\documentclass[natbib,smallextended]{svjour3}       
\smartqed  
\usepackage{graphicx}
\usepackage{hyperref}
\usepackage{longtable}
\usepackage{dpfloat}
\usepackage{booktabs}
\usepackage{threeparttable}
\usepackage{multirow}
\usepackage{pdflscape}
\usepackage{xtab}
\usepackage{geometry}
\usepackage{apj-jour}
\def\oso3{{\em OSO-3\/}}
\def\oso5{{\em OSO-5\/}}
\def\oso6{{\em OSO-6\/}}
\def\oso7{{\em OSO-7\/}}
\def\oso8{{\em OSO-8\/}}
\def\vela{{\em Vela\/}}
\def\exosat{{\em EXOSAT\/}}
\def\einstein{{\em Einstein\/}}
\def\sas1{{\em SAS-1\/}}
\def\uhuru{{\em UHURU\/}}
\def\sas2{{\em SAS-2\/}}
\def\rosat{{\em ROSAT\/}}
\def\asca{{\em ASCA\/}}
\def\chandra{{\em Chandra\/}}
\def\sas3{{\em SAS-3\/}}
\def\ariel5{{\em ARIEL-5\/}}
\def\heao1{{\em HEAO-1\/}}

\def\swift{{\em Swift\/}}
\def\sax{{\em BeppoSAX\/}}
\def\rxte{{\em Rossi XTE\/}}
\def\xmm{{\em XMM-Newton\/}}
\def\suzaku{{\em Suzaku\/}}
\def\cgro{{\em Compton Gamma--Ray Observatory\/}} 
\def\integral{{\em INTEGRAL\/}}  
\def\fermi{{\em Fermi\/}}
\def\nustar{{\em NuSTAR\/}}

\begin{document}

\title{Hard X--ray/Soft gamma--ray experiments and missions: overview and prospects}


\author{Erica Cavallari \and Filippo Frontera
}


\institute{F. Frontera \at
              University of Ferrara, Physics and Earth Sciences Department, Via Saragat, 1, 44122 Ferrara, Italy \\
and \\
National Institute of Astrophysics (INAF), Institute of Space Astrophysics and Cosmic Physics (IASF), Bologna, Italy \\
              Tel.: +39-0532-974254\\
              Fax: +39-0532-974210\\
              \email{frontera@fe.infn.it}           
           \and
           E. Cavallari \at
		University of Ferrara, Physics and Earth Sciences Department, Via Saragat, 1, 44122 Ferrara \\
			  \email{ericacavallari@gmail.com}
}

\date{Received: date / Accepted: date}

\maketitle

\begin{abstract}
Starting from 1960s, a great number of missions and experiments have been performed for the study of the high--energy sky. This review gives a wide vision of the most important space missions and balloon experiments that have operated in the 
10--600 keV band, a crucial window for the study of the most energetic and violent phenomena in the Universe. Thus it is important to take the stock of the achievements to better establish what we have still to do with future missions in order to progress in this field, to establish which are the technologies required to solve the still open issues and to extend our knowledge of the Universe.
 
\end{abstract}

\section{Introduction}
\label{intro}
It is recognized that hard X--ray/soft gamma--ray astronomy (10--600 keV) is a crucial  window for the study of the most energetic and violent events in the Universe. An increasing number of missions devoted to observations in this band or including instruments specifically devoted to perform hard X--ray observations have been performed (e.g., \sax, \rxte) or are still operational (e.g., \integral, \swift, \fermi).

 Actually in the remote past this  was only considered an ancillary energy band. The bands which were considered key for astrophysical observations were the 2--10 keV band, also dubbed classic energy band, and the soft X--ray band ($\le 2$~keV). The major efforts for X--ray observations were concentrated in these bands, e.g., {\em SAS-1} or \uhuru\ satellite, {\em HEAO-2} or \einstein\ satellite, \exosat, \rosat, \asca, \chandra, 
\xmm\ missions.
In these large observatories, at most an extension to harder X-ray energies was foreseen by putting on board small hard X--ray  instruments, e.g., {\em OSO-7}.

The first real attempt to survey the hard X--ray sky was performed with the NASA HEAO-1 observatory in which, in addition to very large instruments working in the classical X-ray astronomy band, a relatively sensitive instrument (HEAO-A4) was on board.  
In order to partially compensate the deficiency of sensitive hard X--ray instruments aboard satellite missions, many balloon experiments were performed. However, due to short flight durations, only studies of peculiar sources were possible.  

With the ESA \integral\ observatory, and the NASA \swift\ satellite, unprecedented sky surveys in the band beyond 20 keV are being performed. 
As a consequence, hundreds of celestial sources have already been discovered, new classes of Galactic sources are being identified, an overview of the extragalactic sky is available, while evidence of extended matter-antimatter annihilation emission from our Galactic center and evidence of Galactic nucleo-synthesis processes have been also reported. 
In order to take full advantage of the extraordinary potential of soft  gamma--ray astronomy, a new generation of telescopes is needed. The current instrumentation, mainly based on direct-viewing detectors,  is penalized by its modest sensitivity, that improves at best as the square root of the detector surface.  The only solution to the limitations of the current generation of gamma--ray instruments is the use of focusing optics. 

To study the hard X-ray continuum spectra from celestial sources up to about 80 keV and their possible emission/absorption lines within this range, multilayer optics based on X-ray diffraction in reflection configuration, are now developed and, for the first time, launched aboard an American  satellite (\nustar).
However, if we want to  study the continuum  emission beyond 80~keV, a further effort is needed. Laue lenses, based on diffraction from crystals in a transmission configuration, are now under development and are possible candidates for extending the focusing energy band.

But, in order to establish how to face the future of the hard X--ray astronomy, it is important to take stock of the current status of the field, to know the instruments used, the observations performed with their limitations, and the results obtained. In this review paper, we outline the history of the main hard X--ray experiments flown aboard satellite missions and stratospheric balloons, their main features, their limitations and their main achievements.
The list of all acronyms and abbreviations used throughout the text can be found in Table~\ref{table:acro}.

\section{The birth of hard X--ray astronomy}
\label{s:birth}

\begin{figure*}
\centering\includegraphics [width=0.80\textwidth]{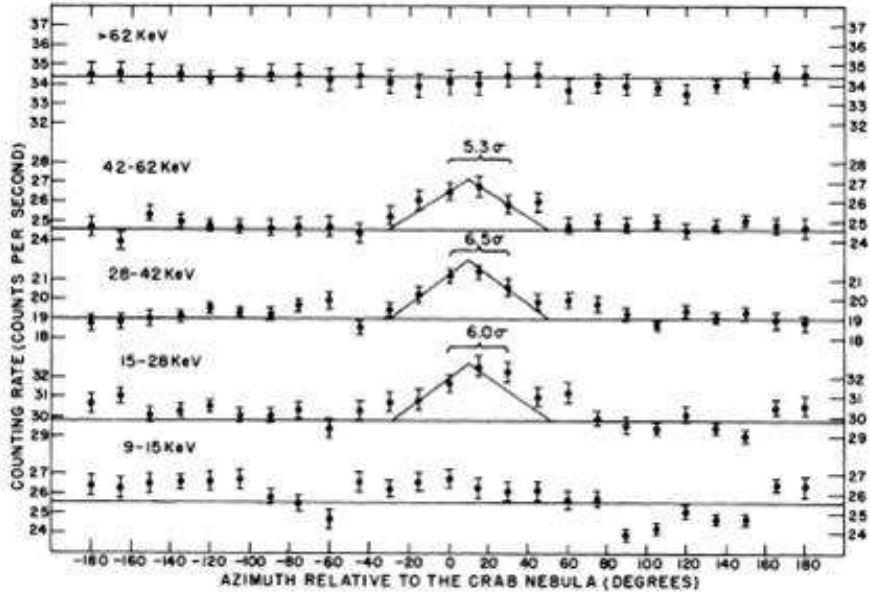}
\caption{The first discovery of an X--ray source (Crab Nebula) at hard X--ray energies. Reprinted from \citet{Clark1965}}
\label{f:Clark1965}  
\end{figure*}

Soon after the serendipitous discovery in 1962 by \citet{Giacconi62} of an extrasolar X-ray source, Sco X-1, X-ray astronomy progressed rapidly. 
The first detection of a hard X--ray extrasolar source occurred in July 21, 1964 with a balloon-borne experiment of the Massachusetts Institute of Technology (MIT) \citep{Clark1965}. The target source was the Crab Nebula, that was previously detected near 4 keV by \citet{Bowyer64}. The search for 
high--energy X--rays was an understandable curiosity. The X--ray detector was a scintillation counter made of a NaI(Tl) crystal of 97~cm$^2$ area and 1~mm thick. A slat collimator made of brass limited the FOV to 32 deg in one direction and 110 deg into perpendicular direction. The scintillation pulses were transmitted in 5 energy channels: 9--15 keV, 15--28 keV, 28--42 keV, 42--62 keV, and $>$62 keV. The balloon was launched from Palestine, Texas and achieved a float of 2.9 mbars. The Crab Nebula was observed during the meridian transit (see Fig.~\ref{f:Clark1965}). The average background level in the energy range 15 to 62 keV was $B = 1.6\times 10^{-2}$~counts/(cm$^2$ s keV), with excess counts due to the source at the level of 
5-7$\sigma$, depending on the energy channel. The spectral distribution was consistent with a power--law. So, thanks to a hard X--ray observation, for the first time it was possible to establish that the X--rays from the Crab Nebula were not due to blackbody emission, as expected from the surface of a neutron star, but were likely due to a non-thermal process.

%

One year later, on September 23, 1965, another balloon experiment was launched from Palestine, under the responsibility of Laurence E. Peterson of the University of California, San Diego (UCSD)
\citep{Peterson66a} for observing the Crab Nebula. The energy band was 16 to 120 keV, the detector material was again 
NaI(Tl), the detector thickness was 5 mm and its area was only 9.4 cm$^2$, a factor 10 lower than that adopted by \citet{Clark1965}. However, the detector was actively shielded with a CsI scintillator and, as a consequence, the background level $B$ was a factor of about 7 lower than that obtained by \citet{Clark1965}. The signal from the Crab was very clear and, for the first time, a 15 channel spectrum was obtained, determining for the first time the power--law dependence of the Crab Nebula photon spectrum 
($F_{\gamma}\propto E^{-1.91\pm0.08}$).

With a similar balloon experiment, again under the responsibility of Laurence E. Peterson, Sco X-1 was observed for the first time at hard X--ray energies \citep{Peterson66b}. The launch was performed on June 18, 1965. It was possible to derive the spectrum of the source up to 50 keV, establishing that the data were well fit with an exponential law 
($\propto \exp(-0.23 E)$) up to 35 keV followed by a non-thermal component visible up to 50 keV.

In the same year, two balloon flights were performed from Hyderabad (India) with a payload which included a hard X--ray detector made of a NaI(Tl) scintillator of 7 inches diameter and 0.5 inch thickness, viewed by a 7 inch photomultiplier (PMT).
The FOV of the telescope was about 20 deg, with its axis inclined with respect to the vertical by 22 deg. The FOV rotated around the vertical with a period of 11.2 min. Thanks to this rotation a region of the celestial sphere including the Cygnus region was scanned during the flight. For the first time Cyg X--1, discovered at low energies in the same year with a rocket experiment \citep{Bowyer65}, was detected at hard X--ray energies from 20 to 58 keV \citep{McCracken65}.

Soon after the above balloon experiments, many others were successfully
performed by X--ray astronomy groups spread over many countries (USA,
France, Italy, The Netherlands, India, Japan). Most of them were devoted
to observations of the Crab, Cygnus region, Sco X--1, Galactic Center
region, and, for the first time, hard spectra and time variability were
observed \citep{Brini67, Bleeker67, Lewin67, Chodil68, Haymes68,
Peterson68, Riegler68, Overbeck68, Rocchia69, Glass69, Bingham69,
Webber70, Agrawal71}, \citep{Deerenberg71, Agrawal72, Matsuoka72}.

The interest for hard X--ray observations was also demonstated by the acceptance of a hard X--ray experiment aboard the 3rd Orbiting Solar Observatory (OSO~3), launched on March 8, 1967 and operational until November 10, 1969. The hard X-ray experiment was led by UCSD \citep{Hudson69a,Hudson69b}, and mounted on the spinning wheel of the satellite with a radial view. It consisted of a single thin NaI(Tl) scintillation crystal, 5 mm thick and 9.57 cm$^2$ area, viewed from a photomultiplier enclosed in a cup-shaped CsI(Tl) anti-coincidence shield 5 cm thick. The instrument operated from 7.7 to 210 keV with 6 channels. The FOV was 23$^\circ$ FWHM. It scanned the entire sky over the course of the mission. 

Aboard OSO~6 there was a hard X--ray experiment (27--189 keV) \citep{Brini71}, devoted to studying both solar X--ray flares and the celestial sky. The detection of GRBs was obtained with this instrument \citep{Pizzichini75} soon after their discovery with the Vela satellites in 1973 \citep{Klebesadel73} 

\begin{figure*}
\centering\includegraphics [width=0.80\textwidth]{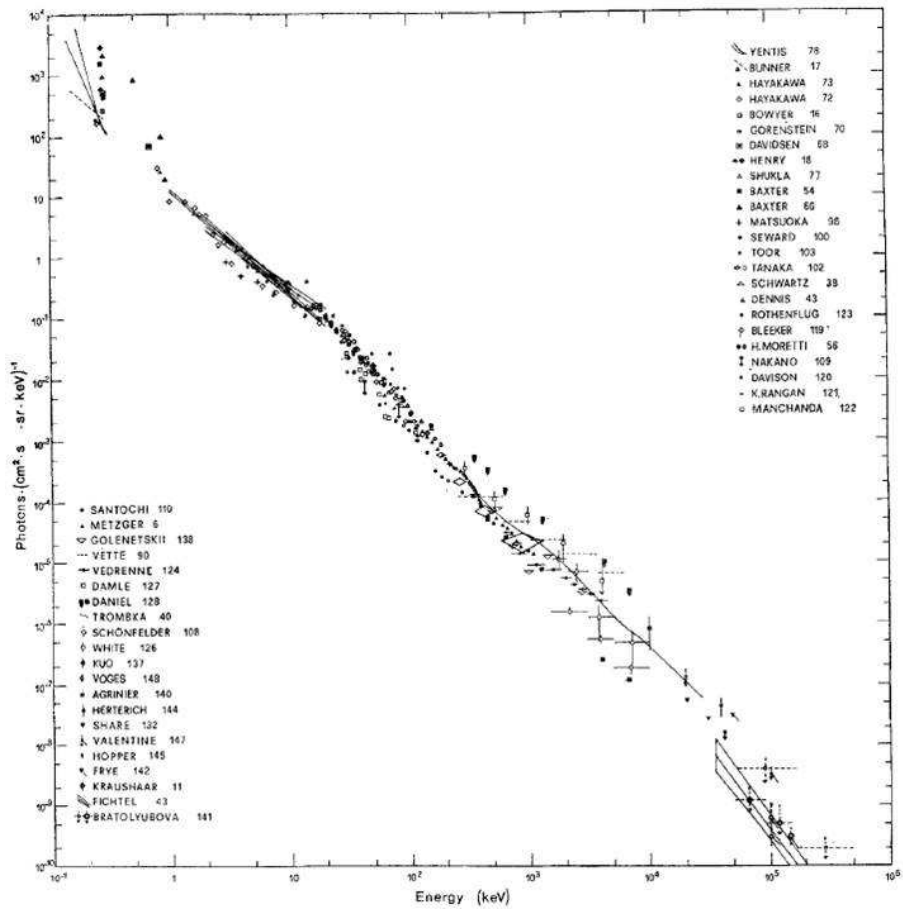}
\caption{The spectrum of the Cosmic X--ray Background as derived from the observations, mainly balloon experiments, performed up to the early 1970s. It could be described by two power--laws, one with index $1.590\pm 0.021$ from 1 to 20 keV, and the other with index $2.040\pm 0.013$ from 20 to 200 keV,  with a break at 20 keV. Reprinted from \citet{Horstman75}.}
\label{f:CXB-Horstman75}  
\end{figure*}

There was also a great interest for understanding the spectrum and thus the emission mechanism of the cosmic diffuse X--ray background (CXB), discovered at low energies by
\citet{Giacconi62}. Many balloon and rocket experiments, but also observations with OSO~3, were performed during the 1960s to observe the CXB \citep[e.g.,][]{Brini70,Schwartz70}. An exhaustive and critical  review of the observations up to the early 1970s was performed by \citet{Horstman75} (see Fig.~\ref{f:CXB-Horstman75}).

%
%

Hard X--ray astronomy was born. Two key features were immediately clear: the determination of the emission mechanism of the celestial X--ray sources, and the detection of  mysterious GRBs. Concerning the former, the spectrum derived in the classical X--ray band (2--10 keV), in spite of its importance, was in many cases insufficient to establish the emission mechanisms in play. Concerning GRBs, hard X--ray astronomy was crucial for their discovery.

\section{Satellite missions and balloon experiments}
\label{pastmissions}

Table~\ref{t:sat-missions} reports the main characteristics  of the hard X--ray experiments aboard the main satellite missions, while Table~\ref{tab:balloons} reports the main features of the balloon experiments devoted to hard X--ray astronomical observations. We discuss them and their most relevant scientific results in the following subsections, subdivided according to their decade of operation.

\newgeometry{left=1cm,bottom=1cm}
\begin{landscape}
\tiny
\begin{longtable}[c]{llllllllll}
\caption{Main satellite missions}
\label{t:sat-missions}
\\
\toprule
	&Launch date & Orbit 	&On board HE 	&Detector & Thickness & Energy range  	&  Energy & Useful or & FOV \\
Mission &Site & Altitude (km) &Instruments 	&type		&
(g/cm$^2$) & (keV) & resolution (\%)&	effective area (${\rm cm}^2$) ($^a$) &  \\
  &Termination date &Inclination(${^\circ}$)&  & 		&		    &	          &		&		& \\
\midrule
\endfirsthead
\caption{Main satellite missions (continued)} \\
\toprule
	&Launch date &Orbit 	&On board HE 	&Detector & Thickness
& Energy range 	& Energy & Useful or & FOV \\
Mission &Site &Altitude (km) &Instruments 	& type	&
(g/cm$^2$) & (keV) &	resolution  & effective area (${\rm cm}^2$) ($^a$)   &  \\
	&Termination date &Inclination(${^\circ}$)&  &	  	& 			&	    &			&			 &    \\
\midrule
\endhead
OSO-3		& 8 March 1967		& 550		& XRT & SD $+$ C	 &
1.835		& 7.7--210 &  45\% @30~keV & 9.57 & 23$^\circ$  \\
		& Cape Canaveral &  33   &		&	&	&	&		&	&	\\
		& 10 November 1969 &		&	&	&	&	&
	&	&	\\
\midrule
OSO-5		& 22 January 1969	& 555		& XRD	& SD $+$ C &
2.86		& 14--254	&	nd	& 70		&	40$^\circ$ \\
		& Cape Canaveral &	35	&	&	&	&	&	&	&	\\
		& July 1975	 &		&	&	&	&	&
	&	&	\\
\midrule
OSO-6		& 9 August 1969		& 465--516 & SXM & SD $+$ C &
4.66		& 25--190	& 28\% @134~keV & 5.1 & $17^\circ \times 23^\circ$ 	\\
		& Cape Canaveral	 & 32.9	&	&	&	&	&		&	&	\\
		& 31 December 1972 &		&	&	&	&	&
	&	&	\\ 
\midrule
OSO-7 	& 29 September 1971 		& 321--572			& HXT 			& SD 	   &  3.67 	& 7--500  & 33\% @60~keV		&
	64			& 6.5${^\circ}$ \\ 
		& Cape Canaveral 		& 33.1 				& XCE 			& PC+C &	0.061	& 1--60 &  nd		& 186 			& 1--3${^\circ}$ \\
		& 9 July 1974 	&		&	&	&	&	&	&     &	\\			
\midrule
Ariel~5	& 15 October 1974			& 500--400
	& ST			& SD	& 18.04	& 20--1200	& 30\% @662keV	& 8	&	8$^\circ$	\\
	& San Marco (Kenya)	& 2.8		&	&	&	&	&	&	&	\\
	& Spring 1980		&		&	&	&	&	&	&	&	\\
\midrule
OSO~8	& 21 June 1975				& 544--559			& CXRS			& PC+C	& nd	& 2--60 &  nd			& 537
	& 3--5${^\circ}$ \\ \cmidrule {4-10}
		& Cape Canaveral, USA		& 32.9				& HECX			& SD+C  &  nd	& 20 keV--3 MeV	& nd &  27.5			& 5${^\circ}$ \\
		& 1 October 1978			& 					& 				& 			& 				& 				& \\
\midrule
HEAO 1	& 1 August 1977				& 445--445			& HED(A2)		& PC+C	& nd	& 3--60	& nd		& 2700
	& various (see text) \\ \cmidrule {4-10}
		& Cape Canaveral		& 22.75			 	& (A4) LED		& PD+C  & 1.10		& 13--180	&25\% @60~keV	& 225			& 1.4${^\circ}$x20${^\circ}$\\*
		& 9 January 1979			& 					& MED			& PD+C & 27.52		& 0.08--2 MeV & 10\% @1~MeV	& 168			& 17${^\circ}$\\* 
		&							&					& HED & PD+C  & 	27.52		& 0.2--10 MeV &10\%@1~MeV	& 125			& 10${^\circ}$\\	
\midrule
Venera 11-14 & 9-14 Sept.1978 & variable 
	& Konus 			& SD (6 units) & 11.01	& 30--300& 
66\%~@200~keV	& 50/unit			& 4$\pi$		\\
		& Baikonur, URSS			& NA						&	&	&	&	&	&	&	\\
		&	March 1983			&						&	&	&	&	&	&	&	\\				
\midrule
Hakucho		& 21 February 1979		& 521--533			& HDX			& SD+C  &  nd		& 10--100	& nd	& 45			& 4.4${^\circ}$x10${^\circ}$\\
		& USC, Japan				& 29.9			 	& 				&	&	&		&				&				&\\
		& 16 April 1985				& 	&	&				&				&			&				& 				&   \\
\midrule
HEAO 3	& 20 September 1979			& 486--505			& HRGRS			& SSD	 & $\sim 24$	& 0.06--10 MeV& 0.2\% @1.46~MeV &    75(@100 keV)	& 27${^\circ}$\\
		& Cape Canaveral, USA		& 43.6			 	&				&			&				&		&	&		&\\*
		& 1 June 1980				&					&				&			&				& 	&	&		&\\*
\midrule
Tenma	& 20 February 1983			& 489--503			& GSPC			& GSPC+C	& 0.04	& 2--60	& 9.5\% @6~keV			& 720			& 3.1--2.5${^\circ}$\\
		& USC, Japan			& 31.5			 	&				&     GSPC+RMC	&	&			&
		&		& 3.8${^\circ}$\\* \cmidrule{4-10}
		& 11 November 1991			&  					&  RBM/GBD		& SD	& nd	& 10--100	& nd	& 14			& 1 sr\\
\midrule
EXOSAT	& 26 May 1983				& 347--191709		& ME			& PC+C	& 0.04	& 1--50	& 18\% @22~keV		& 1800			& 45'\\
		& Vandenberg, USA			& 72.5			 	& 				& 			& 				&	&	&		 	&\\*
		& 9 April 1986				&  					&		&	&		&			&				&				&\\*
\midrule
Ginga	& 5 February 1987			& 517--708			& GBD		& SD	 & 3.67	& 14--400 & 15\% @60~keV	& 60			& no coll\\
		& USC, Japan				& 31.1			 	&	 & PC		& 0.02  & 2--30		& 	nd			& 	63			&  no coll \\
		& 1 November 1991			&  					& 				& 		&	& 				& 		&		&\\
\midrule
KVANT-MIR	& 31 March 1987			&	354--374		&
HEXE		&  PD+C  & 1.17	 & 15--200	&30\% @60~keV	800
1.6$^\circ \times 1.6^\circ$	\\
		& Baikonur, USSR & 51.6	& GSPC & GSPC+C	& nd	& 2--100	&  nd		&   300 & 3$^\circ \times 3^\circ$ \\
		& 23 March 2001	 & 		& Pulsar X--1	& PD+C		& nd		& 30--800	& nd	&  1256	& 
3$^\circ \times 3^\circ$ \\
\midrule
Granat	& 1 December 1989			& 2000--200000		& SIGMA			& PSD+CM	& 4.40   		& 
35 keV--1.3 MeV & 8\% @511~keV		& 794			& 11.4${^\circ}$x10.5${^\circ}$\\ \cmidrule{4-10}
		& Tyuratam, USSR			& 51.6			   	& ART--P		& MWPC+CM	& 0.08	& 3--60	& 25\% @6~keV
		& 1160			& 3.6${^\circ}$x3.6${^\circ}$\\* \cmidrule{4-10}
		& 27 November 1998			& 					& ART--S		& MWPC+RC	& 0.2		& 3--100
		& 11\% @ 60~keV	& 800(@100~keV)	& 2${^\circ}$x2${^\circ}$\\* \cmidrule{4-10}
		&							&					& PHEBUS		& SD	& 85.56	&
		0.075--124 MeV&    nd	      &  280			& no coll \\* \cmidrule{4-10}
		&							&					& WATCH			& SD+RMC	& 0.734	&
		6--180	& 30\% @60~keV	 	& 95			& 9~sr	\\* \cmidrule {4-10}
		&							&					& KONUS--B		& SD	&  18.35		& 10 keV--8 MeV	&	nd	&	2200	& non coll \\
\midrule
CGRO 	& 5 April 1991				& 362--457			& OSSE			& SD+C	& 36.7		& 0.05--10 MeV	& 8\% @662~keV
& 2000(@511~keV)					& 3.8${^\circ}$x11.4${^\circ}$  \\\cmidrule {4-10}
		& Cape Canaveral, USA 		& 28.5			 	& BATSE (LAD)			& SD		& 4.66		& 30 keV--2 MeV	&
		27\%~88~keV &	 1.62 ${\rm m}^2$& no coll \\ \cmidrule{4-10}
		& 4 June 2000 		&					& COMPTEL 		&	CT	& 7.65 (low Z)	& 1-30 MeV	&	8.8\% @1.27~MeV	& 10/50			& 1 sr\\
			&			&		& 31.20 (high Z) &
		&		&		&		&		&	\\
\midrule\pagebreak
Wind 		& 1 November 1994	&L$_1$ Lagrangian point	&	Konus		& SD (2 units)& 27.96		& 10~keV --10~MeV &
66\%~@200~keV	& 126/unit			& 4$\pi$		\\
		& Cape Canaveral		& NA			&		&
	&	&	&	&	&	\\
		& OPERATIONAL	&		&		&		&		&		&		&		&		\\	

\midrule
RXTE	& 30 December 1995			& 409-409			& PCA			& PC+C	& 0.02	& 2--60	& 18\%~@6~keV		&
6250(@6~keV) & 1${^\circ}$\\ 
		& Cape Canaveral, USA		& 28.5			 	& HEXTE			& PD+C	& 1.1		& 17-240	& 15\%~60~keV	& 1600			& 1${^\circ}$\\*
		& 5 January 2012			& 					&		&		&			&				&		&		& \\
\midrule
BeppoSAX& 30 April 1996				& 575--594			& HPGSPC		& GSPC+C	& 0.50 	& 4--120	&
4\%~60~keV	& 240			& 1${^\circ}$x1${^\circ}$\\ 
		& Cape Canaveral, USA		& 4			 		& 				& 			& 				& 				& \\* 
		& 30 April 2002				&  					& PDS			& PD+RC	& 1.10		& 15--300	&  15\%~@60~keV		& 640			& 1.3${^\circ}$\\*
\midrule
HETE 2	& 9 October 1999			& 590--650			& FREGATE		& SD	& 3.67		& 6--400	& 
12\%~@122~keV	& 120			& 3 sr\\
		& Kwajalein, Rep. Marshall Isl.	& 1.95			& 		&		&	&		&				&				&\\*
		& March 2008		&			& 				& 				&			&				&		&		&\\*
\midrule
INTEGRAL& 17 October 2002			& 639--153000		& SPI			& SD+CM	& 37.24	& 20 keV--8 MeV  	& 0.19\%~@1.3~MeV	& 250(through CM)			& 16${^\circ}$ FC\\
		& Tyuratam, Kazakhstan		& 51.7			 	& 		&		& 			&
		&		&		&\\* \cmidrule{4-10}
		& OPERATIONAL				& 					& ISGRI			& SSD+CM	& 1.17		& 15 keV--1 MeV	& 8\%~@60~keV		&1300(through CM)			& 9${^\circ}$x9${^\circ}$ FC\\*
	&		&	 	& PICsIT	& SD+CM	& 13.53		& 175--10000				& 18\%~@511~keV		& 1400 (@500~keV)		& 9${^\circ}$x9${^\circ}$ FC	\\*
\midrule
Swift	& 20 November 2004			& 585--604			& BAT			& SSD+CM &	1.17		& 15--150	& 
2.2\%~@80~keV		& 2620 (through CM)	& 
100${^\circ}$x60${^\circ}$ HF \\
		& Cape Canaveral, USA		& 20.6			 	&			&	&			&				&		&		&\\*
		& OPERATIONAL				& 					&		&		&			&				&		&		&\\
\midrule
SUZAKU & 10 July 2005				& 550--550			& HXD/GSO			& PD+C	& 3.35	& 40--600		& 24\%~@100~keV   & 273(@150 keV) 		& 4.5${^\circ} \times 4.5{^\circ}$ \\
		& USC,Japan					& 31				& HXD/PIN		& SSD+C 	& 0.46 & 10--70			&	3~keV			&	160 (@20~keV)			& 34'$\times$34' ($<$100 keV) \\*
		& 2 September 2015				& 					&		&		&			&				&		&		& \\
\midrule
AGILE	& 23 April 2007				& 524--553			& SuperAGILE	& SSD+CM	& 0.095  		& 18--60	& 8~keV		& 670 (through CM)			& 107${^\circ}$x68${^\circ}$\\
		& Shriharikota, India		& 2.5			 	&		&		&			&		&		&				&\\*
		& OPERATIONAL				& 					&		&		&			&				&		&		&\\
\midrule
\fermi\	& 11 June 2008				& 542--562			& GBM/NaI			& SD	 &  4.66			& 10 keV--1 MeV& 12\%~@511~keV 	&	1200		& 9.5 sr\\
		& Cape Canaveral, USA		& 25.6			 	&  GBM/BGO				& SD		& 90.55	& 150 keV--30~MeV	& 7.5\%~@2~MeV		& 110		&  4$\pi$~sr  \\*
		& OPERATIONAL				& 					&		&		&			&				&		&		&\\
\midrule
NuSTAR	& 13 June 2012				& 610--650			&	XRT			& PSD		&  0.37		& 3--78.4 & 1.3\%~@68~keV		& 847(@9 keV)	& 10'(@10 keV)\\
		& Kwajalen Atoll			& 6			 		&		& 		&		& 		& 				& 60(@78 keV)				& 6'(@68 keV)\\*
		& OPERATIONAL		&	& 					&		&		&			&				&				&\\
\midrule
ASTROSAT	& 28 Sept 2015	&	600	&	LAXPC		& MWPC+C
& 0.37	& 3--80	& 13\%~@60~keV &		6000	& 
$1^\circ \times 1^\circ$ \\
		& SDSC, Sriharikota	& 6	& CZTI		& PSD+CM	& 1.24	& 10--150		& 2\%~@60~keV	& 1000	& $6^\circ \times 6^\circ$ \\ 
	& OPERATIONAL	&	&	&	&	&	&	&	&	\\	
\midrule
\multicolumn{10}{p{1.4\textheight}}{$^a$ With {\em useful area}, only used for non-focusing telescopes, we mean the detector geometrical area exposed to the source through a possible  collimator or coded mask. With {\em effective area}, in the case of non-focusing telescopes we mean the geometrical area of the detector through a possible collimator or coded mask times the detection efficiency. Thus the effective area depends on photon energy. 
In the case of focusing telescopes, with effective area we mean the geometrical area of the mirrors projected on the focal plane times the reflection efficiency at a given energy $E$. When it is possible, we report the effective area in which also the detection efficiency is taken into account.}
\end{longtable}
\end{landscape}

\restoregeometry

\newgeometry{left=1cm,bottom=1cm}
\begin{landscape}
\tiny
\begin{longtable}[c]{llllllll}
\caption{The most significant hard X-ray balloon experiments}
\label{tab:balloons}
 \\
\toprule
Group/ 			&Launch date			&Detector 
&	Thickness	& Energy	& Energy resolution (\%)	& Useful or effective					&FOV(FWHM)			\\
Experiment		&Site					&type	 & (g/cm$^2$)	& range	& (FWHM)	& area (${\rm cm}^2$) ($^a$)	&					\\
				&Float altitude 		&	 &		& (keV) 	&		 &				&					\\
\midrule
\endfirsthead
\caption{The most significant hard X-ray balloon experiments (continued)} \\
\toprule
Group/ 			&Launch date			&Detector		& Thickness	&Energy	& Energy resolution (\%)		&Useful or effective					&FOV(FWHM)			\\
Experiment		&Site					&type		& 
(g/cm$^2$)		&range &	(FWHM)	&area (${\rm cm}^2$) ($^a$)	&					\\
				&Float altitude 		&	&		& (keV) 	&		 	&			&		
		\\
\midrule
\endhead
Rice Un.		&4 June 1967			&SD+C		& 18.64		&35-560	& 9.2\%~@511~keV			&81						&24${^\circ}$		\\
				&Palestine (Texas)		&	&		&			&		&				&					\\
				&3.35-3.65 mbar			&	&		&			&		&				&
				\\	\cmidrule{2-8}
				&25 November 1970		&SD+C		& 18.64 &		23-930	& 9.2\%~@511~keV		&75 (@661 keV)			&24${^\circ}$		\\
				&Paran{\'a} 	(Argentina)		&			&		&	&						&
	&					\\ 
				&3.1-3.4 mbar			&			&	&		&		&				&
				\\	\cmidrule{2-8}
				&1 Aprile 1974 			&SD+C  & 18.35		&0.02-12.27 MeV	& 12\%~@511~keV			&182				&13${^\circ}$		\\
				&Rio Cuarto	(Argentina)	&			&	&		&			&			&					\\
\midrule
Tata Inst.		&April 1968				&SD+C		&1.47		& 22.5--154 	& nd		&97						&18.6${^\circ}$		\\
				&Hyderabad (India)		&			&	&		&			&			&					\\
				&5-7 g/${\rm cm}^2$		&			&			&			&			&
				&			\\	\cmidrule{2-8}
				&1985-1992				&MWPC		& 0.09	& 20-100	& 	nd	&2400					&5${^\circ}$x5${^\circ}$\\
				&Hyderabad (India)		&			&	&		&		&				&					\\
				&4.5 g/${\rm cm}^2$		&	&		&			&			&			&					\\				
\midrule
Bologna 		&23 June 1970			&SD+C		& 4.66	&20-200	& 20\%~@134~keV			&136					&13${^\circ}$		\\
				&Gap Tallard (France)	&			&
	&			&			&			&					\\	
				&5 g/${\rm cm}^2$		&	&		&			&			&			&
				\\	\cmidrule{2-8}
				&1 July 1972			&SD+C		
& 4.66	&	30-200	& 	30\%~@134~keV	 	&280					&13${^\circ}$		\\
				&Gap Tallard (France)	&			&	&		&		&				&		\\
				&6 g/${\rm cm}^2$		&			&			&			&			&
				&			\\	\cmidrule{2-8}
				&29 July 1976 			&SD+C		&4.66		&20-300	&    poor		&525					&14${^\circ}$		\\
				&Trapani (Italy)		&			&	&		&			&			&					\\
				&			&			&			&			&			&				&					\\
\midrule
NRL				&17 October 1973		& PD+C  &	
1.10(NaI)	& 20--160	& 35\%~@60~keV			& 70					& 20$^\circ$		\\ 
				&Palestine (Texas)	& PD+C &  1.35 (CsI)	& 20--160		& 35\%~@60~keV		& 70								& 20${^\circ}$				\\ 
				&2.6 g/${\rm cm}^2$		&			&			&		&				&
				&			\\	\cmidrule {2-8}
				&10 May 1976			&PD+C		
& 1.83		&20-250 	& 24\%@60~keV		&765					&10${^\circ}$		\\
				&Palestine (Texas)		&			&			&			&			&		&			\\
				&2.3 g/${\rm cm}^2$		&			&			&			&			&
				&			\\	\cmidrule {2-8}
				&24 November 1977		&PD+C		&1.83	&	20-250	& 	24\%@60~keV	&765		&5${^\circ}$		\\
				&AS(Australia)&			&			&			&			&		
&			\\
				&			&			&			&			&						&	&				\\
\midrule
INPE			& 1973--1974			&SD	 	& 37.28		&0.3-17 MeV	& 14\%~@511~keV			&32						&		2$\pi$			\\
				&Sao Josè dos Campos (Brazil)&		&			&			&			&			&		\\ 
				&3.5-4 g/${\rm cm}^2$	&			&			&			&			&
				&		\\ \cmidrule {2-8}
			& 1978			&SD	 	& 2.33		&0.1-2 MeV	& 14\%~@511~keV			&32						&		2$\pi$			\\
				&Sao Josè dos Campos (Brazil)&		&			&						&		&	&			\\
				&3.5-4 g/${\rm cm}^2$	&			&			&			&			&		&		\\

\midrule
AIT/MPE			&3 May 1976		&SD+C	  &	1.83	&17-160	& 29\%~@60~keV		&87
&2${^\circ}$x10${^\circ}$\\	\cmidrule {3-8}
				&Palestine (Texas)		&PD+C	 	
& 1.10		&15-135	& 22\%~@60~keV		&108					&2${^\circ}$x10${^\circ}$ \\
				&2.8 g/${\rm cm}^2$		&			&			&			&			&
				&			\\	\cmidrule {2-8}
				&Sept--Oct 1977		&PD+C 	& 1.10	&10-200 &	22\%~@60~keV		&766					&  3${^\circ}$		\\
				&Palestine (Texas)		&			&			&		&			&		&					\\
				&3.1 g/${\rm cm}^2$		&			&			&			&			&
				&			\\	\cmidrule {2-8}
				&1978					&PD+C		& 1.10		&	10-200	& 22\%~@60~keV			&766					&3${^\circ}$		\\
				&AS (Australia)&			&			&			&			&			&		\\
				&3.5 g/${\rm cm}^2$		&			&			&		&				&		&			\\
\midrule
ISAS			&1977-1979				&SD+RMC & nd		& 30-200		& nd		& $\sim 500$		&146${^\circ}$		\\
				&SBC (Japan)			&			&			&			&			&		&			\\
				&5 mbar			&		&			&			&						&
	&				\\
\midrule
Uni. Tasmania	&20 November 1978		&MWPC+C	& 0.11			&20-100		&  nd		&5200					&7${^\circ}$x20${^\circ}$\\
				&AS (Australia)	&		&			&			&			&			&		\\
				&3.5 g/${\rm cm}^2$		&			&			&			&			&		&			\\
\midrule
MISO			&  1977,1978,1979,1980			&CT		&	9 (NE311), 36.7 (NaI)   	&0.1--20 MeV 	& 32\%~@1~MeV		&
560					&3${^\circ}$x3${^\circ}$\\ \cmidrule{3-8}
				&Palestine (Texas)		&  HXD	
&  nd			&	20--280		& nd		
&	600			 		&3${^\circ}$x3${^\circ}$			\\ 
				&	4~mbar		& 		&	&		&		&			&     \\
\midrule\pagebreak
Bell--Sandia		& Sept 1977,April 1979	& GeD		&
			33.52		&	0.1--5~MeV		& 
0.6\%~@511~keV		& 	21	&	15$^\circ$		\\
				& AS(Australia)	&		&		&		&		&		&		\\	
				&	3.6~mbar		&		&		&		&		&		&		\\

\midrule
MPI Compton-Tel		& Oct.1977; May 1979		& NE213+NaI	& 13.11(LS)+28.26(SD)	& 1--10~MeV		&
10\%~@2~MeV	& 24.4 (@1.5--2~MeV)& 40$^\circ$-50$^\circ$ \\
				& Palestine (Texas)	&		&
	&		&		&		&		\\
				& 3.5				&		&			&		&		&		&	\\
\midrule
UCR Compton-Tel		& 10 Nov. 1981			& LS(S1)+LS(S2)		& 10.87 (S1), 17.4 (S2)			& 0.3--30 MeV			& 20\%~@4~MeV	& 100 (@1~MeV, on axis) & 60$^\circ$	\\
				& AS (Australia)		&		&
	& 		&		&		&		\\
				& 4.5				&		&
	&		&		&		&	\\
\midrule
XG				&1980, 1981			&SD+C	
& 3.67	  	&20-200	&  23\%~@60~keV			&1455					&9.2${^\circ}$x9.2${^\circ}$ \\
				&Palestine (Texas)		&			&			&		&				&		&			\\
				&2.8 mbar 	 	&		&			&			&			&			&\\ 
\cmidrule{2-8}
				&1982				& SD+PD+C	
& 3.67+1.10	  	&20-200	&  23\%~@60~keV			&1455					&9.2${^\circ}$x9.2${^\circ}$ \\
				&Milo Base (Sicily)		&			&			&		&				&		&			\\
				&3.5 mbar 	 	&		&			&			&			&			&\\ 
\midrule
POKER			&1981, 1985		&MWPC+C	& $\sim 0.2$		&15-200	&  13\%~@60~keV			&2700x4					&5${^\circ}$x5${^\circ}$\\
			&Milo Base (Italy)		&			&			&			&			&		&			\\ 
				& 2.4~mbar		&			&			&			&			&
				&			\\ \cmidrule {2-8}
				&May 1989			&MWPC+C & 0.21		&15-280 & 13\%~@60~keV		&2500x3					&1.9${^\circ}$x1.9${^\circ}$\\
				&AS (Australia)&			&	&		&			&			&   \\
				&3.48-3.64 mbar 	 	&			&	&		&			&			&  \\
\midrule
FIGARO II		& July 1986, July 1990		&SD+C		
& 18.35		&170-6000 	&  nd		&3600					&50${^\circ}$		\\
				&Milo Base (Italy)		&			&			&			&			& 		&			\\
				&4~mbar			&		&			&			& 						&	&				\\
			& November 1988		&	
& 		& 	&  		&				& 		\\
				&Queensland (Australia)		&			&			&			&			& 		&			\\
				&4--4.5~mbar			&		&			&			& 						&	&				\\
\midrule
MIFRASO			& July 1986, 1987			&SD+C(HED)		&	2.20		&	15-300	& 25\%~@60~keV		&2700
&2.6${^\circ}$	 	\\ \cmidrule{3-8}
				&Milo Base (Italy)		& PC+C(LED)
& $\sim 0.2$	&	10--120  & 12\%~@60~keV		& 	900						&	2.6${^\circ}$
\\
				&3.8-4 mbar	&	&	&	&	
&					&  \\
\midrule
EXITE			&Oct. 1988, May 1989			&PSD+C+CM	&2.20	 	&	20-300 	& 11\%~@122~keV		&934					&3.4${^\circ}$ 		\\
				&Ft.Sumner(NM,USA), AS(Australia)&			&			&			&			&	&				\\
				&3--4~g/(${\rm cm}^2$)		&			&			&			&			&		&			\\
\midrule
GRIS 			& 12 flights 1978--90 		&GeD+C	
& 34.58		&20-8000 	& 0.4\%~@500~keV		&61.5 (@847 keV)		&17${^\circ}$(@500keV)\\
			&Ft.Sumner(NM,USA),AS(Australia)	&	
&		&			&			&			&					\\
			& April, May 1992 			& GeD+C		&	34.58		&	20--8000		&	0.4\%~@511~keV		&	100~(@511~keV)	&	18${^\circ}$(@500keV)		\\
			&	AS(Australia) &			&		&
&		&			& \\
\midrule
UAH-MSFC		&3 flights 1987-1988		&SD+C 	& 4.66		&18--960 	& $\sim$40\%~@80~keV		& 2027					&15.5${^\circ} \times 180^\circ$		\\
				&AS (Australia)& SD+C	 		& 4.66	&	18--960	& $\sim$40\%~@80~keV		&	2027					& 30.75${^\circ} \times 180^\circ$	\\
				&3 mbar					&			&			&		&				&		&			\\
\midrule
GRIP			& May 1987, Nov. 1987, Apr. 1988, Apr.1989		& PSD+CM	& 18.35	& 30--5000	& 16.6\%~@50~keV	& 500~(@200~keV) & 18${^\circ}$ \\
				& AS (Australia)	&		&		&
     &		&		&			\\
			&  nd		&		&		&		&
	&			&		\\					
\midrule
LXeGRIT			&May 1999, Oct 2000				&CT	&	$\sim 21$		&200-25000	& 9\%~@1~MeV			&400					&		1~sr 			\\
				&Ft. Sumner (New Mexico)&			&			&			&			&			&		\\
				&-						&			&			&						&			$2\pi$		\\
\midrule
HERO			&24 May 2001			&FO+PSD	& 0.27 			&20-45	&	5\%		&4 (@40 keV)			&6' (@40keV)		\\
				&Ft. Sumner (New Mexico)&			&			&				&		&			&		\\
				&39 km		&			&			&			&						&
	&					\\
\midrule
CLAIRE			&14 June 2001			&LL+GeD	
& 21.28		&167-173	& 1.4\%~@170~keV		&64 (@170 keV)			&1.5'				\\
				&Gap-Tallard (France)	&			&			&			&			&			&		\\
				&41 km			&		&			&			&						&		&			\\
\midrule
InFoc$\mu$s		& July 2001		&FO+CZT	& 1.24		&  20-40 	& 12.5\%~@32~keV		&49 (@30 keV)			&11'				\\
				&Ft. Sumner(NM,USA)	&			&			&		&			&			&		\\
				&	36~km			&		&			&			&						&		&			\\
\midrule
HEFT			&18 May 2005			&FO+CZT	& 1.24		&6-68		& 1.4\%~@70~keV		&250 (@40 keV)			&17'(@20keV)		\\
				&Ft. Sumner(NM, USA)	&			&			&				&		&				&	\\	
				&39~km			&			&			&			&			&				&	\\
\midrule\pagebreak
protoEXIST1		&9 October 2009			&PSD+CM 	
& 3.1			&30-600  &	10\%~@30~keV		&256					&9${^\circ}$x9${^\circ}$\\
				&Ft.Sumner (NM,USA)	&			&			&			&			&			&		\\
				& 40~km		&			&			&			&			&				&	\\
\midrule
\multicolumn{8}{p{1.4\textheight}}{$^a$ With {\em useful area}, only used for non-focusing telescopes, we mean the detector geometrical area exposed to the source through a possible  collimator or coded mask. With {\em effective area}, in the case of non-focusing telescopes we mean the geometrical area of the detector through a possible collimator or coded mask times the detection efficiency. Thus the effective area depends on photon energy. 
In the case of focusing telescopes, with effective area we mean the geometric area of the mirrors projected on the focal plane times the reflection efficiency at a given energy $E$. When it is possible, we report the effective area in which also the detection efficiency is taken into account.}

\end{longtable}
 
\end{landscape}

\restoregeometry

\normalsize

\subsection{\bf Satellite missions and balloon experiments in the 1970s} 

The seventies were the golden age of the balloon experiments, while most of the performed X--ray satellite missions were optimized for getting the best response and effective area in the classic X--ray energy band (2--10 keV), even if an extension to higher energies was preserved.

In the case of proportional counters (PC) on board satellites, the strategy generally adopted to extend the energy pass-band was that of having stacks of detector arrays one behind the other. The top array, filled with a low Z gas (e.g. Argon), functioned as an X--ray filter of the bottom array. The latter was generally filled with high Z gas (generally Xenon). In this way it was possible to extend the band even up to 60 keV.
 
To get a better efficiency at high energies than that of  proportional counters, hard X--ray detectors based on scintillators of NaI(Tl), capable of efficiently detecting photons up several hundreds of keV, were developed and adopted. A further improvement in scintillator detectors was achieved with the introduction of  phoswich detectors of NaI/CsI, that permitted high  detection efficiency and a very low background \citep[e.g.,][]{Frontera1997;bepposax}. This type of instrument became the work-horse of both satellite missions and balloon-borne experiments until the present epoch. The balloon missions were  devoted to specific studies of peculiar X--ray sources, while the satellite missions were designed also for the survey of the hard X-ray sky.

\subsubsection{Satellite missions}

Hard X-ray instrumentation was aboard almost all satellites dedicated to X--ray astronomy, such as \sas3, \ariel5\ and \heao1, and also aboard satellites whose main goal was the study of the Sun, like the Orbiting Solar Observatories {\em OSO~3}, {\em OSO~5}, {\em OSO~6} (that we have discussed before), {\em OSO~7} and {\em OSO~8}. Most of the hard X--ray experiments in this decade had wide fields of view, allowing them to perform surveys of the sky.
A particular case is that of the \vela\ satellites designed for nuclear test detection. 

\begin{itemize}

\item{\bf Vela series satellites}

The Vela satellites represented a series of 12 military spacecraft with life time of the order of 1 year, launched between 17 October 1963 (Vela 1A and 1B) and 8 April 1970 (Vela 6A and 6B). Goal of the satellites was to monitor the explosion of nuclear bombs in the terrestrial atmosphere, which were vetoed by the  Partial Test Ban Treaty issued on 10 October 1963.

The satellites were spinning and had on board X--ray and gamma--ray detectors. The most sophisticated instrumentation was aboard {\em Vela~5} and {\em Vela~6}. It included a gamma-ray experiment \citep{Klebesadel73} which consisted of six CsI scintillation crystals, with a total volume of 60 cm$^3$,  distributed so to achieve nearly isotropic sensitivity. The energy  passband was 0.2-1 MeV in the case of Vela 5 and 0.3--1.5 MeV in the case of Vela 6. The scintillators were shielded for charged particles, thanks to a high-Z shield. 

On 2 July 1967, the Vela 3 and Vela 4 satellites detected a flash of gamma--ray emission unlike any known nuclear weapon signature. Other similar events were observed with other Vela satellites launched later and with better instruments. By analyzing the different arrival times of the bursts as detected by different satellites, the Los Alamos team led by Ray Klebesadel was able to determine rough estimates for the sky positions of sixteen bursts and to definitively rule out a terrestrial or solar origin \citep{Klebesadel73}. In the time period 1969 to 1979 the Vela spacecraft (5 and 6) recorded 73 gamma-ray bursts. A preliminary catalog of  events was reported by \citet{Strong74}.
\\
\\

\item{\bf OSO~5}

{\em OSO~5} was launched on January 22, 1969 with a Delta rocket from Cape Canaveral (US) and lasted until July 1975. For celestial X--ray observations, the instruments had to be located on the wheel section of the spacecraft, which, rotating, provided overall gyroscopic stability to the satellite. On the rotating wheel of {\em OSO~5} there was, among others, a CsI scintillation crystal \citep{Frost1971}. 
The energy range was 14--254 keV with 9 energy channels. It was primarily designed to measure, in addition to solar X-ray flares, the intensity, spectrum and spatial distribution of the diffuse cosmic background.

The most striking result was indeed the spectrum of the diffuse background in the 14-200 keV energy range 
\citep{Dennis1973}.
\\

\item{\bf OSO-7}

The first relevant mission in the seventies with hard X--ray instrumentation was {\em OSO~7}. It was launched on September 29, 1971 by a Delta rocket and ended on July 9, 1974. The primary objectives were to perform solar physics experiments and to map the celestial sphere.

The rotating wheel carried four X--ray instruments which looked radially outwards, and scanned across the Sun every 2 sec. Two of them were solar observing instruments, and the other two were cosmic X-ray instruments: the MIT X-ray Cosmic Experiment (\textbf{XCE}) and the UCSD Hard X-ray Telescope (\textbf{HXT}). We concentrate on these two instruments.  

The MIT \textbf{XCE} \citep{Clark1973;oso7} (see also 
Table~\ref{t:sat-missions})
 was devoted to measure the positions, spectra and time variations of X-ray sources from 1 keV up to 60 keV. It consisted of two banks of gas proportional counters 
with cilindrical collimators that defined two circular FOVs, one with 1${^\circ}$ FWHM and the other with 3${^\circ}$ FWHM. Each bank had four counters. Starting from the front, the counters in each bank had Ne (1--6 keV), Ar (3--10 keV), Kr (15--40 keV) and Xe (25--60 keV) as the principal filling gas, stacked in such a way  that each would act as a filter for those behind it 
\citep{Markert1979;oso7}. 
As the wheel rotated with a period of about 2 s, the FOVs swept out two circular scan bands in the sky. For each wheel revolution, the counts were accumulated in equal azimuthal bins, the content of which was sequentially telemetered. 
The angular resolution was about 14 deg.  As the spin axis was precessed to maintain an angle near 90 deg from the Sun direction, the scan bands gradually swept over the sky. In this way all the sky could be surveyed.

The UCSD Hard X-ray Telescope (\textbf{HXT}) \citep{Peterson1973;oso7,Cox2000;oso7-8heao1-3} was designed mainly to measure the spectrum and intensity of known and new X-ray sources in the 7-500 keV energy range. It consisted of a NaI(Tl) scintillation crystal (see Table~\ref{t:sat-missions}) 
viewed by a Photomultiplier (PMT). The detector was 
surrounded by a 4 cm thick \citep{Ulmer1972;oso7} CsI(Na) anticoincidence shield, viewed by 6 PMTs, 
with 10 drilled holes to define its FOV (6.5${^\circ}$). Thanks to the wheel rotation, it provided a spatial resolution of 0.2${^\circ}$. 

Among the significant results, in addition to those on solar flares \citep[e.g.,][]{Datlowe74},  there were a hard X--ray scanning of the sky with both the UCSD and MIT instruments \citep{Peterson1973;oso7, Markert1979;oso7}, the first spectral studies extended to hard X--rays of strong Galactic  \citep{Ulmer73, Ulmer73b, Baity73, Baity74, Ulmer74, Ulmer74b, Ulmer75} and extragalactic sources previously discovered with the \uhuru\ satellite \citep[e.g.,][]{Mushotzky75,Mushotzky76}, the discovery of the 8.7-day periodicity from Vela X-1 \citep{Ulmer72} which led to its optical identification as a High Mass X--ray Binary (HMXB). Most of the results in the hard X--ray band came from the UCSD instrument, given its higher efficiency than the proportional counter instrument of MIT \citep[e.g.,][]{Clark72}.
\\
\\
\\

\item{\bf OSO 8}

It was launched on June 21, 1975 by a Delta rocket and ended on October 1, 1978. OSO--8 consisted of a rotating wheel and a non-spinning upper section ("sail"). Four experiments were mounted in the rotating wheel to exclusively observe cosmic X-ray sources. The first three experiments had their fields-of-view either aligned to the spin axis of the spacecraft or at small angles to it. The fourth instrument observed cosmic X-ray sources during the satellite night. Only two of these experiments covered a significant part of the hard X--ray band: the  Cosmic X-ray Spectroscopy (\textbf{CXRS}) experiment and the High-Energy Celestial X-rays (\textbf{HECX}) experiment.

The \textbf{CXRS} experiment \citep{Serlemitsos1976;oso8} was designed to determine the spectra of point-like sources and the spectrum of the diffuse cosmic X-ray background in the 2--60 keV energy range. It consisted of two Xenon (A and C) and one Argon (B) proportional counters. Detectors A and C covered the 2--60 keV energy band. A was located behind a 5${^\circ}$ collimator, with a Beryllium window and oriented antiparallel to the spin axis. 
Detector C was located behind a 5${^\circ}$ FOV collimator with a Mylar window and oriented parallel to the spacecraft spin axis. 
The time resolution ranged from 160 ms down to 1.25 ms \citep{Serlemitsos1976;oso8}.

The \textbf{HECX} experiment \citep{Dennis77,Cox2000;oso7-8heao1-3} was designed to measure the energy spectra of celestial X-ray sources and the primary X--ray background in the 20 keV-3 MeV energy range. The detector consisted of 2 independent CsI(Na) crystals shielded by a large, actively collimated CsI(Na) shielding. 
The instrument axis was offset by 5 deg from the negative spin axis of the wheel. 
One of the two central crystals was completely shielded and served as a monitor of the internal detector background spectrum. 

OSO--8 was devoted mainly to observations of Galactic X--ray binaries, with significant results at hard X--ray energies (the best up to 60 keV) on the brightest ones, e.g. Crab Nebula \citep{Dolan77}, Vela X-1 \citep{Becker78}, Cen X-3 \citep{Dolan84}, Her X-1 \citep{Maurer79}, AM Her (first time HECX and CXRS coincident spectrum over the range 2–-250 keV) \citep{Coe79}, Cyg X-1 \citep{Dolan79}, 4U1700$-$37 \citep{Dolan80}, Cyg X-3 \citep{Dolan82}. 
Also high--energy spectral observations of the brightest  AGNs, like NGC 4151 \citep{Mushotzky78} and Cen A \citep{Beall76}, were performed.  
\\

\item{\bf Ariel V}

It was launched on October 15, 1974 from the San Marco platform (Kenya) and ended in the spring of 1980. The hard X-ray experiment was a High--energy Scintillation Telescope ({\bf ST}).

{\bf ST} was provided by the Imperial College, London University \citep{Engel77,Coe82} and was designed to extend the spectral information on selected X-ray sources in the energy region from 20 keV to 2 MeV. 
The detector was a disk of 
CsI(Na) scintillator actively collimated.
The detector axis was inclined by a few degrees with respect to the satellite spin axis so that it rotated as the satellite spun. With this method the sources could be approximately located and the background removed.

%
%

Given the small useful area (8 cm$^2$, see 
Table~\ref{t:sat-missions}), only strong sources could be detected. Positive results concerned the Galactic Center region \citep{Coe81}, Galactic source spectra and time variability, like Serpens X--1 \citep{Coe78c}; Cen X--3, GX301$-$2, and 3U1254$-$69 \citep{Coe76}, Circinus X--1 in outburst \citep{Coe76b}; Cyg X--1 and the transient A0620$-$00 \citep{Coe76c}; the discovery of the X--ray nova A0535$+$26 in hard X--rays \citep{Coe75}; the detection of Am Her type degenerate dwarfs \citep{Coe78b}; the confirmation of the cyclotron line feature at 64 keV previously discovered by \citet{Trumper77} from Her X--1 (see below) \citep{Coe77}; spectral results on extragalactic sources, like NGC4151 \citep{Coe81b}.
A summary of the hard X--ray observations  can be found in a paper by \citet{Coe82}, where for the majority of the observed sources only upper limits could be given. 
\\

\item{\bf SAS~3}

The Small Astronomical Satellite 3 (\sas3), also known as {\em Explorer~53} or {\em SAS-C}, was launched on May 5, 1975 by a Scout rocket and ended in April 1979. It was the third of a series of small spacecraft intended to survey the X-ray sky locating the sources with an accuracy of 15 arcsec, and to study a selected set of sources over the energy range from 0.1 to 55 keV. 

The payload included four instruments \citep{Buff1977;sas3}, one of which was
%
a set of three slat collimator detectors ({\bf SCDs}), made of  proportional counters, looking out perpendicularly to the spacecraft Z-axis. 
One of the three {\bf SCDs} had an extension up to 60 keV by positioning a Xenon counter behind the Argon counter. Its on--axis useful area was only 75 cm$^2$, three times lower than the Argon counters. 

Few significant results of the many obtained with SAS~3 concerned the high--energy band, in particular observations of the strongest sources \citep[e.g.][]{Remillard84,Doty81}.
\\

\item{\bf HEAO 1}

The High--Energy Astronomy Observatories ({\bf HEAO}) satellite series  marked the epoch of large scientific payloads. The hard X--ray energies were successfully covered by the first satellite of the series, {\bf HEAO~1}.
  
It was launched on August 12, 1977 by an Atlas--Centaur rocket and terminated its operations on January 9, 1979. It was designed to map and survey the celestial sphere for X-- and gamma--ray sources in the 150 eV--10 MeV energy range, to establish the size and precise location of X--ray sources, to determine the contribution of discrete sources to the X--ray background, and to study peculiar X--ray sources and their time variations. The satellite could operate in scanning and pointing modes. In scanning mode it rotated clockwise about the Earth--Sun direction  (z axis) with a period of 33 minutes. As the detector axes were directed perpendicularly to the z axis, each observatory rotation provided a scan of a great circle on the sky. When passing over the South Atlantic Anomaly (SAA), high--voltage supplies were turned off or reduced to prevent damage caused by saturation effects.

 The experiments on board that covered also or only the hard X--ray passband were a Cosmic X--Ray experiment(\textbf{CX}) (also dubbed A--2) and a Low--Energy Gamma--Ray and Hard X--Ray Sky Survey experiment (\textbf{LEGR \& HXSS}) (A--4).

The {\bf A--2} experiment \citep{Rothschild79,Marshall1979;heao1} was designed primarily to measure the diffuse X--ray background in the 0.15--60 keV energy range. It consisted of 6 mechanically collimated, gas--filled, multilayer, multiwire proportional counters to cover 3 broad spectral bands. Three of them were 
high--energy detectors ({\bf HED}) made of Xenon--filled (1 atm) proportional counters that covered the 3--60 keV energy range.
Charged particles were rejected by a top veto layer of 
Propane--Neon. The 3 HED were collimated by the use of a dual FOV collimator which provided the same detector with two co--aligned sections having different FOV in the scan direction. Thus all collimator sections viewed $3^\circ$ (FWHM) normal to the scan plane, while along the scan plane, one of the collimator sections viewed $3^\circ$ FWHM and the other  either $1.5^\circ$  or $6^\circ$ FWHM. This collimator configuration was designed to simultaneously measure both the diffuse X--ray flux and the detector internal background.  

The {\bf A--4} experiment \citep{Jung1989;heao1} was designed to measure point--like sources and the diffuse cosmic background in the 10 keV--10 MeV energy range. It consisted of 7 different collimated NaI(Tl)/CsI(Na) scintillators placed inside CsI(Na) wells: 2 Low Energy Detectors ({\bf LED}), 4 Medium Energy Detectors ({\bf MED}) and one High--Energy Detector ({\bf HED}).

The {\bf LEDs} were sensitive in the 13--180 keV energy range. 
The NaI(Tl)
worked as a primary detector, while CsI(Na) 
as an active shield (phoswich configuration).
Charged particles were rejected with a thin plastic scintillator covering the apertures.  

The 4 {\bf MED} \citep{Bouchet2001;heao1,Cox2000;oso7-8heao1-3} phoswich detectors had an energy range of 80 keV--2 MeV with a small useful area (42~cm$^2$ each). Also in this case, the NaI(Tl)
worked as a primary detector, and CsI(Na) 
as an active shield. 

The {\bf HED} phoswich detector \citep{Cox2000;oso7-8heao1-3} had an energy range of 0.2--10 MeV.
Similarly, the NaI(Tl) 
worked as primary detector and CsI(Na) 
as active shield.

Each of the detectors was equipped with a pulse--shape analyzer and a discriminator that recognized the true events and rejected the CsI(Na) events. The experiment also contained three particle monitors, which measured proton and electron fluxes in three energy ranges. There was a high--resolution timing system that allowed to measure GRBs. 

One of the most important results of \heao1\ was the most complete and sensitive survey of the hard X--ray sky in the 13--180 keV energy band \citep{Levine84}. Forty--four sources were detected in the 40--80 keV energy band, and 14 in the 80--180 keV band. Most of the sources are Galactic; seven are extragalactic (see Fig.~\ref{f:HEAO1-A4}).

%
%
\begin{figure*}
\centering\includegraphics [width=0.80\textwidth]{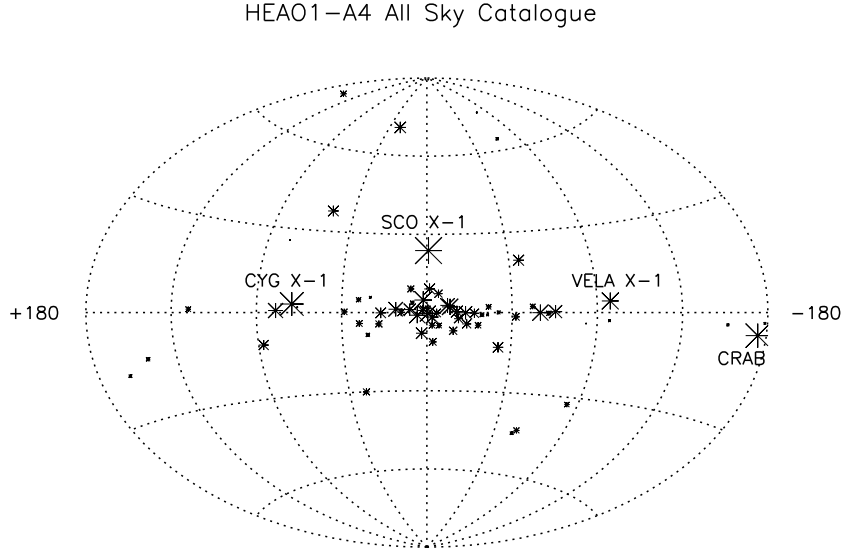}
\vspace*{-3\baselineskip}
\caption{The first view of the hard X--ray sky in Galacttic coordinates obtained with the HEAO--1 A4 experiment. The survey is complete, except in regions of source confusion, down to an intensity level of about 1/75 of the Crab Nebula in the 13--80 keV energy band.  Figure derived from the catalog published by \citet{Levine84}.}
\label{f:HEAO1-A4}  
\end{figure*}
Another key result was the most definite spectrum of the CXB in the 13--180 keV energy band \citep{Gruber99}, previously observed mainly with balloon experiments (see Fig.~\ref{f:CXB-Horstman75}). Results on the CXB in the 80--400 keV band were also reported using the A-4 MED experiment \citep{Kinzer97}.
Another key result was the CXB spectrum obtained with the A-2 experiment in the energy band from 3 to 50 keV. It was found consistent with free-free emission from an optically thin plasma of $40 \pm 5$~keV temperature \citep{Marshall80}.

Many other results in the hard X--ray band, mainly obtained with the A-4 experiment, concerned spectra and time variability studies of single Galactic and extragalactic sources. Among the Galactic X-ray sources, significant results  were obtained from X--ray pulsars like Her X--1 \citep{Gruber80,Gorecki82}, SMC X--1 \citep{Gruber84}, Cen X--3 \citep{Howe83}, 4U1626$-$67 \citep{Pravdo79}, 4U0115$+$63 \citep{Wheaton79}, LMC X--4 \citep{Lang81}; from bright X--ray transients \citep{Cooke84}; from Low Mass X-ray Binaries (LMXBs), like Sco X--1 \citep{Rothschild80,Soong83}.
Also hard X--ray measurements of type I bursts were obtained, e.g., from MXB1728$-$34 \citep{Hoffman79}.
  
A relevant result concerning the Galaxy Clusters was the discovery of a hard X--ray component from the Perseus cluster
 \citep{Primini81}, and from the Centaurus and A1060 clusters
 \citep{Mitchell80}.

Also obtained were the first spectra from the brightest AGNs, like those from Seyfert 1 galaxies up to 50 keV with the A--2 experiment \citep{Mushotzky80}, and above 50 keV with A--4 \citep{Baity84,Dil81}). Results from the radio galaxy Cen-A were obtained up to 2 MeV and from the quasar 3C273 up to 120 keV \citep{Primini79}.
A surprising result with the A-4 MED experiment was the discovery of a transient source near the Galactic Center in the 300--650 energy range, whose spectrum was consistent with a Gaussian \citep{Briggs94}.  
\\

\item{\bf Venera 11-14}

The high interest of the scientific community to unveil the mystery on the origin and nature of GRBs discovered few years before \citep{Klebesadel73} affected also Russian astrophysicists who successfully proposed  a hard X--ray/soft gamma--ray GRB experiment {\bf Konus} aboard the Russian interplanetary missions {\em Venera 11} and {\em Venera 12} launched in Sept. 1978 \citep{Mazets79}. The experiment was developed by the  
Ioffe Physico--Technical Institute in St. Petersburg and consisted of six NaI(Tl) scintillators, which were completely open apart from a shield on the sides and bottom. The detectors were  oriented along 6 different directions and  covered all-sky. The different orientation of the detector axes allowed to get a localization of the GRB sources  $\ge 4$~deg, while, when the mutual large distance of the two missions was also taken into account, a localization in the arcmin range of the source direction was even possible through triangulation. In addition to the localization, it was possible to get temporal structure and photon spectrum of the events \citep{Mazets79,Mazets81}. 

A modified version of the {\bf Konus} experiment flown aboard {\em Venera 11} and {\em Venera 12} was also flown aboard {\em Venera 13 and 14} launched in 1981 October 30 and November 4, respectively \citep{Golenetskii84}. The main differences concerned the number of energy channels and a better time resolution. For example, the temporal accumulation of the photon spectra was 0.5~s instead of 4~s. 

The Konus results were outstanding. Concerning GRBs, we wish to mention the important discovery, within single GRBs,  of a time-resolved correlation between luminosity and peak energy (interpreted as bremsstrahlung temperature) of the $E F(E)$ spectrum \citep{Golenetskii83}, and the earliest evidence  of an isotropic distribution of the GRB positions in the sky \citep{Mazets81b,Mazets88} (see Fig.~\ref{f:Konus-Mazets88}). Other key results concern the discovery  on 5 March 1979 of the first pulsating burst,  later called Soft Gamma Ray Repeater (SGR), from the supernova remnant N49 in the Large Magellanic Cloud  \citep{Mazets79b}: SGR~0526$-$66. Konus detected a total of 16 recurrent bursts from this source \citep{Golenetskii84} and obtained the earliest detections of the famous SGR~1806$-$20 and SGR~1900+14 \citep{Mazets79c}.
\\

%
%
\begin{figure*}
\centering\includegraphics [width=0.80\textwidth]{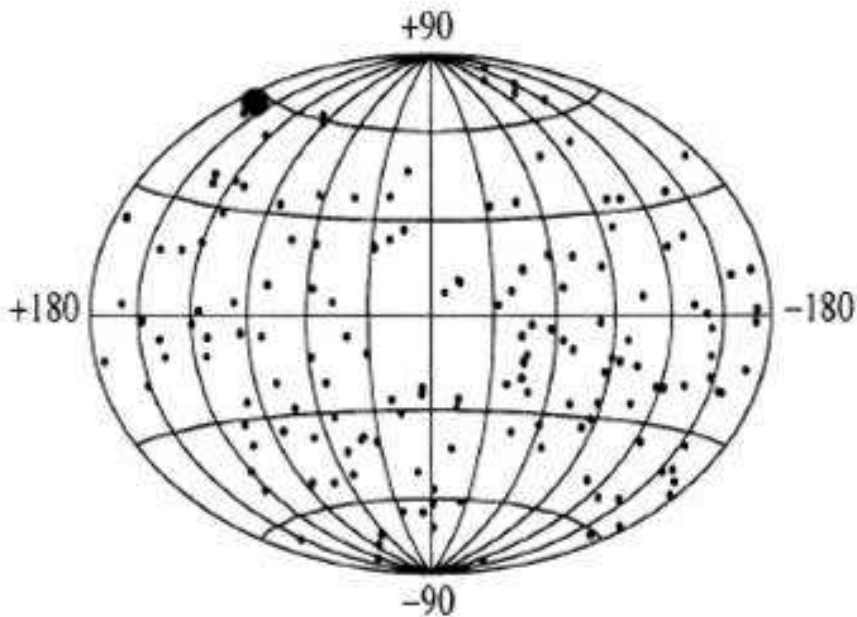}
\caption{Distribution of GRBs in the sky in Galactic coordinates as obtained with the Konus experiment  aboard the Venera 11--14 missions. Reprinted from \citet{Mazets88}.}
\label{f:Konus-Mazets88}  
\end{figure*}

\end{itemize}

\subsubsection{Balloon experiments}
\label {1970balloons}

As discussed above, many X-ray astronomy groups performed balloon experiments soon after the first hard X--ray astronomy discoveries. The interest for balloon experiments continued in the 1970s, given the possibility of designing and performing them in a time much shorter than the satellite experiments and, thanks to the development of large balloon sizes, the possibility of launching large detection areas, much larger and sensitive than the satellite experiments. We discuss the most significant balloon experiments performed in the 1970s (see also Table~\ref{tab:balloons}).

\begin{itemize}

\item{\bf Rice University group}

With a hard X--ray instrumentation developed in the 1960s \citep{Haymes68}, several balloon experiments were performed from USA and from Australia.
Concisely the used detector consisted of a NaI(Tl) crystal 
viewed from a PMT. 
The FOV (24$^\circ$ FWHM) was obtained by means of a 10~inch diameter by a 12~inch long NaI(Tl) well scintillator that surrounded the central detector. The spectra were collected in the energy  band for 35 to 560 keV, with the events beyond 560 keV counted in a single integral channel.
Significant spectral results were obtained from the Crab \citep{Haymes68}, GX~3$+$1 \citep{Haymes69}, the Cygnus region (X--1 plus X--3) \citep{Haymes70}, Sco X--1 \citep{Haymes72}, and the Galactic Center Region \citep{Haymes75}.

Using the experiment above, but with an extended spectral band from 23 to 930 keV, two balloon flights were performed from Paran\'a (Argentina). 
The Galactic Center region was observed in both flights, and, combining the results, for the first time a significant (5.3$\sigma$) emission feature (see Fig.~\ref{f:GCline-Johnson73}) at $476\pm 24$~keV with a flux of $(1.8\pm 0.5)\times10^{-3}$~photons/(cm$^2$~s~keV) was detected \citep{Johnson72,Johnson73}. 

%
%
\begin{figure*}
\centering\includegraphics [width=0.80\textwidth]{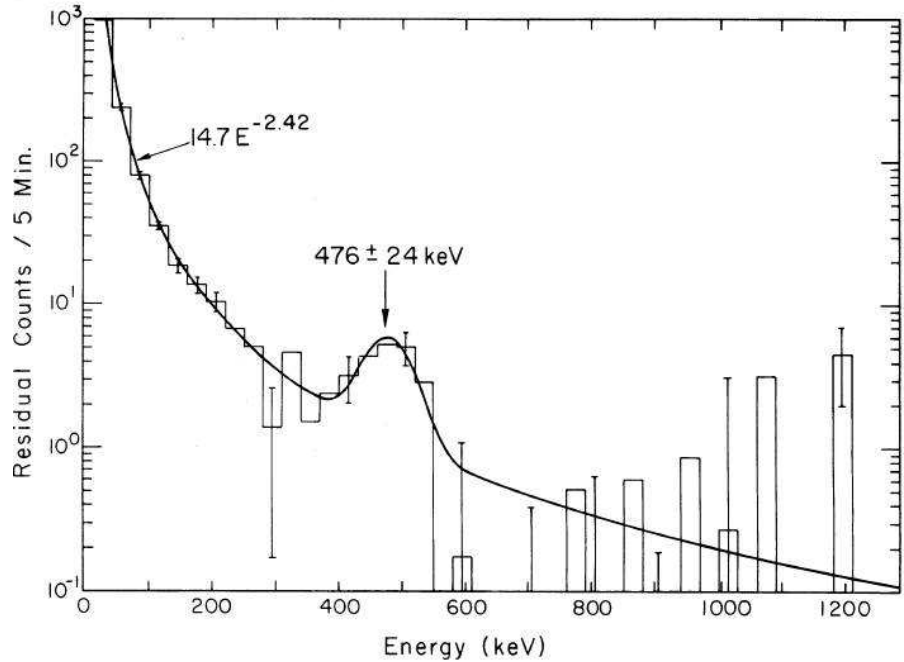}
\caption{The first discovery of a gamma--ray line from the Galactic Center Region, obtained with a balloon experiment. Reprinted from \citet{Johnson73}}
\label{f:GCline-Johnson73}  
\end{figure*}

An improved balloon experiment was launched 
in 1974 from Rio Cuarto (Argentina). 
The experiment \citep{Haymes75} consisted of a NaI(Tl) crystal with a similar thickness and a larger cross section (see Table~\ref{tab:balloons}). 
The FOV 
was defined by a NaI(Tl) collimator. The energy range was 0.02-12.27 MeV.
Among the results there was the observation of Cen A/NGC 5128 and GX 1+4 \citep{Haymes75,Koo80}.

The same experiment was flown in 
1977 from Palestine (Texas). 
It was devoted to the observation of the Seyfert galaxy NGC 4151 and its long term time variability \citep{Meegan79}.
\\

\item{\bf Tata Institute group}

A balloon experiment was developed by the Tata institute of Fundamental Research in Bombay (India).  
The detector consisted of a thin NaI(Tl) crystal 
and a Beryllium entrance window. The FOV was 
obtained by means of a graded shield collimator of Lead, Tin and Copper. A plastic scintillator surrounding the detector acted as an anti-coincidence shield. The energy range was 22.5-154 keV.

With this experiment, several balloon flights were performed starting from April 1968 \citep{Agrawal71, Agrawal72}.
The balloons were launched from Hyderabad (India) and lasted a few hours at the float altitude. 
They were dedicated to study the intensity, energy spectrum and time variations from various sources, such as Sco X-1, Cyg X-1, Crab \citep{Agrawal71,Agrawal72}. In an experiment performed on May 1, 1971, simultaneous hard-X and optical observations of Sco X--1 were performed \citep{Matsuoka72}.
\\

\item{\bf Bologna group}

Toward the end of the 1960s \citep{Brini70a}, the detection system of the Bologna group was made up of two identical detectors of NaI(Tl) with a passband in the 20-200 keV energy range. 
A passive collimator together with semi-active anticoincidence (AC) shields around the central detector gave a triangular response with a FOV of 
13${^\circ}$ FWHM \citep{Brini1971;hxr70}.

The experiment \citep{frontera1972;hxr70} was launched for the first time in 1970
aboard a stabilized platform to study the hard X--ray
pulsating emission from NP 0532, the pulsar in the Crab Nebula \citep{Brini71,Cavani71}. 
Another balloon experiment with the same instrumentation was performed in 1971, devoted to the observation of Cyg X--1 \citep{Frontera75}.

With an improved detection system and same passband, a  balloon experiment was launched in 1972.
The detector consisted of a NaI(Tl) crystal with the same thickness and FOV, but with an increased area.

The goal was the long term variability study of Cyg X-1 and the observation of Cyg X-3 \citep{Frontera75}. Combining the various observations of Cygnus X--1, an outstanding result was obtained (see Fig.~\ref{f:QPO-CygX1-Frontera75}): the first discovery of Quasi-Periodic Oscillations (QPOs) from the source with a frequency centroid of $5.75\times10^{-2}$~Hz \citep{Frontera75b,Vanderklis95}. A confirmation of these QPOs was obtained with BATSE about 20 yrs later \citep{Kouveliotou93}.

%
%
\begin{figure*}
\centering\includegraphics [width=0.80\textwidth]{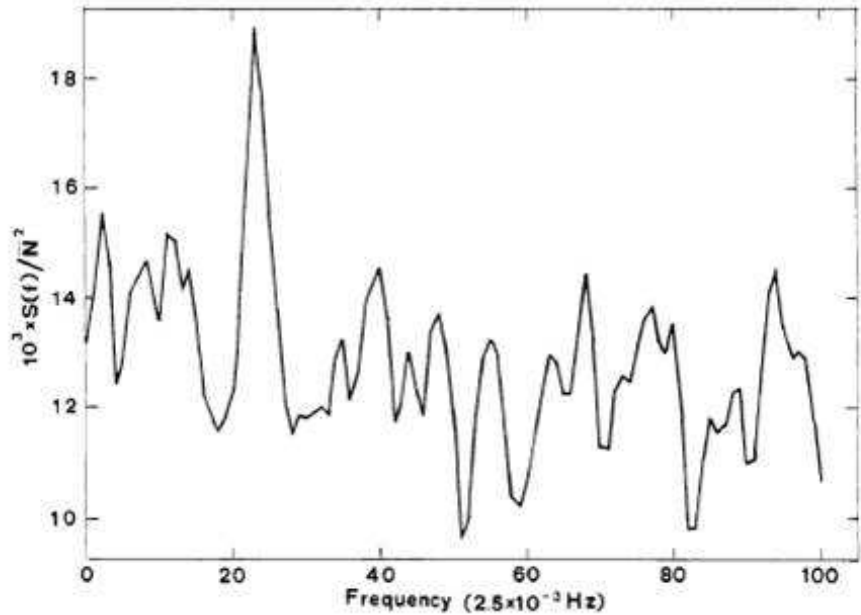}
\caption{Power Density Spectrum of Cygnus X--1, obtained from the results of three balloon experiments performed to study  the source time variability. The first evidence of Quasi-Periodic Oscillations (QPO) was discovered. Reprinted from \citet{Frontera75b}.}
\label{f:QPO-CygX1-Frontera75}  
\end{figure*}

Another balloon experiment of the Bologna group was flown in 1976 from Trapani (Sicily) for a transatlantic flight. The flight terminated before arriving in the East coast of USA. The total useful duration was  about 75 h \citep{Frontera81}.
The experiment consisted of two independent collimated hard X-ray telescopes, 4 m apart, both pointing to the zenith. Each detector consisted of a large NaI(Tl) crystal. 
The nominal energy range was 20 to 300 keV. Due to telemetry limitations, the scientific data transmitted were the counts in 0.83 s in two energy channels (20-150 keV and 150-300 keV) and the 60 channel energy spectra integrated over 106 s. 
The detector background varied as a consequence of the change of both the geomagnetic latitude and the float altitude \citep{Frontera81}.  
Among the relevant results of this experiment there were the detection of pseudo gamma-ray bursts of long duration due to phosporescence in the detector produced by high--energy cosmic rays \citep{Frontera81a}, the detection of a latitude effect in the X--ray counts \citep{Frontera81}, the observation of extragalactic (such as NGC~4151 and 
MCG~8-11-11) \citep{Frontera79a} and Galactic sources, like X~Persei \citep{Frontera79b}, and the discovery of a transient source \citep{Frontera79c}. 
\\

\item{\bf NRL balloon experiments}

The first balloon experiment of the Naval Research Laboratory (NRL) performed in the 1970s was launched in 1973 from Palestine (Texas) \citep{Kinzer78}. 
There were 2 phoswich X-ray detectors (A and B), that differed from each other only because of the CsI and NaI crystal roles interchanged. 
Both scintillators were viewed from a single Photomultiplier Tube (PMT). The energy passband was 20 to 160 keV. 
The two detectors were oriented vertically and were looking towards the north galactic pole. 
The balloon experiment was devoted to the measurement of the CXB spectrum above 20 keV \citep{Kinzer78}.

After this experiment, an improved experiment with a larger detection area and 20-250 keV energy passband, was developed  and, for the first time, launched in May 1976 from Palestine (Texas) \citep{Johnson78,Strickman79}.
The FOV (10${^\circ}$ FWHM) was defined by a graded collimator
and could be oriented according to a programmed observation schedule.
The experiment was devoted to the observation of several sources \citep{Johnson78}, among which the Crab Nebula 
\citep{Strickman79} and Cyg X-1 \citep{Johnson78}, from which significant fluctuations on time scales ranging from 10 s down to 0.1 s were observed.

The same experiment \citep{Strickman80} with a smaller FOV (5${^\circ}$ FWHM) was also flown in 1977
from Alice Springs (Australia), with main objective the observation of GX 1+4 \citep{Strickman80}. 
\\

\item{\bf INPE group, Brazil}

Also the National Institute for Space Research (INPE) in Sao Jos\'e dos Campos (Brazil) developed a balloon experiment that was launched three times \citep{Buivan79} in 1973, 
1974 
and 1978 
with flight durations of a few hours.
The detector was a NaI(Tl) crystal 4~inches diameter by 4~inches thickness for  1973 and 1974 flights and 3 inches diameter times 1/4 inches thickness for the 1978 flight. The energy passsband was 0.9-17 MeV, 0.3-5 MeV and 0.1-2 MeV in 1973, 1974 and 1978, respectively. The flights of 1973 and 1978 were dedicated to the observation of the atmospheric gamma-ray component, while that of 1974 was dedicated to the observation of the Galactic Center region. The most significant result was the spectrum of the atmospheric gamma-ray emission \citep{Buivan79}.
\\

\item{\bf AIT/MPE group}

The first significant balloon experiment of the Astronomical Institute of the Tubingen University (AIT) in collaboration with the Max Planck Institute for Extraterrestrial Physics (MPE), was performed on May 3, 1976 from Palestine (Texas) \citep{Kendziorra77}. Previously a balloon experiment was performed in 1975 \citep{Pietsch76}. 
There were two independent collimated telescopes mounted in parallel: a NaI(Tl) scintillator detector (the same used in the 1975 flight)
shielded by well type CsI crystals, and a NaI/CsI phoswich detector. 
Their passband was 17-160 keV and 15-135 keV, respectively.
The FOV of both telescopes was 2${^\circ}\times$10${^\circ}$   FWHM.
The balloon experiment was devoted to the observation of Her X--1, Cyg X--1, Cyg~X--2 and Cyg~X--3 \citep{Trumper78}.  The most important result was the first discovery of a strong line feature at 58 keV in the pulsed X-ray spectrum of Her X-1 \citep{Trumper77,Trumper78} 
(see Fig.~\ref{f:HerX1-line-Trumper78}). 

%
%
\begin{figure*}
\centering\includegraphics [width=0.80\textwidth]{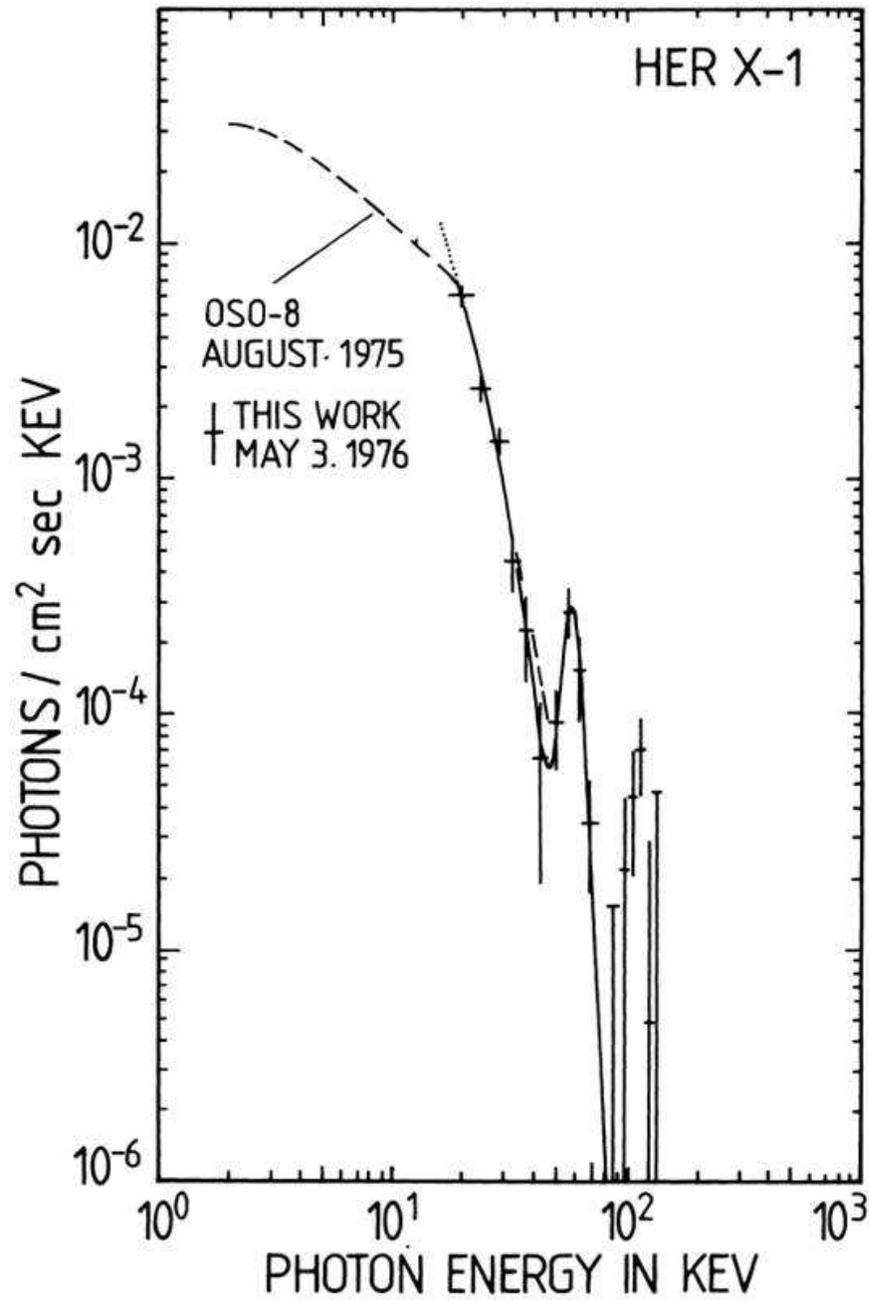}
\caption{X--ray spectrum of the X--ray pulsar Her X--1, with the first discovery of a cyclotron line. The result was obtained with a balloon experiment. Initially the line was interpreted as an emission line at 58 keV. Reprinted from \citet{Trumper78}.}
\label{f:HerX1-line-Trumper78}  
\end{figure*}

An improved balloon experiment was launched two times in 1977, 
from Palestine (Texas \citep{Staubert78,Reppin79,Staubert80}.  
The energy range was 10--200 keV.
%
A passive graded shield (Pb, Sn and Cu) and a plastic anticoincidence scintillator were used to reduce the background. Also a radioactive source of $^{109}$Cd was used to perform an in-flight calibration. During the first flight, the X-ray binary system AM Herculis was observed. By combining the flight results with those obtained with {\em OSO~8}, it was possible to derive an accurate spectrum of the source and find that it was consistent with a thermal bremsstrahlung spectrum \citep{Staubert78}.
During the second flight the time variability of Cyg~X--2 was investigated \citep{Reppin79}.

The same experiment was launched on November 22, 1978 from Alice Springs (Australia) reaching a float altitude of 3.5 g/cm$^2$. It was devoted to the observation of Vela X-1 (4U 0900-40) to study its 283 s pulsation and its hard X--ray spectrum \citep{Staubert80}.
\\

\item{\bf ISAS group, Japan} 

With the goal of detecting GRBs, an ISAS (Institute of Space and Aeronautical Science) group \citep{Yamagami79}  developed a balloon experiment which used a set of 3 NaI(Tl) scintillator detectors with 5 inches diameter, each located below a rotating cross-modulation collimator (RCMC), with a FOV of about 146${^\circ}$ FWHM and a passband of 30-200 keV. The experiment, with the telescope axis directed toward the zenith, was launched three times in 1975 aboard a balloon for a total duration flight of 145 hrs. 

Positive results were obtained with the detection and first accurate localization (within 0.2 deg) of a GRB event \citep{Nishimura78}, two years after the publication of the GRB discovery. Other two long duration flights were performed in 1977 and 1979 (total duration of 138 hrs) with an improved experiment, made of an array of 4 detectors, each surmounted by a RCMC. Two GRBs events were detected and one of them accurately localized \citep{Yamagami79}. 
\\

\item{\bf University of Tasmania group, Australia}

A balloon group, operating at the University of Tasmania in Hobert, developed a balloon experiment that employed a multi-wire proportional counter \citep{Greenhill79} 
filled with Xenon at a pressure of 716 mbar together with 304 mbar of Helium. It was sensitive to the 20-100 keV energy range. The detector was surrounded by veto counters and a graded shield of Sn and Cu.  There was a collimator made of Nickel-plated pewter sheets, separated from the Xenon filled volume by means of an Aluminum window of 0.5 mm thick, but inside the hermetic chamber. 

The balloon was launched in 1978 
on November 20, 1978 from Alice Springs (Australia).
Among its significant results there was the observation of known X-ray sources (Sco X-1, Cir X-1)  and the observation of a transient event believed to be a gamma-ray burst  
\citep{Greenhill79b}.
\\

\item{\bf MISO collaboration}

The MIlan-SOuthampton ({\bf MISO}) collaboration, established between the University of Southampton (UK) and the Institute of Cosmic Physics of the Italian CNR in Milan (Italy), developed a gamma-ray  telescope operating in the 0.1-20 MeV energy range. The telescope consisted of two scintillators (a liquid scintillator NE 311, later replaced by a plastic scintillator, and a NaI(Tl)) that formed a Compton-coincidence detection system.
The FOV (3${^\circ}$ FWHM) was obtained by means of a semi-active shielding system. To extend the band to lower energies, a passively shielded hard X-ray detector (20-280 keV)
was mounted parallel with the main telescope and the same FOV  \citep{Baker1979;miso,Baker1981;miso,Perotti1986;miso}.

The experiment was successfully flown from Palestine (Texas) in May 1977, October 1978, September 1979 and May 1980. Positive results were obtained in the observation of the Seyfert I galaxies NGC~4151 \citep{Perotti79,Butler81} and MCG~8$-$11$-$11 \citep{Baker81}, and in the observation of the sky region containing the bright high--energy gamma--ray source CG135+1 \citep{DiCocco81}.
\\

\item{\bf MPI Compton Telescope experiment}

The Max Planck Institute for Extraterrestrial Physics (MPI) was the first institution that developed a Compton Telescope for soft gamma--ray astronomy (1--10 MeV) \citep{Schonfelder73}. The telescope was based on the use of plastic scintillators for both the Compton scatterer and the absorber, both covered by an anticoincidence plastic scintillator. After this earliest development, the MPI moved to a Compton telescope based on an array of organic liquid scintillators as Compton scatterer and an array of thick NaI(Tl) inorganic crystals as 
absorber \citep{Schonfelder82}. Also in this case, an anticoincidence shield of plastic scintillators was added.

Scientific results were obtained with both telescope configurations. With the earliest, the energy spectrum of the cosmic gamma--ray background (CGB) was measured 
\citep{Schonfelder74}, and, thanks to the telescope half--opening angle of 20$^\circ$,  it was possible establish for the first time that at least 87\% of the CGB was of extragalactic origin.
With the second telescope configuration, two balloon flights had been performed: in 1977 and 1979. The first one was the most fruitful. It was possible to derive a sky map of the anticenter region of our Galaxy \citep{Graser81}, the gamma--ray spectrum of Crab and its pulsar \citep{Penningsfeld79}, and the spectrum of the diffuse (primary and terrestrial albedo) gamma--ray components \citep{Schonfelder80}.
\\

\item{\bf Groups committed to detect Gamma--ray lines}

Following theoretical studies by many authors 
\citep[e.g.,][]{Meneguzzi75, Lingenfelter76, Yoshimori79b}  on the possible production of Galactic gamma-ray lines and their detectability, a high interest was inspired in the search of these lines, in particular of the positron annihilation line at 0.511 MeV.

As discussed above, the first discovery of a spectral feature from the Galactic Center region at $476\pm 24$ keV (see 
Fig.~\ref{f:GCline-Johnson73}) was due to \citet{Johnson72} and \citet{Johnson73}. It was interpreted by \citet{Leventhal73} as a positronium annihilation line. After this discovery,  several balloon experiments were performed devoted to the search of gamma--ray lines \citep[e.g.,][]{Buivan79}, but only a few significant results were found.

When the Crab Nebula was in the field of view of their 
gamma--ray telescope \citet{Yoshimori79} reported the possible detection of a gamma--ray line at $\sim 400$~keV with a flux of $(7.4 \pm5.4)\times 10^{-3}$~photons/(cm$^2$~s), never confirmed. 

The most fruitful collaboration was that established by {\bf Bell and Sandia Laboratories}. They developed a balloon telescope made of a Ge(Li) solid state detector 6.3 cm thick with a detection area of about 21~cm$^2$  and a FOV of 15$^\circ$, working in the 0.1--5 MeV energy range \citep{Leventhal77b,Leventhal78}. The FOV was obtained with a heavy anticoincidence shield of NaI(Tl) around the main detector. The gondola was alt-azimuth stabilized with a pointing accuracy of 1$^\circ$.  

Their most relevant result was the discovery, in a balloon flight performed on 1977 from Alice Spring, and confirmation with another flight performed in 1979, of a line feature at $(510.7\pm0.5)$ keV  \citep{Leventhal78,Leventhal80} from the Galactic Center direction (see Fig.~\ref{f:GCline-Leventhal78}).  Its flux at the top of the atmosphere was $(1.22\pm 0.22)\times10^{-3}$~photons/(cm$^2$~s). 

%
%
\begin{figure*}
\centering\includegraphics [width=0.80\textwidth]{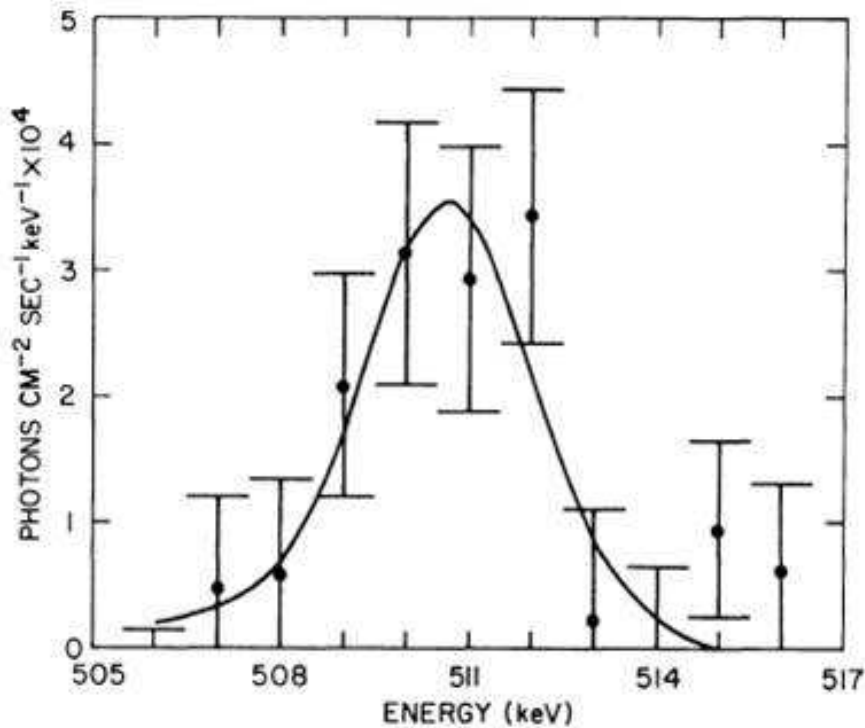}
\caption{First discovery of  a positron annihilation line from the Galactic Center region with the Bell--Sandia balloon experiment performed on 1977 September 11-12. Reprinted from \citet{Leventhal78}.}
\label{f:GCline-Leventhal78}  
\end{figure*}

 \end{itemize}

\subsection{\bf Satellite missions and balloon experiments in the 1980s}

The 1980s saw the second generation of hard X--ray missions and balloon experiments. 
In the case of satellite missions, the highest interest was still devoted to the low energy band ($<20$ keV), but  an extension to higher energies ($\sim$50~keV) was envisaged. In spite of that, in most of these cases no significant results were reported. This was also due to the fact that, in general, the analysis of the data collected in the hard X--ray band was much more problematic than at lower energies, where the 
signal-to-noise ratio was much higher. Only in few cases, {\em HEAO~3} and {\em Mir--Kvant}, significant results at high energies were obtained and reported.
In the case of balloons, the 1980s decade was characterized by fewer but more sensitive experiments.

\subsubsection{Satellite missions}

\begin{itemize}

\item{\bf Ariel~VI}

Ariel~VI was the follow up to Ariel~V. It was launched on June 3, 1979
and operated until February 1982. The only instrument with an energy passband in the hard X--ray range was an X-ray Proportional Counter Spectrometer of the Leicester University. The instrument \citep{Ricketts1982;ariel6} consisted of an array of four multi-layered, Xenon-filled proportional counters with a total area of 300 cm$^2$, an energy passband from 1 to 50 keV, and an approximately circular FOV of 3$^\circ$ FWHM viewing along the satellite spin axis. 
It was designed for detailed measurement of time variability and spectra of both galactic and extragalactic sources, but the scientific output, at least at high--energies, was almost null due to electromagnetic interference from ground-based radar hampering the pointing operations. 
\\

\item{\bf Hakucho}

The first X-ray astronomy Japanese satellite, {\em Hakucho},  was launched on February 21, 1979 and terminated on April 16, 1985. The payload included a Hard X-ray (\textbf{HDX}) experiment \citep{Hayakawa1981;hakucho} consisting of a NaI(Tl) scintillation counter sensitive to the 10-100 keV energy range. 
In spite of the numerous outstanding results obtained at low energies on Galactic X--ray binaries (black hole candidates, bursters, pulsars) \citep[e.g.,][]{Hayakawa1981;hakucho}, supernova remnants \citep[e.g.,][]{Inoue1979;hakucho}, etc, due to the very small useful area of HXD, no significant results were obtained at hard X--ray energies. 
\\

\item{\bf HEAO 3}

The third HEAO mission was launched on September 20, 1979 by an Atlas-Centaur rocket and terminated on May 29, 1981. The experiments on board were a High Resolution Gamma--Ray Spectrometer (HRGRS) and two cosmic--ray experiments. The HRGRS \citep{Mahoney80} was developed to search for gamma-ray line emissions in the energy passband from 50 keV to 10 MeV. It consisted of four cooled, p-type, drifted-Germanium detectors
shielded by a 
thick CsI anticoincidence shield. 
The energy resolution, initially 3 keV FWHM at 1.46 MeV, degraded with time because of radiation damage. 
It operated only for about one half year, until June 1, 1980 when the cryogen for the detectors exhausted.

Among the significant results there were
%
%
the measurement of the Galactic Center 511 keV line, that gave an intensity of $(1.25 \pm 0.18)\times 10^{-3}$~photons/(cm$^2$~s) in the fall of 1979 and $(0.99 \pm 0.18)\times 10^{-3}$~photons/(cm$^2$~s) in the spring of 1980 \citep{Mahoney94}, the discovery of the Al$^{26}$ line emission (1809 keV) in the interstellar medium, the search for the 511 keV annihilation line from active galaxies \citep{Marscher84}, the search for gamma--ray line emission from the the strongest known X--ray sources (e.g., Cyg X--1, Cyg X--3, SS~433), and a final negative response \citep{Mahoney84b} to the issue of a possible line at 73 and 400 keV from the Crab pulsar reported by several authors 
\citep[e.g.,][]{Ling79,Strickman82,Manchanda82,Leventhal77a,Hameury83}.
\\ 

\item{\bf TENMA}

The second Japanese X-ray astronomy satellite ASTRO-B, better known as {\bf TENMA}, was launched on February 20, 1983, and terminated on November 22, 1985. It was designed to study  spectra and temporal variations of X-ray sources, to make an all-sky survey for studying X-ray bursts and transients and to observe soft X-ray sources with a reflecting telescope. The observing efficiency was greatly reduced after a battery failure in July 1984. 
The experiments on board \citep{Tanaka1984;tenma} included a Gas Scintillation Proportional Counter (2--60 keV;
\textbf{GSPC}) and a Radiation Belt Monitor/Gamma-ray Burst Detector (10--100~keV; \textbf{RBM/GBD}).

The {\bf GSPC} experiment with energy passband 2--60 keV \citep{Tanaka1984;tenma} 
consisted of a set of 10 GSPCs, each made of a ceramic gas cell filled with 1 atm of Xenon and 0.2 atm of Helium, covered by a 100 $\mu$m thick Beryllium window and viewed by a suitable PMT. 
Four counters had a FOV of 3.1${^\circ}$ FWHM, while  another four had a FOV of 2.5${^\circ}$. The last two, with a FOV of 3.8${^\circ}$ FOV,  were surmounted by a bigrid rotating modulation collimator to improve the angular resolution. 
On--board calibration was performed using \textsuperscript{109}Cd radioactive sources. 

The {\bf RBM/GBD} with energy passband 10 to 100 keV \citep{Tanaka1984;tenma} consisted of 2 sets of NaI(Tl) scintillation counters. 
One of the counters was viewing in the direction of the spin axis and the other was scanning the sky with a fan-beam FOV. 
Detected GRBs were recorded with a time resolution of 1/8 s.

In spite of the many significant results on X--ray sources obtained at low energies, the extension to hard X--ray energies was limited, at most, to 20-30 keV  for either Galactic 
\citep[e.g.,][]{Nakamura89,Leahy89} or extragalactic sources \cite[e.g.,][]{Miyoshi86}, with detailed spectral studies, inclusive of Iron lines, of the brightest Galactic ones \citep{Inoue85}. No results with the RBM/GBD were reported.
\\

\item{\bf EXOSAT}

The European X-ray Observatory SATellite (EXOSAT) 
\citep{Taylor1981;exosat} was launched on May 26, 1983. On April 9, 1986 a failure in the attitude control system caused the termination of the operations. EXOSAT reentered in the atmosphere on May 6, 1986. 
The experiment on board that covered a small part of the hard X-ray band was the Medium-Energy Cosmic X-ray (\textbf{ME}) experiment.

The ME detector \citep{Turner1981;exosat} consisted of an array of 8 proportional counters sensitive in the 1-50 keV energy range. Each counter comprised 2 multi-wire proportional chambers,
one in front of the FOV and the other on the back. The front chamber was filled with 2 bar Argon-Carbon dioxide, while the rear chamber was filled with Xenon-Carbon dioxide. The chambers were separated by means of a 1.5 mm Beryllium window. The front window was made of 62 $\mu$m Beryllium (32 $\mu$m for only one quadrant). 

In spite of the numerous  and outstanding results obtained with EXOSAT at low energies ($<20$ keV), no significant results were reported at higher energies.
\\
\\

\item{\bf GINGA}

Also known as Astro-C, {\em Ginga} was a three-axis stabilized satellite, launched on February 5, 1987 and reentered the Earth atmosphere on November 1, 1991. It was designed to mainly study X--ray spectra and time variability of celestial, galactic and extragalactic sources. {\em Ginga} carried three scientific instruments \citep{Makino87}: a Large Area proportional Counter ({\bf LAC}) (1.5--30~keV), an All-Sky X-ray Monitor ({\bf ASM}) (1.5--30 keV), and a Gamma-ray Burst Detector (\textbf{GBD}) (1.5--400 keV). So, high energies were covered only by GBD.

{\bf GBD} detector \citep{Murakami1989;ginga} consisted of
a proportional counter ({\bf PC}) and a scintillation spectrometer ({\bf SS}), both pointing to a direction parallel to the Z-axis of the satellite. The energy ranges were 2-30 keV and 14-400 keV for PC and SS, respectively.
The {\bf PC} was filled with Xenon and Carbon dioxide (10$\%$) at 1.16~atm pressure, 
and an X--ray entrance window of 63.5 $\mu$m of Beryllium.  
The {\bf SS} used a
NaI(Tl) crystal
coupled to a PMT via a light guide. The entrance window of SS was an Aluminum sheet of 0.2~mm thickness.
The lateral cylindrical surface of the scintillator was 
covered by a graded passive shield.
%
No collimator limited the FOV of the two detectors.

In addition to detect several GRBs \citep[e.g.,][]{Murakami91}, GBD seemed to show absorption features from some of these events, that immediately were interpreted as cyclotron absorption lines \citep{Murakami89,Murakami91b}. However these lines were never observed later in other GRBs.
\\

\item{\bf Mir--Kvant}

The {\em Kvant} module, launched on 1987 March 31, was attached to the {\em Mir} space station. It operated until fall 1989 and was restarted in October 1990.

There were 4 instruments in the module. Apart from a coded mask imaging spectrometer ({\bf TTM/COMIS}) working in the low energy band (2--30 keV), the other instruments ({\bf HEXE}, {\bf GSPC}, and  {\bf Pulsar X--1}), all coaligned, were working in the hard X--ray energy band.

{\bf HEXE} (High--Energy X--ray Experiment) \citep{Borkus95} consisted of 4 individual phoswich detectors made of NaI(Tl) and CsI(Na) surmounted by a  Tungsten collimator. 

{\bf GSPC} (or {\bf Sirene 2})  was a high-pressure 
(3 atm) gas scintillation proportional counter covering the 2-100 keV energy range.

{\bf Pulsar X-1} consisted of 4 NaI/CsI phoswich detectors covering the 30-800 keV energy range. 

Several observations of Galactic sources were performed, deriving the hard X--ray component of their emission.
We mention here significant broad--band spectral results obtained on  the Vulpecula X--ray nova \citep{Sunyaev88b}, the decay of the hard X--ray counterpart of SN1987A \citep{Sunyaev89}, the Galactic Center region \citep{Sunyaev91a}, bright black-hole candidates, X-ray pulsars and LMXBs \citep{Sunyaev91b}, bright hard X--ray transients \citep{Borkous97,Kaniovsky97}.
 
\end{itemize}

\subsubsection {Balloon experiments}
\label {1980balloons}

We report here those experiments that gave the most significant results.

\begin{itemize}

\item{\bf Bell--Sandia Laboratories experiments} 

The Bell--Sandia Collaboration (see above) continued in the 1980s with other balloon experiments mainly devoted to the monitoring of the Galactic Center (GC), from which a 511 keV positron annihilation line was previously detected (see above). Their balloon experiment was again flown from Alice Springs (Australia) on 1981 November 21 and 1984 November 20. The result was that the 511 keV line was no more detected, with 1$\sigma$ upper limit of $\sim 4.4\times 10^{-4}$~photons/(cm$^2$~s) \citep{Leventhal82,Leventhal86}. This result showed the variability of this line (see Fig.~\ref{f:GCline-vs-time-Leventhal86}).
However, measurements in the same epoch with the wide field ($\sim 130^\circ$) {\bf Gamma Ray Spectrometer} aboard the {\bf Solar Maximum Mission} satellite,
showed a strong 511 keV line ($2.3^{+0.5}_{-0.8} \times 10^{-3}$~photons/(cm$^2$~s)) with  small variability \citep{Share90}. This discrepancy and the positive detection in 1988 with the balloon experiment GRIS (see below) was interpreted by \citet{Lingenfelter89} as due to the presence of a point--like source near the Galactic Center and a diffuse line emission in the Galactic Center region.
\\

%
%
\begin{figure*}
\centering\includegraphics[width=0.80\textwidth]{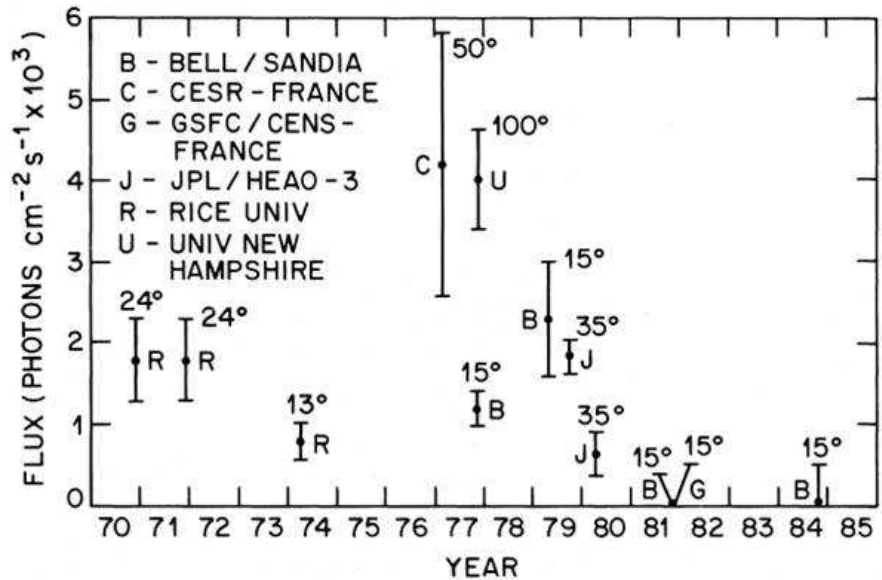}
\caption{The time behaviour of the 511 keV positron annihilation line from the Galactic Center region since its first discovery 
to 1984 November 20. The FOV of the instruments used is also 
reported, because there was the suspect that the line intensity could be correlated with the instrument FOV and thus that it 
could come from an additional source or that the source could be spatially extended \citep{Dunphy83}.  Reprinted from 
\citet{Leventhal86}.
}
\label{f:GCline-vs-time-Leventhal86}  
\end{figure*}

\item{\bf UCR Compton Telescope experiment}

A Compton telescope was also developed by the University of California in Riverside (UCR) \citep{Herzo75}, based on a double array of liquid scintillator tanks, the first array as Compton scatterer and the second one as absorber. An anticoincidence shield of plastic scintillators surrounded both arrays. The telescope exhibited an angular resolution of about 8$^\circ$ (HWHM).
With this instrument, as a result of a flight performed in 1981 from Alice Springs (Australia), it was possible to report for the first time, in the 0.3--30 MeV energy band,  on pulsations from the Vela pulsar PSR~0833$-$45 and its spectrum \citep{Tumer84}.
\\

\item{\bf XG experiment}

The {\bf XG} experiment (from Italian X-Grande, i.e., X-Large) was a large balloon experiment of the Bologna group \citep{Frontera1985;xgrande} with an operational energy band from 20 to 200 keV. 
It consisted of an array of 16 independent square NaI(Tl) crystals, each 
viewed by a PMT and separately collimated by a graded mechanical square collimator. 
The entire telescope was actively shielded by a plastic scintillator. 
Data were transmitted with a time resolution down to 1 ms.

The experiment was flown for two times from Palestine (Texas) in 1980 and 1981. In 1982 it was launched from the Milo base in Sicily. In the last flight four of the units were replaced with NaI/CsI phoswich units with different thickness values in order to select the best configuration for the high--energy instrument {\em PDS} for the BeppoSAX satellite    
 \citep{Frontera85}.

In addition to spectral results obtained on Galactic (Cygnus X-1, Cygnus X--2, Her X--1, X-Persei, Crab Nebula) and extragalactic (NGC5548, Perseus Cluster) sources \citep{Frontera85c,Matt90}, the most significant and outstanding  result was the high-statistics detection of the recurrent transient X-ray pulsar A0535+26 near the maximum of a large outburst. It was the first time in which single periodic pulses  were visible, and, in addition to a study of the timing properties, a very detailed pulse-phase resolved spectral analysis (see Fig.~\ref{f:A0535-Dalfiume88})  was possible to be performed \citep{Frontera85b, Dalfiume88}.
\\
%
%
\begin{figure*}
\centering\includegraphics [width=0.80\textwidth]{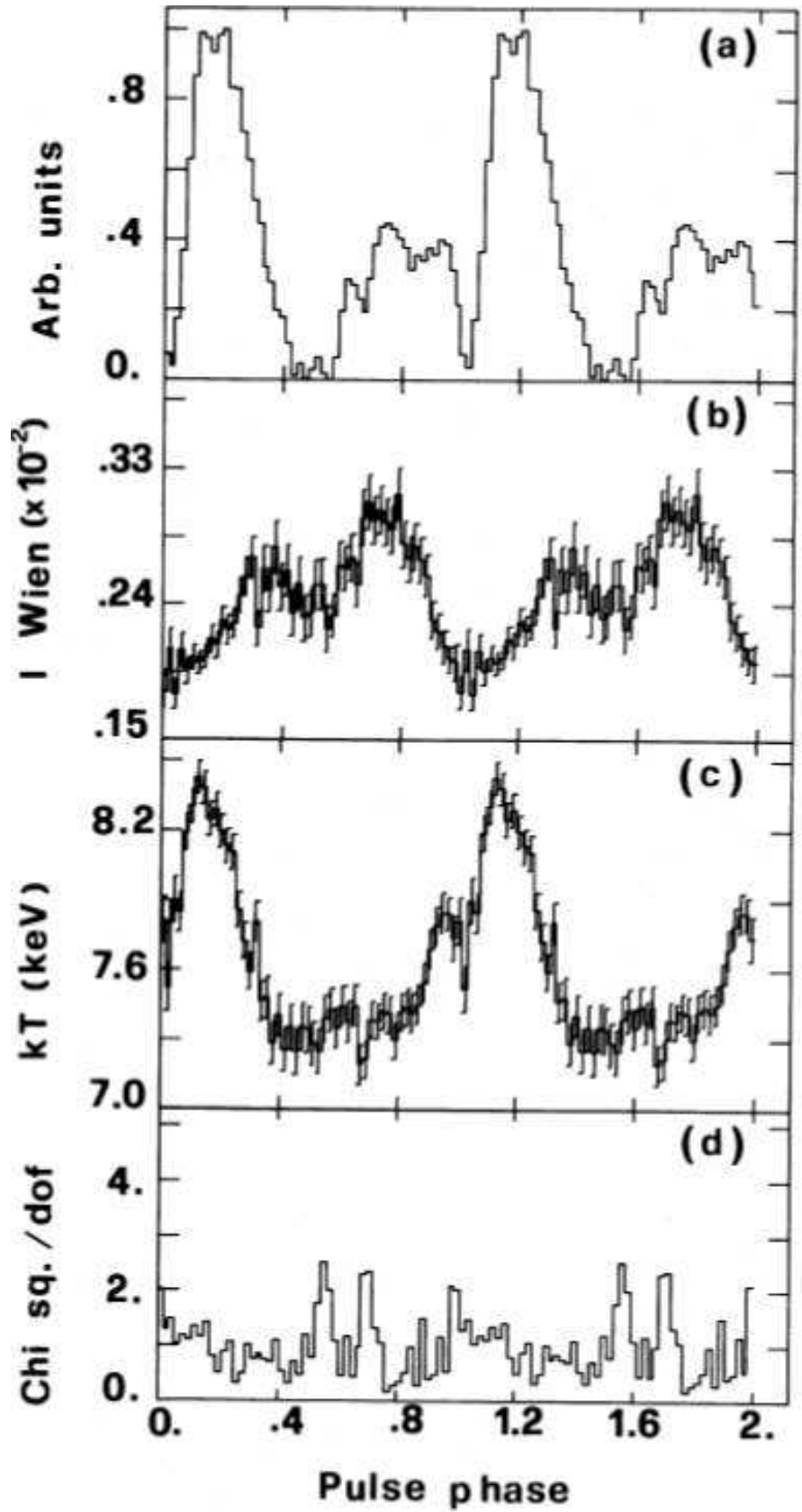}
\caption{Very detailed pulse--phase resolved spectroscopy in 27--100 keV of the X--ray recurrent transient pulsar A~0535$+$26 observed with the XG-experiment during a balloon flight on 5 October 1980 during a source outburst. The best fit spectrum was a Wien law. (a) Average pulse profile. (b)Intensity of the Wien law versus pulse phase. (c) Wien temperature  versus pulse phase. (d) $\chi^2$ per degree of freedom. It was the first time that a so detailed pulse-phase spectroscopy in hard X--rays was possible to be performed. Reprinted from \citet{Dalfiume88}.}
\label{f:A0535-Dalfiume88}  
\end{figure*}

\item{\bf POKER}

{\bf POKER}, a balloon experiment of the Frascati group, was flown in 1981 and in 1985 from the Milo balloon base in Sicily, and again in 1989 from Alice Springs, Australia. 

The instrument \citep{Bazzano83;POKER} was a very large array of Multiwire Proportional Counters (MWPC) filled with a gas mixture of Xenon-Argon-Isobuthane. For the flights of summer 1981 and 1985, it consisted of 4 units of passively collimated MWPC, 
with an efficiency higher than 20$\%$ in the energy range 15-110 keV.
%
%
For the flight of May 1989 the detector consisted of 3 MWPC modules instead of 4.
%
To improve the background rejection, each MWPC and collimator module were surrounded with a plastic scintillator shield \citep{Bazzano1990;poker}.

Results were reported from all flights: for the 1981 flight, the detection of the recurrent transient X--ray pulsar A~0535+26 during one of its off--states \citep{Polcaro83}, a hard X--ray detection of galaxy clusters \citep{Bazzano84} never confirmed; the detection of 3 AGNs (NGC~4151, MCG~8$-$11$-$11, Mkn~421) \citep{Ubertini84}; for the 1985 flight, observation of Cyg X--1 and the Crab pulsar \citep{Ubertini91,Ubertini94}; while for the 1989 flight, observations of  the radio galaxy Cen~A \citep{Ubertini93}, Sco~X--1 \citep{Ubertini92}, and the Galactic Center region were 
reported \citep{Bazzano92,Bazzano93}.
\\

\item{\bf FIGARO II}

The French Italian GAmma Ray Observatory (FIGARO) was specifically designed to observe cosmic sources with a well-established time signature, like pulsars.
The first version of FIGARO, launched the first time from Brazil in 1983, was destroyed following a balloon burst \citep{Agnetta85}. Then it was reconstructed (FIGARO II) and successfully flown first from the Milo base (Trapani, Italy) in 1986, then from Charleville (Queensland, Australia) in 1988, and again from Milo base in 1990.  

The principal detector of FIGARO II 
\citep{Agnetta85,Agrinier1990;figaro2} consisted of a square array of nine NaI(Tl) tiles.
The energy passband was 0.2-6 MeV \citep{Agnetta1989;figaro2}. The detector was actively shielded against the environmental 
background with a wall of 12 NaI(Tl) modules along the four sides and  a block of plastic scintillators from below. To reject charged particles from the entrance window there was a 5 mm thick plastic scintillator on the top of the experiment.

Significant results were obtained from all flights. The main goal of the first and third flight was the study of the Crab pulsar. Both the pulse profile and spectrum of the source in different energy bands were well determined \citep{Agrinier90}. The second flight was mainly devoted to the Vela pulsar PSR~B0833$-$45, the brightest source in high--energy gamma--rays ($>$50~MeV).  Very low upper limits to the source intensity in the instrument passband were reported \citep{Sacco90}. A very intriguing result, but never confirmed, was the almost 3$\sigma$ evidence of a 0.44 MeV line feature in the spectrum of the Crab pulsar \citep{Massaro91} observed during the third flight. During this flight, in addition to Crab, evidence of hard X--ray periodic emission (103.2 s) was reported from the binary pulsar A0535+26 during its "off" transient state \citep{Cusumano92} 
\\

\item{\bf MIFRASO}

The {\bf MIFRASO} experiment (MIlano FRAscati SOuthampton collaboration) consisted of a High--Energy Detector (HED)  and a Low Energy Detector (LED)\citep{Baker1984;mifraso}. The HED was made of an array of eight identical scintillation counters of NaI(Tl) 6 mm thick, 
actively shielded on the bottom by an equal number of much thicker NaI(Tl) crystals (50 mm). The collimator was surrounded by a plastic scintillator as veto system.


The LED was made of two high-pressure Xenon gas proportional counters,
sensitive in the 10-120 keV energy range.
%

MIFRASO was launched from Milo Base, Sicily (Italy) in 1986 and in 1987. The results concerned the hard X--ray detection of the Coma Cluster \citep{Bazzano90}, the quasar 3C273 \citep{Dean90},  and the Seyfert galaxies NGC~4151     \citep{Perotti90b} and MCG~8$-$11$-$11 \citep{Perotti90}. Interesting was also the observation of A0535+26 far from the periastron passage (phase 0.25) \citep{Coe90} when  outbursts were often observed \cite[e.g.,][]{Frontera85b}. 
\\

\item{\bf EXITE}

The Energetic X-ray Imaging Experiment ({\bf EXITE}), developed at the Harvard--Smithsonian Center for Astrophysics,  was one of the first coded mask telescopes \citep{Braga1990;exite} with a 20--300 keV energy passband, flown aboard a stratospheric balloon.  The central detector was a 
NaI(Tl) scintillator, optically coupled to an image intensifier tube. It was surrounded by a graded passive shield and by  
an active shielding of plastic scintillator.
The space resolution was about 6~mm FWHM. The coded mask, with square cell side of 13~mm, was distant 2 m from the detector, yielding an angular resolution of 32 arcminutes FWHM at 100 keV and a location accuracy of about 25~arcmin. Two 1D crossed collimators, placed between the detector and the coded mask, defined a FOV of 3.4${^\circ}$ FWHM. 
The coded mask
had a Uniformly Redundant Array (URA) pattern. 

EXITE was flown three times.
The first flight was carried out in Alice Springs (Australia) in May 1988, with no scientific results due to a mechanical problem. The second balloon flight took place in Fort Sumner (New Mexico) in October 1988.
The third flight took place in Australia, Alice Springs in May 1989, in the context of a NASA balloon campaign devoted to the supernova 1987A.

A variety of both galactic and extragalactic objects were observed. Significant results include the hard X--ray detection of the {\em Einstein} source 1E1740.7-2942 with a possible second source, perhaps a transient source, about 40 arcmin West (EXS1737.9-2952) \citep{Grindlay93}, and the first imaging observation of the black hole candidate GX~339$-$4 at hard X-ray energies (see Fig.~\ref{f:GX339-4-Covault92}) \citep{Covault92}.
\\

%
\begin{figure*}
\centering\includegraphics [width=0.80\textwidth]{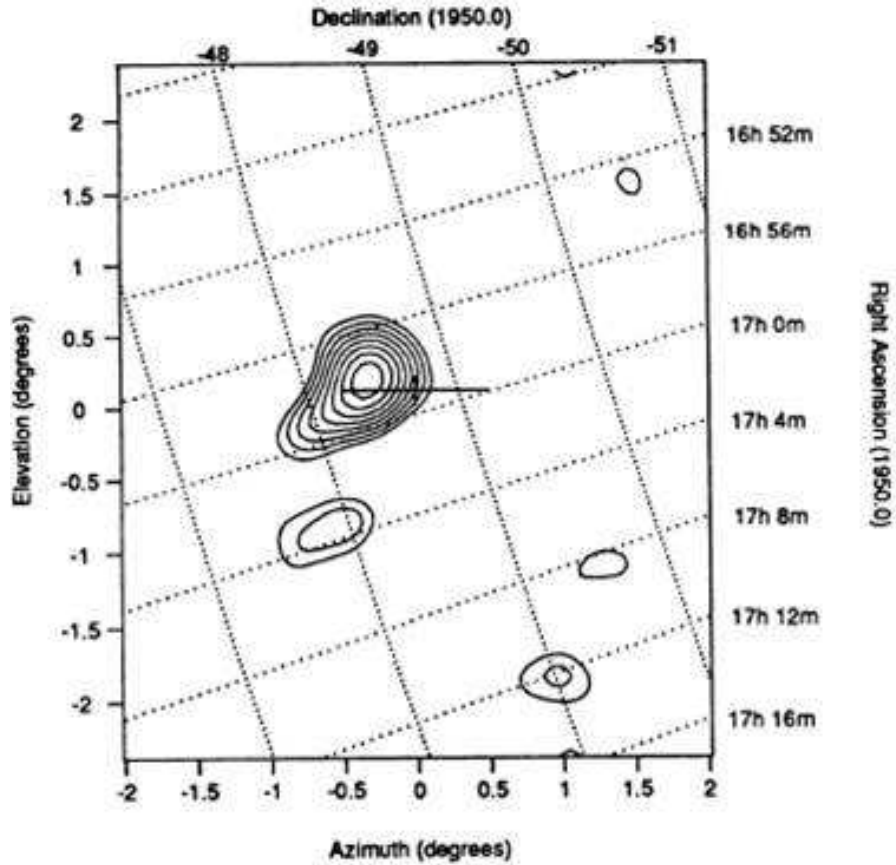}
\caption{One of the first hard X--ray images of a celestial source (GX~339$-$4) obtained with the coded mask aboard the EXITE balloon experiment.  Reprinted from \citet{Covault92}.}
\label{f:GX339-4-Covault92}  
\end{figure*}

\item{\bf GRIS}

The Gamma-Ray Imaging Spectrometer ({\bf GRIS}), developed at the NASA Goddard Space Flight Center, in collaboration with the Bell--Sandia Laboratories 
 \citep{Teegarden1985;gris2,Tueller90}, was a balloon-borne experiment using cooled Germanium detectors for 
high--resolution gamma--ray spectroscopy in the 20 keV--8 MeV energy range.
The basic instrument consisted of an array of 7 coaxial high-purity Germanium detectors surrounded by a thick active NaI(Tl) shield/collimator. An active, uniformly redundant coded mask, was used to generate sky maps in its FOV, with a source positioning accuracy of 0.2$^\circ$. 
Each detector was shielded by a 15 cm thick NaI active anti-coincidence shield. 

GRIS had 9 successful flights (2 from Fort Sumner and 7  from Alice Springs) over 8 years, from 1978 to 1985. Later it was flown two times from Alice Springs (in 1988), where it was devoted to the observation of SN1987A. In 1990 it was again flown from Fort Sumner. 

Among the results obtained from the GRIS campaigns there was the measurement of gamma-ray lines (rest energies of 846.8 and 1238.3 keV) from SN1987A \citep{Tueller90}, and the positive detection, at a flux level of $(11.8\pm1.6)\times 10^{-4}$~photons/(cm$^2$~s), of the 511 keV positron annihilation line from the Galactic Center region \citep{Gehrels91,Cheng97}, after the negative results obtained in the early 1980s by  \citet{Leventhal82,Leventhal86}.
\\

\item{\bf UAH--MSFC balloon experiment}

As a result of a collaboration between the University of Alabama in Huntsville (UAH) and the Marshall Space Flight Center (MSFC), a balloon experiment devoted to the observation of SN1987A was launched three times from Alice Springs (Australia): in October 1987, and in April and November 1988.  

The experiment was based on two large area detection units  of the same size and design as the LADs adopted for the BATSE experiment (see below) \citep{Pendleton95}. 
%
A collimator based on passive slats of lead 
was adopted. The energy passband was 18 to 960 keV, with a different binning in different energy bands.

The results were very positive: in all flights SN1987A was detected and the spectrum determined up to 300 keV. The amount of $^{56}$Co was constrained \citep{Pendleton95}. 
\\

\item{\bf GRIP}

The Gamma--Ray Imaging Payload ({\bf GRIP}) was an experiment developed at the CalTech Institute \citep{Althouse85}. It consisted of a shielded detector system surmounted by a coded--aperture mask at 2.5 m distance. The detector was a position-sensitive NaI(Tl) scintillator viewed by  19 PMTs. A gain control was performed during the balloon flights. The background was minimized by using a side plastic shield and a bottom scintillator detector similar to the primary detector. The coded mask, made of Lead, with half of open pixels, had hexagonal cells in a uniformly redundant array and was rotating at 1~rpm for an unbiased background subtraction.

The most important flights were those performed from Alice Springs (Australia) in 1987 (two times, May and November), 1988, and 1989. The 1987 flights were devoted to observe the supernova SN1987A. It was not detected in the first flight \citep{Witteborn87}, but was clearly detected in the second flight \citep{Cook88}, providing for the first time the gamma--ray image of a supernova. During the second and third flights  gamma--ray imaging of the Galactic Center region was performed, deriving gamma--ray flux and spectrum  of the brightest sources, among which, the {\em Einstein} source 1E~1740.7$-$2942 \citep{Cook91a, Cook91b, Heindl93}.  Another important result was the non-detection of the 511 keV line. Assuming that the line emission was due to 1E~1740.7$-$2942, the 95\% upper limits were $6.8 \times 10^{-4}$~cm$^{-2}$~s$^{-1}$ for the 1988 flight and $3.7 \times 10^{-4}$~cm$^{-2}$~s$^{-1}$ for the 1989 flight, confirming the time variability of this line \citep{Heindl93}.   
\\
 
\end{itemize}

\subsection{\bf Satellite missions and balloon experiments in the 1990s}

In the 1990s, a big effort on satellite missions was performed in order to extend the energy band up to hard X-ray rays. This effort was made possible by the use of new instruments, e.g, coded masks on board {\bf GRANAT}.
Here we discuss, in addition to  GRANAT, {\bf CGRO}, {\bf RXTE} and {\bf BeppoSAX}. We also devote attention to  {\bf COMPTEL} on board CGRO, although this experiment had a passband beyond  hard X-rays.

Concerning balloon experiments, this decade was not characterized by significant experiments. Those launched were classical esperiments, mainly due to new groups entering into stage from emerging countries (Brazil, China, India) \citep[e.g.][]{Braga95}. Most of the efforts of the leading groups were devoted to the design of new imaging experiments, especially for long duration balloon flights, e.g. EXITE2 \citep{Grindlay98}, ALISE \citep{Bazzano91}, AXEL \citep{Sood96}, MARGIE \citep{Cherry95;MARGIE}, LASE \citep{Dsilva98}.
An interesting review on the significant scientific results obtained with balloon experiments was reported by \citet{Teegarden94}.

\subsubsection{Satellite missions}
\label{1990missions}

\begin{itemize}

\item{\bf GRANAT}

Also known as {\bf Astron 2}, {\bf GRANAT} was the result of a collaboration between Russia and European countries. It was launched on December 1, 1989 with a Proton rocket and operated for almost 9 years until November 27, 1998. 

It was designed to study gamma-ray bursts and other transient X-ray sources, and also to image X-ray sources near the Galactic Center.
The most relevant hard X--ray experiments on board were the X--ray/Gamma--ray Imaging Telescope \textbf{SIGMA} (35 keV--1.3~MeV), the X--ray telescopes \textbf{ART-P} (3--60~keV) and \textbf{ART-S} (3--100 keV), the gamma-ray burst monitor \textbf{PHEBUS} (0.075--124~MeV),   the All-sky monitor \textbf{WATCH} (6--180 keV), and the Gamma-ray burst experiment \textbf{KONUS-B} (10~keV--8~MeV).

\textbf{SIGMA} \citep{Gilfanov91,Burenin1999;granat} was the  result of a collaboration between CESR (Toulouse) and CEA (Saclay). It was designed to produce high-resolution images of the hard X--ray/soft gamma--ray sky.
The telescope consisted of a position sensitive detector (PSD) surmounted by a coded aperture mask at 2.5~m distance with 15 arcminutes of angular resolution. 
%
The PSD consisted of a large NaI(Tl) disk (diameter of 57 cm) 
viewed by 61 hexagonal PMTs. 
The total useful area (see Table~\ref{t:sat-missions}) was determined by the 
central rectangular zone of the PSD whose size matched the 
basic mask pattern. 
An in--flight calibrator was made of an \textsuperscript{241}Am radioactive source.
There was also an anticoincidence shield of CsI surrounding the camera, and a thin plastic scintillator located on the top of the PSD to veto the incoming charged particles. 
The  CsI anticoincidence shield was used to detect GRBs \citep{Pelaez1991;granat}.

\textbf{ART-P} (Astrophysical R{\"o}ngten Telescope) 
\citep{Sunyaev1990;granat}, designed by IKI in Moscow, consisted of 4 coaxial, completely independent modules.
Each of the modules included a position-sensitive MWPC surmounted by a coded mask to get an angular resolution of 
5~arcmin and a positional accuracy of 1.5~arcmin. 
The best time resolution was 4 ms.

\textbf{ART-S} (Astrophysical R\"ongten Telescope-Spectrometer) \citep{Sunyaev1990;granat} was also designed by  IKI in Moscow. 
It consisted of 4 detectors based on spectroscopic MWPCs. 

The \textbf{PHEBUS experiment} 
\citep{Talon1993;granat,Vilmer1994;granat}, designed by CESR (Toulouse, France), consisted of 6 independent BGO detectors surrounded by a plastic anticoincidence shield and oriented in such a way to obtain a complete
open field of view. 
Phebus could operate in 2 modes: in the absence of a burst (waiting mode or Normal Mode) detected photons in the 0.1-1.6 MeV energy range with a 64 s time resolution were recorded. If the count-rate exceeded the background level by about 8$\sigma$  the Burst Mode (BM) was activated, and a 31.25 ms time resolution was turned on.

The \textbf{WATCH} (Wide Angle Telescope for Cosmic Hard X-rays)  experiment \citep{Crosby1998;granat}, designed by the Danish Space Research Institute, was composed of 4 detection units mounted in a tetrahedral geometry. The detectors were based on rotation-modulation collimators with the second grid of the collimator replaced by 2 interleaved grids of NaI(Tl) and CsI(Na) detectors, viewed by a single PMT. 
The signals
from the two types of scintillators could be separated electronically, due to the different decay characteristics of the scintillator materials. The modulation grid 
provided a 5.7${^\circ}$ angular resolution. 
%
The instrument could localize bright sources 
within 0.5${^\circ}$. 
During a burst or a transient event, count rates were accumulated with a time resolution of 1 s into 36 energy channels. 

The {\bf KONUS-B experiment} \citep{Golenetskii1991;granat}, designed by the Ioffe Physico-Technical Institute in St. Petersburg to continue their research on 
gamma--ray bursts \citep{Mazets87}, used 7 NaI(Tl) scintillation detectors distributed around the spacecraft.
The lateral surface of the crystals was shielded with 
5 mm thick
Lead. When the counting rate in the 50--200 keV band was rising by  $6\sigma$ over the background level, the energy spectra and time histories were acquired. The first 8 spectra were measured with 1/16 s time resolution while the remaining spectra had adaptive time resolutions depending on the count rate. 
The range of time resolution for the time histories was from 0.25 s to 8 s. 
The instrument operated only from 11 December  1989 to 20 February 1990. Over that period, 60 solar flares and 19 cosmic gamma-ray bursts were detected.

Many relevant results were obtained with the {\em Granat} observatory. Among them, we wish to mention the discovery of new sources, like the X--ray nova GRS~1915+105 with 
WATCH \citep{Castro-Tirado92}, a very deep  image of the Galactic Center region  
\citep[e.g.,][]{Bouchet91, Trudolyubov99} with evidence of a short time (one day) transient bump around 500 keV from the Galactic micro-quasar 1E1740-294, the first discovery of a likely redshifted electron-positron annihilation line (480~keV) still from 1E1740-294 (broad) and from the X-ray Nova in Musca (narrow) (see Fig.~\ref{f:480keVline-Goldwurm92}) \citep{Mandrou94,Goldwurm92,Gilfanov91}, 
the study of spectra and time variability of black hole candidates \citep[e.g.,][]{Grebenev97}, the detection of many GRBs and their spectra \citep[e.g.,][]{Golenetskii91}. The reality of the 480~keV line from Nova Muscae has been contentious, but a new similar line at 511 keV recently observed from the microquasar V404~Cygni (see below) sheds a new light to the line from Nova Muscae.
\\

%
%
\begin{figure*}
\centering\includegraphics [width=0.80\textwidth]{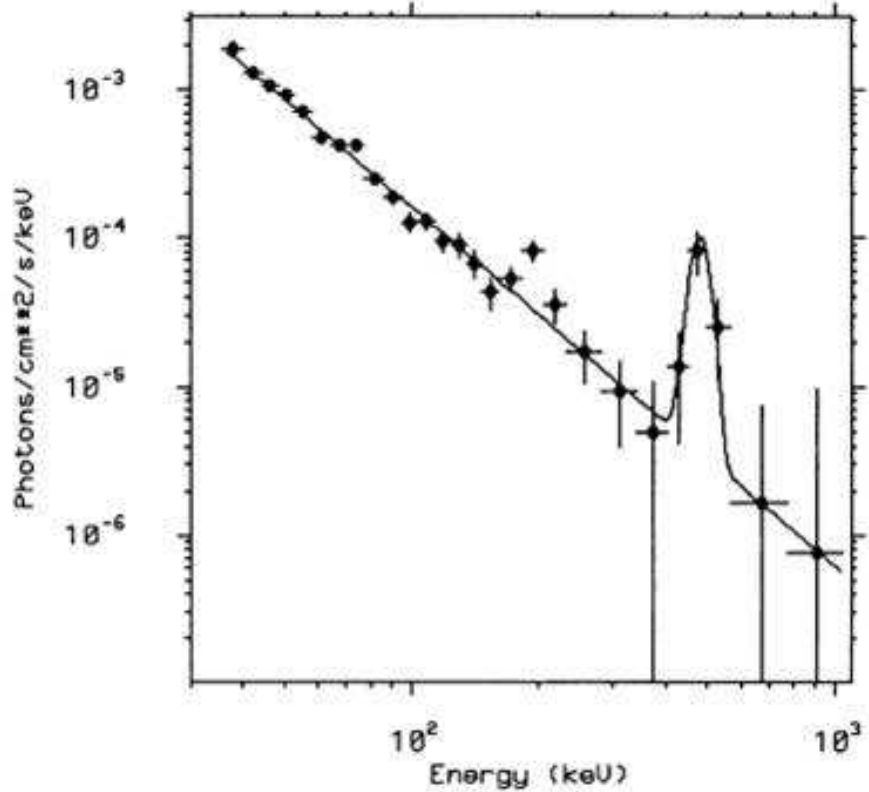}
\caption{First observation of a 480 keV emission line from a point-like source (Nova Muscae) with the SIGMA telescope, interpreted by the observers as a likely  redshifted positron annihilation line. Reprinted from \citet{Goldwurm92}.}
\label{f:480keVline-Goldwurm92}  
\end{figure*}

\item{\bf CGRO}

The {\em Compton} Gamma Ray Observatory ({\bf CGRO}) was launched on April 5, 1991 by the Space Shuttle Atlantis and reentered the Earth atmosphere on June 4, 2000.
The hard X--ray/soft gamma--ray experiments on board were an Oriented Scintillation Spectrometer Experiment (\textbf{OSSE}), a Burst And Transient Source Experiment (\textbf{BATSE}), and a COMPton TELescope (\textbf{COMPTEL}).

\textbf{OSSE} \citep{Johnson1993;cgro} consisted of 4 collimated NaI(Tl)/CsI(Na) phoswich scintillation detectors to provide gamma-ray line and continuum emission detection capability in the 0.05-10 MeV energy range. Each of these detectors could be individually pointed, allowing observations of a gamma-ray source to be alternated with observations of nearby background regions. 
The phoswich was enclosed in an annular shield of NaI(Tl) scintillation crystal in anticoincidence with the gamma-ray interactions in the phoswich. The anticoincidence shield had also the capability to measure the GRB rate. A plastic scintillation detector covered the detector aperture to get
an anticoincidence shield to charged-particles. For most observations, two detectors were pointed at the source while the other two were offset
for simultaneous background measurements. For time-variable phenomena, all four detectors could be pointed at the source for maximizing sensitivity. 

\textbf{BATSE} \citep{Fishman92,Harmon2002;cgro}, which was developed mainly to detect and locate gamma-ray bursts, consisted of 8, completely open, NaI(Tl) Large Area Detectors (LADs) at the
corners of the spacecraft, each sensitive in the 30 keV-2 MeV energy range and with an area of 2025 cm$^2$. 
The GRB histories could be transmitted with different time resolutions down to $\mu$s time scales. For each LAD there was a smaller spectroscopy detector (SD) with a detection area of about 600~cm$^2$ \citep{McNamara95}, optimized for energy resolution and broad energy coverage (10~keV--11~MeV).  Some Earth occultation measurements were performed with the BATSE SDs.

\textbf{COMPTEL} \citep{Schonfelder1993;cgro} explored the 0.75--30 MeV energy range with an angular resolution of 
1--2~deg in its FOV.
It consisted of two detector layers, an upper one of low-Z material (liquid scintillator NE~213A), and a lower one of high-Z material (NaI scintillator),  separated from each other by a distance of 1.5 m. Each detector was entirely surrounded by a thin anticoincidence shield of plastic scintillator to reject charged particles. 
The effective area varied between 10 and 50 cm$^2$ depending on  energy and event selection. 
The continuum and gamma-ray line source sensitivity are shown in \citet{Schonfelder1993;cgro}.

Many significant results were obtained with {\em CGRO}.
Among the many important results obtained with {\bf BATSE}, we wish to mention
the undeniable confirmation of the isotropic distribution of the GRB events (see Fig. ~\ref{f:skydistr-Paciesas99}) \citep[e.g.,][]{Paciesas99}, the first hint of which, as we have seen,  was earlier reported by \citet{Mazets88}. A GRB isotropic distribution was crucial to exclude their origin as being in the disk of our Galaxy, but did not excluded the possibility that they could be either local or have origin in an extended halo of our Galaxy with a typical source distance of about 100~kpc 
\citep[e.g.,][]{Hartmann94}. A significant discovery of BATSE was that of a unique object
GRO~J1744$-$28 called "the Bursting Pulsar" 
\citep{Kouveliotou96}. Broad band spectrum and phase analysis of the source were later measured with OSSE \citep{Strickman96b}.
%
%
\begin{figure*}
\centering\includegraphics [width=0.80\textwidth]{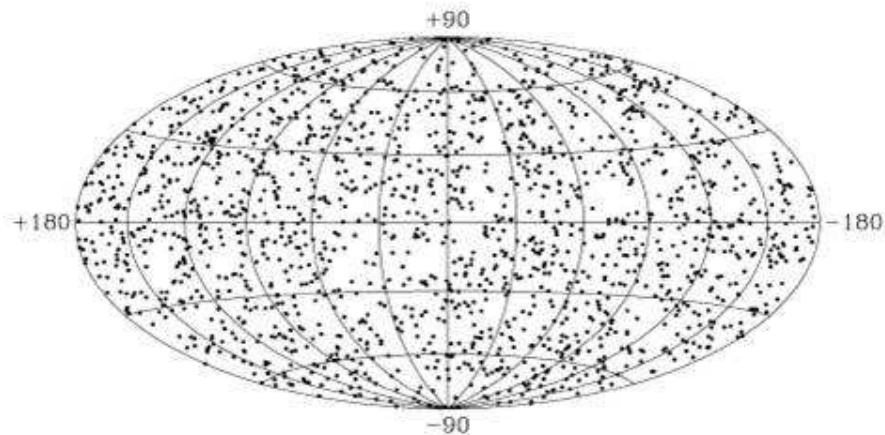}
\caption{Sky distribution in Galactic coordinates of 1637 GRBs detected with BATSE. Reprinted from \citet{Paciesas99}.}
\label{f:skydistr-Paciesas99}  
\end{figure*}

Among the outstanding results obtained with {\bf COMPTEL}, we mention the mapping of the Milky Way using the \textsuperscript{26}Al Gamma-ray line \citep{Diehl94}, the detection of the first soft gamma--ray pulsar PSR~B1509$-$58
\citep{Kuiper99}, and the discovery of only Blazars as AGNs emitters at MeV energies (no Seyfert galaxies), opening the field of gamma--ray astronomy \citep{Collmar99}.  

With {\bf OSSE} numerous results were obtained. We wish to mention the measurement of the 511 keV positron annihilation line from Galactic Center region at a flux level of $(2.5 \pm 0.3)\times 10^{-4}$~photons/(cm$^2$~s) \citep{Purcell93} and its mapping \citep{Purcell97}, and the measurement of the broad band spectrum and time variability from Galactic and extragalactic sources \citep[e.g.][]{Grove96}. Most classes of X--ray sources, compatible with the instrument sensitivity, were investigated,  from X--ray pulsars (e.g., A0535+26 from which an absorption feature at 110 keV was discovered \citep{Grove95}), to supernova remnants (e.g., Vela \citep{DeJager96}), Galactic center sources (e.g., 1E~1740.7$-$2942 \citep{Jung95}), BH candidates \citep[e.g.,][]{Grabelsky95,Phlips96}, radio galaxies \citep{Kinzer95}, Seyfert galaxies \citep[e.g.,][]{Maisack93,Fabian93,Johnson94,Zdziarski95,Bassani95,Johnson97,
Zdziarski00}, starburst galaxies \citep{Bhattacharya94}, blazars \citep[e.g.,][]{McNaron-Brown95}. 
\\
\\
\\

\item{\bf Wind interplanetary mission}

Thanks to an agreement between USA and the Russian Federation, an upgraded version of the {\bf Konus} experiment aboard {\em Venera} 11-14 missions was launched aboard the American {\em Wind} spacecraft, launched in 1994, and still operational. The new Konus consists of two detection units made of NaI(Tl) scintillators completely open apart from a shield on the sides and on the bottom of the detector, similarly to the previous Konus configuration. The two units were oriented along the spin axis of the spacecraft, one looking toward the top and the other toward the bottom, covering in this way the full sky. Currently the energy range is 20~keV to 15 MeV \citep{Aptekar12}.

Several thousands of GRBs have been detected so far, providing for each of them a broad band spectrum. Thanks to its very high distance from the Earth, Konus--Wind provides an almost unbiased rate of GRBs and of other transient events, like SGRs \citep{Aptekar12}. The event localization on the sky can only be obtained by combining the arrival time of the events at Konus--Wind and at other gamma-ray satellites (Interplanetary Network, IPN). Most of the results obtained with Konus--Wind can be found in GCN (GRB Coordinate Network) circulars.
\\

\item{\bf RXTE}

The {\em Rossi} X-ray Timing Explorer ({\bf RXTE}) was launched on December 30, 1995 by a Delta rocket and terminated its operations on January 5, 2012. It was designed to mainly study the temporal and broad-band spectral phenomena associated with stellar and Galactic systems.

The  experiments on board were two narrow field instruments and a wide field detector: the Proportional Counter Array (\textbf{PCA}), the High-Energy X-ray Timing Experiment (\textbf{HEXTE}), and the  All Sky Monitor (\textbf{ASM}). The hard X--ray band was covered by the narrow field instruments: PCA (2--60 keV) and HEXTE (17--240 keV).

The \textbf{PCA} \citep{Bradt1993;rxte}, designed to measure short-term variability down to $\mu$s time scale, consisted of an array of 5 proportional counters. 
Each counter was an extended version of the successful HEAO-1 A2 HED sealed detector. 
There were tubular hexagonal collimators providing 1${^\circ}$ FOV FWHM. The time resolution was 1 $\mu$s. 

\textbf{HEXTE} \citep{Gruber1996;rxte,Rothschild98} consisted of 2 independent clusters of 4 NaI(Tl)/CsI(Na) phoswich scintillation detectors, each viewed by a PMT. 
Around each cluster there were 4 plastic anticoincidence scintillators. 
The 1${^\circ}$ FWHM FOV of each detector was defined by Lead  honeycomb collimators. 
There was a calibration source of \textsuperscript{241}Am. The time resolution was 10 $\mu$s \citep{Bradt1993;rxte}. 

Numerous results were obtained with RXTE, demonstrated by thousands of scientific papers. However, many of them have been obtained from the data in the low energy band of PCA (up to 20-30 keV), like the discovery of the kHz QPOs from Low Mass X--ray Binaries (LMXBs) \citep[e.g.,][]{Vanderklis96,Vanderklis99} that has allowed important astrophysical inferences, like the constraints on mass and radius of neutron stars \citep{Zhang97}. 

Here, we wish to emphasize some results obtained when the 
high--energy data are included, either in the entire PCA passband (2--60 keV) or in the PCA plus HEXTE energy band (2--200 keV). During the contemporary operational life of CGRO, RXTE, \sax\ and INTEGRAL, also coordinated source observations  were performed  (e.g., the transient 198~s X--ray pulsar GRO~J2058+42, \citet{Wilson98}) and very broad band spectra could  be derived (e.g., GRO~J1655$-$40, \citet{Tomsick99}).

The results obtained with only RXTE mainly concern temporal and spectral variability of compact Galactic sources in binary systems (mainly LMXBs), from long times scales, like state transitions of BHCs, e.g., Cyg~X--1, GRO~J1655$-$40, GX~339$-$4 \citep{Belloni96,Mendez98,Smith99}, down to ms time scales or shorter \citep[e.g.,][]{Feng99,Gierlinski03}. Many results concerned the erratic time variability of Black Hole Candidates (BHCs) \citep[e.g.,][]{Lin00}. However, high--energy spectra ($>$40 keV) could not be determined in the case of weak sources \citep[see, e.g., 47~Tucanae,][]{Ferguson99}.
 
In the case of Galactic X--ray sources, several observations concerned the X--ray  pulsars, with the discovery of new cyclotron lines, e.g. EXO~2030$+$375 \citep{Reig99} or their harmonics \citep[e.g. 4U~0115+63, Vela~X--1, GX~301$-$2, MXB~0656$-$072, 4U~1538$-$52,][]{Heindl99,Kreykenbohm99,Kreykenbohm04,McBride06,Rodes-Roca09}.

Many X--ray sources studied with PCA and HEXTE were transient or recurrent transient sources whose outbursts were discovered with the ASM on board. Most of them are compact in binary systems, e.g., XTE~J1550$-$564, whose broad band spectral and temporal behaviour was found similar to that of BHCs \citep{Belloni02}, XTE J1752$-$223  new Galactic BH candidate \citep{Shaposhnikov10}, or the recurrent transient 4U 1608-52, known to be a neutron star in a LMXB \citep{Gierlinski02}. Many Soft X--ray Transients (SXTs) discovered with ASM were followed up with PCA and HEXTE \citep[e.g.,][]{Maccarone03}.

In the case of extragalactic sources, different classes were observed in the hard X--ray band and investigated in their spectrum and, in some cases, in temporal variability,  from Seyfert galaxies \citep[e.g.,][]{McHardy99,Lee99,Benlloch01} to blazars \citep[e.g.][]{Lawson99}, starburst galaxies \citep[e.g.,][]{Gruber99b,Rephaeli02}, galaxy clusters \citep[e.g., Coma Cluster, A754, A3667][]{Rephaeli99,Valinia99,Rephaeli04}. However, only in the case of bright AGNs, spectral shape could be determined up to 100 keV and beyond \citep{Rivers11}. In the other cases, the sensitivity limits prevented to give the high--energy spectral shape \citep[e.g.,][]{Madejski99}. A measurement of the CXB spectrum up to 15 keV was also obtained by \citet{Revnivtsev03}, exploiting three years of RXTE/PCA scanning and slewing observations, and using the Earth--viewing data for the estimate of the instrument background. 
\\

\item{\bf BeppoSAX}

The "Satellite per Astronomia X" ({\bf SAX}) \citep{Boella97}, result of a collaboration between Italy and the Netherlands, was launched on April 30, 1996 by an Atlas-Centaur rocket and operated until April 30, 2002.
After its launch, SAX was renamed {\bf BeppoSAX}, after Giuseppe (diminutive "Beppo") Occhialini. 

The experiments on board included Narrow Field ({\bf NFI}) and Wide Field Instruments ({\bf WFI}). The NFIs were four focusing telescopes, three of which ({\bf MECS}) with a 2--10 keV passband and one ({\bf LECS}) with a 0.1--10 keV passband, a High Pressure Gas Scintillation Proportional Counter (\textbf{HP-GSPC}) with a 4--120 keV passband, and a Phoswich Detection System (\textbf{PDS}) with a 15--200 keV passband. The WFIs were 2 Wide Field Cameras ({\bf WFCs}) with a 2--30 keV passband and a Gamma--Ray Burst Monitor ({\bf GRBM}) with a nominal energy band from 40 to 700 keV. We concentrate on the instruments with passband that covered the hard X--ray band.

\textbf{HP-GSPC} \citep{Manzo1997;bepposax} was filled with a high purity gas mixture of Xenon (90$\%$) and Helium (10$\%$) at 5 atm, with a X--ray entrance window of Beryllium.
On top of the detector there was 
an hexagonal collimator made of Aluminum plus Lead. 
The entire detector was shielded with
Lead plus 
Tin in all directions other than the FOV. Radioactive sources of \textsuperscript{55}Fe and \textsuperscript{109}Cd  were adopted for in--flight calibration. 

\textbf{PDS} \citep{Frontera1997;bepposax}, consisted of a square array of 4 independent and collimated NaI(Tl)/CsI(Na) phoswich scintillation detectors. 
The collimator assembly consisted of two hexagonal X-ray collimators, one per each pair of detectors, that could be independently rocked back and forth to allow the simultaneous monitoring of source and background.
%
Anticoincidence shields of CsI(Na) surrounded the sides, while a thin plastic scintillator covered the X--ray entrance window.
The best time resolution was 16 $\mu$s.  
%
There was a gain control source of \textsuperscript{241}Am and a movable calibrator made of a\textsuperscript{57}Co radioactive source distributed along a line.

The 4 lateral and independent shields of PDS
were also used as GRB monitor ({\bf GRBM}) in the 40-700 keV energy range with temporal resolution down to 0.5 ms \citep{Frontera1997;bepposax,Frontera97b,Costa98}. A trigger system was implemented to identify GRB events. Thanks to the independence of the four scintillators, the instrument was also capable of providing  crude GRB locations (within few degrees) \citep{Pamini90}.  Among other things, this feature was of key importance for deriving the photon spectra of the events.

\sax\ was designed to perform spectroscopic and time variability studies of celestial X-ray sources and to perform a periodic monitoring of Galaxy plane. However, thanks to the GRBM and WFCs,
the GRBM team initiated, since the \sax\ Science Verification Phase, a search for promptly identifying GRBs with GRBM and accurately localizing them (within a few arcmin) with WFCs. This search, as it is well known, resulted to be very fruitful and conducted to the the first discovery, with the MECS telescopes, of the X--ray afterglow of the 1997 February 28 GRB event \citep{Costa97} (see Fig.~\ref{f:970228loc-Frontera98}), and, two months later, to the first determination of the GRB redshift \citep{Metzger97} and thus of the cosmological distance of their progenitors. These discoveries, whose entire story can be found in \citet{Costa11} and \citet{Frontera15}, made \sax\ famous. GRB discoveries with \sax\ continued for its entire operational life time. A summary of the GRB afterglow results can be found in \citet{Frontera03}. A catalog of the 1082 GRBs detected with GRBM  and the spectral determination of the brightest ones were also published \citep{Frontera09,Guidorzi11}.
In addition to the results on GRBs, the GRBM provided other important results. We mention here the spectral and temporal comparative properties of two large flares from the Soft Gamma-ray Repeater SGR 1900+14 occurred
on August 27, 1998 and April 18, 2001 \citep{Guidorzi04}
%
%
\begin{figure*}
\centering\includegraphics [angle=-90,width=0.80\textwidth]{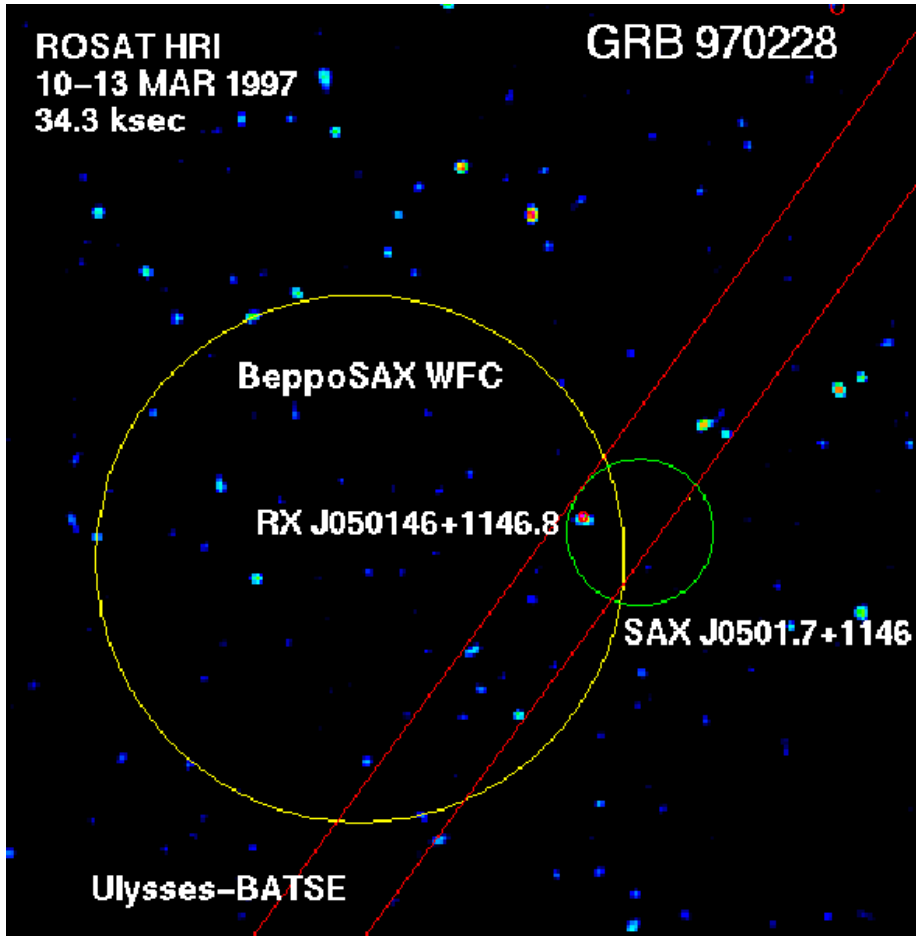}
\caption{Localization improvement ladder of the first afterglow source discovered with \sax. {\em Large circle}: error box of the GRB~970228 event with WFCs, 3~hrs from the event. {\em Red straight lines:} triangulation annulus derived from \sax\ and {\em Ulysses} spacecraft timings \citep{Hurley97}.  {\em Small circle:} error box of the X--ray afterglow source obtained with  MECS 8~hrs from the primary event. {\em Red dot:}  localization uncertainty of the X--ray afterglow source obtained with ROSAT about 12~days from the primary event. The ROSAT position was coincident within 2~arcsec with that of the optical transient associated with GRB970228 \citep{Vanparadijs97}. This result confirmed that X-ray afterglow source and the optical transient were the same object. Reprinted from \citet{Frontera98}.}
\label{f:970228loc-Frontera98}  
\end{figure*}

In addition to GRB discoveries, thanks to the very broad passband of the instrumentation (0.1--200 keV), to the well matched sensitivity over the entire band, and to its almost equatorial orbit (PDS, for example, had the lowest background level among all the already flown high--energy instruments),  \sax\ provided key results also in the observations of Galactic and extragalactic sources, for which it was designed.  For the first time, the hard X--ray band covered by PDS ($>$15 keV) was in numerous cases exploited along with the lower energy band, deriving broad band spectra and time variability properties of Galactic and extragalactic sources.

Concerning Galactic sources, significant spectral results were obtained from different classes of sources, either transient or persistent, from supernova remnants, e.g., Cas~A  \citep{Favata97} from which also the detection of the lines at 67.9 and 78.4 keV associated with the nuclear decay of $^{44}$Ti was obtained \citep{Vink01}, to LMXBs, e.g., the dipping source XB~1916$-$053 \citep{Church98}, the bursting sources 
GS~1826$-$238 \citep{intZand99} and 4U~0614+091 \citep{Piraino99}, the bursting transients 
SAX~J1810.8$-$2609  \citep{Natalucci00} and SAX J1747.0$-$2853  \citep{Natalucci00b}, sources in Globular Clusters, e.g., X~1724$-$308 in Terzan 2 \citep{Guainazzi98}, the  Rapid Burster \citep{Masetti00}; BHCs, e.g., 4U~1630$-$47 \citep{Oosterbroek98}, the superluminal source in outburst GRS~1915+105 \citep{Feroci99}, GX~339$-$4 \citep{Chiappetti99}, GRS~1758$-$258 \citep{Sidoli02}, the transient in outburst 
XTE~J1650$-$500 
(see Fig.~\ref{f:1650-500-Montanari09})  \citep{Montanari09},  the HMXB Cygnus~X--1 in two spectral states \citep{Frontera01}. 

%
%
\begin{figure*}
\centering\includegraphics [angle=-90,width=0.80\textwidth]{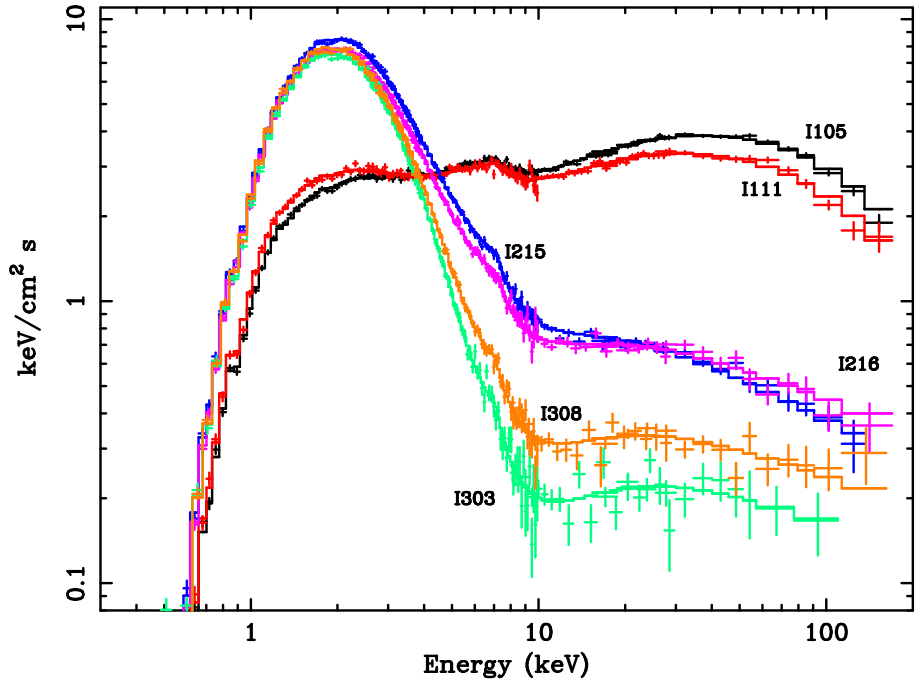}
\caption{An example of broad band spectra from celestial X--ray sources obtained with \sax. The figure shows the spectral evolution of the BHC transient XTE~J1650$-$500 observed with 
\sax\ at different times during an outburst. Reprinted from \citet{Montanari09}.}
\label{f:1650-500-Montanari09}  
\end{figure*}

The detection of cyclotron lines, and, in some cases, of their harmonics were found in several X--ray pulsars, e.g. 4U~1907+09 \citep{Cusumano98}, Vela~X--1 \citep{Orlandini98}, Cen~X--3 \citep{Santangelo98,Burderi00}, 4U~1626$-$67 \citep{Orlandini98b}, X~0115$+$63 from which 4 harmonic features were discovered \citep{Santangelo99}.
A review of the broad--band spectra of accretion--powered X--ray pulsars observed with \sax\ was done by \citet{Dalfiume00}.

Concerning extragalactic sources, significant broad band spectra had been derived from different classes, from BL Lac objects/blazars, e.g. PKS~2155$-$304 \citep{Giommi98}, Mkn~421 \citep{Guainazzi99,Malizia00}, Mkn~501 \citep{Pian98}, NGC~7674 \citep{Malaguti98}, ON~231 \citep{Tagliaferri00}, PKS~0528+134 \citep{Ghisellini99}, to LINER galaxies, e.g. NGC~3998 \citep{Pellegrini00}, to Seyfert 1 galaxies, e.g. NGC~4593 \citep{Guainazzi99b}, NGC~2110 \citep{Malaguti99}, to Seyfert 2 galaxies, e.g. NGC~1068 \citep{Matt97}, NGC~2992 \citep{Gilli00},  MCG~$-$6$-$30$-$15 \citep{Guainazzi99c}, Mkn~3 \citep{Cappi99}, to quasars, e.g. 3C~373 \citep{Grandi97}, the high redshift quasar PKS~2149$-$306 \citep{Elvis00}, to hard X--ray radiation from galaxy clusters, e.g. Coma Cluster \citep{Fusco-Femiano99}, A~2256 \citep{Fusco-Femiano00} and Centaurus Cluster \citep{Molendi02}.

Also, the hard X--ray spectrum up to 60 keV and the absolute intensity of the cosmic X--ray Diffuse Background was investigated with the PDS with significant results \citep{Frontera07}.  

\end{itemize}

\subsubsection {Balloon Experiments}
\label {1990balloons}

\begin{itemize}

\item{\bf GRIS 1992}

We have already discussed the GRIS experiment. A larger configuration  was again launched two times in 
1992 \citep{Leventhal93}. The number of detectors was still 7, but they were  made of the largest available high--purity Ge (detector volume of about 2000~cm$^3$).
The instrument was capable to point to a source with an accuracy of a few tenths of degree.

A full Galactic center transit of 12 hr duration was achieved in both flights. The result was the detection of the electron/positron annihilation  line in both flights, with a 511 keV flux of $(7.7 \pm 1.2) \times 10^{-4}$ and $(8.9 \pm 1.1) \times 10^{-4}$~photons/(cm$^2$~s), respectively. It was the first time that successive high-resolution balloon measurements were achieved on a time scale of days.
\\ 

\item{\bf LXeGRIT}

The Liquid Xenon Gamma-Ray Imaging Telescope (LXeGRIT), a result of a collaboration led by the Columbia University Laboratory, was flown in 1989 and 1999 from Fort Sumner 
\citep{Aprile04}.
The main instrument \citep{Aprile2000;lxegrit,Oberlack2000;lxegrit} consisted in a large-volume liquid Xenon Compton telescope based on full event imaging in a time projection chamber. 
It imaged gamma-rays in the energy range from 200 keV to 25 MeV, with an angular resolution of 3$^\circ$ at 1.8 MeV. 
The flights, part of which was spent to observe the Crab Nebula, were mainly used to estimate the experiment performance (imaging and spectroscopy) \citep{Aprile03,Aprile04}.
\\

\item{\bf Tata Institute experiment}

A balloon experiment successfully flown several times between 1985 and 1992, with the observation of several X--ray Binaries (e.g., 4U~1907$+$09, GX~1$+$4 \citep{Chitnis93}, Cygnus X-3 \citep{Rao91}) was that developed at the Tata Institute for Fundamental Research (TIFR) in Bombay (India) \citep[e.g.,][]{Chitnis93}. The telescope consisted of two collimated 
large--area Xenon--filled multi-anode proportional counters. 
The detection efficiency was $\ge 50$\% between 20 and 80 keV. 

\end{itemize}

\subsection{\bf 2000s missions and balloon experiments}
\label{2000missions}

The satellite missions launched in this decade are characterized by broad energy band and imaging capabilities,  also in hard X--rays (e.g., {\em INTEGRAL}). Two of these missions, {\em HETE 2} and the still operational \swift, were specifically designed for GRB studies. But, also the high--energy gamma--ray missions \fermi\ and {\em AGILE} were designed taking into account specific GRB detections.

Concerning balloon experiments, this decade sees the design and development of hard X--ray polarimeters, e.g., HX-POL \citep{Krawczynski09}, POGOlite \citep{Pearce08}, GRAPE \citep{Bloser06}  and, for the first time, the first test flights of hard X--ray focusing telescopes (see below).

\subsubsection{Satellite missions}

\begin{itemize}

\item{\bf HETE 2}

The High--Energy Transient Explorer 2 (HETE-2) \citep{Ricker03} was launched on October 9, 2000 by a Pegasus rocket and terminated in March 2008. It was designed to detect GRBs, determine their properties with simultaneous observations at soft, medium and high--energy X-rays, and to provide within several seconds their precise localization. The instruments onboard were a Soft X--ray Camera \citep[{\bf SXC}; 0.5--10 keV;][]{Villasenor03}), a Wide-Field X--Ray Monitor \citep[{\bf WXM}; 2–-20~keV;][]{Kawai03} and a FREnch GAmma--ray TElescope \citep[{\bf FREGATE};~6--400~keV][]{Atteia03}. 
We limit our description to FREGATE.

\textbf{FREGATE} was a classical hard X--ray/soft gamma--ray detector with the main task of alerting the other instruments of the occurrence of a GRB. It was made of 4 NaI(Tl) scintillation cleaved crystals. 
The instrument worked well for the entire operational mission life time.

Among the most significant results of HETE 2  there was the discovery of the first GRB (030329) \citep{Vanderspek03} unambiguously associated with a supernova \citep{Stanek03}, confirming the association of the \sax\ GRB~980425 with the supernova SN1998bw \citep{Galama98}. Another outstanding result was the  study of X-Ray Flashes (XRF), a class of events discovered with \sax\ \citep{Heise01}. With HETE~2 it was possible to establish that XRFs and X--ray rich events are less energetic subclasses of GRBs, establishing that all three kinds of bursts (GRB, XRF, XRR) arise from the same phenomenon \citep[e.g.,][]{Sakamoto05}. 
Another important result, soon after confirmed with the 
\swift\ satellite (see below), was the apparent
detection of an extended emission after short GRBs \citep{Villasenor05}, confirming the results found
with BATSE \citep[e.g.,][]{Lazzati01} and  \sax\ GRBM \citep{Montanari05}.
\\

\item {\bf INTEGRAL}

The INTErnational Gamma-Ray Astrophysics Laboratory (INTEGRAL) was launched on October 17, 2002 by a Proton rocket and it is still operational. It was designed to produce a complete map of the sky in the hard X--ray/soft gamma--ray band.

The main experiments on board are a Gamma-ray Spectrometer (\textbf{SPI}) and a Gamma-ray Imager (\textbf{IBIS}).

\textbf{SPI} \citep{Vedrenne2003;integral,Roques2003;integral} consists of an array of 19 n-type Ge cooled detectors with hexagonal shape,   surrounded by an active anticoincidence shield of BGO. The energy passband is 20 keV-8 MeV.
An active cryogenic system guarantees a temperature of the Ge detectors at 85~K. 
A Tungsten coded aperture mask located 1.7 m from the Germanium array provides 
an angular resolution of 2.5${^\circ}$ (FWHM) in its 
FOV. 
The mask geometry is circular with 127 hexagonal pixels, of
which 63 are opaque. 
To further reduce the background, a plastic scintillator is located below the mask. 

\textbf{IBIS} is a coded aperture mask gamma-ray telescope sensitive to the 15 keV-10 MeV energy range 
\citep{Ubertini2003;integral}. The detector assembly has two position-sensitive detection planes: a) a front layer \citep[{\bf ISGRI};][]{Lebrun2003;integral} of CdTe pixels 
with a passband of 15 keV-1 MeV; and b) a back layer 
\citep[{\bf PICsIT};][]{Dicocco2003} of CsI(Tl) pixels with 
a passband of 170 keV-10 MeV. 

The separation between the detecting planes is about 94 mm, so IBIS can also work as a Compton telescope. An anticoincidence shield made of BGO crystals read out by PMTs is around the detector. 
The coded mask
is located 3.2 m above the collimated detection plane  
and  provides an angular resolution of 12~arcmin. 

Many relevant results have already been obtained with INTEGRAL on both Galactic (mainly compact objects) and extragalactic (mainly AGNs) sources. We mention here the most relevant ones. One of them certainly concerns the discovery of highly obscured ($N_H>10^{23}$~cm$^{-2}$) super--giant High Mass X--ray Binaries (sgHMXBs) (see, e.g., the review by \citet{Walter15}). From the same class of sources, INTEGRAL also discovered fast X-ray outbursts lasting less than a day, typically a few hours, now known as Super-giant Fast X--ray Transients (SFXT) \citep{Sguera05,Sguera06}. 

Another very important result obtained with INTEGRAL plus the HEXTE instrument aboard RXTE is  discovery of unexpected hard spectral tails ($>$10 keV) in the total and pulsed spectra of Anomalous X--ray Pulsars (AXPs) and Soft Gamma-ray Repeaters (SGRs) (see, e.g., Fig.~\ref{f:RXSJ1708-4009}) \citep{Kuiper04,Kuiper06,Kuiper08,Gotz06}. Theoretical interpretations have been advanced \citep[e.g.,][]{Beloborodov13}, but, due to sensitivity limitations, the shape of the hard tails is still undefined above 100 keV. 

%
%
\begin{figure*}
\centering\includegraphics [width=0.80\textwidth]{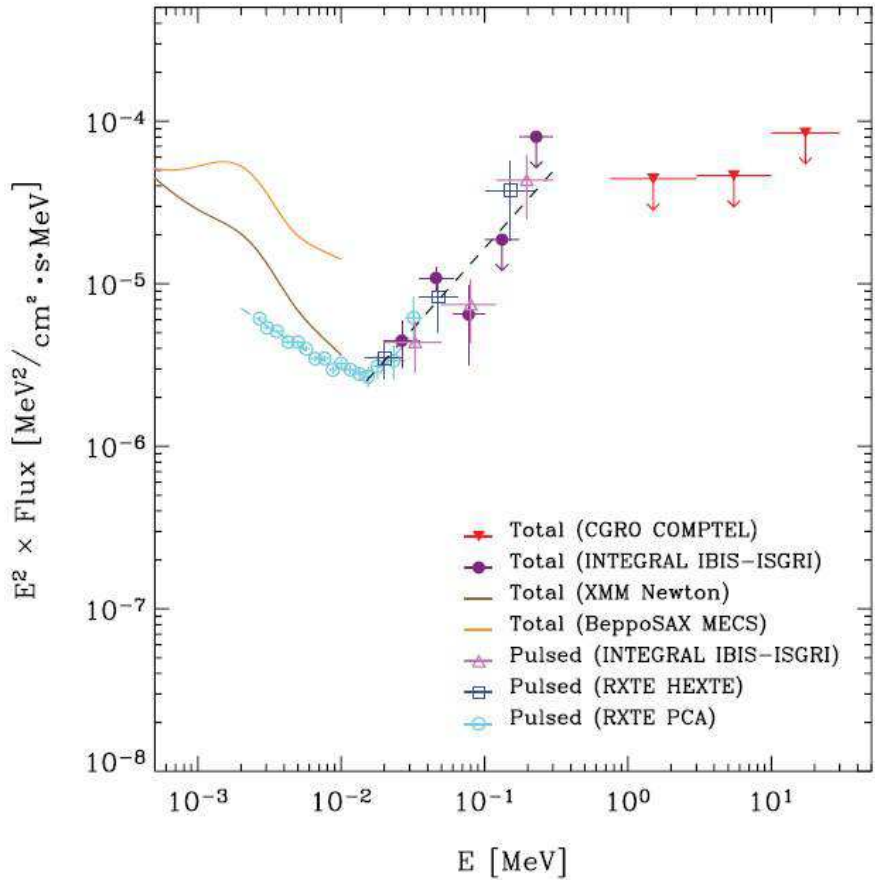}
\caption{Discovery of a high--energy component from Anomalous X--ray Pulsars. The figure shows the hard X--ray energy results obtained with \integral\ ISGRI and \rxte\ HEXTE for the AXP RXS~J1708$-$4009. Note the upper limits above 100 keV up to the energy band covered with \cgro\ COMPTEL.  Reprinted from \citet{Kuiper06}.}
\label{f:RXSJ1708-4009}  
\end{figure*}

Thanks to its wide field of view, its imaging capability and its long operation (more than 10 years), INTEGRAL has allowed the first complete hard X--ray sky survey after that performed with HEAO--1 A4 (see above). The most recent 
survey in the 17--100 keV energy band is that reported by \citet{Bird16}, while at energies beyond 100 keV see \citet{Krivonos15}.

Another outstanding INTEGRAL result is the measurement of polarized radiation. Polarized hard X--ray photons ($>$200~keV) have been measured from the Crab Nebula at an angle parallel to the pulsar rotation axis \citep{Dean08,Forot08}. Strongly polarized radiation (polarization fraction of 67$\pm30$\%) has also been detected in the 400~keV--2~MeV energy  band from Cygnus X--1 (see Fig.~\ref{f:CygX1-polar-Laurent11}), revealing that the MeV emission is probably related to the jet first detected in the radio band \citep{Laurent11,Jourdain12}. Also from several GRBs a significant level of polarized radiation has been detected with INTEGRAL \citep{Gotz09,Gotz13,Gotz14}.
%
%
\begin{figure*}
\centering\includegraphics [width=0.80\textwidth]{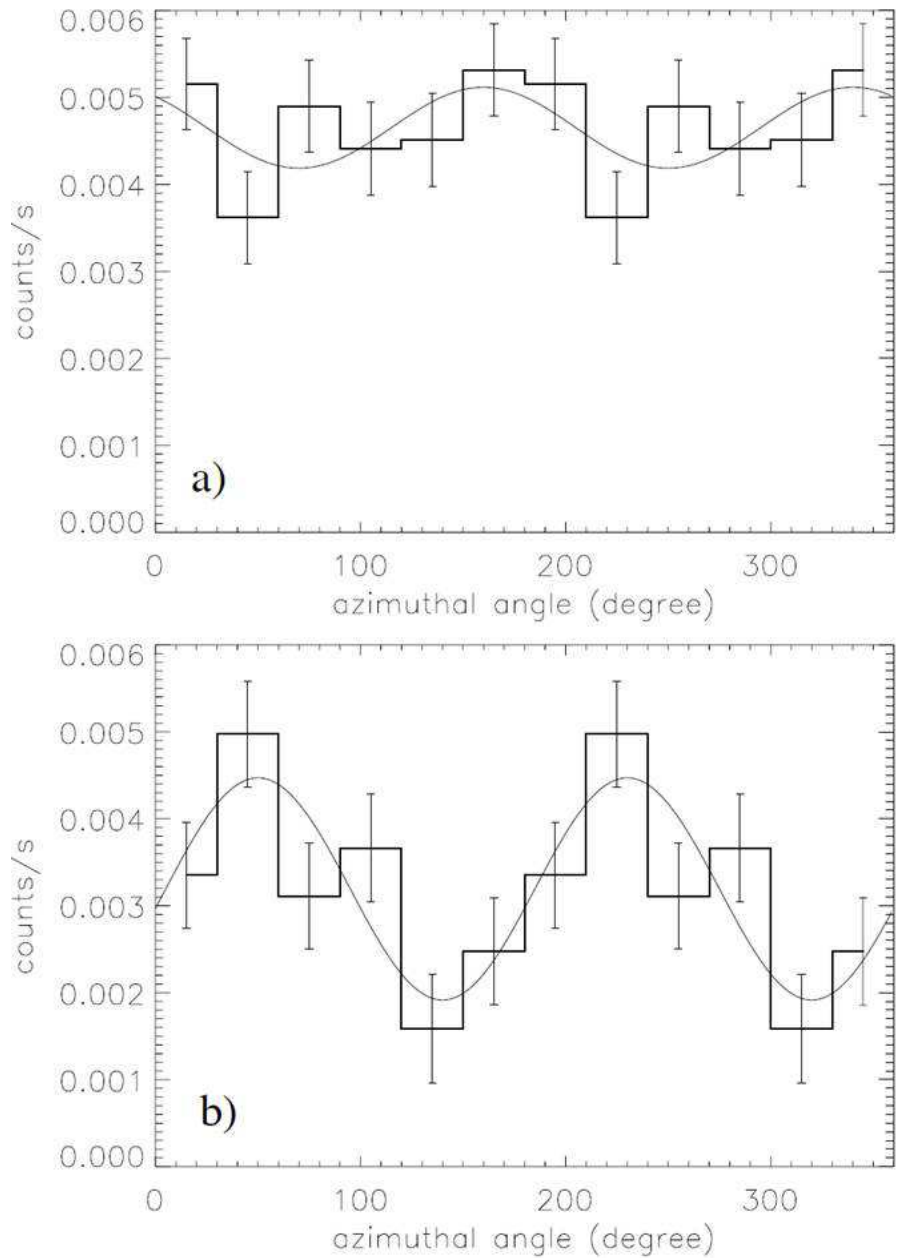}
\caption{Discovery of polarized soft gamma--ray radiation from Cygnus X--1 with \integral. The polarization signal is negligible in the energy in the 250--400 keV energy band (panel a), but it becomes detectable (polarization fraction $= 67 \pm 30$ \%) at higher energies (400--2000 keV; panel b).  Reprinted from \citet{Laurent11b}.}
\label{f:CygX1-polar-Laurent11}  
\end{figure*}

For the first time INTEGRAL allowed to measure the space distribution of the 511 keV positron annihilation line from the Galactic Center region. However, while
  \citet{Weidenspointner08} reported an asymmetric disk emission, \citet{Bouchet09}  showed an emission symmetric with respect to the Galactic Center. 
This result is very important to understand the origin of the Galactic positron population and numerous papers were triggered by these observations. Different potential sources of positrons were suggested, like Sgr A, pulsars, binaries, sources of radioactive elements like $^{26}$Al, $^{56}$Co, $^{44}$Ti, dark matter. 

Detection with INTEGRAL of a 511~keV emission line has also been recently reported by \citet{Siegert16} from the microquasar V404 Cygni. This observation follows the emission line observation around 500 keV from the X--ray Nova in Musca with SIGMA, discussed above  \citep{Goldwurm92}. These results support the conjecture that the diffuse emission of annihilation gamma--rays in the Bulge region of our Galaxy could be due to microquasars. 

Another peculiar result concerns the detection of hard X--ray lines. INTEGRAL detected the lines at 67.9 and 78.4 keV associated with the nuclear decay of $^{44}$Ti from the remnant of the supernova 1987A \citep{Grebenev12}. From the measured fluxes it was possible to establish that this decay provided sufficient energy to power
the remnant at late times. The initial mass of $^{44}$Ti was estimated to be $(3.1 \pm 0.8)\times 10^{-4}$~M$_{\odot}$, which is near the upper bound of theoretical predictions \citep{Grebenev12}.
Another relevant result is the first ever detection of the gamma--ray lines at 847~keV and 1238~keV in the spectrum of the SNIa SN2014J. From the line fluxes it was possible to establish that $0.62\pm 0.13$~M$_\odot$ of radioactive $^{56}$Ni was synthesized during the explosion 
\citep{Churazov14}.
\\

\item{{\bf Swift}}

The \swift\ Gamma-ray burst Explorer was launched on November 20, 2004 by a Delta rocket and it is still operational. It was designed mainly for studying the early afterglow of GRBs. For that purpose it includes a Burst Alert Telescope ({\bf BAT}) for detecting and locating GRBs, an X-ray Telescope ({\bf XRT}, 2--10 keV) and a UV/Visible light telescope 
({\bf UVOT}) for promptly detecting and monitoring their X--ray and optical afterglow. In addition, the GRB coordinates are promptly distributed for giving the opportunity of prompt multi-wavelength observations from the ground. We concentrate on the hard X-ray experiment BAT.

\textbf{BAT} is a hard X--ray telescope with a passband from 15 to 150 keV  that makes use of a coded mask system to locate celestial sources \citep{Barthelmy2005;swift}. In its passband, it achieves 
an angular resolution of 20 arcmin, with a location capability of 1.4 arcmin. The detector is an array of 
Cd(Zn)Te crystals, positioned 1 m below the mask.
The mask is made of 
Lead tiles distributed in a half-filled random pattern. 
To reduce the background a graded shield is located on the side walls between the mask and the detector plane and under the detector plane.  BAT observes 88$\%$ of the sky every day with a detection sensitivity of 5.3 mCrab for a full-day observation \citep{Krimm2013;swift}. 

Among the numerous \swift\ results on GRBs (see, e.g., the review by \citet{Gehrels13}), we wish to mention the detection of the most distant long GRB (090429B, z$\sim$9.4) \citep{Cucchiara11}, the first accurate localization and redshift measurement of a short burst \citep{Gehrels05}, the detection so far of more than 1000 GRBs (1080 GRBs in September 2016), with about 1/3 with known redshifts. 
In addition to the GRB discoveries, BAT has allowed many significant results on other celestial phenomena and sources. Among them, we mention the discovery of Tidal Disruption Flares (TDFs) with relativistic outflow, like that occurred on  2011 March 28 \citep[e.g.][]{Burrows11,Bloom11}. 
BAT has already been a rich source of discovery of new Galactic and
extragalactic sources and has provided light curves for several hundred hard X-ray sources \citep{Krimm2013;swift}. Among the Galactic sources, the most interesting objects discovered and/or monitored and are certainly the several magnetars \citep[see, e.g.,][]{Rea11,Rea12} and SFXTs 
\citep[e.g.][]{Romano14}.
Thanks to the daily survey of 88\%  of the sky, a source catalog  of the entire sky, in particular of the high Galactic latitude sky, has been possible \citep[e.g.,][]{Cusumano10} (see Fig.~\ref{f:Cusumano10}). Also deep surveys, down to $\sim$1mCrab level, have been performed with the goal of establishing the nature of the source population responsible for the CXB at hard X--ray energies \citep[e.g.,][]{Ajello08}. Also, it was possible to derive the CXB spectrum at hard X--ray energies with the BAT \citep{Ajello08b}. 
%
%
\begin{figure*}
\centering\includegraphics [width=0.90\textwidth]{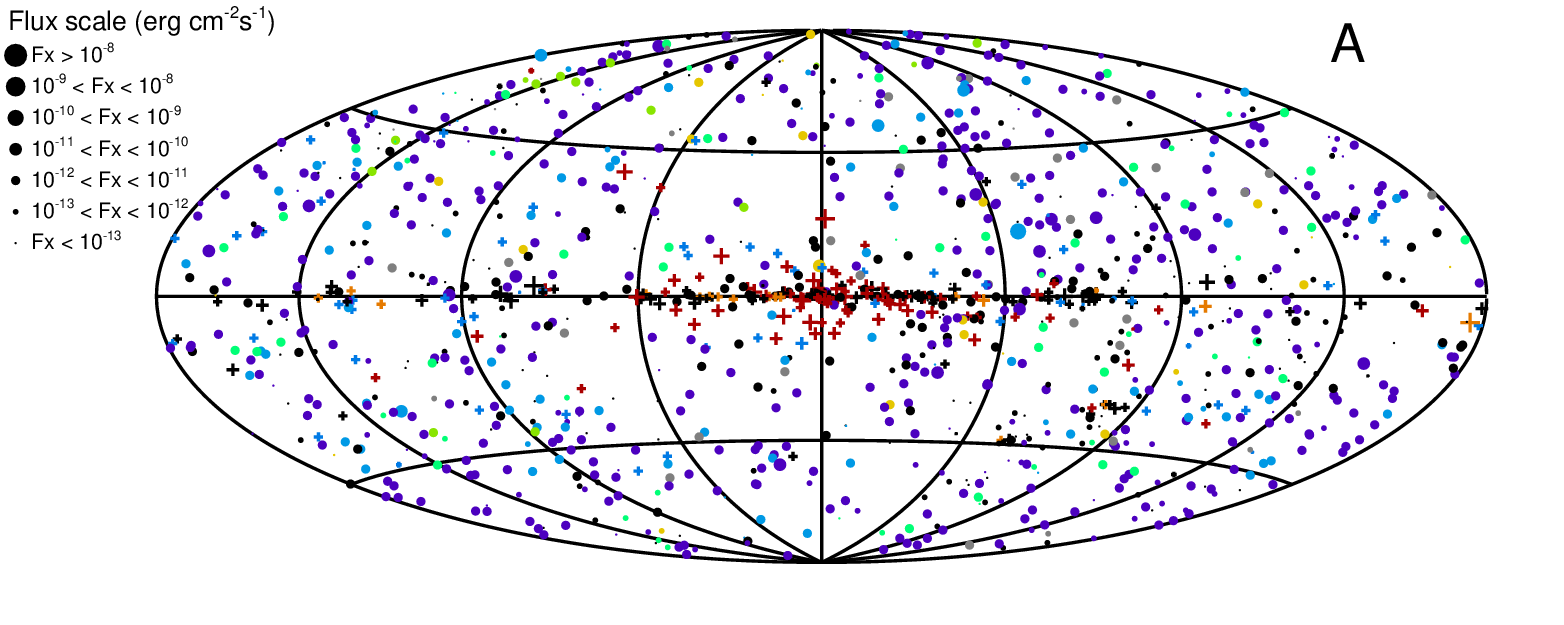}
\includegraphics [width=0.90\textwidth]{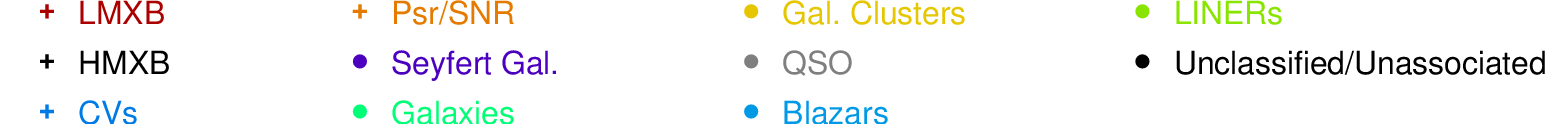}
\caption{The hard X--ray (15--150 keV) sky in Galactic coordinates obtained with the BAT telescope aboard \swift\ down to a flux level of $10^{-13}$~erg~cm$^{-2}$~s$^{-1}$. The size of the symbols is proportional to the flux in 15--150 keV, while the different colours specify the object class according to the legend in the bottom of the figure.  Reprinted from \citet{Cusumano10}.}
\label{f:Cusumano10}  
\end{figure*}
\\

\item{\bf SUZAKU}

Also known as {\bf Astro-E2}, the fifth Japanese X-ray astronomy satellite {\bf SUZAKU} was launched on July 10, 2005 by a M-5 rocket and came to an end on September 2, 2015. As the failed mission Astro-E, it was designed to obtain precise measurements of high-energy processes in stars, supernova remnants, galaxies, clusters of galaxies, and the environments around neutron stars and black holes.

The satellite carried three experiments on board \citep{Inoue03}: an X-ray Imaging Spectrometer ({\bf XIS}) consisting of four imaging CCD cameras sensitive in the 0.2-12.0 keV band, each located at the focal plane of a dedicated X-ray telescope. The second was a non-imaging, collimated Hard X-ray Detector ({\bf HXD}) sensitive in the 10-600 keV band. The third instrument would have been an X-ray micro-calorimeter (X-ray Spectrometer ({\bf XRS}), but it lost all of its cryogen before scientific observations could begin. We concentrate here on the hard X--ray instrument.

\textbf{HXD} \citep{Takahashi07} consisted of an array of 16 (4 $\times$ 4) phoswich counters located under Silicon PIN diodes, which in turn were surrounded by BGO anti-coincidence well-type shields that had also the role of active gross collimators \citep{Takahashi07}. Each phoswich counter was a Gadolinium Silicate crystal (GSO; $Gd_{2}SiO_{5}(Ce)$) optically coupled with a BGO crystal. 
The PIN diodes absorbed X-rays with energies below 70 keV, but gradually became transparent to harder X--rays, which in turn reached and were detected by the GSO detectors. Passive collimators defined the instrument FOV, which was narrow below 100 keV and broader at higher energies (see Table~\ref{t:sat-missions}).
The time resolution was 61 $\mu$s \citep{Mitsuda2007;suzaku}. 

The lateral anticoincidence shield counters of BGO were also used as a Wide-band All-sky Monitor (\textbf{WAM}) to detect GRBs in the 50~keV-5 MeV range. The WAM FOV was about 2$\pi$ with an effective area of 800  cm\textsuperscript{2} at 100 keV and 400  cm\textsuperscript{2} at 1 MeV 
\citep{Mitsuda2007;suzaku}. The WAM had a time resolution of 31.25 ms when it was triggered by a GRB and of 1 s in the waiting time \citep{Mitsuda2007;suzaku}.

Numerous interesting results were obtained with \suzaku. Many of them include data from the {\bf HXD}, providing broad band spectra from 0.7~keV up to hard X--ray energies. They concern Galactic and extragalactic objects. Among the latter, significant results were obtained from AGNs, like the Seyfert 1 galaxy NGC~3516 \citep{Markowitz06} and MCG$-$6$-$30$-$15 \citep{Miniutti07}, the obscured Seyfert 1.9 MCG~$-$5$-$23$-$16 \citep{Reeves07}, the very obscured Seyfert 2 NGC~4945 \citep{Itoh08}, NGC~2273 (first detection of hard X--rays) \citep{Awaki09}, and NGC~4388 \citep{Shirai08},  the quasars PKS~1510$-$089 \citep{Kataoka08}, the TeV blazars
1ES~1101$-$232 and 1ES~1553+113 \citep{Reimer08}, the high redshift ($z= 2.69$) flat-spectrum radio quasar (FSRQ) RBS~315 \citep{Tavecchio07}, the Compton-thick (column density $>10^{24}$~cm$^{-2}$) AGN NGC~1365 \citep{Risaliti09}. 

Also, spectra of galaxy clusters extended up to hard X--ray energies were reported, like that from Ophiuchus, one of the hottest cool core clusters \citep{Fujita08}, that from the merging cluster Abell~3667 \citep{Nakazawa09} and that from Abell cluster A2163, found to be the hottest ($kT\sim14$~keV) \citep{Ota14}.

Significant diffuse hard X--ray emission was detected for the first time from the Galactic Center \citep{Yuasa08}.
Among the broad band spectra of Galactic X--ray sources extending up to hard X--ray energies, we mention the nice results obtained from the black-hole binaries GRO~J1655$-$40 \citep{Takahashi08} and GRS~1915$+$105 \citep{Ueda10}, those obtained from the supernova remant RX~J1713.7$-$3946 \citep{Takahashi08b,Tanaka08}, the discovery of photons up to 70 keV  from a classical nova (V2491~Cygni) \citep{Takei09}, the broad--band spectra from Anomalous X--ray Pulsars (AXP) and Soft Gamma--ray Repeaters (SGR), with the strong  confirmation  of a hard X--ray tail in the persistent spectra from AXP 1E~1547.0$-$5408 \citep{Enoto10} after its discovery with the INTEGRAL satellite \citep{Baldovin09}, the discovery of a hard tail during short bursts from SGR~0501$+$4516 \citep{Nakagawa11}, and the first detection of hard X--rays from an Ultra Luminous Source (ULX) \citep[M82~X--1,][]{Miyawaki09}.  

Also cyclotron resonance scattering features (CRSF) in X--ray pulsars were investigated, like that from A~0535$+$26 during a very low luminosity
state of the source \citep{Terada06}, the discovery of a second harmonic at 72 keV from Her~X--1 \citep{Enoto08}, the discovery, in collaboration with \rxte, of a CRSF at 54 keV from GX~304$-$1 \citep{Yamamoto11}, the first discovery of an absorption feature at 15 keV from a SFXT \citep[AX~J1841.0$-$0536,][]{Kawabata12}, the first firm detection of a CRSF at 76 keV in the X-ray spectrum of the Be X-ray binary pulsar 
GRO~J1008-57 \citep{Yamamoto14}.
  
Also, hard X--ray spectral properties from symbiotic stars have been derived, e.g., SS73~17 \citep{Smith08}. 

Interesting observations concern the identification of the nature of sources found with all--sky surveys performed with \integral\ IBIS  and \swift\ BAT \citep[e.g.,][]{Morris09}. An interesting case is that of XSS~J12270$-$4859 that was initially classified as an Intermediate Polar, while, thanks to the broad band spectra obtained with \suzaku, the source turned out to be a LMXB harbouring a milli-second pulsar, which switches between an accretion-powered state (LMXB) and a rotation-powered pulsar state (a so-called transitional pulsar).  

Also the WAM instrument provided significant hard X-ray spectral results on GRB prompt emission \citep[e.g.,][]{Ohno08}. 
\\

\item{\bf AGILE}

AGILE (Astrorivelatore Gamma a Immagini LEggero) is an Italian mission, launched on 2007 April 23 by a PSLV-CA indian rocket and it is still operational.
The mission is devoted to high--energy gamma-rays ($>$100 MeV), but it includes a hard X-ray monitor, \textbf{SuperAGILE} \citep{Feroci2007;agile}, devoted to monitor the hard  X--ray sky in the 15--45 keV energy band  with an on-axis angular resolution of 6 arcmin.
SuperAGILE is composed by 4 Silicon-microstrip detectors coupled with a set of mutually orthogonal one-dimensional coded masks \citep{Feroci2007;agile}. 
Mask and detector have the same size and are 142 mm far away. The instrument has a source location accuracy of 2-3 arcminutes.
A segmented anti-coincidence system made of plastic shield surrounds all detectors. 

Various results of the SuperAGILE monitor have been reported. We mention here the review of  various classes of sources detected in the first two years \citep{Feroci10}, the one-year continuum monitoring of the X--ray pulsar of GX 301$-$2 
\citep{Evangelista10}.
\\
\\

\item{\bf \fermi\ gamma--ray space telescope}

The {\bf \fermi\ gamma--ray space telescope}, formerly called {\bf GLAST} ({\bf Gamma-ray LArge Space Telescope}), was launched on June 11, 2008 by a Delta II rocket and it is still operational. It was designed mainly to 
survey high--energy gamma--rays (20~MeV--300~GeV) 
from astronomical sources. It also includes a hard X-ray experiment, the Gamma-ray Burst Monitor (\textbf{GBM}), designed to detect GRBs in the 10 keV-30 MeV energy range, with the goal of extending down the energy band covered by the Large Area Telescope ({\bf LAT}) and of computing prompt burst locations onboard to allow re-orienting of the LAT to observe the delayed emission from the located GRBs.

The {\bf GBM} \citep{Meegan09} consists of an array of 12 thin disks of NaI(Tl) oriented in 12 different directions and 2 BGO scintillation detectors. The main features of GBM are given in Table~\ref{t:sat-missions}.
Each NaI crystal is packed in an hermetically sealed Aluminum-housing with a thin window (0.2 mm thick Beryllium). 
The BGOs are mounted on opposite sides. 
The best time resolution is 2 $\mu$s in time-tagged event (TTE) mode.

Many GRBs are being detected with GBM, of which the brightest ones could be crudely localized (within a few degrees), too coarse for follow-up at longer wavelengths. Instead GBM has the advantage, with respect to Swift/BAT, of a broad passband. In addition to GRB catalogs \citep[e.g.][]{Vonkienlin14}, many accurate spectral studies of the GRB prompt emission have been performed \citep[e.g.,][]{Guiriec10,Gruber11,Nava11}, with significant results, like the detection of a blackbody component in the spectra of GRBs  in addition to the non-thermal one \citep[e.g.,][]{Guiriec11}.
Other significant results of GBM concern the spectral and temporal properties of Terrestrial Gamma--ray Flashes (TGFs) \citep[e.g.,][]{Briggs10}.

In addition to the results obtained on GRBs and TGFs, important results obtained with GBM concern the long monitoring and discovery of variable sources like X--ray pulsars and magnetars (AXPs, SGRs). In addition, through the Earth occultation technique it is possible to perform a long duration monitoring of persistent hard X--ray sources. Examples of these results are the discovery of a new SGR \citep{Vanderhorst10}, the discovery of a new torque reversal and spin-up of the X--ray pulsar 4U~1626$-$67 when GBM data are combined with those from Swift/BAT \citep{Camero-Arranz10},  the spectral and temporal studies of SGRs, like SGR~J1550$-$5418 ($=$1E~1547.0$-$5408) \citep{Kaneko10}, the  light-curve of hundreds of  strong X--ray/soft gamma--ray Galactic and extragalactic sources \citep{Wilson-Hodge12}, the hard X--ray study of thermonuclear bursts, e.g. from the NS low-mass X-ray binary 4U 0614+09 \citep{Linares12}. Very important was a 7\% decline of the Crab Nebula flux (nebular component) in the 15--50 keV band, observed over two years, from 2008 to 2010 and confirmed by other instruments \citep{Wilson-Hodge11}.  

\end{itemize}

\subsubsection {Balloon experiments}
\label {2000balloons}

The most significant balloon experiments launched in the 2000s decade are reported below.
\\
\\

\begin{itemize}

\item{{\bf HERO}}

The High-Energy Replicated Optics (HERO) experiment 
\citep{Ramsey2002;hero,Iniewsky2010;hero} was the first experiment flown with focusing optics (FO). It was flown in 
2001 on a stratospheric balloon from Fort Sumner for 17 h. 

It consisted of 2 co-aligned mirror modules, each module containing 3 nested mirror shells, sensitive in the 20-45 keV energy range, providing
an angular resolution better than 45 arcseconds Half-Power Diameter (HPD).
The focal-plane detector was a Gas Scintillation Proportional Counter (GSPC), filled with Xenon-Helium mixture at a total pressure of 10 atm, with 350 $\mu$m spatial resolution. 

The hard X-ray optics were made of full-shell electroformed-nickel-replicated  mirrors (0.25 mm thick) coated with 50 nm Iridium to enhance the high-energy reflectivity. The mirrors were conical approximations to a Wolter I geometry, with a monolithic shell structure containing both parabolic and hyperbolic segments. The mirrors had a 3 m focal length.

The balloon experiment was successful and, for the first time, hard X--ray images of the Crab Nebula plus its pulsar, Cygnus X-1 and 
GRS~1915+105, were obtained \citep{Ramsey2002;hero}. 
\\

\item{\bf CLAIRE}

CLAIRE was an experiment flown on 2000 June 15 and on 2001 June 14 from Gap-Tallard (France), with the primary objective of testing a narrow band gamma--ray lens under space conditions. Scientific results were  obtained in the second flight \citep{Halloin03, vonBallmoos2005;claire}. 

The lens had a diameter of 45 cm (8 crystal rings) and a focal length of 279 cm. It consisted of 556 Ge$_{0.98}$Si$_{0.02}$ flat Germanium-rich mosaic crystals with two different sizes (1$\times$1~cm$^2$ and 1$\times$ 0.7~cm$^2$) disposed on 8 concentric rings. The crystals were properly oriented to focus only photons within a bandwidth of 3 keV centred at 170 keV. The expected focal spot was 1.5 cm diameter. In the focal plane there was a detector made of a 3$\times$3 array of High Purity Germanium (HPGe) crystals 
with 1.5$\times$1.5~cm$^2$ cross section,
cooled by a liquid Nitrogen criostat.
The detector was actively shielded by a CsI(Tl) side shield and BGO collimators. 

The 2001 experiment pointed the Crab nebula for 72 minutes, collecting 33 photons in the 3 keV bandpass of the lens, consistently with those expected
 \citep{vonBallmoos2005;claire}.
\\

\item{\bf InFOC$\mu$S}

Another balloon experiment with a hard X--ray focusing telescope was the International Focusing Optics Collaboration for $\mu$Crab Sensitivity (InFOC$\mu$S), led by Nagoya University in Japan \citep{Tawara2001;infocus}. It was flown from Fort Sumner (USA) in 
2004 for 22.5 hrs.

The hard X-ray telescope consisted of a depth-graded Platinum/Carbon multilayer mirror and a CdZnTe (CZT) focal plane detector. The telescope, using the conical approximation to the Wolter I geometry, consisted of 255 nested mirrors with a focal length of 8 m \citep{Tawara2001;infocus}. It was sensitive in the 20-40 keV energy range, with
an angular resolution of 2.4 arcminutes (HPD).
 The pixellated planar CdZnTe detector was configured with a 12$\times$12 segmented array of detection pixels of 2~mm$^2$ cross section \citep{Baumgartner2003;infocus}. 
The detector was surrounded by a 3 cm thick CsI anticoincidence active shield to reduce background from particles and photons. 

The Cygnus X-1 image and spectrum were obtained in spite of an instability of the pointing system  \citep{Baumgartner2003;infocus}.
\\

\item{\bf HEFT}

Also the High--Energy Focusing Telescope ({\em HEFT}) was a balloon-borne experiment with a hard X--ray focusing telescope on board. It was the result of an international collaboration led by the Columbia University and CalTech. It was flown on May 18, 2005 from Fort Sumner (New Mexico) for a total of 25 hrs at a floating altitude of about 40 km.
It consisted of an array of 3 co-aligned conical-approximation Wolter I mirrors \citep{Koglin2006;heft}, each of which focusing hard X--rays (6--68 keV)  onto a focal plane well shielded, solid-state CZT pixel detector. Each telescope module contained 72 mirror shells, each divided into 5 mirror segments. The mirror modules were made of a glass substrate with depth-graded W/Si and Ni/Si multilayers 
\citep{Harrison2000;heft}. The focal length was of 6~m, and the angular resolution (HPD) in its FOV  was 1 arcmin.

During the flight, successful observations of celestial objects as Crab Nebula, Cyg X-1, Her~X-1, GRS~1915 and X-Persei \citep{Koglin2006;heft}, and the simultaneous observation with Swift of the Blazar 3C454.3, were performed. However, the results were only presented at conferences with no proceedings \citep[e.g.,][]{Chen06}. The successful flight of {\em HEFT} was important for  the design of the {\bf NuSTAR} mission (see below).
\\

\item{\bf ProtoEXIST1}

{\em ProtoEXIST1} was a balloon experiment devoted to qualify the technology required for the High--Energy Telescope (HET) of the Energetic X-ray Imaging Survey Telescope (EXIST) mission proposal. It was launched from the Columbia Scientific Balloon Facility at Ft. Sumner in
2009 
\citep{Hong11}.

It was a wide-field hard X-ray coded-aperture telescope in the 30-600 keV band. The detector plane consisted of an 8$\times$8 array of 
pixellated CZT crystals
mounted on a set of readout electronics boards.
A tungsten mask, mounted at 90 cm above the detector, provided shadowgrams of the X--ray sources with an angular resolution of 20 arcmin in its FCFOV of 
9$^\circ \times 9^\circ$  (see Table~\ref{tab:balloons}). 

In order to reduce the background radiation, the detector was surrounded by 
passive shields on the four sides all the way to the mask. On the back side of the detector, a 
CsI(Na) active anticoincidence shield provided signals to tag charged particle induced events as well as $\ge$100~keV background photons from below. 

During the flight the BHC Cygnus X--1 was observed for 1 hr, obtaining its image and hard X--ray spectrum \citep{Hong11}.

\end{itemize}

\subsection{\bf 2010s satellites and balloon experiments}
\label{2010satellites}

The today effort in hard X--ray/soft gamma--rays is mainly devoted to significantly increase the instrument sensitivity and its angular resolution. This can be obtained by means of focusing techniques. {\bf NuSTAR} is the first launched space mission devoted to this objective although the energy band is limited to 80 keV. A focusing optics telescope sensitive up to 80 keV and a and a non-focusing telescope sensitive up to 600 keV were also part of the {\bf ASTRO--H} payload \citep{Takahashi14}. Unfortunately, after a successful Science Verification Phase, this mission, launched on Feb. 17, 2016 and later renamed to {\bf Hitomi}, was lost on March 29, 2016 at the beginning of the Operational Phase due to a S/W problem. In addition to these focusing missions, an Indian satellite, {\bf ASTROSAT},  with 
non-focusing hard X--ray instrumentation, has been launched, which can be considered as a {\em RXTE} follow-up. In addition to these missions, it merits to mention the very small (3.8 kg) GAmma--ray burst Polarimeter ({\bf GAP}) devoted to measure the polarization of bright GRBs in the 50--300 keV band. GAP has been launched in 2010 aboard the Japanese solar--power sail demonstrator {\bf IKAROS} \citep{Yonetoku11a}, with some significant results already obtained  \citep{Yonetoku11,Yonetoku12}.

In this epoch there are very few balloon experiments. We mention here the {\bf PoGOLite Pathfinder} experiment devoted to test the performance of the {\bf PoGOlite} polarimeter \citep{Chauvin16}.

\subsubsection{Satellite missions}

We discuss  {\em NuSTAR} and the Indian mission {\em ASTROSAT}.

\begin{itemize}

\item{\bf NuSTAR}

The NUclear Spectroscopic Telescope ARray ({\em NuSTAR}) 
\citep{Harrison2013;nustar} was launched on June 13, 2012 by a Pegasus-XL rocket. It is the first satellite mission that makes use of focusing telescopes to image the sky in the 3-79 keV energy range. Based, in large part, on the technologies developed for the HEFT balloon experiment (see above), it is still operational.

It consists of 2 co-aligned grazing incidence hard X-ray telescopes focusing onto position sensitive detectors, with 18 arcseconds FWHM angular resolution. Their focal length is 10.14~m obtained with an extendable mast. Each optic module contains 133 nested multilayer-coated grazing incidence shells in a conical approximation to a Wolter-I geometry. Each shell is coated with depth-graded multilayer structures which increase the grazing angle with a significant reflectivity. 
The inner 89 shells are coated with depth-graded Pt/C multilayers that reflect efficiently below the Pt K--absorption edge at 78.4 keV. The outer 44 shells are coated with depth-graded W/Si multilayers that reflect efficiently below the W K--absorption edge at 69.5 keV. 
Focal plane detectors, surrounded by a CsI anti-coincidence shield, are made of a $2\times 2$ array of CZT counters. 
The best time resolution is $2\mu$s. 

With NuSTAR, for the first time, most classes of Galactic and extragalactic sources become visible in hard X--rays. Among the Galactic sources, we mention the first detection of high-energy X-ray emission from the Galactic center non-thermal filament G359.89$-$0.08 (Sgr A-E) \citep{Zhang14}, the first hard X--ray image of the inner 5'$\times$5' (12~pc$\times$12~pc) of the Galaxy with an angular resolution of 18" \citep{Perez15}, the first detection of hard X--ray emission from the Arches star cluster (11 arcmin from Sgr~A$^*$) \citep{Krivonos14}, the observation of the first energy-dependent hard phase lag (up to 4~s) in the pulse profile of the transient Be/X-ray binary pulsar  ($P\sim 12.29$~s) GS 0834$-$430, never reported before from a high--mass X--ray binary pulsar \citep{Miyasaka13}, the observation at soft and hard X--rays of magnetar sources, e.g. 1E~1841$-$045 \citep{HAn13} confirming the non-thermal component at high-energies but also the need to extend the energy band of the focusing telescopes to higher energies in order to fully test magnetar models \citep[e.g.][]{Beloborodov13}, the discovery of the transient magnetar SGR~J1745$-$29 with a hard spectral tail located at the Galactic centre \citep{Mori13}, the first hard X--ray image of the Pulsar Wind Nebula (PWN) MSH~15$-$52 and its total spectrum up to 20 keV \citep{An14}, the first sensitive hard X--ray spectrum of a neutron star transient in quiescence, Cen~X--4, finding, unexpectedly, that the hard X--ray component is consistent with a thermal bremsstrahlung model with $kT_e = 18$~keV \citep{Chakrabarty14}, the first hard X-ray detection of transitional pulsars, like PSR~J1023$+$0038, which switch between an accretion--powered state (LMXB) and a rotation--powered state (radio ms-pulsar) \citep{Li14}.

Concerning extragalactic sources, we mention the first detection
of a SN (SN~2010jl) outside the Local Group in the hard X-ray band \citep{Ofek14}, the first detection of a young extragalactic stripped-envelope SN (2014C) out to 40 keV \citep{Margutti16}, the hard X--ray spectral detection of Ultraluminous X--ray sources, e.g. ULX~5 in the Circinus galaxy \citep{Walton13} and ULX~2 in M82 from which a 1.37~s pulsation was discovered, implying a magnetized neutron star as emitting source \citep{Bachetti14}, the capability to study extremely active
supermassive black holes located at very high redshift, like the blazar B2 1023+25 ($z = 5.3$) \citep{Sbarrato13}, the best signal-to-noise ratio obtained to date from active galactic nuclei over the 3–-79 keV band, as in the case of the Seyfert 1 galaxy IC~4329A, whose high--energy cutoff \cite[$E_{cut} = 178^{+74}_{-40}$~keV,][]{Brenneman14}) could be determined, the inference of a maximally rotating supermassive black hole (NGC~4051) from the spectral property of hard X--ray radiation \citep{Risaliti14}.

Also deep extragalactic surveys have been performed at hard X--ray energies to characterize the source population that contributes to the high--energy CXB \citep[e.g.,][]{Alexander13,Civano15}. 

Another very important result is the detection  of the hard X--ray afterglow (up to 79 keV) (see Fig.~\ref{f:130427A-Kouv13}) from a very bright GRB (130427A) \citep{Kouveliotou13}, after the first detection of a X--ray afterglow (from GRB~990123) obtained with the \sax\ PDS instrument \citep{Maiorano05}. These hard X--ray afterglow observations are  crucial to establish the emission mechanism of the GRB afterglow and characterize the circumburst environment. They should be extended to  higher energies. 
%
%
\begin{figure*}
\centering\includegraphics [width=0.80\textwidth]{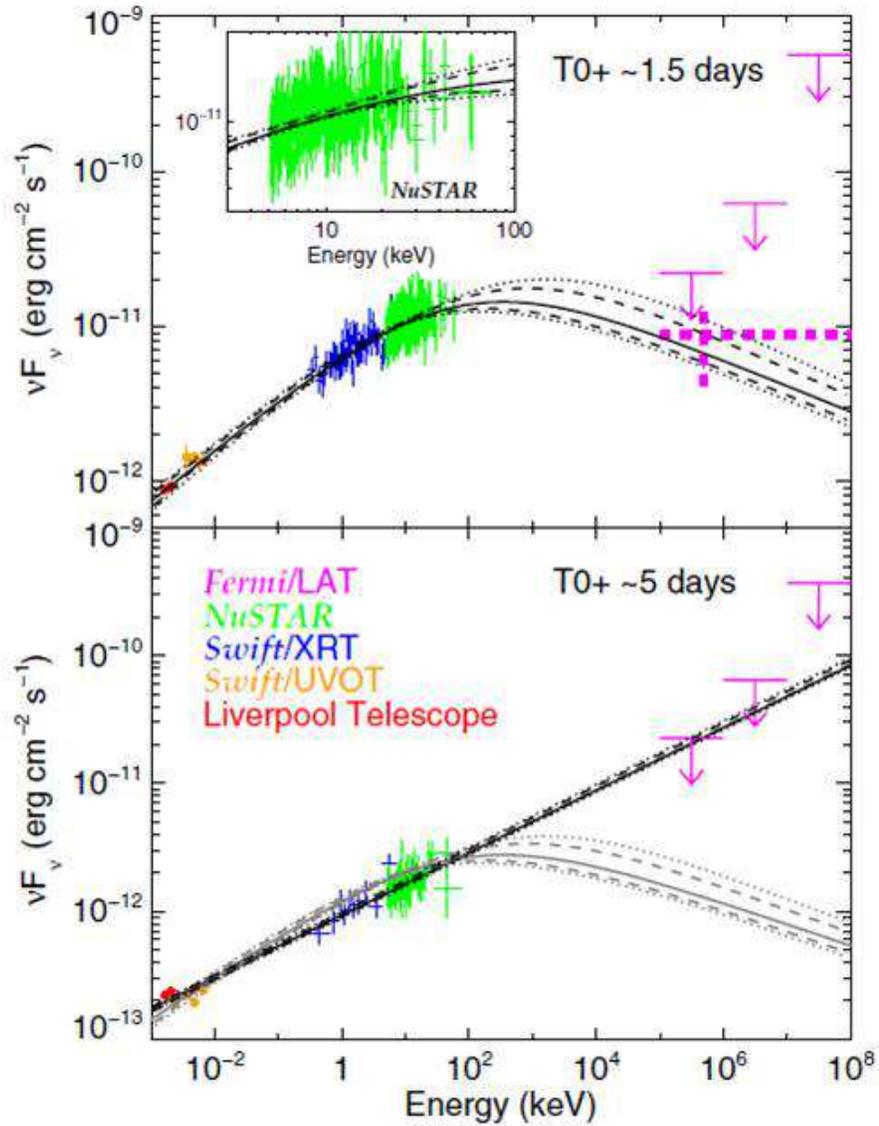}
\caption{Afterglow spectrum of GRB~130427A measured up to 79 keV with NuSTAR, after 1.5 and 5 days after the event. The extension of the measurements at high energies is crucial to establish the emission mechanism of the afterglow radiation. Reprinted from \citet{Kouveliotou13}.}
\label{f:130427A-Kouv13}  
\end{figure*}

Several results also concern CRSFs in X--ray pulsars. Some of them concern the discovery of further properties of the CRSFs  already known, e.g. Her~X--1 \citep{Furst13} and Vela~X--1 \citep{Furst14a}. Another significant result concerns the confirmation of a line feature at 78 keV from the high-mass X-ray binary GRO J1008$-$57 \citep{Bellm14}. Also new CRSFs have been  discovered, like a variable line at 12.5 keV from the transient Be--neutron star binary KS~1947$+$300 
\citep{Furst14b}, and a CRSF at 31.3 keV from the ROSAT X--ray pulsar RX~J05205$-$6932 during a high luminosity ($3.6\times10^{38}$~erg~s$^{-1}$) outburst \citep{Tendulkar14}. Important is the first measurement of a cyclotron line at 17 keV from a SFXT (IGR~J17544$-$2619), from which the surface magnetic field could be derived ($B = (1.45 \pm 0.03)\times 10^{12} (1+z)$~G, $z$ being the gravitational redshift), demonstrating the NS nature of these sources \citep{Bhalerao15}.
\\

\item{\bf ASTROSAT}

{\bf ASTROSAT} is the first indian satellite devoted to multiwavelength observations from space. It was launched by the Indian launch vehicle PSLV from the Satish Dhawan Space Centre, Sriharikota, on 2015 September 28. 

The five instruments on board \citep{Singh14} cover the visible, near UV, far UV, soft X-ray band and hard X-ray (3–-80 and 10–-150 keV) band. In the 3--80 keV passband, {\em ASTROSAT} makes use of a cluster of 3 co-aligned identical Large Area X-ray Proportional Counters ({\bf LAXPCs}), each with a multi-wire-multi-layer configuration and a FOV, obtained through mechanical collimators, of $1^{\circ}\times 1^{\circ}$ FWHM \citep{Yadav16a}.

In the 10--150 keV passband, {\em ASTROSAT} uses an imaging telescope ({\bf CZTI}) with
an angular resolution of 8 arcmin. It consists of a pixellated CZT detector array surmounted by a 
two--dimensional coded mask. 
CZTI is expected to provide measurements of the X-ray polarization level of bright X-ray sources in the energy range of 100-300 keV \citep{Vadawale15} to constrain, with an exposure time  of 100 ks,  any intrinsic polarization greater than $\sim$40\% ($>$500 mCrab) at 
$3\sigma$ confidence level.
 
At the time of this paper, in addition to GCNs on the detection of GRBs \cite[e.g.,][]{Vadawale16}, scientific results are being published. We wish to mention
the polarization study of the first GRB (151006A) observed with {\em ASTROSAT} \citep{Rao16a}, and the observation of the micro--quasar GRS~1915$+$105, with the study of the its power spectrum with energy and  2--8 Hz QPO \citep{Yadav16}. A review of the first results obtained with {\em ASTROSAT} is given by \citet{Rao16b}. 

\end{itemize}

\section{Achievements, open problems and future prospects}
\label{comparison future}

\begin{figure*}
\centering\includegraphics [width=0.90\textwidth]{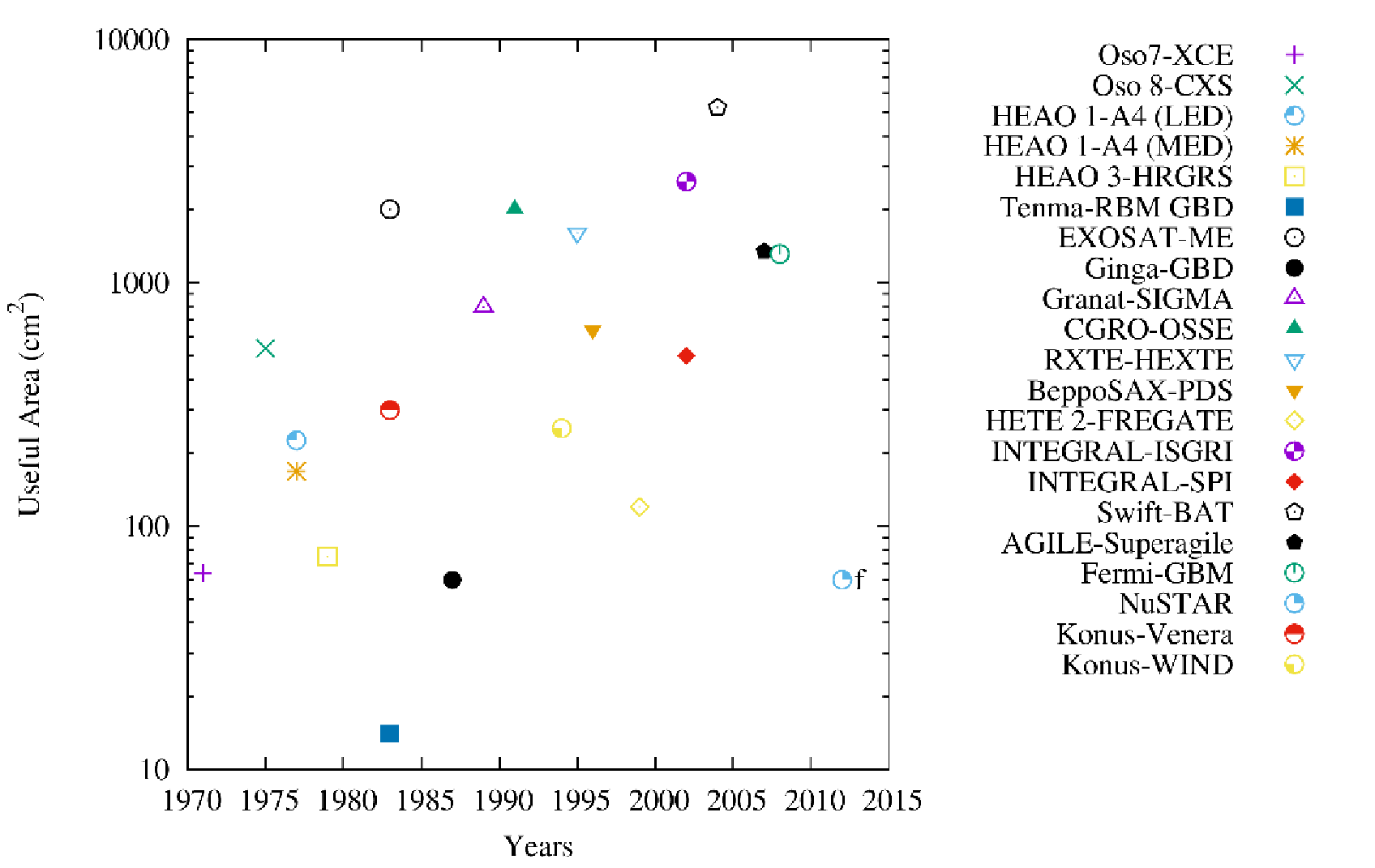}
\includegraphics [width=0.90\textwidth]{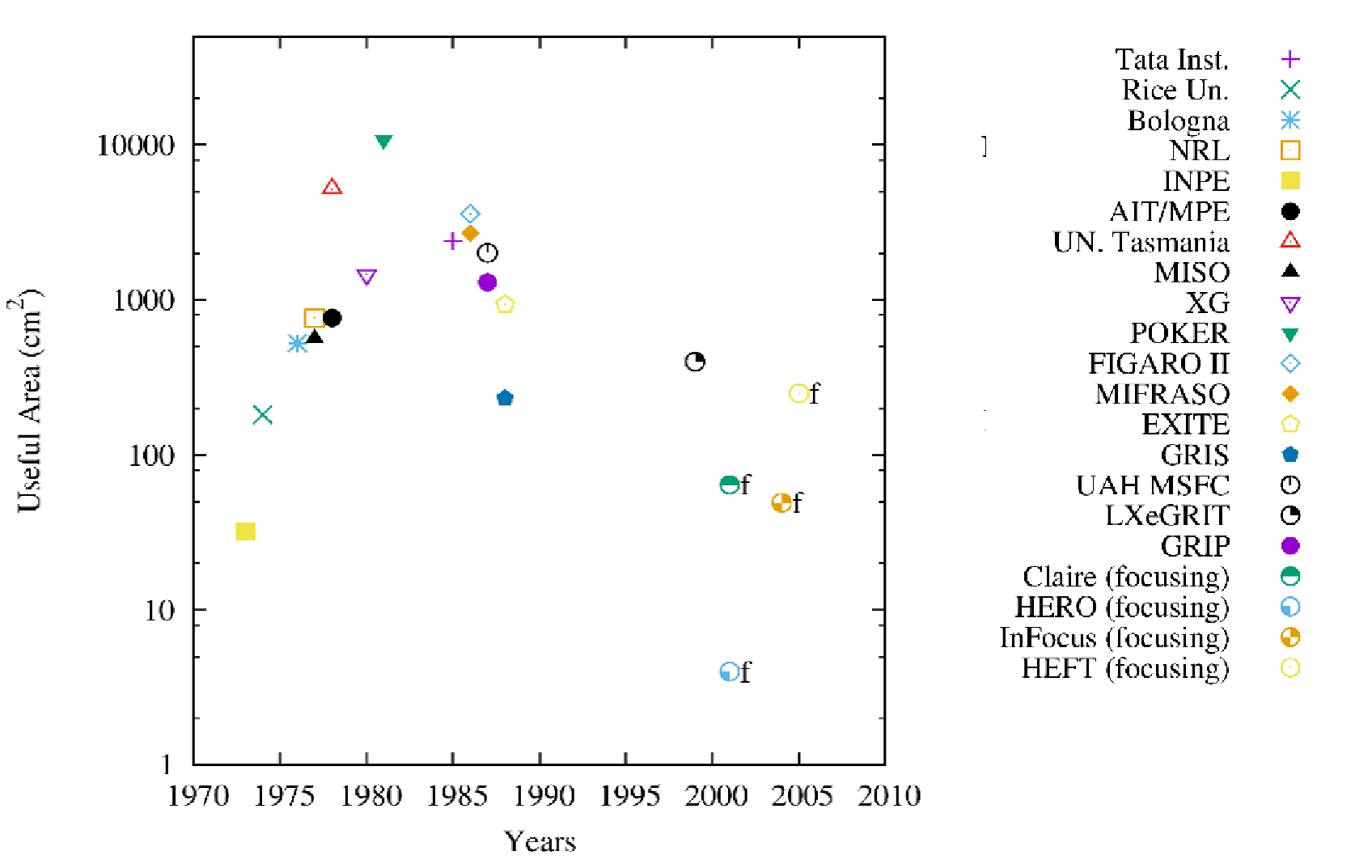}
\caption{Trend with time of the useful area, defined as the geometrical area through the collimator or coded mask, of the hard X--ray non-focusing instruments. In the case of focusing instruments (marked with "f"), it is shown the trend with time of the effective area. {\em Top panel}: Satellite instruments. Data are taken from Table~\ref{t:sat-missions}. {\em Bottom panel}: balloon instruments. Data are taken from 
Table~\ref{tab:balloons}.}
\label{f:useful-area}  
\end{figure*}

\begin{figure*}
\centering\includegraphics [width=0.80\textwidth]{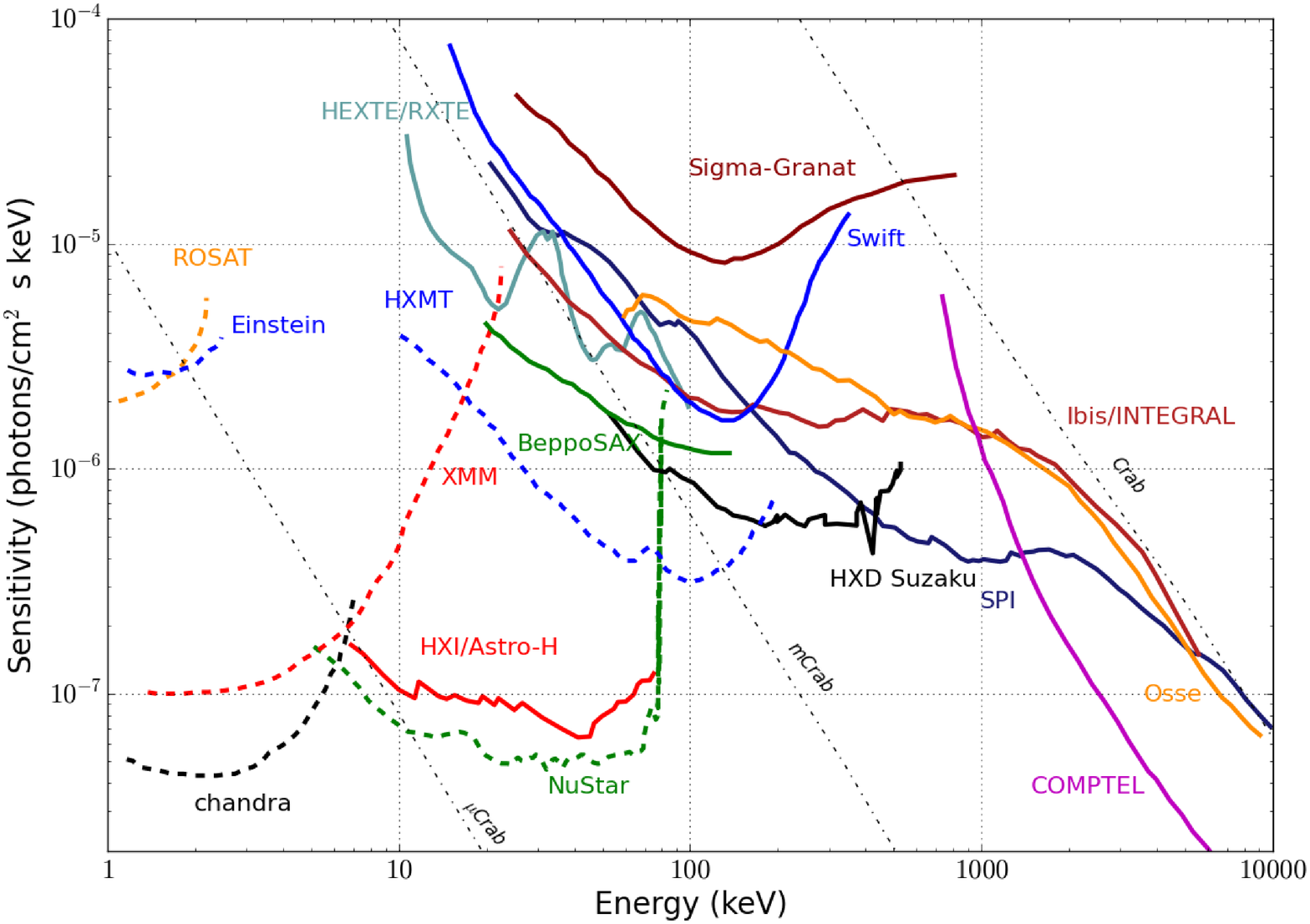}
\caption{Three sigma sensitivity, for an exposure time $T = 10^5$~s and a bandwidth  $\Delta E = E/2$, of the most relevant missions in the hard X--ray/soft gamma--ray band. Adapted from \citet{Virgilli17}.}
\label{f:sensitivity}  
\end{figure*}

From the review of the experiments discussed above, it appears clear that hard X--ray astronomy has reached a very advanced stage of development. Since the beginning 
(see Section~\ref{s:birth}), its key role has been the determination of the emission mechanisms at work in X--ray celestial sources. In the 1970s/1980s, this was possible only for the brightest sources. With the increase of the experiment useful or effective  area (see Fig.~\ref{f:useful-area}) and thus of sensitivity (see Fig.~\ref{f:sensitivity}), the emission mechanisms of weaker sources could be investigated. Thanks to satellites, like \sax, \rxte\ and \integral, we have seen that, in order to investigate the emission mechanisms at play in X--ray sources, the measurement of broad--band spectra from soft X--rays to hard X--rays/soft gamma--rays, is crucial. This broad--band study should be now extended to several hundreds of sources, Galactic and extragalactic, of different classes, that are known to be hard X--ray emitters also on the basis of the INTEGRAL/IBIS and Swift/BAT sky surveys discussed above. To get broad band spectra up to hard X--ray/soft gamma--ray energies  for the weakest sources, direct--viewing telescopes, in spite of their very exciting results obtained so far,  are inadequate owing to their limited sensitivity (see Fig.~\ref{f:sensitivity}). The solution is the use of broad band X--ray focusing telescopes. At low X--ray energies ($\le$10 keV), it is already many years that focusing is a mature technology, while at hard X--ray energies up to about 80 keV the technology has become mature only recenty, as brilliantly demonstrated by the \nustar\ mission. However, even the \nustar\ results have demonstrated that an extension of the focusing to higher energies and a better sensitivity in the upper part of the \nustar\ telescope passband 
($>$30-40 keV) are a must.

%
%
%

%
%

Laue lenses offer a viable  solution to the implementation of a focusing telescope up to  600 keV and beyond \citep[see, e.g.,][]{Frontera2013;laue}. The Laue lens development is in an advanced stage and Laue lenses are expected to achieve sensitivities 2--3 orders of magnitude better than the \integral\ sensitivity at the same energies  and a much better angular resolution ($\sim 20$~arcsec) \citep{Virgilli17}.

In the mean time, a Chinese mission, the Hard X-ray Modulation Telescope (\textbf{HXMT}) \citep{li07, ZhangS14}, just launched (June 15, 2017) and renamed {\bf Insight}, could do an important job for broad--band studies. Indeed it includes a low--energy ({\bf LE}; 1--15 keV), a medium--energy ({\bf ME}; 5--30~keV) and a high--energy ({\bf HE}; 20--250~keV) instrument. The {\em HE} telescope consists of 18 NaI(Tl)/CsI(Na) phoswich crystals, with a design similar to that of the BeppoSAX/PDS, but with a much larger total geometric area (about 5000 cm$^2$). Thanks to the collimator configuration, {\bf HE} is also capable to perform, with an angular  resolution of about 5 arcmin, imaging of the sky region in its FOV ($5.7^\circ \times 5.7^\circ$ FWHM). The telescope is expected to achieve a continuum sensitivity of about
3$\times 10^{-7}$~ph~cm$^{-2}$~s$^{-1}$~keV$^{-1}$ at 100 keV (3$\sigma$, 100~ks).  
Main goals of HXMT include a sensitive monitoring of the Galaxy plane and broad band (1--250~keV) pointed observations  of peculiar objects.

Almost contemporarily, a Gamma-Ray Burst polarimeter 
{\bf POLAR} was launched on September 15, 2016 aboard the Chinese space laboratory Tiangong-2 (TG-2), and is still in the Science Verification Phase, for polarization measurements of the GRB prompt emission in the 50--500 keV band \citep{Kole16}. 

Also devoted to GRB studies in a broad passband is the {\bf SVOM} Sino--French mission, whose launch is foreseen in 2021. It includes a coded-mask telescope operating in the 4-150 keV energy range for real-time GRB detection and prompt localization, and a non-imaging gamma-ray monitor for GRB spectroscopy from 15 keV to 5~MeV \citep{Wei16}.

No other hard X--ray instruments are at this time approved for satellite missions. We expect that, once the technological developments are accomplished,  future missions with broad band optics, hopefully inclusive of Laue lenses, can eventually  perform not only deep studies of the many hard X--ray sources discovered with INTEGRAL and  Swift/BAT  and of those expected to be discovered  in the next years with HXMT, but also solve fundamental issues still open in astrophysics. 

Among them, we wish to mention the origin of the 511 keV annihilation line from the Galactic Center region.
We have seen that the presence of this line, initially discovered with balloon experiments \citep{Leventhal78}, is now well established. The monitoring of the line with balloon experiments found it variable with  upper limits to its intensity \citep[e.g.,][]{Leventhal86} in the same years in which the {\em SMM} Gamma--Ray Spectrometer with a broader FOV detected it \citep{Share90}. These seemingly contradictory results  opened the possibility that this line was partly due to a diffuse emission, and partly to point--like sources \citep{Lingenfelter89}. INTEGRAL measured the space distribution of the emission, but, due to the still low angular resolution (12 arcmin), has left the origin of this radiation unsolved, whether it is the result of a superposition of point--like sources, or it is intrinsically diffuse or a combination of both types.
Only a focusing telescope with much higher sensitivity than INTEGRAL and with a much better angular resolution could settle the issue, and establish the nature of the emission. An exciting possibility is that this line is due to dark matter annihilation.
  
Related to the previous issue, it is of great relevance to confirm the result found with INTEGRAL in the case of V404 Cygni \citep{Siegert16}, and, previously, with SIGMA aboard GRANAT in the case of Nova Muscae \citep{Goldwurm92}, that the 511~keV positron annihilation line  has origin in microquasars. This can be done by studying a significant sample of  microquasars with much higher sensitivity.

Another still open issue is the systematic study of explosive events, like thermonuclear or core--collapse Supernovae, through the detection of nuclear lines produced in the explosion, in particular the 158 keV line from $^{56}$Ni, and the 67.9 and 78.4 keV lines from $^{44}$Ti.

Another relevant open issue is the determination of the still mysterious shape of the high--energy spectra of anomalous X--ray pulsars and SGRs. We have seen that a hard tail from these sources has been determined with INTEGRAL plus RXTE up to 100 keV. However, in order to establish its origin
 \citep[e.g.,][]{Beloborodov13}, the detection should be extended beyond this energy, where now only upper limits can be given \citep{Kuiper06,Kuiper08,Gotz06}.
   
In spite of their importance for establishing the physics and their contribution to the high--energy Cosmic X--ray Background, the high--energy spectra of AGNs are still poorly known beyond 100 keV. This knowledge is important to determine their cutoff energies and, in the case of Blazars, the determination of their spectra in the critical band around 400 keV where the spectra are still undetected.

As we have seen, GRBs have been discovered in the hard X--ray band. This band, extended to the MeV region from one side, and to low energy X--rays from the other side,  is crucial not only to investigate the still debated physics of the GRB prompt emission but also its afterglow. The few detections of hard X--ray afterglow emission available so far (one with BeppSAX \citep{Maiorano05} another with NuSTAR \citep{Kouveliotou13}) have shown the importance of the high--energy spectra in order to establish the afterglow emission mechanisms. Unfortunately, these observations are very rare due to the limitations of the instruments and the parallel need of prompt follow up of promptly localized GRBs. Also in this case, much more sensitive hard X--ray/soft gamma--ray instruments are requested for the study of the GRB afterglow.   

Another important issue is the polarization level of the 
high--energy emission from different classes of sources. Thus far we have very few polarization results: those obtained with the Gamma--Ray Burst Polarimeter (GAP) aboard the solar-power sail demonstrator IKAROS 
\citep{Yonetoku11, Yonetoku12}, and those obtained with INTEGRAL from Crab \citep{Dean08}, Cygnus X--1 \citep{Jourdain12} and some GRBs \citep{Gotz09,Gotz13,Gotz14} we have  discussed above. These results have shown the importance of the polarization measurements for establishing the emission mechanisms. A much  better sensitivity is requested to extend this search to different classes of sources, taking into account that the polarization properties can change with the time scale of the X--ray source variability and destroy the polarization level if the observation time requested to achieve the needed sensitivity is longer. 

We expect that only focusing telescopes extending the band at least up to 600 keV can solve the above and many other open issues. For the first time a new window will be open beyond 100 keV ushering in a large discovery space and making a great breakthrough in the knowledge of the high--energy Universe.

In addition to the studies of peculiar objects or sky regions, also a sensitive soft gamma--ray monitoring of wide regions of the sky is important in a band, beyond 100 keV,  yet scarcely explored. This could allow, among others, to discover new persistent and transient gamma--ray sources, inclusive of supernovae explosions, to study GRBs and other transient phenomena, like Tidal Disruption Events, Soft Gamma--Ray Repeaters, electromagnetic counterparts of Gravitational Wave Events, possible high--energy emission associated to Fast Radio Bursts \citep{DeLaunay16}, to study their polarization properties, and to discover new diffuse line emission of nuclear origin.  
To do that, a new generation of telescopes is required with a significantly better sensitivity than that of the INTEGRAL mission at the same energies. A possibility that could have been offered by the wide field (90$^\circ \times 70^\circ$) 
High--Energy  Telescope (5--600~keV) proposed for the EXIST mission \citep[e.g.,][]{Grindlay10}, unfortunately not approved by NASA.  Another possibility is the Pair And Compton Telescope (PACT; 100 keV--100 MeV) \citep{Laurent14}, with a wide FOV ($>$3~sr),  but with a still poorer angular resolution (of the order of 1 deg below 10 MeV). The development status of this instrument is advanced. 

The combination of both wide field hard X--ray/soft gamma--ray instruments and broad--band narrow field telescopes up to hard X--rays/soft gamma--rays on board of one or more satellites could be a winning strategy of the next step of the 
hard X--ray/soft gamma--ray astronomy for an unprecedented study of the high--energy Universe.

\begin{acknowledgements}
First of all we wish to thank the referee, Dr. Lucien Kuiper, for his careful reading of the manuscipt and his very precious comments and suggestions. We would like to thank also Lorenzo Amati, Mauro Orlandini and John B. Stephen from INAF, IASF Bologna, Cristiano Guidorzi and Piero Rosati from University of Ferrara,  for their useful comments. Last, but not least, we wish to thank Enrico Virgilli from University of Ferrara for his help in the figure preparation.
\end{acknowledgements}

\newpage


\begin{longtable}{ll}
\caption{Acronyms}\\
AGILE		& Astrorivelatore Gamma a Immagini LEggero \\
AGN		& Active Galactic Nuclei\\
AIT		& Astronomisches Institut der Universitat Tubingen\\
ART-P		& Astrophysical Roentgen Telescope\\
ART-S		& Astrophysical Roentgen Telescope-Spectrometer\\	
AS		& Alice Springs\\ 
ASM		& All-Sky X-ray Monitor\\
AXP		& Anomalous X-ray Pulsar\\
BAT		& Burst Alert Telescope\\
BATSE		& Burst And Transient Source Experiment\\
BGO		& Bismuth Germanate\\
BHC		& Black Hole Candidate\\
BM		& Burst Mode\\
C		& mechanical Collimator\\
CAL		& CALibration system\\
CEA		& Commissariat à l'énergie atomique et aux énergies alternatives\\
CESR		& Centre d'Etude Spatiale des Rayonnements\\
CGRO		& Compton Gamma-Ray Observatory\\
CM		& Coded Mask\\
COMPTEL	& COMPton TELescope\\
CRSF		& Cyclotron Resonance Scattering Feature\\
CT		& Compton Telescope\\
CXB		& Cosmic X-ray Background\\
CXRS		& Cosmic X-Ray Spectroscopy\\
CZTI		& Cadmium Zinc Telluride Imager\\
EXIST		& Energetic X-ray Imaging Survey Telescope\\
EXITE		& Energetic X-ray Imaging Telescope Experiment\\
EXOSAT	& European X-ray Observatory SATellite\\
FO		& Focusing Optics \\
FCFOV		& Fully Coded Field of View\\
FIGARO	& French Italian GAmma Ray Observatory\\
FOV		& Field Of View\\
FREGATE	& FREnch GAmma-ray TElescope\\
FWHM		& Full Width at Half Maximum\\
GBD		& Gamma-ray Burst Detector\\
GC		& Galactic Center\\
GeD		& Germanium Detector\\
GLAST		& Gamma-ray LArge Space Telescope\\
GRB		& Gamma-Ray Burst\\
GRBM		& Gamma-Ray Burst Monitor\\
GRIS		& Gamma-Ray Imaging Spectrometer\\
GSPC		& Gas Scintillation Proportional Counter\\
HDX		& Hard X-ray Detector\\
HEAO		& High-Energy Astronomy Observatory\\
HED		& High-Energy Detector\\
HECX		& High-Energy Celestial X-rays\\
HEFT		& High-Energy Focusing Telescope\\
HERO		& High-Energy Replicated Optics\\
HET		& High-Energy Telescope\\
HETE		& High-Energy Transient Explorer\\
HEXE		& High-Energy EXperiment\\
HEXTE		& High-Energy X-ray Timing Experiment\\
HMXRB		& High-Mass X-Ray Binary\\
HPD		& Half Power Diameter\\
HPGSPC	& High Pressure Gas Scintillation Proportional Counter\\
HRGRS		& High Resolution Gamma-Ray Spectrometer\\
HWHM		& Half Width at Half Maximum\\
HXD		& Hard X-ray Detector\\
HXI		& Hard X-ray Imager\\
HXMT		& Hard X-ray Modulation Telescope\\
HXT		& Hard X-ray Telescope\\
IBIS		& Imager on-Board the INTEGRAL Satellite\\
InFOC$\nu$S	& International Focusing Optics Collaboration for muCrab Sensitivity\\
INPE		& National Institute for Space Research\\
INTEGRAL	& INTErnational Gamma-Ray Astrophysical Laboratory\\
ISGRI		& INTEGRAL Soft Gamma-Ray Imager\\
ISAS		& Institute of Space and Aeronautical Science\\
LAC		& Large Area proportional Counter\\
LAD		& Large Area Detector\\
LAT		& Large Area Telescope\\
LAXPCs	& Large Area X-ray Proportional Counter\\
LED		& Low Energy Detectors\\
LEGR HXSS	& Low-Energy Gamma-Ray and Hard X-ray Sky Survey\\
LL		&Laue Lenses\\
LMXB		& Low Mass X-ray Binary\\
LS		& Liquid Scintillator\\
LXeGRIT	& Liquid Xenon Gamma-Ray Imaging Telescope\\
ME		& Medium Energy\\
MED		& Medium Energy Detector\\
MIFRASO	& MIlano FRAscati SOuthampton\\
MISO		& MIlan SOuthampton\\
MPE		& Max Planck Institute for Extraterrestrial Physics\\  
MPI		& Max Planck Institute\\
MSFC		& NASA Marshall Space Flight Center\\
MSP		& MilliSecond Pulsar\\
MWPC		& Multi-Wire Proportional Counter\\
NA		& Not Applicable\\
nd		& no data\\
NM		& Normal Mode\\
NRL		& Naval Research Laboratory\\
OSO		& Orbiting Solar Observatory\\
OSSE		& Oriented Scintillation Spectrometer Experiment\\
PC		& Proportional Counter\\
PCA		& Proportional Counter Array\\
PCFOV		& Partially Coded Field Of View\\
PD		& Phoswich Detector \\
PDS		& BeppoSAX Phoswich Detection System\\
PS		& Plastic Scintillator\\
PMT		& Photo Multiplier Tube\\
PSD		& Position Sensitive Detector\\
PWN		& Pulsar Wind Nebula\\
QPOs		& Quasi Periodic Oscillation\\
RBM/GBD	& Radiation Belt Monitor-Gamma-ray Burst Detector\\
RC		& Rocking Collimator \\
RMC		& Rotation Modulation Collimator\\
RXTE		& Rossi X-ray Timing Explorer\\
SAA		& South Atlantic Anomaly\\
SAS		& Small Astronomical Satellite\\
SAX		& Satellite per Astronomia X\\
SBC		& Sanriku Balloon Center\\
SCD		& Slat Collimator Detector\\
SD		& Scintillation crystal Detector\\
SFXT		& Supergiant Fast X-Ray Transient\\
SGD		& Soft Gamma-ray Detector\\
SGR		& Soft Gamma-ray Repeater\\
SSD		& Silicon Detector\\
SXC		& Soft X-ray Camera\\
SXI		& Soft X-ray Imager\\
SXS		& Soft X-ray Spectrometer\\
TCD		& Tube Collimated Detector\\
TDF		& Tidal Disruption Flare\\
TGF		& Terrestrial Gamma-ray Flash\\
TIFR		& Tata Institute of Fundamental Research\\
UAH		& University of Alabama in Huntsville\\
UCR		& University of California Riverside\\
UCSD		& University of California San Diego\\
ULX		& Ultra Luminous X-ray Source\\
URA		& Uniformly Randomized\\
USC		& Uchinoura Space Center\\
UVOT		& UV and Optical Telescope\\
XCE		& X-ray Cosmic Experiment\\
XG		& X-Grande\\
XIS		& X-ray Imaging Spectrometer\\
XRF		& X-Ray Flash\\
XRS		& X-Ray Spectrometer\\
XRT		& focusing X-Ray Telescope\\
WAM		& Wide-band All-sky Monitor\\
WATCH		& Wide Angle Telescope for Cosmic Hard X-rays\\
WFC		& Wide Field Camera\\
\label{table:acro} 
\end{longtable}

%
%

%

\bibliographystyle{aps-nameyear}
\bibliography{bibliography-ff}

\begin{thebibliography}{572}
\ifx \bisbn   \undefined \def \bisbn  #1{ISBN #1}\fi
\ifx \binits  \undefined \def \binits#1{#1} \fi
\ifx \bauthor  \undefined \def \bauthor#1{#1} \fi
\ifx \bjtitle  \undefined \def \bjtitle#1{\textrm{#1}}\fi
\ifx \batitle  \undefined \def \batitle#1{#1} \fi
\ifx \bctitle  \undefined \def \bctitle#1{#1} \fi
\ifx \bvolume  \undefined \def \bvolume#1{\textbf{#1}}\fi
\ifx \byear  \undefined \def \byear#1{#1} \fi
\ifx \bissue  \undefined \def \bissue#1{#1} \fi
\ifx \bfpage  \undefined \def \bfpage#1{#1} \fi
\ifx \blpage  \undefined \def \blpage #1{#1} \fi
\ifx \burl  \undefined \def \burl#1{#1} \fi
\ifx \doiurl  \undefined \def \doiurl#1{#1} \fi
\ifx \betal  \undefined \def \betal{et al.} \fi
\ifx \binstitute  \undefined \def \binstitute#1{#1} \fi
\ifx \beditor  \undefined \def \beditor#1{#1} \fi
\ifx \bpublisher  \undefined \def \bpublisher#1{#1} \fi
\ifx \bbtitle  \undefined \def \bbtitle#1{\textit{#1}} \fi
\ifx \bedition  \undefined \def \bedition#1{#1} \fi
\ifx \bseriesno  \undefined \def \bseriesno#1{#1} \fi
\ifx \blocation  \undefined \def \blocation#1{#1} \fi
\ifx \bsertitle  \undefined \def \bsertitle#1{#1} \fi
\ifx \bsnm \undefined \def \bsnm#1{#1} \fi
\ifx \bsuffix \undefined \def \bsuffix#1{#1} \fi
\ifx \bparticle \undefined \def \bparticle#1{#1} \fi
\ifx \barticle \undefined \def \barticle#1{#1} \fi
\ifx \botherref \undefined \def \botherref #1{#1} \fi
\ifx \url \undefined \def \url#1{#1} \fi
\ifx \bchapter \undefined \def \bchapter#1{#1} \fi
\ifx \bbook \undefined \def \bbook#1{#1} \fi
\ifx \bcomment \undefined \def \bcomment#1{#1} \fi
\ifx \oauthor \undefined \def \oauthor#1{#1} \fi
\ifx \citeauthoryear \undefined \def \citeauthoryear#1{#1} \fi
\ifx \texttildelow  \undefined \def \texttildelow{\symbol{126}} \fi
\def \endbibitem {}
\ifx \bconflocation  \undefined \def \bconflocation#1{#1} \fi

\bibitem[\protect\citeauthoryear{{Agnetta} et~al.}{1985}]{Agnetta85}
\begin{barticle}
\bauthor{\binits{G.} \bsnm{{Agnetta}}},
\bauthor{\binits{B.} \bsnm{{Agrinier}}},
\bauthor{\binits{J.P.} \bsnm{{Chabaud}}},
\bauthor{\binits{E.} \bsnm{{Costa}}},
\bauthor{\binits{R.} \bsnm{{Diraffaele}}},
\bauthor{\binits{P.} \bsnm{{Frabel}}},
\bauthor{\binits{G.} \bsnm{{Gerardi}}},
\bauthor{\binits{C.} \bsnm{{Gouiffes}}},
\bauthor{\binits{M.F.} \bsnm{{Landrea}}},
\bauthor{\binits{P.} \bsnm{{Mandrou}}},
\batitle{{The FIGARO experiment for the observation of time marked sources in
  the low energy gamma-ray range}}.
\bjtitle{International Cosmic Ray Conference}
\bvolume{3},
\bfpage{334}--\blpage{337}
(\byear{1985})
\end{barticle}
\endbibitem

\bibitem[\protect\citeauthoryear{{Agnetta} et~al.}{1989}]{Agnetta1989;figaro2}
\begin{barticle}
\bauthor{\binits{G.} \bsnm{{Agnetta}}},
\bauthor{\binits{R.} \bsnm{{di Raffaele}}},
\bauthor{\binits{T.} \bsnm{{Mineo}}},
\bauthor{\binits{B.} \bsnm{{Sacco}}},
\bauthor{\binits{L.} \bsnm{{Scarsi}}},
\bauthor{\binits{B.} \bsnm{{Agrinier}}},
\bauthor{\binits{J.C.} \bsnm{{Christy}}},
\bauthor{\binits{B.} \bsnm{{Parlier}}},
\bauthor{\binits{J.P.} \bsnm{{Chabaud}}},
\bauthor{\binits{P.} \bsnm{{Frabel}}},
\bauthor{\binits{P.} \bsnm{{Mandrou}}},
\bauthor{\binits{M.} \bsnm{{Niel}}},
\bauthor{\binits{G.} \bsnm{{Rouaix}}},
\bauthor{\binits{G.} \bsnm{{Vedrenne}}},
\bauthor{\binits{E.} \bsnm{{Costa}}},
\bauthor{\binits{G.} \bsnm{{Gerardi}}},
\bauthor{\binits{J.L.} \bsnm{{Masnou}}},
\bauthor{\binits{E.} \bsnm{{Massaro}}},
\bauthor{\binits{M.} \bsnm{{Salvati}}},
\batitle{{Figaro II experiment: description and technical performance}}.
\bjtitle{Nuclear Instruments and Methods in Physics Research A}
\bvolume{281},
\bfpage{197}--\blpage{206}
(\byear{1989}).
doi:\doiurl{10.1016/0168-9002(89)91235-7}
\end{barticle}
\endbibitem

\bibitem[\protect\citeauthoryear{{Agrawal} et~al.}{1971}]{Agrawal71}
\begin{barticle}
\bauthor{\binits{P.C.} \bsnm{{Agrawal}}},
\bauthor{\binits{S.} \bsnm{{Biswas}}},
\bauthor{\binits{G.S.} \bsnm{{Gokhale}}},
\bauthor{\binits{V.S.} \bsnm{{Iyengar}}},
\bauthor{\binits{P.K.} \bsnm{{Kunte}}},
\bauthor{\binits{R.K.} \bsnm{{Manchanda}}},
\bauthor{\binits{B.V.} \bsnm{{Sreekantan}}},
\batitle{{Intensity and Energy Spectra of Several X-Ray Sources in the 20-150
  KeV Energy Band.}}
\bjtitle{International Cosmic Ray Conference}
\bvolume{1},
\bfpage{20}
(\byear{1971})
\end{barticle}
\endbibitem

\bibitem[\protect\citeauthoryear{{Agrawal} et~al.}{1972}]{Agrawal72}
\begin{barticle}
\bauthor{\binits{P.C.} \bsnm{{Agrawal}}},
\bauthor{\binits{G.S.} \bsnm{{Gokhale}}},
\bauthor{\binits{V.S.} \bsnm{{Iyengar}}},
\bauthor{\binits{P.K.} \bsnm{{Kunte}}},
\bauthor{\binits{R.K.} \bsnm{{Manchanda}}},
\bauthor{\binits{B.V.} \bsnm{{Sreekantan}}},
\batitle{{Energy Spectrum and Time Variations of Hard X-Rays from Cyg X-1}}.
\bjtitle{\apss}
\bvolume{18},
\bfpage{408}--\blpage{424}
(\byear{1972}).
doi:\doiurl{10.1007/BF00645405}
\end{barticle}
\endbibitem

\bibitem[\protect\citeauthoryear{{Agrinier}
  et~al.}{1990a}]{Agrinier1990;figaro2}
\begin{barticle}
\bauthor{\binits{B.} \bsnm{{Agrinier}}},
\bauthor{\binits{J.L.} \bsnm{{Masnou}}},
\bauthor{\binits{B.} \bsnm{{Parlier}}},
\bauthor{\binits{M.} \bsnm{{Niel}}},
\bauthor{\binits{P.} \bsnm{{Mandrou}}},
\bauthor{\binits{G.} \bsnm{{Vedrenne}}},
\bauthor{\binits{G.} \bsnm{{Gerardi}}},
\bauthor{\binits{T.} \bsnm{{Mineo}}},
\bauthor{\binits{B.} \bsnm{{Sacco}}},
\bauthor{\binits{L.} \bsnm{{Scarsi}}},
\bauthor{\binits{E.} \bsnm{{Costa}}},
\bauthor{\binits{E.} \bsnm{{Massaro}}},
\bauthor{\binits{G.} \bsnm{{Matt}}},
\bauthor{\binits{M.} \bsnm{{Salvati}}},
\bauthor{\binits{A.G.} \bsnm{{Lyne}}},
\batitle{{Observation of the Crab pulsar, PSR 0531 + 21, at 0.2-6.0 MeV with
  the FIGARO II experiment}}.
\bjtitle{\apj}
\bvolume{355},
\bfpage{645}--\blpage{650}
(\byear{1990}a).
doi:\doiurl{10.1086/168797}
\end{barticle}
\endbibitem

\bibitem[\protect\citeauthoryear{{Agrinier} et~al.}{1990b}]{Agrinier90}
\begin{barticle}
\bauthor{\binits{B.} \bsnm{{Agrinier}}},
\bauthor{\binits{J.L.} \bsnm{{Masnou}}},
\bauthor{\binits{B.} \bsnm{{Parlier}}},
\bauthor{\binits{M.} \bsnm{{Niel}}},
\bauthor{\binits{P.} \bsnm{{Mandrou}}},
\bauthor{\binits{G.} \bsnm{{Vedrenne}}},
\bauthor{\binits{G.} \bsnm{{Gerardi}}},
\bauthor{\binits{T.} \bsnm{{Mineo}}},
\bauthor{\binits{B.} \bsnm{{Sacco}}},
\bauthor{\binits{L.} \bsnm{{Scarsi}}},
\bauthor{\binits{E.} \bsnm{{Costa}}},
\bauthor{\binits{E.} \bsnm{{Massaro}}},
\bauthor{\binits{G.} \bsnm{{Matt}}},
\bauthor{\binits{M.} \bsnm{{Salvati}}},
\bauthor{\binits{A.G.} \bsnm{{Lyne}}},
\batitle{{Observation of the Crab pulsar, PSR 0531 + 21, at 0.2-6.0 MeV with
  the FIGARO II experiment}}.
\bjtitle{\apj}
\bvolume{355},
\bfpage{645}--\blpage{650}
(\byear{1990}b).
doi:\doiurl{10.1086/168797}
\end{barticle}
\endbibitem

\bibitem[\protect\citeauthoryear{{Ajello} et~al.}{2008a}]{Ajello08}
\begin{barticle}
\bauthor{\binits{M.} \bsnm{{Ajello}}},
\bauthor{\binits{J.} \bsnm{{Greiner}}},
\bauthor{\binits{G.} \bsnm{{Kanbach}}},
\bauthor{\binits{A.} \bsnm{{Rau}}},
\bauthor{\binits{A.W.} \bsnm{{Strong}}},
\bauthor{\binits{J.A.} \bsnm{{Kennea}}},
\batitle{{BAT X-Ray Survey. I. Methodology and X-Ray Identification}}.
\bjtitle{\apj}
\bvolume{678},
\bfpage{102}--\blpage{115}
(\byear{2008}a).
doi:\doiurl{10.1086/529418}
\end{barticle}
\endbibitem

\bibitem[\protect\citeauthoryear{{Ajello} et~al.}{2008b}]{Ajello08b}
\begin{barticle}
\bauthor{\binits{M.} \bsnm{{Ajello}}},
\bauthor{\binits{J.} \bsnm{{Greiner}}},
\bauthor{\binits{G.} \bsnm{{Sato}}},
\bauthor{\binits{D.R.} \bsnm{{Willis}}},
\bauthor{\binits{G.} \bsnm{{Kanbach}}},
\bauthor{\binits{A.W.} \bsnm{{Strong}}},
\bauthor{\binits{R.} \bsnm{{Diehl}}},
\bauthor{\binits{G.} \bsnm{{Hasinger}}},
\bauthor{\binits{N.} \bsnm{{Gehrels}}},
\bauthor{\binits{C.B.} \bsnm{{Markwardt}}},
\bauthor{\binits{J.} \bsnm{{Tueller}}},
\batitle{{Cosmic X-Ray Background and Earth Albedo Spectra with Swift BAT}}.
\bjtitle{\apj}
\bvolume{689},
\bfpage{666}--\blpage{677}
(\byear{2008}b).
doi:\doiurl{10.1086/592595}
\end{barticle}
\endbibitem

\bibitem[\protect\citeauthoryear{{Alexander} et~al.}{2013}]{Alexander13}
\begin{barticle}
\bauthor{\binits{D.M.} \bsnm{{Alexander}}},
\bauthor{\binits{D.} \bsnm{{Stern}}},
\bauthor{\binits{A.} \bsnm{{Del Moro}}},
\bauthor{\binits{G.B.} \bsnm{{Lansbury}}},
\bauthor{\binits{R.J.} \bsnm{{Assef}}},
\bauthor{\binits{J.} \bsnm{{Aird}}},
\bauthor{\binits{M.} \bsnm{{Ajello}}},
\bauthor{\binits{D.R.} \bsnm{{Ballantyne}}},
\bauthor{\binits{F.E.} \bsnm{{Bauer}}},
\bauthor{\binits{S.E.} \bsnm{{Boggs}}},
\bauthor{\binits{W.N.} \bsnm{{Brandt}}},
\bauthor{\binits{F.E.} \bsnm{{Christensen}}},
\bauthor{\binits{F.} \bsnm{{Civano}}},
\bauthor{\binits{A.} \bsnm{{Comastri}}},
\bauthor{\binits{W.W.} \bsnm{{Craig}}},
\bauthor{\binits{M.} \bsnm{{Elvis}}},
\bauthor{\binits{B.W.} \bsnm{{Grefenstette}}},
\bauthor{\binits{C.J.} \bsnm{{Hailey}}},
\bauthor{\binits{F.A.} \bsnm{{Harrison}}},
\bauthor{\binits{R.C.} \bsnm{{Hickox}}},
\bauthor{\binits{B.} \bsnm{{Luo}}},
\bauthor{\binits{K.K.} \bsnm{{Madsen}}},
\bauthor{\binits{J.R.} \bsnm{{Mullaney}}},
\bauthor{\binits{M.} \bsnm{{Perri}}},
\bauthor{\binits{S.} \bsnm{{Puccetti}}},
\bauthor{\binits{C.} \bsnm{{Saez}}},
\bauthor{\binits{E.} \bsnm{{Treister}}},
\bauthor{\binits{C.M.} \bsnm{{Urry}}},
\bauthor{\binits{W.W.} \bsnm{{Zhang}}},
\bauthor{\binits{C.R.} \bsnm{{Bridge}}},
\bauthor{\binits{P.R.M.} \bsnm{{Eisenhardt}}},
\bauthor{\binits{A.H.} \bsnm{{Gonzalez}}},
\bauthor{\binits{S.H.} \bsnm{{Miller}}},
\bauthor{\binits{C.W.} \bsnm{{Tsai}}},
\batitle{{The NuSTAR Extragalactic Survey: A First Sensitive Look at the
  High-energy Cosmic X-Ray Background Population}}.
\bjtitle{\apj}
\bvolume{773},
\bfpage{125}
(\byear{2013}).
doi:\doiurl{10.1088/0004-637X/773/2/125}
\end{barticle}
\endbibitem

\bibitem[\protect\citeauthoryear{{Althouse} et~al.}{1985}]{Althouse85}
\begin{botherref}
\oauthor{\binits{W.E.} \bsnm{{Althouse}}},
\oauthor{\binits{W.R.} \bsnm{{Cook}}},
\oauthor{\binits{A.C.} \bsnm{{Cummings}}},
\oauthor{\binits{M.H.} \bsnm{{Finger}}},
\oauthor{\binits{T.A.} \bsnm{{Prince}}},
\oauthor{\binits{S.M.} \bsnm{{Schindler}}},
\oauthor{\binits{C.H.} \bsnm{{Starr}}},
\oauthor{\binits{E.C.} \bsnm{{Stone}}},
{A balloon-borne imaging gamma-ray telescope}.
International Cosmic Ray Conference
\textbf{3}
(1985)
\end{botherref}
\endbibitem

\bibitem[\protect\citeauthoryear{{An} et~al.}{2013}]{HAn13}
\begin{barticle}
\bauthor{\binits{H.} \bsnm{{An}}},
\bauthor{\binits{R.} \bsnm{{Hasco{\"e}t}}},
\bauthor{\binits{V.M.} \bsnm{{Kaspi}}},
\bauthor{\binits{A.M.} \bsnm{{Beloborodov}}},
\bauthor{\binits{F.} \bsnm{{Dufour}}},
\bauthor{\binits{E.V.} \bsnm{{Gotthelf}}},
\bauthor{\binits{R.} \bsnm{{Archibald}}},
\bauthor{\binits{M.} \bsnm{{Bachetti}}},
\bauthor{\binits{S.E.} \bsnm{{Boggs}}},
\bauthor{\binits{F.E.} \bsnm{{Christensen}}},
\bauthor{\binits{W.W.} \bsnm{{Craig}}},
\bauthor{\binits{B.W.} \bsnm{{Greffenstette}}},
\bauthor{\binits{C.J.} \bsnm{{Hailey}}},
\bauthor{\binits{F.A.} \bsnm{{Harrison}}},
\bauthor{\binits{T.} \bsnm{{Kitaguchi}}},
\bauthor{\binits{C.} \bsnm{{Kouveliotou}}},
\bauthor{\binits{K.K.} \bsnm{{Madsen}}},
\bauthor{\binits{C.B.} \bsnm{{Markwardt}}},
\bauthor{\binits{D.} \bsnm{{Stern}}},
\bauthor{\binits{J.K.} \bsnm{{Vogel}}},
\bauthor{\binits{W.W.} \bsnm{{Zhang}}},
\batitle{{NuSTAR Observations of Magnetar 1E 1841-045}}.
\bjtitle{\apj}
\bvolume{779},
\bfpage{163}
(\byear{2013}).
doi:\doiurl{10.1088/0004-637X/779/2/163}
\end{barticle}
\endbibitem

\bibitem[\protect\citeauthoryear{{An} et~al.}{2014}]{An14}
\begin{barticle}
\bauthor{\binits{H.} \bsnm{{An}}},
\bauthor{\binits{K.K.} \bsnm{{Madsen}}},
\bauthor{\binits{S.P.} \bsnm{{Reynolds}}},
\bauthor{\binits{V.M.} \bsnm{{Kaspi}}},
\bauthor{\binits{F.A.} \bsnm{{Harrison}}},
\bauthor{\binits{S.E.} \bsnm{{Boggs}}},
\bauthor{\binits{F.E.} \bsnm{{Christensen}}},
\bauthor{\binits{W.W.} \bsnm{{Craig}}},
\bauthor{\binits{C.L.} \bsnm{{Fryer}}},
\bauthor{\binits{B.W.} \bsnm{{Grefenstette}}},
\bauthor{\binits{C.J.} \bsnm{{Hailey}}},
\bauthor{\binits{K.} \bsnm{{Mori}}},
\bauthor{\binits{D.} \bsnm{{Stern}}},
\bauthor{\binits{W.W.} \bsnm{{Zhang}}},
\batitle{{High-energy X-Ray Imaging of the Pulsar Wind Nebula MSH 15-52:
  Constraints on Particle Acceleration and Transport}}.
\bjtitle{\apj}
\bvolume{793},
\bfpage{90}
(\byear{2014}).
doi:\doiurl{10.1088/0004-637X/793/2/90}
\end{barticle}
\endbibitem

\bibitem[\protect\citeauthoryear{{Aprile} et~al.}{2000}]{Aprile2000;lxegrit}
\begin{bchapter}
\bauthor{\binits{E.} \bsnm{{Aprile}}},
\bauthor{\binits{A.} \bsnm{{Curioni}}},
\bauthor{\binits{V.} \bsnm{{Egorov}}},
\bauthor{\binits{K.-L.} \bsnm{{Giboni}}},
\bauthor{\binits{U.G.} \bsnm{{Oberlack}}},
\bauthor{\binits{S.} \bsnm{{Ventura}}},
\bauthor{\binits{T.} \bsnm{{Doke}}},
\bauthor{\binits{J.} \bsnm{{Kikuchi}}},
\bauthor{\binits{K.} \bsnm{{Takizawa}}},
\bauthor{\binits{E.L.} \bsnm{{Chupp}}},
\bauthor{\binits{P.P.} \bsnm{{Dunphy}}},
\bctitle{{Spectroscopy and imaging performance of the Liquid Xenon Gamma-Ray
  Imaging Telescope (LXeGRIT)}},
in \bbtitle{X-Ray and Gamma-Ray Instrumentation for Astronomy XI},
ed. by \beditor{\binits{K.A.} \bsnm{{Flanagan}}},
\beditor{\binits{O.H.} \bsnm{{Siegmund}}}
\bsertitle{\procspie},
vol. \bseriesno{4140},
\byear{2000},
pp. \bfpage{333}--\blpage{343}
\end{bchapter}
\endbibitem

\bibitem[\protect\citeauthoryear{{Aprile} et~al.}{2003}]{Aprile03}
\begin{bchapter}
\bauthor{\binits{E.} \bsnm{{Aprile}}},
\bauthor{\binits{A.} \bsnm{{Curioni}}},
\bauthor{\binits{K.-L.} \bsnm{{Giboni}}},
\bauthor{\binits{M.} \bsnm{{Kobayashi}}},
\bauthor{\binits{U.G.} \bsnm{{Oberlack}}},
\bauthor{\binits{E.L.} \bsnm{{Chupp}}},
\bauthor{\binits{P.P.} \bsnm{{Dunphy}}},
\bauthor{\binits{S.} \bsnm{{Ventura}}},
\bauthor{\binits{T.} \bsnm{{Doke}}},
\bauthor{\binits{J.} \bsnm{{Kikuchi}}},
\bctitle{{Observation of 1-10 MeV Gamma-Rays from the Crab with the
  Balloon-Borne LXeGRIT Compton Telescope}},
in \bbtitle{AAS/High Energy Astrophysics Division \#7}.
\bsertitle{Bulletin of the American Astronomical Society},
vol. \bseriesno{35},
\byear{2003},
p. \bfpage{640}
\end{bchapter}
\endbibitem

\bibitem[\protect\citeauthoryear{{Aprile} et~al.}{2004}]{Aprile04}
\begin{barticle}
\bauthor{\binits{E.} \bsnm{{Aprile}}},
\bauthor{\binits{A.} \bsnm{{Curioni}}},
\bauthor{\binits{K.L.} \bsnm{{Giboni}}},
\bauthor{\binits{M.} \bsnm{{Kobayashi}}},
\bauthor{\binits{U.G.} \bsnm{{Oberlack}}},
\bauthor{\binits{S.} \bsnm{{Ventura}}},
\bauthor{\binits{E.L.} \bsnm{{Chupp}}},
\bauthor{\binits{P.P.} \bsnm{{Dunphy}}},
\bauthor{\binits{T.} \bsnm{{Doke}}},
\bauthor{\binits{J.} \bsnm{{Kikuchi}}},
\batitle{{Calibration and in-flight performance of the Compton telescope
  prototype LXeGRIT}}.
\bjtitle{\nar}
\bvolume{48},
\bfpage{257}--\blpage{262}
(\byear{2004}).
doi:\doiurl{10.1016/j.newar.2003.11.053}
\end{barticle}
\endbibitem

\bibitem[\protect\citeauthoryear{{Aptekar} et~al.}{2012}]{Aptekar12}
\begin{bchapter}
\bauthor{\binits{R.} \bsnm{{Aptekar}}},
\bauthor{\binits{S.V.} \bsnm{{Golenetskii}}},
\bauthor{\binits{D.D.} \bsnm{{Frederiks}}},
\bauthor{\binits{E.P.} \bsnm{{Mazets}}},
\bauthor{\binits{V.D.} \bsnm{{Palshin}}},
\bctitle{{Cosmic gamma-ray bursts studies with Ioffe Institute Konus
  experiments}},
in \bbtitle{-Ray Bursts 2012 Conference (GRB 2012)},
\byear{2012},
p. \bfpage{118}
\end{bchapter}
\endbibitem

\bibitem[\protect\citeauthoryear{{Atteia} et~al.}{2003}]{Atteia03}
\begin{bchapter}
\bauthor{\binits{J.-L.} \bsnm{{Atteia}}},
\bauthor{\binits{M.} \bsnm{{Boer}}},
\bauthor{\binits{F.} \bsnm{{Cotin}}},
\bauthor{\binits{J.} \bsnm{{Couteret}}},
\bauthor{\binits{J.-P.} \bsnm{{Dezalay}}},
\bauthor{\binits{M.} \bsnm{{Ehanno}}},
\bauthor{\binits{J.} \bsnm{{Evrard}}},
\bauthor{\binits{D.} \bsnm{{Lagrange}}},
\bauthor{\binits{M.} \bsnm{{Niel}}},
\bauthor{\binits{J.-F.} \bsnm{{Olive}}},
\bauthor{\binits{G.} \bsnm{{Rouaix}}},
\bauthor{\binits{P.} \bsnm{{Souleille}}},
\bauthor{\binits{G.} \bsnm{{Vedrenne}}},
\bauthor{\binits{K.} \bsnm{{Hurley}}},
\bauthor{\binits{G.} \bsnm{{Ricker}}},
\bauthor{\binits{R.} \bsnm{{Vanderspek}}},
\bauthor{\binits{G.} \bsnm{{Crew}}},
\bauthor{\binits{J.} \bsnm{{Doty}}},
\bauthor{\binits{N.} \bsnm{{Butler}}},
\bctitle{{In-Flight Performance and First Results of FREGATE}},
in \bbtitle{Gamma-Ray Burst and Afterglow Astronomy 2001: A Workshop
  Celebrating the First Year of the HETE Mission},
ed. by \beditor{\binits{G.R.} \bsnm{{Ricker}}},
\beditor{\binits{R.K.} \bsnm{{Vanderspek}}}
\bsertitle{American Institute of Physics Conference Series},
vol. \bseriesno{662},
\byear{2003},
pp. \bfpage{17}--\blpage{24}.
doi:\doiurl{10.1063/1.1579292}
\end{bchapter}
\endbibitem

\bibitem[\protect\citeauthoryear{{Awaki} et~al.}{2009}]{Awaki09}
\begin{barticle}
\bauthor{\binits{H.} \bsnm{{Awaki}}},
\bauthor{\binits{Y.} \bsnm{{Terashima}}},
\bauthor{\binits{Y.} \bsnm{{Higaki}}},
\bauthor{\binits{Y.} \bsnm{{Fukazawa}}},
\batitle{{Detection of Hard X-Rays from the Compton-Thick Seyfert 2 Galaxy NGC
  2273 with Suzaku}}.
\bjtitle{\pasj}
\bvolume{61},
\bfpage{317}--\blpage{325}
(\byear{2009}).
doi:\doiurl{10.1093/pasj/61.sp1.S317}
\end{barticle}
\endbibitem

\bibitem[\protect\citeauthoryear{{Bachetti} et~al.}{2014}]{Bachetti14}
\begin{barticle}
\bauthor{\binits{M.} \bsnm{{Bachetti}}},
\bauthor{\binits{F.A.} \bsnm{{Harrison}}},
\bauthor{\binits{D.J.} \bsnm{{Walton}}},
\bauthor{\binits{B.W.} \bsnm{{Grefenstette}}},
\bauthor{\binits{D.} \bsnm{{Chakrabarty}}},
\bauthor{\binits{F.} \bsnm{{F{\"u}rst}}},
\bauthor{\binits{D.} \bsnm{{Barret}}},
\bauthor{\binits{A.} \bsnm{{Beloborodov}}},
\bauthor{\binits{S.E.} \bsnm{{Boggs}}},
\bauthor{\binits{F.E.} \bsnm{{Christensen}}},
\bauthor{\binits{W.W.} \bsnm{{Craig}}},
\bauthor{\binits{A.C.} \bsnm{{Fabian}}},
\bauthor{\binits{C.J.} \bsnm{{Hailey}}},
\bauthor{\binits{A.} \bsnm{{Hornschemeier}}},
\bauthor{\binits{V.} \bsnm{{Kaspi}}},
\bauthor{\binits{S.R.} \bsnm{{Kulkarni}}},
\bauthor{\binits{T.} \bsnm{{Maccarone}}},
\bauthor{\binits{J.M.} \bsnm{{Miller}}},
\bauthor{\binits{V.} \bsnm{{Rana}}},
\bauthor{\binits{D.} \bsnm{{Stern}}},
\bauthor{\binits{S.P.} \bsnm{{Tendulkar}}},
\bauthor{\binits{J.} \bsnm{{Tomsick}}},
\bauthor{\binits{N.A.} \bsnm{{Webb}}},
\bauthor{\binits{W.W.} \bsnm{{Zhang}}},
\batitle{{An ultraluminous X-ray source powered by an accreting neutron star}}.
\bjtitle{\nat}
\bvolume{514},
\bfpage{202}--\blpage{204}
(\byear{2014}).
doi:\doiurl{10.1038/nature13791}
\end{barticle}
\endbibitem

\bibitem[\protect\citeauthoryear{{Baity} et~al.}{1973}]{Baity73}
\begin{barticle}
\bauthor{\binits{W.A.} \bsnm{{Baity}}},
\bauthor{\binits{M.P.} \bsnm{{Ulmer}}},
\bauthor{\binits{W.A.} \bsnm{{Wheaton}}},
\bauthor{\binits{L.E.} \bsnm{{Peterson}}},
\batitle{{Observations of Cyg X-1 and Cyg X-3 Above 7 keV from OSO-7}}.
\bjtitle{Nature Physical Science}
\bvolume{245},
\bfpage{90}--\blpage{92}
(\byear{1973}).
doi:\doiurl{10.1038/physci245090a0}
\end{barticle}
\endbibitem

\bibitem[\protect\citeauthoryear{{Baity} et~al.}{1974}]{Baity74}
\begin{barticle}
\bauthor{\binits{W.A.} \bsnm{{Baity}}},
\bauthor{\binits{M.P.} \bsnm{{Ulmer}}},
\bauthor{\binits{W.A.} \bsnm{{Wheaton}}},
\bauthor{\binits{L.E.} \bsnm{{Peterson}}},
\batitle{{Extended Observations of $>$ 7-keV X-Rays from Centaurus X-3 by the
  OSO-7 Satellite}}.
\bjtitle{\apj}
\bvolume{187},
\bfpage{341}--\blpage{344}
(\byear{1974}).
doi:\doiurl{10.1086/152634}
\end{barticle}
\endbibitem

\bibitem[\protect\citeauthoryear{{Baity} et~al.}{1984}]{Baity84}
\begin{barticle}
\bauthor{\binits{W.A.} \bsnm{{Baity}}},
\bauthor{\binits{D.M.} \bsnm{{Worrall}}},
\bauthor{\binits{R.E.} \bsnm{{Rothschild}}},
\bauthor{\binits{R.F.} \bsnm{{Mushotzky}}},
\bauthor{\binits{A.F.} \bsnm{{Tennant}}},
\bauthor{\binits{F.A.} \bsnm{{Primini}}},
\batitle{{Observations of NGC 4151 at 2 keV to 2 MeV from HEAO 1}}.
\bjtitle{\apj}
\bvolume{279},
\bfpage{555}--\blpage{562}
(\byear{1984}).
doi:\doiurl{10.1086/161920}
\end{barticle}
\endbibitem

\bibitem[\protect\citeauthoryear{{Baker} et~al.}{1979}]{Baker1979;miso}
\begin{barticle}
\bauthor{\binits{R.E.} \bsnm{{Baker}}},
\bauthor{\binits{R.C.} \bsnm{{Butler}}},
\bauthor{\binits{A.J.} \bsnm{{Dean}}},
\bauthor{\binits{G.} \bsnm{{Di Cocco}}},
\bauthor{\binits{N.A.} \bsnm{{Dipper}}},
\bauthor{\binits{S.J.} \bsnm{{Martin}}},
\bauthor{\binits{K.E.} \bsnm{{Mount}}},
\bauthor{\binits{D.} \bsnm{{Ramsden}}},
\bauthor{\binits{G.} \bsnm{{Barbaglia}}},
\bauthor{\binits{L.} \bsnm{{Barbareschi}}},
\bauthor{\binits{G.} \bsnm{{Boella}}},
\bauthor{\binits{A.} \bsnm{{Bussini}}},
\bauthor{\binits{A.} \bsnm{{Igiuni}}},
\bauthor{\binits{P.} \bsnm{{Inzani}}},
\bauthor{\binits{F.} \bsnm{{Perotti}}},
\bauthor{\binits{G.} \bsnm{{Villa}}},
\batitle{{The MISO low energy gamma -ray telescope.}}
\bjtitle{Nuclear Instruments and Methods}
\bvolume{158},
\bfpage{595}--\blpage{604}
(\byear{1979}).
doi:\doiurl{10.1016/S0029-554X(79)96375-4}
\end{barticle}
\endbibitem

\bibitem[\protect\citeauthoryear{{Baker} et~al.}{1981a}]{Baker1981;miso}
\begin{barticle}
\bauthor{\binits{R.E.} \bsnm{{Baker}}},
\bauthor{\binits{L.} \bsnm{{Bassani}}},
\bauthor{\binits{A.J.} \bsnm{{Dean}}},
\bauthor{\binits{D.} \bsnm{{Ramsden}}},
\bauthor{\binits{R.C.} \bsnm{{Butler}}},
\bauthor{\binits{G.} \bsnm{{Di Cocco}}},
\bauthor{\binits{G.} \bsnm{{Boella}}},
\bauthor{\binits{A.} \bsnm{{della Ventura}}},
\bauthor{\binits{F.} \bsnm{{Perotti}}},
\bauthor{\binits{G.} \bsnm{{Villa}}},
\batitle{{Low energy gamma ray observations with the MISO telescope}}.
\bjtitle{International Cosmic Ray Conference}
\bvolume{1},
\bfpage{222}--\blpage{225}
(\byear{1981}a)
\end{barticle}
\endbibitem

\bibitem[\protect\citeauthoryear{{Baker} et~al.}{1981b}]{Baker81}
\begin{barticle}
\bauthor{\binits{R.E.} \bsnm{{Baker}}},
\bauthor{\binits{L.} \bsnm{{Bassani}}},
\bauthor{\binits{A.J.} \bsnm{{Dean}}},
\bauthor{\binits{D.} \bsnm{{Ramsden}}},
\bauthor{\binits{R.C.} \bsnm{{Butler}}},
\bauthor{\binits{G.} \bsnm{{Di Cocco}}},
\bauthor{\binits{G.} \bsnm{{Boella}}},
\bauthor{\binits{A.} \bsnm{{della Ventura}}},
\bauthor{\binits{F.} \bsnm{{Perotti}}},
\bauthor{\binits{G.} \bsnm{{Villa}}},
\batitle{{Low energy gamma ray observations with the MISO telescope}}.
\bjtitle{International Cosmic Ray Conference}
\bvolume{1},
\bfpage{222}--\blpage{225}
(\byear{1981}b)
\end{barticle}
\endbibitem

\bibitem[\protect\citeauthoryear{{Baker} et~al.}{1984}]{Baker1984;mifraso}
\begin{barticle}
\bauthor{\binits{R.E.} \bsnm{{Baker}}},
\bauthor{\binits{G.} \bsnm{{Barbaglia}}},
\bauthor{\binits{A.} \bsnm{{Bazzano}}},
\bauthor{\binits{L.} \bsnm{{Boccaccini}}},
\bauthor{\binits{A.} \bsnm{{Bussini}}},
\bauthor{\binits{A.} \bsnm{{Carzaniga}}},
\bauthor{\binits{A.} \bsnm{{Court}}},
\bauthor{\binits{A.J.} \bsnm{{Dean}}},
\bauthor{\binits{N.A.} \bsnm{{Dipper}}},
\bauthor{\binits{G.} \bsnm{{Ferrandi}}},
\bauthor{\binits{N.} \bsnm{{Haskell}}},
\bauthor{\binits{C.} \bsnm{{La Padula}}},
\bauthor{\binits{R.A.} \bsnm{{Lewis}}},
\bauthor{\binits{D.} \bsnm{{Maccagni}}},
\bauthor{\binits{M.} \bsnm{{Mastropietro}}},
\bauthor{\binits{R.} \bsnm{{Patriarca}}},
\bauthor{\binits{F.} \bsnm{{Perotti}}},
\bauthor{\binits{V.F.} \bsnm{{Polcaro}}},
\bauthor{\binits{E.} \bsnm{{Quadrini}}},
\bauthor{\binits{D.} \bsnm{{Ramsden}}},
\bauthor{\binits{S.} \bsnm{{Sembay}}},
\bauthor{\binits{R.} \bsnm{{Spicer}}},
\bauthor{\binits{P.} \bsnm{{Ubertini}}},
\bauthor{\binits{G.} \bsnm{{Villa}}},
\bauthor{\binits{D.} \bsnm{{Whatley}}},
\batitle{{A large area telescope for balloon-borne hard X-ray astronomy}}.
\bjtitle{Nuclear Instruments and Methods in Physics Research A}
\bvolume{228},
\bfpage{183}--\blpage{192}
(\byear{1984}).
doi:\doiurl{10.1016/0168-9002(84)90030-5}
\end{barticle}
\endbibitem

\bibitem[\protect\citeauthoryear{{Baldovin} et~al.}{2009}]{Baldovin09}
\begin{botherref}
\oauthor{\binits{C.} \bsnm{{Baldovin}}},
\oauthor{\binits{V.} \bsnm{{Savchenko}}},
\oauthor{\binits{V.} \bsnm{{Beckmann}}},
\oauthor{\binits{A.} \bsnm{{Neronov}}},
\oauthor{\binits{D.} \bsnm{{Goetz}}},
\oauthor{\binits{P.} \bsnm{{den Hartog}}},
\oauthor{\binits{W.} \bsnm{{Hermsen}}},
\oauthor{\binits{L.} \bsnm{{Kuiper}}},
\oauthor{\binits{S.} \bsnm{{Mereghetti}}},
\oauthor{\binits{K.} \bsnm{{Hurley}}},
{INTEGRAL observes continued activity from AXP 1E1547.0-5408}.
The A stronomer's Telegram
\textbf{1908}
(2009)
\end{botherref}
\endbibitem

\bibitem[\protect\citeauthoryear{{Barthelmy}
  et~al.}{2005}]{Barthelmy2005;swift}
\begin{barticle}
\bauthor{\binits{S.D.} \bsnm{{Barthelmy}}},
\bauthor{\binits{L.M.} \bsnm{{Barbier}}},
\bauthor{\binits{J.R.} \bsnm{{Cummings}}},
\bauthor{\binits{E.E.} \bsnm{{Fenimore}}},
\bauthor{\binits{N.} \bsnm{{Gehrels}}},
\bauthor{\binits{D.} \bsnm{{Hullinger}}},
\bauthor{\binits{H.A.} \bsnm{{Krimm}}},
\bauthor{\binits{C.B.} \bsnm{{Markwardt}}},
\bauthor{\binits{D.M.} \bsnm{{Palmer}}},
\bauthor{\binits{A.} \bsnm{{Parsons}}},
\bauthor{\binits{G.} \bsnm{{Sato}}},
\bauthor{\binits{M.} \bsnm{{Suzuki}}},
\bauthor{\binits{T.} \bsnm{{Takahashi}}},
\bauthor{\binits{M.} \bsnm{{Tashiro}}},
\bauthor{\binits{J.} \bsnm{{Tueller}}},
\batitle{{The Burst Alert Telescope (BAT) on the SWIFT Midex Mission}}.
\bjtitle{\ssr}
\bvolume{120},
\bfpage{143}--\blpage{164}
(\byear{2005}).
doi:\doiurl{10.1007/s11214-005-5096-3}
\end{barticle}
\endbibitem

\bibitem[\protect\citeauthoryear{{Bassani} et~al.}{1995}]{Bassani95}
\begin{barticle}
\bauthor{\binits{L.} \bsnm{{Bassani}}},
\bauthor{\binits{G.} \bsnm{{Malaguti}}},
\bauthor{\binits{E.} \bsnm{{Jourdain}}},
\bauthor{\binits{J.P.} \bsnm{{Roques}}},
\bauthor{\binits{W.N.} \bsnm{{Johnson}}},
\batitle{{Detection of soft gamma-ray emission from the Seyfert 2 galaxy NGC
  4507 by the OSSE telescope}}.
\bjtitle{\apjl}
\bvolume{444},
\bfpage{73}--\blpage{76}
(\byear{1995}).
doi:\doiurl{10.1086/187863}
\end{barticle}
\endbibitem

\bibitem[\protect\citeauthoryear{{Baumgartner}
  et~al.}{2003}]{Baumgartner2003;infocus}
\begin{bchapter}
\bauthor{\binits{W.H.} \bsnm{{Baumgartner}}},
\bauthor{\binits{J.} \bsnm{{Tueller}}},
\bauthor{\binits{H.} \bsnm{{Krimm}}},
\bauthor{\binits{S.D.} \bsnm{{Barthelmy}}},
\bauthor{\binits{F.} \bsnm{{Berendse}}},
\bauthor{\binits{L.} \bsnm{{Ryan}}},
\bauthor{\binits{F.B.} \bsnm{{Birsa}}},
\bauthor{\binits{T.} \bsnm{{Okajima}}},
\bauthor{\binits{H.} \bsnm{{Kunieda}}},
\bauthor{\binits{Y.} \bsnm{{Ogasaka}}},
\bauthor{\binits{Y.} \bsnm{{Tawara}}},
\bauthor{\binits{K.} \bsnm{{Tamura}}},
\bctitle{{InFOCuS hard x-ray telescope: pixellated CZT detector/shield
  performance and flight results}},
in \bbtitle{X-Ray and Gamma-Ray Telescopes and Instruments for Astronomy.},
ed. by \beditor{\binits{J.E.} \bsnm{{Tr{\"u}mper}}},
\beditor{\binits{H.D.} \bsnm{{Tananbaum}}}
\bsertitle{Society of Photo-Optical Instrumentation Engineers (SPIE) Conference
  Series},
vol. \bseriesno{4851},
\byear{2003},
pp. \bfpage{945}--\blpage{956}.
doi:\doiurl{10.1117/12.461491}
\end{bchapter}
\endbibitem

\bibitem[\protect\citeauthoryear{{Bazzano} et~al.}{1983}]{Bazzano83;POKER}
\begin{barticle}
\bauthor{\binits{A.} \bsnm{{Bazzano}}},
\bauthor{\binits{L.} \bsnm{{Boccaccini}}},
\bauthor{\binits{C.} \bsnm{{La Padula}}},
\bauthor{\binits{M.} \bsnm{{Mastropietro}}},
\bauthor{\binits{R.} \bsnm{{Patriarca}}},
\bauthor{\binits{F.} \bsnm{{Polcaro}}},
\bauthor{\binits{P.} \bsnm{{Ubertini}}},
\bauthor{\binits{R.} \bsnm{{Manchanda}}},
\batitle{{Design and performance of a one square meter proportional counter
  system for hard x-ray astronomy}}.
\bjtitle{Nuclear Instruments and Methods in Physics Research}
\bvolume{214},
\bfpage{481}--\blpage{490}
(\byear{1983}).
doi:\doiurl{10.1016/0167-5087(83)90620-8}
\end{barticle}
\endbibitem

\bibitem[\protect\citeauthoryear{{Bazzano} et~al.}{1984}]{Bazzano84}
\begin{barticle}
\bauthor{\binits{A.} \bsnm{{Bazzano}}},
\bauthor{\binits{R.} \bsnm{{Fusco-Femiano}}},
\bauthor{\binits{C.} \bsnm{{La Padula}}},
\bauthor{\binits{V.F.} \bsnm{{Polcaro}}},
\bauthor{\binits{P.} \bsnm{{Ubertini}}},
\bauthor{\binits{R.K.} \bsnm{{Manchanda}}},
\batitle{{Evidence of hard X-ray emission from three clusters of galaxies}}.
\bjtitle{\apj}
\bvolume{279},
\bfpage{515}--\blpage{520}
(\byear{1984}).
doi:\doiurl{10.1086/161915}
\end{barticle}
\endbibitem

\bibitem[\protect\citeauthoryear{{Bazzano} et~al.}{1990a}]{Bazzano90}
\begin{barticle}
\bauthor{\binits{A.} \bsnm{{Bazzano}}},
\bauthor{\binits{R.} \bsnm{{Fusco-Femiano}}},
\bauthor{\binits{P.} \bsnm{{Ubertini}}},
\bauthor{\binits{F.} \bsnm{{Perotti}}},
\bauthor{\binits{E.} \bsnm{{Quadrini}}},
\bauthor{\binits{A.J.} \bsnm{{Court}}},
\bauthor{\binits{N.A.} \bsnm{{Dean}}},
\bauthor{\binits{A.J.} \bsnm{{Dipper}}},
\bauthor{\binits{R.} \bsnm{{Lewis}}},
\bauthor{\binits{J.B.} \bsnm{{Stephen}}},
\batitle{{Hard X-rays from Coma Cluster region}}.
\bjtitle{\apjl}
\bvolume{362},
\bfpage{51}--\blpage{54}
(\byear{1990}a).
doi:\doiurl{10.1086/185845}
\end{barticle}
\endbibitem

\bibitem[\protect\citeauthoryear{{Bazzano} et~al.}{1990b}]{Bazzano1990;poker}
\begin{barticle}
\bauthor{\binits{A.} \bsnm{{Bazzano}}},
\bauthor{\binits{L.} \bsnm{{Boccaccini}}},
\bauthor{\binits{M.} \bsnm{{Federici}}},
\bauthor{\binits{C.} \bsnm{{La Padula}}},
\bauthor{\binits{M.} \bsnm{{Mastropietro}}},
\bauthor{\binits{R.} \bsnm{{Patriarca}}},
\bauthor{\binits{P.} \bsnm{{Ubertini}}},
\bauthor{\binits{R.} \bsnm{{Sood}}},
\bauthor{\binits{Z.} \bsnm{{Ye}}},
\batitle{{In-Flight Performance of the Hard X-Ray Balloon-Borne Experiment
  ''Poker''}}.
\bjtitle{International Cosmic Ray Conference}
\bvolume{4},
\bfpage{216}
(\byear{1990}b)
\end{barticle}
\endbibitem

\bibitem[\protect\citeauthoryear{{Bazzano} et~al.}{1991}]{Bazzano91}
\begin{barticle}
\bauthor{\binits{A.} \bsnm{{Bazzano}}},
\bauthor{\binits{R.} \bsnm{{Bellisario}}},
\bauthor{\binits{L.} \bsnm{{Boccaccini}}},
\bauthor{\binits{M.} \bsnm{{Federici}}},
\bauthor{\binits{C.} \bsnm{{La Padula}}},
\bauthor{\binits{R.} \bsnm{{Patriarca}}},
\bauthor{\binits{P.} \bsnm{{Ubertini}}},
\batitle{{ALISE: A Balloon Borne Hard X-Ray Imager}}.
\bjtitle{International Cosmic Ray Conference}
\bvolume{2},
\bfpage{555}
(\byear{1991})
\end{barticle}
\endbibitem

\bibitem[\protect\citeauthoryear{{Bazzano} et~al.}{1992}]{Bazzano92}
\begin{barticle}
\bauthor{\binits{A.} \bsnm{{Bazzano}}},
\bauthor{\binits{C.} \bsnm{{La Padula}}},
\bauthor{\binits{P.} \bsnm{{Ubertini}}},
\bauthor{\binits{R.K.} \bsnm{{Sood}}},
\batitle{{Variability of the Galactic center in the hard X-ray range}}.
\bjtitle{\apjl}
\bvolume{385},
\bfpage{17}--\blpage{20}
(\byear{1992}).
doi:\doiurl{10.1086/186267}
\end{barticle}
\endbibitem

\bibitem[\protect\citeauthoryear{{Bazzano} et~al.}{1993}]{Bazzano93}
\begin{barticle}
\bauthor{\binits{A.} \bsnm{{Bazzano}}},
\bauthor{\binits{M.} \bsnm{{Cocchi}}},
\bauthor{\binits{C.} \bsnm{{La Padula}}},
\bauthor{\binits{R.} \bsnm{{Sood}}},
\bauthor{\binits{P.} \bsnm{{Ubertini}}},
\batitle{{Hard X-ray observation of GRS 1758-258}}.
\bjtitle{\aaps}
\bvolume{97},
\bfpage{169}--\blpage{171}
(\byear{1993})
\end{barticle}
\endbibitem

\bibitem[\protect\citeauthoryear{{Beall} et~al.}{1976}]{Beall76}
\begin{bchapter}
\bauthor{\binits{J.} \bsnm{{Beall}}},
\bauthor{\binits{B.R.} \bsnm{{Dennis}}},
\bauthor{\binits{C.J.} \bsnm{{Crannell}}},
\bauthor{\binits{J.F.} \bsnm{{Dolan}}},
\bauthor{\binits{K.J.} \bsnm{{Frost}}},
\bauthor{\binits{L.E.} \bsnm{{Orwig}}},
\bctitle{{The 30 to 200 keV X-Ray Spectrum of Cen A Measured from OSO-8}},
in \bbtitle{Bulletin of the American Astronomical Society}.
\bsertitle{Bulletin of the American Astronomical Society},
vol. \bseriesno{8},
\byear{1976},
p. \bfpage{363}
\end{bchapter}
\endbibitem

\bibitem[\protect\citeauthoryear{{Becker} et~al.}{1978}]{Becker78}
\begin{barticle}
\bauthor{\binits{R.H.} \bsnm{{Becker}}},
\bauthor{\binits{R.E.} \bsnm{{Rothschild}}},
\bauthor{\binits{E.A.} \bsnm{{Boldt}}},
\bauthor{\binits{S.S.} \bsnm{{Holt}}},
\bauthor{\binits{S.H.} \bsnm{{Pravdo}}},
\bauthor{\binits{P.J.} \bsnm{{Serlemitsos}}},
\bauthor{\binits{J.H.} \bsnm{{Swank}}},
\batitle{{Extended observations of VELA X-1 by OSO 8}}.
\bjtitle{\apj}
\bvolume{221},
\bfpage{912}--\blpage{916}
(\byear{1978}).
doi:\doiurl{10.1086/156094}
\end{barticle}
\endbibitem

\bibitem[\protect\citeauthoryear{{Bellm} et~al.}{2014}]{Bellm14}
\begin{barticle}
\bauthor{\binits{E.C.} \bsnm{{Bellm}}},
\bauthor{\binits{F.} \bsnm{{F{\"u}rst}}},
\bauthor{\binits{K.} \bsnm{{Pottschmidt}}},
\bauthor{\binits{J.A.} \bsnm{{Tomsick}}},
\bauthor{\binits{S.E.} \bsnm{{Boggs}}},
\bauthor{\binits{D.} \bsnm{{Chakrabarty}}},
\bauthor{\binits{F.E.} \bsnm{{Christensen}}},
\bauthor{\binits{W.W.} \bsnm{{Craig}}},
\bauthor{\binits{C.J.} \bsnm{{Hailey}}},
\bauthor{\binits{F.A.} \bsnm{{Harrison}}},
\bauthor{\binits{D.} \bsnm{{Stern}}},
\bauthor{\binits{D.J.} \bsnm{{Walton}}},
\bauthor{\binits{J.} \bsnm{{Wilms}}},
\bauthor{\binits{W.W.} \bsnm{{Zhang}}},
\batitle{{Confirmation of a High Magnetic Field in GRO J1008-57}}.
\bjtitle{\apj}
\bvolume{792},
\bfpage{108}
(\byear{2014}).
doi:\doiurl{10.1088/0004-637X/792/2/108}
\end{barticle}
\endbibitem

\bibitem[\protect\citeauthoryear{{Belloni} et~al.}{1996}]{Belloni96}
\begin{barticle}
\bauthor{\binits{T.} \bsnm{{Belloni}}},
\bauthor{\binits{M.} \bsnm{{Mendez}}},
\bauthor{\binits{M.} \bsnm{{van der Klis}}},
\bauthor{\binits{G.} \bsnm{{Hasinger}}},
\bauthor{\binits{W.H.G.} \bsnm{{Lewin}}},
\bauthor{\binits{J.} \bsnm{{van Paradijs}}},
\batitle{{An Intermediate State of Cygnus X-1}}.
\bjtitle{\apjl}
\bvolume{472},
\bfpage{107}
(\byear{1996}).
doi:\doiurl{10.1086/310369}
\end{barticle}
\endbibitem

\bibitem[\protect\citeauthoryear{{Belloni} et~al.}{2002}]{Belloni02}
\begin{barticle}
\bauthor{\binits{T.} \bsnm{{Belloni}}},
\bauthor{\binits{A.P.} \bsnm{{Colombo}}},
\bauthor{\binits{J.} \bsnm{{Homan}}},
\bauthor{\binits{S.} \bsnm{{Campana}}},
\bauthor{\binits{M.} \bsnm{{van der Klis}}},
\batitle{{A low/hard state outburst of XTE J1550-564}}.
\bjtitle{\aap}
\bvolume{390},
\bfpage{199}--\blpage{204}
(\byear{2002}).
doi:\doiurl{10.1051/0004-6361:20020703}
\end{barticle}
\endbibitem

\bibitem[\protect\citeauthoryear{{Beloborodov}}{2013}]{Beloborodov13}
\begin{barticle}
\bauthor{\binits{A.M.} \bsnm{{Beloborodov}}},
\batitle{{On the Mechanism of Hard X-Ray Emission from Magnetars}}.
\bjtitle{\apj}
\bvolume{762},
\bfpage{13}
(\byear{2013}).
doi:\doiurl{10.1088/0004-637X/762/1/13}
\end{barticle}
\endbibitem

\bibitem[\protect\citeauthoryear{{Benlloch} et~al.}{2001}]{Benlloch01}
\begin{barticle}
\bauthor{\binits{S.} \bsnm{{Benlloch}}},
\bauthor{\binits{R.E.} \bsnm{{Rothschild}}},
\bauthor{\binits{J.} \bsnm{{Wilms}}},
\bauthor{\binits{C.S.} \bsnm{{Reynolds}}},
\bauthor{\binits{W.A.} \bsnm{{Heindl}}},
\bauthor{\binits{R.} \bsnm{{Staubert}}},
\batitle{{RXTE monitoring of Centaurus A}}.
\bjtitle{\aap}
\bvolume{371},
\bfpage{858}--\blpage{864}
(\byear{2001}).
doi:\doiurl{10.1051/0004-6361:20010438}
\end{barticle}
\endbibitem

\bibitem[\protect\citeauthoryear{{Bhalerao} et~al.}{2015}]{Bhalerao15}
\begin{barticle}
\bauthor{\binits{V.} \bsnm{{Bhalerao}}},
\bauthor{\binits{P.} \bsnm{{Romano}}},
\bauthor{\binits{J.} \bsnm{{Tomsick}}},
\bauthor{\binits{L.} \bsnm{{Natalucci}}},
\bauthor{\binits{D.M.} \bsnm{{Smith}}},
\bauthor{\binits{E.} \bsnm{{Bellm}}},
\bauthor{\binits{S.E.} \bsnm{{Boggs}}},
\bauthor{\binits{D.} \bsnm{{Chakrabarty}}},
\bauthor{\binits{F.E.} \bsnm{{Christensen}}},
\bauthor{\binits{W.W.} \bsnm{{Craig}}},
\bauthor{\binits{F.} \bsnm{{Fuerst}}},
\bauthor{\binits{C.J.} \bsnm{{Hailey}}},
\bauthor{\binits{F.A.} \bsnm{{Harrison}}},
\bauthor{\binits{R.A.} \bsnm{{Krivonos}}},
\bauthor{\binits{T.-N.} \bsnm{{Lu}}},
\bauthor{\binits{K.} \bsnm{{Madsen}}},
\bauthor{\binits{D.} \bsnm{{Stern}}},
\bauthor{\binits{G.} \bsnm{{Younes}}},
\bauthor{\binits{W.} \bsnm{{Zhang}}},
\batitle{{NuSTAR detection of a cyclotron line in the supergiant fast X-ray
  transient IGR J17544-2619}}.
\bjtitle{\mnras}
\bvolume{447},
\bfpage{2274}--\blpage{2281}
(\byear{2015}).
doi:\doiurl{10.1093/mnras/stu2495}
\end{barticle}
\endbibitem

\bibitem[\protect\citeauthoryear{{Bhattacharya} et~al.}{1994}]{Bhattacharya94}
\begin{barticle}
\bauthor{\binits{D.} \bsnm{{Bhattacharya}}},
\bauthor{\binits{L.-S.} \bsnm{{The}}},
\bauthor{\binits{J.D.} \bsnm{{Kurfess}}},
\bauthor{\binits{D.D.} \bsnm{{Clayton}}},
\bauthor{\binits{N.} \bsnm{{Gehrels}}},
\bauthor{\binits{M.D.} \bsnm{{Leising}}},
\bauthor{\binits{D.A.} \bsnm{{Grabelsky}}},
\bauthor{\binits{W.N.} \bsnm{{Johnson}}},
\bauthor{\binits{G.V.} \bsnm{{Jung}}},
\bauthor{\binits{R.L.} \bsnm{{Kinzer}}},
\bauthor{\binits{W.R.} \bsnm{{Purcell}}},
\bauthor{\binits{M.S.} \bsnm{{Strickman}}},
\bauthor{\binits{M.P.} \bsnm{{Ulmer}}},
\batitle{{Gamma-ray observations of NGC 253 and M82 with OSSE}}.
\bjtitle{\apj}
\bvolume{437},
\bfpage{173}--\blpage{178}
(\byear{1994}).
doi:\doiurl{10.1086/174985}
\end{barticle}
\endbibitem

\bibitem[\protect\citeauthoryear{{Bingham} and {Clark}}{1969}]{Bingham69}
\begin{barticle}
\bauthor{\binits{R.G.} \bsnm{{Bingham}}},
\bauthor{\binits{C.D.} \bsnm{{Clark}}},
\batitle{{High-Energy X-Rays from Cygnus XR-1}}.
\bjtitle{\apj}
\bvolume{158},
\bfpage{207}
(\byear{1969}).
doi:\doiurl{10.1086/150184}
\end{barticle}
\endbibitem

\bibitem[\protect\citeauthoryear{{Bird} et~al.}{2016}]{Bird16}
\begin{barticle}
\bauthor{\binits{A.J.} \bsnm{{Bird}}},
\bauthor{\binits{A.} \bsnm{{Bazzano}}},
\bauthor{\binits{A.} \bsnm{{Malizia}}},
\bauthor{\binits{M.} \bsnm{{Fiocchi}}},
\bauthor{\binits{V.} \bsnm{{Sguera}}},
\bauthor{\binits{L.} \bsnm{{Bassani}}},
\bauthor{\binits{A.B.} \bsnm{{Hill}}},
\bauthor{\binits{P.} \bsnm{{Ubertini}}},
\bauthor{\binits{C.} \bsnm{{Winkler}}},
\batitle{{The IBIS Soft Gamma-Ray Sky after 1000 Integral Orbits}}.
\bjtitle{\apjs}
\bvolume{223},
\bfpage{15}
(\byear{2016}).
doi:\doiurl{10.3847/0067-0049/223/1/15}
\end{barticle}
\endbibitem

\bibitem[\protect\citeauthoryear{{Bleeker} et~al.}{1967}]{Bleeker67}
\begin{barticle}
\bauthor{\binits{J.A.M.} \bsnm{{Bleeker}}},
\bauthor{\binits{J.J.} \bsnm{{Burger}}},
\bauthor{\binits{A.J.M.} \bsnm{{Deerenberg}}},
\bauthor{\binits{A.} \bsnm{{Scheepmaker}}},
\bauthor{\binits{B.N.} \bsnm{{Swanenburg}}},
\bauthor{\binits{Y.} \bsnm{{Tanaka}}},
\batitle{{Balloon Observation of the X-Ray Sources in the Cygnus Region in the
  Energy Range 20-130 KEV}}.
\bjtitle{\apj}
\bvolume{147},
\bfpage{391}
(\byear{1967}).
doi:\doiurl{10.1086/149019}
\end{barticle}
\endbibitem

\bibitem[\protect\citeauthoryear{{Bloom} et~al.}{2011}]{Bloom11}
\begin{barticle}
\bauthor{\binits{J.S.} \bsnm{{Bloom}}},
\bauthor{\binits{D.} \bsnm{{Giannios}}},
\bauthor{\binits{B.D.} \bsnm{{Metzger}}},
\bauthor{\binits{S.B.} \bsnm{{Cenko}}},
\bauthor{\binits{D.A.} \bsnm{{Perley}}},
\bauthor{\binits{N.R.} \bsnm{{Butler}}},
\bauthor{\binits{N.R.} \bsnm{{Tanvir}}},
\bauthor{\binits{A.J.} \bsnm{{Levan}}},
\bauthor{\binits{P.T.} \bsnm{{O'Brien}}},
\bauthor{\binits{L.E.} \bsnm{{Strubbe}}},
\bauthor{\binits{F.} \bsnm{{De Colle}}},
\bauthor{\binits{E.} \bsnm{{Ramirez-Ruiz}}},
\bauthor{\binits{W.H.} \bsnm{{Lee}}},
\bauthor{\binits{S.} \bsnm{{Nayakshin}}},
\bauthor{\binits{E.} \bsnm{{Quataert}}},
\bauthor{\binits{A.R.} \bsnm{{King}}},
\bauthor{\binits{A.} \bsnm{{Cucchiara}}},
\bauthor{\binits{J.} \bsnm{{Guillochon}}},
\bauthor{\binits{G.C.} \bsnm{{Bower}}},
\bauthor{\binits{A.S.} \bsnm{{Fruchter}}},
\bauthor{\binits{A.N.} \bsnm{{Morgan}}},
\bauthor{\binits{A.J.} \bsnm{{van der Horst}}},
\batitle{{A Possible Relativistic Jetted Outburst from a Massive Black Hole Fed
  by a Tidally Disrupted Star}}.
\bjtitle{Science}
\bvolume{333},
\bfpage{203}
(\byear{2011}).
doi:\doiurl{10.1126/science.1207150}
\end{barticle}
\endbibitem

\bibitem[\protect\citeauthoryear{{Bloser} et~al.}{2006}]{Bloser06}
\begin{barticle}
\bauthor{\binits{P.F.} \bsnm{{Bloser}}},
\bauthor{\binits{J.S.} \bsnm{{Legere}}},
\bauthor{\binits{J.R.} \bsnm{{Macri}}},
\bauthor{\binits{M.L.} \bsnm{{McConnell}}},
\bauthor{\binits{T.} \bsnm{{Narita}}},
\bauthor{\binits{J.M.} \bsnm{{Ryan}}},
\batitle{{GRAPE - A Balloon-Borne Gamma-Ray Polarimeter Experiment}}.
\bjtitle{Chinese Journal of Astronomy and Astrophysics Supplement}
\bvolume{6}(\bissue{1}),
\bfpage{393}--\blpage{397}
(\byear{2006}).
doi:\doiurl{10.1088/1009-9271/6/S1/54}
\end{barticle}
\endbibitem

\bibitem[\protect\citeauthoryear{{Boella} et~al.}{1997}]{Boella97}
\begin{botherref}
\oauthor{\binits{G.} \bsnm{{Boella}}},
\oauthor{\binits{R.C.} \bsnm{{Butler}}},
\oauthor{\binits{G.C.} \bsnm{{Perola}}},
\oauthor{\binits{L.} \bsnm{{Piro}}},
\oauthor{\binits{L.} \bsnm{{Scarsi}}},
\oauthor{\binits{J.A.M.} \bsnm{{Bleeker}}},
{BeppoSAX, the wide band mission for X-ray astronomy}.
\aaps
\textbf{122}
(1997).
doi:\doiurl{10.1051/aas:1997136}
\end{botherref}
\endbibitem

\bibitem[\protect\citeauthoryear{{Borkous} et~al.}{1997}]{Borkous97}
\begin{bchapter}
\bauthor{\binits{V.V.} \bsnm{{Borkous}}},
\bauthor{\binits{A.S.} \bsnm{{Kaniovsky}}},
\bauthor{\binits{R.A.} \bsnm{{Sunyaev}}},
\bauthor{\binits{V.V.} \bsnm{{Efremov}}},
\bauthor{\binits{A.} \bsnm{{Kendziorra}}},
\bauthor{\binits{P.} \bsnm{{Kretschmar}}},
\bauthor{\binits{M.} \bsnm{{Kunz}}},
\bauthor{\binits{M.} \bsnm{{Maisack}}},
\bauthor{\binits{R.} \bsnm{{Staubert}}},
\bauthor{\binits{J.} \bsnm{{Englhauser}}},
\bauthor{\binits{W.} \bsnm{{Pietsch}}},
\bauthor{\binits{C.} \bsnm{{Reppin}}},
\bauthor{\binits{J.} \bsnm{{Tr{\"u}mper}}},
\bctitle{{Hard X-Ray Observations of the Bursting Pulsar GRO J1744-28 by HEXE
  onboard MIR-KVANT}},
in \bbtitle{The Transparent Universe},
ed. by \beditor{\binits{C.} \bsnm{{Winkler}}},
\beditor{\binits{T.J.-L.} \bsnm{{Courvoisier}}},
\beditor{\binits{P.} \bsnm{{Durouchoux}}}
\bsertitle{ESA Special Publication},
vol. \bseriesno{382},
\byear{1997},
p. \bfpage{299}
\end{bchapter}
\endbibitem

\bibitem[\protect\citeauthoryear{{Borkus} et~al.}{1995}]{Borkus95}
\begin{barticle}
\bauthor{\binits{V.V.} \bsnm{{Borkus}}},
\bauthor{\binits{A.S.} \bsnm{{Kaniovsky}}},
\bauthor{\binits{V.V.} \bsnm{{Efremov}}},
\bauthor{\binits{K.N.} \bsnm{{Borozdin}}},
\bauthor{\binits{R.A.} \bsnm{{Sunyaev}}},
\bauthor{\binits{N.L.} \bsnm{{Aleksandrovich}}},
\bauthor{\binits{V.A.} \bsnm{{Arefev}}},
\bauthor{\binits{P.} \bsnm{{Kretschmar}}},
\bauthor{\binits{M.} \bsnm{{Kunz}}},
\bauthor{\binits{M.} \bsnm{{Maisack}}},
\bauthor{\binits{R.} \bsnm{{Staubert}}},
\bauthor{\binits{J.} \bsnm{{Englhauser}}},
\bauthor{\binits{W.} \bsnm{{Pietsch}}},
\bauthor{\binits{C.} \bsnm{{Reppin}}},
\bauthor{\binits{J.} \bsnm{{Tr{\"u}mper}}},
\bauthor{\binits{G.K.} \bsnm{{Skinner}}},
\batitle{{Observations of the X-ray source Cygnus X-1 from the Roentgen
  Observatory aboard the orbiting Mir-Kvant module}}.
\bjtitle{Astronomy Letters}
\bvolume{21},
\bfpage{794}--\blpage{803}
(\byear{1995})
\end{barticle}
\endbibitem

\bibitem[\protect\citeauthoryear{{Bouchet} et~al.}{1991}]{Bouchet91}
\begin{barticle}
\bauthor{\binits{L.} \bsnm{{Bouchet}}},
\bauthor{\binits{P.} \bsnm{{Mandrou}}},
\bauthor{\binits{J.P.} \bsnm{{Roques}}},
\bauthor{\binits{G.} \bsnm{{Vedrenne}}},
\bauthor{\binits{B.} \bsnm{{Cordier}}},
\bauthor{\binits{A.} \bsnm{{Goldwurm}}},
\bauthor{\binits{F.} \bsnm{{Lebrun}}},
\bauthor{\binits{J.} \bsnm{{Paul}}},
\bauthor{\binits{R.} \bsnm{{Sunyaev}}},
\bauthor{\binits{E.} \bsnm{{Churazov}}},
\bauthor{\binits{M.} \bsnm{{Gilfanov}}},
\bauthor{\binits{M.} \bsnm{{Pavlinsky}}},
\bauthor{\binits{S.} \bsnm{{Grebenev}}},
\bauthor{\binits{G.} \bsnm{{Babalyan}}},
\bauthor{\binits{I.} \bsnm{{Dekhanov}}},
\bauthor{\binits{N.} \bsnm{{Khavenson}}},
\batitle{{Sigma discovery of variable e(+)-e(-) annihilation radiation from the
  near Galactic center variable compact source 1E 1740.7 - 2942}}.
\bjtitle{\apjl}
\bvolume{383},
\bfpage{45}--\blpage{48}
(\byear{1991}).
doi:\doiurl{10.1086/186237}
\end{barticle}
\endbibitem

\bibitem[\protect\citeauthoryear{{Bouchet} et~al.}{2001}]{Bouchet2001;heao1}
\begin{barticle}
\bauthor{\binits{L.} \bsnm{{Bouchet}}},
\bauthor{\binits{J.P.} \bsnm{{Roques}}},
\bauthor{\binits{J.} \bsnm{{Ballet}}},
\bauthor{\binits{A.} \bsnm{{Goldwurm}}},
\bauthor{\binits{J.} \bsnm{{Paul}}},
\batitle{{The SIGMA/Granat Telescope: Calibration and Data Reduction}}.
\bjtitle{\apj}
\bvolume{548},
\bfpage{990}--\blpage{1009}
(\byear{2001}).
doi:\doiurl{10.1086/318997}
\end{barticle}
\endbibitem

\bibitem[\protect\citeauthoryear{{Bouchet} et~al.}{2009}]{Bouchet09}
\begin{bchapter}
\bauthor{\binits{L.} \bsnm{{Bouchet}}},
\bauthor{\binits{A.} \bsnm{{Strong}}},
\bauthor{\binits{E.} \bsnm{{Jourdain}}},
\bauthor{\binits{J.P.} \bsnm{{Roques}}},
\bauthor{\binits{T.} \bsnm{{Porter}}},
\bauthor{\binits{I.} \bsnm{{Moskalenko}}},
\bauthor{\binits{R.} \bsnm{{Diehl}}},
\bauthor{\binits{E.} \bsnm{{Orlando}}},
\bctitle{{A complete all-sky survey with INTEGRAL/SPI: sources census, hard
  X-ray diffuse emission and annihilation line}},
in \bbtitle{The Extreme Sky: Sampling the Universe above 10 keV},
\byear{2009},
p. \bfpage{16}
\end{bchapter}
\endbibitem

\bibitem[\protect\citeauthoryear{{Bowyer} et~al.}{1964}]{Bowyer64}
\begin{barticle}
\bauthor{\binits{S.} \bsnm{{Bowyer}}},
\bauthor{\binits{E.T.} \bsnm{{Byram}}},
\bauthor{\binits{T.A.} \bsnm{{Chubb}}},
\bauthor{\binits{H.} \bsnm{{Friedman}}},
\batitle{{X-ray Sources in the Galaxy}}.
\bjtitle{\nat}
\bvolume{201},
\bfpage{1307}--\blpage{1308}
(\byear{1964}).
doi:\doiurl{10.1038/2011307a0}
\end{barticle}
\endbibitem

\bibitem[\protect\citeauthoryear{{Bowyer} et~al.}{1965}]{Bowyer65}
\begin{barticle}
\bauthor{\binits{S.} \bsnm{{Bowyer}}},
\bauthor{\binits{E.T.} \bsnm{{Byram}}},
\bauthor{\binits{T.A.} \bsnm{{Chubb}}},
\bauthor{\binits{H.} \bsnm{{Friedman}}},
\batitle{{Observational results of X-ray astronomy}}.
\bjtitle{Annales d'Astrophysique}
\bvolume{28},
\bfpage{791}
(\byear{1965})
\end{barticle}
\endbibitem

\bibitem[\protect\citeauthoryear{{Bradt} et~al.}{1993}]{Bradt1993;rxte}
\begin{barticle}
\bauthor{\binits{H.V.} \bsnm{{Bradt}}},
\bauthor{\binits{R.E.} \bsnm{{Rothschild}}},
\bauthor{\binits{J.H.} \bsnm{{Swank}}},
\batitle{{X-ray timing explorer mission}}.
\bjtitle{\aaps}
\bvolume{97},
\bfpage{355}--\blpage{360}
(\byear{1993})
\end{barticle}
\endbibitem

\bibitem[\protect\citeauthoryear{{Braga} et~al.}{1995}]{Braga95}
\begin{botherref}
\oauthor{\binits{J.} \bsnm{{Braga}}},
\oauthor{\binits{F.} \bsnm{{D'Amico}}},
\oauthor{\binits{T.} \bsnm{{Villela}}},
{Development of the balloon-borne hard X-ray imaging experiment TIMAX}.
Advances in Space Research
\textbf{15}
(1995)
\end{botherref}
\endbibitem

\bibitem[\protect\citeauthoryear{{Braga} et~al.}{1990}]{Braga1990;exite}
\begin{barticle}
\bauthor{\binits{J.} \bsnm{{Braga}}},
\bauthor{\binits{C.E.} \bsnm{{Covault}}},
\bauthor{\binits{R.} \bsnm{{Manandhar}}},
\bauthor{\binits{J.E.} \bsnm{{Grindlay}}},
\batitle{{Hard X-ray arcmin imaging with the EXITE telescope.}}
\bjtitle{\rmxaa}
\bvolume{21},
\bfpage{633}--\blpage{637}
(\byear{1990})
\end{barticle}
\endbibitem

\bibitem[\protect\citeauthoryear{{Brenneman} et~al.}{2014}]{Brenneman14}
\begin{barticle}
\bauthor{\binits{L.W.} \bsnm{{Brenneman}}},
\bauthor{\binits{G.} \bsnm{{Madejski}}},
\bauthor{\binits{F.} \bsnm{{Fuerst}}},
\bauthor{\binits{G.} \bsnm{{Matt}}},
\bauthor{\binits{M.} \bsnm{{Elvis}}},
\bauthor{\binits{F.A.} \bsnm{{Harrison}}},
\bauthor{\binits{D.R.} \bsnm{{Ballantyne}}},
\bauthor{\binits{S.E.} \bsnm{{Boggs}}},
\bauthor{\binits{F.E.} \bsnm{{Christensen}}},
\bauthor{\binits{W.W.} \bsnm{{Craig}}},
\bauthor{\binits{A.C.} \bsnm{{Fabian}}},
\bauthor{\binits{B.W.} \bsnm{{Grefenstette}}},
\bauthor{\binits{C.J.} \bsnm{{Hailey}}},
\bauthor{\binits{K.K.} \bsnm{{Madsen}}},
\bauthor{\binits{A.} \bsnm{{Marinucci}}},
\bauthor{\binits{E.} \bsnm{{Rivers}}},
\bauthor{\binits{D.} \bsnm{{Stern}}},
\bauthor{\binits{D.J.} \bsnm{{Walton}}},
\bauthor{\binits{W.W.} \bsnm{{Zhang}}},
\batitle{{Measuring the Coronal Properties of IC 4329A with NuSTAR}}.
\bjtitle{\apj}
\bvolume{781},
\bfpage{83}
(\byear{2014}).
doi:\doiurl{10.1088/0004-637X/781/2/83}
\end{barticle}
\endbibitem

\bibitem[\protect\citeauthoryear{{Briggs} et~al.}{1994}]{Briggs94}
\begin{bchapter}
\bauthor{\binits{M.S.} \bsnm{{Briggs}}},
\bauthor{\binits{D.E.} \bsnm{{Gruber}}},
\bauthor{\binits{J.L.} \bsnm{{Matteson}}},
\bauthor{\binits{L.E.} \bsnm{{Peterson}}},
\bctitle{{A transient MeV range gamma-ray sources observed by HEAO-1.}},
in \bbtitle{American Institute of Physics Conference Series},
ed. by \beditor{\binits{C.E.} \bsnm{{Fichtel}}},
\beditor{\binits{N.} \bsnm{{Gehrels}}},
\beditor{\binits{J.P.} \bsnm{{Norris}}}
\bsertitle{American Institute of Physics Conference Series},
vol. \bseriesno{304},
\byear{1994},
pp. \bfpage{255}--\blpage{259}
\end{bchapter}
\endbibitem

\bibitem[\protect\citeauthoryear{{Briggs} et~al.}{2010}]{Briggs10}
\begin{barticle}
\bauthor{\binits{M.S.} \bsnm{{Briggs}}},
\bauthor{\binits{G.J.} \bsnm{{Fishman}}},
\bauthor{\binits{V.} \bsnm{{Connaughton}}},
\bauthor{\binits{P.N.} \bsnm{{Bhat}}},
\bauthor{\binits{W.S.} \bsnm{{Paciesas}}},
\bauthor{\binits{R.D.} \bsnm{{Preece}}},
\bauthor{\binits{C.} \bsnm{{Wilson-Hodge}}},
\bauthor{\binits{V.L.} \bsnm{{Chaplin}}},
\bauthor{\binits{R.M.} \bsnm{{Kippen}}},
\bauthor{\binits{A.} \bsnm{{von Kienlin}}},
\bauthor{\binits{C.A.} \bsnm{{Meegan}}},
\bauthor{\binits{E.} \bsnm{{Bissaldi}}},
\bauthor{\binits{J.R.} \bsnm{{Dwyer}}},
\bauthor{\binits{D.M.} \bsnm{{Smith}}},
\bauthor{\binits{R.H.} \bsnm{{Holzworth}}},
\bauthor{\binits{J.E.} \bsnm{{Grove}}},
\bauthor{\binits{A.} \bsnm{{Chekhtman}}},
\batitle{{First results on terrestrial gamma ray flashes from the Fermi
  Gamma-ray Burst Monitor}}.
\bjtitle{Journal of Geophysical Research (Space Physics)}
\bvolume{115},
\bfpage{07323}
(\byear{2010}).
doi:\doiurl{10.1029/2009JA015242}
\end{barticle}
\endbibitem

\bibitem[\protect\citeauthoryear{{Brini} et~al.}{1970a}]{Brini70a}
\begin{barticle}
\bauthor{\binits{D.} \bsnm{{Brini}}},
\bauthor{\binits{F.} \bsnm{{Frontera}}},
\bauthor{\binits{F.} \bsnm{{Fuligni}}},
\batitle{{Upper limits of X-ray fluxes from some quasars}}.
\bjtitle{Nuovo Cimento B Serie}
\bvolume{65},
\bfpage{181}--\blpage{186}
(\byear{1970}a).
doi:\doiurl{10.1007/BF02711194}
\end{barticle}
\endbibitem

\bibitem[\protect\citeauthoryear{{Brini} et~al.}{1970b}]{Brini70}
\begin{bchapter}
\bauthor{\binits{D.} \bsnm{{Brini}}},
\bauthor{\binits{F.} \bsnm{{Fuligni}}},
\bauthor{\binits{E.} \bsnm{{Horstman-Moretti}}},
\bctitle{{Measurement of the Cosmic X-Ray Background in the 25-200 KE V
  Range}},
in \bbtitle{Non-Solar X- and Gamma-Ray Astronomy},
ed. by \beditor{\binits{L.} \bsnm{{Gratton}}}
\bsertitle{IAU Symposium},
vol. \bseriesno{37},
\byear{1970}b,
p. \bfpage{321}
\end{bchapter}
\endbibitem

\bibitem[\protect\citeauthoryear{{Brini} et~al.}{1967}]{Brini67}
\begin{barticle}
\bauthor{\binits{D.} \bsnm{{Brini}}},
\bauthor{\binits{U.} \bsnm{{Ciriegi}}},
\bauthor{\binits{F.} \bsnm{{Fuligni}}},
\bauthor{\binits{E.} \bsnm{{Moretti}}},
\bauthor{\binits{G.} \bsnm{{Vespignani}}},
\batitle{{Cosmic X-Ray Sources in the 20-180 KEV Energy Range}}.
\bjtitle{\apj}
\bvolume{149},
\bfpage{429}
(\byear{1967}).
doi:\doiurl{10.1086/149267}
\end{barticle}
\endbibitem

\bibitem[\protect\citeauthoryear{{Brini} et~al.}{1971a}]{Brini71}
\begin{barticle}
\bauthor{\binits{D.} \bsnm{{Brini}}},
\bauthor{\binits{C.} \bsnm{{Cavani}}},
\bauthor{\binits{F.} \bsnm{{Frontera}}},
\bauthor{\binits{F.} \bsnm{{Fuligni}}},
\batitle{{Pulsars-X-ray emission from NP 0532 in the 20-200 keV range}}.
\bjtitle{Nature Physical Science}
\bvolume{232},
\bfpage{80}
(\byear{1971}a)
\end{barticle}
\endbibitem

\bibitem[\protect\citeauthoryear{{Brini} et~al.}{1971b}]{Brini1971;hxr70}
\begin{barticle}
\bauthor{\binits{D.} \bsnm{{Brini}}},
\bauthor{\binits{C.} \bsnm{{Cavani}}},
\bauthor{\binits{F.} \bsnm{{Frontera}}},
\bauthor{\binits{F.} \bsnm{{Fuligni}}},
\batitle{{Pulsed X-ray Emission from NP 0532 in the 20-200 keV Range}}.
\bjtitle{Nature Physical Science}
\bvolume{232},
\bfpage{79}--\blpage{81}
(\byear{1971}b).
doi:\doiurl{10.1038/physci232079b0}
\end{barticle}
\endbibitem

\bibitem[\protect\citeauthoryear{{Buff} et~al.}{1977}]{Buff1977;sas3}
\begin{barticle}
\bauthor{\binits{J.} \bsnm{{Buff}}},
\bauthor{\binits{G.} \bsnm{{Jernigan}}},
\bauthor{\binits{B.} \bsnm{{Laufer}}},
\bauthor{\binits{H.} \bsnm{{Bradt}}},
\bauthor{\binits{G.W.} \bsnm{{Clark}}},
\bauthor{\binits{W.H.G.} \bsnm{{Lewin}}},
\bauthor{\binits{T.} \bsnm{{Matilsky}}},
\bauthor{\binits{W.} \bsnm{{Mayer}}},
\bauthor{\binits{F.} \bsnm{{Primini}}},
\batitle{{Intense X-ray flares from Aquila X-1 and Circinus X-1}}.
\bjtitle{\apj}
\bvolume{212},
\bfpage{768}--\blpage{773}
(\byear{1977}).
doi:\doiurl{10.1086/155102}
\end{barticle}
\endbibitem

\bibitem[\protect\citeauthoryear{{Buivan} et~al.}{1979}]{Buivan79}
\begin{barticle}
\bauthor{\binits{N.A.} \bsnm{{Buivan}}},
\bauthor{\binits{K.R.} \bsnm{{Rao}}},
\bauthor{\binits{I.M.} \bsnm{{Martin}}},
\batitle{{Gamma-ray lines observed in balloon flights at high rigidity}}.
\bjtitle{\apss}
\bvolume{64},
\bfpage{339}--\blpage{346}
(\byear{1979}).
doi:\doiurl{10.1007/BF00639513}
\end{barticle}
\endbibitem

\bibitem[\protect\citeauthoryear{{Burderi} et~al.}{2000}]{Burderi00}
\begin{barticle}
\bauthor{\binits{L.} \bsnm{{Burderi}}},
\bauthor{\binits{T.} \bsnm{{Di Salvo}}},
\bauthor{\binits{N.R.} \bsnm{{Robba}}},
\bauthor{\binits{A.} \bsnm{{La Barbera}}},
\bauthor{\binits{M.} \bsnm{{Guainazzi}}},
\batitle{{The 0.1-100 KEV Spectrum of Centaurus X-3: Pulse Phase Spectroscopy
  of the Cyclotron Line and Magnetic Field Structure}}.
\bjtitle{\apj}
\bvolume{530},
\bfpage{429}--\blpage{440}
(\byear{2000}).
doi:\doiurl{10.1086/308336}
\end{barticle}
\endbibitem

\bibitem[\protect\citeauthoryear{{Burenin} et~al.}{1999}]{Burenin1999;granat}
\begin{barticle}
\bauthor{\binits{R.A.} \bsnm{{Burenin}}},
\bauthor{\binits{A.A.} \bsnm{{Vikhlinin}}},
\bauthor{\binits{M.R.} \bsnm{{Gilfanov}}},
\bauthor{\binits{O.V.} \bsnm{{Terekhov}}},
\bauthor{\binits{A.Y.} \bsnm{{Tkachenko}}},
\bauthor{\binits{S.Y.} \bsnm{{Sazonov}}},
\bauthor{\binits{E.M.} \bsnm{{Churazov}}},
\bauthor{\binits{R.A.} \bsnm{{Sunyaev}}},
\bauthor{\binits{P.} \bsnm{{Goldoni}}},
\bauthor{\binits{A.} \bsnm{{Claret}}},
\bauthor{\binits{A.} \bsnm{{Goldwurm}}},
\bauthor{\binits{J.} \bsnm{{Paul}}},
\bauthor{\binits{J.P.} \bsnm{{Roques}}},
\bauthor{\binits{E.} \bsnm{{Jourdain}}},
\bauthor{\binits{F.} \bsnm{{Pelaez}}},
\bauthor{\binits{G.} \bsnm{{Vedrenne}}},
\batitle{{GRANAT/SIGMA observation of the early afterglow from GRB 920723 in
  soft gamma-rays}}.
\bjtitle{\aap}
\bvolume{344},
\bfpage{53}--\blpage{56}
(\byear{1999})
\end{barticle}
\endbibitem

\bibitem[\protect\citeauthoryear{{Burrows} et~al.}{2011}]{Burrows11}
\begin{barticle}
\bauthor{\binits{D.N.} \bsnm{{Burrows}}},
\bauthor{\binits{J.A.} \bsnm{{Kennea}}},
\bauthor{\binits{G.} \bsnm{{Ghisellini}}},
\bauthor{\binits{V.} \bsnm{{Mangano}}},
\bauthor{\binits{B.} \bsnm{{Zhang}}},
\bauthor{\binits{K.L.} \bsnm{{Page}}},
\bauthor{\binits{M.} \bsnm{{Eracleous}}},
\bauthor{\binits{P.} \bsnm{{Romano}}},
\bauthor{\binits{T.} \bsnm{{Sakamoto}}},
\bauthor{\binits{A.D.} \bsnm{{Falcone}}},
\bauthor{\binits{J.P.} \bsnm{{Osborne}}},
\bauthor{\binits{S.} \bsnm{{Campana}}},
\bauthor{\binits{A.P.} \bsnm{{Beardmore}}},
\bauthor{\binits{A.A.} \bsnm{{Breeveld}}},
\bauthor{\binits{M.M.} \bsnm{{Chester}}},
\bauthor{\binits{R.} \bsnm{{Corbet}}},
\bauthor{\binits{S.} \bsnm{{Covino}}},
\bauthor{\binits{J.R.} \bsnm{{Cummings}}},
\bauthor{\binits{P.} \bsnm{{D'Avanzo}}},
\bauthor{\binits{V.} \bsnm{{D'Elia}}},
\bauthor{\binits{P.} \bsnm{{Esposito}}},
\bauthor{\binits{P.A.} \bsnm{{Evans}}},
\bauthor{\binits{D.} \bsnm{{Fugazza}}},
\bauthor{\binits{J.M.} \bsnm{{Gelbord}}},
\bauthor{\binits{K.} \bsnm{{Hiroi}}},
\bauthor{\binits{S.T.} \bsnm{{Holland}}},
\bauthor{\binits{K.Y.} \bsnm{{Huang}}},
\bauthor{\binits{M.} \bsnm{{Im}}},
\bauthor{\binits{G.} \bsnm{{Israel}}},
\bauthor{\binits{Y.} \bsnm{{Jeon}}},
\bauthor{\binits{Y.-B.} \bsnm{{Jeon}}},
\bauthor{\binits{H.D.} \bsnm{{Jun}}},
\bauthor{\binits{N.} \bsnm{{Kawai}}},
\bauthor{\binits{J.H.} \bsnm{{Kim}}},
\bauthor{\binits{H.A.} \bsnm{{Krimm}}},
\bauthor{\binits{F.E.} \bsnm{{Marshall}}},
\bauthor{\bsnm{{P.~M{\'e}sz{\'a}ros}}},
\bauthor{\binits{H.} \bsnm{{Negoro}}},
\bauthor{\binits{N.} \bsnm{{Omodei}}},
\bauthor{\binits{W.-K.} \bsnm{{Park}}},
\bauthor{\binits{J.S.} \bsnm{{Perkins}}},
\bauthor{\binits{M.} \bsnm{{Sugizaki}}},
\bauthor{\binits{H.-I.} \bsnm{{Sung}}},
\bauthor{\binits{G.} \bsnm{{Tagliaferri}}},
\bauthor{\binits{E.} \bsnm{{Troja}}},
\bauthor{\binits{Y.} \bsnm{{Ueda}}},
\bauthor{\binits{Y.} \bsnm{{Urata}}},
\bauthor{\binits{R.} \bsnm{{Usui}}},
\bauthor{\binits{L.A.} \bsnm{{Antonelli}}},
\bauthor{\binits{S.D.} \bsnm{{Barthelmy}}},
\bauthor{\binits{G.} \bsnm{{Cusumano}}},
\bauthor{\binits{P.} \bsnm{{Giommi}}},
\bauthor{\binits{A.} \bsnm{{Melandri}}},
\bauthor{\binits{M.} \bsnm{{Perri}}},
\bauthor{\binits{J.L.} \bsnm{{Racusin}}},
\bauthor{\binits{B.} \bsnm{{Sbarufatti}}},
\bauthor{\binits{M.H.} \bsnm{{Siegel}}},
\bauthor{\binits{N.} \bsnm{{Gehrels}}},
\batitle{{Relativistic jet activity from the tidal disruption of a star by a
  massive black hole}}.
\bjtitle{\nat}
\bvolume{476},
\bfpage{421}--\blpage{424}
(\byear{2011}).
doi:\doiurl{10.1038/nature10374}
\end{barticle}
\endbibitem

\bibitem[\protect\citeauthoryear{{Butler} et~al.}{1981}]{Butler81}
\begin{barticle}
\bauthor{\binits{R.C.} \bsnm{{Butler}}},
\bauthor{\binits{G.} \bsnm{{Di Cocco}}},
\bauthor{\binits{G.} \bsnm{{Boella}}},
\bauthor{\binits{A.} \bsnm{{della Ventura}}},
\bauthor{\binits{F.} \bsnm{{Perotti}}},
\bauthor{\binits{G.} \bsnm{{Villa}}},
\bauthor{\binits{R.E.} \bsnm{{Baker}}},
\bauthor{\binits{L.} \bsnm{{Bassani}}},
\bauthor{\binits{A.J.} \bsnm{{Dean}}},
\bauthor{\binits{D.} \bsnm{{Ramsden}}},
\batitle{{Low energy gamma ray observations of NGC 4151}}.
\bjtitle{International Cosmic Ray Conference}
\bvolume{1},
\bfpage{226}--\blpage{229}
(\byear{1981})
\end{barticle}
\endbibitem

\bibitem[\protect\citeauthoryear{{Camero-Arranz}
  et~al.}{2010}]{Camero-Arranz10}
\begin{barticle}
\bauthor{\binits{A.} \bsnm{{Camero-Arranz}}},
\bauthor{\binits{M.H.} \bsnm{{Finger}}},
\bauthor{\binits{N.R.} \bsnm{{Ikhsanov}}},
\bauthor{\binits{C.A.} \bsnm{{Wilson-Hodge}}},
\bauthor{\binits{E.} \bsnm{{Beklen}}},
\batitle{{New Torque Reversal and Spin-up of 4U 1626-67 Observed by
  Fermi/Gamma-ray Burst Monitor and Swift/Burst Alert Telescope}}.
\bjtitle{\apj}
\bvolume{708},
\bfpage{1500}--\blpage{1506}
(\byear{2010}).
doi:\doiurl{10.1088/0004-637X/708/2/1500}
\end{barticle}
\endbibitem

\bibitem[\protect\citeauthoryear{{Cappi} et~al.}{1999}]{Cappi99}
\begin{barticle}
\bauthor{\binits{M.} \bsnm{{Cappi}}},
\bauthor{\binits{L.} \bsnm{{Bassani}}},
\bauthor{\binits{A.} \bsnm{{Comastri}}},
\bauthor{\binits{M.} \bsnm{{Guainazzi}}},
\bauthor{\binits{T.} \bsnm{{Maccacaro}}},
\bauthor{\binits{G.} \bsnm{{Malaguti}}},
\bauthor{\binits{G.} \bsnm{{Matt}}},
\bauthor{\binits{G.G.C.} \bsnm{{Palumbo}}},
\bauthor{\binits{P.} \bsnm{{Blanco}}},
\bauthor{\binits{M.} \bsnm{{Dadina}}},
\bauthor{\binits{D.} \bsnm{{dal Fiume}}},
\bauthor{\binits{G.} \bsnm{{Di Cocco}}},
\bauthor{\binits{A.C.} \bsnm{{Fabian}}},
\bauthor{\binits{F.} \bsnm{{Frontera}}},
\bauthor{\binits{R.} \bsnm{{Maiolino}}},
\bauthor{\binits{L.} \bsnm{{Piro}}},
\bauthor{\binits{M.} \bsnm{{Trifoglio}}},
\bauthor{\binits{N.} \bsnm{{Zhang}}},
\batitle{{BeppoSAX observations of MKN 3: Piercing through the torus of a
  Seyfert 2 galaxy}}.
\bjtitle{\aap}
\bvolume{344},
\bfpage{857}--\blpage{867}
(\byear{1999})
\end{barticle}
\endbibitem

\bibitem[\protect\citeauthoryear{{Castro-Tirado}
  et~al.}{1992}]{Castro-Tirado92}
\begin{botherref}
\oauthor{\binits{A.J.} \bsnm{{Castro-Tirado}}},
\oauthor{\binits{S.} \bsnm{{Brandt}}},
\oauthor{\binits{N.} \bsnm{{Lund}}},
{GRS 1915+105}.
\iaucirc
\textbf{5590}
(1992)
\end{botherref}
\endbibitem

\bibitem[\protect\citeauthoryear{{Cavani} et~al.}{1971}]{Cavani71}
\begin{barticle}
\bauthor{\binits{C.} \bsnm{{Cavani}}},
\bauthor{\binits{F.} \bsnm{{Frontera}}},
\bauthor{\binits{F.} \bsnm{{Fuligni}}},
\bauthor{\binits{D.} \bsnm{{Brini}}},
\batitle{{Hard X-ray Spectrum of NP 0532}}.
\bjtitle{Nature Physical Science}
\bvolume{233},
\bfpage{153}--\blpage{155}
(\byear{1971}).
doi:\doiurl{10.1038/physci233153a0}
\end{barticle}
\endbibitem

\bibitem[\protect\citeauthoryear{{Chakrabarty} et~al.}{2014}]{Chakrabarty14}
\begin{barticle}
\bauthor{\binits{D.} \bsnm{{Chakrabarty}}},
\bauthor{\binits{J.A.} \bsnm{{Tomsick}}},
\bauthor{\binits{B.W.} \bsnm{{Grefenstette}}},
\bauthor{\binits{D.} \bsnm{{Psaltis}}},
\bauthor{\binits{M.} \bsnm{{Bachetti}}},
\bauthor{\binits{D.} \bsnm{{Barret}}},
\bauthor{\binits{S.E.} \bsnm{{Boggs}}},
\bauthor{\binits{F.E.} \bsnm{{Christensen}}},
\bauthor{\binits{W.W.} \bsnm{{Craig}}},
\bauthor{\binits{F.} \bsnm{{F{\"u}rst}}},
\bauthor{\binits{C.J.} \bsnm{{Hailey}}},
\bauthor{\binits{F.A.} \bsnm{{Harrison}}},
\bauthor{\binits{V.M.} \bsnm{{Kaspi}}},
\bauthor{\binits{J.M.} \bsnm{{Miller}}},
\bauthor{\binits{M.A.} \bsnm{{Nowak}}},
\bauthor{\binits{V.} \bsnm{{Rana}}},
\bauthor{\binits{D.} \bsnm{{Stern}}},
\bauthor{\binits{D.R.} \bsnm{{Wik}}},
\bauthor{\binits{J.} \bsnm{{Wilms}}},
\bauthor{\binits{W.W.} \bsnm{{Zhang}}},
\batitle{{A Hard X-Ray Power-law Spectral Cutoff in Centaurus X-4}}.
\bjtitle{\apj}
\bvolume{797},
\bfpage{92}
(\byear{2014}).
doi:\doiurl{10.1088/0004-637X/797/2/92}
\end{barticle}
\endbibitem

\bibitem[\protect\citeauthoryear{{Chauvin} et~al.}{2016}]{Chauvin16}
\begin{barticle}
\bauthor{\binits{M.} \bsnm{{Chauvin}}},
\bauthor{\binits{H.-G.} \bsnm{{Flor{\'e}n}}},
\bauthor{\binits{M.} \bsnm{{Jackson}}},
\bauthor{\binits{T.} \bsnm{{Kamae}}},
\bauthor{\binits{T.} \bsnm{{Kawano}}},
\bauthor{\binits{M.} \bsnm{{Kiss}}},
\bauthor{\binits{M.} \bsnm{{Kole}}},
\bauthor{\binits{V.} \bsnm{{Mikhalev}}},
\bauthor{\binits{E.} \bsnm{{Moretti}}},
\bauthor{\binits{G.} \bsnm{{Olofsson}}},
\bauthor{\binits{S.} \bsnm{{Rydstr{\"o}m}}},
\bauthor{\binits{H.} \bsnm{{Takahashi}}},
\bauthor{\binits{J.} \bsnm{{Lind}}},
\bauthor{\binits{J.-E.} \bsnm{{Str{\"o}mberg}}},
\bauthor{\binits{O.} \bsnm{{Welin}}},
\bauthor{\binits{A.} \bsnm{{Iyudin}}},
\bauthor{\binits{D.} \bsnm{{Shifrin}}},
\bauthor{\binits{M.} \bsnm{{Pearce}}},
\batitle{{The design and flight performance of the PoGOLite Pathfinder
  balloon-borne hard X-ray polarimeter}}.
\bjtitle{Experimental Astronomy}
\bvolume{41},
\bfpage{17}--\blpage{41}
(\byear{2016}).
doi:\doiurl{10.1007/s10686-015-9474-x}
\end{barticle}
\endbibitem

\bibitem[\protect\citeauthoryear{{Chen} et~al.}{2006}]{Chen06}
\begin{bchapter}
\bauthor{\binits{C.M.H.} \bsnm{{Chen}}},
\bauthor{\binits{W.H.} \bsnm{{Baumgartner}}},
\bauthor{\binits{J.C.} \bsnm{{Chonko}}},
\bauthor{\binits{F.E.} \bsnm{{Christiansen}}},
\bauthor{\binits{W.R.} \bsnm{{Cook}}},
\bauthor{\binits{W.W.} \bsnm{{Craig}}},
\bauthor{\binits{C.J.} \bsnm{{Hailey}}},
\bauthor{\binits{F.A.} \bsnm{{Harrison}}},
\bauthor{\binits{C.P.} \bsnm{{Jensen}}},
\bauthor{\binits{J.E.} \bsnm{{Koglin}}},
\bauthor{\binits{K.} \bsnm{{Kruse Madsen}}},
\bauthor{\binits{R.} \bsnm{{McLean}}},
\bauthor{\binits{M.J.} \bsnm{{Pivovaroff}}},
\bauthor{\binits{K.P.} \bsnm{{Ziock}}},
\bctitle{{In-flight Performance of the Balloon-borne High Energy Focusing
  Telescope}},
in \bbtitle{AAS/High Energy Astrophysics Division \#9}.
\bsertitle{Bulletin of the American Astronomical Society},
vol. \bseriesno{38},
\byear{2006},
p. \bfpage{383}
\end{bchapter}
\endbibitem

\bibitem[\protect\citeauthoryear{{Cheng} et~al.}{1997}]{Cheng97}
\begin{bchapter}
\bauthor{\binits{L.} \bsnm{{Cheng}}},
\bauthor{\binits{M.} \bsnm{{Leventhal}}},
\bauthor{\binits{J.} \bsnm{{Tueller}}},
\bauthor{\binits{N.} \bsnm{{Gehrels}}},
\bauthor{\binits{A.} \bsnm{{Parsons}}},
\bauthor{\binits{B.J.} \bsnm{{Teegarden}}},
\bauthor{\binits{S.D.} \bsnm{{Barthelmy}}},
\bauthor{\binits{J.E.} \bsnm{{Naya}}},
\bauthor{\binits{L.M.} \bsnm{{Bartlett}}},
\bctitle{{GRIS Detection of Positron Annihilation Radiation from the Milky-Way
  Galaxy}},
in \bbtitle{APS April Meeting Abstracts},
\byear{1997}
\end{bchapter}
\endbibitem

\bibitem[\protect\citeauthoryear{{Cherry} et~al.}{1995}]{Cherry95;MARGIE}
\begin{barticle}
\bauthor{\binits{M.L.} \bsnm{{Cherry}}},
\bauthor{\binits{P.P.} \bsnm{{Altice}}},
\bauthor{\binits{M.B.} \bsnm{{Barakat}}},
\bauthor{\binits{C.X.} \bsnm{{Chen}}},
\bauthor{\binits{B.K.} \bsnm{{Dann}}},
\bauthor{\binits{A.} \bsnm{{Drake}}},
\bauthor{\binits{S.B.} \bsnm{{Ellison}}},
\bauthor{\binits{C.J.} \bsnm{{Gagne}}},
\bauthor{\binits{J.} \bsnm{{Gordon}}},
\bauthor{\binits{T.G.} \bsnm{{Guzik}}},
\bauthor{\binits{R.} \bsnm{{Lockwood}}},
\bauthor{\binits{K.} \bsnm{{Johnston}}},
\bauthor{\binits{J.} \bsnm{{Macri}}},
\bauthor{\binits{M.L.} \bsnm{{McConnell}}},
\bauthor{\binits{V.} \bsnm{{Nagarkar}}},
\bauthor{\binits{J.M.} \bsnm{{Ryan}}},
\bauthor{\binits{S.} \bsnm{{Vasile}}},
\batitle{{A New Balloon-Borne Detector for High Angular Resolution Hard X-Ray
  Astronomy}}.
\bjtitle{International Cosmic Ray Conference}
\bvolume{2},
\bfpage{45}
(\byear{1995})
\end{barticle}
\endbibitem

\bibitem[\protect\citeauthoryear{{Chiappetti} et~al.}{1999}]{Chiappetti99}
\begin{barticle}
\bauthor{\binits{L.} \bsnm{{Chiappetti}}},
\bauthor{\binits{F.} \bsnm{{Haardt}}},
\bauthor{\binits{A.} \bsnm{{Treves}}},
\bauthor{\binits{D.} \bsnm{{Ricci}}},
\bauthor{\binits{A.} \bsnm{{Santangelo}}},
\bauthor{\binits{S.} \bsnm{{Mereghetti}}},
\bauthor{\binits{T.} \bsnm{{Belloni}}},
\batitle{{BeppoSAX observations of the exotic black hole candidate GX 339-4}}.
\bjtitle{Nuclear Physics B Proceedings Supplements}
\bvolume{69},
\bfpage{340}--\blpage{343}
(\byear{1999}).
doi:\doiurl{10.1016/S0920-5632(98)00236-9}
\end{barticle}
\endbibitem

\bibitem[\protect\citeauthoryear{{Chitnis} et~al.}{1993}]{Chitnis93}
\begin{barticle}
\bauthor{\binits{V.R.} \bsnm{{Chitnis}}},
\bauthor{\binits{P.C.} \bsnm{{Agrawal}}},
\bauthor{\binits{R.K.} \bsnm{{Manchanda}}},
\bauthor{\binits{A.R.} \bsnm{{Rao}}},
\batitle{{Hard X-ray spectra of some X-ray binaries from balloon-borne
  observations}}.
\bjtitle{Bulletin of the Astronomical Society of India}
\bvolume{21},
\bfpage{555}--\blpage{559}
(\byear{1993})
\end{barticle}
\endbibitem

\bibitem[\protect\citeauthoryear{{Chodil} et~al.}{1968}]{Chodil68}
\begin{barticle}
\bauthor{\binits{G.} \bsnm{{Chodil}}},
\bauthor{\binits{H.} \bsnm{{Mark}}},
\bauthor{\binits{R.} \bsnm{{Rodrigues}}},
\bauthor{\binits{C.D.} \bsnm{{Swift}}},
\batitle{{Observation of the Cygnus Region with a Balloon-Borne X-Ray
  Detector}}.
\bjtitle{\apjl}
\bvolume{151},
\bfpage{1}
(\byear{1968}).
doi:\doiurl{10.1086/180127}
\end{barticle}
\endbibitem

\bibitem[\protect\citeauthoryear{{Churazov} et~al.}{2014}]{Churazov14}
\begin{barticle}
\bauthor{\binits{E.} \bsnm{{Churazov}}},
\bauthor{\binits{R.} \bsnm{{Sunyaev}}},
\bauthor{\binits{J.} \bsnm{{Isern}}},
\bauthor{\binits{J.} \bsnm{{Kn{\"o}dlseder}}},
\bauthor{\binits{P.} \bsnm{{Jean}}},
\bauthor{\binits{F.} \bsnm{{Lebrun}}},
\bauthor{\binits{N.} \bsnm{{Chugai}}},
\bauthor{\binits{S.} \bsnm{{Grebenev}}},
\bauthor{\binits{E.} \bsnm{{Bravo}}},
\bauthor{\binits{S.} \bsnm{{Sazonov}}},
\bauthor{\binits{M.} \bsnm{{Renaud}}},
\batitle{{Cobalt-56 {$\gamma$}-ray emission lines from the type Ia supernova
  2014J}}.
\bjtitle{\nat}
\bvolume{512},
\bfpage{406}--\blpage{408}
(\byear{2014}).
doi:\doiurl{10.1038/nature13672}
\end{barticle}
\endbibitem

\bibitem[\protect\citeauthoryear{{Church} et~al.}{1998}]{Church98}
\begin{barticle}
\bauthor{\binits{M.J.} \bsnm{{Church}}},
\bauthor{\binits{A.N.} \bsnm{{Parmar}}},
\bauthor{\binits{M.} \bsnm{{Balucinska-Church}}},
\bauthor{\binits{T.} \bsnm{{Oosterbroek}}},
\bauthor{\binits{D.} \bsnm{{dal Fiume}}},
\bauthor{\binits{M.} \bsnm{{Orlandini}}},
\batitle{{Progressive covering in dipping and Comptonization in the spectrum of
  XB 1916-053 from the BeppoSAX observation}}.
\bjtitle{\aap}
\bvolume{338},
\bfpage{556}--\blpage{562}
(\byear{1998})
\end{barticle}
\endbibitem

\bibitem[\protect\citeauthoryear{{Civano} et~al.}{2015}]{Civano15}
\begin{barticle}
\bauthor{\binits{F.} \bsnm{{Civano}}},
\bauthor{\binits{R.C.} \bsnm{{Hickox}}},
\bauthor{\binits{S.} \bsnm{{Puccetti}}},
\bauthor{\binits{A.} \bsnm{{Comastri}}},
\bauthor{\binits{J.R.} \bsnm{{Mullaney}}},
\bauthor{\binits{L.} \bsnm{{Zappacosta}}},
\bauthor{\binits{S.M.} \bsnm{{LaMassa}}},
\bauthor{\binits{J.} \bsnm{{Aird}}},
\bauthor{\binits{D.M.} \bsnm{{Alexander}}},
\bauthor{\binits{D.R.} \bsnm{{Ballantyne}}},
\bauthor{\binits{F.E.} \bsnm{{Bauer}}},
\bauthor{\binits{W.N.} \bsnm{{Brandt}}},
\bauthor{\binits{S.E.} \bsnm{{Boggs}}},
\bauthor{\binits{F.E.} \bsnm{{Christensen}}},
\bauthor{\binits{W.W.} \bsnm{{Craig}}},
\bauthor{\binits{A.} \bsnm{{Del-Moro}}},
\bauthor{\binits{M.} \bsnm{{Elvis}}},
\bauthor{\binits{K.} \bsnm{{Forster}}},
\bauthor{\binits{P.} \bsnm{{Gandhi}}},
\bauthor{\binits{B.W.} \bsnm{{Grefenstette}}},
\bauthor{\binits{C.J.} \bsnm{{Hailey}}},
\bauthor{\binits{F.A.} \bsnm{{Harrison}}},
\bauthor{\binits{G.B.} \bsnm{{Lansbury}}},
\bauthor{\binits{B.} \bsnm{{Luo}}},
\bauthor{\binits{K.} \bsnm{{Madsen}}},
\bauthor{\binits{C.} \bsnm{{Saez}}},
\bauthor{\binits{D.} \bsnm{{Stern}}},
\bauthor{\binits{E.} \bsnm{{Treister}}},
\bauthor{\binits{M.C.} \bsnm{{Urry}}},
\bauthor{\binits{D.R.} \bsnm{{Wik}}},
\bauthor{\binits{W.} \bsnm{{Zhang}}},
\batitle{{The Nustar Extragalactic Surveys: Overview and Catalog from the
  COSMOS Field}}.
\bjtitle{\apj}
\bvolume{808},
\bfpage{185}
(\byear{2015}).
doi:\doiurl{10.1088/0004-637X/808/2/185}
\end{barticle}
\endbibitem

\bibitem[\protect\citeauthoryear{{Clark}}{1965}]{Clark1965}
\begin{barticle}
\bauthor{\binits{G.W.} \bsnm{{Clark}}},
\batitle{{Balloon Observation of the X-Ray Spectrum of the Crab Nebula Above 15
  keV}}.
\bjtitle{Physical Review Letters}
\bvolume{14},
\bfpage{91}--\blpage{94}
(\byear{1965}).
doi:\doiurl{10.1103/PhysRevLett.14.91}
\end{barticle}
\endbibitem

\bibitem[\protect\citeauthoryear{{Clark} et~al.}{1972}]{Clark72}
\begin{barticle}
\bauthor{\binits{G.W.} \bsnm{{Clark}}},
\bauthor{\binits{H.V.} \bsnm{{Bradt}}},
\bauthor{\binits{W.H.G.} \bsnm{{Lewin}}},
\bauthor{\binits{T.H.} \bsnm{{Markert}}},
\bauthor{\binits{H.W.} \bsnm{{Schnopper}}},
\bauthor{\binits{G.F.} \bsnm{{Sprott}}},
\batitle{{Measurement of the Position and Spectrum of Hercules X-1 from the
  OSO-7 Satellite}}.
\bjtitle{\apjl}
\bvolume{177},
\bfpage{109}
(\byear{1972}).
doi:\doiurl{10.1086/181062}
\end{barticle}
\endbibitem

\bibitem[\protect\citeauthoryear{{Clark} et~al.}{1973}]{Clark1973;oso7}
\begin{barticle}
\bauthor{\binits{G.W.} \bsnm{{Clark}}},
\bauthor{\binits{H.V.} \bsnm{{Bradt}}},
\bauthor{\binits{W.H.G.} \bsnm{{Lewin}}},
\bauthor{\binits{T.H.} \bsnm{{Markert}}},
\bauthor{\binits{H.W.} \bsnm{{Schnopper}}},
\bauthor{\binits{G.F.} \bsnm{{Sprott}}},
\batitle{{Observations of Taurus X-I by the 1-60 keV X-Ray Detector on the
  OSO-7}}.
\bjtitle{\apj}
\bvolume{179},
\bfpage{263}--\blpage{268}
(\byear{1973}).
doi:\doiurl{10.1086/151866}
\end{barticle}
\endbibitem

\bibitem[\protect\citeauthoryear{{Coe} et~al.}{1976a}]{Coe76c}
\begin{barticle}
\bauthor{\binits{M.J.} \bsnm{{Coe}}},
\bauthor{\binits{A.R.} \bsnm{{Engel}}},
\bauthor{\binits{J.J.} \bsnm{{Quenby}}},
\batitle{{Anti-correlated hard and soft X-ray intensity variations of the
  black-hole candidates CYG X-1 and A0620-00}}.
\bjtitle{\nat}
\bvolume{259},
\bfpage{544}
(\byear{1976}a).
doi:\doiurl{10.1038/259544a0}
\end{barticle}
\endbibitem

\bibitem[\protect\citeauthoryear{{Coe} et~al.}{1976b}]{Coe76}
\begin{barticle}
\bauthor{\binits{M.J.} \bsnm{{Coe}}},
\bauthor{\binits{A.R.} \bsnm{{Engel}}},
\bauthor{\binits{J.J.} \bsnm{{Quenby}}},
\batitle{{Hard X-ray observations in the region of Centaurus}}.
\bjtitle{\mnras}
\bvolume{177},
\bfpage{31}--\blpage{36}
(\byear{1976}b)
\end{barticle}
\endbibitem

\bibitem[\protect\citeauthoryear{{Coe} et~al.}{1976c}]{Coe76b}
\begin{barticle}
\bauthor{\binits{M.J.} \bsnm{{Coe}}},
\bauthor{\binits{A.R.} \bsnm{{Engel}}},
\bauthor{\binits{J.J.} \bsnm{{Quenby}}},
\batitle{{Hard X-ray observations of Circinus X-1 during an outburst}}.
\bjtitle{\nat}
\bvolume{262},
\bfpage{563}
(\byear{1976}c).
doi:\doiurl{10.1038/262563a0}
\end{barticle}
\endbibitem

\bibitem[\protect\citeauthoryear{{Coe} et~al.}{1978}]{Coe78b}
\begin{barticle}
\bauthor{\binits{M.J.} \bsnm{{Coe}}},
\bauthor{\binits{A.R.} \bsnm{{Engel}}},
\bauthor{\binits{J.J.} \bsnm{{Quenby}}},
\batitle{{Hard X-ray observations of white dwarf binary systems}}.
\bjtitle{\nat}
\bvolume{272},
\bfpage{37}
(\byear{1978}).
doi:\doiurl{10.1038/272037a0}
\end{barticle}
\endbibitem

\bibitem[\protect\citeauthoryear{{Coe} et~al.}{1975}]{Coe75}
\begin{barticle}
\bauthor{\binits{M.J.} \bsnm{{Coe}}},
\bauthor{\binits{G.F.} \bsnm{{Carpenter}}},
\bauthor{\binits{A.R.} \bsnm{{Engel}}},
\bauthor{\binits{J.J.} \bsnm{{Quenby}}},
\batitle{{Hard X-ray measurements of nova A0535+26 in Taurus}}.
\bjtitle{\nat}
\bvolume{256},
\bfpage{630}
(\byear{1975}).
doi:\doiurl{10.1038/256630a0}
\end{barticle}
\endbibitem

\bibitem[\protect\citeauthoryear{{Coe} et~al.}{1977}]{Coe77}
\begin{barticle}
\bauthor{\binits{M.J.} \bsnm{{Coe}}},
\bauthor{\binits{A.R.} \bsnm{{Engel}}},
\bauthor{\binits{J.J.} \bsnm{{Quenby}}},
\bauthor{\binits{C.S.} \bsnm{{Dyer}}},
\batitle{{A line feature at 64 keV in the X-ray spectrum of HER X-1}}.
\bjtitle{\nat}
\bvolume{268},
\bfpage{508}
(\byear{1977}).
doi:\doiurl{10.1038/268508a0}
\end{barticle}
\endbibitem

\bibitem[\protect\citeauthoryear{{Coe} et~al.}{1978}]{Coe78c}
\begin{barticle}
\bauthor{\binits{M.J.} \bsnm{{Coe}}},
\bauthor{\binits{A.R.} \bsnm{{Engel}}},
\bauthor{\binits{J.J.} \bsnm{{Quenby}}},
\bauthor{\binits{G.F.} \bsnm{{Carpenter}}},
\bauthor{\binits{S.J.} \bsnm{{Bell-Burnell}}},
\bauthor{\binits{P.J.N.} \bsnm{{Davison}}},
\batitle{{The X-ray source Serpens X-1 - Ariel 5 observations and discussion of
  models for the spectrum and time variability}}.
\bjtitle{\mnras}
\bvolume{184},
\bfpage{355}--\blpage{364}
(\byear{1978})
\end{barticle}
\endbibitem

\bibitem[\protect\citeauthoryear{{Coe} et~al.}{1979}]{Coe79}
\begin{barticle}
\bauthor{\binits{M.J.} \bsnm{{Coe}}},
\bauthor{\binits{B.R.} \bsnm{{Dennis}}},
\bauthor{\binits{J.F.} \bsnm{{Dolan}}},
\bauthor{\binits{C.J.} \bsnm{{Crannell}}},
\bauthor{\binits{K.J.} \bsnm{{Frost}}},
\bauthor{\binits{L.E.} \bsnm{{Orwig}}},
\batitle{{X-ray observations of AM Herculis from OSO 8}}.
\bjtitle{\nat}
\bvolume{279},
\bfpage{509}
(\byear{1979}).
doi:\doiurl{10.1038/279509a0}
\end{barticle}
\endbibitem

\bibitem[\protect\citeauthoryear{{Coe} et~al.}{1981a}]{Coe81}
\begin{barticle}
\bauthor{\binits{M.J.} \bsnm{{Coe}}},
\bauthor{\binits{A.R.} \bsnm{{Engel}}},
\bauthor{\binits{A.J.} \bsnm{{Evans}}},
\bauthor{\binits{J.J.} \bsnm{{Quenby}}},
\batitle{{Ariel 5 hard X-ray studies of the galactic center region}}.
\bjtitle{\apj}
\bvolume{243},
\bfpage{155}--\blpage{160}
(\byear{1981}a).
doi:\doiurl{10.1086/158578}
\end{barticle}
\endbibitem

\bibitem[\protect\citeauthoryear{{Coe} et~al.}{1981b}]{Coe81b}
\begin{barticle}
\bauthor{\binits{M.J.} \bsnm{{Coe}}},
\bauthor{\binits{L.} \bsnm{{Bassani}}},
\bauthor{\binits{A.R.} \bsnm{{Engel}}},
\bauthor{\binits{J.J.} \bsnm{{Quenby}}},
\batitle{{The hard X-ray spectrum of NGC 4151}}.
\bjtitle{\mnras}
\bvolume{195},
\bfpage{241}--\blpage{244}
(\byear{1981}b)
\end{barticle}
\endbibitem

\bibitem[\protect\citeauthoryear{{Coe} et~al.}{1982}]{Coe82}
\begin{barticle}
\bauthor{\binits{M.J.} \bsnm{{Coe}}},
\bauthor{\binits{G.F.} \bsnm{{Carpenter}}},
\bauthor{\binits{A.R.} \bsnm{{Engel}}},
\bauthor{\binits{J.J.} \bsnm{{Quenby}}},
\batitle{{Journal of high energy X-ray observations from the Ariel V
  satellite}}.
\bjtitle{\mnras}
\bvolume{200},
\bfpage{385}--\blpage{390}
(\byear{1982})
\end{barticle}
\endbibitem

\bibitem[\protect\citeauthoryear{{Coe} et~al.}{1990}]{Coe90}
\begin{barticle}
\bauthor{\binits{M.J.} \bsnm{{Coe}}},
\bauthor{\binits{I.R.} \bsnm{{Carstairs}}},
\bauthor{\binits{A.J.} \bsnm{{Court}}},
\bauthor{\binits{S.R.} \bsnm{{Davies}}},
\bauthor{\binits{A.J.} \bsnm{{Dean}}},
\bauthor{\binits{N.A.} \bsnm{{Dipper}}},
\bauthor{\binits{R.A.} \bsnm{{Lewis}}},
\bauthor{\binits{F.} \bsnm{{Perotti}}},
\bauthor{\binits{E.} \bsnm{{Quadrini}}},
\bauthor{\binits{A.} \bsnm{{Bazzano}}},
\bauthor{\binits{P.} \bsnm{{Ubertini}}},
\bauthor{\binits{J.B.} \bsnm{{Stephen}}},
\batitle{{High-energy X-ray observations of A0535 + 26}}.
\bjtitle{\mnras}
\bvolume{243},
\bfpage{475}--\blpage{479}
(\byear{1990})
\end{barticle}
\endbibitem

\bibitem[\protect\citeauthoryear{{Collmar} et~al.}{1999}]{Collmar99}
\begin{barticle}
\bauthor{\binits{W.} \bsnm{{Collmar}}},
\bauthor{\binits{K.} \bsnm{{Bennett}}},
\bauthor{\binits{H.} \bsnm{{Bloemen}}},
\bauthor{\binits{J.J.} \bsnm{{Blom}}},
\bauthor{\binits{W.} \bsnm{{Hermsen}}},
\bauthor{\binits{G.G.} \bsnm{{Lichti}}},
\bauthor{\binits{J.} \bsnm{{Ryan}}},
\bauthor{\binits{V.} \bsnm{{Sch{\"o}nfelder}}},
\bauthor{\binits{J.G.} \bsnm{{Stacy}}},
\bauthor{\binits{H.} \bsnm{{Steinle}}},
\bauthor{\binits{O.R.} \bsnm{{Williams}}},
\bauthor{\binits{C.} \bsnm{{Winkler}}},
\batitle{{COMPTEL Observations of AGN at Mev-Energies}}.
\bjtitle{Astrophysical Letters and Communications}
\bvolume{39},
\bfpage{57}
(\byear{1999})
\end{barticle}
\endbibitem

\bibitem[\protect\citeauthoryear{{Cook} et~al.}{1988}]{Cook88}
\begin{barticle}
\bauthor{\binits{W.R.} \bsnm{{Cook}}},
\bauthor{\binits{D.M.} \bsnm{{Palmer}}},
\bauthor{\binits{T.A.} \bsnm{{Prince}}},
\bauthor{\binits{S.M.} \bsnm{{Schindler}}},
\bauthor{\binits{C.H.} \bsnm{{Starr}}},
\bauthor{\binits{E.C.} \bsnm{{Stone}}},
\batitle{{An imaging observation of SN 1987A at gamma-ray energies}}.
\bjtitle{\apjl}
\bvolume{334},
\bfpage{87}--\blpage{90}
(\byear{1988}).
doi:\doiurl{10.1086/185318}
\end{barticle}
\endbibitem

\bibitem[\protect\citeauthoryear{{Cook} et~al.}{1991a}]{Cook91a}
\begin{barticle}
\bauthor{\binits{W.R.} \bsnm{{Cook}}},
\bauthor{\binits{J.M.} \bsnm{{Grunsfeld}}},
\bauthor{\binits{W.A.} \bsnm{{Heindl}}},
\bauthor{\binits{D.M.} \bsnm{{Palmer}}},
\bauthor{\binits{T.A.} \bsnm{{Prince}}},
\bauthor{\binits{S.M.} \bsnm{{Schindler}}},
\bauthor{\binits{E.C.} \bsnm{{Stone}}},
\batitle{{Coded-aperture imaging of the Galactic center region at gamma-ray
  energies}}.
\bjtitle{\apjl}
\bvolume{372},
\bfpage{75}--\blpage{78}
(\byear{1991}a).
doi:\doiurl{10.1086/186027}
\end{barticle}
\endbibitem

\bibitem[\protect\citeauthoryear{{Cook} et~al.}{1991b}]{Cook91b}
\begin{barticle}
\bauthor{\binits{W.R.} \bsnm{{Cook}}},
\bauthor{\binits{J.M.} \bsnm{{Grunsfeld}}},
\bauthor{\binits{W.A.} \bsnm{{Heindl}}},
\bauthor{\binits{D.M.} \bsnm{{Palmer}}},
\bauthor{\binits{T.A.} \bsnm{{Prince}}},
\bauthor{\binits{S.M.} \bsnm{{Schindler}}},
\bauthor{\binits{C.H.} \bsnm{{Starr}}},
\bauthor{\binits{E.C.} \bsnm{{Stone}}},
\batitle{{Recent results of gamma-ray imaging observations of the Galactic
  center and Crab/A0535 + 26 regions}}.
\bjtitle{Advances in Space Research}
\bvolume{11},
\bfpage{191}--\blpage{202}
(\byear{1991}b).
doi:\doiurl{10.1016/0273-1177(91)90171-F}
\end{barticle}
\endbibitem

\bibitem[\protect\citeauthoryear{{Cooke} et~al.}{1984}]{Cooke84}
\begin{barticle}
\bauthor{\binits{B.A.} \bsnm{{Cooke}}},
\bauthor{\binits{A.M.} \bsnm{{Levine}}},
\bauthor{\binits{F.L.} \bsnm{{Lang}}},
\bauthor{\binits{F.A.} \bsnm{{Primini}}},
\bauthor{\binits{W.H.G.} \bsnm{{Lewin}}},
\batitle{{HEAO 1 high-energy X-ray observations of three bright transient X-ray
  sources H1705-250 (Nova Ophiuchi), H1743-322, and H1833-077 (Scutum X-1)}}.
\bjtitle{\apj}
\bvolume{285},
\bfpage{258}--\blpage{263}
(\byear{1984}).
doi:\doiurl{10.1086/162500}
\end{barticle}
\endbibitem

\bibitem[\protect\citeauthoryear{{Costa} and {Frontera}}{2011}]{Costa11}
\begin{botherref}
\oauthor{\binits{E.} \bsnm{{Costa}}},
\oauthor{\binits{F.} \bsnm{{Frontera}}},
{Gamma Ray Burst origin and their afterglow: story of a discovery and more}.
ArXiv e-prints
(2011)
\end{botherref}
\endbibitem

\bibitem[\protect\citeauthoryear{{Costa} et~al.}{1997}]{Costa97}
\begin{barticle}
\bauthor{\binits{E.} \bsnm{{Costa}}},
\bauthor{\binits{F.} \bsnm{{Frontera}}},
\bauthor{\binits{J.} \bsnm{{Heise}}},
\bauthor{\binits{M.} \bsnm{{Feroci}}},
\bauthor{\binits{J.} \bsnm{{in't Zand}}},
\bauthor{\binits{F.} \bsnm{{Fiore}}},
\bauthor{\binits{M.N.} \bsnm{{Cinti}}},
\bauthor{\binits{D.} \bsnm{{Dal Fiume}}},
\bauthor{\binits{L.} \bsnm{{Nicastro}}},
\bauthor{\binits{M.} \bsnm{{Orlandini}}},
\bauthor{\binits{E.} \bsnm{{Palazzi}}},
\bauthor{\binits{M.} \bsnm{{Rapisarda\#}}},
\bauthor{\binits{G.} \bsnm{{Zavattini}}},
\bauthor{\binits{R.} \bsnm{{Jager}}},
\bauthor{\binits{A.} \bsnm{{Parmar}}},
\bauthor{\binits{A.} \bsnm{{Owens}}},
\bauthor{\binits{S.} \bsnm{{Molendi}}},
\bauthor{\binits{G.} \bsnm{{Cusumano}}},
\bauthor{\binits{M.C.} \bsnm{{Maccarone}}},
\bauthor{\binits{S.} \bsnm{{Giarrusso}}},
\bauthor{\binits{A.} \bsnm{{Coletta}}},
\bauthor{\binits{L.A.} \bsnm{{Antonelli}}},
\bauthor{\binits{P.} \bsnm{{Giommi}}},
\bauthor{\binits{J.M.} \bsnm{{Muller}}},
\bauthor{\binits{L.} \bsnm{{Piro}}},
\bauthor{\binits{R.C.} \bsnm{{Butler}}},
\batitle{{Discovery of an X-ray afterglow associated with the {$\gamma$}-ray
  burst of 28 February 1997}}.
\bjtitle{\nat}
\bvolume{387},
\bfpage{783}--\blpage{785}
(\byear{1997}).
doi:\doiurl{10.1038/42885}
\end{barticle}
\endbibitem

\bibitem[\protect\citeauthoryear{{Costa} et~al.}{1998}]{Costa98}
\begin{barticle}
\bauthor{\binits{E.} \bsnm{{Costa}}},
\bauthor{\binits{F.} \bsnm{{Frontera}}},
\bauthor{\binits{D.} \bsnm{{dal Fiume}}},
\bauthor{\binits{L.} \bsnm{{Amati}}},
\bauthor{\binits{M.N.} \bsnm{{Cinti}}},
\bauthor{\binits{P.} \bsnm{{Collina}}},
\bauthor{\binits{M.} \bsnm{{Feroci}}},
\bauthor{\binits{L.} \bsnm{{Nicastro}}},
\bauthor{\binits{M.} \bsnm{{Orlandini}}},
\bauthor{\binits{E.} \bsnm{{Palazzi}}},
\bauthor{\binits{M.} \bsnm{{Rapisarda}}},
\bauthor{\binits{G.} \bsnm{{Zavattini}}},
\batitle{{The gamma-ray bursts monitor onboard SAX}}.
\bjtitle{Advances in Space Research}
\bvolume{22},
\bfpage{1129}--\blpage{1132}
(\byear{1998}).
doi:\doiurl{10.1016/S0273-1177(98)00208-7}
\end{barticle}
\endbibitem

\bibitem[\protect\citeauthoryear{{Covault} et~al.}{1992}]{Covault92}
\begin{barticle}
\bauthor{\binits{C.E.} \bsnm{{Covault}}},
\bauthor{\binits{J.E.} \bsnm{{Grindlay}}},
\bauthor{\binits{R.P.} \bsnm{{Manandhar}}},
\batitle{{Hard X-ray imaging of the Galactic black hole candidate GX 339 - 4}}.
\bjtitle{\apjl}
\bvolume{388},
\bfpage{65}--\blpage{67}
(\byear{1992}).
doi:\doiurl{10.1086/186331}
\end{barticle}
\endbibitem

\bibitem[\protect\citeauthoryear{{Cox}}{2000}]{Cox2000;oso7-8heao1-3}
\begin{bbook}
\bauthor{\binits{A.N.} \bsnm{{Cox}}},
\bbtitle{{Allen's astrophysical quantities}}
\byear{2000}
\end{bbook}
\endbibitem

\bibitem[\protect\citeauthoryear{{Crosby} et~al.}{1998}]{Crosby1998;granat}
\begin{barticle}
\bauthor{\binits{N.} \bsnm{{Crosby}}},
\bauthor{\binits{N.} \bsnm{{Vilmer}}},
\bauthor{\binits{N.} \bsnm{{Lund}}},
\bauthor{\binits{R.} \bsnm{{Sunyaev}}},
\batitle{{Deka-keV X-ray observations of solar bursts with WATCH/GRANAT:
  frequency distributions of burst parameters}}.
\bjtitle{\aap}
\bvolume{334},
\bfpage{299}--\blpage{313}
(\byear{1998})
\end{barticle}
\endbibitem

\bibitem[\protect\citeauthoryear{{Cucchiara} et~al.}{2011}]{Cucchiara11}
\begin{barticle}
\bauthor{\binits{A.} \bsnm{{Cucchiara}}},
\bauthor{\binits{A.J.} \bsnm{{Levan}}},
\bauthor{\binits{D.B.} \bsnm{{Fox}}},
\bauthor{\binits{N.R.} \bsnm{{Tanvir}}},
\bauthor{\binits{T.N.} \bsnm{{Ukwatta}}},
\bauthor{\binits{E.} \bsnm{{Berger}}},
\bauthor{\binits{T.} \bsnm{{Kr{\"u}hler}}},
\bauthor{\binits{A.} \bsnm{{K{\"u}pc{\"u} Yolda{\c s}}}},
\bauthor{\binits{X.F.} \bsnm{{Wu}}},
\bauthor{\binits{K.} \bsnm{{Toma}}},
\bauthor{\binits{J.} \bsnm{{Greiner}}},
\bauthor{\binits{F.E.} \bsnm{{Olivares}}},
\bauthor{\binits{A.} \bsnm{{Rowlinson}}},
\bauthor{\binits{L.} \bsnm{{Amati}}},
\bauthor{\binits{T.} \bsnm{{Sakamoto}}},
\bauthor{\binits{K.} \bsnm{{Roth}}},
\bauthor{\binits{A.} \bsnm{{Stephens}}},
\bauthor{\binits{A.} \bsnm{{Fritz}}},
\bauthor{\binits{J.P.U.} \bsnm{{Fynbo}}},
\bauthor{\binits{J.} \bsnm{{Hjorth}}},
\bauthor{\binits{D.} \bsnm{{Malesani}}},
\bauthor{\binits{P.} \bsnm{{Jakobsson}}},
\bauthor{\binits{K.} \bsnm{{Wiersema}}},
\bauthor{\binits{P.T.} \bsnm{{O'Brien}}},
\bauthor{\binits{A.M.} \bsnm{{Soderberg}}},
\bauthor{\binits{R.J.} \bsnm{{Foley}}},
\bauthor{\binits{A.S.} \bsnm{{Fruchter}}},
\bauthor{\binits{J.} \bsnm{{Rhoads}}},
\bauthor{\binits{R.E.} \bsnm{{Rutledge}}},
\bauthor{\binits{B.P.} \bsnm{{Schmidt}}},
\bauthor{\binits{M.A.} \bsnm{{Dopita}}},
\bauthor{\binits{P.} \bsnm{{Podsiadlowski}}},
\bauthor{\binits{R.} \bsnm{{Willingale}}},
\bauthor{\binits{C.} \bsnm{{Wolf}}},
\bauthor{\binits{S.R.} \bsnm{{Kulkarni}}},
\bauthor{\binits{P.} \bsnm{{D'Avanzo}}},
\batitle{{A Photometric Redshift of z \~{} 9.4 for GRB 090429B}}.
\bjtitle{\apj}
\bvolume{736},
\bfpage{7}
(\byear{2011}).
doi:\doiurl{10.1088/0004-637X/736/1/7}
\end{barticle}
\endbibitem

\bibitem[\protect\citeauthoryear{{Cusumano} et~al.}{1992}]{Cusumano92}
\begin{barticle}
\bauthor{\binits{G.} \bsnm{{Cusumano}}},
\bauthor{\binits{T.} \bsnm{{Mineo}}},
\bauthor{\binits{B.} \bsnm{{Sacco}}},
\bauthor{\binits{L.} \bsnm{{Scarsi}}},
\bauthor{\binits{G.} \bsnm{{Gerardi}}},
\bauthor{\binits{B.} \bsnm{{Agrinier}}},
\bauthor{\binits{E.} \bsnm{{Barouch}}},
\bauthor{\binits{R.} \bsnm{{Comte}}},
\bauthor{\binits{B.} \bsnm{{Parlier}}},
\bauthor{\binits{J.L.} \bsnm{{Masnou}}},
\bauthor{\binits{E.} \bsnm{{Massaro}}},
\bauthor{\binits{G.} \bsnm{{Matt}}},
\bauthor{\binits{E.} \bsnm{{Costa}}},
\bauthor{\binits{M.} \bsnm{{Salvati}}},
\bauthor{\binits{P.} \bsnm{{Mandrou}}},
\bauthor{\binits{M.} \bsnm{{Niel}}},
\bauthor{\binits{J.F.} \bsnm{{Olive}}},
\batitle{{Observation of A0535 + 26 at energies above 150 keV with the FIGARO
  II experiment}}.
\bjtitle{\apjl}
\bvolume{398},
\bfpage{103}--\blpage{106}
(\byear{1992}).
doi:\doiurl{10.1086/186587}
\end{barticle}
\endbibitem

\bibitem[\protect\citeauthoryear{{Cusumano} et~al.}{1998}]{Cusumano98}
\begin{barticle}
\bauthor{\binits{G.} \bsnm{{Cusumano}}},
\bauthor{\binits{T.} \bsnm{{di Salvo}}},
\bauthor{\binits{L.} \bsnm{{Burderi}}},
\bauthor{\binits{M.} \bsnm{{Orlandini}}},
\bauthor{\binits{S.} \bsnm{{Piraino}}},
\bauthor{\binits{N.} \bsnm{{Robba}}},
\bauthor{\binits{A.} \bsnm{{Santangelo}}},
\batitle{{Detection of a cyclotron line and its second harmonic in 4U1907+09}}.
\bjtitle{\aap}
\bvolume{338},
\bfpage{79}--\blpage{82}
(\byear{1998})
\end{barticle}
\endbibitem

\bibitem[\protect\citeauthoryear{{Cusumano} et~al.}{2010}]{Cusumano10}
\begin{barticle}
\bauthor{\binits{G.} \bsnm{{Cusumano}}},
\bauthor{\binits{V.} \bsnm{{La Parola}}},
\bauthor{\binits{A.} \bsnm{{Segreto}}},
\bauthor{\binits{C.} \bsnm{{Ferrigno}}},
\bauthor{\binits{A.} \bsnm{{Maselli}}},
\bauthor{\binits{B.} \bsnm{{Sbarufatti}}},
\bauthor{\binits{P.} \bsnm{{Romano}}},
\bauthor{\binits{G.} \bsnm{{Chincarini}}},
\bauthor{\binits{P.} \bsnm{{Giommi}}},
\bauthor{\binits{N.} \bsnm{{Masetti}}},
\bauthor{\binits{A.} \bsnm{{Moretti}}},
\bauthor{\binits{P.} \bsnm{{Parisi}}},
\bauthor{\binits{G.} \bsnm{{Tagliaferri}}},
\batitle{{The Palermo Swift-BAT hard X-ray catalogue. III. Results after 54
  months of sky survey}}.
\bjtitle{\aap}
\bvolume{524},
\bfpage{64}
(\byear{2010}).
doi:\doiurl{10.1051/0004-6361/201015249}
\end{barticle}
\endbibitem

\bibitem[\protect\citeauthoryear{{dal Fiume} et~al.}{1988}]{Dalfiume88}
\begin{barticle}
\bauthor{\binits{D.} \bsnm{{dal Fiume}}},
\bauthor{\binits{F.} \bsnm{{Frontera}}},
\bauthor{\binits{E.} \bsnm{{Morelli}}},
\batitle{{The X-ray pulsar A0535+26 in hard X-rays - Average spectrum,
  pulse-phase spectroscopy, and spectral time variability}}.
\bjtitle{\apj}
\bvolume{331},
\bfpage{313}--\blpage{320}
(\byear{1988}).
doi:\doiurl{10.1086/166555}
\end{barticle}
\endbibitem

\bibitem[\protect\citeauthoryear{{dal Fiume} et~al.}{2000}]{Dalfiume00}
\begin{barticle}
\bauthor{\binits{D.} \bsnm{{dal Fiume}}},
\bauthor{\binits{M.} \bsnm{{Orlandini}}},
\bauthor{\binits{S.} \bsnm{{del Sordo}}},
\bauthor{\binits{F.} \bsnm{{Frontera}}},
\bauthor{\binits{T.} \bsnm{{Oosterbroek}}},
\bauthor{\binits{E.} \bsnm{{Palazzi}}},
\bauthor{\binits{A.N.} \bsnm{{Parmar}}},
\bauthor{\binits{S.} \bsnm{{Piraino}}},
\bauthor{\binits{A.} \bsnm{{Santangelo}}},
\bauthor{\binits{A.} \bsnm{{Segreto}}},
\batitle{{The Broad Band Spectral Properties of Binary X-ray Pulsars}}.
\bjtitle{Advances in Space Research}
\bvolume{25},
\bfpage{399}--\blpage{408}
(\byear{2000}).
doi:\doiurl{10.1016/S0273-1177(99)00767-X}
\end{barticle}
\endbibitem

\bibitem[\protect\citeauthoryear{{Datlowe} et~al.}{1974}]{Datlowe74}
\begin{barticle}
\bauthor{\binits{D.W.} \bsnm{{Datlowe}}},
\bauthor{\binits{M.J.} \bsnm{{Elcan}}},
\bauthor{\binits{H.S.} \bsnm{{Hudson}}},
\batitle{{OSO-7 observations of solar X-rays in the energy range 10-100 keV}}.
\bjtitle{\solphys}
\bvolume{39},
\bfpage{155}--\blpage{174}
(\byear{1974}).
doi:\doiurl{10.1007/BF00154978}
\end{barticle}
\endbibitem

\bibitem[\protect\citeauthoryear{{de Jager} et~al.}{1996}]{DeJager96}
\begin{barticle}
\bauthor{\binits{O.C.} \bsnm{{de Jager}}},
\bauthor{\binits{A.K.} \bsnm{{Harding}}},
\bauthor{\binits{M.S.} \bsnm{{Strickman}}},
\batitle{{OSSE Detection of Gamma Rays from the VELA Synchrotron Nebula}}.
\bjtitle{\apj}
\bvolume{460},
\bfpage{729}
(\byear{1996}).
doi:\doiurl{10.1086/177005}
\end{barticle}
\endbibitem

\bibitem[\protect\citeauthoryear{{Dean} et~al.}{1990}]{Dean90}
\begin{barticle}
\bauthor{\binits{A.J.} \bsnm{{Dean}}},
\bauthor{\binits{A.} \bsnm{{Bazzano}}},
\bauthor{\binits{A.J.} \bsnm{{Court}}},
\bauthor{\binits{N.A.} \bsnm{{Dipper}}},
\bauthor{\binits{R.A.} \bsnm{{Lewis}}},
\bauthor{\binits{P.} \bsnm{{Maggioli}}},
\bauthor{\binits{F.} \bsnm{{Perotti}}},
\bauthor{\binits{M.} \bsnm{{Quadrini}}},
\bauthor{\binits{J.B.} \bsnm{{Stephen}}},
\bauthor{\binits{P.} \bsnm{{Ubertini}}},
\batitle{{Nonthermal X-ray emission from 3C 273 - The core of a knotty
  problem?}}
\bjtitle{\apj}
\bvolume{349},
\bfpage{41}--\blpage{44}
(\byear{1990}).
doi:\doiurl{10.1086/168291}
\end{barticle}
\endbibitem

\bibitem[\protect\citeauthoryear{{Dean} et~al.}{2008}]{Dean08}
\begin{barticle}
\bauthor{\binits{A.J.} \bsnm{{Dean}}},
\bauthor{\binits{D.J.} \bsnm{{Clark}}},
\bauthor{\binits{J.B.} \bsnm{{Stephen}}},
\bauthor{\binits{V.A.} \bsnm{{McBride}}},
\bauthor{\binits{L.} \bsnm{{Bassani}}},
\bauthor{\binits{A.} \bsnm{{Bazzano}}},
\bauthor{\binits{A.J.} \bsnm{{Bird}}},
\bauthor{\binits{A.B.} \bsnm{{Hill}}},
\bauthor{\binits{S.E.} \bsnm{{Shaw}}},
\bauthor{\binits{P.} \bsnm{{Ubertini}}},
\batitle{{Polarized Gamma-Ray Emission from the Crab}}.
\bjtitle{Science}
\bvolume{321},
\bfpage{1183}
(\byear{2008}).
doi:\doiurl{10.1126/science.1149056}
\end{barticle}
\endbibitem

\bibitem[\protect\citeauthoryear{{Deerenberg} and
  {Bleeker}}{1971}]{Deerenberg71}
\begin{barticle}
\bauthor{\binits{A.J.M.} \bsnm{{Deerenberg}}},
\bauthor{\binits{J.A.M.} \bsnm{{Bleeker}}},
\batitle{{Measurement of the Pulsation Ratio of the Crab Nebula in the X-ray
  Region from 20 to 80 keV}}.
\bjtitle{Nature Physical Science}
\bvolume{229},
\bfpage{113}--\blpage{114}
(\byear{1971}).
doi:\doiurl{10.1038/physci229113a0}
\end{barticle}
\endbibitem

\bibitem[\protect\citeauthoryear{{DeLaunay} et~al.}{2016}]{DeLaunay16}
\begin{barticle}
\bauthor{\binits{J.J.} \bsnm{{DeLaunay}}},
\bauthor{\binits{D.B.} \bsnm{{Fox}}},
\bauthor{\binits{K.} \bsnm{{Murase}}},
\bauthor{\binits{P.} \bsnm{{M{\'e}sz{\'a}ros}}},
\bauthor{\binits{A.} \bsnm{{Keivani}}},
\bauthor{\binits{C.} \bsnm{{Messick}}},
\bauthor{\binits{M.A.} \bsnm{{Mostaf{\'a}}}},
\bauthor{\binits{F.} \bsnm{{Oikonomou}}},
\bauthor{\binits{G.} \bsnm{{Te{\v s}i{\'c}}}},
\bauthor{\binits{C.F.} \bsnm{{Turley}}},
\batitle{{Discovery of a Transient Gamma-Ray Counterpart to FRB 131104}}.
\bjtitle{\apjl}
\bvolume{832},
\bfpage{1}
(\byear{2016}).
doi:\doiurl{10.3847/2041-8205/832/1/L1}
\end{barticle}
\endbibitem

\bibitem[\protect\citeauthoryear{{Dennis} et~al.}{1973}]{Dennis1973}
\begin{barticle}
\bauthor{\binits{B.R.} \bsnm{{Dennis}}},
\bauthor{\binits{A.N.} \bsnm{{Suri}}},
\bauthor{\binits{K.J.} \bsnm{{Frost}}},
\batitle{{The Diffuse X-Ray Spectrum from 14 TO 200 keV as Measured on OSO-5}}.
\bjtitle{\apj}
\bvolume{186},
\bfpage{97}--\blpage{103}
(\byear{1973}).
doi:\doiurl{10.1086/152480}
\end{barticle}
\endbibitem

\bibitem[\protect\citeauthoryear{{Dennis} et~al.}{1977}]{Dennis77}
\begin{barticle}
\bauthor{\binits{B.R.} \bsnm{{Dennis}}},
\bauthor{\binits{K.J.} \bsnm{{Frost}}},
\bauthor{\binits{R.J.} \bsnm{{Lencho}}},
\bauthor{\binits{L.E.} \bsnm{{Orwig}}},
\batitle{{The high-energy celestial X-ray instrument on board OSO-8}}.
\bjtitle{Space Science Instrumentation}
\bvolume{3},
\bfpage{325}--\blpage{342}
(\byear{1977})
\end{barticle}
\endbibitem

\bibitem[\protect\citeauthoryear{{Di Cocco} et~al.}{1981}]{DiCocco81}
\begin{barticle}
\bauthor{\binits{G.} \bsnm{{Di Cocco}}},
\bauthor{\binits{G.} \bsnm{{Boella}}},
\bauthor{\binits{A.} \bsnm{{della Ventura}}},
\bauthor{\binits{G.} \bsnm{{Mangia}}},
\bauthor{\binits{F.} \bsnm{{Perotti}}},
\bauthor{\binits{G.} \bsnm{{Villa}}},
\bauthor{\binits{R.E.} \bsnm{{Baker}}},
\bauthor{\binits{R.C.} \bsnm{{Butler}}},
\bauthor{\binits{A.J.} \bsnm{{Dean}}},
\bauthor{\binits{R.I.} \bsnm{{Hayles}}},
\batitle{{Low energy gamma-ray observations of CG135+1 and CG195+4}}.
\bjtitle{Advances in Space Research}
\bvolume{1},
\bfpage{231}--\blpage{234}
(\byear{1981}).
doi:\doiurl{10.1016/0273-1177(81)90200-3}
\end{barticle}
\endbibitem

\bibitem[\protect\citeauthoryear{{Di Cocco} et~al.}{2003}]{Dicocco2003}
\begin{barticle}
\bauthor{\binits{G.} \bsnm{{Di Cocco}}},
\bauthor{\binits{E.} \bsnm{{Caroli}}},
\bauthor{\binits{E.} \bsnm{{Celesti}}},
\bauthor{\binits{L.} \bsnm{{Foschini}}},
\bauthor{\binits{F.} \bsnm{{Gianotti}}},
\bauthor{\binits{C.} \bsnm{{Labanti}}},
\bauthor{\binits{G.} \bsnm{{Malaguti}}},
\bauthor{\binits{A.} \bsnm{{Mauri}}},
\bauthor{\binits{E.} \bsnm{{Rossi}}},
\bauthor{\binits{F.} \bsnm{{Schiavone}}},
\bauthor{\binits{A.} \bsnm{{Spizzichino}}},
\bauthor{\binits{J.B.} \bsnm{{Stephen}}},
\bauthor{\binits{A.} \bsnm{{Traci}}},
\bauthor{\binits{M.} \bsnm{{Trifoglio}}},
\batitle{{IBIS/PICsIT in-flight performances}}.
\bjtitle{\aap}
\bvolume{411},
\bfpage{189}--\blpage{195}
(\byear{2003}).
doi:\doiurl{10.1051/0004-6361:20031346}
\end{barticle}
\endbibitem

\bibitem[\protect\citeauthoryear{{Diehl} et~al.}{1994}]{Diehl94}
\begin{barticle}
\bauthor{\binits{R.} \bsnm{{Diehl}}},
\bauthor{\binits{C.} \bsnm{{Dupraz}}},
\bauthor{\binits{K.} \bsnm{{Bennett}}},
\bauthor{\binits{H.} \bsnm{{Bloemen}}},
\bauthor{\binits{H.} \bsnm{{de Boer}}},
\bauthor{\binits{W.} \bsnm{{Hermsen}}},
\bauthor{\binits{G.G.} \bsnm{{Lichti}}},
\bauthor{\binits{M.} \bsnm{{McConnell}}},
\bauthor{\binits{D.} \bsnm{{Morris}}},
\bauthor{\binits{J.} \bsnm{{Ryan}}},
\bauthor{\binits{V.} \bsnm{{Sch{\"o}nfelder}}},
\bauthor{\binits{H.} \bsnm{{Steinle}}},
\bauthor{\binits{A.W.} \bsnm{{Strong}}},
\bauthor{\binits{B.N.} \bsnm{{Swanenburg}}},
\bauthor{\binits{M.} \bsnm{{Varendorff}}},
\bauthor{\binits{C.} \bsnm{{Winkler}}},
\batitle{{COMPTEL observations of the 1.809 MeV gamma-ray line from galactic
  Al-26}}.
\bjtitle{\apjs}
\bvolume{92},
\bfpage{429}--\blpage{432}
(\byear{1994}).
doi:\doiurl{10.1086/191990}
\end{barticle}
\endbibitem

\bibitem[\protect\citeauthoryear{{Dil} et~al.}{1981}]{Dil81}
\begin{barticle}
\bauthor{\binits{S.} \bsnm{{Dil}}},
\bauthor{\binits{F.A.} \bsnm{{Primini}}},
\bauthor{\binits{E.} \bsnm{{Basinska}}},
\bauthor{\binits{M.} \bsnm{{Bautz}}},
\bauthor{\binits{S.K.} \bsnm{{Howe}}},
\bauthor{\binits{F.} \bsnm{{Lang}}},
\bauthor{\binits{A.M.} \bsnm{{Levine}}},
\bauthor{\binits{W.H.G.} \bsnm{{Lewin}}},
\bauthor{\binits{D.W.} \bsnm{{Worrall}}},
\bauthor{\binits{P.L.} \bsnm{{Nolan}}},
\bauthor{\binits{J.L.} \bsnm{{Matteson}}},
\batitle{{HEAO 1 observations of high-energy X-rays from the Seyfert I galaxy
  MKN 509}}.
\bjtitle{\apj}
\bvolume{250},
\bfpage{513}--\blpage{516}
(\byear{1981}).
doi:\doiurl{10.1086/159397}
\end{barticle}
\endbibitem

\bibitem[\protect\citeauthoryear{{Dolan} et~al.}{1977}]{Dolan77}
\begin{barticle}
\bauthor{\binits{J.F.} \bsnm{{Dolan}}},
\bauthor{\binits{C.J.} \bsnm{{Crannell}}},
\bauthor{\binits{B.R.} \bsnm{{Dennis}}},
\bauthor{\binits{K.J.} \bsnm{{Frost}}},
\bauthor{\binits{L.E.} \bsnm{{Orwig}}},
\bauthor{\binits{G.S.} \bsnm{{Maurer}}},
\batitle{{The high-energy X-ray spectrum of the Crab Nebula observed from
  OSO-8}}.
\bjtitle{\apj}
\bvolume{217},
\bfpage{809}--\blpage{814}
(\byear{1977}).
doi:\doiurl{10.1086/155628}
\end{barticle}
\endbibitem

\bibitem[\protect\citeauthoryear{{Dolan} et~al.}{1979}]{Dolan79}
\begin{barticle}
\bauthor{\binits{J.F.} \bsnm{{Dolan}}},
\bauthor{\binits{C.J.} \bsnm{{Crannell}}},
\bauthor{\binits{B.R.} \bsnm{{Dennis}}},
\bauthor{\binits{K.J.} \bsnm{{Frost}}},
\bauthor{\binits{L.E.} \bsnm{{Orwig}}},
\batitle{{High-energy X-ray spectra of Cygnus XR-1 observed from OSO 8}}.
\bjtitle{\apj}
\bvolume{230},
\bfpage{551}--\blpage{557}
(\byear{1979}).
doi:\doiurl{10.1086/157111}
\end{barticle}
\endbibitem

\bibitem[\protect\citeauthoryear{{Dolan} et~al.}{1980}]{Dolan80}
\begin{barticle}
\bauthor{\binits{J.F.} \bsnm{{Dolan}}},
\bauthor{\binits{M.J.} \bsnm{{Coe}}},
\bauthor{\binits{C.J.} \bsnm{{Crannell}}},
\bauthor{\binits{B.R.} \bsnm{{Dennis}}},
\bauthor{\binits{K.J.} \bsnm{{Frost}}},
\bauthor{\binits{L.E.} \bsnm{{Orwig}}},
\bauthor{\binits{G.S.} \bsnm{{Maurer}}},
\batitle{{The high energy X-ray spectrum of 4U 1700-37 observed from OSO 8}}.
\bjtitle{\apj}
\bvolume{238},
\bfpage{238}--\blpage{243}
(\byear{1980}).
doi:\doiurl{10.1086/157980}
\end{barticle}
\endbibitem

\bibitem[\protect\citeauthoryear{{Dolan} et~al.}{1982}]{Dolan82}
\begin{barticle}
\bauthor{\binits{J.F.} \bsnm{{Dolan}}},
\bauthor{\binits{C.J.} \bsnm{{Crannell}}},
\bauthor{\binits{B.R.} \bsnm{{Dennis}}},
\bauthor{\binits{K.J.} \bsnm{{Frost}}},
\bauthor{\binits{L.E.} \bsnm{{Orwig}}},
\batitle{{High energy X-ray observations of CYG X-3 from OSO-8 - Further
  evidence of a 34.1 day period}}.
\bjtitle{\aplett}
\bvolume{22},
\bfpage{147}--\blpage{152}
(\byear{1982})
\end{barticle}
\endbibitem

\bibitem[\protect\citeauthoryear{{Dolan} et~al.}{1984}]{Dolan84}
\begin{barticle}
\bauthor{\binits{J.F.} \bsnm{{Dolan}}},
\bauthor{\binits{C.J.} \bsnm{{Crannell}}},
\bauthor{\binits{B.R.} \bsnm{{Dennis}}},
\bauthor{\binits{K.J.} \bsnm{{Frost}}},
\bauthor{\binits{L.E.} \bsnm{{Orwig}}},
\batitle{{The high-energy X-ray spectrum of Centaurus XR-3 observed from OSO
  8}}.
\bjtitle{\apj}
\bvolume{278},
\bfpage{266}--\blpage{269}
(\byear{1984}).
doi:\doiurl{10.1086/161790}
\end{barticle}
\endbibitem

\bibitem[\protect\citeauthoryear{{Doty} et~al.}{1981}]{Doty81}
\begin{barticle}
\bauthor{\binits{J.P.} \bsnm{{Doty}}},
\bauthor{\binits{W.H.G.} \bsnm{{Lewin}}},
\bauthor{\binits{J.A.} \bsnm{{Hoffman}}},
\batitle{{SAS 3 observations of GX 1 + 4}}.
\bjtitle{\apj}
\bvolume{243},
\bfpage{257}--\blpage{262}
(\byear{1981}).
doi:\doiurl{10.1086/158592}
\end{barticle}
\endbibitem

\bibitem[\protect\citeauthoryear{{D'Silva} et~al.}{1998}]{Dsilva98}
\begin{barticle}
\bauthor{\binits{J.A.R.} \bsnm{{D'Silva}}},
\bauthor{\binits{P.P.} \bsnm{{Madhwani}}},
\bauthor{\binits{N.} \bsnm{{Tembhurne}}},
\bauthor{\binits{R.K.} \bsnm{{Manchanda}}},
\batitle{{A high sensitivity payload for balloon-borne hard X-ray astronomy.}}
\bjtitle{Nuclear Instruments and Methods in Physics Research A}
\bvolume{412},
\bfpage{342}--\blpage{354}
(\byear{1998}).
doi:\doiurl{10.1016/S0168-9002(98)00496-3}
\end{barticle}
\endbibitem

\bibitem[\protect\citeauthoryear{{Dunphy} et~al.}{1983}]{Dunphy83}
\begin{bchapter}
\bauthor{\binits{P.P.} \bsnm{{Dunphy}}},
\bauthor{\binits{E.L.} \bsnm{{Chupp}}},
\bauthor{\binits{D.J.} \bsnm{{Forrest}}},
\bctitle{{Effects of line width and spatial extent on measurements of the 0.51
  MeV Galactic center line}},
in \bbtitle{Positron-Electron Pairs in Astrophysics},
ed. by \beditor{\binits{M.L.} \bsnm{{Burns}}},
\beditor{\binits{A.K.} \bsnm{{Harding}}},
\beditor{\binits{R.} \bsnm{{Ramaty}}}
\bsertitle{American Institute of Physics Conference Series},
vol. \bseriesno{101},
\byear{1983},
pp. \bfpage{237}--\blpage{241}.
doi:\doiurl{10.1063/1.34094}
\end{bchapter}
\endbibitem

\bibitem[\protect\citeauthoryear{{Elvis} et~al.}{2000}]{Elvis00}
\begin{barticle}
\bauthor{\binits{M.} \bsnm{{Elvis}}},
\bauthor{\binits{F.} \bsnm{{Fiore}}},
\bauthor{\binits{A.} \bsnm{{Siemiginowska}}},
\bauthor{\binits{J.} \bsnm{{Bechtold}}},
\bauthor{\binits{S.} \bsnm{{Mathur}}},
\bauthor{\binits{J.} \bsnm{{McDowell}}},
\batitle{{150 KEV Emission from PKS 2149-306 with BEPPOSAX}}.
\bjtitle{\apj}
\bvolume{543},
\bfpage{545}--\blpage{551}
(\byear{2000}).
doi:\doiurl{10.1086/317117}
\end{barticle}
\endbibitem

\bibitem[\protect\citeauthoryear{{Engel} and {Coe}}{1977}]{Engel77}
\begin{barticle}
\bauthor{\binits{A.R.} \bsnm{{Engel}}},
\bauthor{\binits{M.J.} \bsnm{{Coe}}},
\batitle{{The high energy X--ray detector on the Ariel--5 satellite}}.
\bjtitle{Space Science Instrumentation}
\bvolume{3},
\bfpage{407}--\blpage{421}
(\byear{1977})
\end{barticle}
\endbibitem

\bibitem[\protect\citeauthoryear{{Enoto} et~al.}{2008}]{Enoto08}
\begin{barticle}
\bauthor{\binits{T.} \bsnm{{Enoto}}},
\bauthor{\binits{K.} \bsnm{{Makishima}}},
\bauthor{\binits{Y.} \bsnm{{Terada}}},
\bauthor{\binits{T.} \bsnm{{Mihara}}},
\bauthor{\binits{K.} \bsnm{{Nakazawa}}},
\bauthor{\binits{T.} \bsnm{{Ueda}}},
\bauthor{\binits{T.} \bsnm{{Dotani}}},
\bauthor{\binits{M.} \bsnm{{Kokubun}}},
\bauthor{\binits{F.} \bsnm{{Nagase}}},
\bauthor{\binits{S.} \bsnm{{Naik}}},
\bauthor{\binits{M.} \bsnm{{Suzuki}}},
\bauthor{\binits{M.} \bsnm{{Nakajima}}},
\bauthor{\binits{H.} \bsnm{{Takahashi}}},
\batitle{{Suzaku Observations of Hercules X-1: Measurements of the Two
  Cyclotron Harmonics}}.
\bjtitle{\pasj}
\bvolume{60},
\bfpage{57}--\blpage{68}
(\byear{2008})
\end{barticle}
\endbibitem

\bibitem[\protect\citeauthoryear{{Enoto} et~al.}{2010}]{Enoto10}
\begin{barticle}
\bauthor{\binits{T.} \bsnm{{Enoto}}},
\bauthor{\binits{K.} \bsnm{{Nakazawa}}},
\bauthor{\binits{K.} \bsnm{{Makishima}}},
\bauthor{\binits{Y.E.} \bsnm{{Nakagawa}}},
\bauthor{\binits{T.} \bsnm{{Sakamoto}}},
\bauthor{\binits{M.} \bsnm{{Ohno}}},
\bauthor{\binits{T.} \bsnm{{Takahashi}}},
\bauthor{\binits{Y.} \bsnm{{Terada}}},
\bauthor{\binits{K.} \bsnm{{Yamaoka}}},
\bauthor{\binits{T.} \bsnm{{Murakami}}},
\bauthor{\binits{H.} \bsnm{{Takahashi}}},
\batitle{{Suzaku Discovery of a Hard X-Ray Tail in the Persistent Spectra from
  the Magnetar 1E 1547.0-5408 during its 2009 Activity}}.
\bjtitle{\pasj}
\bvolume{62},
\bfpage{475}--\blpage{485}
(\byear{2010}).
doi:\doiurl{10.1093/pasj/62.2.475}
\end{barticle}
\endbibitem

\bibitem[\protect\citeauthoryear{{Evangelista} et~al.}{2010}]{Evangelista10}
\begin{barticle}
\bauthor{\binits{Y.} \bsnm{{Evangelista}}},
\bauthor{\binits{M.} \bsnm{{Feroci}}},
\bauthor{\binits{E.} \bsnm{{Costa}}},
\bauthor{\binits{E.} \bsnm{{Del Monte}}},
\bauthor{\binits{I.} \bsnm{{Donnarumma}}},
\bauthor{\binits{I.} \bsnm{{Lapshov}}},
\bauthor{\binits{F.} \bsnm{{Lazzarotto}}},
\bauthor{\binits{L.} \bsnm{{Pacciani}}},
\bauthor{\binits{M.} \bsnm{{Rapisarda}}},
\bauthor{\binits{P.} \bsnm{{Soffitta}}},
\bauthor{\binits{A.} \bsnm{{Argan}}},
\bauthor{\binits{G.} \bsnm{{Barbiellini}}},
\bauthor{\binits{F.} \bsnm{{Boffelli}}},
\bauthor{\binits{A.} \bsnm{{Bulgarelli}}},
\bauthor{\binits{P.} \bsnm{{Caraveo}}},
\bauthor{\binits{P.W.} \bsnm{{Cattaneo}}},
\bauthor{\binits{A.} \bsnm{{Chen}}},
\bauthor{\binits{F.} \bsnm{{D'Ammando}}},
\bauthor{\binits{G.} \bsnm{{Di Cocco}}},
\bauthor{\binits{F.} \bsnm{{Fuschino}}},
\bauthor{\binits{M.} \bsnm{{Galli}}},
\bauthor{\binits{F.} \bsnm{{Gianotti}}},
\bauthor{\binits{A.} \bsnm{{Giuliani}}},
\bauthor{\binits{C.} \bsnm{{Labanti}}},
\bauthor{\binits{P.} \bsnm{{Lipari}}},
\bauthor{\binits{F.} \bsnm{{Longo}}},
\bauthor{\binits{M.} \bsnm{{Marisaldi}}},
\bauthor{\binits{S.} \bsnm{{Mereghetti}}},
\bauthor{\binits{E.} \bsnm{{Moretti}}},
\bauthor{\binits{A.} \bsnm{{Morselli}}},
\bauthor{\binits{A.} \bsnm{{Pellizzoni}}},
\bauthor{\binits{F.} \bsnm{{Perotti}}},
\bauthor{\binits{G.} \bsnm{{Piano}}},
\bauthor{\binits{P.} \bsnm{{Picozza}}},
\bauthor{\binits{M.} \bsnm{{Pilia}}},
\bauthor{\binits{M.} \bsnm{{Prest}}},
\bauthor{\binits{G.} \bsnm{{Pucella}}},
\bauthor{\binits{A.} \bsnm{{Rappoldi}}},
\bauthor{\binits{S.} \bsnm{{Sabatini}}},
\bauthor{\binits{E.} \bsnm{{Striani}}},
\bauthor{\binits{M.} \bsnm{{Tavani}}},
\bauthor{\binits{M.} \bsnm{{Trifoglio}}},
\bauthor{\binits{A.} \bsnm{{Trois}}},
\bauthor{\binits{E.} \bsnm{{Vallazza}}},
\bauthor{\binits{S.} \bsnm{{Vercellone}}},
\bauthor{\binits{V.} \bsnm{{Vittorini}}},
\bauthor{\binits{A.} \bsnm{{Zambra}}},
\bauthor{\binits{L.A.} \bsnm{{Antonelli}}},
\bauthor{\binits{S.} \bsnm{{Cutini}}},
\bauthor{\binits{C.} \bsnm{{Pittori}}},
\bauthor{\binits{B.} \bsnm{{Preger}}},
\bauthor{\binits{P.} \bsnm{{Santolamazza}}},
\bauthor{\binits{F.} \bsnm{{Verrecchia}}},
\bauthor{\binits{P.} \bsnm{{Giommi}}},
\bauthor{\binits{L.} \bsnm{{Salotti}}},
\batitle{{Temporal Properties of GX 301-2 Over a Year-long Observation with
  SuperAGILE}}.
\bjtitle{\apj}
\bvolume{708},
\bfpage{1663}--\blpage{1673}
(\byear{2010}).
doi:\doiurl{10.1088/0004-637X/708/2/1663}
\end{barticle}
\endbibitem

\bibitem[\protect\citeauthoryear{{Fabian} et~al.}{1993}]{Fabian93}
\begin{barticle}
\bauthor{\binits{A.C.} \bsnm{{Fabian}}},
\bauthor{\binits{K.} \bsnm{{Nandra}}},
\bauthor{\binits{A.} \bsnm{{Celotti}}},
\bauthor{\binits{M.J.} \bsnm{{Rees}}},
\bauthor{\binits{J.E.} \bsnm{{Grove}}},
\bauthor{\binits{W.N.} \bsnm{{Johnson}}},
\batitle{{OSSE Observations of the Bright Seyfert 1 Galaxy IC 4329A}}.
\bjtitle{\apjl}
\bvolume{416},
\bfpage{57}
(\byear{1993}).
doi:\doiurl{10.1086/187070}
\end{barticle}
\endbibitem

\bibitem[\protect\citeauthoryear{{Favata} et~al.}{1997}]{Favata97}
\begin{barticle}
\bauthor{\binits{F.} \bsnm{{Favata}}},
\bauthor{\binits{J.} \bsnm{{Vink}}},
\bauthor{\binits{D.} \bsnm{{dal Fiume}}},
\bauthor{\binits{A.N.} \bsnm{{Parmar}}},
\bauthor{\binits{A.} \bsnm{{Santangelo}}},
\bauthor{\binits{T.} \bsnm{{Mineo}}},
\bauthor{\binits{A.} \bsnm{{Preite-Martinez}}},
\bauthor{\binits{J.S.} \bsnm{{Kaastra}}},
\bauthor{\binits{J.A.M.} \bsnm{{Bleeker}}},
\batitle{{The broad-band X-ray spectrum of the CAS A supernova remnant as seen
  by the BeppoSAX observatory.}}
\bjtitle{\aap}
\bvolume{324},
\bfpage{49}--\blpage{52}
(\byear{1997})
\end{barticle}
\endbibitem

\bibitem[\protect\citeauthoryear{{Feng} et~al.}{1999}]{Feng99}
\begin{barticle}
\bauthor{\binits{Y.X.} \bsnm{{Feng}}},
\bauthor{\binits{T.P.} \bsnm{{Li}}},
\bauthor{\binits{L.} \bsnm{{Chen}}},
\batitle{{X-Ray Shots of Cygnus X-1}}.
\bjtitle{\apj}
\bvolume{514},
\bfpage{373}--\blpage{382}
(\byear{1999}).
doi:\doiurl{10.1086/306946}
\end{barticle}
\endbibitem

\bibitem[\protect\citeauthoryear{{Ferguson} et~al.}{1999}]{Ferguson99}
\begin{barticle}
\bauthor{\binits{C.} \bsnm{{Ferguson}}},
\bauthor{\binits{F.} \bsnm{{Lei}}},
\bauthor{\binits{A.J.} \bsnm{{Dean}}},
\bauthor{\binits{A.J.} \bsnm{{Bird}}},
\bauthor{\binits{J.J.} \bsnm{{Lockley}}},
\batitle{{New observational constraints on hard X-/gamma -ray millisecond
  pulsar emission from 47 Tucanae}}.
\bjtitle{\aap}
\bvolume{350},
\bfpage{847}--\blpage{851}
(\byear{1999})
\end{barticle}
\endbibitem

\bibitem[\protect\citeauthoryear{{Feroci} et~al.}{1999}]{Feroci99}
\begin{barticle}
\bauthor{\binits{M.} \bsnm{{Feroci}}},
\bauthor{\binits{G.} \bsnm{{Matt}}},
\bauthor{\binits{G.} \bsnm{{Pooley}}},
\bauthor{\binits{E.} \bsnm{{Costa}}},
\bauthor{\binits{M.} \bsnm{{Tavani}}},
\bauthor{\binits{T.} \bsnm{{Belloni}}},
\batitle{{Inner accretion disk disappearance during a radio flare in GRS
  1915+105}}.
\bjtitle{\aap}
\bvolume{351},
\bfpage{985}--\blpage{992}
(\byear{1999})
\end{barticle}
\endbibitem

\bibitem[\protect\citeauthoryear{{Feroci} et~al.}{2007}]{Feroci2007;agile}
\begin{barticle}
\bauthor{\binits{M.} \bsnm{{Feroci}}},
\bauthor{\binits{E.} \bsnm{{Costa}}},
\bauthor{\binits{P.} \bsnm{{Soffitta}}},
\bauthor{\binits{E.} \bsnm{{Del Monte}}},
\bauthor{\binits{G.} \bsnm{{di Persio}}},
\bauthor{\binits{I.} \bsnm{{Donnarumma}}},
\bauthor{\binits{Y.} \bsnm{{Evangelista}}},
\bauthor{\binits{M.} \bsnm{{Frutti}}},
\bauthor{\binits{I.} \bsnm{{Lapshov}}},
\bauthor{\binits{F.} \bsnm{{Lazzarotto}}},
\bauthor{\binits{M.} \bsnm{{Mastropietro}}},
\bauthor{\binits{E.} \bsnm{{Morelli}}},
\bauthor{\binits{L.} \bsnm{{Pacciani}}},
\bauthor{\binits{G.} \bsnm{{Porrovecchio}}},
\bauthor{\binits{M.} \bsnm{{Rapisarda}}},
\bauthor{\binits{A.} \bsnm{{Rubini}}},
\bauthor{\binits{M.} \bsnm{{Tavani}}},
\bauthor{\binits{A.} \bsnm{{Argan}}},
\batitle{{SuperAGILE: The hard X-ray imager for the AGILE space mission}}.
\bjtitle{Nuclear Instruments and Methods in Physics Research A}
\bvolume{581},
\bfpage{728}--\blpage{754}
(\byear{2007}).
doi:\doiurl{10.1016/j.nima.2007.07.147}
\end{barticle}
\endbibitem

\bibitem[\protect\citeauthoryear{{Feroci} et~al.}{2010}]{Feroci10}
\begin{barticle}
\bauthor{\binits{M.} \bsnm{{Feroci}}},
\bauthor{\binits{E.} \bsnm{{Costa}}},
\bauthor{\binits{E.} \bsnm{{Del Monte}}},
\bauthor{\binits{I.} \bsnm{{Donnarumma}}},
\bauthor{\binits{Y.} \bsnm{{Evangelista}}},
\bauthor{\binits{I.} \bsnm{{Lapshov}}},
\bauthor{\binits{F.} \bsnm{{Lazzarotto}}},
\bauthor{\binits{L.} \bsnm{{Pacciani}}},
\bauthor{\binits{M.} \bsnm{{Rapisarda}}},
\bauthor{\binits{P.} \bsnm{{Soffitta}}},
\bauthor{\binits{G.} \bsnm{{di Persio}}},
\bauthor{\binits{M.} \bsnm{{Frutti}}},
\bauthor{\binits{M.} \bsnm{{Mastropietro}}},
\bauthor{\binits{E.} \bsnm{{Morelli}}},
\bauthor{\binits{G.} \bsnm{{Porrovecchio}}},
\bauthor{\binits{A.} \bsnm{{Rubini}}},
\bauthor{\binits{A.} \bsnm{{Antonelli}}},
\bauthor{\binits{A.} \bsnm{{Argan}}},
\bauthor{\binits{G.} \bsnm{{Barbiellini}}},
\bauthor{\binits{F.} \bsnm{{Boffelli}}},
\bauthor{\binits{A.} \bsnm{{Bulgarelli}}},
\bauthor{\binits{P.} \bsnm{{Caraveo}}},
\bauthor{\binits{P.W.} \bsnm{{Cattaneo}}},
\bauthor{\binits{A.W.} \bsnm{{Chen}}},
\bauthor{\binits{V.} \bsnm{{Cocco}}},
\bauthor{\binits{S.} \bsnm{{Colafrancesco}}},
\bauthor{\binits{S.} \bsnm{{Cutini}}},
\bauthor{\binits{F.} \bsnm{{D'Ammando}}},
\bauthor{\binits{G.} \bsnm{{de Paris}}},
\bauthor{\binits{G.} \bsnm{{Di Cocco}}},
\bauthor{\binits{G.} \bsnm{{Fanari}}},
\bauthor{\binits{A.} \bsnm{{Ferrari}}},
\bauthor{\binits{M.} \bsnm{{Fiorini}}},
\bauthor{\binits{F.} \bsnm{{Fornari}}},
\bauthor{\binits{F.} \bsnm{{Fuschino}}},
\bauthor{\binits{T.} \bsnm{{Froysland}}},
\bauthor{\binits{M.} \bsnm{{Galli}}},
\bauthor{\binits{D.} \bsnm{{Gasparrini}}},
\bauthor{\binits{F.} \bsnm{{Gianotti}}},
\bauthor{\binits{P.} \bsnm{{Giommi}}},
\bauthor{\binits{A.} \bsnm{{Giuliani}}},
\bauthor{\binits{C.} \bsnm{{Labanti}}},
\bauthor{\binits{F.} \bsnm{{Liello}}},
\bauthor{\binits{P.} \bsnm{{Lipari}}},
\bauthor{\binits{F.} \bsnm{{Longo}}},
\bauthor{\binits{E.} \bsnm{{Mattaini}}},
\bauthor{\binits{M.} \bsnm{{Marisaldi}}},
\bauthor{\binits{A.} \bsnm{{Mauri}}},
\bauthor{\binits{F.} \bsnm{{Mauri}}},
\bauthor{\binits{S.} \bsnm{{Mereghetti}}},
\bauthor{\binits{E.} \bsnm{{Moretti}}},
\bauthor{\binits{A.} \bsnm{{Morselli}}},
\bauthor{\binits{A.} \bsnm{{Pellizzoni}}},
\bauthor{\binits{F.} \bsnm{{Perotti}}},
\bauthor{\binits{G.} \bsnm{{Piano}}},
\bauthor{\binits{P.} \bsnm{{Picozza}}},
\bauthor{\binits{M.} \bsnm{{Pilia}}},
\bauthor{\binits{C.} \bsnm{{Pittori}}},
\bauthor{\binits{C.} \bsnm{{Pontoni}}},
\bauthor{\binits{B.} \bsnm{{Preger}}},
\bauthor{\binits{M.} \bsnm{{Prest}}},
\bauthor{\binits{R.} \bsnm{{Primavera}}},
\bauthor{\binits{G.} \bsnm{{Pucella}}},
\bauthor{\binits{A.} \bsnm{{Rappoldi}}},
\bauthor{\binits{E.} \bsnm{{Rossi}}},
\bauthor{\binits{S.} \bsnm{{Sabatini}}},
\bauthor{\binits{P.} \bsnm{{Santolamazza}}},
\bauthor{\binits{M.} \bsnm{{Tavani}}},
\bauthor{\binits{S.} \bsnm{{Stellato}}},
\bauthor{\binits{F.} \bsnm{{Tamburelli}}},
\bauthor{\binits{A.} \bsnm{{Traci}}},
\bauthor{\binits{M.} \bsnm{{Trifoglio}}},
\bauthor{\binits{A.} \bsnm{{Trois}}},
\bauthor{\binits{E.} \bsnm{{Vallazza}}},
\bauthor{\binits{S.} \bsnm{{Vercellone}}},
\bauthor{\binits{F.} \bsnm{{Verrecchia}}},
\bauthor{\binits{V.} \bsnm{{Vittorini}}},
\bauthor{\binits{A.} \bsnm{{Zambra}}},
\bauthor{\binits{D.} \bsnm{{Zanello}}},
\bauthor{\binits{L.} \bsnm{{Salotti}}},
\batitle{{Monitoring the hard X-ray sky with SuperAGILE}}.
\bjtitle{\aap}
\bvolume{510},
\bfpage{9}
(\byear{2010}).
doi:\doiurl{10.1051/0004-6361/200912972}
\end{barticle}
\endbibitem

\bibitem[\protect\citeauthoryear{{Fishman} et~al.}{1992}]{Fishman92}
\begin{bchapter}
\bauthor{\binits{G.J.} \bsnm{{Fishman}}},
\bauthor{\binits{C.A.} \bsnm{{Meegan}}},
\bauthor{\binits{R.B.} \bsnm{{Wilson}}},
\bauthor{\binits{W.S.} \bsnm{{Paciesas}}},
\bauthor{\binits{G.N.} \bsnm{{Pendleton}}},
\bctitle{{The BATSE experiment on the Compton Gamma Ray Observatory: Status and
  some early results}},
in \bbtitle{NASA Conference Publication},
ed. by \beditor{\binits{C.R.} \bsnm{{Shrader}}},
\beditor{\binits{N.} \bsnm{{Gehrels}}},
\beditor{\binits{B.} \bsnm{{Dennis}}}
\bsertitle{NASA Conference Publication},
vol. \bseriesno{3137},
\byear{1992}
\end{bchapter}
\endbibitem

\bibitem[\protect\citeauthoryear{{Forot} et~al.}{2008}]{Forot08}
\begin{barticle}
\bauthor{\binits{M.} \bsnm{{Forot}}},
\bauthor{\binits{P.} \bsnm{{Laurent}}},
\bauthor{\binits{I.A.} \bsnm{{Grenier}}},
\bauthor{\binits{C.} \bsnm{{Gouiff{\`e}s}}},
\bauthor{\binits{F.} \bsnm{{Lebrun}}},
\batitle{{Polarization of the Crab Pulsar and Nebula as Observed by the
  INTEGRAL/IBIS Telescope}}.
\bjtitle{\apjl}
\bvolume{688},
\bfpage{29}
(\byear{2008}).
doi:\doiurl{10.1086/593974}
\end{barticle}
\endbibitem

\bibitem[\protect\citeauthoryear{{Frontera}}{2003}]{Frontera03}
\begin{bchapter}
\bauthor{\binits{F.} \bsnm{{Frontera}}},
\bctitle{{X-Ray Observations of gamma-Ray Burst Afterglows}},
in \bbtitle{Supernovae and Gamma-Ray Bursters},
ed. by \beditor{\binits{K.} \bsnm{{Weiler}}}
\bsertitle{Lecture Notes in Physics, Berlin Springer Verlag},
vol. \bseriesno{598},
\byear{2003},
pp. \bfpage{317}--\blpage{342}
\end{bchapter}
\endbibitem

\bibitem[\protect\citeauthoryear{{Frontera}}{2015}]{Frontera15}
\begin{bchapter}
\bauthor{\binits{F.} \bsnm{{Frontera}}},
\bctitle{{GRB Afterglow Discovery with Bepposax: its Story 15 Years Later}},
in \bbtitle{Thirteenth Marcel Grossmann Meeting: On Recent Developments in
  Theoretical and Experimental General Relativity, Astrophysics and
  Relativistic Field Theories},
ed. by \beditor{\binits{K.} \bsnm{{Rosquist}}},
\byear{2015},
pp. \bfpage{33}--\blpage{53}
\end{bchapter}
\endbibitem

\bibitem[\protect\citeauthoryear{{Frontera} and {Fuligni}}{1975}]{Frontera75b}
\begin{barticle}
\bauthor{\binits{F.} \bsnm{{Frontera}}},
\bauthor{\binits{F.} \bsnm{{Fuligni}}},
\batitle{{Evidence for long-period sporadic pulsations in the hard X-ray flux
  of CYG X-1}}.
\bjtitle{\apjl}
\bvolume{198},
\bfpage{105}--\blpage{108}
(\byear{1975}).
doi:\doiurl{10.1086/181823}
\end{barticle}
\endbibitem

\bibitem[\protect\citeauthoryear{{Frontera} et~al.}{1975}]{Frontera75}
\begin{barticle}
\bauthor{\binits{F.} \bsnm{{Frontera}}},
\bauthor{\binits{F.} \bsnm{{Fuligni}}},
\bauthor{\binits{C.} \bsnm{{Cavani}}},
\batitle{{Long term variability of CYG X-1 in hard X-rays}}.
\bjtitle{\apss}
\bvolume{32},
\bfpage{197}--\blpage{203}
(\byear{1975}).
doi:\doiurl{10.1007/BF00646225}
\end{barticle}
\endbibitem

\bibitem[\protect\citeauthoryear{{Frontera} et~al.}{1972}]{frontera1972;hxr70}
\begin{barticle}
\bauthor{\binits{F.} \bsnm{{Frontera}}},
\bauthor{\binits{F.} \bsnm{{Fuligni}}},
\bauthor{\binits{D.} \bsnm{{Brini}}},
\bauthor{\binits{C.} \bsnm{{Cavani}}},
\batitle{{Search for long-period X-ray pulsations in the NP 0527 region.}}
\bjtitle{Nuovo Cimento Lettere}
\bvolume{5},
\bfpage{131}--\blpage{134}
(\byear{1972})
\end{barticle}
\endbibitem

\bibitem[\protect\citeauthoryear{{Frontera} et~al.}{1979a}]{Frontera79a}
\begin{barticle}
\bauthor{\binits{F.} \bsnm{{Frontera}}},
\bauthor{\binits{F.} \bsnm{{Fuligni}}},
\bauthor{\binits{E.} \bsnm{{Morelli}}},
\bauthor{\binits{G.} \bsnm{{Ventura}}},
\batitle{{Hard X-ray emission from NGC 4151 and MCG 8-11-11 regions}}.
\bjtitle{\apj}
\bvolume{234},
\bfpage{477}--\blpage{480}
(\byear{1979}a).
doi:\doiurl{10.1086/157517}
\end{barticle}
\endbibitem

\bibitem[\protect\citeauthoryear{{Frontera} et~al.}{1979b}]{Frontera79b}
\begin{barticle}
\bauthor{\binits{F.} \bsnm{{Frontera}}},
\bauthor{\binits{F.} \bsnm{{Fuligni}}},
\bauthor{\binits{E.} \bsnm{{Morelli}}},
\bauthor{\binits{G.} \bsnm{{Ventura}}},
\batitle{{Observation of hard X-ray flux variability of 3U 0352+30 = X
  Persei}}.
\bjtitle{\apj}
\bvolume{229},
\bfpage{291}--\blpage{293}
(\byear{1979}b).
doi:\doiurl{10.1086/156953}
\end{barticle}
\endbibitem

\bibitem[\protect\citeauthoryear{{Frontera} et~al.}{1981a}]{Frontera81a}
\begin{barticle}
\bauthor{\binits{F.} \bsnm{{Frontera}}},
\bauthor{\binits{F.} \bsnm{{Fuligni}}},
\bauthor{\binits{E.} \bsnm{{Morelli}}},
\bauthor{\binits{G.} \bsnm{{Pizzichini}}},
\bauthor{\binits{G.} \bsnm{{Ventura}}},
\batitle{{Detection of pseudo gamma-ray bursts of long duration}}.
\bjtitle{\apss}
\bvolume{75},
\bfpage{211}--\blpage{217}
(\byear{1981}a).
doi:\doiurl{10.1007/BF00651397}
\end{barticle}
\endbibitem

\bibitem[\protect\citeauthoryear{{Frontera} et~al.}{1981b}]{Frontera81}
\begin{barticle}
\bauthor{\binits{F.} \bsnm{{Frontera}}},
\bauthor{\binits{F.} \bsnm{{Fuligni}}},
\bauthor{\binits{E.} \bsnm{{Morelli}}},
\bauthor{\binits{G.} \bsnm{{Ventura}}},
\batitle{{Hard X-ray latitude effect measured during a transatlantic balloon
  flight}}.
\bjtitle{Advances in Space Research}
\bvolume{1},
\bfpage{107}--\blpage{110}
(\byear{1981}b).
doi:\doiurl{10.1016/0273-1177(81)90456-7}
\end{barticle}
\endbibitem

\bibitem[\protect\citeauthoryear{{Frontera} et~al.}{1985a}]{Frontera85c}
\begin{barticle}
\bauthor{\binits{F.} \bsnm{{Frontera}}},
\bauthor{\binits{D.} \bsnm{{dal Fiume}}},
\bauthor{\binits{W.} \bsnm{{Dusi}}},
\bauthor{\binits{E.} \bsnm{{Morelli}}},
\bauthor{\binits{G.} \bsnm{{Spada}}},
\batitle{{Hard X-ray observation of galactic X-ray sources}}.
\bjtitle{Advances in Space Research}
\bvolume{5},
\bfpage{125}--\blpage{128}
(\byear{1985}a).
doi:\doiurl{10.1016/0273-1177(85)90464-8}
\end{barticle}
\endbibitem

\bibitem[\protect\citeauthoryear{{Frontera} et~al.}{1985b}]{Frontera85}
\begin{barticle}
\bauthor{\binits{F.} \bsnm{{Frontera}}},
\bauthor{\binits{O.} \bsnm{{Catani}}},
\bauthor{\binits{E.} \bsnm{{Costa}}},
\bauthor{\binits{D.} \bsnm{{dal Fiume}}},
\bauthor{\binits{G.} \bsnm{{Landini}}},
\bauthor{\binits{E.} \bsnm{{Morelli}}},
\bauthor{\binits{A.} \bsnm{{Rubini}}},
\bauthor{\binits{S.} \bsnm{{Silvestri}}},
\bauthor{\binits{G.} \bsnm{{Spada}}},
\bauthor{\binits{M.} \bsnm{{Trifoglio}}},
\batitle{{Performance of different phoswich configurations in a balloon flight
  experiment}}.
\bjtitle{Nuclear Instruments and Methods in Physics Research A}
\bvolume{235},
\bfpage{573}--\blpage{581}
(\byear{1985}b).
doi:\doiurl{10.1016/0168-9002(85)90110-X}
\end{barticle}
\endbibitem

\bibitem[\protect\citeauthoryear{{Frontera}
  et~al.}{1985c}]{Frontera1985;xgrande}
\begin{barticle}
\bauthor{\binits{F.} \bsnm{{Frontera}}},
\bauthor{\binits{D.} \bsnm{{dal Fiume}}},
\bauthor{\binits{E.} \bsnm{{Morelli}}},
\bauthor{\binits{G.} \bsnm{{Spada}}},
\batitle{{The X-ray pulsar A0535 + 26 - Pulse profile and its time variability
  in hard X-rays}}.
\bjtitle{\apj}
\bvolume{298},
\bfpage{585}--\blpage{595}
(\byear{1985}c).
doi:\doiurl{10.1086/163643}
\end{barticle}
\endbibitem

\bibitem[\protect\citeauthoryear{{Frontera} et~al.}{1985d}]{Frontera85b}
\begin{barticle}
\bauthor{\binits{F.} \bsnm{{Frontera}}},
\bauthor{\binits{D.} \bsnm{{dal Fiume}}},
\bauthor{\binits{E.} \bsnm{{Morelli}}},
\bauthor{\binits{G.} \bsnm{{Spada}}},
\batitle{{The X-ray pulsar A0535 + 26 - Pulse profile and its time variability
  in hard X-rays}}.
\bjtitle{\apj}
\bvolume{298},
\bfpage{585}--\blpage{595}
(\byear{1985}d).
doi:\doiurl{10.1086/163643}
\end{barticle}
\endbibitem

\bibitem[\protect\citeauthoryear{{Frontera} et~al.}{1997a}]{Frontera97b}
\begin{barticle}
\bauthor{\binits{F.} \bsnm{{Frontera}}},
\bauthor{\binits{E.} \bsnm{{Costa}}},
\bauthor{\binits{D.} \bsnm{{Dal Fiume}}},
\bauthor{\binits{M.} \bsnm{{Feroci}}},
\bauthor{\binits{L.} \bsnm{{Amati}}},
\bauthor{\binits{M.N.} \bsnm{{Cinti}}},
\bauthor{\binits{A.} \bsnm{{Coletta}}},
\bauthor{\binits{P.} \bsnm{{Collina}}},
\bauthor{\binits{L.} \bsnm{{Nicastro}}},
\bauthor{\binits{M.} \bsnm{{Orlandini}}},
\bauthor{\binits{E.} \bsnm{{Palazzi}}},
\bauthor{\binits{L.} \bsnm{{Piro}}},
\bauthor{\binits{M.} \bsnm{{Rapisarda}}},
\bauthor{\binits{G.} \bsnm{{Zavattini}}},
\batitle{{Initial Results from the Gamma-ray Burst Monitor aboard the X-ray
  Astronomy Satellite BEPPOSAX}}.
\bjtitle{International Cosmic Ray Conference}
\bvolume{3},
\bfpage{25}
(\byear{1997}a)
\end{barticle}
\endbibitem

\bibitem[\protect\citeauthoryear{{Frontera}
  et~al.}{1997b}]{Frontera1997;bepposax}
\begin{barticle}
\bauthor{\binits{F.} \bsnm{{Frontera}}},
\bauthor{\binits{E.} \bsnm{{Costa}}},
\bauthor{\binits{D.} \bsnm{{dal Fiume}}},
\bauthor{\binits{M.} \bsnm{{Feroci}}},
\bauthor{\binits{L.} \bsnm{{Nicastro}}},
\bauthor{\binits{M.} \bsnm{{Orlandini}}},
\bauthor{\binits{E.} \bsnm{{Palazzi}}},
\bauthor{\binits{G.} \bsnm{{Zavattini}}},
\batitle{{The high energy instrument PDS on-board the BeppoSAX X--ray astronomy
  satellite}}.
\bjtitle{\aaps}
\bvolume{122},
\bfpage{357}--\blpage{369}
(\byear{1997}b).
doi:\doiurl{10.1051/aas:1997140}
\end{barticle}
\endbibitem

\bibitem[\protect\citeauthoryear{{Frontera} et~al.}{1998}]{Frontera98}
\begin{barticle}
\bauthor{\binits{F.} \bsnm{{Frontera}}},
\bauthor{\binits{J.} \bsnm{{Greiner}}},
\bauthor{\binits{L.A.} \bsnm{{Antonelli}}},
\bauthor{\binits{E.} \bsnm{{Costa}}},
\bauthor{\binits{F.} \bsnm{{Fiore}}},
\bauthor{\binits{A.N.} \bsnm{{Parmar}}},
\bauthor{\binits{L.} \bsnm{{Piro}}},
\bauthor{\binits{T.} \bsnm{{Boller}}},
\bauthor{\binits{W.} \bsnm{{Voges}}},
\batitle{{High resolution imaging of the X-ray afterglow of GRB970228 with
  ROSAT}}.
\bjtitle{\aap}
\bvolume{334},
\bfpage{69}--\blpage{72}
(\byear{1998})
\end{barticle}
\endbibitem

\bibitem[\protect\citeauthoryear{{Frontera} et~al.}{2001}]{Frontera01}
\begin{barticle}
\bauthor{\binits{F.} \bsnm{{Frontera}}},
\bauthor{\binits{E.} \bsnm{{Palazzi}}},
\bauthor{\binits{A.A.} \bsnm{{Zdziarski}}},
\bauthor{\binits{F.} \bsnm{{Haardt}}},
\bauthor{\binits{G.C.} \bsnm{{Perola}}},
\bauthor{\binits{L.} \bsnm{{Chiappetti}}},
\bauthor{\binits{G.} \bsnm{{Cusumano}}},
\bauthor{\binits{D.} \bsnm{{Dal Fiume}}},
\bauthor{\binits{S.} \bsnm{{Del Sordo}}},
\bauthor{\binits{M.} \bsnm{{Orlandini}}},
\bauthor{\binits{A.N.} \bsnm{{Parmar}}},
\bauthor{\binits{L.} \bsnm{{Piro}}},
\bauthor{\binits{A.} \bsnm{{Santangelo}}},
\bauthor{\binits{A.} \bsnm{{Segreto}}},
\bauthor{\binits{A.} \bsnm{{Treves}}},
\bauthor{\binits{M.} \bsnm{{Trifoglio}}},
\batitle{{Broadband Spectrum of Cygnus X-1 in Two Spectral States with
  BeppoSAX}}.
\bjtitle{\apj}
\bvolume{546},
\bfpage{1027}--\blpage{1037}
(\byear{2001}).
doi:\doiurl{10.1086/318304}
\end{barticle}
\endbibitem

\bibitem[\protect\citeauthoryear{{Frontera} et~al.}{2007}]{Frontera07}
\begin{barticle}
\bauthor{\binits{F.} \bsnm{{Frontera}}},
\bauthor{\binits{M.} \bsnm{{Orlandini}}},
\bauthor{\binits{R.} \bsnm{{Landi}}},
\bauthor{\binits{A.} \bsnm{{Comastri}}},
\bauthor{\binits{F.} \bsnm{{Fiore}}},
\bauthor{\binits{G.} \bsnm{{Setti}}},
\bauthor{\binits{L.} \bsnm{{Amati}}},
\bauthor{\binits{E.} \bsnm{{Costa}}},
\bauthor{\binits{N.} \bsnm{{Masetti}}},
\bauthor{\binits{E.} \bsnm{{Palazzi}}},
\batitle{{The Cosmic X-Ray Background and the Population of the Most Heavily
  Obscured AGNs}}.
\bjtitle{\apj}
\bvolume{666},
\bfpage{86}--\blpage{95}
(\byear{2007}).
doi:\doiurl{10.1086/519985}
\end{barticle}
\endbibitem

\bibitem[\protect\citeauthoryear{{Frontera} et~al.}{2009}]{Frontera09}
\begin{barticle}
\bauthor{\binits{F.} \bsnm{{Frontera}}},
\bauthor{\binits{C.} \bsnm{{Guidorzi}}},
\bauthor{\binits{E.} \bsnm{{Montanari}}},
\bauthor{\binits{F.} \bsnm{{Rossi}}},
\bauthor{\binits{E.} \bsnm{{Costa}}},
\bauthor{\binits{M.} \bsnm{{Feroci}}},
\bauthor{\binits{F.} \bsnm{{Calura}}},
\bauthor{\binits{M.} \bsnm{{Rapisarda}}},
\bauthor{\binits{L.} \bsnm{{Amati}}},
\bauthor{\binits{D.} \bsnm{{Carturan}}},
\bauthor{\binits{M.R.} \bsnm{{Cinti}}},
\bauthor{\binits{D.D.} \bsnm{{Fiume}}},
\bauthor{\binits{L.} \bsnm{{Nicastro}}},
\bauthor{\binits{M.} \bsnm{{Orlandini}}},
\batitle{{The Gamma-Ray Burst Catalog Obtained with the Gamma-Ray Burst Monitor
  Aboard BeppoSAX}}.
\bjtitle{\apjs}
\bvolume{180},
\bfpage{192}--\blpage{223}
(\byear{2009}).
doi:\doiurl{10.1088/0067-0049/180/1/192}
\end{barticle}
\endbibitem

\bibitem[\protect\citeauthoryear{{Frontera} et~al.}{2013}]{Frontera2013;laue}
\begin{bchapter}
\bauthor{\binits{F.} \bsnm{{Frontera}}},
\bauthor{\binits{E.} \bsnm{{Virgilli}}},
\bauthor{\binits{V.} \bsnm{{Valsan}}},
\bauthor{\binits{V.} \bsnm{{Liccardo}}},
\bauthor{\binits{V.} \bsnm{{Carassiti}}},
\bauthor{\binits{E.} \bsnm{{Caroli}}},
\bauthor{\binits{F.} \bsnm{{Cassese}}},
\bauthor{\binits{C.} \bsnm{{Ferrari}}},
\bauthor{\binits{V.} \bsnm{{Guidi}}},
\bauthor{\binits{S.} \bsnm{{Mottini}}},
\bauthor{\binits{M.} \bsnm{{Pecora}}},
\bauthor{\binits{B.} \bsnm{{Negri}}},
\bauthor{\binits{L.} \bsnm{{Recanatesi}}},
\bauthor{\binits{L.} \bsnm{{Amati}}},
\bauthor{\binits{N.} \bsnm{{Auricchio}}},
\bauthor{\binits{L.} \bsnm{{Bassani}}},
\bauthor{\binits{R.} \bsnm{{Campana}}},
\bauthor{\binits{R.} \bsnm{{Farinelli}}},
\bauthor{\binits{C.} \bsnm{{Guidorzi}}},
\bauthor{\binits{C.} \bsnm{{Labanti}}},
\bauthor{\binits{R.} \bsnm{{Landi}}},
\bauthor{\binits{A.} \bsnm{{Malizia}}},
\bauthor{\binits{M.} \bsnm{{Orlandini}}},
\bauthor{\binits{P.} \bsnm{{Rosati}}},
\bauthor{\binits{V.} \bsnm{{Sguera}}},
\bauthor{\binits{J.} \bsnm{{Stephen}}},
\bauthor{\binits{L.} \bsnm{{Titarchuk}}},
\bctitle{{Scientific prospects in soft gamma-ray astronomy enabled by the LAUE
  project}},
in \bbtitle{Society of Photo-Optical Instrumentation Engineers (SPIE)
  Conference Series}.
\bsertitle{Society of Photo-Optical Instrumentation Engineers (SPIE) Conference
  Series},
vol. \bseriesno{8861},
\byear{2013},
p. \bfpage{6}.
doi:\doiurl{10.1117/12.2023589}
\end{bchapter}
\endbibitem

\bibitem[\protect\citeauthoryear{{Frost} et~al.}{1971}]{Frost1971}
\begin{bchapter}
\bauthor{\binits{K.J.} \bsnm{{Frost}}},
\bauthor{\binits{B.R.} \bsnm{{Dennis}}},
\bauthor{\binits{R.J.} \bsnm{{Lencho}}},
\bctitle{{Experiment to Measure Hard Solar and Celestial X-Rays from the Fifth
  Orbiting Solar Observatory}},
in \bbtitle{New techniques in Space Astronomy},
ed. by \beditor{\binits{F.} \bsnm{{Labuhn}}},
\beditor{\binits{R.} \bsnm{{Lust}}}
\bsertitle{IAU Symposium},
vol. \bseriesno{41},
\byear{1971},
p. \bfpage{185}
\end{bchapter}
\endbibitem

\bibitem[\protect\citeauthoryear{{Fujita} et~al.}{2008}]{Fujita08}
\begin{barticle}
\bauthor{\binits{Y.} \bsnm{{Fujita}}},
\bauthor{\binits{K.} \bsnm{{Hayashida}}},
\bauthor{\binits{M.} \bsnm{{Nagai}}},
\bauthor{\binits{S.} \bsnm{{Inoue}}},
\bauthor{\binits{H.} \bsnm{{Matsumoto}}},
\bauthor{\binits{N.} \bsnm{{Okabe}}},
\bauthor{\binits{T.H.} \bsnm{{Reiprich}}},
\bauthor{\binits{C.L.} \bsnm{{Sarazin}}},
\bauthor{\binits{M.} \bsnm{{Takizawa}}},
\batitle{{Suzaku Observation of the Ophiuchus Galaxy Cluster: One of the
  Hottest Cool Core Clusters}}.
\bjtitle{\pasj}
\bvolume{60},
\bfpage{1133}--\blpage{1142}
(\byear{2008}).
doi:\doiurl{10.1093/pasj/60.5.1133}
\end{barticle}
\endbibitem

\bibitem[\protect\citeauthoryear{{Fuligni} et~al.}{1979}]{Frontera79c}
\begin{bchapter}
\bauthor{\binits{F.} \bsnm{{Fuligni}}},
\bauthor{\binits{W.} \bsnm{{Dusi}}},
\bauthor{\binits{F.} \bsnm{{Frontera}}},
\bauthor{\binits{E.} \bsnm{{Morelli}}},
\bauthor{\binits{G.} \bsnm{{Ventura}}},
\bctitle{{Hard X-Ray Transient Source Observed During the 1976 Transatlantic
  Flight}},
in \bbtitle{X-ray Astronomy},
ed. by \beditor{\binits{W.A.} \bsnm{{Baity}}},
\beditor{\binits{L.E.} \bsnm{{Peterson}}},
\byear{1979},
p. \bfpage{505}
\end{bchapter}
\endbibitem

\bibitem[\protect\citeauthoryear{{F{\"u}rst} et~al.}{2013}]{Furst13}
\begin{barticle}
\bauthor{\binits{F.} \bsnm{{F{\"u}rst}}},
\bauthor{\binits{B.W.} \bsnm{{Grefenstette}}},
\bauthor{\binits{R.} \bsnm{{Staubert}}},
\bauthor{\binits{J.A.} \bsnm{{Tomsick}}},
\bauthor{\binits{M.} \bsnm{{Bachetti}}},
\bauthor{\binits{D.} \bsnm{{Barret}}},
\bauthor{\binits{E.C.} \bsnm{{Bellm}}},
\bauthor{\binits{S.E.} \bsnm{{Boggs}}},
\bauthor{\binits{J.} \bsnm{{Chenevez}}},
\bauthor{\binits{F.E.} \bsnm{{Christensen}}},
\bauthor{\binits{W.W.} \bsnm{{Craig}}},
\bauthor{\binits{C.J.} \bsnm{{Hailey}}},
\bauthor{\binits{F.} \bsnm{{Harrison}}},
\bauthor{\binits{D.} \bsnm{{Klochkov}}},
\bauthor{\binits{K.K.} \bsnm{{Madsen}}},
\bauthor{\binits{K.} \bsnm{{Pottschmidt}}},
\bauthor{\binits{D.} \bsnm{{Stern}}},
\bauthor{\binits{D.J.} \bsnm{{Walton}}},
\bauthor{\binits{J.} \bsnm{{Wilms}}},
\bauthor{\binits{W.} \bsnm{{Zhang}}},
\batitle{{The Smooth Cyclotron Line in Her X-1 as Seen with Nuclear
  Spectroscopic Telescope Array}}.
\bjtitle{\apj}
\bvolume{779},
\bfpage{69}
(\byear{2013}).
doi:\doiurl{10.1088/0004-637X/779/1/69}
\end{barticle}
\endbibitem

\bibitem[\protect\citeauthoryear{{F{\"u}rst} et~al.}{2014a}]{Furst14b}
\begin{barticle}
\bauthor{\binits{F.} \bsnm{{F{\"u}rst}}},
\bauthor{\binits{K.} \bsnm{{Pottschmidt}}},
\bauthor{\binits{J.} \bsnm{{Wilms}}},
\bauthor{\binits{J.} \bsnm{{Kennea}}},
\bauthor{\binits{M.} \bsnm{{Bachetti}}},
\bauthor{\binits{E.} \bsnm{{Bellm}}},
\bauthor{\binits{S.E.} \bsnm{{Boggs}}},
\bauthor{\binits{D.} \bsnm{{Chakrabarty}}},
\bauthor{\binits{F.E.} \bsnm{{Christensen}}},
\bauthor{\binits{W.W.} \bsnm{{Craig}}},
\bauthor{\binits{C.J.} \bsnm{{Hailey}}},
\bauthor{\binits{F.} \bsnm{{Harrison}}},
\bauthor{\binits{D.} \bsnm{{Stern}}},
\bauthor{\binits{J.A.} \bsnm{{Tomsick}}},
\bauthor{\binits{D.J.} \bsnm{{Walton}}},
\bauthor{\binits{W.} \bsnm{{Zhang}}},
\batitle{{NuSTAR Discovery of a Cyclotron Line in KS 1947+300}}.
\bjtitle{\apjl}
\bvolume{784},
\bfpage{40}
(\byear{2014}a).
doi:\doiurl{10.1088/2041-8205/784/2/L40}
\end{barticle}
\endbibitem

\bibitem[\protect\citeauthoryear{{F{\"u}rst} et~al.}{2014b}]{Furst14a}
\begin{barticle}
\bauthor{\binits{F.} \bsnm{{F{\"u}rst}}},
\bauthor{\binits{K.} \bsnm{{Pottschmidt}}},
\bauthor{\binits{J.} \bsnm{{Wilms}}},
\bauthor{\binits{J.A.} \bsnm{{Tomsick}}},
\bauthor{\binits{M.} \bsnm{{Bachetti}}},
\bauthor{\binits{S.E.} \bsnm{{Boggs}}},
\bauthor{\binits{F.E.} \bsnm{{Christensen}}},
\bauthor{\binits{W.W.} \bsnm{{Craig}}},
\bauthor{\binits{B.W.} \bsnm{{Grefenstette}}},
\bauthor{\binits{C.J.} \bsnm{{Hailey}}},
\bauthor{\binits{F.} \bsnm{{Harrison}}},
\bauthor{\binits{K.K.} \bsnm{{Madsen}}},
\bauthor{\binits{J.M.} \bsnm{{Miller}}},
\bauthor{\binits{D.} \bsnm{{Stern}}},
\bauthor{\binits{D.J.} \bsnm{{Walton}}},
\bauthor{\binits{W.} \bsnm{{Zhang}}},
\batitle{{NuSTAR Discovery of a Luminosity Dependent Cyclotron Line Energy in
  Vela X-1}}.
\bjtitle{\apj}
\bvolume{780},
\bfpage{133}
(\byear{2014}b).
doi:\doiurl{10.1088/0004-637X/780/2/133}
\end{barticle}
\endbibitem

\bibitem[\protect\citeauthoryear{{Fusco-Femiano}
  et~al.}{1999}]{Fusco-Femiano99}
\begin{barticle}
\bauthor{\binits{R.} \bsnm{{Fusco-Femiano}}},
\bauthor{\binits{D.} \bsnm{{Dal Fiume}}},
\bauthor{\binits{L.} \bsnm{{Feretti}}},
\bauthor{\binits{G.} \bsnm{{Giovannini}}},
\bauthor{\binits{P.} \bsnm{{Grandi}}},
\bauthor{\binits{G.} \bsnm{{Matt}}},
\bauthor{\binits{S.} \bsnm{{Molendi}}},
\bauthor{\binits{A.} \bsnm{{Santangelo}}},
\batitle{{Hard X-Ray Radiation in the Coma Cluster Spectrum}}.
\bjtitle{\apjl}
\bvolume{513},
\bfpage{21}--\blpage{24}
(\byear{1999}).
doi:\doiurl{10.1086/311902}
\end{barticle}
\endbibitem

\bibitem[\protect\citeauthoryear{{Fusco-Femiano}
  et~al.}{2000}]{Fusco-Femiano00}
\begin{barticle}
\bauthor{\binits{R.} \bsnm{{Fusco-Femiano}}},
\bauthor{\binits{D.} \bsnm{{Dal Fiume}}},
\bauthor{\binits{S.} \bsnm{{De Grandi}}},
\bauthor{\binits{L.} \bsnm{{Feretti}}},
\bauthor{\binits{G.} \bsnm{{Giovannini}}},
\bauthor{\binits{P.} \bsnm{{Grandi}}},
\bauthor{\binits{A.} \bsnm{{Malizia}}},
\bauthor{\binits{G.} \bsnm{{Matt}}},
\bauthor{\binits{S.} \bsnm{{Molendi}}},
\batitle{{Hard X-Ray Emission from the Galaxy Cluster A2256}}.
\bjtitle{\apjl}
\bvolume{534},
\bfpage{7}--\blpage{10}
(\byear{2000}).
doi:\doiurl{10.1086/312639}
\end{barticle}
\endbibitem

\bibitem[\protect\citeauthoryear{{Galama} et~al.}{1998}]{Galama98}
\begin{barticle}
\bauthor{\binits{T.J.} \bsnm{{Galama}}},
\bauthor{\binits{P.M.} \bsnm{{Vreeswijk}}},
\bauthor{\binits{J.} \bsnm{{van Paradijs}}},
\bauthor{\binits{C.} \bsnm{{Kouveliotou}}},
\bauthor{\binits{T.} \bsnm{{Augusteijn}}},
\bauthor{\binits{H.} \bsnm{{B{\"o}hnhardt}}},
\bauthor{\binits{J.P.} \bsnm{{Brewer}}},
\bauthor{\binits{V.} \bsnm{{Doublier}}},
\bauthor{\binits{J.-F.} \bsnm{{Gonzalez}}},
\bauthor{\binits{B.} \bsnm{{Leibundgut}}},
\bauthor{\binits{C.} \bsnm{{Lidman}}},
\bauthor{\binits{O.R.} \bsnm{{Hainaut}}},
\bauthor{\binits{F.} \bsnm{{Patat}}},
\bauthor{\binits{J.} \bsnm{{Heise}}},
\bauthor{\binits{J.} \bsnm{{in't Zand}}},
\bauthor{\binits{K.} \bsnm{{Hurley}}},
\bauthor{\binits{P.J.} \bsnm{{Groot}}},
\bauthor{\binits{R.G.} \bsnm{{Strom}}},
\bauthor{\binits{P.A.} \bsnm{{Mazzali}}},
\bauthor{\binits{K.} \bsnm{{Iwamoto}}},
\bauthor{\binits{K.} \bsnm{{Nomoto}}},
\bauthor{\binits{H.} \bsnm{{Umeda}}},
\bauthor{\binits{T.} \bsnm{{Nakamura}}},
\bauthor{\binits{T.R.} \bsnm{{Young}}},
\bauthor{\binits{T.} \bsnm{{Suzuki}}},
\bauthor{\binits{T.} \bsnm{{Shigeyama}}},
\bauthor{\binits{T.} \bsnm{{Koshut}}},
\bauthor{\binits{M.} \bsnm{{Kippen}}},
\bauthor{\binits{C.} \bsnm{{Robinson}}},
\bauthor{\binits{P.} \bsnm{{de Wildt}}},
\bauthor{\binits{R.A.M.J.} \bsnm{{Wijers}}},
\bauthor{\binits{N.} \bsnm{{Tanvir}}},
\bauthor{\binits{J.} \bsnm{{Greiner}}},
\bauthor{\binits{E.} \bsnm{{Pian}}},
\bauthor{\binits{E.} \bsnm{{Palazzi}}},
\bauthor{\binits{F.} \bsnm{{Frontera}}},
\bauthor{\binits{N.} \bsnm{{Masetti}}},
\bauthor{\binits{L.} \bsnm{{Nicastro}}},
\bauthor{\binits{M.} \bsnm{{Feroci}}},
\bauthor{\binits{E.} \bsnm{{Costa}}},
\bauthor{\binits{L.} \bsnm{{Piro}}},
\bauthor{\binits{B.A.} \bsnm{{Peterson}}},
\bauthor{\binits{C.} \bsnm{{Tinney}}},
\bauthor{\binits{B.} \bsnm{{Boyle}}},
\bauthor{\binits{R.} \bsnm{{Cannon}}},
\bauthor{\binits{R.} \bsnm{{Stathakis}}},
\bauthor{\binits{E.} \bsnm{{Sadler}}},
\bauthor{\binits{M.C.} \bsnm{{Begam}}},
\bauthor{\binits{P.} \bsnm{{Ianna}}},
\batitle{{An unusual supernova in the error box of the {$\gamma$}-ray burst of
  25 April 1998}}.
\bjtitle{\nat}
\bvolume{395},
\bfpage{670}--\blpage{672}
(\byear{1998}).
doi:\doiurl{10.1038/27150}
\end{barticle}
\endbibitem

\bibitem[\protect\citeauthoryear{{Gehrels} and {Razzaque}}{2013}]{Gehrels13}
\begin{barticle}
\bauthor{\binits{N.} \bsnm{{Gehrels}}},
\bauthor{\binits{S.} \bsnm{{Razzaque}}},
\batitle{{Gamma-ray bursts in the swift-Fermi era}}.
\bjtitle{Frontiers of Physics}
\bvolume{8},
\bfpage{661}--\blpage{678}
(\byear{2013}).
doi:\doiurl{10.1007/s11467-013-0282-3}
\end{barticle}
\endbibitem

\bibitem[\protect\citeauthoryear{{Gehrels} et~al.}{1991}]{Gehrels91}
\begin{barticle}
\bauthor{\binits{N.} \bsnm{{Gehrels}}},
\bauthor{\binits{S.D.} \bsnm{{Barthelmy}}},
\bauthor{\binits{B.J.} \bsnm{{Teegarden}}},
\bauthor{\binits{J.} \bsnm{{Tueller}}},
\bauthor{\binits{M.} \bsnm{{Leventhal}}},
\bauthor{\binits{C.J.} \bsnm{{MacCallum}}},
\batitle{{GRIS observations of positron annihilation radiation from the
  Galactic center}}.
\bjtitle{\apjl}
\bvolume{375},
\bfpage{13}--\blpage{16}
(\byear{1991}).
doi:\doiurl{10.1086/186077}
\end{barticle}
\endbibitem

\bibitem[\protect\citeauthoryear{{Gehrels} et~al.}{2005}]{Gehrels05}
\begin{barticle}
\bauthor{\binits{N.} \bsnm{{Gehrels}}},
\bauthor{\binits{C.L.} \bsnm{{Sarazin}}},
\bauthor{\binits{P.T.} \bsnm{{O'Brien}}},
\bauthor{\binits{B.} \bsnm{{Zhang}}},
\bauthor{\binits{L.} \bsnm{{Barbier}}},
\bauthor{\binits{S.D.} \bsnm{{Barthelmy}}},
\bauthor{\binits{A.} \bsnm{{Blustin}}},
\bauthor{\binits{D.N.} \bsnm{{Burrows}}},
\bauthor{\binits{J.} \bsnm{{Cannizzo}}},
\bauthor{\binits{J.R.} \bsnm{{Cummings}}},
\bauthor{\binits{M.} \bsnm{{Goad}}},
\bauthor{\binits{S.T.} \bsnm{{Holland}}},
\bauthor{\binits{C.P.} \bsnm{{Hurkett}}},
\bauthor{\binits{J.A.} \bsnm{{Kennea}}},
\bauthor{\binits{A.} \bsnm{{Levan}}},
\bauthor{\binits{C.B.} \bsnm{{Markwardt}}},
\bauthor{\binits{K.O.} \bsnm{{Mason}}},
\bauthor{\binits{P.} \bsnm{{Meszaros}}},
\bauthor{\binits{M.} \bsnm{{Page}}},
\bauthor{\binits{D.M.} \bsnm{{Palmer}}},
\bauthor{\binits{E.} \bsnm{{Rol}}},
\bauthor{\binits{T.} \bsnm{{Sakamoto}}},
\bauthor{\binits{R.} \bsnm{{Willingale}}},
\bauthor{\binits{L.} \bsnm{{Angelini}}},
\bauthor{\binits{A.} \bsnm{{Beardmore}}},
\bauthor{\binits{P.T.} \bsnm{{Boyd}}},
\bauthor{\binits{A.} \bsnm{{Breeveld}}},
\bauthor{\binits{S.} \bsnm{{Campana}}},
\bauthor{\binits{M.M.} \bsnm{{Chester}}},
\bauthor{\binits{G.} \bsnm{{Chincarini}}},
\bauthor{\binits{L.R.} \bsnm{{Cominsky}}},
\bauthor{\binits{G.} \bsnm{{Cusumano}}},
\bauthor{\binits{M.} \bsnm{{de Pasquale}}},
\bauthor{\binits{E.E.} \bsnm{{Fenimore}}},
\bauthor{\binits{P.} \bsnm{{Giommi}}},
\bauthor{\binits{C.} \bsnm{{Gronwall}}},
\bauthor{\binits{D.} \bsnm{{Grupe}}},
\bauthor{\binits{J.E.} \bsnm{{Hill}}},
\bauthor{\binits{D.} \bsnm{{Hinshaw}}},
\bauthor{\binits{J.} \bsnm{{Hjorth}}},
\bauthor{\binits{D.} \bsnm{{Hullinger}}},
\bauthor{\binits{K.C.} \bsnm{{Hurley}}},
\bauthor{\binits{S.} \bsnm{{Klose}}},
\bauthor{\binits{S.} \bsnm{{Kobayashi}}},
\bauthor{\binits{C.} \bsnm{{Kouveliotou}}},
\bauthor{\binits{H.A.} \bsnm{{Krimm}}},
\bauthor{\binits{V.} \bsnm{{Mangano}}},
\bauthor{\binits{F.E.} \bsnm{{Marshall}}},
\bauthor{\binits{K.} \bsnm{{McGowan}}},
\bauthor{\binits{A.} \bsnm{{Moretti}}},
\bauthor{\binits{R.F.} \bsnm{{Mushotzky}}},
\bauthor{\binits{K.} \bsnm{{Nakazawa}}},
\bauthor{\binits{J.P.} \bsnm{{Norris}}},
\bauthor{\binits{J.A.} \bsnm{{Nousek}}},
\bauthor{\binits{J.P.} \bsnm{{Osborne}}},
\bauthor{\binits{K.} \bsnm{{Page}}},
\bauthor{\binits{A.M.} \bsnm{{Parsons}}},
\bauthor{\binits{S.} \bsnm{{Patel}}},
\bauthor{\binits{M.} \bsnm{{Perri}}},
\bauthor{\binits{T.} \bsnm{{Poole}}},
\bauthor{\binits{P.} \bsnm{{Romano}}},
\bauthor{\binits{P.W.A.} \bsnm{{Roming}}},
\bauthor{\binits{S.} \bsnm{{Rosen}}},
\bauthor{\binits{G.} \bsnm{{Sato}}},
\bauthor{\binits{P.} \bsnm{{Schady}}},
\bauthor{\binits{A.P.} \bsnm{{Smale}}},
\bauthor{\binits{J.} \bsnm{{Sollerman}}},
\bauthor{\binits{R.} \bsnm{{Starling}}},
\bauthor{\binits{M.} \bsnm{{Still}}},
\bauthor{\binits{M.} \bsnm{{Suzuki}}},
\bauthor{\binits{G.} \bsnm{{Tagliaferri}}},
\bauthor{\binits{T.} \bsnm{{Takahashi}}},
\bauthor{\binits{M.} \bsnm{{Tashiro}}},
\bauthor{\binits{J.} \bsnm{{Tueller}}},
\bauthor{\binits{A.A.} \bsnm{{Wells}}},
\bauthor{\binits{N.E.} \bsnm{{White}}},
\bauthor{\binits{R.A.M.J.} \bsnm{{Wijers}}},
\batitle{{A short {$\gamma$}-ray burst apparently associated with an elliptical
  galaxy at redshift z = 0.225}}.
\bjtitle{\nat}
\bvolume{437},
\bfpage{851}--\blpage{854}
(\byear{2005}).
doi:\doiurl{10.1038/nature04142}
\end{barticle}
\endbibitem

\bibitem[\protect\citeauthoryear{{Ghisellini} et~al.}{1999}]{Ghisellini99}
\begin{barticle}
\bauthor{\binits{G.} \bsnm{{Ghisellini}}},
\bauthor{\binits{L.} \bsnm{{Costamante}}},
\bauthor{\binits{G.} \bsnm{{Tagliaferri}}},
\bauthor{\binits{L.} \bsnm{{Maraschi}}},
\bauthor{\binits{A.} \bsnm{{Celotti}}},
\bauthor{\binits{G.} \bsnm{{Fossati}}},
\bauthor{\binits{E.} \bsnm{{Pian}}},
\bauthor{\binits{A.} \bsnm{{Comastri}}},
\bauthor{\binits{G.} \bsnm{{de Francesco}}},
\bauthor{\binits{L.} \bsnm{{Lanteri}}},
\bauthor{\binits{C.M.} \bsnm{{Raiteri}}},
\bauthor{\binits{G.} \bsnm{{Sobrito}}},
\bauthor{\binits{M.} \bsnm{{Villata}}},
\bauthor{\binits{I.S.} \bsnm{{Glass}}},
\bauthor{\binits{P.} \bsnm{{Grandi}}},
\bauthor{\binits{C.} \bsnm{{Perola}}},
\bauthor{\binits{A.} \bsnm{{Treves}}},
\batitle{{The blazar PKS 0528+134: new results from BeppoSAX observations}}.
\bjtitle{\aap}
\bvolume{348},
\bfpage{63}--\blpage{70}
(\byear{1999})
\end{barticle}
\endbibitem

\bibitem[\protect\citeauthoryear{{Giacconi} et~al.}{1962}]{Giacconi62}
\begin{barticle}
\bauthor{\binits{R.} \bsnm{{Giacconi}}},
\bauthor{\binits{H.} \bsnm{{Gursky}}},
\bauthor{\binits{F.R.} \bsnm{{Paolini}}},
\bauthor{\binits{B.B.} \bsnm{{Rossi}}},
\batitle{{Evidence for x Rays From Sources Outside the Solar System}}.
\bjtitle{Physical Review Letters}
\bvolume{9},
\bfpage{439}--\blpage{443}
(\byear{1962}).
doi:\doiurl{10.1103/PhysRevLett.9.439}
\end{barticle}
\endbibitem

\bibitem[\protect\citeauthoryear{{Gierli{\'n}ski} and
  {Done}}{2002}]{Gierlinski02}
\begin{barticle}
\bauthor{\binits{M.} \bsnm{{Gierli{\'n}ski}}},
\bauthor{\binits{C.} \bsnm{{Done}}},
\batitle{{The X-ray spectrum of the atoll source 4U 1608-52}}.
\bjtitle{\mnras}
\bvolume{337},
\bfpage{1373}--\blpage{1380}
(\byear{2002}).
doi:\doiurl{10.1046/j.1365-8711.2002.06009.x}
\end{barticle}
\endbibitem

\bibitem[\protect\citeauthoryear{{Gierli{\'n}ski} and
  {Zdziarski}}{2003}]{Gierlinski03}
\begin{barticle}
\bauthor{\binits{M.} \bsnm{{Gierli{\'n}ski}}},
\bauthor{\binits{A.A.} \bsnm{{Zdziarski}}},
\batitle{{Discovery of powerful millisecond flares from Cygnus X-1}}.
\bjtitle{\mnras}
\bvolume{343},
\bfpage{84}--\blpage{88}
(\byear{2003}).
doi:\doiurl{10.1046/j.1365-8711.2003.06890.x}
\end{barticle}
\endbibitem

\bibitem[\protect\citeauthoryear{{Gilfanov} et~al.}{1991}]{Gilfanov91}
\begin{barticle}
\bauthor{\binits{M.} \bsnm{{Gilfanov}}},
\bauthor{\binits{R.} \bsnm{{Sunyaev}}},
\bauthor{\binits{E.} \bsnm{{Churazov}}},
\bauthor{\binits{M.} \bsnm{{Pavlinskii}}},
\bauthor{\binits{S.} \bsnm{{Grebenev}}},
\bauthor{\binits{R.} \bsnm{{Kremnev}}},
\bauthor{\binits{K.} \bsnm{{Sukhanov}}},
\bauthor{\binits{N.} \bsnm{{Kuleshova}}},
\bauthor{\binits{A.} \bsnm{{Goldwurm}}},
\bauthor{\binits{J.} \bsnm{{Ballet}}},
\bauthor{\binits{B.} \bsnm{{Cordier}}},
\bauthor{\binits{J.} \bsnm{{Paul}}},
\bauthor{\binits{M.} \bsnm{{Denis}}},
\bauthor{\binits{L.} \bsnm{{Bouchet}}},
\bauthor{\binits{D.} \bsnm{{Barret}}},
\bauthor{\binits{J.P.} \bsnm{{Roques}}},
\batitle{{Observations of Nova MUSCAE with the Sigma Telescope on the GRANAT
  Observatory - Spectroscopic Properties in Hard X-Rays and Discovery of the
  Annihilation Line in the Spectrum}}.
\bjtitle{Soviet Astronomy Letters}
\bvolume{17},
\bfpage{437}
(\byear{1991})
\end{barticle}
\endbibitem

\bibitem[\protect\citeauthoryear{{Gilli} et~al.}{2000}]{Gilli00}
\begin{barticle}
\bauthor{\binits{R.} \bsnm{{Gilli}}},
\bauthor{\binits{R.} \bsnm{{Maiolino}}},
\bauthor{\binits{A.} \bsnm{{Marconi}}},
\bauthor{\binits{G.} \bsnm{{Risaliti}}},
\bauthor{\binits{M.} \bsnm{{Dadina}}},
\bauthor{\binits{K.A.} \bsnm{{Weaver}}},
\bauthor{\binits{E.J.M.} \bsnm{{Colbert}}},
\batitle{{The variability of the Seyfert galaxy NGC 2992: the case for a
  revived AGN}}.
\bjtitle{\aap}
\bvolume{355},
\bfpage{485}--\blpage{498}
(\byear{2000})
\end{barticle}
\endbibitem

\bibitem[\protect\citeauthoryear{{Giommi} et~al.}{1998}]{Giommi98}
\begin{barticle}
\bauthor{\binits{P.} \bsnm{{Giommi}}},
\bauthor{\binits{F.} \bsnm{{Fiore}}},
\bauthor{\binits{M.} \bsnm{{Guainazzi}}},
\bauthor{\binits{M.} \bsnm{{Feroci}}},
\bauthor{\binits{F.} \bsnm{{Frontera}}},
\bauthor{\binits{G.} \bsnm{{Ghisellini}}},
\bauthor{\binits{P.} \bsnm{{Grandi}}},
\bauthor{\binits{L.} \bsnm{{Maraschi}}},
\bauthor{\binits{T.} \bsnm{{Mineo}}},
\bauthor{\binits{S.} \bsnm{{Molendi}}},
\bauthor{\binits{A.} \bsnm{{Orr}}},
\bauthor{\binits{S.} \bsnm{{Piraino}}},
\bauthor{\binits{A.} \bsnm{{Segreto}}},
\bauthor{\binits{G.} \bsnm{{Tagliaferri}}},
\bauthor{\binits{A.} \bsnm{{Treves}}},
\batitle{{The complex 0.1-100 keV X-ray spectrum of PKS 2155-304}}.
\bjtitle{\aap}
\bvolume{333},
\bfpage{5}--\blpage{8}
(\byear{1998})
\end{barticle}
\endbibitem

\bibitem[\protect\citeauthoryear{{Glass}}{1969}]{Glass69}
\begin{barticle}
\bauthor{\binits{I.S.} \bsnm{{Glass}}},
\batitle{{Observations of X-Rays from Taurus X-1 and Cygnus X-1}}.
\bjtitle{\apj}
\bvolume{157},
\bfpage{215}
(\byear{1969}).
doi:\doiurl{10.1086/150061}
\end{barticle}
\endbibitem

\bibitem[\protect\citeauthoryear{{Goldwurm} et~al.}{1992}]{Goldwurm92}
\begin{barticle}
\bauthor{\binits{A.} \bsnm{{Goldwurm}}},
\bauthor{\binits{J.} \bsnm{{Ballet}}},
\bauthor{\binits{B.} \bsnm{{Cordier}}},
\bauthor{\binits{J.} \bsnm{{Paul}}},
\bauthor{\binits{L.} \bsnm{{Bouchet}}},
\bauthor{\binits{J.P.} \bsnm{{Roques}}},
\bauthor{\binits{D.} \bsnm{{Barret}}},
\bauthor{\binits{P.} \bsnm{{Mandrou}}},
\bauthor{\binits{R.} \bsnm{{Sunyaev}}},
\bauthor{\binits{E.} \bsnm{{Churazov}}},
\bauthor{\binits{M.} \bsnm{{Gilfanov}}},
\bauthor{\binits{A.} \bsnm{{Dyachkov}}},
\bauthor{\binits{N.} \bsnm{{Khavenson}}},
\bauthor{\binits{V.} \bsnm{{Kovtunenko}}},
\bauthor{\binits{R.} \bsnm{{Kremnev}}},
\bauthor{\binits{K.} \bsnm{{Sukhanov}}},
\batitle{{Sigma/GRANAT soft gamma-ray observations of the X-ray nova in Musca -
  Discovery of positron annihilation emission line}}.
\bjtitle{\apjl}
\bvolume{389},
\bfpage{79}--\blpage{82}
(\byear{1992}).
doi:\doiurl{10.1086/186353}
\end{barticle}
\endbibitem

\bibitem[\protect\citeauthoryear{{Golenetskii} et~al.}{1984}]{Golenetskii84}
\begin{barticle}
\bauthor{\binits{S.V.} \bsnm{{Golenetskii}}},
\bauthor{\binits{V.N.} \bsnm{{Ilinskii}}},
\bauthor{\binits{E.P.} \bsnm{{Mazets}}},
\batitle{{Recurrent bursts in GBS0526 - 66, the source of the 5 March 1979
  gamma-ray burst}}.
\bjtitle{\nat}
\bvolume{307},
\bfpage{41}--\blpage{43}
(\byear{1984}).
doi:\doiurl{10.1038/307041a0}
\end{barticle}
\endbibitem

\bibitem[\protect\citeauthoryear{{Golenetskii} et~al.}{1983}]{Golenetskii83}
\begin{barticle}
\bauthor{\binits{S.V.} \bsnm{{Golenetskii}}},
\bauthor{\binits{E.P.} \bsnm{{Mazets}}},
\bauthor{\binits{R.L.} \bsnm{{Aptekar}}},
\bauthor{\binits{V.N.} \bsnm{{Ilinskii}}},
\batitle{{Correlation between luminosity and temperature in gamma-ray burst
  sources}}.
\bjtitle{\nat}
\bvolume{306},
\bfpage{451}--\blpage{453}
(\byear{1983}).
doi:\doiurl{10.1038/306451a0}
\end{barticle}
\endbibitem

\bibitem[\protect\citeauthoryear{{Golenetskii}
  et~al.}{1991a}]{Golenetskii1991;granat}
\begin{barticle}
\bauthor{\binits{S.V.} \bsnm{{Golenetskii}}},
\bauthor{\binits{R.L.} \bsnm{{Aptekar}}},
\bauthor{\binits{Y.A.} \bsnm{{Guryan}}},
\bauthor{\binits{I.V.} \bsnm{{Dementev}}},
\bauthor{\binits{V.N.} \bsnm{{Ilinskii}}},
\bauthor{\binits{E.P.} \bsnm{{Mazets}}},
\bauthor{\binits{V.N.} \bsnm{{Panov}}},
\bauthor{\binits{Z.Y.} \bsnm{{Sokolova}}},
\bauthor{\binits{D.D.} \bsnm{{Frederiks}}},
\bauthor{\binits{T.V.} \bsnm{{Kharitonova}}},
\bauthor{\binits{L.O.} \bsnm{{Sheshin}}},
\batitle{{Observations of Gamma-Ray Bursts with the Konus-B Instrument on the
  GRANAT Station}}.
\bjtitle{Soviet Astronomy Letters}
\bvolume{17},
\bfpage{83}
(\byear{1991}a)
\end{barticle}
\endbibitem

\bibitem[\protect\citeauthoryear{{Golenetskii} et~al.}{1991b}]{Golenetskii91}
\begin{barticle}
\bauthor{\binits{S.V.} \bsnm{{Golenetskii}}},
\bauthor{\binits{R.L.} \bsnm{{Aptekar}}},
\bauthor{\binits{Y.A.} \bsnm{{Guryan}}},
\bauthor{\binits{I.V.} \bsnm{{Dementev}}},
\bauthor{\binits{V.N.} \bsnm{{Ilinskii}}},
\bauthor{\binits{E.P.} \bsnm{{Mazets}}},
\bauthor{\binits{V.N.} \bsnm{{Panov}}},
\bauthor{\binits{Z.Y.} \bsnm{{Sokolova}}},
\bauthor{\binits{D.D.} \bsnm{{Frederiks}}},
\bauthor{\binits{T.V.} \bsnm{{Kharitonova}}},
\bauthor{\binits{L.O.} \bsnm{{Sheshin}}},
\batitle{{Observations of Gamma-Ray Bursts with the Konus-B Instrument on the
  GRANAT Station}}.
\bjtitle{Soviet Astronomy Letters}
\bvolume{17},
\bfpage{83}
(\byear{1991}b)
\end{barticle}
\endbibitem

\bibitem[\protect\citeauthoryear{{Gorecki} et~al.}{1982}]{Gorecki82}
\begin{barticle}
\bauthor{\binits{A.} \bsnm{{Gorecki}}},
\bauthor{\binits{A.} \bsnm{{Levine}}},
\bauthor{\binits{M.} \bsnm{{Bautz}}},
\bauthor{\binits{F.} \bsnm{{Lang}}},
\bauthor{\binits{F.A.} \bsnm{{Primini}}},
\bauthor{\binits{W.H.G.} \bsnm{{Lewin}}},
\bauthor{\binits{W.A.} \bsnm{{Baity}}},
\bauthor{\binits{D.E.} \bsnm{{Gruber}}},
\bauthor{\binits{R.E.} \bsnm{{Rothschild}}},
\batitle{{HEAO 1 observations of the long-term variability of Hercules X-1}}.
\bjtitle{\apj}
\bvolume{256},
\bfpage{234}--\blpage{237}
(\byear{1982}).
doi:\doiurl{10.1086/159900}
\end{barticle}
\endbibitem

\bibitem[\protect\citeauthoryear{{G{\"o}tz} et~al.}{2006}]{Gotz06}
\begin{barticle}
\bauthor{\binits{D.} \bsnm{{G{\"o}tz}}},
\bauthor{\binits{S.} \bsnm{{Mereghetti}}},
\bauthor{\binits{A.} \bsnm{{Tiengo}}},
\bauthor{\binits{P.} \bsnm{{Esposito}}},
\batitle{{Magnetars as persistent hard X-ray sources: INTEGRAL discovery of a
  hard tail in SGR 1900+14}}.
\bjtitle{\aap}
\bvolume{449},
\bfpage{31}--\blpage{34}
(\byear{2006}).
doi:\doiurl{10.1051/0004-6361:20064870}
\end{barticle}
\endbibitem

\bibitem[\protect\citeauthoryear{{G{\"o}tz} et~al.}{2009}]{Gotz09}
\begin{barticle}
\bauthor{\binits{D.} \bsnm{{G{\"o}tz}}},
\bauthor{\binits{P.} \bsnm{{Laurent}}},
\bauthor{\binits{F.} \bsnm{{Lebrun}}},
\bauthor{\binits{F.} \bsnm{{Daigne}}},
\bauthor{\binits{{\v Z}.} \bsnm{{Bo{\v s}njak}}},
\batitle{{Variable Polarization Measured in the Prompt Emission of GRB 041219A
  Using IBIS on Board INTEGRAL}}.
\bjtitle{\apjl}
\bvolume{695},
\bfpage{208}--\blpage{212}
(\byear{2009}).
doi:\doiurl{10.1088/0004-637X/695/2/L208}
\end{barticle}
\endbibitem

\bibitem[\protect\citeauthoryear{{G{\"o}tz} et~al.}{2013}]{Gotz13}
\begin{barticle}
\bauthor{\binits{D.} \bsnm{{G{\"o}tz}}},
\bauthor{\binits{S.} \bsnm{{Covino}}},
\bauthor{\binits{A.} \bsnm{{Fern{\'a}ndez-Soto}}},
\bauthor{\binits{P.} \bsnm{{Laurent}}},
\bauthor{\binits{{\v Z}.} \bsnm{{Bo{\v s}njak}}},
\batitle{{The polarized gamma-ray burst GRB 061122}}.
\bjtitle{\mnras}
\bvolume{431},
\bfpage{3550}--\blpage{3556}
(\byear{2013}).
doi:\doiurl{10.1093/mnras/stt439}
\end{barticle}
\endbibitem

\bibitem[\protect\citeauthoryear{{G{\"o}tz} et~al.}{2014}]{Gotz14}
\begin{barticle}
\bauthor{\binits{D.} \bsnm{{G{\"o}tz}}},
\bauthor{\binits{P.} \bsnm{{Laurent}}},
\bauthor{\binits{S.} \bsnm{{Antier}}},
\bauthor{\binits{S.} \bsnm{{Covino}}},
\bauthor{\binits{P.} \bsnm{{D'Avanzo}}},
\bauthor{\binits{V.} \bsnm{{D'Elia}}},
\bauthor{\binits{A.} \bsnm{{Melandri}}},
\batitle{{GRB 140206A: the most distant polarized gamma-ray burst}}.
\bjtitle{\mnras}
\bvolume{444},
\bfpage{2776}--\blpage{2782}
(\byear{2014}).
doi:\doiurl{10.1093/mnras/stu1634}
\end{barticle}
\endbibitem

\bibitem[\protect\citeauthoryear{{Grabelsky} et~al.}{1995}]{Grabelsky95}
\begin{barticle}
\bauthor{\binits{D.A.} \bsnm{{Grabelsky}}},
\bauthor{\binits{S.M.} \bsnm{{Maltz}}},
\bauthor{\binits{W.R.} \bsnm{{Purcell}}},
\bauthor{\binits{M.P.} \bsnm{{Ulmer}}},
\bauthor{\binits{J.E.} \bsnm{{Grove}}},
\bauthor{\binits{W.N.} \bsnm{{Johnson}}},
\bauthor{\binits{R.L.} \bsnm{{Kinzer}}},
\bauthor{\binits{J.D.} \bsnm{{Kurfess}}},
\bauthor{\binits{M.S.} \bsnm{{Strickman}}},
\bauthor{\binits{G.V.} \bsnm{{Jung}}},
\batitle{{OSSE observations of GX 339-4}}.
\bjtitle{\apj}
\bvolume{441},
\bfpage{800}--\blpage{805}
(\byear{1995}).
doi:\doiurl{10.1086/175403}
\end{barticle}
\endbibitem

\bibitem[\protect\citeauthoryear{{Grandi} et~al.}{1997}]{Grandi97}
\begin{barticle}
\bauthor{\binits{P.} \bsnm{{Grandi}}},
\bauthor{\binits{M.} \bsnm{{Guainazzi}}},
\bauthor{\binits{T.} \bsnm{{Mineo}}},
\bauthor{\binits{A.N.} \bsnm{{Parmar}}},
\bauthor{\binits{F.} \bsnm{{Fiore}}},
\bauthor{\binits{A.} \bsnm{{Matteuzzi}}},
\bauthor{\binits{F.} \bsnm{{Nicastro}}},
\bauthor{\binits{G.C.} \bsnm{{Perola}}},
\bauthor{\binits{L.} \bsnm{{Piro}}},
\bauthor{\binits{M.} \bsnm{{Cappi}}},
\bauthor{\binits{G.} \bsnm{{Cusumano}}},
\bauthor{\binits{S.} \bsnm{{Frontera}}},
\bauthor{\binits{F.} \bsnm{{Giarrusso}}},
\bauthor{\binits{E.} \bsnm{{Palazzi}}},
\bauthor{\binits{S.} \bsnm{{Piraino}}},
\batitle{{BeppoSAX observation of 3C 273: broadband spectrum and detection of a
  low-energy absorption feature.}}
\bjtitle{\aap}
\bvolume{325},
\bfpage{17}--\blpage{20}
(\byear{1997})
\end{barticle}
\endbibitem

\bibitem[\protect\citeauthoryear{{Graser} and
  {Sch{\"o}nfelder}}{1981}]{Graser81}
\begin{barticle}
\bauthor{\binits{U.} \bsnm{{Graser}}},
\bauthor{\binits{V.} \bsnm{{Sch{\"o}nfelder}}},
\batitle{{Sky map of the galactic anticenter at MeV-gamma-ray energies}}.
\bjtitle{International Cosmic Ray Conference}
\bvolume{9},
\bfpage{76}--\blpage{79}
(\byear{1981})
\end{barticle}
\endbibitem

\bibitem[\protect\citeauthoryear{{Grebenev} et~al.}{1997}]{Grebenev97}
\begin{barticle}
\bauthor{\binits{S.A.} \bsnm{{Grebenev}}},
\bauthor{\binits{R.A.} \bsnm{{Sunyaev}}},
\bauthor{\binits{M.N.} \bsnm{{Pavlinsky}}},
\batitle{{Spectral states of galactic black hole candidates: results of
  observations with ART-P/Granat}}.
\bjtitle{Advances in Space Research}
\bvolume{19},
\bfpage{15}--\blpage{23}
(\byear{1997}).
doi:\doiurl{10.1016/S0273-1177(97)00031-8}
\end{barticle}
\endbibitem

\bibitem[\protect\citeauthoryear{{Grebenev} et~al.}{2012}]{Grebenev12}
\begin{barticle}
\bauthor{\binits{S.A.} \bsnm{{Grebenev}}},
\bauthor{\binits{A.A.} \bsnm{{Lutovinov}}},
\bauthor{\binits{S.S.} \bsnm{{Tsygankov}}},
\bauthor{\binits{C.} \bsnm{{Winkler}}},
\batitle{{Hard-X-ray emission lines from the decay of $^{44}$Ti in the remnant
  of supernova 1987A}}.
\bjtitle{\nat}
\bvolume{490},
\bfpage{373}--\blpage{375}
(\byear{2012}).
doi:\doiurl{10.1038/nature11473}
\end{barticle}
\endbibitem

\bibitem[\protect\citeauthoryear{{Greenhill} et~al.}{1979a}]{Greenhill79}
\begin{barticle}
\bauthor{\binits{J.G.} \bsnm{{Greenhill}}},
\bauthor{\binits{A.G.} \bsnm{{Fenton}}},
\bauthor{\binits{K.B.} \bsnm{{Fenton}}},
\bauthor{\binits{R.M.} \bsnm{{Thomas}}},
\bauthor{\binits{M.L.} \bsnm{{Duldig}}},
\bauthor{\binits{M.W.} \bsnm{{Emery}}},
\bauthor{\binits{D.J.} \bsnm{{Cooke}}},
\bauthor{\binits{J.} \bsnm{{Phillips}}},
\bauthor{\binits{D.J.} \bsnm{{Watts}}},
\bauthor{\binits{R.M.} \bsnm{{Hudson}}},
\bauthor{\binits{E.} \bsnm{{Middleton}}},
\bauthor{\binits{G.} \bsnm{{Salmon}}},
\batitle{{a Large Area Proportional Counter for Balloon-Borne X-Ray
  Astronomy}}.
\bjtitle{International Cosmic Ray Conference}
\bvolume{11},
\bfpage{8}
(\byear{1979}a)
\end{barticle}
\endbibitem

\bibitem[\protect\citeauthoryear{{Greenhill} et~al.}{1979b}]{Greenhill79b}
\begin{barticle}
\bauthor{\binits{J.G.} \bsnm{{Greenhill}}},
\bauthor{\binits{M.L.} \bsnm{{Duldig}}},
\bauthor{\binits{M.W.} \bsnm{{Emery}}},
\bauthor{\binits{A.G.} \bsnm{{Fenton}}},
\bauthor{\binits{K.B.} \bsnm{{Fenton}}},
\bauthor{\binits{R.M.} \bsnm{{Thomas}}},
\bauthor{\binits{D.J.} \bsnm{{Watts}}},
\batitle{{Balloon observations of several southern X-ray sources}}.
\bjtitle{Proceedings of the Astronomical Society of Australia}
\bvolume{3},
\bfpage{349}
(\byear{1979}b)
\end{barticle}
\endbibitem

\bibitem[\protect\citeauthoryear{{Grindlay}}{1998}]{Grindlay98}
\begin{barticle}
\bauthor{\binits{J.E.} \bsnm{{Grindlay}}},
\batitle{{Balloon-borne hard x-ray imaging and future surveys}}.
\bjtitle{Advances in Space Research}
\bvolume{21},
\bfpage{999}--\blpage{1008}
(\byear{1998}).
doi:\doiurl{10.1016/S0273-1177(97)01088-0}
\end{barticle}
\endbibitem

\bibitem[\protect\citeauthoryear{{Grindlay} et~al.}{1993}]{Grindlay93}
\begin{barticle}
\bauthor{\binits{J.E.} \bsnm{{Grindlay}}},
\bauthor{\binits{C.E.} \bsnm{{Covault}}},
\bauthor{\binits{R.P.} \bsnm{{Manandhar}}},
\batitle{{EXITE observation of the Galactic center - A new transient?}}
\bjtitle{\aaps}
\bvolume{97},
\bfpage{155}--\blpage{158}
(\byear{1993})
\end{barticle}
\endbibitem

\bibitem[\protect\citeauthoryear{{Grindlay} et~al.}{2010}]{Grindlay10}
\begin{bchapter}
\bauthor{\binits{J.} \bsnm{{Grindlay}}},
\bauthor{\binits{N.} \bsnm{{Gehrels}}},
\bauthor{\binits{J.} \bsnm{{Bloom}}},
\bauthor{\binits{P.} \bsnm{{Coppi}}},
\bauthor{\binits{A.} \bsnm{{Soderberg}}},
\bauthor{\binits{J.} \bsnm{{Hong}}},
\bauthor{\binits{B.} \bsnm{{Allen}}},
\bauthor{\binits{S.} \bsnm{{Barthelmy}}},
\bauthor{\binits{G.} \bsnm{{Tagliaferri}}},
\bauthor{\binits{H.} \bsnm{{Moseley}}},
\bauthor{\binits{A.} \bsnm{{Kutyrev}}},
\bauthor{\binits{G.} \bsnm{{Fabbiano}}},
\bauthor{\binits{G.} \bsnm{{Fishman}}},
\bauthor{\binits{B.} \bsnm{{Ramsey}}},
\bauthor{\binits{R.} \bsnm{{Della Ceca}}},
\bauthor{\binits{L.} \bsnm{{Natalucci}}},
\bauthor{\binits{P.} \bsnm{{Ubertini}} \bsuffix{III}},
\bctitle{{Overview of EXIST mission science and implementation}},
in \bbtitle{Space Telescopes and Instrumentation 2010: Ultraviolet to Gamma
  Ray}.
\bsertitle{\procspie},
vol. \bseriesno{7732},
\byear{2010},
p. \bfpage{77321}.
doi:\doiurl{10.1117/12.857895}
\end{bchapter}
\endbibitem

\bibitem[\protect\citeauthoryear{{Grove}}{1996}]{Grove96}
\begin{barticle}
\bauthor{\binits{J.E.} \bsnm{{Grove}}},
\batitle{{OSSE highlights of the low-energy gamma-ray sky}}.
\bjtitle{\memsai}
\bvolume{67},
\bfpage{127}
(\byear{1996})
\end{barticle}
\endbibitem

\bibitem[\protect\citeauthoryear{{Grove} et~al.}{1995}]{Grove95}
\begin{barticle}
\bauthor{\binits{J.E.} \bsnm{{Grove}}},
\bauthor{\binits{M.S.} \bsnm{{Strickman}}},
\bauthor{\binits{W.N.} \bsnm{{Johnson}}},
\bauthor{\binits{J.D.} \bsnm{{Kurfess}}},
\bauthor{\binits{R.L.} \bsnm{{Kinzer}}},
\bauthor{\binits{C.H.} \bsnm{{Starr}}},
\bauthor{\binits{G.V.} \bsnm{{Jung}}},
\bauthor{\binits{E.} \bsnm{{Kendziorra}}},
\bauthor{\binits{P.} \bsnm{{Kretschmar}}},
\bauthor{\binits{M.} \bsnm{{Maisack}}},
\bauthor{\binits{R.} \bsnm{{Staubert}}},
\batitle{{The soft gamma-ray spectrum of A0535+26: Detection of an absorption
  feature at 110 keV by OSSE}}.
\bjtitle{\apjl}
\bvolume{438},
\bfpage{25}--\blpage{28}
(\byear{1995}).
doi:\doiurl{10.1086/187706}
\end{barticle}
\endbibitem

\bibitem[\protect\citeauthoryear{{Gruber} and {Rephaeli}}{1999}]{Gruber99b}
\begin{barticle}
\bauthor{\binits{D.E.} \bsnm{{Gruber}}},
\bauthor{\binits{Y.} \bsnm{{Rephaeli}}},
\batitle{{RXTE observations of the starburst galaxy M82}}.
\bjtitle{Nuclear Physics B Proceedings Supplements}
\bvolume{69},
\bfpage{550}--\blpage{553}
(\byear{1999}).
doi:\doiurl{10.1016/S0920-5632(98)00285-0}
\end{barticle}
\endbibitem

\bibitem[\protect\citeauthoryear{{Gruber} and {Rothschild}}{1984}]{Gruber84}
\begin{barticle}
\bauthor{\binits{D.E.} \bsnm{{Gruber}}},
\bauthor{\binits{R.E.} \bsnm{{Rothschild}}},
\batitle{{SMC X-1 variability observed from HEAO 1}}.
\bjtitle{\apj}
\bvolume{283},
\bfpage{546}--\blpage{551}
(\byear{1984}).
doi:\doiurl{10.1086/162338}
\end{barticle}
\endbibitem

\bibitem[\protect\citeauthoryear{{Gruber} et~al.}{1980}]{Gruber80}
\begin{barticle}
\bauthor{\binits{D.E.} \bsnm{{Gruber}}},
\bauthor{\binits{J.L.} \bsnm{{Matteson}}},
\bauthor{\binits{P.L.} \bsnm{{Nolan}}},
\bauthor{\binits{F.K.} \bsnm{{Knight}}},
\bauthor{\binits{W.A.} \bsnm{{Baity}}},
\bauthor{\binits{R.E.} \bsnm{{Rothschild}}},
\bauthor{\binits{L.E.} \bsnm{{Peterson}}},
\bauthor{\binits{J.A.} \bsnm{{Hoffman}}},
\bauthor{\binits{A.} \bsnm{{Scheepmaker}}},
\bauthor{\binits{W.A.} \bsnm{{Wheaton}}},
\bauthor{\binits{F.A.} \bsnm{{Primini}}},
\bauthor{\binits{A.M.} \bsnm{{Levine}}},
\bauthor{\binits{W.H.G.} \bsnm{{Lewin}}},
\batitle{{Hercules X-1 hard X-ray pulsations observed from HEAO 1}}.
\bjtitle{\apjl}
\bvolume{240},
\bfpage{127}--\blpage{131}
(\byear{1980}).
doi:\doiurl{10.1086/183338}
\end{barticle}
\endbibitem

\bibitem[\protect\citeauthoryear{{Gruber} et~al.}{1996}]{Gruber1996;rxte}
\begin{barticle}
\bauthor{\binits{D.E.} \bsnm{{Gruber}}},
\bauthor{\binits{P.R.} \bsnm{{Blanco}}},
\bauthor{\binits{W.A.} \bsnm{{Heindl}}},
\bauthor{\binits{M.R.} \bsnm{{Pelling}}},
\bauthor{\binits{R.E.} \bsnm{{Rothschild}}},
\bauthor{\binits{P.L.} \bsnm{{Hink}}},
\batitle{{The high energy X-ray timing experiment on XTE.}}
\bjtitle{\aaps}
\bvolume{120},
\bfpage{641}
(\byear{1996})
\end{barticle}
\endbibitem

\bibitem[\protect\citeauthoryear{{Gruber} et~al.}{1999}]{Gruber99}
\begin{barticle}
\bauthor{\binits{D.E.} \bsnm{{Gruber}}},
\bauthor{\binits{J.L.} \bsnm{{Matteson}}},
\bauthor{\binits{L.E.} \bsnm{{Peterson}}},
\bauthor{\binits{G.V.} \bsnm{{Jung}}},
\batitle{{The Spectrum of Diffuse Cosmic Hard X-Rays Measured with HEAO 1}}.
\bjtitle{\apj}
\bvolume{520},
\bfpage{124}--\blpage{129}
(\byear{1999}).
doi:\doiurl{10.1086/307450}
\end{barticle}
\endbibitem

\bibitem[\protect\citeauthoryear{{Gruber} et~al.}{2011}]{Gruber11}
\begin{barticle}
\bauthor{\binits{D.} \bsnm{{Gruber}}},
\bauthor{\binits{J.} \bsnm{{Greiner}}},
\bauthor{\binits{A.} \bsnm{{von Kienlin}}},
\bauthor{\binits{A.} \bsnm{{Rau}}},
\bauthor{\binits{M.S.} \bsnm{{Briggs}}},
\bauthor{\binits{V.} \bsnm{{Connaughton}}},
\bauthor{\binits{A.} \bsnm{{Goldstein}}},
\bauthor{\binits{A.J.} \bsnm{{van der Horst}}},
\bauthor{\binits{M.} \bsnm{{Nardini}}},
\bauthor{\binits{P.N.} \bsnm{{Bhat}}},
\bauthor{\binits{E.} \bsnm{{Bissaldi}}},
\bauthor{\binits{J.M.} \bsnm{{Burgess}}},
\bauthor{\binits{V.L.} \bsnm{{Chaplin}}},
\bauthor{\binits{R.} \bsnm{{Diehl}}},
\bauthor{\binits{G.J.} \bsnm{{Fishman}}},
\bauthor{\binits{G.} \bsnm{{Fitzpatrick}}},
\bauthor{\binits{S.} \bsnm{{Foley}}},
\bauthor{\binits{M.H.} \bsnm{{Gibby}}},
\bauthor{\binits{M.M.} \bsnm{{Giles}}},
\bauthor{\binits{S.} \bsnm{{Guiriec}}},
\bauthor{\binits{R.M.} \bsnm{{Kippen}}},
\bauthor{\binits{C.} \bsnm{{Kouveliotou}}},
\bauthor{\binits{L.} \bsnm{{Lin}}},
\bauthor{\binits{S.} \bsnm{{McBreen}}},
\bauthor{\binits{C.A.} \bsnm{{Meegan}}},
\bauthor{\binits{F.} \bsnm{{Olivares E.}}},
\bauthor{\binits{W.S.} \bsnm{{Paciesas}}},
\bauthor{\binits{R.D.} \bsnm{{Preece}}},
\bauthor{\binits{D.} \bsnm{{Tierney}}},
\bauthor{\binits{C.} \bsnm{{Wilson-Hodge}}},
\batitle{{Rest-frame properties of 32 gamma-ray bursts observed by the Fermi
  Gamma-ray Burst Monitor}}.
\bjtitle{\aap}
\bvolume{531},
\bfpage{20}
(\byear{2011}).
doi:\doiurl{10.1051/0004-6361/201116953}
\end{barticle}
\endbibitem

\bibitem[\protect\citeauthoryear{{Guainazzi} et~al.}{1998}]{Guainazzi98}
\begin{barticle}
\bauthor{\binits{M.} \bsnm{{Guainazzi}}},
\bauthor{\binits{A.N.} \bsnm{{Parmar}}},
\bauthor{\binits{A.} \bsnm{{Segreto}}},
\bauthor{\binits{L.} \bsnm{{Stella}}},
\bauthor{\binits{D.} \bsnm{{dal Fiume}}},
\bauthor{\binits{T.} \bsnm{{Oosterbroek}}},
\batitle{{The comptonized X-ray source X 1724-308 in the globular cluster
  Terzan 2}}.
\bjtitle{\aap}
\bvolume{339},
\bfpage{802}--\blpage{810}
(\byear{1998})
\end{barticle}
\endbibitem

\bibitem[\protect\citeauthoryear{{Guainazzi} et~al.}{1999a}]{Guainazzi99c}
\begin{barticle}
\bauthor{\binits{M.} \bsnm{{Guainazzi}}},
\bauthor{\binits{G.} \bsnm{{Matt}}},
\bauthor{\binits{S.} \bsnm{{Molendi}}},
\bauthor{\binits{A.} \bsnm{{Orr}}},
\bauthor{\binits{F.} \bsnm{{Fiore}}},
\bauthor{\binits{P.} \bsnm{{Grandi}}},
\bauthor{\binits{A.} \bsnm{{Matteuzzi}}},
\bauthor{\binits{T.} \bsnm{{Mineo}}},
\bauthor{\binits{G.C.} \bsnm{{Perola}}},
\bauthor{\binits{A.N.} \bsnm{{Parmar}}},
\bauthor{\binits{L.} \bsnm{{Piro}}},
\batitle{{BeppoSAX confirms extreme relativistic effects in the X-ray spectrum
  of MCG-6-30-15}}.
\bjtitle{\aap}
\bvolume{341},
\bfpage{27}--\blpage{30}
(\byear{1999}a)
\end{barticle}
\endbibitem

\bibitem[\protect\citeauthoryear{{Guainazzi} et~al.}{1999b}]{Guainazzi99b}
\begin{barticle}
\bauthor{\binits{M.} \bsnm{{Guainazzi}}},
\bauthor{\binits{G.C.} \bsnm{{Perola}}},
\bauthor{\binits{G.} \bsnm{{Matt}}},
\bauthor{\binits{F.} \bsnm{{Nicastro}}},
\bauthor{\binits{L.} \bsnm{{Bassani}}},
\bauthor{\binits{F.} \bsnm{{Fiore}}},
\bauthor{\binits{D.} \bsnm{{dal Fiume}}},
\bauthor{\binits{L.} \bsnm{{Piro}}},
\batitle{{The complex 0.1-200 keV spectrum of the Seyfert 1 Galaxy NGC 4593}}.
\bjtitle{\aap}
\bvolume{346},
\bfpage{407}--\blpage{414}
(\byear{1999}b)
\end{barticle}
\endbibitem

\bibitem[\protect\citeauthoryear{{Guainazzi} et~al.}{1999c}]{Guainazzi99}
\begin{barticle}
\bauthor{\binits{M.} \bsnm{{Guainazzi}}},
\bauthor{\binits{G.} \bsnm{{Vacanti}}},
\bauthor{\binits{A.} \bsnm{{Malizia}}},
\bauthor{\binits{K.S.} \bsnm{{O'Flaherty}}},
\bauthor{\binits{E.} \bsnm{{Palazzi}}},
\bauthor{\binits{A.N.} \bsnm{{Parmar}}},
\batitle{{The quiescent state broadband X-ray spectrum and variability of MKN
  421}}.
\bjtitle{\aap}
\bvolume{342},
\bfpage{124}--\blpage{130}
(\byear{1999}c)
\end{barticle}
\endbibitem

\bibitem[\protect\citeauthoryear{{Guidorzi} et~al.}{2004}]{Guidorzi04}
\begin{barticle}
\bauthor{\binits{C.} \bsnm{{Guidorzi}}},
\bauthor{\binits{F.} \bsnm{{Frontera}}},
\bauthor{\binits{E.} \bsnm{{Montanari}}},
\bauthor{\binits{M.} \bsnm{{Feroci}}},
\bauthor{\binits{L.} \bsnm{{Amati}}},
\bauthor{\binits{E.} \bsnm{{Costa}}},
\bauthor{\binits{M.} \bsnm{{Orlandini}}},
\batitle{{Comparative study of the two large flares from SGR1900+14 with the
  BeppoSAX Gamma-Ray Burst Monitor}}.
\bjtitle{\aap}
\bvolume{416},
\bfpage{297}--\blpage{310}
(\byear{2004}).
doi:\doiurl{10.1051/0004-6361:20034511}
\end{barticle}
\endbibitem

\bibitem[\protect\citeauthoryear{{Guidorzi} et~al.}{2011}]{Guidorzi11}
\begin{barticle}
\bauthor{\binits{C.} \bsnm{{Guidorzi}}},
\bauthor{\binits{M.} \bsnm{{Lacapra}}},
\bauthor{\binits{F.} \bsnm{{Frontera}}},
\bauthor{\binits{E.} \bsnm{{Montanari}}},
\bauthor{\binits{L.} \bsnm{{Amati}}},
\bauthor{\binits{F.} \bsnm{{Calura}}},
\bauthor{\binits{L.} \bsnm{{Nicastro}}},
\bauthor{\binits{M.} \bsnm{{Orlandini}}},
\batitle{{Spectral catalogue of bright gamma-ray bursts detected with the
  BeppoSAX/GRBM}}.
\bjtitle{\aap}
\bvolume{526},
\bfpage{49}
(\byear{2011}).
doi:\doiurl{10.1051/0004-6361/201015752}
\end{barticle}
\endbibitem

\bibitem[\protect\citeauthoryear{{Guiriec} et~al.}{2010}]{Guiriec10}
\begin{barticle}
\bauthor{\binits{S.} \bsnm{{Guiriec}}},
\bauthor{\binits{M.S.} \bsnm{{Briggs}}},
\bauthor{\binits{V.} \bsnm{{Connaugthon}}},
\bauthor{\binits{E.} \bsnm{{Kara}}},
\bauthor{\binits{F.} \bsnm{{Daigne}}},
\bauthor{\binits{C.} \bsnm{{Kouveliotou}}},
\bauthor{\binits{A.J.} \bsnm{{van der Horst}}},
\bauthor{\binits{W.} \bsnm{{Paciesas}}},
\bauthor{\binits{C.A.} \bsnm{{Meegan}}},
\bauthor{\binits{P.N.} \bsnm{{Bhat}}},
\bauthor{\binits{S.} \bsnm{{Foley}}},
\bauthor{\binits{E.} \bsnm{{Bissaldi}}},
\bauthor{\binits{M.} \bsnm{{Burgess}}},
\bauthor{\binits{V.} \bsnm{{Chaplin}}},
\bauthor{\binits{R.} \bsnm{{Diehl}}},
\bauthor{\binits{G.} \bsnm{{Fishman}}},
\bauthor{\binits{M.} \bsnm{{Gibby}}},
\bauthor{\binits{M.M.} \bsnm{{Giles}}},
\bauthor{\binits{A.} \bsnm{{Goldstein}}},
\bauthor{\binits{J.} \bsnm{{Greiner}}},
\bauthor{\binits{D.} \bsnm{{Gruber}}},
\bauthor{\binits{A.} \bsnm{{von Kienlin}}},
\bauthor{\binits{M.} \bsnm{{Kippen}}},
\bauthor{\binits{S.} \bsnm{{McBreen}}},
\bauthor{\binits{R.} \bsnm{{Preece}}},
\bauthor{\binits{A.} \bsnm{{Rau}}},
\bauthor{\binits{D.} \bsnm{{Tierney}}},
\bauthor{\binits{C.} \bsnm{{Wilson-Hodge}}},
\batitle{{Time-resolved Spectroscopy of the Three Brightest and Hardest Short
  Gamma-ray Bursts Observed with the Fermi Gamma-ray Burst Monitor}}.
\bjtitle{\apj}
\bvolume{725},
\bfpage{225}--\blpage{241}
(\byear{2010}).
doi:\doiurl{10.1088/0004-637X/725/1/225}
\end{barticle}
\endbibitem

\bibitem[\protect\citeauthoryear{{Guiriec} et~al.}{2011}]{Guiriec11}
\begin{barticle}
\bauthor{\binits{S.} \bsnm{{Guiriec}}},
\bauthor{\binits{V.} \bsnm{{Connaughton}}},
\bauthor{\binits{M.S.} \bsnm{{Briggs}}},
\bauthor{\binits{M.} \bsnm{{Burgess}}},
\bauthor{\binits{F.} \bsnm{{Ryde}}},
\bauthor{\binits{F.} \bsnm{{Daigne}}},
\bauthor{\binits{P.} \bsnm{{M{\'e}sz{\'a}ros}}},
\bauthor{\binits{A.} \bsnm{{Goldstein}}},
\bauthor{\binits{J.} \bsnm{{McEnery}}},
\bauthor{\binits{N.} \bsnm{{Omodei}}},
\bauthor{\binits{P.N.} \bsnm{{Bhat}}},
\bauthor{\binits{E.} \bsnm{{Bissaldi}}},
\bauthor{\binits{A.} \bsnm{{Camero-Arranz}}},
\bauthor{\binits{V.} \bsnm{{Chaplin}}},
\bauthor{\binits{R.} \bsnm{{Diehl}}},
\bauthor{\binits{G.} \bsnm{{Fishman}}},
\bauthor{\binits{S.} \bsnm{{Foley}}},
\bauthor{\binits{M.} \bsnm{{Gibby}}},
\bauthor{\binits{M.M.} \bsnm{{Giles}}},
\bauthor{\binits{J.} \bsnm{{Greiner}}},
\bauthor{\binits{D.} \bsnm{{Gruber}}},
\bauthor{\binits{A.} \bsnm{{von Kienlin}}},
\bauthor{\binits{M.} \bsnm{{Kippen}}},
\bauthor{\binits{C.} \bsnm{{Kouveliotou}}},
\bauthor{\binits{S.} \bsnm{{McBreen}}},
\bauthor{\binits{C.A.} \bsnm{{Meegan}}},
\bauthor{\binits{W.} \bsnm{{Paciesas}}},
\bauthor{\binits{R.} \bsnm{{Preece}}},
\bauthor{\binits{A.} \bsnm{{Rau}}},
\bauthor{\binits{D.} \bsnm{{Tierney}}},
\bauthor{\binits{A.J.} \bsnm{{van der Horst}}},
\bauthor{\binits{C.} \bsnm{{Wilson-Hodge}}},
\batitle{{Detection of a Thermal Spectral Component in the Prompt Emission of
  GRB 100724B}}.
\bjtitle{\apjl}
\bvolume{727},
\bfpage{33}
(\byear{2011}).
doi:\doiurl{10.1088/2041-8205/727/2/L33}
\end{barticle}
\endbibitem

\bibitem[\protect\citeauthoryear{{Halloin} et~al.}{2003}]{Halloin03}
\begin{barticle}
\bauthor{\binits{H.} \bsnm{{Halloin}}},
\bauthor{\binits{P.} \bsnm{{von Ballmoos}}},
\bauthor{\binits{J.} \bsnm{{Evrard}}},
\bauthor{\binits{G.K.} \bsnm{{Skinner}}},
\bauthor{\binits{N.} \bsnm{{Abrosimov}}},
\bauthor{\binits{P.} \bsnm{{Bastie}}},
\bauthor{\binits{G.} \bsnm{{Di Cocco}}},
\bauthor{\binits{M.} \bsnm{{George}}},
\bauthor{\binits{B.} \bsnm{{Hamelin}}},
\bauthor{\binits{P.} \bsnm{{Jean}}},
\bauthor{\binits{J.} \bsnm{{Kn{\"o}dleseder}}},
\bauthor{\binits{P.} \bsnm{{Laporte}}},
\bauthor{\binits{C.} \bsnm{{Badenes}}},
\bauthor{\binits{P.} \bsnm{{Laurent}}},
\bauthor{\binits{R.K.} \bsnm{{Smither}}},
\batitle{{Performance of CLAIRE, the first balloon-borne {$\gamma$}-ray lens
  telescope}}.
\bjtitle{Nuclear Instruments and Methods in Physics Research A}
\bvolume{504},
\bfpage{120}--\blpage{125}
(\byear{2003}).
doi:\doiurl{10.1016/S0168-9002(03)00807-6}
\end{barticle}
\endbibitem

\bibitem[\protect\citeauthoryear{{Hameury} et~al.}{1983}]{Hameury83}
\begin{barticle}
\bauthor{\binits{J.M.} \bsnm{{Hameury}}},
\bauthor{\binits{D.} \bsnm{{Boclet}}},
\bauthor{\binits{P.} \bsnm{{Durouchoux}}},
\bauthor{\binits{T.L.} \bsnm{{Cline}}},
\bauthor{\binits{B.J.} \bsnm{{Teegarden}}},
\bauthor{\binits{J.} \bsnm{{Tueller}}},
\bauthor{\binits{W.S.} \bsnm{{Paciesas}}},
\bauthor{\binits{R.C.} \bsnm{{Haymes}}},
\batitle{{Hard X-ray observations of the Crab Nebula and A0535+26 with a high
  energy resolution spectrometer}}.
\bjtitle{\apj}
\bvolume{270},
\bfpage{144}--\blpage{149}
(\byear{1983}).
doi:\doiurl{10.1086/161105}
\end{barticle}
\endbibitem

\bibitem[\protect\citeauthoryear{{Harmon} et~al.}{2002}]{Harmon2002;cgro}
\begin{barticle}
\bauthor{\binits{B.A.} \bsnm{{Harmon}}},
\bauthor{\binits{G.J.} \bsnm{{Fishman}}},
\bauthor{\binits{C.A.} \bsnm{{Wilson}}},
\bauthor{\binits{W.S.} \bsnm{{Paciesas}}},
\bauthor{\binits{S.N.} \bsnm{{Zhang}}},
\bauthor{\binits{M.H.} \bsnm{{Finger}}},
\bauthor{\binits{T.M.} \bsnm{{Koshut}}},
\bauthor{\binits{M.L.} \bsnm{{McCollough}}},
\bauthor{\binits{C.R.} \bsnm{{Robinson}}},
\bauthor{\binits{B.C.} \bsnm{{Rubin}}},
\batitle{{The Burst and Transient Source Experiment Earth Occultation
  Technique}}.
\bjtitle{\apjs}
\bvolume{138},
\bfpage{149}--\blpage{183}
(\byear{2002}).
doi:\doiurl{10.1086/324018}
\end{barticle}
\endbibitem

\bibitem[\protect\citeauthoryear{{Harrison} et~al.}{2000}]{Harrison2000;heft}
\begin{bchapter}
\bauthor{\binits{F.A.} \bsnm{{Harrison}}},
\bauthor{\binits{S.E.} \bsnm{{Boggs}}},
\bauthor{\binits{A.E.} \bsnm{{Bolotnikov}}},
\bauthor{\binits{F.E.} \bsnm{{Christensen}}},
\bauthor{\binits{W.R.} \bsnm{{Cook}}},
\bauthor{\binits{W.W.} \bsnm{{Craig}}},
\bauthor{\binits{C.J.} \bsnm{{Hailey}}},
\bauthor{\binits{M.A.} \bsnm{{Jimenez-Garate}}},
\bauthor{\binits{P.H.} \bsnm{{Mao}}},
\bauthor{\binits{S.M.} \bsnm{{Schindler}}},
\bauthor{\binits{D.L.} \bsnm{{Windt}}},
\bctitle{{Development of the High-Energy Focusing Telescope (HEFT) balloon
  experiment}},
in \bbtitle{X-Ray Optics, Instruments, and Missions III},
ed. by \beditor{\binits{J.E.} \bsnm{{Tr{\"u}mper}}},
\beditor{\binits{B.} \bsnm{{Aschenbach}}}
\bsertitle{Society of Photo-Optical Instrumentation Engineers (SPIE) Conference
  Series},
vol. \bseriesno{4012},
\byear{2000},
pp. \bfpage{693}--\blpage{699}
\end{bchapter}
\endbibitem

\bibitem[\protect\citeauthoryear{{Harrison} et~al.}{2013}]{Harrison2013;nustar}
\begin{barticle}
\bauthor{\binits{F.A.} \bsnm{{Harrison}}},
\bauthor{\binits{W.W.} \bsnm{{Craig}}},
\bauthor{\binits{F.E.} \bsnm{{Christensen}}},
\bauthor{\binits{C.J.} \bsnm{{Hailey}}},
\bauthor{\binits{W.W.} \bsnm{{Zhang}}},
\bauthor{\binits{S.E.} \bsnm{{Boggs}}},
\bauthor{\binits{D.} \bsnm{{Stern}}},
\bauthor{\binits{W.R.} \bsnm{{Cook}}},
\bauthor{\binits{K.} \bsnm{{Forster}}},
\bauthor{\binits{P.} \bsnm{{Giommi}}},
\bauthor{\binits{B.W.} \bsnm{{Grefenstette}}},
\bauthor{\binits{Y.} \bsnm{{Kim}}},
\bauthor{\binits{T.} \bsnm{{Kitaguchi}}},
\bauthor{\binits{J.E.} \bsnm{{Koglin}}},
\bauthor{\binits{K.K.} \bsnm{{Madsen}}},
\bauthor{\binits{P.H.} \bsnm{{Mao}}},
\bauthor{\binits{H.} \bsnm{{Miyasaka}}},
\bauthor{\binits{K.} \bsnm{{Mori}}},
\bauthor{\binits{M.} \bsnm{{Perri}}},
\bauthor{\binits{M.J.} \bsnm{{Pivovaroff}}},
\bauthor{\binits{S.} \bsnm{{Puccetti}}},
\bauthor{\binits{V.R.} \bsnm{{Rana}}},
\bauthor{\binits{N.J.} \bsnm{{Westergaard}}},
\bauthor{\binits{J.} \bsnm{{Willis}}},
\bauthor{\binits{A.} \bsnm{{Zoglauer}}},
\bauthor{\binits{H.} \bsnm{{An}}},
\bauthor{\binits{M.} \bsnm{{Bachetti}}},
\bauthor{\binits{N.M.} \bsnm{{Barri{\`e}re}}},
\bauthor{\binits{E.C.} \bsnm{{Bellm}}},
\bauthor{\binits{V.} \bsnm{{Bhalerao}}},
\bauthor{\binits{N.F.} \bsnm{{Brejnholt}}},
\bauthor{\binits{F.} \bsnm{{Fuerst}}},
\bauthor{\binits{C.C.} \bsnm{{Liebe}}},
\bauthor{\binits{C.B.} \bsnm{{Markwardt}}},
\bauthor{\binits{M.} \bsnm{{Nynka}}},
\bauthor{\binits{J.K.} \bsnm{{Vogel}}},
\bauthor{\binits{D.J.} \bsnm{{Walton}}},
\bauthor{\binits{D.R.} \bsnm{{Wik}}},
\bauthor{\binits{D.M.} \bsnm{{Alexander}}},
\bauthor{\binits{L.R.} \bsnm{{Cominsky}}},
\bauthor{\binits{A.E.} \bsnm{{Hornschemeier}}},
\bauthor{\binits{A.} \bsnm{{Hornstrup}}},
\bauthor{\binits{V.M.} \bsnm{{Kaspi}}},
\bauthor{\binits{G.M.} \bsnm{{Madejski}}},
\bauthor{\binits{G.} \bsnm{{Matt}}},
\bauthor{\binits{S.} \bsnm{{Molendi}}},
\bauthor{\binits{D.M.} \bsnm{{Smith}}},
\bauthor{\binits{J.A.} \bsnm{{Tomsick}}},
\bauthor{\binits{M.} \bsnm{{Ajello}}},
\bauthor{\binits{D.R.} \bsnm{{Ballantyne}}},
\bauthor{\binits{M.} \bsnm{{Balokovi{\'c}}}},
\bauthor{\binits{D.} \bsnm{{Barret}}},
\bauthor{\binits{F.E.} \bsnm{{Bauer}}},
\bauthor{\binits{R.D.} \bsnm{{Blandford}}},
\bauthor{\binits{W.N.} \bsnm{{Brandt}}},
\bauthor{\binits{L.W.} \bsnm{{Brenneman}}},
\bauthor{\binits{J.} \bsnm{{Chiang}}},
\bauthor{\binits{D.} \bsnm{{Chakrabarty}}},
\bauthor{\binits{J.} \bsnm{{Chenevez}}},
\bauthor{\binits{A.} \bsnm{{Comastri}}},
\bauthor{\binits{F.} \bsnm{{Dufour}}},
\bauthor{\binits{M.} \bsnm{{Elvis}}},
\bauthor{\binits{A.C.} \bsnm{{Fabian}}},
\bauthor{\binits{D.} \bsnm{{Farrah}}},
\bauthor{\binits{C.L.} \bsnm{{Fryer}}},
\bauthor{\binits{E.V.} \bsnm{{Gotthelf}}},
\bauthor{\binits{J.E.} \bsnm{{Grindlay}}},
\bauthor{\binits{D.J.} \bsnm{{Helfand}}},
\bauthor{\binits{R.} \bsnm{{Krivonos}}},
\bauthor{\binits{D.L.} \bsnm{{Meier}}},
\bauthor{\binits{J.M.} \bsnm{{Miller}}},
\bauthor{\binits{L.} \bsnm{{Natalucci}}},
\bauthor{\binits{P.} \bsnm{{Ogle}}},
\bauthor{\binits{E.O.} \bsnm{{Ofek}}},
\bauthor{\binits{A.} \bsnm{{Ptak}}},
\bauthor{\binits{S.P.} \bsnm{{Reynolds}}},
\bauthor{\binits{J.R.} \bsnm{{Rigby}}},
\bauthor{\binits{G.} \bsnm{{Tagliaferri}}},
\bauthor{\binits{S.E.} \bsnm{{Thorsett}}},
\bauthor{\binits{E.} \bsnm{{Treister}}},
\bauthor{\binits{C.M.} \bsnm{{Urry}}},
\batitle{{The Nuclear Spectroscopic Telescope Array (NuSTAR) High-energy X-Ray
  Mission}}.
\bjtitle{\apj}
\bvolume{770},
\bfpage{103}
(\byear{2013}).
doi:\doiurl{10.1088/0004-637X/770/2/103}
\end{barticle}
\endbibitem

\bibitem[\protect\citeauthoryear{{Hartmann} et~al.}{1994}]{Hartmann94}
\begin{botherref}
\oauthor{\binits{D.H.} \bsnm{{Hartmann}}},
\oauthor{\binits{L.E.} \bsnm{{Brown}}},
\oauthor{\binits{L.-S.} \bsnm{{The}}},
\oauthor{\binits{E.V.} \bsnm{{Linder}}},
\oauthor{\binits{V.} \bsnm{{Petrosian}}},
\oauthor{\binits{G.R.} \bsnm{{Blumenthal}}},
\oauthor{\binits{K.C.} \bsnm{{Hurley}}},
{Do gamma-ray bursts originate from an extended Galactic Halo of high-velocity
  neutron stars?}
\apjs
\textbf{90}
(1994).
doi:\doiurl{10.1086/191921}
\end{botherref}
\endbibitem

\bibitem[\protect\citeauthoryear{{Hayakawa}}{1981}]{Hayakawa1981;hakucho}
\begin{barticle}
\bauthor{\binits{S.} \bsnm{{Hayakawa}}},
\batitle{{Galactic X-rays observed with X-ray astronomy satellite 'Hakucho'}}.
\bjtitle{\ssr}
\bvolume{29},
\bfpage{221}--\blpage{290}
(\byear{1981}).
doi:\doiurl{10.1007/BF00229297}
\end{barticle}
\endbibitem

\bibitem[\protect\citeauthoryear{{Haymes} and {Harnden}}{1970}]{Haymes70}
\begin{barticle}
\bauthor{\binits{R.C.} \bsnm{{Haymes}}},
\bauthor{\binits{F.R.} \bsnm{{Harnden}} \bsuffix{Jr.}},
\batitle{{Low-Energy Gamma Radiation from Cygnus}}.
\bjtitle{\apj}
\bvolume{159},
\bfpage{1111}
(\byear{1970}).
doi:\doiurl{10.1086/150391}
\end{barticle}
\endbibitem

\bibitem[\protect\citeauthoryear{{Haymes} et~al.}{1968}]{Haymes68}
\begin{barticle}
\bauthor{\binits{R.C.} \bsnm{{Haymes}}},
\bauthor{\binits{D.V.} \bsnm{{Ellis}}},
\bauthor{\binits{G.J.} \bsnm{{Fishman}}},
\bauthor{\binits{J.D.} \bsnm{{Kurfess}}},
\bauthor{\binits{W.H.} \bsnm{{Tucker}}},
\batitle{{Observation of Gamma Radiation from the Crab Nebula}}.
\bjtitle{\apjl}
\bvolume{151},
\bfpage{9}
(\byear{1968}).
doi:\doiurl{10.1086/180129}
\end{barticle}
\endbibitem

\bibitem[\protect\citeauthoryear{{Haymes} et~al.}{1969}]{Haymes69}
\begin{barticle}
\bauthor{\binits{R.C.} \bsnm{{Haymes}}},
\bauthor{\binits{D.V.} \bsnm{{Ellis}}},
\bauthor{\binits{G.J.} \bsnm{{Fishman}}},
\bauthor{\binits{S.W.} \bsnm{{Glenn}}},
\bauthor{\binits{J.D.} \bsnm{{Kurfess}}},
\batitle{{Observation of Hard Radiation from the Region of the Galactic
  Center}}.
\bjtitle{\apj}
\bvolume{157},
\bfpage{1455}
(\byear{1969}).
doi:\doiurl{10.1086/150164}
\end{barticle}
\endbibitem

\bibitem[\protect\citeauthoryear{{Haymes} et~al.}{1972}]{Haymes72}
\begin{barticle}
\bauthor{\binits{R.C.} \bsnm{{Haymes}}},
\bauthor{\binits{F.R.} \bsnm{{Harnden}} \bsuffix{Jr.}},
\bauthor{\binits{W.N.} \bsnm{{Johnson}} \bsuffix{III}},
\bauthor{\binits{H.M.} \bsnm{{Prichard}}},
\bauthor{\binits{H.E.} \bsnm{{Bosch}}},
\batitle{{The Low-Energy Gamma-Ray Spectrum of Scorpius X-1.}}
\bjtitle{\apjl}
\bvolume{172},
\bfpage{47}
(\byear{1972}).
doi:\doiurl{10.1086/180888}
\end{barticle}
\endbibitem

\bibitem[\protect\citeauthoryear{{Haymes} et~al.}{1975}]{Haymes75}
\begin{barticle}
\bauthor{\binits{R.C.} \bsnm{{Haymes}}},
\bauthor{\binits{G.D.} \bsnm{{Walraven}}},
\bauthor{\binits{C.A.} \bsnm{{Meegan}}},
\bauthor{\binits{R.D.} \bsnm{{Hall}}},
\bauthor{\binits{F.T.} \bsnm{{Djuth}}},
\bauthor{\binits{D.H.} \bsnm{{Shelton}}},
\batitle{{Detection of nuclear gamma rays from the galactic center region}}.
\bjtitle{\apj}
\bvolume{201},
\bfpage{593}--\blpage{602}
(\byear{1975}).
doi:\doiurl{10.1086/153925}
\end{barticle}
\endbibitem

\bibitem[\protect\citeauthoryear{{Heindl} et~al.}{1993}]{Heindl93}
\begin{barticle}
\bauthor{\binits{W.A.} \bsnm{{Heindl}}},
\bauthor{\binits{W.R.} \bsnm{{Cook}}},
\bauthor{\binits{J.M.} \bsnm{{Grunsfeld}}},
\bauthor{\binits{D.M.} \bsnm{{Palmer}}},
\bauthor{\binits{T.A.} \bsnm{{Prince}}},
\bauthor{\binits{S.M.} \bsnm{{Schindler}}},
\bauthor{\binits{E.C.} \bsnm{{Stone}}},
\batitle{{An observation of the Galactic center hard X-ray source, 1E
  1740.7-2942, with the Caltech coded-aperture telescope}}.
\bjtitle{\apj}
\bvolume{408},
\bfpage{507}--\blpage{513}
(\byear{1993}).
doi:\doiurl{10.1086/172608}
\end{barticle}
\endbibitem

\bibitem[\protect\citeauthoryear{{Heindl} et~al.}{1999}]{Heindl99}
\begin{barticle}
\bauthor{\binits{W.A.} \bsnm{{Heindl}}},
\bauthor{\binits{W.} \bsnm{{Coburn}}},
\bauthor{\binits{D.E.} \bsnm{{Gruber}}},
\bauthor{\binits{M.R.} \bsnm{{Pelling}}},
\bauthor{\binits{R.E.} \bsnm{{Rothschild}}},
\bauthor{\binits{J.} \bsnm{{Wilms}}},
\bauthor{\binits{K.} \bsnm{{Pottschmidt}}},
\bauthor{\binits{R.} \bsnm{{Staubert}}},
\batitle{{Discovery of a Third Harmonic Cyclotron Resonance Scattering Feature
  in the X-Ray Spectrum of 4U 0115+63}}.
\bjtitle{\apjl}
\bvolume{521},
\bfpage{49}--\blpage{53}
(\byear{1999}).
doi:\doiurl{10.1086/312172}
\end{barticle}
\endbibitem

\bibitem[\protect\citeauthoryear{{Heise} et~al.}{2001}]{Heise01}
\begin{bchapter}
\bauthor{\binits{J.} \bsnm{{Heise}}},
\bauthor{\binits{J.I.} \bsnm{{Zand}}},
\bauthor{\binits{R.M.} \bsnm{{Kippen}}},
\bauthor{\binits{P.M.} \bsnm{{Woods}}},
\bctitle{{X-Ray Flashes and X-Ray Rich Gamma Ray Bursts}},
in \bbtitle{Gamma-ray Bursts in the Afterglow Era},
ed. by \beditor{\binits{E.} \bsnm{{Costa}}},
\beditor{\binits{F.} \bsnm{{Frontera}}},
\beditor{\binits{J.} \bsnm{{Hjorth}}},
\byear{2001},
p. \bfpage{16}
\end{bchapter}
\endbibitem

\bibitem[\protect\citeauthoryear{{Herzo}}{1975}]{Herzo75}
\begin{barticle}
\bauthor{\binits{D.} \bsnm{{Herzo}}},
\batitle{{A large double scatter telescope for gamma rays and neutrons}}.
\bjtitle{Nuclear Instruments and Methods}
\bvolume{123},
\bfpage{583}--\blpage{597}
(\byear{1975}).
doi:\doiurl{10.1016/0029-554X(75)90215-3}
\end{barticle}
\endbibitem

\bibitem[\protect\citeauthoryear{{Hoffman} et~al.}{1979}]{Hoffman79}
\begin{barticle}
\bauthor{\binits{J.A.} \bsnm{{Hoffman}}},
\bauthor{\binits{W.H.G.} \bsnm{{Lewin}}},
\bauthor{\binits{F.A.} \bsnm{{Primini}}},
\bauthor{\binits{W.A.} \bsnm{{Wheaton}}},
\bauthor{\binits{J.H.} \bsnm{{Swank}}},
\bauthor{\binits{E.A.} \bsnm{{Boldt}}},
\bauthor{\binits{S.S.} \bsnm{{Holt}}},
\bauthor{\binits{P.J.} \bsnm{{Serlemitsos}}},
\bauthor{\binits{G.H.} \bsnm{{Share}}},
\bauthor{\binits{K.} \bsnm{{Wood}}},
\bauthor{\binits{D.} \bsnm{{Yentis}}},
\bauthor{\binits{W.D.} \bsnm{{Evans}}},
\bauthor{\binits{J.L.} \bsnm{{Matteson}}},
\bauthor{\binits{D.E.} \bsnm{{Gruber}}},
\bauthor{\binits{L.E.} \bsnm{{Peterson}}},
\batitle{{HEAO 1 observation of a type I burst from MXB 1728-34}}.
\bjtitle{\apjl}
\bvolume{233},
\bfpage{51}--\blpage{55}
(\byear{1979}).
doi:\doiurl{10.1086/183075}
\end{barticle}
\endbibitem

\bibitem[\protect\citeauthoryear{{Hong} et~al.}{2011}]{Hong11}
\begin{barticle}
\bauthor{\binits{J.} \bsnm{{Hong}}},
\bauthor{\binits{B.} \bsnm{{Allen}}},
\bauthor{\binits{J.} \bsnm{{Grindlay}}},
\bauthor{\binits{S.} \bsnm{{Barthelemy}}},
\bauthor{\binits{R.} \bsnm{{Baker}}},
\bauthor{\binits{A.} \bsnm{{Garson}}},
\bauthor{\binits{H.} \bsnm{{Krawczynski}}},
\bauthor{\binits{J.} \bsnm{{Apple}}},
\bauthor{\binits{W.H.} \bsnm{{Cleveland}}},
\batitle{{Flight performance of an advanced CZT imaging detector in a
  balloon-borne wide-field hard X-ray telescope --ProtoEXIST1}}.
\bjtitle{Nuclear Instruments and Methods in Physics Research A}
\bvolume{654},
\bfpage{361}--\blpage{372}
(\byear{2011}).
doi:\doiurl{10.1016/j.nima.2011.07.025}
\end{barticle}
\endbibitem

\bibitem[\protect\citeauthoryear{{Horstman} et~al.}{1975}]{Horstman75}
\begin{barticle}
\bauthor{\binits{H.M.} \bsnm{{Horstman}}},
\bauthor{\binits{G.} \bsnm{{Cavallo}}},
\bauthor{\binits{E.} \bsnm{{Moretti-Horstman}}},
\batitle{{The X and gamma diffuse background}}.
\bjtitle{Nuovo Cimento Rivista Serie}
\bvolume{5},
\bfpage{255}--\blpage{311}
(\byear{1975}).
doi:\doiurl{10.1007/BF02747042}
\end{barticle}
\endbibitem

\bibitem[\protect\citeauthoryear{{Howe} et~al.}{1983}]{Howe83}
\begin{barticle}
\bauthor{\binits{S.K.} \bsnm{{Howe}}},
\bauthor{\binits{F.A.} \bsnm{{Primini}}},
\bauthor{\binits{M.W.} \bsnm{{Bautz}}},
\bauthor{\binits{F.L.} \bsnm{{Lang}}},
\bauthor{\binits{A.M.} \bsnm{{Levine}}},
\bauthor{\binits{W.H.G.} \bsnm{{Lewin}}},
\batitle{{HEAO 1 high-energy X-ray observations of Centaurus X-3}}.
\bjtitle{\apj}
\bvolume{272},
\bfpage{678}--\blpage{686}
(\byear{1983}).
doi:\doiurl{10.1086/161330}
\end{barticle}
\endbibitem

\bibitem[\protect\citeauthoryear{{Hudson} et~al.}{1969a}]{Hudson69a}
\begin{barticle}
\bauthor{\binits{H.S.} \bsnm{{Hudson}}},
\bauthor{\binits{L.E.} \bsnm{{Peterson}}},
\bauthor{\binits{D.A.} \bsnm{{Schwartz}}},
\batitle{{Solar and Cosmic X-Rays above 7.7 keV}}.
\bjtitle{\solphys}
\bvolume{6},
\bfpage{205}--\blpage{215}
(\byear{1969}a).
doi:\doiurl{10.1007/BF00150945}
\end{barticle}
\endbibitem

\bibitem[\protect\citeauthoryear{{Hudson} et~al.}{1969b}]{Hudson69b}
\begin{barticle}
\bauthor{\binits{H.S.} \bsnm{{Hudson}}},
\bauthor{\binits{L.E.} \bsnm{{Peterson}}},
\bauthor{\binits{D.A.} \bsnm{{Schwartz}}},
\batitle{{The Hard Solar X-Ray Spectrum Observed from the Third Orbiting Solar
  Observatory}}.
\bjtitle{\apj}
\bvolume{157},
\bfpage{389}
(\byear{1969}b).
doi:\doiurl{10.1086/150075}
\end{barticle}
\endbibitem

\bibitem[\protect\citeauthoryear{{Hurley} et~al.}{1997}]{Hurley97}
\begin{barticle}
\bauthor{\binits{K.} \bsnm{{Hurley}}},
\bauthor{\binits{E.} \bsnm{{Costa}}},
\bauthor{\binits{M.} \bsnm{{Feroci}}},
\bauthor{\binits{F.} \bsnm{{Frontera}}},
\bauthor{\binits{T.} \bsnm{{Cline}}},
\bauthor{\binits{D.} \bsnm{{Dal Fiume}}},
\bauthor{\binits{M.} \bsnm{{Orlandini}}},
\bauthor{\binits{M.} \bsnm{{Boer}}},
\bauthor{\binits{E.} \bsnm{{Mazets}}},
\bauthor{\binits{R.} \bsnm{{Aptekar}}},
\bauthor{\binits{S.} \bsnm{{Golenetskii}}},
\bauthor{\binits{M.} \bsnm{{Terekhov}}},
\batitle{{Third Interplanetary Network Localization, Time History, Fluence,
  Peak Flux, and Distance Lower Limit of the 1997 February 28 Gamma-Ray
  Burst}}.
\bjtitle{\apjl}
\bvolume{485},
\bfpage{1}--\blpage{3}
(\byear{1997}).
doi:\doiurl{10.1086/310808}
\end{barticle}
\endbibitem

\bibitem[\protect\citeauthoryear{{in 't Zand} et~al.}{1999}]{intZand99}
\begin{barticle}
\bauthor{\binits{J.J.M.} \bsnm{{in 't Zand}}},
\bauthor{\binits{J.} \bsnm{{Heise}}},
\bauthor{\binits{E.} \bsnm{{Kuulkers}}},
\bauthor{\binits{A.} \bsnm{{Bazzano}}},
\bauthor{\binits{M.} \bsnm{{Cocchi}}},
\bauthor{\binits{P.} \bsnm{{Ubertini}}},
\batitle{{Broad-band X-ray measurements of GS 1826-238}}.
\bjtitle{\aap}
\bvolume{347},
\bfpage{891}--\blpage{896}
(\byear{1999})
\end{barticle}
\endbibitem

\bibitem[\protect\citeauthoryear{Iniewsky}{2010}]{Iniewsky2010;hero}
\begin{bbook}
\bauthor{\binits{K.} \bsnm{Iniewsky}},
\bbtitle{Semiconductor Radiation Detection Systems}
(\bpublisher{CRC Press}, \blocation{???}, \byear{2010})
\end{bbook}
\endbibitem

\bibitem[\protect\citeauthoryear{{Inoue}}{1985}]{Inoue85}
\begin{barticle}
\bauthor{\binits{H.} \bsnm{{Inoue}}},
\batitle{{TENMA observations of bright binary X-ray sources}}.
\bjtitle{\ssr}
\bvolume{40},
\bfpage{317}--\blpage{338}
(\byear{1985}).
doi:\doiurl{10.1007/BF00212905}
\end{barticle}
\endbibitem

\bibitem[\protect\citeauthoryear{{Inoue}}{2003}]{Inoue03}
\begin{barticle}
\bauthor{\binits{H.} \bsnm{{Inoue}}},
\batitle{{The Astro-E mission}}.
\bjtitle{Advances in Space Research}
\bvolume{32},
\bfpage{2089}--\blpage{2090}
(\byear{2003}).
doi:\doiurl{10.1016/S0273-1177(03)90649-1}
\end{barticle}
\endbibitem

\bibitem[\protect\citeauthoryear{{Inoue} et~al.}{1979}]{Inoue1979;hakucho}
\begin{barticle}
\bauthor{\binits{H.} \bsnm{{Inoue}}},
\bauthor{\binits{K.} \bsnm{{Koyama}}},
\bauthor{\binits{M.} \bsnm{{Matsuoka}}},
\bauthor{\binits{T.} \bsnm{{Ohashi}}},
\bauthor{\binits{Y.} \bsnm{{Tanaka}}},
\bauthor{\binits{H.} \bsnm{{Tsunemi}}},
\batitle{{Emission Line Features in the Soft X-Ray Spectra of the North Polar
  Spur and the Cygnus Loop}}.
\bjtitle{International Cosmic Ray Conference}
\bvolume{1},
\bfpage{24}
(\byear{1979})
\end{barticle}
\endbibitem

\bibitem[\protect\citeauthoryear{{Itoh} et~al.}{2008}]{Itoh08}
\begin{barticle}
\bauthor{\binits{T.} \bsnm{{Itoh}}},
\bauthor{\binits{C.} \bsnm{{Done}}},
\bauthor{\binits{K.} \bsnm{{Makishima}}},
\bauthor{\binits{G.} \bsnm{{Madejski}}},
\bauthor{\binits{H.} \bsnm{{Awaki}}},
\bauthor{\binits{P.} \bsnm{{Gandhi}}},
\bauthor{\binits{N.} \bsnm{{Isobe}}},
\bauthor{\binits{G.C.} \bsnm{{Dewangan}}},
\bauthor{\binits{R.E.} \bsnm{{Griffthis}}},
\bauthor{\binits{N.} \bsnm{{Anabuki}}},
\bauthor{\binits{T.} \bsnm{{Okajima}}},
\bauthor{\binits{J.N.} \bsnm{{Reeves}}},
\bauthor{\binits{T.} \bsnm{{Takahashi}}},
\bauthor{\binits{Y.} \bsnm{{Ueda}}},
\bauthor{\binits{S.} \bsnm{{Eguchi}}},
\bauthor{\binits{T.} \bsnm{{Yaqoob}}},
\batitle{{Suzaku Wide-Band X-Ray Spectroscopy of the Seyfert2 AGN in NGC
  4945}}.
\bjtitle{\pasj}
\bvolume{60},
\bfpage{251}--\blpage{262}
(\byear{2008})
\end{barticle}
\endbibitem

\bibitem[\protect\citeauthoryear{{Johnson} et~al.}{1978}]{Johnson78}
\begin{bchapter}
\bauthor{\binits{W.N.} \bsnm{{Johnson}}},
\bauthor{\binits{J.D.} \bsnm{{Kurfess}}},
\bauthor{\binits{D.M.} \bsnm{{Saulnier}}},
\bctitle{{A hard X-ray experiment for long-duration balloon flights.}},
in \bbtitle{New Instrumentation for Space Astronomy},
ed. by \beditor{\binits{K.A.} \bsnm{{van der Hucht}}},
\beditor{\binits{G.} \bsnm{{Vaiana}}},
\byear{1978},
pp. \bfpage{169}--\blpage{172}
\end{bchapter}
\endbibitem

\bibitem[\protect\citeauthoryear{{Johnson} and {Haymes}}{1973}]{Johnson73}
\begin{barticle}
\bauthor{\binits{W.N.} \bsnm{{Johnson}} \bsuffix{III}},
\bauthor{\binits{R.C.} \bsnm{{Haymes}}},
\batitle{{Detection of a Gamma-Ray Spectral Line from the Galactic-Center
  Region}}.
\bjtitle{\apj}
\bvolume{184},
\bfpage{103}--\blpage{126}
(\byear{1973}).
doi:\doiurl{10.1086/152309}
\end{barticle}
\endbibitem

\bibitem[\protect\citeauthoryear{{Johnson} et~al.}{1972}]{Johnson72}
\begin{barticle}
\bauthor{\binits{W.N.} \bsnm{{Johnson}} \bsuffix{III}},
\bauthor{\binits{F.R.} \bsnm{{Harnden}} \bsuffix{Jr.}},
\bauthor{\binits{R.C.} \bsnm{{Haymes}}},
\batitle{{The Spectrum of Low-Energy Gamma Radiation from the Galactic-Center
  Region.}}
\bjtitle{\apjl}
\bvolume{172},
\bfpage{1}
(\byear{1972}).
doi:\doiurl{10.1086/180878}
\end{barticle}
\endbibitem

\bibitem[\protect\citeauthoryear{{Johnson} et~al.}{1993}]{Johnson1993;cgro}
\begin{barticle}
\bauthor{\binits{W.N.} \bsnm{{Johnson}}},
\bauthor{\binits{R.L.} \bsnm{{Kinzer}}},
\bauthor{\binits{J.D.} \bsnm{{Kurfess}}},
\bauthor{\binits{M.S.} \bsnm{{Strickman}}},
\bauthor{\binits{W.R.} \bsnm{{Purcell}}},
\bauthor{\binits{D.A.} \bsnm{{Grabelsky}}},
\bauthor{\binits{M.P.} \bsnm{{Ulmer}}},
\bauthor{\binits{D.A.} \bsnm{{Hillis}}},
\bauthor{\binits{G.V.} \bsnm{{Jung}}},
\bauthor{\binits{R.A.} \bsnm{{Cameron}}},
\batitle{{The Oriented Scintillation Spectrometer Experiment - Instrument
  description}}.
\bjtitle{\apjs}
\bvolume{86},
\bfpage{693}--\blpage{712}
(\byear{1993}).
doi:\doiurl{10.1086/191795}
\end{barticle}
\endbibitem

\bibitem[\protect\citeauthoryear{{Johnson} et~al.}{1994}]{Johnson94}
\begin{bchapter}
\bauthor{\binits{W.N.} \bsnm{{Johnson}}},
\bauthor{\binits{J.E.} \bsnm{{Grove}}},
\bauthor{\binits{R.L.} \bsnm{{Kinzer}}},
\bauthor{\binits{R.A.} \bsnm{{Kroeger}}},
\bauthor{\binits{J.D.} \bsnm{{Kurfess}}},
\bauthor{\binits{M.S.} \bsnm{{Strickman}}},
\bauthor{\binits{K.} \bsnm{{McNaron-Brown}}},
\bauthor{\binits{D.A.} \bsnm{{Grabelsky}}},
\bauthor{\binits{W.R.} \bsnm{{Purcell}}},
\bauthor{\binits{M.P.} \bsnm{{Ulmer}}},
\bauthor{\binits{G.V.} \bsnm{{Jung}}},
\bctitle{{The canonical Seyfert spectrum: the implications of OSSE
  observations.}},
in \bbtitle{American Institute of Physics Conference Series},
ed. by \beditor{\binits{C.E.} \bsnm{{Fichtel}}},
\beditor{\binits{N.} \bsnm{{Gehrels}}},
\beditor{\binits{J.P.} \bsnm{{Norris}}}
\bsertitle{American Institute of Physics Conference Series},
vol. \bseriesno{304},
\byear{1994},
pp. \bfpage{515}--\blpage{524}
\end{bchapter}
\endbibitem

\bibitem[\protect\citeauthoryear{{Johnson} et~al.}{1997}]{Johnson97}
\begin{barticle}
\bauthor{\binits{W.N.} \bsnm{{Johnson}}},
\bauthor{\binits{K.} \bsnm{{McNaron-Brown}}},
\bauthor{\binits{J.D.} \bsnm{{Kurfess}}},
\bauthor{\binits{A.A.} \bsnm{{Zdziarski}}},
\bauthor{\binits{P.} \bsnm{{Magdziarz}}},
\bauthor{\binits{N.} \bsnm{{Gehrels}}},
\batitle{{Long-Term Monitoring of NGC 4151 by OSSE}}.
\bjtitle{\apj}
\bvolume{482},
\bfpage{173}--\blpage{177}
(\byear{1997})
\end{barticle}
\endbibitem

\bibitem[\protect\citeauthoryear{{Jourdain} et~al.}{2012}]{Jourdain12}
\begin{barticle}
\bauthor{\binits{E.} \bsnm{{Jourdain}}},
\bauthor{\binits{J.P.} \bsnm{{Roques}}},
\bauthor{\binits{M.} \bsnm{{Chauvin}}},
\bauthor{\binits{D.J.} \bsnm{{Clark}}},
\batitle{{Separation of Two Contributions to the High Energy Emission of Cygnus
  X-1: Polarization Measurements with INTEGRAL SPI}}.
\bjtitle{\apj}
\bvolume{761},
\bfpage{27}
(\byear{2012}).
doi:\doiurl{10.1088/0004-637X/761/1/27}
\end{barticle}
\endbibitem

\bibitem[\protect\citeauthoryear{{Jung}}{1989}]{Jung1989;heao1}
\begin{barticle}
\bauthor{\binits{G.V.} \bsnm{{Jung}}},
\batitle{{The hard X-ray to low-energy gamma-ray spectrum of the Crab Nebula}}.
\bjtitle{\apj}
\bvolume{338},
\bfpage{972}--\blpage{982}
(\byear{1989}).
doi:\doiurl{10.1086/167249}
\end{barticle}
\endbibitem

\bibitem[\protect\citeauthoryear{{Jung} et~al.}{1995}]{Jung95}
\begin{barticle}
\bauthor{\binits{G.V.} \bsnm{{Jung}}},
\bauthor{\binits{D.J.} \bsnm{{Kurfess}}},
\bauthor{\binits{W.N.} \bsnm{{Johnson}}},
\bauthor{\binits{R.L.} \bsnm{{Kinzer}}},
\bauthor{\binits{J.E.} \bsnm{{Grove}}},
\bauthor{\binits{M.S.} \bsnm{{Strickman}}},
\bauthor{\binits{W.R.} \bsnm{{Purcell}}},
\bauthor{\binits{D.A.} \bsnm{{Grabelsky}}},
\bauthor{\binits{M.P.} \bsnm{{Ulmer}}},
\batitle{{OSSE observations of 1E 1740.7-2942 in 1992 September}}.
\bjtitle{\aap}
\bvolume{295},
\bfpage{23}--\blpage{26}
(\byear{1995})
\end{barticle}
\endbibitem

\bibitem[\protect\citeauthoryear{{Kaneko} et~al.}{2010}]{Kaneko10}
\begin{barticle}
\bauthor{\binits{Y.} \bsnm{{Kaneko}}},
\bauthor{\binits{E.} \bsnm{{G{\"o}{\v g}{\"u}{\c s}}}},
\bauthor{\binits{C.} \bsnm{{Kouveliotou}}},
\bauthor{\binits{J.} \bsnm{{Granot}}},
\bauthor{\binits{E.} \bsnm{{Ramirez-Ruiz}}},
\bauthor{\binits{A.J.} \bsnm{{van der Horst}}},
\bauthor{\binits{A.L.} \bsnm{{Watts}}},
\bauthor{\binits{M.H.} \bsnm{{Finger}}},
\bauthor{\binits{N.} \bsnm{{Gehrels}}},
\bauthor{\binits{A.} \bsnm{{Pe'er}}},
\bauthor{\binits{M.} \bsnm{{van der Klis}}},
\bauthor{\binits{A.} \bsnm{{von Kienlin}}},
\bauthor{\binits{S.} \bsnm{{Wachter}}},
\bauthor{\binits{C.A.} \bsnm{{Wilson-Hodge}}},
\bauthor{\binits{P.M.} \bsnm{{Woods}}},
\batitle{{Magnetar Twists: Fermi/Gamma-Ray Burst Monitor Detection of SGR
  J1550-5418}}.
\bjtitle{\apj}
\bvolume{710},
\bfpage{1335}--\blpage{1342}
(\byear{2010}).
doi:\doiurl{10.1088/0004-637X/710/2/1335}
\end{barticle}
\endbibitem

\bibitem[\protect\citeauthoryear{{Kaniovsky} et~al.}{1997}]{Kaniovsky97}
\begin{barticle}
\bauthor{\binits{A.S.} \bsnm{{Kaniovsky}}},
\bauthor{\binits{V.A.} \bsnm{{Arefiev}}},
\bauthor{\binits{N.L.} \bsnm{{Aleksandrovich}}},
\bauthor{\binits{V.V.} \bsnm{{Borkous}}},
\bauthor{\binits{K.N.} \bsnm{{Borozdin}}},
\bauthor{\binits{V.V.} \bsnm{{Efremov}}},
\bauthor{\binits{R.A.} \bsnm{{Sunyaev}}},
\bauthor{\binits{E.} \bsnm{{Kendziorra}}},
\bauthor{\binits{P.} \bsnm{{Kretschmar}}},
\bauthor{\binits{M.} \bsnm{{Kunz}}},
\bauthor{\binits{M.} \bsnm{{Maisack}}},
\bauthor{\binits{R.} \bsnm{{Staubert}}},
\bauthor{\binits{S.} \bsnm{{Doebereiner}}},
\bauthor{\binits{J.} \bsnm{{Englhauser}}},
\bauthor{\binits{W.} \bsnm{{Pietsch}}},
\bauthor{\binits{C.} \bsnm{{Reppin}}},
\bauthor{\binits{J.} \bsnm{{Tr{\"u}mper}}},
\bauthor{\binits{G.K.} \bsnm{{Skinner}}},
\bauthor{\binits{A.P.} \bsnm{{Willmore}}},
\bauthor{\binits{A.C.} \bsnm{{Brinkman}}},
\bauthor{\binits{J.} \bsnm{{Heise}}},
\bauthor{\binits{R.} \bsnm{{Jager}}},
\batitle{{Three hard X-ray transients: GRO J0422+32, GRS 1716-24, GRS 1009-45.
  Broad band observations by Roentgen-MIR-KVANT observatory}}.
\bjtitle{Advances in Space Research}
\bvolume{19},
\bfpage{29}--\blpage{34}
(\byear{1997}).
doi:\doiurl{10.1016/S0273-1177(97)00033-1}
\end{barticle}
\endbibitem

\bibitem[\protect\citeauthoryear{{Kataoka} et~al.}{2008}]{Kataoka08}
\begin{barticle}
\bauthor{\binits{J.} \bsnm{{Kataoka}}},
\bauthor{\binits{G.} \bsnm{{Madejski}}},
\bauthor{\binits{M.} \bsnm{{Sikora}}},
\bauthor{\binits{P.} \bsnm{{Roming}}},
\bauthor{\binits{M.M.} \bsnm{{Chester}}},
\bauthor{\binits{D.} \bsnm{{Grupe}}},
\bauthor{\binits{Y.} \bsnm{{Tsubuku}}},
\bauthor{\binits{R.} \bsnm{{Sato}}},
\bauthor{\binits{N.} \bsnm{{Kawai}}},
\bauthor{\binits{G.} \bsnm{{Tosti}}},
\bauthor{\binits{D.} \bsnm{{Impiombato}}},
\bauthor{\binits{Y.Y.} \bsnm{{Kovalev}}},
\bauthor{\binits{Y.A.} \bsnm{{Kovalev}}},
\bauthor{\binits{P.G.} \bsnm{{Edwards}}},
\bauthor{\binits{S.J.} \bsnm{{Wagner}}},
\bauthor{\binits{R.} \bsnm{{Moderski}}},
\bauthor{\binits{{\L}.} \bsnm{{Stawarz}}},
\bauthor{\binits{T.} \bsnm{{Takahashi}}},
\bauthor{\binits{S.} \bsnm{{Watanabe}}},
\batitle{{Multiwavelength Observations of the Powerful Gamma-Ray Quasar PKS
  1510-089: Clues on the Jet Composition}}.
\bjtitle{\apj}
\bvolume{672},
\bfpage{787}--\blpage{799}
(\byear{2008}).
doi:\doiurl{10.1086/523093}
\end{barticle}
\endbibitem

\bibitem[\protect\citeauthoryear{{Kawabata Nobukawa} et~al.}{2012}]{Kawabata12}
\begin{botherref}
\oauthor{\binits{K.} \bsnm{{Kawabata Nobukawa}}},
\oauthor{\binits{M.} \bsnm{{Nobukawa}}},
\oauthor{\binits{T.G.} \bsnm{{Tsuru}}},
\oauthor{\binits{K.} \bsnm{{Koyama}}},
{Suzaku Observation of the Supergiant Fast X-Ray Transient AX J1841.0-0536}.
\pasj
\textbf{64}
(2012).
doi:\doiurl{10.1093/pasj/64.5.99}
\end{botherref}
\endbibitem

\bibitem[\protect\citeauthoryear{{Kawai} et~al.}{2003}]{Kawai03}
\begin{bchapter}
\bauthor{\binits{N.} \bsnm{{Kawai}}},
\bauthor{\binits{A.} \bsnm{{Yoshida}}},
\bauthor{\binits{M.} \bsnm{{Matsuoka}}},
\bauthor{\binits{Y.} \bsnm{{Shirasaki}}},
\bauthor{\binits{T.} \bsnm{{Tamagawa}}},
\bauthor{\binits{K.} \bsnm{{Torii}}},
\bauthor{\binits{T.} \bsnm{{Sakamoto}}},
\bauthor{\binits{D.} \bsnm{{Takahashi}}},
\bauthor{\binits{E.} \bsnm{{Fenimore}}},
\bauthor{\binits{M.} \bsnm{{Galassi}}},
\bauthor{\binits{T.} \bsnm{{Tavenner}}},
\bauthor{\binits{D.Q.} \bsnm{{Lamb}}},
\bauthor{\binits{C.} \bsnm{{Graziani}}},
\bauthor{\binits{T.} \bsnm{{Donaghy}}},
\bauthor{\binits{R.} \bsnm{{Vanderspek}}},
\bauthor{\binits{M.} \bsnm{{Yamauchi}}},
\bauthor{\binits{K.} \bsnm{{Takagishi}}},
\bauthor{\binits{I.} \bsnm{{Hatsukade}}},
\bctitle{{In-Orbit Performance of WXM (Wide-Field X-Ray Monitor)}},
in \bbtitle{Gamma-Ray Burst and Afterglow Astronomy 2001: A Workshop
  Celebrating the First Year of the HETE Mission},
ed. by \beditor{\binits{G.R.} \bsnm{{Ricker}}},
\beditor{\binits{R.K.} \bsnm{{Vanderspek}}}
\bsertitle{American Institute of Physics Conference Series},
vol. \bseriesno{662},
\byear{2003},
pp. \bfpage{25}--\blpage{32}.
doi:\doiurl{10.1063/1.1579293}
\end{bchapter}
\endbibitem

\bibitem[\protect\citeauthoryear{{Kendziorra} et~al.}{1977}]{Kendziorra77}
\begin{barticle}
\bauthor{\binits{E.} \bsnm{{Kendziorra}}},
\bauthor{\binits{R.} \bsnm{{Staubert}}},
\bauthor{\binits{W.} \bsnm{{Pietsch}}},
\bauthor{\binits{C.} \bsnm{{Reppin}}},
\bauthor{\binits{B.} \bsnm{{Sacco}}},
\bauthor{\binits{J.} \bsnm{{Tr{\"u}mper}}},
\batitle{{Hercules X-1 - The 1.24 second pulsation in hard X-rays}}.
\bjtitle{\apjl}
\bvolume{217},
\bfpage{93}--\blpage{96}
(\byear{1977}).
doi:\doiurl{10.1086/182546}
\end{barticle}
\endbibitem

\bibitem[\protect\citeauthoryear{{Kinzer} et~al.}{1978}]{Kinzer78}
\begin{barticle}
\bauthor{\binits{R.L.} \bsnm{{Kinzer}}},
\bauthor{\binits{W.N.} \bsnm{{Johnson}}},
\bauthor{\binits{J.D.} \bsnm{{Kurfess}}},
\batitle{{A balloon observation of the diffuse cosmic X-radiation above 20
  keV}}.
\bjtitle{\apj}
\bvolume{222},
\bfpage{370}--\blpage{378}
(\byear{1978}).
doi:\doiurl{10.1086/156150}
\end{barticle}
\endbibitem

\bibitem[\protect\citeauthoryear{{Kinzer} et~al.}{1995}]{Kinzer95}
\begin{barticle}
\bauthor{\binits{R.L.} \bsnm{{Kinzer}}},
\bauthor{\binits{W.N.} \bsnm{{Johnson}}},
\bauthor{\binits{C.D.} \bsnm{{Dermer}}},
\bauthor{\binits{J.D.} \bsnm{{Kurfess}}},
\bauthor{\binits{M.S.} \bsnm{{Strickman}}},
\bauthor{\binits{J.E.} \bsnm{{Grove}}},
\bauthor{\binits{R.A.} \bsnm{{Kroeger}}},
\bauthor{\binits{D.A.} \bsnm{{Grabelsky}}},
\bauthor{\binits{W.R.} \bsnm{{Purcell}}},
\bauthor{\binits{M.P.} \bsnm{{Ulmer}}},
\bauthor{\binits{G.V.} \bsnm{{Jung}}},
\bauthor{\binits{K.} \bsnm{{McNaron-Brown}}},
\batitle{{OSSE Observations of Gamma-Ray Emission from Centaurus A}}.
\bjtitle{\apj}
\bvolume{449},
\bfpage{105}
(\byear{1995}).
doi:\doiurl{10.1086/176036}
\end{barticle}
\endbibitem

\bibitem[\protect\citeauthoryear{{Kinzer} et~al.}{1997}]{Kinzer97}
\begin{barticle}
\bauthor{\binits{R.L.} \bsnm{{Kinzer}}},
\bauthor{\binits{G.V.} \bsnm{{Jung}}},
\bauthor{\binits{D.E.} \bsnm{{Gruber}}},
\bauthor{\binits{J.L.} \bsnm{{Matteson}}},
\bauthor{\bsnm{{Peterson}}},
\bauthor{\bsnm{{L.~E.}}},
\batitle{{Diffuse Cosmic Gamma Radiation Measured by HEAO 1}}.
\bjtitle{\apj}
\bvolume{475},
\bfpage{361}--\blpage{372}
(\byear{1997})
\end{barticle}
\endbibitem

\bibitem[\protect\citeauthoryear{{Klebesadel} et~al.}{1973}]{Klebesadel73}
\begin{barticle}
\bauthor{\binits{R.W.} \bsnm{{Klebesadel}}},
\bauthor{\binits{I.B.} \bsnm{{Strong}}},
\bauthor{\binits{R.A.} \bsnm{{Olson}}},
\batitle{{Observations of Gamma-Ray Bursts of Cosmic Origin}}.
\bjtitle{\apjl}
\bvolume{182},
\bfpage{85}
(\byear{1973}).
doi:\doiurl{10.1086/181225}
\end{barticle}
\endbibitem

\bibitem[\protect\citeauthoryear{{Koglin} et~al.}{2006}]{Koglin2006;heft}
\begin{bchapter}
\bauthor{\binits{J.E.} \bsnm{{Koglin}}},
\bauthor{\binits{W.H.} \bsnm{{Baumgartner}}},
\bauthor{\binits{C.M.H.} \bsnm{{Chen}}},
\bauthor{\binits{J.C.} \bsnm{{Chonko}}},
\bauthor{\binits{F.E.} \bsnm{{Christensen}}},
\bauthor{\binits{W.W.} \bsnm{{Craig}}},
\bauthor{\binits{T.R.} \bsnm{{Decker}}},
\bauthor{\binits{C.J.} \bsnm{{Hailey}}},
\bauthor{\binits{F.A.} \bsnm{{Harrison}}},
\bauthor{\binits{C.P.} \bsnm{{Jensen}}},
\bauthor{\binits{K.K.} \bsnm{{Madsen}}},
\bauthor{\binits{M.J.} \bsnm{{Pivovaroff}}},
\bauthor{\binits{G.} \bsnm{{Tajiri}}},
\bctitle{{Calibration of HEFT Hard X-ray Optics}},
in \bbtitle{The X-ray Universe 2005},
ed. by \beditor{\binits{A.} \bsnm{{Wilson}}}
\bsertitle{ESA Special Publication},
vol. \bseriesno{604},
\byear{2006},
p. \bfpage{955}
\end{bchapter}
\endbibitem

\bibitem[\protect\citeauthoryear{{Kole} et~al.}{2016}]{Kole16}
\begin{botherref}
\oauthor{\binits{M.} \bsnm{{Kole}}},
\oauthor{\binits{T.W.} \bsnm{{Bao}}},
\oauthor{\binits{T.} \bsnm{{Batsch}}},
\oauthor{\binits{T.} \bsnm{{Bernasconi}}},
\oauthor{\binits{F.} \bsnm{{Cadoux}}},
\oauthor{\binits{J.Y.} \bsnm{{Chai}}},
\oauthor{\binits{Y.W.} \bsnm{{Dong}}},
\oauthor{\binits{N.} \bsnm{{Gauvin}}},
\oauthor{\binits{W.} \bsnm{{Hajdas}}},
\oauthor{\binits{J.J.} \bsnm{{He}}},
\oauthor{\binits{M.N.} \bsnm{{Kong}}},
\oauthor{\binits{S.W.} \bsnm{{Kong}}},
\oauthor{\binits{C.} \bsnm{{Lechanoine-Leluc}}},
\oauthor{\binits{L.} \bsnm{{Li}}},
\oauthor{\binits{Z.H.} \bsnm{{Li}}},
\oauthor{\binits{J.T.} \bsnm{{Liu}}},
\oauthor{\binits{X.} \bsnm{{Liu}}},
\oauthor{\binits{R.} \bsnm{{Marcinkowski}}},
\oauthor{\binits{S.} \bsnm{{Orsi}}},
\oauthor{\binits{M.} \bsnm{{Pohl}}},
\oauthor{\binits{N.} \bsnm{{Produit}}},
\oauthor{\binits{D.} \bsnm{{Rapin}}},
\oauthor{\binits{A.} \bsnm{{Rutczynska}}},
\oauthor{\binits{D.} \bsnm{{Rybka}}},
\oauthor{\binits{H.L.} \bsnm{{Shi}}},
\oauthor{\binits{L.M.} \bsnm{{Song}}},
\oauthor{\binits{J.C.} \bsnm{{Sun}}},
\oauthor{\binits{J.} \bsnm{{Szabelski}}},
\oauthor{\binits{R.J.} \bsnm{{Wang}}},
\oauthor{\binits{Y.J.} \bsnm{{Wang}}},
\oauthor{\binits{X.} \bsnm{{Wen}}},
\oauthor{\binits{B.B.} \bsnm{{Wu}}},
\oauthor{\binits{X.} \bsnm{{Wu}}},
\oauthor{\binits{H.L.} \bsnm{{Xiao}}},
\oauthor{\binits{S.L.} \bsnm{{Xiong}}},
\oauthor{\binits{H.H.} \bsnm{{Xu}}},
\oauthor{\binits{M.} \bsnm{{Xu}}},
\oauthor{\binits{J.} \bsnm{{Zhang}}},
\oauthor{\binits{L.} \bsnm{{Zhang}}},
\oauthor{\binits{L.Y.} \bsnm{{Zhang}}},
\oauthor{\binits{S.N.} \bsnm{{Zhang}}},
\oauthor{\binits{X.F.} \bsnm{{Zhang}}},
\oauthor{\binits{Y.J.} \bsnm{{Zhang}}},
\oauthor{\binits{A.} \bsnm{{Zwolinska}}},
{POLAR: Final Calibration and In-Flight Performance of a Dedicated GRB
  Polarimeter}.
ArXiv e-prints
(2016)
\end{botherref}
\endbibitem

\bibitem[\protect\citeauthoryear{{Koo} and {Haymes}}{1980}]{Koo80}
\begin{barticle}
\bauthor{\binits{J.-W.C.} \bsnm{{Koo}}},
\bauthor{\binits{R.C.} \bsnm{{Haymes}}},
\batitle{{The hard X-ray periodicity of GX 1 + 4}}.
\bjtitle{\apjl}
\bvolume{239},
\bfpage{57}--\blpage{60}
(\byear{1980}).
doi:\doiurl{10.1086/183291}
\end{barticle}
\endbibitem

\bibitem[\protect\citeauthoryear{{Kouveliotou} et~al.}{1993}]{Kouveliotou93}
\begin{bchapter}
\bauthor{\binits{C.} \bsnm{{Kouveliotou}}},
\bauthor{\binits{M.H.} \bsnm{{Finger}}},
\bauthor{\binits{G.J.} \bsnm{{Fishman}}},
\bauthor{\binits{C.A.} \bsnm{{Meegan}}},
\bauthor{\binits{R.B.} \bsnm{{Wilson}}},
\bauthor{\binits{W.S.} \bsnm{{Paciesas}}},
\bauthor{\binits{T.} \bsnm{{Minamitani}}},
\bauthor{\binits{J.} \bsnm{{van Paradijs}}},
\bctitle{{Detection of quasi-periodic oscillations (QSO) from Cyg X-1 and GRO
  J0422+32.}},
in \bbtitle{American Institute of Physics Conference Series},
ed. by \beditor{\binits{M.} \bsnm{{Friedlander}}},
\beditor{\binits{N.} \bsnm{{Gehrels}}},
\beditor{\binits{D.J.} \bsnm{{Macomb}}}
\bsertitle{American Institute of Physics Conference Series},
vol. \bseriesno{280},
\byear{1993},
pp. \bfpage{319}--\blpage{323}
\end{bchapter}
\endbibitem

\bibitem[\protect\citeauthoryear{{Kouveliotou} et~al.}{1996}]{Kouveliotou96}
\begin{botherref}
\oauthor{\binits{C.} \bsnm{{Kouveliotou}}},
\oauthor{\binits{J.} \bsnm{{Kommers}}},
\oauthor{\binits{W.H.G.} \bsnm{{Lewin}}},
\oauthor{\binits{J.} \bsnm{{van Paradijs}}},
\oauthor{\binits{G.J.} \bsnm{{Fishman}}},
\oauthor{\binits{M.S.} \bsnm{{Briggs}}},
\oauthor{\binits{K.} \bsnm{{Hurley}}},
\oauthor{\binits{A.} \bsnm{{Harmon}}},
\oauthor{\binits{M.H.} \bsnm{{Finger}}},
\oauthor{\binits{R.B.} \bsnm{{Wilson}}},
{GRO J1744-28}.
\iaucirc
(1996)
\end{botherref}
\endbibitem

\bibitem[\protect\citeauthoryear{{Kouveliotou} et~al.}{2013}]{Kouveliotou13}
\begin{barticle}
\bauthor{\binits{C.} \bsnm{{Kouveliotou}}},
\bauthor{\binits{J.} \bsnm{{Granot}}},
\bauthor{\binits{J.L.} \bsnm{{Racusin}}},
\bauthor{\binits{E.} \bsnm{{Bellm}}},
\bauthor{\binits{G.} \bsnm{{Vianello}}},
\bauthor{\binits{S.} \bsnm{{Oates}}},
\bauthor{\binits{C.L.} \bsnm{{Fryer}}},
\bauthor{\binits{S.E.} \bsnm{{Boggs}}},
\bauthor{\binits{F.E.} \bsnm{{Christensen}}},
\bauthor{\binits{W.W.} \bsnm{{Craig}}},
\bauthor{\binits{C.D.} \bsnm{{Dermer}}},
\bauthor{\binits{N.} \bsnm{{Gehrels}}},
\bauthor{\binits{C.J.} \bsnm{{Hailey}}},
\bauthor{\binits{F.A.} \bsnm{{Harrison}}},
\bauthor{\binits{A.} \bsnm{{Melandri}}},
\bauthor{\binits{J.E.} \bsnm{{McEnery}}},
\bauthor{\binits{C.G.} \bsnm{{Mundell}}},
\bauthor{\binits{D.K.} \bsnm{{Stern}}},
\bauthor{\binits{G.} \bsnm{{Tagliaferri}}},
\bauthor{\binits{W.W.} \bsnm{{Zhang}}},
\batitle{{NuSTAR Observations of GRB 130427A Establish a Single Component
  Synchrotron Afterglow Origin for the Late Optical to Multi-GeV Emission}}.
\bjtitle{\apjl}
\bvolume{779},
\bfpage{1}
(\byear{2013}).
doi:\doiurl{10.1088/2041-8205/779/1/L1}
\end{barticle}
\endbibitem

\bibitem[\protect\citeauthoryear{{Krawczynski} et~al.}{2009}]{Krawczynski09}
\begin{barticle}
\bauthor{\binits{H.} \bsnm{{Krawczynski}}},
\bauthor{\binits{A.} \bsnm{{Garson}} \bsuffix{III}},
\bauthor{\binits{J.} \bsnm{{Martin}}},
\bauthor{\binits{Q.} \bsnm{{Li}}},
\bauthor{\binits{M.} \bsnm{{Beilicke}}},
\bauthor{\binits{P.} \bsnm{{Dowkontt}}},
\bauthor{\binits{K.} \bsnm{{Lee}}},
\bauthor{\binits{E.} \bsnm{{Wulf}}},
\bauthor{\binits{J.} \bsnm{{Kurfess}}},
\bauthor{\binits{E.I.} \bsnm{{Novikova}}},
\bauthor{\binits{G.} \bsnm{{de Geronimo}}},
\bauthor{\binits{M.G.} \bsnm{{Baring}}},
\bauthor{\binits{A.K.} \bsnm{{Harding}}},
\bauthor{\binits{J.} \bsnm{{Grindlay}}},
\bauthor{\binits{J.S.} \bsnm{{Hong}}},
\batitle{{HX-POL -- A Balloon-Borne Hard X-Ray Polarimeter}}.
\bjtitle{IEEE Transactions on Nuclear Science}
\bvolume{56},
\bfpage{3607}--\blpage{3613}
(\byear{2009}).
doi:\doiurl{10.1109/TNS.2009.2034523}
\end{barticle}
\endbibitem

\bibitem[\protect\citeauthoryear{{Kreykenbohm} et~al.}{1999}]{Kreykenbohm99}
\begin{barticle}
\bauthor{\binits{I.} \bsnm{{Kreykenbohm}}},
\bauthor{\binits{P.} \bsnm{{Kretschmar}}},
\bauthor{\binits{J.} \bsnm{{Wilms}}},
\bauthor{\binits{R.} \bsnm{{Staubert}}},
\bauthor{\binits{E.} \bsnm{{Kendziorra}}},
\bauthor{\binits{D.E.} \bsnm{{Gruber}}},
\bauthor{\binits{W.A.} \bsnm{{Heindl}}},
\bauthor{\binits{R.E.} \bsnm{{Rothschild}}},
\batitle{{VELA X-1 as seen by RXTE}}.
\bjtitle{\aap}
\bvolume{341},
\bfpage{141}--\blpage{150}
(\byear{1999})
\end{barticle}
\endbibitem

\bibitem[\protect\citeauthoryear{{Kreykenbohm} et~al.}{2004}]{Kreykenbohm04}
\begin{barticle}
\bauthor{\binits{I.} \bsnm{{Kreykenbohm}}},
\bauthor{\binits{J.} \bsnm{{Wilms}}},
\bauthor{\binits{W.} \bsnm{{Coburn}}},
\bauthor{\binits{M.} \bsnm{{Kuster}}},
\bauthor{\binits{R.E.} \bsnm{{Rothschild}}},
\bauthor{\binits{W.A.} \bsnm{{Heindl}}},
\bauthor{\binits{P.} \bsnm{{Kretschmar}}},
\bauthor{\binits{R.} \bsnm{{Staubert}}},
\batitle{{The variable cyclotron line in GX 301-2}}.
\bjtitle{\aap}
\bvolume{427},
\bfpage{975}--\blpage{986}
(\byear{2004}).
doi:\doiurl{10.1051/0004-6361:20035836}
\end{barticle}
\endbibitem

\bibitem[\protect\citeauthoryear{{Krimm} et~al.}{2013}]{Krimm2013;swift}
\begin{barticle}
\bauthor{\binits{H.A.} \bsnm{{Krimm}}},
\bauthor{\binits{S.T.} \bsnm{{Holland}}},
\bauthor{\binits{R.H.D.} \bsnm{{Corbet}}},
\bauthor{\binits{A.B.} \bsnm{{Pearlman}}},
\bauthor{\binits{P.} \bsnm{{Romano}}},
\bauthor{\binits{J.A.} \bsnm{{Kennea}}},
\bauthor{\binits{J.S.} \bsnm{{Bloom}}},
\bauthor{\binits{S.D.} \bsnm{{Barthelmy}}},
\bauthor{\binits{W.H.} \bsnm{{Baumgartner}}},
\bauthor{\binits{J.R.} \bsnm{{Cummings}}},
\bauthor{\binits{N.} \bsnm{{Gehrels}}},
\bauthor{\binits{A.Y.} \bsnm{{Lien}}},
\bauthor{\binits{C.B.} \bsnm{{Markwardt}}},
\bauthor{\binits{D.M.} \bsnm{{Palmer}}},
\bauthor{\binits{T.} \bsnm{{Sakamoto}}},
\bauthor{\binits{M.} \bsnm{{Stamatikos}}},
\bauthor{\binits{T.N.} \bsnm{{Ukwatta}}},
\batitle{{The Swift/BAT Hard X-Ray Transient Monitor}}.
\bjtitle{\apjs}
\bvolume{209},
\bfpage{14}
(\byear{2013}).
doi:\doiurl{10.1088/0067-0049/209/1/14}
\end{barticle}
\endbibitem

\bibitem[\protect\citeauthoryear{{Krivonos} et~al.}{2014}]{Krivonos14}
\begin{barticle}
\bauthor{\binits{R.A.} \bsnm{{Krivonos}}},
\bauthor{\binits{J.A.} \bsnm{{Tomsick}}},
\bauthor{\binits{F.E.} \bsnm{{Bauer}}},
\bauthor{\binits{F.K.} \bsnm{{Baganoff}}},
\bauthor{\binits{N.M.} \bsnm{{Barriere}}},
\bauthor{\binits{A.} \bsnm{{Bodaghee}}},
\bauthor{\binits{S.E.} \bsnm{{Boggs}}},
\bauthor{\binits{F.E.} \bsnm{{Christensen}}},
\bauthor{\binits{W.W.} \bsnm{{Craig}}},
\bauthor{\binits{B.W.} \bsnm{{Grefenstette}}},
\bauthor{\binits{C.J.} \bsnm{{Hailey}}},
\bauthor{\binits{F.A.} \bsnm{{Harrison}}},
\bauthor{\binits{J.} \bsnm{{Hong}}},
\bauthor{\binits{K.K.} \bsnm{{Madsen}}},
\bauthor{\binits{K.} \bsnm{{Mori}}},
\bauthor{\binits{M.} \bsnm{{Nynka}}},
\bauthor{\binits{D.} \bsnm{{Stern}}},
\bauthor{\binits{W.W.} \bsnm{{Zhang}}},
\batitle{{First Hard X-Ray Detection of the Non-thermal Emission around the
  Arches Cluster: Morphology and Spectral Studies with NuSTAR}}.
\bjtitle{\apj}
\bvolume{781},
\bfpage{107}
(\byear{2014}).
doi:\doiurl{10.1088/0004-637X/781/2/107}
\end{barticle}
\endbibitem

\bibitem[\protect\citeauthoryear{{Krivonos} et~al.}{2015}]{Krivonos15}
\begin{barticle}
\bauthor{\binits{R.} \bsnm{{Krivonos}}},
\bauthor{\binits{S.} \bsnm{{Tsygankov}}},
\bauthor{\binits{A.} \bsnm{{Lutovinov}}},
\bauthor{\binits{M.} \bsnm{{Revnivtsev}}},
\bauthor{\binits{E.} \bsnm{{Churazov}}},
\bauthor{\binits{R.} \bsnm{{Sunyaev}}},
\batitle{{INTEGRAL 11-year hard X-ray survey above 100 keV}}.
\bjtitle{\mnras}
\bvolume{448},
\bfpage{3766}--\blpage{3774}
(\byear{2015}).
doi:\doiurl{10.1093/mnras/stv150}
\end{barticle}
\endbibitem

\bibitem[\protect\citeauthoryear{{Kuiper} et~al.}{2008}]{Kuiper08}
\begin{botherref}
\oauthor{\binits{L.} \bsnm{{Kuiper}}},
\oauthor{\binits{P.R.} \bsnm{{den Hartog}}},
\oauthor{\binits{W.} \bsnm{{Hermsen}}},
{Hard X-ray/soft gamma-ray Characteristics of the Persistent Emission from
  Magnetars}.
ArXiv e-prints
(2008)
\end{botherref}
\endbibitem

\bibitem[\protect\citeauthoryear{{Kuiper} et~al.}{2004}]{Kuiper04}
\begin{barticle}
\bauthor{\binits{L.} \bsnm{{Kuiper}}},
\bauthor{\binits{W.} \bsnm{{Hermsen}}},
\bauthor{\binits{M.} \bsnm{{Mendez}}},
\batitle{{Discovery of Hard Nonthermal Pulsed X-Ray Emission from the Anomalous
  X-Ray Pulsar 1E 1841-045}}.
\bjtitle{\apj}
\bvolume{613},
\bfpage{1173}--\blpage{1178}
(\byear{2004}).
doi:\doiurl{10.1086/423129}
\end{barticle}
\endbibitem

\bibitem[\protect\citeauthoryear{{Kuiper} et~al.}{1999}]{Kuiper99}
\begin{barticle}
\bauthor{\binits{L.} \bsnm{{Kuiper}}},
\bauthor{\binits{W.} \bsnm{{Hermsen}}},
\bauthor{\binits{J.M.} \bsnm{{Krijger}}},
\bauthor{\binits{K.} \bsnm{{Bennett}}},
\bauthor{\binits{A.} \bsnm{{Carrami{\~n}ana}}},
\bauthor{\binits{V.} \bsnm{{Sch{\"o}nfelder}}},
\bauthor{\binits{M.} \bsnm{{Bailes}}},
\bauthor{\binits{R.N.} \bsnm{{Manchester}}},
\batitle{{COMPTEL detection of pulsed gamma -ray emission from PSR B1509-58 up
  to at least 10 MeV}}.
\bjtitle{\aap}
\bvolume{351},
\bfpage{119}--\blpage{132}
(\byear{1999})
\end{barticle}
\endbibitem

\bibitem[\protect\citeauthoryear{{Kuiper} et~al.}{2006}]{Kuiper06}
\begin{barticle}
\bauthor{\binits{L.} \bsnm{{Kuiper}}},
\bauthor{\binits{W.} \bsnm{{Hermsen}}},
\bauthor{\binits{P.R.} \bsnm{{den Hartog}}},
\bauthor{\binits{W.} \bsnm{{Collmar}}},
\batitle{{Discovery of Luminous Pulsed Hard X-Ray Emission from Anomalous X-Ray
  Pulsars 1RXS J1708-4009, 4U 0142+61, and 1E 2259+586 by INTEGRAL and RXTE}}.
\bjtitle{\apj}
\bvolume{645},
\bfpage{556}--\blpage{575}
(\byear{2006}).
doi:\doiurl{10.1086/504317}
\end{barticle}
\endbibitem

\bibitem[\protect\citeauthoryear{{Lang} et~al.}{1981}]{Lang81}
\begin{barticle}
\bauthor{\binits{F.L.} \bsnm{{Lang}}},
\bauthor{\binits{A.M.} \bsnm{{Levine}}},
\bauthor{\binits{M.} \bsnm{{Bautz}}},
\bauthor{\binits{S.} \bsnm{{Hauskins}}},
\bauthor{\binits{S.} \bsnm{{Howe}}},
\bauthor{\binits{F.A.} \bsnm{{Primini}}},
\bauthor{\binits{W.H.G.} \bsnm{{Lewin}}},
\bauthor{\binits{W.A.} \bsnm{{Baity}}},
\bauthor{\binits{F.K.} \bsnm{{Knight}}},
\bauthor{\binits{R.E.} \bsnm{{Rotschild}}},
\bauthor{\binits{J.A.} \bsnm{{Petterson}}},
\batitle{{Discovery of a 30.5 day periodicity in LMC X-4}}.
\bjtitle{\apjl}
\bvolume{246},
\bfpage{21}--\blpage{25}
(\byear{1981}).
doi:\doiurl{10.1086/183545}
\end{barticle}
\endbibitem

\bibitem[\protect\citeauthoryear{{Laurent} et~al.}{2011a}]{Laurent11b}
\begin{barticle}
\bauthor{\binits{P.} \bsnm{{Laurent}}},
\bauthor{\binits{D.} \bsnm{{Gotz}}},
\bauthor{\binits{P.} \bsnm{{Binetruy}}},
\bauthor{\binits{S.} \bsnm{{Covino}}},
\bauthor{\binits{A.} \bsnm{{Fernandez-Soto}}},
\batitle{{Constraints on Lorentz Invariance Violation using integral/IBIS
  observations of GRB041219A}}.
\bjtitle{\prd}
\bvolume{83}(\bissue{12}),
\bfpage{121301}
(\byear{2011}a).
doi:\doiurl{10.1103/PhysRevD.83.121301}
\end{barticle}
\endbibitem

\bibitem[\protect\citeauthoryear{{Laurent} et~al.}{2011b}]{Laurent11}
\begin{barticle}
\bauthor{\binits{P.} \bsnm{{Laurent}}},
\bauthor{\binits{J.} \bsnm{{Rodriguez}}},
\bauthor{\binits{J.} \bsnm{{Wilms}}},
\bauthor{\binits{M.} \bsnm{{Cadolle Bel}}},
\bauthor{\binits{K.} \bsnm{{Pottschmidt}}},
\bauthor{\binits{V.} \bsnm{{Grinberg}}},
\batitle{{Polarized Gamma-Ray Emission from the Galactic Black Hole Cygnus
  X-1}}.
\bjtitle{Science}
\bvolume{332},
\bfpage{438}
(\byear{2011}b).
doi:\doiurl{10.1126/science.1200848}
\end{barticle}
\endbibitem

\bibitem[\protect\citeauthoryear{{Laurent} et~al.}{2014}]{Laurent14}
\begin{bchapter}
\bauthor{\binits{P.} \bsnm{{Laurent}}},
\bauthor{\binits{V.} \bsnm{{Tatischeff}}},
\bauthor{\binits{N.} \bsnm{{de Ser{\'e}ville}}},
\bauthor{\binits{O.} \bsnm{{Limousin}}},
\bauthor{\binits{W.} \bsnm{{Bertoli}}},
\bauthor{\binits{E.} \bsnm{{Br{\'e}elle}}},
\bauthor{\binits{Y.} \bsnm{{Dolgorouky}}},
\bauthor{\binits{A.} \bsnm{{Gostojic}}},
\bauthor{\binits{C.} \bsnm{{Hamadache}}},
\bauthor{\binits{M.} \bsnm{{Khalil}}},
\bauthor{\binits{J.} \bsnm{{Kiener}}},
\bctitle{{PACT: a sensitive 100 keV-10 MeV all sky pairs and Compton
  telescope}},
in \bbtitle{Space Telescopes and Instrumentation 2014: Ultraviolet to Gamma
  Ray}.
\bsertitle{\procspie},
vol. \bseriesno{9144},
\byear{2014},
p. \bfpage{91440}.
doi:\doiurl{10.1117/12.2055898}
\end{bchapter}
\endbibitem

\bibitem[\protect\citeauthoryear{{Lawson} et~al.}{1999}]{Lawson99}
\begin{barticle}
\bauthor{\binits{A.J.} \bsnm{{Lawson}}},
\bauthor{\binits{I.M.} \bsnm{{McHardy}}},
\bauthor{\binits{A.M.} \bsnm{{Newsam}}},
\batitle{{RXTE monitoring of the blazars 3C 279 and 3C 273}}.
\bjtitle{Nuclear Physics B Proceedings Supplements}
\bvolume{69},
\bfpage{439}--\blpage{444}
(\byear{1999}).
doi:\doiurl{10.1016/S0920-5632(98)00258-8}
\end{barticle}
\endbibitem

\bibitem[\protect\citeauthoryear{{Lazzati} et~al.}{2001}]{Lazzati01}
\begin{barticle}
\bauthor{\binits{D.} \bsnm{{Lazzati}}},
\bauthor{\binits{E.} \bsnm{{Ramirez-Ruiz}}},
\bauthor{\binits{G.} \bsnm{{Ghisellini}}},
\batitle{{Possible detection of hard X-ray afterglows of short gamma -ray
  bursts}}.
\bjtitle{\aap}
\bvolume{379},
\bfpage{39}--\blpage{43}
(\byear{2001}).
doi:\doiurl{10.1051/0004-6361:20011485}
\end{barticle}
\endbibitem

\bibitem[\protect\citeauthoryear{{Leahy} et~al.}{1989}]{Leahy89}
\begin{barticle}
\bauthor{\binits{D.A.} \bsnm{{Leahy}}},
\bauthor{\binits{M.} \bsnm{{Matsuoka}}},
\bauthor{\binits{N.} \bsnm{{Kawai}}},
\bauthor{\binits{F.} \bsnm{{Makino}}},
\batitle{{TENMA observations of iron line emission from GX301-2}}.
\bjtitle{\mnras}
\bvolume{237},
\bfpage{269}--\blpage{276}
(\byear{1989}).
doi:\doiurl{10.1093/mnras/237.1.269}
\end{barticle}
\endbibitem

\bibitem[\protect\citeauthoryear{{Lebrun} et~al.}{2003}]{Lebrun2003;integral}
\begin{barticle}
\bauthor{\binits{F.} \bsnm{{Lebrun}}},
\bauthor{\binits{J.P.} \bsnm{{Leray}}},
\bauthor{\binits{P.} \bsnm{{Lavocat}}},
\bauthor{\binits{J.} \bsnm{{Cr{\'e}tolle}}},
\bauthor{\binits{M.} \bsnm{{Arqu{\`e}s}}},
\bauthor{\binits{C.} \bsnm{{Blondel}}},
\bauthor{\binits{C.} \bsnm{{Bonnin}}},
\bauthor{\binits{A.} \bsnm{{Bou{\`e}re}}},
\bauthor{\binits{C.} \bsnm{{Cara}}},
\bauthor{\binits{T.} \bsnm{{Chaleil}}},
\bauthor{\binits{F.} \bsnm{{Daly}}},
\bauthor{\binits{F.} \bsnm{{Desages}}},
\bauthor{\binits{H.} \bsnm{{Dzitko}}},
\bauthor{\binits{B.} \bsnm{{Horeau}}},
\bauthor{\binits{P.} \bsnm{{Laurent}}},
\bauthor{\binits{O.} \bsnm{{Limousin}}},
\bauthor{\binits{F.} \bsnm{{Mathy}}},
\bauthor{\binits{V.} \bsnm{{Mauguen}}},
\bauthor{\binits{F.} \bsnm{{Meignier}}},
\bauthor{\binits{F.} \bsnm{{Molini{\'e}}}},
\bauthor{\binits{E.} \bsnm{{Poindron}}},
\bauthor{\binits{M.} \bsnm{{Rouger}}},
\bauthor{\binits{A.} \bsnm{{Sauvageon}}},
\bauthor{\binits{T.} \bsnm{{Tourrette}}},
\batitle{{ISGRI: The INTEGRAL Soft Gamma-Ray Imager}}.
\bjtitle{\aap}
\bvolume{411},
\bfpage{141}--\blpage{148}
(\byear{2003}).
doi:\doiurl{10.1051/0004-6361:20031367}
\end{barticle}
\endbibitem

\bibitem[\protect\citeauthoryear{{Lee} et~al.}{1999}]{Lee99}
\begin{barticle}
\bauthor{\binits{J.C.} \bsnm{{Lee}}},
\bauthor{\binits{A.C.} \bsnm{{Fabian}}},
\bauthor{\binits{W.N.} \bsnm{{Brandt}}},
\bauthor{\binits{C.S.} \bsnm{{Reynolds}}},
\bauthor{\binits{K.} \bsnm{{Iwasawa}}},
\batitle{{First constraints on iron abundance versus reflection fraction from
  the Seyfert 1 galaxy MCG-6-30-15}}.
\bjtitle{\mnras}
\bvolume{310},
\bfpage{973}--\blpage{981}
(\byear{1999}).
doi:\doiurl{10.1046/j.1365-8711.1999.02999.x}
\end{barticle}
\endbibitem

\bibitem[\protect\citeauthoryear{{Leventhal}}{1973}]{Leventhal73}
\begin{barticle}
\bauthor{\binits{M.} \bsnm{{Leventhal}}},
\batitle{{Positronium-Formation Redshift of the 511 keV Annihilation Line}}.
\bjtitle{\apjl}
\bvolume{183},
\bfpage{147}
(\byear{1973}).
doi:\doiurl{10.1086/181274}
\end{barticle}
\endbibitem

\bibitem[\protect\citeauthoryear{{Leventhal} et~al.}{1977}]{Leventhal77b}
\begin{barticle}
\bauthor{\binits{M.} \bsnm{{Leventhal}}},
\bauthor{\binits{C.} \bsnm{{MacCallum}}},
\bauthor{\binits{A.} \bsnm{{Watts}}},
\batitle{{A search for gamma-ray lines from Nova Cygni 1975, Nova Serpentis
  1970, and the Crab Nebula}}.
\bjtitle{\apj}
\bvolume{216},
\bfpage{491}--\blpage{502}
(\byear{1977}).
doi:\doiurl{10.1086/155490}
\end{barticle}
\endbibitem

\bibitem[\protect\citeauthoryear{{Leventhal} et~al.}{1978}]{Leventhal78}
\begin{barticle}
\bauthor{\binits{M.} \bsnm{{Leventhal}}},
\bauthor{\binits{C.J.} \bsnm{{MacCallum}}},
\bauthor{\binits{P.D.} \bsnm{{Stang}}},
\batitle{{Detection of 511 keV positron annihilation radiation from the
  galactic center direction}}.
\bjtitle{\apjl}
\bvolume{225},
\bfpage{11}--\blpage{14}
(\byear{1978}).
doi:\doiurl{10.1086/182782}
\end{barticle}
\endbibitem

\bibitem[\protect\citeauthoryear{{Leventhal} et~al.}{1977}]{Leventhal77a}
\begin{barticle}
\bauthor{\binits{M.} \bsnm{{Leventhal}}},
\bauthor{\binits{C.J.} \bsnm{{MacCallum}}},
\bauthor{\binits{A.C.} \bsnm{{Watts}}},
\batitle{{Possible gamma-ray line from the Crab Nebula}}.
\bjtitle{\nat}
\bvolume{266},
\bfpage{696}--\blpage{698}
(\byear{1977}).
doi:\doiurl{10.1038/266696a0}
\end{barticle}
\endbibitem

\bibitem[\protect\citeauthoryear{{Leventhal} et~al.}{1980}]{Leventhal80}
\begin{barticle}
\bauthor{\binits{M.} \bsnm{{Leventhal}}},
\bauthor{\binits{C.J.} \bsnm{{MacCallum}}},
\bauthor{\binits{A.F.} \bsnm{{Huters}}},
\bauthor{\binits{P.D.} \bsnm{{Stang}}},
\batitle{{Gamma-ray lines and continuum radiation from the galactic center
  direction}}.
\bjtitle{\apj}
\bvolume{240},
\bfpage{338}--\blpage{343}
(\byear{1980}).
doi:\doiurl{10.1086/158237}
\end{barticle}
\endbibitem

\bibitem[\protect\citeauthoryear{{Leventhal} et~al.}{1982}]{Leventhal82}
\begin{barticle}
\bauthor{\binits{M.} \bsnm{{Leventhal}}},
\bauthor{\binits{C.J.} \bsnm{{MacCallum}}},
\bauthor{\binits{A.F.} \bsnm{{Huters}}},
\bauthor{\binits{P.D.} \bsnm{{Stang}}},
\batitle{{Time-variable positron annihilation radiation from the galactic
  center direction}}.
\bjtitle{\apjl}
\bvolume{260},
\bfpage{1}--\blpage{5}
(\byear{1982}).
doi:\doiurl{10.1086/183858}
\end{barticle}
\endbibitem

\bibitem[\protect\citeauthoryear{{Leventhal} et~al.}{1986}]{Leventhal86}
\begin{barticle}
\bauthor{\binits{M.} \bsnm{{Leventhal}}},
\bauthor{\binits{C.J.} \bsnm{{MacCallum}}},
\bauthor{\binits{A.F.} \bsnm{{Huters}}},
\bauthor{\binits{P.D.} \bsnm{{Stang}}},
\batitle{{Current status of the galactic center positron-annihilation source}}.
\bjtitle{\apj}
\bvolume{302},
\bfpage{459}--\blpage{461}
(\byear{1986}).
doi:\doiurl{10.1086/164004}
\end{barticle}
\endbibitem

\bibitem[\protect\citeauthoryear{{Leventhal} et~al.}{1993}]{Leventhal93}
\begin{barticle}
\bauthor{\binits{M.} \bsnm{{Leventhal}}},
\bauthor{\binits{S.D.} \bsnm{{Barthelmy}}},
\bauthor{\binits{N.} \bsnm{{Gehrels}}},
\bauthor{\binits{B.J.} \bsnm{{Teegarden}}},
\bauthor{\binits{J.} \bsnm{{Tueller}}},
\bauthor{\binits{L.M.} \bsnm{{Bartlett}}},
\batitle{{GRIS detections of the 511 keV line from the Galactic center region
  in 1992}}.
\bjtitle{\apjl}
\bvolume{405},
\bfpage{25}--\blpage{28}
(\byear{1993}).
doi:\doiurl{10.1086/186757}
\end{barticle}
\endbibitem

\bibitem[\protect\citeauthoryear{{Levine} et~al.}{1984}]{Levine84}
\begin{barticle}
\bauthor{\binits{A.M.} \bsnm{{Levine}}},
\bauthor{\binits{F.L.} \bsnm{{Lang}}},
\bauthor{\binits{W.H.G.} \bsnm{{Lewin}}},
\bauthor{\binits{F.A.} \bsnm{{Primini}}},
\bauthor{\binits{C.A.} \bsnm{{Dobson}}},
\bauthor{\binits{J.P.} \bsnm{{Doty}}},
\bauthor{\binits{J.A.} \bsnm{{Hoffman}}},
\bauthor{\binits{S.K.} \bsnm{{Howe}}},
\bauthor{\binits{A.} \bsnm{{Scheepmaker}}},
\bauthor{\binits{W.A.} \bsnm{{Wheaton}}},
\bauthor{\binits{J.L.} \bsnm{{Matteson}}},
\bauthor{\binits{W.A.} \bsnm{{Baity}}},
\bauthor{\binits{D.E.} \bsnm{{Gruber}}},
\bauthor{\binits{F.K.} \bsnm{{Knight}}},
\bauthor{\binits{P.L.} \bsnm{{Nolan}}},
\bauthor{\binits{R.M.} \bsnm{{Pelling}}},
\bauthor{\binits{R.E.} \bsnm{{Rothschild}}},
\bauthor{\binits{L.E.} \bsnm{{Peterson}}},
\batitle{{The HEAO 1 A-4 catalog of high-energy X-ray sources}}.
\bjtitle{\apjs}
\bvolume{54},
\bfpage{581}--\blpage{617}
(\byear{1984}).
doi:\doiurl{10.1086/190944}
\end{barticle}
\endbibitem

\bibitem[\protect\citeauthoryear{{Lewin} et~al.}{1967}]{Lewin67}
\begin{barticle}
\bauthor{\binits{W.H.G.} \bsnm{{Lewin}}},
\bauthor{\binits{G.W.} \bsnm{{Clark}}},
\bauthor{\binits{W.B.} \bsnm{{Smith}}},
\batitle{{Spectral Data on SCO X-1 in the Energy Range from 20 TO 100 KEV}}.
\bjtitle{\apjl}
\bvolume{150},
\bfpage{153}
(\byear{1967}).
doi:\doiurl{10.1086/180115}
\end{barticle}
\endbibitem

\bibitem[\protect\citeauthoryear{{Li} et~al.}{2014}]{Li14}
\begin{barticle}
\bauthor{\binits{K.L.} \bsnm{{Li}}},
\bauthor{\binits{A.K.H.} \bsnm{{Kong}}},
\bauthor{\binits{J.} \bsnm{{Takata}}},
\bauthor{\binits{K.S.} \bsnm{{Cheng}}},
\bauthor{\binits{P.H.T.} \bsnm{{Tam}}},
\bauthor{\binits{C.Y.} \bsnm{{Hui}}},
\bauthor{\binits{R.} \bsnm{{Jin}}},
\batitle{{NuSTAR Observations and Broadband Spectral Energy Distribution
  Modeling of the Millisecond Pulsar Binary PSR J1023+0038}}.
\bjtitle{\apj}
\bvolume{797},
\bfpage{111}
(\byear{2014}).
doi:\doiurl{10.1088/0004-637X/797/2/111}
\end{barticle}
\endbibitem

\bibitem[\protect\citeauthoryear{{Li}}{2007}]{li07}
\begin{barticle}
\bauthor{\binits{T.-P.} \bsnm{{Li}}},
\batitle{{HXMT: A Chinese High-Energy Astrophysics Mission}}.
\bjtitle{Nuclear Physics B Proceedings Supplements}
\bvolume{166},
\bfpage{131}--\blpage{139}
(\byear{2007}).
doi:\doiurl{10.1016/j.nuclphysbps.2006.12.070}
\end{barticle}
\endbibitem

\bibitem[\protect\citeauthoryear{{Lin} et~al.}{2000}]{Lin00}
\begin{barticle}
\bauthor{\binits{D.} \bsnm{{Lin}}},
\bauthor{\binits{I.A.} \bsnm{{Smith}}},
\bauthor{\binits{M.} \bsnm{{B{\"o}ttcher}}},
\bauthor{\binits{E.P.} \bsnm{{Liang}}},
\batitle{{The Energy Dependence of the Aperiodic Variability for Cygnus X-1, GX
  339-4, GRS 1758-258, and 1E 1740.7-2942}}.
\bjtitle{\apj}
\bvolume{531},
\bfpage{963}--\blpage{970}
(\byear{2000}).
doi:\doiurl{10.1086/308524}
\end{barticle}
\endbibitem

\bibitem[\protect\citeauthoryear{{Linares} et~al.}{2012}]{Linares12}
\begin{barticle}
\bauthor{\binits{M.} \bsnm{{Linares}}},
\bauthor{\binits{V.} \bsnm{{Connaughton}}},
\bauthor{\binits{P.} \bsnm{{Jenke}}},
\bauthor{\binits{A.J.} \bsnm{{van der Horst}}},
\bauthor{\binits{A.} \bsnm{{Camero-Arranz}}},
\bauthor{\binits{C.} \bsnm{{Kouveliotou}}},
\bauthor{\binits{D.} \bsnm{{Chakrabarty}}},
\bauthor{\binits{E.} \bsnm{{Beklen}}},
\bauthor{\binits{P.N.} \bsnm{{Bhat}}},
\bauthor{\binits{M.S.} \bsnm{{Briggs}}},
\bauthor{\binits{M.} \bsnm{{Finger}}},
\bauthor{\binits{W.S.} \bsnm{{Paciesas}}},
\bauthor{\binits{R.} \bsnm{{Preece}}},
\bauthor{\binits{A.} \bsnm{{von Kienlin}}},
\bauthor{\binits{C.A.} \bsnm{{Wilson-Hodge}}},
\batitle{{The Fermi-GBM X-Ray Burst Monitor: Thermonuclear Bursts from 4U
  0614+09}}.
\bjtitle{\apj}
\bvolume{760},
\bfpage{133}
(\byear{2012}).
doi:\doiurl{10.1088/0004-637X/760/2/133}
\end{barticle}
\endbibitem

\bibitem[\protect\citeauthoryear{{Ling} et~al.}{1979}]{Ling79}
\begin{barticle}
\bauthor{\binits{J.C.} \bsnm{{Ling}}},
\bauthor{\binits{W.A.} \bsnm{{Mahoney}}},
\bauthor{\binits{J.B.} \bsnm{{Willett}}},
\bauthor{\binits{A.S.} \bsnm{{Jacobson}}},
\batitle{{A possible line feature at 73 keV from the Crab Nebula}}.
\bjtitle{\apj}
\bvolume{231},
\bfpage{896}--\blpage{905}
(\byear{1979}).
doi:\doiurl{10.1086/157252}
\end{barticle}
\endbibitem

\bibitem[\protect\citeauthoryear{{Lingenfelter} and
  {Ramaty}}{1976}]{Lingenfelter76}
\begin{barticle}
\bauthor{\binits{R.E.} \bsnm{{Lingenfelter}}},
\bauthor{\binits{R.} \bsnm{{Ramaty}}},
\batitle{{Gamma ray lines from interstellar grains}}.
\bjtitle{NASA STI/Recon Technical Report N}
\bvolume{76},
\bfpage{33117}
(\byear{1976})
\end{barticle}
\endbibitem

\bibitem[\protect\citeauthoryear{{Lingenfelter} and
  {Ramaty}}{1989}]{Lingenfelter89}
\begin{barticle}
\bauthor{\binits{R.E.} \bsnm{{Lingenfelter}}},
\bauthor{\binits{R.} \bsnm{{Ramaty}}},
\batitle{{The nature of the annihilation radiation and gamma-ray continuum from
  the Galactic center region}}.
\bjtitle{\apj}
\bvolume{343},
\bfpage{686}--\blpage{695}
(\byear{1989}).
doi:\doiurl{10.1086/167740}
\end{barticle}
\endbibitem

\bibitem[\protect\citeauthoryear{{Maccarone} and {Coppi}}{2003}]{Maccarone03}
\begin{barticle}
\bauthor{\binits{T.J.} \bsnm{{Maccarone}}},
\bauthor{\binits{P.S.} \bsnm{{Coppi}}},
\batitle{{Hysteresis in the light curves of soft X-ray transients}}.
\bjtitle{\mnras}
\bvolume{338},
\bfpage{189}--\blpage{196}
(\byear{2003}).
doi:\doiurl{10.1046/j.1365-8711.2003.06040.x}
\end{barticle}
\endbibitem

\bibitem[\protect\citeauthoryear{{Madejski} et~al.}{1999}]{Madejski99}
\begin{barticle}
\bauthor{\binits{G.M.} \bsnm{{Madejski}}},
\bauthor{\binits{M.} \bsnm{{Sikora}}},
\bauthor{\binits{T.} \bsnm{{Jaffe}}},
\bauthor{\binits{M.} \bsnm{{B{\L}a{\.z}ejowski}}},
\bauthor{\binits{K.} \bsnm{{Jahoda}}},
\bauthor{\binits{R.} \bsnm{{Moderski}}},
\batitle{{X-Ray Observations of BL Lacertae during the 1997 Outburst and
  Association with Quasar-like Characteristics}}.
\bjtitle{\apj}
\bvolume{521},
\bfpage{145}--\blpage{154}
(\byear{1999}).
doi:\doiurl{10.1086/307524}
\end{barticle}
\endbibitem

\bibitem[\protect\citeauthoryear{{Mahoney} et~al.}{1984}]{Mahoney84b}
\begin{barticle}
\bauthor{\binits{W.A.} \bsnm{{Mahoney}}},
\bauthor{\binits{J.C.} \bsnm{{Ling}}},
\bauthor{\binits{A.S.} \bsnm{{Jacobson}}},
\batitle{{HEAO 3 observations of the Crab pulsar}}.
\bjtitle{\apj}
\bvolume{278},
\bfpage{784}--\blpage{790}
(\byear{1984}).
doi:\doiurl{10.1086/161848}
\end{barticle}
\endbibitem

\bibitem[\protect\citeauthoryear{{Mahoney} et~al.}{1994}]{Mahoney94}
\begin{barticle}
\bauthor{\binits{W.A.} \bsnm{{Mahoney}}},
\bauthor{\binits{J.C.} \bsnm{{Ling}}},
\bauthor{\binits{W.A.} \bsnm{{Wheaton}}},
\batitle{{HEAO 3 observations of the Galactic center 511 keV line}}.
\bjtitle{\apjs}
\bvolume{92},
\bfpage{387}--\blpage{391}
(\byear{1994}).
doi:\doiurl{10.1086/191983}
\end{barticle}
\endbibitem

\bibitem[\protect\citeauthoryear{{Mahoney} et~al.}{1980}]{Mahoney80}
\begin{barticle}
\bauthor{\binits{W.A.} \bsnm{{Mahoney}}},
\bauthor{\binits{J.C.} \bsnm{{Ling}}},
\bauthor{\binits{A.S.} \bsnm{{Jacobson}}},
\bauthor{\binits{R.M.} \bsnm{{Tapphorn}}},
\batitle{{The HEAO 3 gamma-ray spectrometer}}.
\bjtitle{Nuclear Instruments and Methods}
\bvolume{178},
\bfpage{363}--\blpage{381}
(\byear{1980}).
doi:\doiurl{10.1016/0029-554X(80)90815-0}
\end{barticle}
\endbibitem

\bibitem[\protect\citeauthoryear{{Maiorano} et~al.}{2005}]{Maiorano05}
\begin{barticle}
\bauthor{\binits{E.} \bsnm{{Maiorano}}},
\bauthor{\binits{N.} \bsnm{{Masetti}}},
\bauthor{\binits{E.} \bsnm{{Palazzi}}},
\bauthor{\binits{F.} \bsnm{{Frontera}}},
\bauthor{\binits{P.} \bsnm{{Grandi}}},
\bauthor{\binits{E.} \bsnm{{Pian}}},
\bauthor{\binits{L.} \bsnm{{Amati}}},
\bauthor{\binits{L.} \bsnm{{Nicastro}}},
\bauthor{\binits{P.} \bsnm{{Soffitta}}},
\bauthor{\binits{C.} \bsnm{{Guidorzi}}},
\bauthor{\binits{R.} \bsnm{{Landi}}},
\bauthor{\binits{E.} \bsnm{{Montanari}}},
\bauthor{\binits{M.} \bsnm{{Orlandini}}},
\bauthor{\binits{A.} \bsnm{{Corsi}}},
\bauthor{\binits{L.} \bsnm{{Piro}}},
\bauthor{\binits{L.A.} \bsnm{{Antonelli}}},
\bauthor{\binits{E.} \bsnm{{Costa}}},
\bauthor{\binits{M.} \bsnm{{Feroci}}},
\bauthor{\binits{J.} \bsnm{{Heise}}},
\bauthor{\binits{E.} \bsnm{{Kuulkers}}},
\bauthor{\binits{J.J.M.} \bsnm{{in't Zand}}},
\batitle{{The puzzling case of GRB 990123: multiwavelength afterglow study}}.
\bjtitle{\aap}
\bvolume{438},
\bfpage{821}--\blpage{827}
(\byear{2005}).
doi:\doiurl{10.1051/0004-6361:20042534}
\end{barticle}
\endbibitem

\bibitem[\protect\citeauthoryear{{Maisack} et~al.}{1993}]{Maisack93}
\begin{barticle}
\bauthor{\binits{M.} \bsnm{{Maisack}}},
\bauthor{\binits{W.N.} \bsnm{{Johnson}}},
\bauthor{\binits{R.L.} \bsnm{{Kinzer}}},
\bauthor{\binits{M.S.} \bsnm{{Strickman}}},
\bauthor{\binits{J.D.} \bsnm{{Kurfess}}},
\bauthor{\binits{R.A.} \bsnm{{Cameron}}},
\bauthor{\binits{G.V.} \bsnm{{Jung}}},
\bauthor{\binits{D.A.} \bsnm{{Grabelsky}}},
\bauthor{\binits{W.R.} \bsnm{{Purcell}}},
\bauthor{\binits{M.P.} \bsnm{{Ulmer}}},
\batitle{{OSSE observations of NGC 4151}}.
\bjtitle{\apjl}
\bvolume{407},
\bfpage{61}--\blpage{64}
(\byear{1993}).
doi:\doiurl{10.1086/186806}
\end{barticle}
\endbibitem

\bibitem[\protect\citeauthoryear{{Makino}}{1987}]{Makino87}
\begin{barticle}
\bauthor{\binits{F.} \bsnm{{Makino}}},
\batitle{{The X-ray astronomy satellite Astro-C}}.
\bjtitle{\aplett}
\bvolume{25},
\bfpage{223}--\blpage{233}
(\byear{1987})
\end{barticle}
\endbibitem

\bibitem[\protect\citeauthoryear{{Malaguti} et~al.}{1998}]{Malaguti98}
\begin{barticle}
\bauthor{\binits{G.} \bsnm{{Malaguti}}},
\bauthor{\binits{G.G.C.} \bsnm{{Palumbo}}},
\bauthor{\binits{M.} \bsnm{{Cappi}}},
\bauthor{\binits{A.} \bsnm{{Comastri}}},
\bauthor{\binits{C.} \bsnm{{Otani}}},
\bauthor{\binits{M.} \bsnm{{Matsuoka}}},
\bauthor{\binits{M.} \bsnm{{Guainazzi}}},
\bauthor{\binits{L.} \bsnm{{Bassani}}},
\bauthor{\binits{F.} \bsnm{{Frontera}}},
\batitle{{BeppoSAX observation of NGC 7674: a new reflection-dominated Seyfert
  2 galaxy}}.
\bjtitle{\aap}
\bvolume{331},
\bfpage{519}--\blpage{523}
(\byear{1998})
\end{barticle}
\endbibitem

\bibitem[\protect\citeauthoryear{{Malaguti} et~al.}{1999}]{Malaguti99}
\begin{barticle}
\bauthor{\binits{G.} \bsnm{{Malaguti}}},
\bauthor{\binits{L.} \bsnm{{Bassani}}},
\bauthor{\binits{M.} \bsnm{{Cappi}}},
\bauthor{\binits{A.} \bsnm{{Comastri}}},
\bauthor{\binits{G.} \bsnm{{Di Cocco}}},
\bauthor{\binits{A.C.} \bsnm{{Fabian}}},
\bauthor{\binits{G.G.C.} \bsnm{{Palumbo}}},
\bauthor{\binits{T.} \bsnm{{Maccacaro}}},
\bauthor{\binits{R.} \bsnm{{Maiolino}}},
\bauthor{\binits{P.} \bsnm{{Blanco}}},
\bauthor{\binits{M.} \bsnm{{Dadina}}},
\bauthor{\binits{D.} \bsnm{{dal Fiume}}},
\bauthor{\binits{F.} \bsnm{{Frontera}}},
\bauthor{\binits{M.} \bsnm{{Trifoglio}}},
\batitle{{BeppoSAX uncovers the hidden Seyfert 1 nucleus in the Seyfert 2
  galaxy NGC 2110}}.
\bjtitle{\aap}
\bvolume{342},
\bfpage{41}--\blpage{44}
(\byear{1999})
\end{barticle}
\endbibitem

\bibitem[\protect\citeauthoryear{{Malizia} et~al.}{2000}]{Malizia00}
\begin{barticle}
\bauthor{\binits{A.} \bsnm{{Malizia}}},
\bauthor{\binits{M.} \bsnm{{Capalbi}}},
\bauthor{\binits{F.} \bsnm{{Fiore}}},
\bauthor{\binits{P.} \bsnm{{Giommi}}},
\bauthor{\binits{G.} \bsnm{{Gandolfi}}},
\bauthor{\binits{A.} \bsnm{{Tesseri}}},
\bauthor{\binits{L.A.} \bsnm{{Antonelli}}},
\bauthor{\binits{R.C.} \bsnm{{Butler}}},
\bauthor{\binits{G.} \bsnm{{Celidonio}}},
\bauthor{\binits{A.} \bsnm{{Coletta}}},
\bauthor{\binits{L.} \bsnm{{Di Ciolo}}},
\bauthor{\binits{J.M.} \bsnm{{Muller}}},
\bauthor{\binits{L.} \bsnm{{Piro}}},
\bauthor{\binits{S.} \bsnm{{Rebecchi}}},
\bauthor{\binits{D.} \bsnm{{Ricci}}},
\bauthor{\binits{R.} \bsnm{{Ricci}}},
\bauthor{\binits{M.} \bsnm{{Smith}}},
\bauthor{\binits{V.} \bsnm{{Torroni}}},
\batitle{{The 0.1-100keV spectrum and variability of Mrk 421 in a high state}}.
\bjtitle{\mnras}
\bvolume{312},
\bfpage{123}--\blpage{129}
(\byear{2000}).
doi:\doiurl{10.1046/j.1365-8711.2000.03106.x}
\end{barticle}
\endbibitem

\bibitem[\protect\citeauthoryear{{Manchanda} et~al.}{1982}]{Manchanda82}
\begin{barticle}
\bauthor{\binits{R.K.} \bsnm{{Manchanda}}},
\bauthor{\binits{A.} \bsnm{{Bazzano}}},
\bauthor{\binits{C.D.} \bsnm{{La Padula}}},
\bauthor{\binits{V.F.} \bsnm{{Polcaro}}},
\bauthor{\binits{P.} \bsnm{{Ubertini}}},
\batitle{{Line feature around 73 keV from the Crab Nebula}}.
\bjtitle{\apj}
\bvolume{252},
\bfpage{172}--\blpage{178}
(\byear{1982}).
doi:\doiurl{10.1086/159543}
\end{barticle}
\endbibitem

\bibitem[\protect\citeauthoryear{{Mandrou} et~al.}{1994}]{Mandrou94}
\begin{barticle}
\bauthor{\binits{P.} \bsnm{{Mandrou}}},
\bauthor{\binits{J.P.} \bsnm{{Roques}}},
\bauthor{\binits{L.} \bsnm{{Bouchet}}},
\bauthor{\binits{M.} \bsnm{{Niel}}},
\bauthor{\binits{J.} \bsnm{{Paul}}},
\bauthor{\binits{J.P.} \bsnm{{Leray}}},
\bauthor{\binits{F.} \bsnm{{Lebrun}}},
\bauthor{\binits{J.} \bsnm{{Ballet}}},
\bauthor{\binits{E.} \bsnm{{Churazov}}},
\bauthor{\binits{M.} \bsnm{{Gilfanov}}},
\bauthor{\binits{R.} \bsnm{{Sunyaev}}},
\bauthor{\binits{B.} \bsnm{{Novikov}}},
\bauthor{\binits{N.} \bsnm{{Khavenson}}},
\bauthor{\binits{N.} \bsnm{{Kuleshova}}},
\bauthor{\binits{A.} \bsnm{{Sheikhet}}},
\bauthor{\binits{I.} \bsnm{{Tserenin}}},
\batitle{{Review of 3 years of observations with the low-energy gamma-ray
  telescope SIGMA onboard GRANAT}}.
\bjtitle{\apjs}
\bvolume{92},
\bfpage{343}--\blpage{349}
(\byear{1994}).
doi:\doiurl{10.1086/191977}
\end{barticle}
\endbibitem

\bibitem[\protect\citeauthoryear{{Manzo} et~al.}{1997}]{Manzo1997;bepposax}
\begin{barticle}
\bauthor{\binits{G.} \bsnm{{Manzo}}},
\bauthor{\binits{S.} \bsnm{{Giarrusso}}},
\bauthor{\binits{A.} \bsnm{{Santangelo}}},
\bauthor{\binits{F.} \bsnm{{Ciralli}}},
\bauthor{\binits{G.} \bsnm{{Fazio}}},
\bauthor{\binits{S.} \bsnm{{Piraino}}},
\bauthor{\binits{A.} \bsnm{{Segreto}}},
\batitle{{The high pressure gas scintillation proportional counter on-board the
  BeppoSAX X-ray astronomy satellite}}.
\bjtitle{\aaps}
\bvolume{122},
\bfpage{341}--\blpage{356}
(\byear{1997}).
doi:\doiurl{10.1051/aas:1997139}
\end{barticle}
\endbibitem

\bibitem[\protect\citeauthoryear{{Margutti} et~al.}{2016}]{Margutti16}
\begin{botherref}
\oauthor{\binits{R.} \bsnm{{Margutti}}},
\oauthor{\binits{A.} \bsnm{{Kamble}}},
\oauthor{\binits{D.} \bsnm{{Milisavljevic}}},
\oauthor{\binits{S.} \bsnm{{De Mink}}},
\oauthor{\binits{E.} \bsnm{{Zapartas}}},
\oauthor{\binits{M.} \bsnm{{Drout}}},
\oauthor{\binits{R.} \bsnm{{Chornock}}},
\oauthor{\binits{G.} \bsnm{{Risaliti}}},
\oauthor{\binits{B.A.} \bsnm{{Zauderer}}},
\oauthor{\binits{M.} \bsnm{{Bietenholz}}},
\oauthor{\binits{M.} \bsnm{{Cantiello}}},
\oauthor{\binits{S.} \bsnm{{Chakraborti}}},
\oauthor{\binits{L.} \bsnm{{Chomiuk}}},
\oauthor{\binits{W.} \bsnm{{Fong}}},
\oauthor{\binits{B.} \bsnm{{Grefenstette}}},
\oauthor{\binits{C.} \bsnm{{Guidorzi}}},
\oauthor{\binits{R.} \bsnm{{Kirshner}}},
\oauthor{\binits{J.T.} \bsnm{{Parrent}}},
\oauthor{\binits{D.} \bsnm{{Patnaude}}},
\oauthor{\binits{A.M.} \bsnm{{Soderberg}}},
\oauthor{\binits{N.C.} \bsnm{{Gehrels}}},
\oauthor{\binits{F.} \bsnm{{Harrison}}},
{Ejection of the massive Hydrogen-rich envelope timed with the collapse of the
  stripped SN2014C}.
ArXiv e-prints
(2016)
\end{botherref}
\endbibitem

\bibitem[\protect\citeauthoryear{{Markert} et~al.}{1979}]{Markert1979;oso7}
\begin{barticle}
\bauthor{\binits{T.H.} \bsnm{{Markert}}},
\bauthor{\binits{F.N.} \bsnm{{Laird}}},
\bauthor{\binits{G.W.} \bsnm{{Clark}}},
\bauthor{\binits{D.R.} \bsnm{{Hearn}}},
\bauthor{\binits{G.F.} \bsnm{{Sprott}}},
\bauthor{\binits{F.K.} \bsnm{{Li}}},
\bauthor{\binits{H.V.} \bsnm{{Bradt}}},
\bauthor{\binits{W.H.G.} \bsnm{{Lewin}}},
\bauthor{\binits{H.W.} \bsnm{{Schnopper}}},
\bauthor{\binits{P.F.} \bsnm{{Winkler}}},
\batitle{{The MIT/OSO 7 catalog of X-ray sources - Intensities, spectra, and
  long-term variability}}.
\bjtitle{\apjs}
\bvolume{39},
\bfpage{573}--\blpage{632}
(\byear{1979}).
doi:\doiurl{10.1086/190587}
\end{barticle}
\endbibitem

\bibitem[\protect\citeauthoryear{{Markowitz} et~al.}{2006}]{Markowitz06}
\begin{barticle}
\bauthor{\binits{A.} \bsnm{{Markowitz}}},
\bauthor{\binits{J.N.} \bsnm{{Reeves}}},
\bauthor{\binits{P.} \bsnm{{Serlemitsos}}},
\bauthor{\binits{T.} \bsnm{{Yaqoob}}},
\bauthor{\binits{H.} \bsnm{{Awaki}}},
\bauthor{\binits{A.} \bsnm{{Fabian}}},
\bauthor{\binits{L.} \bsnm{{Gallo}}},
\bauthor{\binits{R.E.} \bsnm{{Griffiths}}},
\bauthor{\binits{H.} \bsnm{{Kunieda}}},
\bauthor{\binits{G.} \bsnm{{Miniutti}}},
\bauthor{\binits{R.} \bsnm{{Mushotzky}}},
\bauthor{\binits{T.} \bsnm{{Okajima}}},
\batitle{{Suzaku observation of NGC 3516: complex absorption and the broad and
  narrow Fe K lines}}.
\bjtitle{Astronomische Nachrichten}
\bvolume{327},
\bfpage{1087}
(\byear{2006}).
doi:\doiurl{10.1002/asna.200610697}
\end{barticle}
\endbibitem

\bibitem[\protect\citeauthoryear{{Marscher} et~al.}{1984}]{Marscher84}
\begin{barticle}
\bauthor{\binits{A.P.} \bsnm{{Marscher}}},
\bauthor{\binits{K.} \bsnm{{Brecher}}},
\bauthor{\binits{W.A.} \bsnm{{Wheaton}}},
\bauthor{\binits{J.C.} \bsnm{{Ling}}},
\bauthor{\binits{W.A.} \bsnm{{Mahoney}}},
\bauthor{\binits{A.S.} \bsnm{{Jacobson}}},
\batitle{{Search for 511 keV electron-positron annihilation radiation from
  mildly active galaxies using the HEAO 3 gamma-ray spectrometer}}.
\bjtitle{\apj}
\bvolume{281},
\bfpage{566}--\blpage{569}
(\byear{1984}).
doi:\doiurl{10.1086/162130}
\end{barticle}
\endbibitem

\bibitem[\protect\citeauthoryear{{Marshall} et~al.}{1979}]{Marshall1979;heao1}
\begin{barticle}
\bauthor{\binits{F.E.} \bsnm{{Marshall}}},
\bauthor{\binits{E.A.} \bsnm{{Boldt}}},
\bauthor{\binits{S.S.} \bsnm{{Holt}}},
\bauthor{\binits{R.F.} \bsnm{{Mushotzky}}},
\bauthor{\binits{R.E.} \bsnm{{Rothschild}}},
\bauthor{\binits{P.J.} \bsnm{{Serlemitsos}}},
\bauthor{\binits{S.H.} \bsnm{{Pravdo}}},
\batitle{{New hard X-ray sources observed with HEAO A-2}}.
\bjtitle{\apjs}
\bvolume{40},
\bfpage{657}--\blpage{665}
(\byear{1979}).
doi:\doiurl{10.1086/190600}
\end{barticle}
\endbibitem

\bibitem[\protect\citeauthoryear{{Marshall} et~al.}{1980}]{Marshall80}
\begin{barticle}
\bauthor{\binits{F.E.} \bsnm{{Marshall}}},
\bauthor{\binits{E.A.} \bsnm{{Boldt}}},
\bauthor{\binits{S.S.} \bsnm{{Holt}}},
\bauthor{\binits{R.B.} \bsnm{{Miller}}},
\bauthor{\binits{R.F.} \bsnm{{Mushotzky}}},
\bauthor{\binits{L.A.} \bsnm{{Rose}}},
\bauthor{\binits{R.E.} \bsnm{{Rothschild}}},
\bauthor{\binits{P.J.} \bsnm{{Serlemitsos}}},
\batitle{{The diffuse X-ray background spectrum from 3 to 50 keV}}.
\bjtitle{\apj}
\bvolume{235},
\bfpage{4}--\blpage{10}
(\byear{1980}).
doi:\doiurl{10.1086/157601}
\end{barticle}
\endbibitem

\bibitem[\protect\citeauthoryear{{Masetti} et~al.}{2000}]{Masetti00}
\begin{barticle}
\bauthor{\binits{N.} \bsnm{{Masetti}}},
\bauthor{\binits{F.} \bsnm{{Frontera}}},
\bauthor{\binits{L.} \bsnm{{Stella}}},
\bauthor{\binits{M.} \bsnm{{Orlandini}}},
\bauthor{\binits{A.N.} \bsnm{{Parmar}}},
\bauthor{\binits{S.} \bsnm{{Del Sordo}}},
\bauthor{\binits{L.} \bsnm{{Amati}}},
\bauthor{\binits{E.} \bsnm{{Palazzi}}},
\bauthor{\binits{D.} \bsnm{{Dal Fiume}}},
\bauthor{\binits{G.} \bsnm{{Cusumano}}},
\bauthor{\binits{G.} \bsnm{{Pareschi}}},
\bauthor{\binits{I.} \bsnm{{Lapidus}}},
\bauthor{\binits{R.A.} \bsnm{{Remillard}}},
\batitle{{Hard X-rays from Type II bursts of the Rapid Burster and its
  transition toward quiescence}}.
\bjtitle{\aap}
\bvolume{363},
\bfpage{188}--\blpage{197}
(\byear{2000})
\end{barticle}
\endbibitem

\bibitem[\protect\citeauthoryear{{Massaro} et~al.}{1991}]{Massaro91}
\begin{barticle}
\bauthor{\binits{E.} \bsnm{{Massaro}}},
\bauthor{\binits{G.} \bsnm{{Matt}}},
\bauthor{\binits{M.} \bsnm{{Salvati}}},
\bauthor{\binits{E.} \bsnm{{Costa}}},
\bauthor{\binits{P.} \bsnm{{Mandrou}}},
\bauthor{\binits{M.} \bsnm{{Niel}}},
\bauthor{\binits{J.F.} \bsnm{{Olive}}},
\bauthor{\binits{T.} \bsnm{{Mineo}}},
\bauthor{\binits{B.} \bsnm{{Sacco}}},
\bauthor{\binits{L.} \bsnm{{Scarsi}}},
\bauthor{\binits{G.} \bsnm{{Gerardi}}},
\bauthor{\binits{B.} \bsnm{{Agrinier}}},
\bauthor{\binits{E.} \bsnm{{Barouch}}},
\bauthor{\binits{R.} \bsnm{{Comte}}},
\bauthor{\binits{B.} \bsnm{{Parlier}}},
\bauthor{\binits{J.L.} \bsnm{{Masnou}}},
\batitle{{Detection of a feature at 0.44 MeV in the Crab pulsar spectrum with
  FIGARO II - A redshifted positron annihilation line?}}
\bjtitle{\apjl}
\bvolume{376},
\bfpage{11}--\blpage{15}
(\byear{1991}).
doi:\doiurl{10.1086/186091}
\end{barticle}
\endbibitem

\bibitem[\protect\citeauthoryear{{Matsuoka} et~al.}{1972}]{Matsuoka72}
\begin{barticle}
\bauthor{\binits{M.} \bsnm{{Matsuoka}}},
\bauthor{\binits{M.} \bsnm{{Fujii}}},
\bauthor{\binits{S.} \bsnm{{Miyamoto}}},
\bauthor{\binits{J.} \bsnm{{Nishimura}}},
\bauthor{\binits{M.} \bsnm{{Oda}}},
\bauthor{\binits{Y.} \bsnm{{Ogawara}}},
\bauthor{\binits{S.} \bsnm{{Hayakawa}}},
\bauthor{\binits{I.} \bsnm{{Kasahara}}},
\bauthor{\binits{F.} \bsnm{{Makino}}},
\bauthor{\binits{Y.} \bsnm{{Tanaka}}},
\bauthor{\binits{P.C.} \bsnm{{Agrawal}}},
\bauthor{\binits{B.V.} \bsnm{{Sreekantan}}},
\batitle{{Time Variations of Hard X-Rays from Sco X-1}}.
\bjtitle{\apss}
\bvolume{18},
\bfpage{472}--\blpage{490}
(\byear{1972}).
doi:\doiurl{10.1007/BF00645411}
\end{barticle}
\endbibitem

\bibitem[\protect\citeauthoryear{{Matt} et~al.}{1990}]{Matt90}
\begin{barticle}
\bauthor{\binits{G.} \bsnm{{Matt}}},
\bauthor{\binits{E.} \bsnm{{Costa}}},
\bauthor{\binits{D.} \bsnm{{dal Fiume}}},
\bauthor{\binits{W.} \bsnm{{Dusi}}},
\bauthor{\binits{F.} \bsnm{{Frontera}}},
\bauthor{\binits{E.} \bsnm{{Morelli}}},
\batitle{{Observations of galactic and extragalactic high-energy X-ray
  sources}}.
\bjtitle{\apj}
\bvolume{355},
\bfpage{468}--\blpage{472}
(\byear{1990}).
doi:\doiurl{10.1086/168781}
\end{barticle}
\endbibitem

\bibitem[\protect\citeauthoryear{{Matt} et~al.}{1997}]{Matt97}
\begin{barticle}
\bauthor{\binits{G.} \bsnm{{Matt}}},
\bauthor{\binits{M.} \bsnm{{Guainazzi}}},
\bauthor{\binits{F.} \bsnm{{Frontera}}},
\bauthor{\binits{L.} \bsnm{{Bassani}}},
\bauthor{\binits{W.N.} \bsnm{{Brandt}}},
\bauthor{\binits{A.C.} \bsnm{{Fabian}}},
\bauthor{\binits{F.} \bsnm{{Fiore}}},
\bauthor{\binits{F.} \bsnm{{Haardt}}},
\bauthor{\binits{K.} \bsnm{{Iwasawa}}},
\bauthor{\binits{R.} \bsnm{{Maiolino}}},
\bauthor{\binits{G.} \bsnm{{Malaguti}}},
\bauthor{\binits{A.} \bsnm{{Marconi}}},
\bauthor{\binits{A.} \bsnm{{Matteuzzi}}},
\bauthor{\binits{S.} \bsnm{{Molendi}}},
\bauthor{\binits{G.C.} \bsnm{{Perola}}},
\bauthor{\binits{S.} \bsnm{{Piraino}}},
\bauthor{\binits{L.} \bsnm{{Piro}}},
\batitle{{Hard X-ray detection of NGC 1068 with BeppoSAX.}}
\bjtitle{\aap}
\bvolume{325},
\bfpage{13}--\blpage{16}
(\byear{1997})
\end{barticle}
\endbibitem

\bibitem[\protect\citeauthoryear{{Maurer} et~al.}{1979}]{Maurer79}
\begin{barticle}
\bauthor{\binits{G.S.} \bsnm{{Maurer}}},
\bauthor{\binits{B.R.} \bsnm{{Dennis}}},
\bauthor{\binits{M.J.} \bsnm{{Coe}}},
\bauthor{\binits{C.J.} \bsnm{{Crannell}}},
\bauthor{\binits{J.F.} \bsnm{{Dolan}}},
\bauthor{\binits{K.J.} \bsnm{{Frost}}},
\bauthor{\binits{L.E.} \bsnm{{Orwig}}},
\bauthor{\binits{E.P.} \bsnm{{Cutler}}},
\batitle{{The high-energy pulsed X-ray spectrum of Hercules X-1 as observed
  with OSO 8}}.
\bjtitle{\apj}
\bvolume{231},
\bfpage{906}--\blpage{911}
(\byear{1979}).
doi:\doiurl{10.1086/157253}
\end{barticle}
\endbibitem

\bibitem[\protect\citeauthoryear{{Mazets} and {Golenetskii}}{1981}]{Mazets81}
\begin{barticle}
\bauthor{\binits{E.P.} \bsnm{{Mazets}}},
\bauthor{\binits{S.V.} \bsnm{{Golenetskii}}},
\batitle{{Recent results from the gamma-ray burst studies in the KONUS
  experiment}}.
\bjtitle{\apss}
\bvolume{75},
\bfpage{47}--\blpage{81}
(\byear{1981}).
doi:\doiurl{10.1007/BF00651384}
\end{barticle}
\endbibitem

\bibitem[\protect\citeauthoryear{{Mazets} and {Golenetskii}}{1988}]{Mazets88}
\begin{barticle}
\bauthor{\binits{E.P.} \bsnm{{Mazets}}},
\bauthor{\binits{S.V.} \bsnm{{Golenetskii}}},
\batitle{{Observations of cosmic gamma-ray bursts.}}
\bjtitle{Astrophysics and Space Physics Reviews}
\bvolume{6},
\bfpage{281}--\blpage{311}
(\byear{1988})
\end{barticle}
\endbibitem

\bibitem[\protect\citeauthoryear{{Mazets} and {Golenetskij}}{1987}]{Mazets87}
\begin{barticle}
\bauthor{\binits{E.P.} \bsnm{{Mazets}}},
\bauthor{\binits{S.V.} \bsnm{{Golenetskij}}},
\batitle{{Observational properties of gamma-ray bursts.}}
\bjtitle{Itogi Nauki i Tekhniki Seriia Astronomiia}
\bvolume{32},
\bfpage{16}--\blpage{42}
(\byear{1987})
\end{barticle}
\endbibitem

\bibitem[\protect\citeauthoryear{{Mazets} et~al.}{1979a}]{Mazets79c}
\begin{barticle}
\bauthor{\binits{E.P.} \bsnm{{Mazets}}},
\bauthor{\binits{S.V.} \bsnm{{Golenetskij}}},
\bauthor{\binits{Y.A.} \bsnm{{Guryan}}},
\batitle{{Soft gamma-ray bursts from the source B1900+14}}.
\bjtitle{Soviet Astronomy Letters}
\bvolume{5},
\bfpage{641}--\blpage{643}
(\byear{1979}a)
\end{barticle}
\endbibitem

\bibitem[\protect\citeauthoryear{{Mazets} et~al.}{1979b}]{Mazets79b}
\begin{barticle}
\bauthor{\binits{E.P.} \bsnm{{Mazets}}},
\bauthor{\binits{S.V.} \bsnm{{Golentskii}}},
\bauthor{\binits{V.N.} \bsnm{{Ilinskii}}},
\bauthor{\binits{R.L.} \bsnm{{Aptekar}}},
\bauthor{\binits{I.A.} \bsnm{{Guryan}}},
\batitle{{Observations of a flaring X-ray pulsar in Dorado}}.
\bjtitle{\nat}
\bvolume{282},
\bfpage{587}--\blpage{589}
(\byear{1979}b).
doi:\doiurl{10.1038/282587a0}
\end{barticle}
\endbibitem

\bibitem[\protect\citeauthoryear{{Mazets} et~al.}{1979c}]{Mazets79}
\begin{barticle}
\bauthor{\binits{E.P.} \bsnm{{Mazets}}},
\bauthor{\binits{S.V.} \bsnm{{Golenetskii}}},
\bauthor{\binits{V.N.} \bsnm{{Ilinskii}}},
\bauthor{\binits{V.N.} \bsnm{{Panov}}},
\bauthor{\binits{R.L.} \bsnm{{Aptekar}}},
\bauthor{\binits{I.A.} \bsnm{{Gurian}}},
\bauthor{\binits{I.A.} \bsnm{{Sokolov}}},
\bauthor{\binits{Z.I.} \bsnm{{Sokolova}}},
\bauthor{\binits{T.V.} \bsnm{{Kharitonova}}},
\batitle{{Venera 11 and 12 observations of gamma-ray bursts - The Cone
  experiment}}.
\bjtitle{Soviet Astronomy Letters}
\bvolume{5},
\bfpage{163}--\blpage{167}
(\byear{1979}c)
\end{barticle}
\endbibitem

\bibitem[\protect\citeauthoryear{{Mazets} et~al.}{1981}]{Mazets81b}
\begin{barticle}
\bauthor{\binits{E.P.} \bsnm{{Mazets}}},
\bauthor{\binits{S.V.} \bsnm{{Golenetskii}}},
\bauthor{\binits{V.N.} \bsnm{{Ilinskii}}},
\bauthor{\binits{V.N.} \bsnm{{Panov}}},
\bauthor{\binits{R.L.} \bsnm{{Aptekar}}},
\bauthor{\binits{I.A.} \bsnm{{Gurian}}},
\bauthor{\binits{M.P.} \bsnm{{Proskura}}},
\bauthor{\binits{I.A.} \bsnm{{Sokolov}}},
\bauthor{\binits{Z.I.} \bsnm{{Sokolova}}},
\bauthor{\binits{T.V.} \bsnm{{Kharitonova}}},
\batitle{{Catalog of cosmic gamma-ray bursts from the KONUS experiment data.
  I.}}
\bjtitle{\apss}
\bvolume{80},
\bfpage{3}--\blpage{83}
(\byear{1981}).
doi:\doiurl{10.1007/BF00649140}
\end{barticle}
\endbibitem

\bibitem[\protect\citeauthoryear{{McBride} et~al.}{2006}]{McBride06}
\begin{barticle}
\bauthor{\binits{V.A.} \bsnm{{McBride}}},
\bauthor{\binits{J.} \bsnm{{Wilms}}},
\bauthor{\binits{M.J.} \bsnm{{Coe}}},
\bauthor{\binits{I.} \bsnm{{Kreykenbohm}}},
\bauthor{\binits{R.E.} \bsnm{{Rothschild}}},
\bauthor{\binits{W.} \bsnm{{Coburn}}},
\bauthor{\binits{J.L.} \bsnm{{Galache}}},
\bauthor{\binits{P.} \bsnm{{Kretschmar}}},
\bauthor{\binits{W.R.T.} \bsnm{{Edge}}},
\bauthor{\binits{R.} \bsnm{{Staubert}}},
\batitle{{Study of the cyclotron feature in <ASTROBJ>MXB 0656-072</ASTROBJ>}}.
\bjtitle{\aap}
\bvolume{451},
\bfpage{267}--\blpage{272}
(\byear{2006}).
doi:\doiurl{10.1051/0004-6361:20054239}
\end{barticle}
\endbibitem

\bibitem[\protect\citeauthoryear{{McCracken}}{1965}]{McCracken65}
\begin{barticle}
\bauthor{\binits{K.G.} \bsnm{{McCracken}}},
\batitle{{A celestial source of X rays in the energy range 20 to 58 keV}}.
\bjtitle{International Cosmic Ray Conference}
\bvolume{1},
\bfpage{449}
(\byear{1965})
\end{barticle}
\endbibitem

\bibitem[\protect\citeauthoryear{{McHardy} et~al.}{1999}]{McHardy99}
\begin{barticle}
\bauthor{\binits{I.M.} \bsnm{{McHardy}}},
\bauthor{\binits{I.E.} \bsnm{{Papadakis}}},
\bauthor{\binits{P.} \bsnm{{Uttley}}},
\batitle{{Temporal and spectral variability of AGN with RXTE}}.
\bjtitle{Nuclear Physics B Proceedings Supplements}
\bvolume{69},
\bfpage{509}--\blpage{514}
(\byear{1999}).
doi:\doiurl{10.1016/S0920-5632(98)00272-2}
\end{barticle}
\endbibitem

\bibitem[\protect\citeauthoryear{{McNamara} et~al.}{1995}]{McNamara95}
\begin{barticle}
\bauthor{\binits{B.J.} \bsnm{{McNamara}}},
\bauthor{\binits{B.A.} \bsnm{{Harmon}}},
\bauthor{\binits{T.E.} \bsnm{{Harrison}}},
\batitle{{The extraction of discrete flux measurements and continuous light
  curves from the CGRO/BATSE Spectroscopy Detector datasets.}}
\bjtitle{\aaps}
\bvolume{111},
\bfpage{587}
(\byear{1995})
\end{barticle}
\endbibitem

\bibitem[\protect\citeauthoryear{{McNaron-Brown}
  et~al.}{1995}]{McNaron-Brown95}
\begin{barticle}
\bauthor{\binits{K.} \bsnm{{McNaron-Brown}}},
\bauthor{\binits{W.N.} \bsnm{{Johnson}}},
\bauthor{\binits{G.V.} \bsnm{{Jung}}},
\bauthor{\binits{R.L.} \bsnm{{Kinzer}}},
\bauthor{\binits{J.D.} \bsnm{{Kurfess}}},
\bauthor{\binits{M.S.} \bsnm{{Strickman}}},
\bauthor{\binits{C.D.} \bsnm{{Dermer}}},
\bauthor{\binits{D.A.} \bsnm{{Grabelsky}}},
\bauthor{\binits{W.R.} \bsnm{{Purcell}}},
\bauthor{\binits{M.P.} \bsnm{{Ulmer}}},
\bauthor{\binits{M.} \bsnm{{Kafatos}}},
\bauthor{\binits{P.A.} \bsnm{{Becker}}},
\bauthor{\binits{R.} \bsnm{{Staubert}}},
\bauthor{\binits{M.} \bsnm{{Maisack}}},
\batitle{{OSSE Observations of Blazars}}.
\bjtitle{\apj}
\bvolume{451},
\bfpage{575}
(\byear{1995}).
doi:\doiurl{10.1086/176245}
\end{barticle}
\endbibitem

\bibitem[\protect\citeauthoryear{{Meegan} and {Haymes}}{1979}]{Meegan79}
\begin{barticle}
\bauthor{\binits{C.A.} \bsnm{{Meegan}}},
\bauthor{\binits{R.C.} \bsnm{{Haymes}}},
\batitle{{Evidence for variability of hard X-rays from NGC 4151}}.
\bjtitle{\apj}
\bvolume{233},
\bfpage{510}--\blpage{513}
(\byear{1979}).
doi:\doiurl{10.1086/157411}
\end{barticle}
\endbibitem

\bibitem[\protect\citeauthoryear{{Meegan} et~al.}{2009}]{Meegan09}
\begin{barticle}
\bauthor{\binits{C.} \bsnm{{Meegan}}},
\bauthor{\binits{G.} \bsnm{{Lichti}}},
\bauthor{\binits{P.N.} \bsnm{{Bhat}}},
\bauthor{\binits{E.} \bsnm{{Bissaldi}}},
\bauthor{\binits{M.S.} \bsnm{{Briggs}}},
\bauthor{\binits{V.} \bsnm{{Connaughton}}},
\bauthor{\binits{R.} \bsnm{{Diehl}}},
\bauthor{\binits{G.} \bsnm{{Fishman}}},
\bauthor{\binits{J.} \bsnm{{Greiner}}},
\bauthor{\binits{A.S.} \bsnm{{Hoover}}},
\bauthor{\binits{A.J.} \bsnm{{van der Horst}}},
\bauthor{\binits{A.} \bsnm{{von Kienlin}}},
\bauthor{\binits{R.M.} \bsnm{{Kippen}}},
\bauthor{\binits{C.} \bsnm{{Kouveliotou}}},
\bauthor{\binits{S.} \bsnm{{McBreen}}},
\bauthor{\binits{W.S.} \bsnm{{Paciesas}}},
\bauthor{\binits{R.} \bsnm{{Preece}}},
\bauthor{\binits{H.} \bsnm{{Steinle}}},
\bauthor{\binits{M.S.} \bsnm{{Wallace}}},
\bauthor{\binits{R.B.} \bsnm{{Wilson}}},
\bauthor{\binits{C.} \bsnm{{Wilson-Hodge}}},
\batitle{{The Fermi Gamma-ray Burst Monitor}}.
\bjtitle{\apj}
\bvolume{702},
\bfpage{791}--\blpage{804}
(\byear{2009}).
doi:\doiurl{10.1088/0004-637X/702/1/791}
\end{barticle}
\endbibitem

\bibitem[\protect\citeauthoryear{{M{\'e}ndez} et~al.}{1998}]{Mendez98}
\begin{barticle}
\bauthor{\binits{M.} \bsnm{{M{\'e}ndez}}},
\bauthor{\binits{T.} \bsnm{{Belloni}}},
\bauthor{\binits{M.} \bsnm{{van der Klis}}},
\batitle{{``Canonical'' Black Hole States in the Superluminal Source GRO
  J1655-40}}.
\bjtitle{\apjl}
\bvolume{499},
\bfpage{187}--\blpage{190}
(\byear{1998}).
doi:\doiurl{10.1086/311368}
\end{barticle}
\endbibitem

\bibitem[\protect\citeauthoryear{{Meneguzzi} and {Reeves}}{1975}]{Meneguzzi75}
\begin{barticle}
\bauthor{\binits{M.} \bsnm{{Meneguzzi}}},
\bauthor{\binits{H.} \bsnm{{Reeves}}},
\batitle{{Nuclear gamma ray production by cosmic rays}}.
\bjtitle{\aap}
\bvolume{40},
\bfpage{91}--\blpage{98}
(\byear{1975})
\end{barticle}
\endbibitem

\bibitem[\protect\citeauthoryear{{Metzger} et~al.}{1997}]{Metzger97}
\begin{barticle}
\bauthor{\binits{M.R.} \bsnm{{Metzger}}},
\bauthor{\binits{S.G.} \bsnm{{Djorgovski}}},
\bauthor{\binits{S.R.} \bsnm{{Kulkarni}}},
\bauthor{\binits{C.C.} \bsnm{{Steidel}}},
\bauthor{\binits{K.L.} \bsnm{{Adelberger}}},
\bauthor{\binits{D.A.} \bsnm{{Frail}}},
\bauthor{\binits{E.} \bsnm{{Costa}}},
\bauthor{\binits{F.} \bsnm{{Frontera}}},
\batitle{{Spectral constraints on the redshift of the optical counterpart to
  the {$\gamma$}-ray burst of 8 May 1997}}.
\bjtitle{\nat}
\bvolume{387},
\bfpage{878}--\blpage{880}
(\byear{1997}).
doi:\doiurl{10.1038/43132}
\end{barticle}
\endbibitem

\bibitem[\protect\citeauthoryear{{Miniutti} et~al.}{2007}]{Miniutti07}
\begin{barticle}
\bauthor{\binits{G.} \bsnm{{Miniutti}}},
\bauthor{\binits{A.C.} \bsnm{{Fabian}}},
\bauthor{\binits{N.} \bsnm{{Anabuki}}},
\bauthor{\binits{J.} \bsnm{{Crummy}}},
\bauthor{\binits{Y.} \bsnm{{Fukazawa}}},
\bauthor{\binits{L.} \bsnm{{Gallo}}},
\bauthor{\binits{Y.} \bsnm{{Haba}}},
\bauthor{\binits{K.} \bsnm{{Hayashida}}},
\bauthor{\binits{S.} \bsnm{{Holt}}},
\bauthor{\binits{H.} \bsnm{{Kunieda}}},
\bauthor{\binits{J.} \bsnm{{Larsson}}},
\bauthor{\binits{A.} \bsnm{{Markowitz}}},
\bauthor{\binits{C.} \bsnm{{Matsumoto}}},
\bauthor{\binits{M.} \bsnm{{Ohno}}},
\bauthor{\binits{J.N.} \bsnm{{Reeves}}},
\bauthor{\binits{T.} \bsnm{{Takahashi}}},
\bauthor{\binits{Y.} \bsnm{{Tanaka}}},
\bauthor{\binits{Y.} \bsnm{{Terashima}}},
\bauthor{\binits{K.} \bsnm{{Torii}}},
\bauthor{\binits{Y.} \bsnm{{Ueda}}},
\bauthor{\binits{M.} \bsnm{{Ushio}}},
\bauthor{\binits{S.} \bsnm{{Watanabe}}},
\bauthor{\binits{M.} \bsnm{{Yamauchi}}},
\bauthor{\binits{T.} \bsnm{{Yaqoob}}},
\batitle{{Suzaku Observations of the Hard X-Ray Variability of MCG -6-30-15:
  the Effects of Strong Gravity around a Kerr Black Hole}}.
\bjtitle{\pasj}
\bvolume{59},
\bfpage{315}--\blpage{325}
(\byear{2007}).
doi:\doiurl{10.1093/pasj/59.sp1.S315}
\end{barticle}
\endbibitem

\bibitem[\protect\citeauthoryear{{Mitchell} and {Mushotzky}}{1980}]{Mitchell80}
\begin{barticle}
\bauthor{\binits{R.} \bsnm{{Mitchell}}},
\bauthor{\binits{R.} \bsnm{{Mushotzky}}},
\batitle{{HEAO A-2 observations of the X-ray spectra of the Centaurus and A1060
  clusters of galaxies}}.
\bjtitle{\apj}
\bvolume{236},
\bfpage{730}--\blpage{737}
(\byear{1980}).
doi:\doiurl{10.1086/157796}
\end{barticle}
\endbibitem

\bibitem[\protect\citeauthoryear{{Mitsuda} et~al.}{2007}]{Mitsuda2007;suzaku}
\begin{barticle}
\bauthor{\binits{K.} \bsnm{{Mitsuda}}},
\bauthor{\binits{M.} \bsnm{{Bautz}}},
\bauthor{\binits{H.} \bsnm{{Inoue}}},
\bauthor{\binits{R.L.} \bsnm{{Kelley}}},
\bauthor{\binits{K.} \bsnm{{Koyama}}},
\bauthor{\binits{H.} \bsnm{{Kunieda}}},
\bauthor{\binits{K.} \bsnm{{Makishima}}},
\bauthor{\binits{Y.} \bsnm{{Ogawara}}},
\bauthor{\binits{R.} \bsnm{{Petre}}},
\bauthor{\binits{T.} \bsnm{{Takahashi}}},
\bauthor{\binits{H.} \bsnm{{Tsunemi}}},
\bauthor{\binits{N.E.} \bsnm{{White}}},
\bauthor{\binits{N.} \bsnm{{Anabuki}}},
\bauthor{\binits{L.} \bsnm{{Angelini}}},
\bauthor{\binits{K.} \bsnm{{Arnaud}}},
\bauthor{\binits{H.} \bsnm{{Awaki}}},
\bauthor{\binits{A.} \bsnm{{Bamba}}},
\bauthor{\binits{K.} \bsnm{{Boyce}}},
\bauthor{\binits{G.V.} \bsnm{{Brown}}},
\bauthor{\binits{K.-W.} \bsnm{{Chan}}},
\bauthor{\binits{J.} \bsnm{{Cottam}}},
\bauthor{\binits{T.} \bsnm{{Dotani}}},
\bauthor{\binits{J.} \bsnm{{Doty}}},
\bauthor{\binits{K.} \bsnm{{Ebisawa}}},
\bauthor{\binits{Y.} \bsnm{{Ezoe}}},
\bauthor{\binits{A.C.} \bsnm{{Fabian}}},
\bauthor{\binits{E.} \bsnm{{Figueroa}}},
\bauthor{\binits{R.} \bsnm{{Fujimoto}}},
\bauthor{\binits{Y.} \bsnm{{Fukazawa}}},
\bauthor{\binits{T.} \bsnm{{Furusho}}},
\bauthor{\binits{A.} \bsnm{{Furuzawa}}},
\bauthor{\binits{K.} \bsnm{{Gendreau}}},
\bauthor{\binits{R.E.} \bsnm{{Griffiths}}},
\bauthor{\binits{Y.} \bsnm{{Haba}}},
\bauthor{\binits{K.} \bsnm{{Hamaguchi}}},
\bauthor{\binits{I.} \bsnm{{Harrus}}},
\bauthor{\binits{G.} \bsnm{{Hasinger}}},
\bauthor{\binits{I.} \bsnm{{Hatsukade}}},
\bauthor{\binits{K.} \bsnm{{Hayashida}}},
\bauthor{\binits{P.J.} \bsnm{{Henry}}},
\bauthor{\binits{J.S.} \bsnm{{Hiraga}}},
\bauthor{\binits{S.S.} \bsnm{{Holt}}},
\bauthor{\binits{A.} \bsnm{{Hornschemeier}}},
\bauthor{\binits{J.P.} \bsnm{{Hughes}}},
\bauthor{\binits{U.} \bsnm{{Hwang}}},
\bauthor{\binits{M.} \bsnm{{Ishida}}},
\bauthor{\binits{Y.} \bsnm{{Ishisaki}}},
\bauthor{\binits{N.} \bsnm{{Isobe}}},
\bauthor{\binits{M.} \bsnm{{Itoh}}},
\bauthor{\binits{N.} \bsnm{{Iyomoto}}},
\bauthor{\binits{S.M.} \bsnm{{Kahn}}},
\bauthor{\binits{T.} \bsnm{{Kamae}}},
\bauthor{\binits{H.} \bsnm{{Katagiri}}},
\bauthor{\binits{J.} \bsnm{{Kataoka}}},
\bauthor{\binits{H.} \bsnm{{Katayama}}},
\bauthor{\binits{N.} \bsnm{{Kawai}}},
\bauthor{\binits{C.} \bsnm{{Kilbourne}}},
\bauthor{\binits{K.} \bsnm{{Kinugasa}}},
\bauthor{\binits{S.} \bsnm{{Kissel}}},
\bauthor{\binits{S.} \bsnm{{Kitamoto}}},
\bauthor{\binits{M.} \bsnm{{Kohama}}},
\bauthor{\binits{T.} \bsnm{{Kohmura}}},
\bauthor{\binits{M.} \bsnm{{Kokubun}}},
\bauthor{\binits{T.} \bsnm{{Kotani}}},
\bauthor{\binits{J.} \bsnm{{Kotoku}}},
\bauthor{\binits{A.} \bsnm{{Kubota}}},
\bauthor{\binits{G.M.} \bsnm{{Madejski}}},
\bauthor{\binits{Y.} \bsnm{{Maeda}}},
\bauthor{\binits{F.} \bsnm{{Makino}}},
\bauthor{\binits{A.} \bsnm{{Markowitz}}},
\bauthor{\binits{C.} \bsnm{{Matsumoto}}},
\bauthor{\binits{H.} \bsnm{{Matsumoto}}},
\bauthor{\binits{M.} \bsnm{{Matsuoka}}},
\bauthor{\binits{K.} \bsnm{{Matsushita}}},
\bauthor{\binits{D.} \bsnm{{McCammon}}},
\bauthor{\binits{T.} \bsnm{{Mihara}}},
\bauthor{\binits{K.} \bsnm{{Misaki}}},
\bauthor{\binits{E.} \bsnm{{Miyata}}},
\bauthor{\binits{T.} \bsnm{{Mizuno}}},
\bauthor{\binits{K.} \bsnm{{Mori}}},
\bauthor{\binits{H.} \bsnm{{Mori}}},
\bauthor{\binits{M.} \bsnm{{Morii}}},
\bauthor{\binits{H.} \bsnm{{Moseley}}},
\bauthor{\binits{K.} \bsnm{{Mukai}}},
\bauthor{\binits{H.} \bsnm{{Murakami}}},
\bauthor{\binits{T.} \bsnm{{Murakami}}},
\bauthor{\binits{R.} \bsnm{{Mushotzky}}},
\bauthor{\binits{F.} \bsnm{{Nagase}}},
\bauthor{\binits{M.} \bsnm{{Namiki}}},
\bauthor{\binits{H.} \bsnm{{Negoro}}},
\bauthor{\binits{K.} \bsnm{{Nakazawa}}},
\bauthor{\binits{J.A.} \bsnm{{Nousek}}},
\bauthor{\binits{T.} \bsnm{{Okajima}}},
\bauthor{\binits{Y.} \bsnm{{Ogasaka}}},
\bauthor{\binits{T.} \bsnm{{Ohashi}}},
\bauthor{\binits{T.} \bsnm{{Oshima}}},
\bauthor{\binits{N.} \bsnm{{Ota}}},
\bauthor{\binits{M.} \bsnm{{Ozaki}}},
\bauthor{\binits{H.} \bsnm{{Ozawa}}},
\bauthor{\binits{A.N.} \bsnm{{Parmar}}},
\bauthor{\binits{W.D.} \bsnm{{Pence}}},
\bauthor{\binits{F.S.} \bsnm{{Porter}}},
\bauthor{\binits{J.N.} \bsnm{{Reeves}}},
\bauthor{\binits{G.R.} \bsnm{{Ricker}}},
\bauthor{\binits{I.} \bsnm{{Sakurai}}},
\bauthor{\binits{W.T.} \bsnm{{Sanders}}},
\bauthor{\binits{A.} \bsnm{{Senda}}},
\bauthor{\binits{P.} \bsnm{{Serlemitsos}}},
\bauthor{\binits{R.} \bsnm{{Shibata}}},
\bauthor{\binits{Y.} \bsnm{{Soong}}},
\bauthor{\binits{R.} \bsnm{{Smith}}},
\bauthor{\binits{M.} \bsnm{{Suzuki}}},
\bauthor{\binits{A.E.} \bsnm{{Szymkowiak}}},
\bauthor{\binits{H.} \bsnm{{Takahashi}}},
\bauthor{\binits{T.} \bsnm{{Tamagawa}}},
\bauthor{\binits{K.} \bsnm{{Tamura}}},
\bauthor{\binits{T.} \bsnm{{Tamura}}},
\bauthor{\binits{Y.} \bsnm{{Tanaka}}},
\bauthor{\binits{M.} \bsnm{{Tashiro}}},
\bauthor{\binits{Y.} \bsnm{{Tawara}}},
\bauthor{\binits{Y.} \bsnm{{Terada}}},
\bauthor{\binits{Y.} \bsnm{{Terashima}}},
\bauthor{\binits{H.} \bsnm{{Tomida}}},
\bauthor{\binits{K.} \bsnm{{Torii}}},
\bauthor{\binits{Y.} \bsnm{{Tsuboi}}},
\bauthor{\binits{M.} \bsnm{{Tsujimoto}}},
\bauthor{\binits{T.G.} \bsnm{{Tsuru}}},
\bauthor{\binits{M.J.L..} \bsnm{{Turner}}},
\bauthor{\binits{Y.} \bsnm{{Ueda}}},
\bauthor{\binits{S.} \bsnm{{Ueno}}},
\bauthor{\binits{M.} \bsnm{{Ueno}}},
\bauthor{\binits{S.} \bsnm{{Uno}}},
\bauthor{\binits{Y.} \bsnm{{Urata}}},
\bauthor{\binits{S.} \bsnm{{Watanabe}}},
\bauthor{\binits{N.} \bsnm{{Yamamoto}}},
\bauthor{\binits{K.} \bsnm{{Yamaoka}}},
\bauthor{\binits{N.Y.} \bsnm{{Yamasaki}}},
\bauthor{\binits{K.} \bsnm{{Yamashita}}},
\bauthor{\binits{M.} \bsnm{{Yamauchi}}},
\bauthor{\binits{S.} \bsnm{{Yamauchi}}},
\bauthor{\binits{T.} \bsnm{{Yaqoob}}},
\bauthor{\binits{D.} \bsnm{{Yonetoku}}},
\bauthor{\binits{A.} \bsnm{{Yoshida}}},
\batitle{{The X-Ray Observatory Suzaku}}.
\bjtitle{\pasj}
\bvolume{59},
\bfpage{1}--\blpage{7}
(\byear{2007}).
doi:\doiurl{10.1093/pasj/59.sp1.S1}
\end{barticle}
\endbibitem

\bibitem[\protect\citeauthoryear{{Miyasaka} et~al.}{2013}]{Miyasaka13}
\begin{barticle}
\bauthor{\binits{H.} \bsnm{{Miyasaka}}},
\bauthor{\binits{M.} \bsnm{{Bachetti}}},
\bauthor{\binits{F.A.} \bsnm{{Harrison}}},
\bauthor{\binits{F.} \bsnm{{F{\"u}rst}}},
\bauthor{\binits{D.} \bsnm{{Barret}}},
\bauthor{\binits{E.C.} \bsnm{{Bellm}}},
\bauthor{\binits{S.E.} \bsnm{{Boggs}}},
\bauthor{\binits{D.} \bsnm{{Chakrabarty}}},
\bauthor{\binits{J.} \bsnm{{Chenevez}}},
\bauthor{\binits{F.E.} \bsnm{{Christensen}}},
\bauthor{\binits{W.W.} \bsnm{{Craig}}},
\bauthor{\binits{B.W.} \bsnm{{Grefenstette}}},
\bauthor{\binits{C.J.} \bsnm{{Hailey}}},
\bauthor{\binits{K.K.} \bsnm{{Madsen}}},
\bauthor{\binits{L.} \bsnm{{Natalucci}}},
\bauthor{\binits{K.} \bsnm{{Pottschmidt}}},
\bauthor{\binits{D.} \bsnm{{Stern}}},
\bauthor{\binits{J.A.} \bsnm{{Tomsick}}},
\bauthor{\binits{D.J.} \bsnm{{Walton}}},
\bauthor{\binits{J.} \bsnm{{Wilms}}},
\bauthor{\binits{W.} \bsnm{{Zhang}}},
\batitle{{NuSTAR Detection of Hard X-Ray Phase Lags from the Accreting Pulsar
  GS 0834-430}}.
\bjtitle{\apj}
\bvolume{775},
\bfpage{65}
(\byear{2013}).
doi:\doiurl{10.1088/0004-637X/775/1/65}
\end{barticle}
\endbibitem

\bibitem[\protect\citeauthoryear{{Miyawaki} et~al.}{2009}]{Miyawaki09}
\begin{barticle}
\bauthor{\binits{R.} \bsnm{{Miyawaki}}},
\bauthor{\binits{K.} \bsnm{{Makishima}}},
\bauthor{\binits{S.} \bsnm{{Yamada}}},
\bauthor{\binits{P.} \bsnm{{Gandhi}}},
\bauthor{\binits{T.} \bsnm{{Mizuno}}},
\bauthor{\binits{A.} \bsnm{{Kubota}}},
\bauthor{\binits{T.G.} \bsnm{{Tsuru}}},
\bauthor{\binits{H.} \bsnm{{Matsumoto}}},
\batitle{{Suzaku Observations of M 82 X-1 : Detection of a Curved Hard X-Ray
  Spectrum}}.
\bjtitle{\pasj}
\bvolume{61},
\bfpage{263}--\blpage{278}
(\byear{2009}).
doi:\doiurl{10.1093/pasj/61.sp1.S263}
\end{barticle}
\endbibitem

\bibitem[\protect\citeauthoryear{{Miyoshi} et~al.}{1986}]{Miyoshi86}
\begin{barticle}
\bauthor{\binits{S.} \bsnm{{Miyoshi}}},
\bauthor{\binits{S.} \bsnm{{Hayakawa}}},
\bauthor{\binits{H.} \bsnm{{Kunieda}}},
\bauthor{\binits{F.} \bsnm{{Nagase}}},
\bauthor{\binits{Y.} \bsnm{{Tawara}}},
\batitle{{X-ray observation of AGN's from TENMA}}.
\bjtitle{\apss}
\bvolume{119},
\bfpage{185}--\blpage{190}
(\byear{1986}).
doi:\doiurl{10.1007/BF00648843}
\end{barticle}
\endbibitem

\bibitem[\protect\citeauthoryear{{Molendi} et~al.}{2002}]{Molendi02}
\begin{barticle}
\bauthor{\binits{S.} \bsnm{{Molendi}}},
\bauthor{\binits{S.} \bsnm{{De Grandi}}},
\bauthor{\binits{M.} \bsnm{{Guainazzi}}},
\batitle{{A BeppoSAX view of the Centaurus Cluster}}.
\bjtitle{\aap}
\bvolume{392},
\bfpage{13}--\blpage{17}
(\byear{2002}).
doi:\doiurl{10.1051/0004-6361:20020696}
\end{barticle}
\endbibitem

\bibitem[\protect\citeauthoryear{{Montanari} et~al.}{2009}]{Montanari09}
\begin{barticle}
\bauthor{\binits{E.} \bsnm{{Montanari}}},
\bauthor{\binits{L.} \bsnm{{Titarchuk}}},
\bauthor{\binits{F.} \bsnm{{Frontera}}},
\batitle{{BeppoSAX Observations of the Power and Energy Spectral Evolution in
  the Black Hole Candidate XTE J1650-500}}.
\bjtitle{\apj}
\bvolume{692},
\bfpage{1597}--\blpage{1608}
(\byear{2009}).
doi:\doiurl{10.1088/0004-637X/692/2/1597}
\end{barticle}
\endbibitem

\bibitem[\protect\citeauthoryear{{Montanari} et~al.}{2005}]{Montanari05}
\begin{barticle}
\bauthor{\binits{E.} \bsnm{{Montanari}}},
\bauthor{\binits{F.} \bsnm{{Frontera}}},
\bauthor{\binits{C.} \bsnm{{Guidorzi}}},
\bauthor{\binits{M.} \bsnm{{Rapisarda}}},
\batitle{{Evidence of a Long-Duration Component in the Prompt Emission of Short
  Gamma-Ray Bursts Detected with BeppoSAX}}.
\bjtitle{\apjl}
\bvolume{625},
\bfpage{17}--\blpage{21}
(\byear{2005}).
doi:\doiurl{10.1086/430759}
\end{barticle}
\endbibitem

\bibitem[\protect\citeauthoryear{{Mori} et~al.}{2013}]{Mori13}
\begin{barticle}
\bauthor{\binits{K.} \bsnm{{Mori}}},
\bauthor{\binits{E.V.} \bsnm{{Gotthelf}}},
\bauthor{\binits{S.} \bsnm{{Zhang}}},
\bauthor{\binits{H.} \bsnm{{An}}},
\bauthor{\binits{F.K.} \bsnm{{Baganoff}}},
\bauthor{\binits{N.M.} \bsnm{{Barri{\`e}re}}},
\bauthor{\binits{A.M.} \bsnm{{Beloborodov}}},
\bauthor{\binits{S.E.} \bsnm{{Boggs}}},
\bauthor{\binits{F.E.} \bsnm{{Christensen}}},
\bauthor{\binits{W.W.} \bsnm{{Craig}}},
\bauthor{\binits{F.} \bsnm{{Dufour}}},
\bauthor{\binits{B.W.} \bsnm{{Grefenstette}}},
\bauthor{\binits{C.J.} \bsnm{{Hailey}}},
\bauthor{\binits{F.A.} \bsnm{{Harrison}}},
\bauthor{\binits{J.} \bsnm{{Hong}}},
\bauthor{\binits{V.M.} \bsnm{{Kaspi}}},
\bauthor{\binits{J.A.} \bsnm{{Kennea}}},
\bauthor{\binits{K.K.} \bsnm{{Madsen}}},
\bauthor{\binits{C.B.} \bsnm{{Markwardt}}},
\bauthor{\binits{M.} \bsnm{{Nynka}}},
\bauthor{\binits{D.} \bsnm{{Stern}}},
\bauthor{\binits{J.A.} \bsnm{{Tomsick}}},
\bauthor{\binits{W.W.} \bsnm{{Zhang}}},
\batitle{{NuSTAR Discovery of a 3.76 s Transient Magnetar Near Sagittarius
  A*}}.
\bjtitle{\apjl}
\bvolume{770},
\bfpage{23}
(\byear{2013}).
doi:\doiurl{10.1088/2041-8205/770/2/L23}
\end{barticle}
\endbibitem

\bibitem[\protect\citeauthoryear{{Morris} et~al.}{2009}]{Morris09}
\begin{barticle}
\bauthor{\binits{D.C.} \bsnm{{Morris}}},
\bauthor{\binits{R.K.} \bsnm{{Smith}}},
\bauthor{\binits{C.B.} \bsnm{{Markwardt}}},
\bauthor{\binits{R.F.} \bsnm{{Mushotzky}}},
\bauthor{\binits{J.} \bsnm{{Tueller}}},
\bauthor{\binits{T.R.} \bsnm{{Kallman}}},
\bauthor{\binits{K.S.} \bsnm{{Dhuga}}},
\batitle{{Suzaku Observations of Four Heavily Absorbed HMXBs}}.
\bjtitle{\apj}
\bvolume{699},
\bfpage{892}--\blpage{901}
(\byear{2009}).
doi:\doiurl{10.1088/0004-637X/699/1/892}
\end{barticle}
\endbibitem

\bibitem[\protect\citeauthoryear{{Murakami}}{1989}]{Murakami89}
\begin{bchapter}
\bauthor{\binits{T.} \bsnm{{Murakami}}},
\bctitle{{Cyclotron absorption in gamma ray bursts seen with GINGA}},
in \bbtitle{Two Topics in X-Ray Astronomy, Volume 1: X Ray Binaries. Volume 2:
  AGN and the X Ray Background},
ed. by \beditor{\binits{J.} \bsnm{{Hunt}}},
\beditor{\binits{B.} \bsnm{{Battrick}}}
\bsertitle{ESA Special Publication},
vol. \bseriesno{296},
\byear{1989},
pp. \bfpage{173}--\blpage{177}
\end{bchapter}
\endbibitem

\bibitem[\protect\citeauthoryear{{Murakami}}{1991}]{Murakami91}
\begin{barticle}
\bauthor{\binits{T.} \bsnm{{Murakami}}},
\batitle{{The most recent results of GRB observations with GINGA}}.
\bjtitle{Advances in Space Research}
\bvolume{11},
\bfpage{119}--\blpage{123}
(\byear{1991}).
doi:\doiurl{10.1016/0273-1177(91)90160-L}
\end{barticle}
\endbibitem

\bibitem[\protect\citeauthoryear{{Murakami} et~al.}{1989}]{Murakami1989;ginga}
\begin{barticle}
\bauthor{\binits{T.} \bsnm{{Murakami}}},
\bauthor{\binits{M.} \bsnm{{Fujii}}},
\bauthor{\binits{K.} \bsnm{{Hayashida}}},
\bauthor{\binits{M.} \bsnm{{Itoh}}},
\bauthor{\binits{J.} \bsnm{{Nishimura}}},
\bauthor{\binits{T.} \bsnm{{Yamagami}}},
\bauthor{\binits{J.P.} \bsnm{{Conner}}},
\bauthor{\binits{W.D.} \bsnm{{Evans}}},
\bauthor{\binits{E.E.} \bsnm{{Fenimore}}},
\bauthor{\binits{R.W.} \bsnm{{Klebesadel}}},
\bauthor{\binits{K.M.} \bsnm{{Spencer}}},
\bauthor{\binits{H.} \bsnm{{Murakami}}},
\bauthor{\binits{N.} \bsnm{{Kawai}}},
\bauthor{\binits{I.} \bsnm{{Kondo}}},
\bauthor{\binits{M.} \bsnm{{Katoh}}},
\batitle{{The Gamma-ray Burst Detector system on board Ginga}}.
\bjtitle{\pasj}
\bvolume{41},
\bfpage{405}--\blpage{426}
(\byear{1989})
\end{barticle}
\endbibitem

\bibitem[\protect\citeauthoryear{{Murakami} et~al.}{1991}]{Murakami91b}
\begin{bchapter}
\bauthor{\binits{T.} \bsnm{{Murakami}}},
\bauthor{\binits{Y.} \bsnm{{Ogasaka}}},
\bauthor{\binits{A.} \bsnm{{Yoshida}}},
\bauthor{\binits{E.F.} \bsnm{{Fenimore}}},
\bctitle{{Ginga observations of gamma-ray bursts}},
in \bbtitle{American Institute of Physics Conference Series}.
\bsertitle{American Institute of Physics Conference Series},
vol. \bseriesno{265},
\byear{1991},
pp. \bfpage{28}--\blpage{31}.
doi:\doiurl{10.1063/1.42796}
\end{bchapter}
\endbibitem

\bibitem[\protect\citeauthoryear{{Mushotzky} et~al.}{1978}]{Mushotzky78}
\begin{barticle}
\bauthor{\binits{R.F.} \bsnm{{Mushotzky}}},
\bauthor{\binits{S.S.} \bsnm{{Holt}}},
\bauthor{\binits{P.J.} \bsnm{{Serlemitsos}}},
\batitle{{X-ray observations of a flare in NGC 4151 from OSO 8}}.
\bjtitle{\apjl}
\bvolume{225},
\bfpage{115}--\blpage{118}
(\byear{1978}).
doi:\doiurl{10.1086/182806}
\end{barticle}
\endbibitem

\bibitem[\protect\citeauthoryear{{Mushotzky} et~al.}{1975}]{Mushotzky75}
\begin{bchapter}
\bauthor{\binits{R.F.} \bsnm{{Mushotzky}}},
\bauthor{\binits{W.A.} \bsnm{{Baity}}},
\bauthor{\binits{W.A.} \bsnm{{Wheaton}}},
\bauthor{\binits{L.E.} \bsnm{{Peterson}}},
\bctitle{{USCD OSO-7 Observations of Four Extragalactic X-Ray Sources}},
in \bbtitle{Bulletin of the American Astronomical Society}.
\bsertitle{Bulletin of the American Astronomical Society},
vol. \bseriesno{7},
\byear{1975},
p. \bfpage{461}
\end{bchapter}
\endbibitem

\bibitem[\protect\citeauthoryear{{Mushotzky} et~al.}{1976}]{Mushotzky76}
\begin{barticle}
\bauthor{\binits{R.F.} \bsnm{{Mushotzky}}},
\bauthor{\binits{W.A.} \bsnm{{Baity}}},
\bauthor{\binits{W.A.} \bsnm{{Wheaton}}},
\bauthor{\binits{L.E.} \bsnm{{Peterson}}},
\batitle{{UCSD OSO-7 observations of the hard X-ray spectrum and variability of
  Centaurus A}}.
\bjtitle{\apjl}
\bvolume{206},
\bfpage{45}--\blpage{48}
(\byear{1976}).
doi:\doiurl{10.1086/182129}
\end{barticle}
\endbibitem

\bibitem[\protect\citeauthoryear{{Mushotzky} et~al.}{1980}]{Mushotzky80}
\begin{barticle}
\bauthor{\binits{R.F.} \bsnm{{Mushotzky}}},
\bauthor{\binits{F.E.} \bsnm{{Marshall}}},
\bauthor{\binits{E.A.} \bsnm{{Boldt}}},
\bauthor{\binits{S.S.} \bsnm{{Holt}}},
\bauthor{\binits{P.J.} \bsnm{{Serlemitsos}}},
\batitle{{HEAO 1 spectra of X-ray emitting Seyfert 1 galaxies}}.
\bjtitle{\apj}
\bvolume{235},
\bfpage{377}--\blpage{385}
(\byear{1980}).
doi:\doiurl{10.1086/157641}
\end{barticle}
\endbibitem

\bibitem[\protect\citeauthoryear{{Nakagawa} et~al.}{2011}]{Nakagawa11}
\begin{barticle}
\bauthor{\binits{Y.E.} \bsnm{{Nakagawa}}},
\bauthor{\binits{K.} \bsnm{{Makishima}}},
\bauthor{\binits{T.} \bsnm{{Enoto}}},
\batitle{{The Suzaku Discovery of A Hard Power-Law Component in the Spectra of
  Short Bursts from SGR 0501+4516}}.
\bjtitle{\pasj}
\bvolume{63},
\bfpage{813}--\blpage{820}
(\byear{2011}).
doi:\doiurl{10.1093/pasj/63.sp3.S813}
\end{barticle}
\endbibitem

\bibitem[\protect\citeauthoryear{{Nakamura} et~al.}{1989}]{Nakamura89}
\begin{barticle}
\bauthor{\binits{N.} \bsnm{{Nakamura}}},
\bauthor{\binits{T.} \bsnm{{Dotani}}},
\bauthor{\binits{H.} \bsnm{{Inoue}}},
\bauthor{\binits{K.} \bsnm{{Mitsuda}}},
\bauthor{\binits{Y.} \bsnm{{Tanaka}}},
\bauthor{\binits{M.} \bsnm{{Matsuoka}}},
\batitle{{TENMA observation of X-ray bursts from X1608-52}}.
\bjtitle{\pasj}
\bvolume{41},
\bfpage{617}--\blpage{639}
(\byear{1989})
\end{barticle}
\endbibitem

\bibitem[\protect\citeauthoryear{{Nakazawa} et~al.}{2009}]{Nakazawa09}
\begin{barticle}
\bauthor{\binits{K.} \bsnm{{Nakazawa}}},
\bauthor{\binits{C.L.} \bsnm{{Sarazin}}},
\bauthor{\binits{M.} \bsnm{{Kawaharada}}},
\bauthor{\binits{T.} \bsnm{{Kitaguchi}}},
\bauthor{\binits{S.} \bsnm{{Okuyama}}},
\bauthor{\binits{K.} \bsnm{{Makishima}}},
\bauthor{\binits{N.} \bsnm{{Kawano}}},
\bauthor{\binits{Y.} \bsnm{{Fukazawa}}},
\bauthor{\binits{S.} \bsnm{{Inoue}}},
\bauthor{\binits{M.} \bsnm{{Takizawa}}},
\bauthor{\binits{D.R.} \bsnm{{Wik}}},
\bauthor{\binits{A.} \bsnm{{Finoguenov}}},
\bauthor{\binits{T.E.} \bsnm{{Clarke}}},
\batitle{{Hard X-Ray Properties of the Merging Cluster Abell 3667 as Observed
  with Suzaku}}.
\bjtitle{\pasj}
\bvolume{61},
\bfpage{339}--\blpage{355}
(\byear{2009}).
doi:\doiurl{10.1093/pasj/61.2.339}
\end{barticle}
\endbibitem

\bibitem[\protect\citeauthoryear{{Natalucci} et~al.}{2000a}]{Natalucci00b}
\begin{barticle}
\bauthor{\binits{L.} \bsnm{{Natalucci}}},
\bauthor{\binits{A.} \bsnm{{Bazzano}}},
\bauthor{\binits{M.} \bsnm{{Cocchi}}},
\bauthor{\binits{P.} \bsnm{{Ubertini}}},
\bauthor{\binits{J.} \bsnm{{Heise}}},
\bauthor{\binits{E.} \bsnm{{Kuulkers}}},
\bauthor{\binits{J.J.M.} \bsnm{{in't Zand}}},
\batitle{{Broadband Observations of the New X-Ray Burster SAX J1747.0-2853
  during the 1998 March Outburst}}.
\bjtitle{\apjl}
\bvolume{543},
\bfpage{73}--\blpage{76}
(\byear{2000}a).
doi:\doiurl{10.1086/318173}
\end{barticle}
\endbibitem

\bibitem[\protect\citeauthoryear{{Natalucci} et~al.}{2000b}]{Natalucci00}
\begin{barticle}
\bauthor{\binits{L.} \bsnm{{Natalucci}}},
\bauthor{\binits{A.} \bsnm{{Bazzano}}},
\bauthor{\binits{M.} \bsnm{{Cocchi}}},
\bauthor{\binits{P.} \bsnm{{Ubertini}}},
\bauthor{\binits{J.} \bsnm{{Heise}}},
\bauthor{\binits{E.} \bsnm{{Kuulkers}}},
\bauthor{\binits{J.J.M.} \bsnm{{in 't Zand}}},
\bauthor{\binits{M.J.S.} \bsnm{{Smith}}},
\batitle{{SAX J1810.8-2609: A New Hard X-Ray Bursting Transient}}.
\bjtitle{\apj}
\bvolume{536},
\bfpage{891}--\blpage{895}
(\byear{2000}b).
doi:\doiurl{10.1086/308987}
\end{barticle}
\endbibitem

\bibitem[\protect\citeauthoryear{{Nava} et~al.}{2011}]{Nava11}
\begin{barticle}
\bauthor{\binits{L.} \bsnm{{Nava}}},
\bauthor{\binits{G.} \bsnm{{Ghirlanda}}},
\bauthor{\binits{G.} \bsnm{{Ghisellini}}},
\bauthor{\binits{A.} \bsnm{{Celotti}}},
\batitle{{Spectral properties of 438 GRBs detected by Fermi/GBM}}.
\bjtitle{\aap}
\bvolume{530},
\bfpage{21}
(\byear{2011}).
doi:\doiurl{10.1051/0004-6361/201016270}
\end{barticle}
\endbibitem

\bibitem[\protect\citeauthoryear{{Nishimura} et~al.}{1978}]{Nishimura78}
\begin{barticle}
\bauthor{\binits{J.} \bsnm{{Nishimura}}},
\bauthor{\binits{M.} \bsnm{{Fujii}}},
\bauthor{\binits{Y.} \bsnm{{Tawara}}},
\bauthor{\binits{M.} \bsnm{{Oda}}},
\bauthor{\binits{Y.} \bsnm{{Ogawara}}},
\bauthor{\binits{T.} \bsnm{{Yamagami}}},
\bauthor{\binits{S.} \bsnm{{Miyamoto}}},
\bauthor{\binits{M.} \bsnm{{Kajiwara}}},
\bauthor{\binits{H.} \bsnm{{Murakami}}},
\bauthor{\binits{M.} \bsnm{{Yoshimori}}},
\batitle{{Gamma-ray burst observed at balloon altitude}}.
\bjtitle{\nat}
\bvolume{272},
\bfpage{337}
(\byear{1978}).
doi:\doiurl{10.1038/272337a0}
\end{barticle}
\endbibitem

\bibitem[\protect\citeauthoryear{{Oberlack}
  et~al.}{2000}]{Oberlack2000;lxegrit}
\begin{bchapter}
\bauthor{\binits{U.G.} \bsnm{{Oberlack}}},
\bauthor{\binits{E.} \bsnm{{Aprile}}},
\bauthor{\binits{A.} \bsnm{{Curioni}}},
\bauthor{\binits{V.} \bsnm{{Egorov}}},
\bauthor{\binits{K.-L.} \bsnm{{Giboni}}},
\bctitle{{Compton scattering sequence reconstruction algorithm for the liquid
  xenon gamma-ray imaging telescope (LXeGRIT)}},
in \bbtitle{Hard X-Ray, Gamma-Ray, and Neutron Detector Physics II},
ed. by \beditor{\binits{R.B.} \bsnm{{James}}},
\beditor{\binits{R.C.} \bsnm{{Schirato}}}
\bsertitle{Society of Photo-Optical Instrumentation Engineers (SPIE) Conference
  Series},
vol. \bseriesno{4141},
\byear{2000},
pp. \bfpage{168}--\blpage{177}
\end{bchapter}
\endbibitem

\bibitem[\protect\citeauthoryear{{Ofek} et~al.}{2014}]{Ofek14}
\begin{barticle}
\bauthor{\binits{E.O.} \bsnm{{Ofek}}},
\bauthor{\binits{A.} \bsnm{{Zoglauer}}},
\bauthor{\binits{S.E.} \bsnm{{Boggs}}},
\bauthor{\binits{N.M.} \bsnm{{Barri{\'e}re}}},
\bauthor{\binits{S.P.} \bsnm{{Reynolds}}},
\bauthor{\binits{C.L.} \bsnm{{Fryer}}},
\bauthor{\binits{F.A.} \bsnm{{Harrison}}},
\bauthor{\binits{S.B.} \bsnm{{Cenko}}},
\bauthor{\binits{S.R.} \bsnm{{Kulkarni}}},
\bauthor{\binits{A.} \bsnm{{Gal-Yam}}},
\bauthor{\binits{I.} \bsnm{{Arcavi}}},
\bauthor{\binits{E.} \bsnm{{Bellm}}},
\bauthor{\binits{J.S.} \bsnm{{Bloom}}},
\bauthor{\binits{F.} \bsnm{{Christensen}}},
\bauthor{\binits{W.W.} \bsnm{{Craig}}},
\bauthor{\binits{W.} \bsnm{{Even}}},
\bauthor{\binits{A.V.} \bsnm{{Filippenko}}},
\bauthor{\binits{B.} \bsnm{{Grefenstette}}},
\bauthor{\binits{C.J.} \bsnm{{Hailey}}},
\bauthor{\binits{R.} \bsnm{{Laher}}},
\bauthor{\binits{K.} \bsnm{{Madsen}}},
\bauthor{\binits{E.} \bsnm{{Nakar}}},
\bauthor{\binits{P.E.} \bsnm{{Nugent}}},
\bauthor{\binits{D.} \bsnm{{Stern}}},
\bauthor{\binits{M.} \bsnm{{Sullivan}}},
\bauthor{\binits{J.} \bsnm{{Surace}}},
\bauthor{\binits{W.W.} \bsnm{{Zhang}}},
\batitle{{SN 2010jl: Optical to Hard X-Ray Observations Reveal an Explosion
  Embedded in a Ten Solar Mass Cocoon}}.
\bjtitle{\apj}
\bvolume{781},
\bfpage{42}
(\byear{2014}).
doi:\doiurl{10.1088/0004-637X/781/1/42}
\end{barticle}
\endbibitem

\bibitem[\protect\citeauthoryear{{Ohno} et~al.}{2008}]{Ohno08}
\begin{barticle}
\bauthor{\binits{M.} \bsnm{{Ohno}}},
\bauthor{\binits{Y.} \bsnm{{Fukazawa}}},
\bauthor{\binits{T.} \bsnm{{Takahashi}}},
\bauthor{\binits{K.} \bsnm{{Yamaoka}}},
\bauthor{\binits{S.} \bsnm{{Sugita}}},
\bauthor{\binits{V.} \bsnm{{Pal'Shin}}},
\bauthor{\binits{D.} \bsnm{{Frederiks}}},
\bauthor{\binits{P.} \bsnm{{Oleynik}}},
\bauthor{\binits{M.} \bsnm{{Ulanov}}},
\bauthor{\binits{T.} \bsnm{{Sakamoto}}},
\bauthor{\binits{G.} \bsnm{{Sato}}},
\bauthor{\binits{K.} \bsnm{{Hurley}}},
\bauthor{\binits{M.S.} \bsnm{{Tashiro}}},
\bauthor{\binits{Y.} \bsnm{{Urata}}},
\bauthor{\binits{K.} \bsnm{{Onda}}},
\bauthor{\binits{T.} \bsnm{{Tamagawa}}},
\bauthor{\binits{Y.} \bsnm{{Terada}}},
\bauthor{\binits{M.} \bsnm{{Suzuki}}},
\bauthor{\binits{H.} \bsnm{{Soojing}}},
\batitle{{Spectral Properties of Prompt Emission of Four Short Gamma-Ray Bursts
  Observed by the Suzaku-WAM and the Konus-Wind}}.
\bjtitle{\pasj}
\bvolume{60},
\bfpage{361}--\blpage{374}
(\byear{2008})
\end{barticle}
\endbibitem

\bibitem[\protect\citeauthoryear{{Oosterbroek} et~al.}{1998}]{Oosterbroek98}
\begin{barticle}
\bauthor{\binits{T.} \bsnm{{Oosterbroek}}},
\bauthor{\binits{A.N.} \bsnm{{Parmar}}},
\bauthor{\binits{E.} \bsnm{{Kuulkers}}},
\bauthor{\binits{T.} \bsnm{{Belloni}}},
\bauthor{\binits{M.} \bsnm{{van der Klis}}},
\bauthor{\binits{F.} \bsnm{{Frontera}}},
\bauthor{\binits{A.} \bsnm{{Santangelo}}},
\batitle{{The 1998 outburst of the X-ray transient 4U 1630-47 observed with it
  BeppoSAX}}.
\bjtitle{\aap}
\bvolume{340},
\bfpage{431}--\blpage{436}
(\byear{1998})
\end{barticle}
\endbibitem

\bibitem[\protect\citeauthoryear{{Orlandini} et~al.}{1998a}]{Orlandini98b}
\begin{barticle}
\bauthor{\binits{M.} \bsnm{{Orlandini}}},
\bauthor{\binits{D.} \bsnm{{Dal Fiume}}},
\bauthor{\binits{F.} \bsnm{{Frontera}}},
\bauthor{\binits{S.} \bsnm{{Del Sordo}}},
\bauthor{\binits{S.} \bsnm{{Piraino}}},
\bauthor{\binits{A.} \bsnm{{Santangelo}}},
\bauthor{\binits{A.} \bsnm{{Segreto}}},
\bauthor{\binits{T.} \bsnm{{Oosterbroek}}},
\bauthor{\binits{A.N.} \bsnm{{Parmar}}},
\batitle{{BEPPOSAX Observation of 4U 1626-67: Discovery of an Absorption
  Cyclotron Resonance Feature}}.
\bjtitle{\apjl}
\bvolume{500},
\bfpage{163}--\blpage{166}
(\byear{1998}a).
doi:\doiurl{10.1086/311404}
\end{barticle}
\endbibitem

\bibitem[\protect\citeauthoryear{{Orlandini} et~al.}{1998b}]{Orlandini98}
\begin{barticle}
\bauthor{\binits{M.} \bsnm{{Orlandini}}},
\bauthor{\binits{D.} \bsnm{{dal Fiume}}},
\bauthor{\binits{F.} \bsnm{{Frontera}}},
\bauthor{\binits{G.} \bsnm{{Cusumano}}},
\bauthor{\binits{S.} \bsnm{{del Sordo}}},
\bauthor{\binits{S.} \bsnm{{Giarrusso}}},
\bauthor{\binits{S.} \bsnm{{Piraino}}},
\bauthor{\binits{A.} \bsnm{{Segreto}}},
\bauthor{\binits{M.} \bsnm{{Guainazzi}}},
\bauthor{\binits{L.} \bsnm{{Piro}}},
\batitle{{The VELA X-1 pulse-averaged spectrum as observed by BeppoSAX}}.
\bjtitle{\aap}
\bvolume{332},
\bfpage{121}--\blpage{126}
(\byear{1998}b)
\end{barticle}
\endbibitem

\bibitem[\protect\citeauthoryear{{Ota} et~al.}{2014}]{Ota14}
\begin{barticle}
\bauthor{\binits{N.} \bsnm{{Ota}}},
\bauthor{\binits{K.} \bsnm{{Nagayoshi}}},
\bauthor{\binits{G.W.} \bsnm{{Pratt}}},
\bauthor{\binits{T.} \bsnm{{Kitayama}}},
\bauthor{\binits{T.} \bsnm{{Oshima}}},
\bauthor{\binits{T.H.} \bsnm{{Reiprich}}},
\batitle{{Investigating the hard X-ray emission from the hottest Abell cluster
  A2163 with Suzaku}}.
\bjtitle{\aap}
\bvolume{562},
\bfpage{60}
(\byear{2014}).
doi:\doiurl{10.1051/0004-6361/201322878}
\end{barticle}
\endbibitem

\bibitem[\protect\citeauthoryear{{Overbeck} and {Tananbaum}}{1968}]{Overbeck68}
\begin{barticle}
\bauthor{\binits{J.W.} \bsnm{{Overbeck}}},
\bauthor{\binits{H.D.} \bsnm{{Tananbaum}}},
\batitle{{Time Variations in Scorpius X-1 and Cygnus XR-1}}.
\bjtitle{\apj}
\bvolume{153},
\bfpage{899}
(\byear{1968}).
doi:\doiurl{10.1086/149714}
\end{barticle}
\endbibitem

\bibitem[\protect\citeauthoryear{{Paciesas} et~al.}{1999}]{Paciesas99}
\begin{barticle}
\bauthor{\binits{W.S.} \bsnm{{Paciesas}}},
\bauthor{\binits{C.A.} \bsnm{{Meegan}}},
\bauthor{\binits{G.N.} \bsnm{{Pendleton}}},
\bauthor{\binits{M.S.} \bsnm{{Briggs}}},
\bauthor{\binits{C.} \bsnm{{Kouveliotou}}},
\bauthor{\binits{T.M.} \bsnm{{Koshut}}},
\bauthor{\binits{J.P.} \bsnm{{Lestrade}}},
\bauthor{\binits{M.L.} \bsnm{{McCollough}}},
\bauthor{\binits{J.J.} \bsnm{{Brainerd}}},
\bauthor{\binits{J.} \bsnm{{Hakkila}}},
\bauthor{\binits{W.} \bsnm{{Henze}}},
\bauthor{\binits{R.D.} \bsnm{{Preece}}},
\bauthor{\binits{V.} \bsnm{{Connaughton}}},
\bauthor{\binits{R.M.} \bsnm{{Kippen}}},
\bauthor{\binits{R.S.} \bsnm{{Mallozzi}}},
\bauthor{\binits{G.J.} \bsnm{{Fishman}}},
\bauthor{\binits{G.A.} \bsnm{{Richardson}}},
\bauthor{\binits{M.} \bsnm{{Sahi}}},
\batitle{{The Fourth BATSE Gamma-Ray Burst Catalog (Revised)}}.
\bjtitle{\apjs}
\bvolume{122},
\bfpage{465}--\blpage{495}
(\byear{1999}).
doi:\doiurl{10.1086/313224}
\end{barticle}
\endbibitem

\bibitem[\protect\citeauthoryear{{Pamini} et~al.}{1990}]{Pamini90}
\begin{barticle}
\bauthor{\binits{M.} \bsnm{{Pamini}}},
\bauthor{\binits{L.} \bsnm{{Natalucci}}},
\bauthor{\binits{D.} \bsnm{{dal Fiume}}},
\bauthor{\binits{F.} \bsnm{{Frontera}}},
\bauthor{\binits{E.} \bsnm{{Costa}}},
\batitle{{The gamma-ray burst monitor on board the SAX satellite}}.
\bjtitle{Nuovo Cimento C Geophysics Space Physics C}
\bvolume{13},
\bfpage{337}--\blpage{344}
(\byear{1990}).
doi:\doiurl{10.1007/BF02507071}
\end{barticle}
\endbibitem

\bibitem[\protect\citeauthoryear{{Pearce} et~al.}{2008}]{Pearce08}
\begin{barticle}
\bauthor{\binits{M.} \bsnm{{Pearce}}},
\bauthor{\binits{M.} \bsnm{{Arimoto}}},
\bauthor{\binits{M.} \bsnm{{Axelsson}}},
\bauthor{\binits{C.-I.} \bsnm{{Bj{\"o}rnsson}}},
\bauthor{\binits{G.} \bsnm{{Bogaert}}},
\bauthor{\binits{P.} \bsnm{{Carlson}}},
\bauthor{\binits{W.} \bsnm{{Craig}}},
\bauthor{\binits{Y.} \bsnm{{Fukazawa}}},
\bauthor{\binits{S.} \bsnm{{Gunji}}},
\bauthor{\binits{L.} \bsnm{{Hjalmarsdotter}}},
\bauthor{\binits{T.} \bsnm{{Kamae}}},
\bauthor{\binits{Y.} \bsnm{{Kanai}}},
\bauthor{\binits{J.} \bsnm{{Kataoka}}},
\bauthor{\binits{J.} \bsnm{{Katsuta}}},
\bauthor{\binits{N.} \bsnm{{Kawai}}},
\bauthor{\binits{J.} \bsnm{{Kazejev}}},
\bauthor{\binits{M.} \bsnm{{Kiss}}},
\bauthor{\binits{W.} \bsnm{{Klamra}}},
\bauthor{\binits{S.} \bsnm{{Larsson}}},
\bauthor{\binits{G.} \bsnm{{Madejski}}},
\bauthor{\binits{C.} \bsnm{{Marini Bettolo}}},
\bauthor{\binits{T.} \bsnm{{Mizuno}}},
\bauthor{\binits{J.} \bsnm{{Ng}}},
\bauthor{\binits{M.} \bsnm{{Nomachi}}},
\bauthor{\binits{H.} \bsnm{{Odaka}}},
\bauthor{\binits{F.} \bsnm{{Ryde}}},
\bauthor{\binits{H.} \bsnm{{Tajima}}},
\bauthor{\binits{H.} \bsnm{{Takahashi}}},
\bauthor{\binits{T.} \bsnm{{Takahashi}}},
\bauthor{\binits{T.} \bsnm{{Tanaka}}},
\bauthor{\binits{T.} \bsnm{{Thurston}}},
\bauthor{\binits{M.} \bsnm{{Ueno}}},
\bauthor{\binits{G.} \bsnm{{Varner}}},
\bauthor{\binits{H.} \bsnm{{Yoshida}}},
\bauthor{\binits{T.} \bsnm{{Yuasa}}},
\batitle{{PoGOLite: a balloon-borne soft gamma-ray polarimeter}}.
\bjtitle{International Cosmic Ray Conference}
\bvolume{2},
\bfpage{479}--\blpage{482}
(\byear{2008})
\end{barticle}
\endbibitem

\bibitem[\protect\citeauthoryear{{Pelaez} et~al.}{1991}]{Pelaez1991;granat}
\begin{barticle}
\bauthor{\binits{F.} \bsnm{{Pelaez}}},
\bauthor{\binits{J.-L.} \bsnm{{Atteia}}},
\bauthor{\binits{B.} \bsnm{{Mena}}},
\bauthor{\binits{M.} \bsnm{{Niel}}},
\bauthor{\binits{J.} \bsnm{{Ballet}}},
\bauthor{\binits{B.} \bsnm{{Cordier}}},
\bauthor{\binits{A.} \bsnm{{Lambert}}},
\bauthor{\binits{J.} \bsnm{{Paul}}},
\bauthor{\binits{R.A.} \bsnm{{Sunyaev}}},
\bauthor{\binits{D.} \bsnm{{Stepanov}}},
\batitle{{First gamma-ray burst observations with the Sigma telescope}}.
\bjtitle{Advances in Space Research}
\bvolume{11},
\bfpage{135}--\blpage{141}
(\byear{1991}).
doi:\doiurl{10.1016/0273-1177(91)90163-E}
\end{barticle}
\endbibitem

\bibitem[\protect\citeauthoryear{{Pellegrini} et~al.}{2000}]{Pellegrini00}
\begin{barticle}
\bauthor{\binits{S.} \bsnm{{Pellegrini}}},
\bauthor{\binits{M.} \bsnm{{Cappi}}},
\bauthor{\binits{L.} \bsnm{{Bassani}}},
\bauthor{\binits{R.} \bsnm{{della Ceca}}},
\bauthor{\binits{G.G.C.} \bsnm{{Palumbo}}},
\batitle{{The 0.1-100 keV view of NGC3998: an AGN origin for the LINER
  activity}}.
\bjtitle{\aap}
\bvolume{360},
\bfpage{878}--\blpage{886}
(\byear{2000})
\end{barticle}
\endbibitem

\bibitem[\protect\citeauthoryear{{Pendleton} et~al.}{1995}]{Pendleton95}
\begin{barticle}
\bauthor{\binits{G.N.} \bsnm{{Pendleton}}},
\bauthor{\binits{W.S.} \bsnm{{Paciesas}}},
\bauthor{\binits{G.J.} \bsnm{{Fishman}}},
\bauthor{\binits{C.A.} \bsnm{{Meegan}}},
\bauthor{\binits{R.B.} \bsnm{{Wilson}}},
\batitle{{Balloon-borne measurements of the SN 1987A hard X-ray continuum}}.
\bjtitle{\apj}
\bvolume{439},
\bfpage{963}--\blpage{975}
(\byear{1995}).
doi:\doiurl{10.1086/175233}
\end{barticle}
\endbibitem

\bibitem[\protect\citeauthoryear{{Penningsfeld} et~al.}{1979}]{Penningsfeld79}
\begin{barticle}
\bauthor{\binits{F.-P.} \bsnm{{Penningsfeld}}},
\bauthor{\binits{U.} \bsnm{{Graser}}},
\bauthor{\binits{V.} \bsnm{{Sch{\"o}nfelder}}},
\batitle{{The Energy Spectrum of the Crab Nebula and the Pulsar NP 0532 AT
  Mev-Energies}}.
\bjtitle{International Cosmic Ray Conference}
\bvolume{1},
\bfpage{101}
(\byear{1979})
\end{barticle}
\endbibitem

\bibitem[\protect\citeauthoryear{{Perez} et~al.}{2015}]{Perez15}
\begin{barticle}
\bauthor{\binits{K.} \bsnm{{Perez}}},
\bauthor{\binits{C.J.} \bsnm{{Hailey}}},
\bauthor{\binits{F.E.} \bsnm{{Bauer}}},
\bauthor{\binits{R.A.} \bsnm{{Krivonos}}},
\bauthor{\binits{K.} \bsnm{{Mori}}},
\bauthor{\binits{F.K.} \bsnm{{Baganoff}}},
\bauthor{\binits{N.M.} \bsnm{{Barri{\`e}re}}},
\bauthor{\binits{S.E.} \bsnm{{Boggs}}},
\bauthor{\binits{F.E.} \bsnm{{Christensen}}},
\bauthor{\binits{W.W.} \bsnm{{Craig}}},
\bauthor{\binits{B.W.} \bsnm{{Grefenstette}}},
\bauthor{\binits{J.E.} \bsnm{{Grindlay}}},
\bauthor{\binits{F.A.} \bsnm{{Harrison}}},
\bauthor{\binits{J.} \bsnm{{Hong}}},
\bauthor{\binits{K.K.} \bsnm{{Madsen}}},
\bauthor{\binits{M.} \bsnm{{Nynka}}},
\bauthor{\binits{D.} \bsnm{{Stern}}},
\bauthor{\binits{J.A.} \bsnm{{Tomsick}}},
\bauthor{\binits{D.R.} \bsnm{{Wik}}},
\bauthor{\binits{S.} \bsnm{{Zhang}}},
\bauthor{\binits{W.W.} \bsnm{{Zhang}}},
\bauthor{\binits{A.} \bsnm{{Zoglauer}}},
\batitle{{Extended hard-X-ray emission in the inner few parsecs of the
  Galaxy}}.
\bjtitle{\nat}
\bvolume{520},
\bfpage{646}--\blpage{649}
(\byear{2015}).
doi:\doiurl{10.1038/nature14353}
\end{barticle}
\endbibitem

\bibitem[\protect\citeauthoryear{{Perotti} et~al.}{1979}]{Perotti79}
\begin{barticle}
\bauthor{\binits{F.} \bsnm{{Perotti}}},
\bauthor{\binits{A.D.} \bsnm{{Ventura}}},
\bauthor{\binits{G.} \bsnm{{Sechi}}},
\bauthor{\binits{G.} \bsnm{{Villa}}},
\bauthor{\binits{G.} \bsnm{{di Cocco}}},
\bauthor{\binits{R.E.} \bsnm{{Baker}}},
\bauthor{\binits{R.C.} \bsnm{{Butler}}},
\bauthor{\binits{A.J.} \bsnm{{Dean}}},
\bauthor{\binits{S.J.} \bsnm{{Martin}}},
\bauthor{\binits{D.} \bsnm{{Ramsden}}},
\batitle{{Low energy gamma-ray spectrum of NGC4151}}.
\bjtitle{\nat}
\bvolume{282},
\bfpage{484}--\blpage{486}
(\byear{1979}).
doi:\doiurl{10.1038/282484a0}
\end{barticle}
\endbibitem

\bibitem[\protect\citeauthoryear{{Perotti} et~al.}{1986}]{Perotti1986;miso}
\begin{barticle}
\bauthor{\binits{F.} \bsnm{{Perotti}}},
\bauthor{\binits{A.} \bsnm{{Della Ventura}}},
\bauthor{\binits{G.} \bsnm{{Villa}}},
\bauthor{\binits{L.} \bsnm{{Bassani}}},
\bauthor{\binits{R.C.} \bsnm{{Butler}}},
\bauthor{\binits{G.} \bsnm{{Di Cocco}}},
\bauthor{\binits{R.E.} \bsnm{{Baker}}},
\bauthor{\binits{A.J.} \bsnm{{Dean}}},
\bauthor{\binits{C.G.} \bsnm{{Hanson}}},
\batitle{{Hard X-ray observations of Cygnus X-1 with the MISO telescope}}.
\bjtitle{\apj}
\bvolume{300},
\bfpage{297}--\blpage{303}
(\byear{1986}).
doi:\doiurl{10.1086/163803}
\end{barticle}
\endbibitem

\bibitem[\protect\citeauthoryear{{Perotti} et~al.}{1990a}]{Perotti90b}
\begin{barticle}
\bauthor{\binits{F.} \bsnm{{Perotti}}},
\bauthor{\binits{R.} \bsnm{{Buratti}}},
\bauthor{\binits{P.} \bsnm{{Maggioli}}},
\bauthor{\binits{E.} \bsnm{{Quadrini}}},
\bauthor{\binits{A.} \bsnm{{Bazzano}}},
\bauthor{\binits{P.} \bsnm{{Ubertini}}},
\bauthor{\binits{L.} \bsnm{{Bassani}}},
\bauthor{\binits{J.B.} \bsnm{{Stephen}}},
\bauthor{\binits{A.J.} \bsnm{{Court}}},
\bauthor{\binits{A.J.} \bsnm{{Dean}}},
\bauthor{\binits{N.A.} \bsnm{{Dipper}}},
\bauthor{\binits{R.A.} \bsnm{{Lewis}}},
\batitle{{Hard X-ray observation of NGC 4151 with the MIFRASO telescope}}.
\bjtitle{\apj}
\bvolume{356},
\bfpage{467}--\blpage{471}
(\byear{1990}a).
doi:\doiurl{10.1086/168854}
\end{barticle}
\endbibitem

\bibitem[\protect\citeauthoryear{{Perotti} et~al.}{1990b}]{Perotti90}
\begin{barticle}
\bauthor{\binits{F.} \bsnm{{Perotti}}},
\bauthor{\binits{L.} \bsnm{{Bassani}}},
\bauthor{\binits{A.} \bsnm{{Bazzano}}},
\bauthor{\binits{A.J.} \bsnm{{Court}}},
\bauthor{\binits{A.J.} \bsnm{{Dean}}},
\bauthor{\binits{N.A.} \bsnm{{Dipper}}},
\bauthor{\binits{R.A.} \bsnm{{Lewis}}},
\bauthor{\binits{P.} \bsnm{{Maggioli}}},
\bauthor{\binits{M.} \bsnm{{Quadrini}}},
\bauthor{\binits{J.B.} \bsnm{{Stephen}}},
\bauthor{\binits{P.} \bsnm{{Ubertini}}},
\batitle{{MCG8-11-11 - A powerful source of X- and gamma-rays}}.
\bjtitle{\aap}
\bvolume{234},
\bfpage{106}--\blpage{108}
(\byear{1990}b)
\end{barticle}
\endbibitem

\bibitem[\protect\citeauthoryear{{Peterson}}{1973}]{Peterson1973;oso7}
\begin{bchapter}
\bauthor{\binits{L.E.} \bsnm{{Peterson}}},
\bctitle{{Hard Cosmic X-Ray Sources}},
in \bbtitle{X- and Gamma-Ray Astronomy},
ed. by \beditor{\binits{H.} \bsnm{{Bradt}}},
\beditor{\binits{R.} \bsnm{{Giacconi}}}
\bsertitle{IAU Symposium},
vol. \bseriesno{55},
\byear{1973},
p. \bfpage{51}
\end{bchapter}
\endbibitem

\bibitem[\protect\citeauthoryear{{Peterson} and {Jacobson}}{1966}]{Peterson66a}
\begin{barticle}
\bauthor{\binits{L.E.} \bsnm{{Peterson}}},
\bauthor{\binits{A.S.} \bsnm{{Jacobson}}},
\batitle{{The Spectrum of Scorpius XR-1 to 50 KEV}}.
\bjtitle{\apj}
\bvolume{145},
\bfpage{962}
(\byear{1966}).
doi:\doiurl{10.1086/148848}
\end{barticle}
\endbibitem

\bibitem[\protect\citeauthoryear{{Peterson} et~al.}{1966}]{Peterson66b}
\begin{barticle}
\bauthor{\binits{L.E.} \bsnm{{Peterson}}},
\bauthor{\binits{A.S.} \bsnm{{Jacobson}}},
\bauthor{\binits{R.M.} \bsnm{{Pelling}}},
\batitle{{Spectrum of Crab Nebula X Rays to 120 keV}}.
\bjtitle{Physical Review Letters}
\bvolume{16},
\bfpage{142}--\blpage{144}
(\byear{1966}).
doi:\doiurl{10.1103/PhysRevLett.16.142}
\end{barticle}
\endbibitem

\bibitem[\protect\citeauthoryear{{Peterson} et~al.}{1968}]{Peterson68}
\begin{barticle}
\bauthor{\binits{L.E.} \bsnm{{Peterson}}},
\bauthor{\binits{A.S.} \bsnm{{Jacobson}}},
\bauthor{\binits{R.M.} \bsnm{{Pelling}}},
\bauthor{\binits{D.A.} \bsnm{{Schwartz}}},
\batitle{{Observations of cosmic X-ray source in the 10 - 250 keV range.}}
\bjtitle{Canadian Journal of Physics}
\bvolume{46},
\bfpage{437}
(\byear{1968})
\end{barticle}
\endbibitem

\bibitem[\protect\citeauthoryear{{Phlips} et~al.}{1996}]{Phlips96}
\begin{barticle}
\bauthor{\binits{B.F.} \bsnm{{Phlips}}},
\bauthor{\binits{G.V.} \bsnm{{Jung}}},
\bauthor{\binits{M.D.} \bsnm{{Leising}}},
\bauthor{\binits{J.E.} \bsnm{{Grove}}},
\bauthor{\binits{W.N.} \bsnm{{Johnson}}},
\bauthor{\binits{R.L.} \bsnm{{Kinzer}}},
\bauthor{\binits{R.A.} \bsnm{{Kroeger}}},
\bauthor{\binits{J.D.} \bsnm{{Kurfess}}},
\bauthor{\binits{M.S.} \bsnm{{Strickman}}},
\bauthor{\binits{D.A.} \bsnm{{Grabelsky}}},
\bauthor{\binits{S.M.} \bsnm{{Matz}}},
\bauthor{\binits{W.R.} \bsnm{{Purcell}}},
\bauthor{\binits{M.P.} \bsnm{{Ulmer}}},
\bauthor{\binits{K.} \bsnm{{McNaron-Brown}}},
\batitle{{Gamma ray observations of Cygnus X-1 with OSSE.}}
\bjtitle{\aaps}
\bvolume{120},
\bfpage{155}--\blpage{158}
(\byear{1996})
\end{barticle}
\endbibitem

\bibitem[\protect\citeauthoryear{{Pian} et~al.}{1998}]{Pian98}
\begin{barticle}
\bauthor{\binits{E.} \bsnm{{Pian}}},
\bauthor{\binits{G.} \bsnm{{Vacanti}}},
\bauthor{\binits{G.} \bsnm{{Tagliaferri}}},
\bauthor{\binits{G.} \bsnm{{Ghisellini}}},
\bauthor{\binits{L.} \bsnm{{Maraschi}}},
\bauthor{\binits{A.} \bsnm{{Treves}}},
\bauthor{\binits{C.M.} \bsnm{{Urry}}},
\bauthor{\binits{F.} \bsnm{{Fiore}}},
\bauthor{\binits{P.} \bsnm{{Giommi}}},
\bauthor{\binits{E.} \bsnm{{Palazzi}}},
\bauthor{\binits{L.} \bsnm{{Chiappetti}}},
\bauthor{\binits{R.M.} \bsnm{{Sambruna}}},
\batitle{{BeppoSAX Observations of Unprecedented Synchrotron Activity in the BL
  Lacertae Object Markarian 501}}.
\bjtitle{\apjl}
\bvolume{492},
\bfpage{17}--\blpage{20}
(\byear{1998}).
doi:\doiurl{10.1086/311083}
\end{barticle}
\endbibitem

\bibitem[\protect\citeauthoryear{{Pietsch} et~al.}{1976}]{Pietsch76}
\begin{barticle}
\bauthor{\binits{W.} \bsnm{{Pietsch}}},
\bauthor{\binits{E.} \bsnm{{Kendziorra}}},
\bauthor{\binits{R.} \bsnm{{Staubert}}},
\bauthor{\binits{J.} \bsnm{{Tr{\"u}mper}}},
\batitle{{The 4.8 hour variation of Cygnus X-3 at high X-ray energies}}.
\bjtitle{\apjl}
\bvolume{203},
\bfpage{67}--\blpage{69}
(\byear{1976}).
doi:\doiurl{10.1086/182021}
\end{barticle}
\endbibitem

\bibitem[\protect\citeauthoryear{{Piraino} et~al.}{1999}]{Piraino99}
\begin{barticle}
\bauthor{\binits{S.} \bsnm{{Piraino}}},
\bauthor{\binits{A.} \bsnm{{Santangelo}}},
\bauthor{\binits{E.C.} \bsnm{{Ford}}},
\bauthor{\binits{P.} \bsnm{{Kaaret}}},
\batitle{{BeppoSAX observations of the atoll X-ray binary 4U 0614+091}}.
\bjtitle{\aap}
\bvolume{349},
\bfpage{77}--\blpage{81}
(\byear{1999})
\end{barticle}
\endbibitem

\bibitem[\protect\citeauthoryear{{Pizzichini} et~al.}{1975}]{Pizzichini75}
\begin{barticle}
\bauthor{\binits{G.} \bsnm{{Pizzichini}}},
\bauthor{\binits{G.G.C.} \bsnm{{Palumbo}}},
\bauthor{\binits{A.} \bsnm{{Spizzichino}}},
\batitle{{Gamma-ray bursts observed by a hard X-ray experiment aboard OSO-6}}.
\bjtitle{\apjl}
\bvolume{195},
\bfpage{1}--\blpage{5}
(\byear{1975}).
doi:\doiurl{10.1086/181695}
\end{barticle}
\endbibitem

\bibitem[\protect\citeauthoryear{{Polcaro} et~al.}{1983}]{Polcaro83}
\begin{barticle}
\bauthor{\binits{V.F.} \bsnm{{Polcaro}}},
\bauthor{\binits{A.} \bsnm{{Bazzano}}},
\bauthor{\binits{C.} \bsnm{{La Padula}}},
\bauthor{\binits{P.} \bsnm{{Ubertini}}},
\bauthor{\binits{R.K.} \bsnm{{Manchanda}}},
\batitle{{Low state hard X-ray emission from A0535 + 26}}.
\bjtitle{\aap}
\bvolume{127},
\bfpage{333}--\blpage{336}
(\byear{1983})
\end{barticle}
\endbibitem

\bibitem[\protect\citeauthoryear{{Pravdo} et~al.}{1979}]{Pravdo79}
\begin{barticle}
\bauthor{\binits{S.H.} \bsnm{{Pravdo}}},
\bauthor{\binits{N.E.} \bsnm{{White}}},
\bauthor{\binits{E.A.} \bsnm{{Boldt}}},
\bauthor{\binits{S.S.} \bsnm{{Holt}}},
\bauthor{\binits{P.J.} \bsnm{{Serlemitsos}}},
\bauthor{\binits{J.H.} \bsnm{{Swank}}},
\bauthor{\binits{A.E.} \bsnm{{Szymkowiak}}},
\bauthor{\binits{I.} \bsnm{{Tuohy}}},
\bauthor{\binits{G.} \bsnm{{Garmire}}},
\batitle{{HEAO 1 observations of the X-ray pulsar 4U 1626-67}}.
\bjtitle{\apj}
\bvolume{231},
\bfpage{912}--\blpage{918}
(\byear{1979}).
doi:\doiurl{10.1086/157254}
\end{barticle}
\endbibitem

\bibitem[\protect\citeauthoryear{{Primini} et~al.}{1979}]{Primini79}
\begin{barticle}
\bauthor{\binits{F.A.} \bsnm{{Primini}}},
\bauthor{\binits{B.A.} \bsnm{{Cooke}}},
\bauthor{\binits{C.A.} \bsnm{{Dobson}}},
\bauthor{\binits{S.K.} \bsnm{{Howe}}},
\bauthor{\binits{A.} \bsnm{{Scheepmaker}}},
\bauthor{\binits{W.A.} \bsnm{{Wheaton}}},
\bauthor{\binits{W.H.G.} \bsnm{{Lewin}}},
\bauthor{\binits{W.A.} \bsnm{{Baity}}},
\bauthor{\binits{D.E.} \bsnm{{Gruber}}},
\bauthor{\binits{J.L.} \bsnm{{Matteson}}},
\batitle{{HEAO 1 observations of high-energy X-rays from 3C273}}.
\bjtitle{\nat}
\bvolume{278},
\bfpage{234}
(\byear{1979}).
doi:\doiurl{10.1038/278234a0}
\end{barticle}
\endbibitem

\bibitem[\protect\citeauthoryear{{Primini} et~al.}{1981}]{Primini81}
\begin{barticle}
\bauthor{\binits{F.A.} \bsnm{{Primini}}},
\bauthor{\binits{E.} \bsnm{{Basinska}}},
\bauthor{\binits{S.K.} \bsnm{{Howe}}},
\bauthor{\binits{F.} \bsnm{{Lang}}},
\bauthor{\binits{A.M.} \bsnm{{Levine}}},
\bauthor{\binits{W.H.G.} \bsnm{{Lewin}}},
\bauthor{\binits{R.} \bsnm{{Rothschild}}},
\bauthor{\binits{W.A.} \bsnm{{Baity}}},
\bauthor{\binits{D.E.} \bsnm{{Gruber}}},
\bauthor{\binits{F.K.} \bsnm{{Knight}}},
\bauthor{\binits{J.L.} \bsnm{{Matteson}}},
\bauthor{\binits{S.M.} \bsnm{{Lea}}},
\bauthor{\binits{G.A.} \bsnm{{Reichert}}},
\batitle{{HEAO 1 observations of the Perseus cluster above 10 keV}}.
\bjtitle{\apjl}
\bvolume{243},
\bfpage{13}--\blpage{17}
(\byear{1981}).
doi:\doiurl{10.1086/183433}
\end{barticle}
\endbibitem

\bibitem[\protect\citeauthoryear{{Purcell} et~al.}{1993}]{Purcell93}
\begin{barticle}
\bauthor{\binits{W.R.} \bsnm{{Purcell}}},
\bauthor{\binits{D.A.} \bsnm{{Grabelsky}}},
\bauthor{\binits{M.P.} \bsnm{{Ulmer}}},
\bauthor{\binits{W.N.} \bsnm{{Johnson}}},
\bauthor{\binits{R.L.} \bsnm{{Kinzer}}},
\bauthor{\binits{J.D.} \bsnm{{Kurfess}}},
\bauthor{\binits{M.S.} \bsnm{{Strickman}}},
\bauthor{\binits{G.V.} \bsnm{{Jung}}},
\batitle{{OSSE observations of Galactic 511 keV positron annihilation radiation
  - Initial phase 1 results}}.
\bjtitle{\apjl}
\bvolume{413},
\bfpage{85}--\blpage{88}
(\byear{1993}).
doi:\doiurl{10.1086/186965}
\end{barticle}
\endbibitem

\bibitem[\protect\citeauthoryear{{Purcell} et~al.}{1997}]{Purcell97}
\begin{barticle}
\bauthor{\binits{W.R.} \bsnm{{Purcell}}},
\bauthor{\binits{L.-X.} \bsnm{{Cheng}}},
\bauthor{\binits{D.D.} \bsnm{{Dixon}}},
\bauthor{\binits{R.L.} \bsnm{{Kinzer}}},
\bauthor{\binits{J.D.} \bsnm{{Kurfess}}},
\bauthor{\binits{M.} \bsnm{{Leventhal}}},
\bauthor{\binits{M.A.} \bsnm{{Saunders}}},
\bauthor{\binits{J.G.} \bsnm{{Skibo}}},
\bauthor{\binits{D.M.} \bsnm{{Smith}}},
\bauthor{\binits{J.} \bsnm{{Tueller}}},
\batitle{{OSSE Mapping of Galactic 511 keV Positron Annihilation Line
  Emission}}.
\bjtitle{\apj}
\bvolume{491},
\bfpage{725}--\blpage{748}
(\byear{1997})
\end{barticle}
\endbibitem

\bibitem[\protect\citeauthoryear{{Ramsey} et~al.}{2002}]{Ramsey2002;hero}
\begin{barticle}
\bauthor{\binits{B.D.} \bsnm{{Ramsey}}},
\bauthor{\binits{C.D.} \bsnm{{Alexander}}},
\bauthor{\binits{J.A.} \bsnm{{Apple}}},
\bauthor{\binits{C.M.} \bsnm{{Benson}}},
\bauthor{\binits{K.L.} \bsnm{{Dietz}}},
\bauthor{\binits{R.F.} \bsnm{{Elsner}}},
\bauthor{\binits{D.E.} \bsnm{{Engelhaupt}}},
\bauthor{\binits{K.K.} \bsnm{{Ghosh}}},
\bauthor{\binits{J.J.} \bsnm{{Kolodziejczak}}},
\bauthor{\binits{S.L.} \bsnm{{O'Dell}}},
\bauthor{\binits{C.O.} \bsnm{{Speegle}}},
\bauthor{\binits{D.A.} \bsnm{{Swartz}}},
\bauthor{\binits{M.C.} \bsnm{{Weisskopf}}},
\batitle{{First Images from HERO, a Hard X-Ray Focusing Telescope}}.
\bjtitle{\apj}
\bvolume{568},
\bfpage{432}--\blpage{435}
(\byear{2002}).
doi:\doiurl{10.1086/338801}
\end{barticle}
\endbibitem

\bibitem[\protect\citeauthoryear{{Rao} et~al.}{1991}]{Rao91}
\begin{barticle}
\bauthor{\binits{A.R.} \bsnm{{Rao}}},
\bauthor{\binits{P.C.} \bsnm{{Agrawal}}},
\bauthor{\binits{R.K.} \bsnm{{Manchanda}}},
\batitle{{Hard X-ray observations of Cygnus X-3}}.
\bjtitle{\aap}
\bvolume{241},
\bfpage{127}--\blpage{130}
(\byear{1991})
\end{barticle}
\endbibitem

\bibitem[\protect\citeauthoryear{{Rao} et~al.}{2016a}]{Rao16b}
\begin{botherref}
\oauthor{\binits{A.R.} \bsnm{{Rao}}},
\oauthor{\binits{K.P.} \bsnm{{Singh}}},
\oauthor{\binits{D.} \bsnm{{Bhattacharya}}},
{AstroSat - a multi-wavelength astronomy satellite}.
ArXiv e-prints
(2016a)
\end{botherref}
\endbibitem

\bibitem[\protect\citeauthoryear{{Rao} et~al.}{2016b}]{Rao16a}
\begin{barticle}
\bauthor{\binits{A.R.} \bsnm{{Rao}}},
\bauthor{\binits{V.} \bsnm{{Chand}}},
\bauthor{\binits{M.K.} \bsnm{{Hingar}}},
\bauthor{\binits{S.} \bsnm{{Iyyani}}},
\bauthor{\binits{R.} \bsnm{{Khanna}}},
\bauthor{\binits{A.P.K.} \bsnm{{Kutty}}},
\bauthor{\binits{J.P.} \bsnm{{Malkar}}},
\bauthor{\binits{D.} \bsnm{{Paul}}},
\bauthor{\binits{V.B.} \bsnm{{Bhalerao}}},
\bauthor{\binits{D.} \bsnm{{Bhattacharya}}},
\bauthor{\binits{G.C.} \bsnm{{Dewangan}}},
\bauthor{\binits{P.} \bsnm{{Pawar}}},
\bauthor{\binits{A.M.} \bsnm{{Vibhute}}},
\bauthor{\binits{T.} \bsnm{{Chattopadhyay}}},
\bauthor{\binits{N.P.S.} \bsnm{{Mithun}}},
\bauthor{\binits{S.V.} \bsnm{{Vadawale}}},
\bauthor{\binits{N.} \bsnm{{Vagshette}}},
\bauthor{\binits{R.} \bsnm{{Basak}}},
\bauthor{\binits{P.} \bsnm{{Pradeep}}},
\bauthor{\binits{E.} \bsnm{{Samuel}}},
\bauthor{\binits{S.} \bsnm{{Sreekumar}}},
\bauthor{\binits{P.} \bsnm{{Vinod}}},
\bauthor{\binits{K.H.} \bsnm{{Navalgund}}},
\bauthor{\binits{R.} \bsnm{{Pandiyan}}},
\bauthor{\binits{K.S.} \bsnm{{Sarma}}},
\bauthor{\binits{S.} \bsnm{{Seetha}}},
\bauthor{\binits{K.} \bsnm{{Subbarao}}},
\batitle{{AstroSat CZT Imager Observations of GRB 151006A: Timing,
  Spectroscopy, and Polarization Study}}.
\bjtitle{\apj}
\bvolume{833},
\bfpage{86}
(\byear{2016}b).
doi:\doiurl{10.3847/1538-4357/833/1/86}
\end{barticle}
\endbibitem

\bibitem[\protect\citeauthoryear{{Rea} and {Esposito}}{2011}]{Rea11}
\begin{barticle}
\bauthor{\binits{N.} \bsnm{{Rea}}},
\bauthor{\binits{P.} \bsnm{{Esposito}}},
\batitle{{Magnetar outbursts: an observational review}}.
\bjtitle{Astrophysics and Space Science Proceedings}
\bvolume{21},
\bfpage{247}
(\byear{2011})
\end{barticle}
\endbibitem

\bibitem[\protect\citeauthoryear{{Rea} et~al.}{2012}]{Rea12}
\begin{barticle}
\bauthor{\binits{N.} \bsnm{{Rea}}},
\bauthor{\binits{G.L.} \bsnm{{Israel}}},
\bauthor{\binits{P.} \bsnm{{Esposito}}},
\bauthor{\binits{J.A.} \bsnm{{Pons}}},
\bauthor{\binits{A.} \bsnm{{Camero-Arranz}}},
\bauthor{\binits{R.P.} \bsnm{{Mignani}}},
\bauthor{\binits{R.} \bsnm{{Turolla}}},
\bauthor{\binits{S.} \bsnm{{Zane}}},
\bauthor{\binits{M.} \bsnm{{Burgay}}},
\bauthor{\binits{A.} \bsnm{{Possenti}}},
\bauthor{\binits{S.} \bsnm{{Campana}}},
\bauthor{\binits{T.} \bsnm{{Enoto}}},
\bauthor{\binits{N.} \bsnm{{Gehrels}}},
\bauthor{\binits{E.} \bsnm{{G{\"o}{\v g}{\"u}{\c s}}}},
\bauthor{\binits{D.} \bsnm{{G{\"o}tz}}},
\bauthor{\binits{C.} \bsnm{{Kouveliotou}}},
\bauthor{\binits{K.} \bsnm{{Makishima}}},
\bauthor{\binits{S.} \bsnm{{Mereghetti}}},
\bauthor{\binits{S.R.} \bsnm{{Oates}}},
\bauthor{\binits{D.M.} \bsnm{{Palmer}}},
\bauthor{\binits{R.} \bsnm{{Perna}}},
\bauthor{\binits{L.} \bsnm{{Stella}}},
\bauthor{\binits{A.} \bsnm{{Tiengo}}},
\batitle{{A New Low Magnetic Field Magnetar: The 2011 Outburst of Swift
  J1822.3-1606}}.
\bjtitle{\apj}
\bvolume{754},
\bfpage{27}
(\byear{2012}).
doi:\doiurl{10.1088/0004-637X/754/1/27}
\end{barticle}
\endbibitem

\bibitem[\protect\citeauthoryear{{Reeves} et~al.}{2007}]{Reeves07}
\begin{barticle}
\bauthor{\binits{J.N.} \bsnm{{Reeves}}},
\bauthor{\binits{H.} \bsnm{{Awaki}}},
\bauthor{\binits{G.C.} \bsnm{{Dewangan}}},
\bauthor{\binits{A.C.} \bsnm{{Fabian}}},
\bauthor{\binits{Y.} \bsnm{{Fukazawa}}},
\bauthor{\binits{L.} \bsnm{{Gallo}}},
\bauthor{\binits{R.} \bsnm{{Griffiths}}},
\bauthor{\binits{H.} \bsnm{{Inoue}}},
\bauthor{\binits{H.} \bsnm{{Kunieda}}},
\bauthor{\binits{A.} \bsnm{{Markowitz}}},
\bauthor{\binits{G.} \bsnm{{Miniutti}}},
\bauthor{\binits{T.} \bsnm{{Mizuno}}},
\bauthor{\binits{R.} \bsnm{{Mushotzky}}},
\bauthor{\binits{T.} \bsnm{{Okajima}}},
\bauthor{\binits{A.} \bsnm{{Ptak}}},
\bauthor{\binits{T.} \bsnm{{Takahashi}}},
\bauthor{\binits{Y.} \bsnm{{Terashima}}},
\bauthor{\binits{M.} \bsnm{{Ushio}}},
\bauthor{\binits{S.} \bsnm{{Watanabe}}},
\bauthor{\binits{T.} \bsnm{{Yamasaki}}},
\bauthor{\binits{M.} \bsnm{{Yamauchi}}},
\bauthor{\binits{T.} \bsnm{{Yaqoob}}},
\batitle{{Revealing the High Energy Emission from the Obscured Seyfert Galaxy
  MCG-5-23-16 with Suzaku}}.
\bjtitle{\pasj}
\bvolume{59},
\bfpage{301}--\blpage{314}
(\byear{2007}).
doi:\doiurl{10.1093/pasj/59.sp1.S301}
\end{barticle}
\endbibitem

\bibitem[\protect\citeauthoryear{{Reig} and {Coe}}{1999}]{Reig99}
\begin{barticle}
\bauthor{\binits{P.} \bsnm{{Reig}}},
\bauthor{\binits{M.J.} \bsnm{{Coe}}},
\batitle{{X-ray spectral properties of the pulsar EXO 2030+375 during an
  outburst}}.
\bjtitle{\mnras}
\bvolume{302},
\bfpage{700}--\blpage{706}
(\byear{1999}).
doi:\doiurl{10.1046/j.1365-8711.1999.02179.x}
\end{barticle}
\endbibitem

\bibitem[\protect\citeauthoryear{{Reimer} et~al.}{2008}]{Reimer08}
\begin{barticle}
\bauthor{\binits{A.} \bsnm{{Reimer}}},
\bauthor{\binits{L.} \bsnm{{Costamante}}},
\bauthor{\binits{G.} \bsnm{{Madejski}}},
\bauthor{\binits{O.} \bsnm{{Reimer}}},
\bauthor{\binits{D.} \bsnm{{Dorner}}},
\batitle{{A Hard X-Ray View of Two Distant VHE Blazars: 1ES 1101-232 and 1ES
  1553+113}}.
\bjtitle{\apj}
\bvolume{682},
\bfpage{775}--\blpage{783}
(\byear{2008}).
doi:\doiurl{10.1086/589641}
\end{barticle}
\endbibitem

\bibitem[\protect\citeauthoryear{{Remillard} and
  {Canizares}}{1984}]{Remillard84}
\begin{barticle}
\bauthor{\binits{R.A.} \bsnm{{Remillard}}},
\bauthor{\binits{C.R.} \bsnm{{Canizares}}},
\batitle{{SAS 3 observations of Cygnus X-1 - The intensity dips}}.
\bjtitle{\apj}
\bvolume{278},
\bfpage{761}--\blpage{768}
(\byear{1984}).
doi:\doiurl{10.1086/161846}
\end{barticle}
\endbibitem

\bibitem[\protect\citeauthoryear{{Rephaeli} and {Gruber}}{2002}]{Rephaeli02}
\begin{barticle}
\bauthor{\binits{Y.} \bsnm{{Rephaeli}}},
\bauthor{\binits{D.} \bsnm{{Gruber}}},
\batitle{{RXTE view of the starburst galaxies M 82 and NGC 253}}.
\bjtitle{\aap}
\bvolume{389},
\bfpage{752}--\blpage{760}
(\byear{2002}).
doi:\doiurl{10.1051/0004-6361:20020635}
\end{barticle}
\endbibitem

\bibitem[\protect\citeauthoryear{{Rephaeli} and {Gruber}}{2004}]{Rephaeli04}
\begin{barticle}
\bauthor{\binits{Y.} \bsnm{{Rephaeli}}},
\bauthor{\binits{D.} \bsnm{{Gruber}}},
\batitle{{Spectral Analysis of Rossi X-Ray Timing Explorer Observations of
  A3667}}.
\bjtitle{\apj}
\bvolume{606},
\bfpage{825}--\blpage{828}
(\byear{2004}).
doi:\doiurl{10.1086/383123}
\end{barticle}
\endbibitem

\bibitem[\protect\citeauthoryear{{Rephaeli} et~al.}{1999}]{Rephaeli99}
\begin{barticle}
\bauthor{\binits{Y.} \bsnm{{Rephaeli}}},
\bauthor{\binits{D.} \bsnm{{Gruber}}},
\bauthor{\binits{P.} \bsnm{{Blanco}}},
\batitle{{Rossi X-Ray Timing Explorer Observations of the Coma Cluster}}.
\bjtitle{\apjl}
\bvolume{511},
\bfpage{21}--\blpage{24}
(\byear{1999}).
doi:\doiurl{10.1086/311828}
\end{barticle}
\endbibitem

\bibitem[\protect\citeauthoryear{{Reppin} et~al.}{1979}]{Reppin79}
\begin{barticle}
\bauthor{\binits{C.} \bsnm{{Reppin}}},
\bauthor{\binits{W.} \bsnm{{Pietsch}}},
\bauthor{\binits{J.} \bsnm{{Tr{\"u}mper}}},
\bauthor{\binits{W.} \bsnm{{Voges}}},
\bauthor{\binits{E.} \bsnm{{Kendziorra}}},
\bauthor{\binits{R.} \bsnm{{Staubert}}},
\batitle{{Hard X-ray observation of Cygnus X-3}}.
\bjtitle{\apj}
\bvolume{234},
\bfpage{329}--\blpage{332}
(\byear{1979}).
doi:\doiurl{10.1086/157500}
\end{barticle}
\endbibitem

\bibitem[\protect\citeauthoryear{{Revnivtsev} et~al.}{2003}]{Revnivtsev03}
\begin{barticle}
\bauthor{\binits{M.} \bsnm{{Revnivtsev}}},
\bauthor{\binits{M.} \bsnm{{Gilfanov}}},
\bauthor{\binits{R.} \bsnm{{Sunyaev}}},
\bauthor{\binits{K.} \bsnm{{Jahoda}}},
\bauthor{\binits{C.} \bsnm{{Markwardt}}},
\batitle{{The spectrum of the cosmic X-ray background observed by RTXE/PCA}}.
\bjtitle{\aap}
\bvolume{411},
\bfpage{329}--\blpage{334}
(\byear{2003}).
doi:\doiurl{10.1051/0004-6361:20031386}
\end{barticle}
\endbibitem

\bibitem[\protect\citeauthoryear{{Ricker} et~al.}{2003}]{Ricker03}
\begin{bchapter}
\bauthor{\binits{G.R.} \bsnm{{Ricker}}},
\bauthor{\binits{J.-L.} \bsnm{{Atteia}}},
\bauthor{\binits{G.B.} \bsnm{{Crew}}},
\bauthor{\binits{J.P.} \bsnm{{Doty}}},
\bauthor{\binits{E.E.} \bsnm{{Fenimore}}},
\bauthor{\binits{M.} \bsnm{{Galassi}}},
\bauthor{\binits{C.} \bsnm{{Graziani}}},
\bauthor{\binits{K.} \bsnm{{Hurley}}},
\bauthor{\binits{J.G.} \bsnm{{Jernigan}}},
\bauthor{\binits{N.} \bsnm{{Kawai}}},
\bauthor{\binits{D.Q.} \bsnm{{Lamb}}},
\bauthor{\binits{M.} \bsnm{{Matsuoka}}},
\bauthor{\binits{G.} \bsnm{{Pizzichini}}},
\bauthor{\binits{Y.} \bsnm{{Shirasaki}}},
\bauthor{\binits{T.} \bsnm{{Tamagawa}}},
\bauthor{\binits{R.} \bsnm{{Vanderspek}}},
\bauthor{\binits{G.} \bsnm{{Vedrenne}}},
\bauthor{\binits{J.} \bsnm{{Villasenor}}},
\bauthor{\binits{S.E.} \bsnm{{Woosley}}},
\bauthor{\binits{A.} \bsnm{{Yoshida}}},
\bctitle{{The High Energy Transient Explorer (HETE): Mission and Science
  Overview}},
in \bbtitle{Gamma-Ray Burst and Afterglow Astronomy 2001: A Workshop
  Celebrating the First Year of the HETE Mission},
ed. by \beditor{\binits{G.R.} \bsnm{{Ricker}}},
\beditor{\binits{R.K.} \bsnm{{Vanderspek}}}
\bsertitle{American Institute of Physics Conference Series},
vol. \bseriesno{662},
\byear{2003},
pp. \bfpage{3}--\blpage{16}.
doi:\doiurl{10.1063/1.1579291}
\end{bchapter}
\endbibitem

\bibitem[\protect\citeauthoryear{{Ricketts} et~al.}{1982}]{Ricketts1982;ariel6}
\begin{barticle}
\bauthor{\binits{M.J.} \bsnm{{Ricketts}}},
\bauthor{\binits{R.} \bsnm{{Hall}}},
\bauthor{\binits{C.G.} \bsnm{{Page}}},
\bauthor{\binits{C.H.} \bsnm{{Whitford}}},
\bauthor{\binits{K.A.} \bsnm{{Pounds}}},
\batitle{{GX 1+4: pulse period measurement and detection of phase-variable iron
  line emission}}.
\bjtitle{\mnras}
\bvolume{201},
\bfpage{759}--\blpage{768}
(\byear{1982})
\end{barticle}
\endbibitem

\bibitem[\protect\citeauthoryear{{Riegler} et~al.}{1968}]{Riegler68}
\begin{barticle}
\bauthor{\binits{G.R.} \bsnm{{Riegler}}},
\bauthor{\binits{E.} \bsnm{{Boldt}}},
\bauthor{\binits{P.} \bsnm{{Serlemitsos}}},
\batitle{{A Comparison of the X-Ray Spectra from Tau XR-1 and the Vicinity of
  SGR XR-1}}.
\bjtitle{\apjl}
\bvolume{153},
\bfpage{95}
(\byear{1968}).
doi:\doiurl{10.1086/180229}
\end{barticle}
\endbibitem

\bibitem[\protect\citeauthoryear{{Risaliti}}{2014}]{Risaliti14}
\begin{bchapter}
\bauthor{\binits{G.} \bsnm{{Risaliti}}},
\bctitle{{NuSTAR measurement of the high spin of the supermassive black hole in
  NGC 4051}},
in \bbtitle{AAS/High Energy Astrophysics Division}.
\bsertitle{AAS/High Energy Astrophysics Division},
vol. \bseriesno{14},
\byear{2014},
pp. \bfpage{106}--\blpage{05}
\end{bchapter}
\endbibitem

\bibitem[\protect\citeauthoryear{{Risaliti} et~al.}{2009}]{Risaliti09}
\begin{barticle}
\bauthor{\binits{G.} \bsnm{{Risaliti}}},
\bauthor{\binits{V.} \bsnm{{Braito}}},
\bauthor{\binits{V.} \bsnm{{Laparola}}},
\bauthor{\binits{S.} \bsnm{{Bianchi}}},
\bauthor{\binits{M.} \bsnm{{Elvis}}},
\bauthor{\binits{G.} \bsnm{{Fabbiano}}},
\bauthor{\binits{R.} \bsnm{{Maiolino}}},
\bauthor{\binits{G.} \bsnm{{Matt}}},
\bauthor{\binits{J.} \bsnm{{Reeves}}},
\bauthor{\binits{M.} \bsnm{{Salvati}}},
\bauthor{\binits{J.} \bsnm{{Wang}}},
\batitle{{A Strong Excess in the 20-100 keV Emission of NGC 1365}}.
\bjtitle{\apjl}
\bvolume{705},
\bfpage{1}--\blpage{5}
(\byear{2009}).
doi:\doiurl{10.1088/0004-637X/705/1/L1}
\end{barticle}
\endbibitem

\bibitem[\protect\citeauthoryear{{Rivers} et~al.}{2011}]{Rivers11}
\begin{barticle}
\bauthor{\binits{E.} \bsnm{{Rivers}}},
\bauthor{\binits{A.} \bsnm{{Markowitz}}},
\bauthor{\binits{R.} \bsnm{{Rothschild}}},
\batitle{{Spectral Survey of X-ray Bright Active Galactic Nuclei from the Rossi
  X-ray Timing Explorer}}.
\bjtitle{\apjs}
\bvolume{193},
\bfpage{3}
(\byear{2011}).
doi:\doiurl{10.1088/0067-0049/193/1/3}
\end{barticle}
\endbibitem

\bibitem[\protect\citeauthoryear{{Rocchia} et~al.}{1969}]{Rocchia69}
\begin{barticle}
\bauthor{\binits{R.} \bsnm{{Rocchia}}},
\bauthor{\binits{R.} \bsnm{{Rothenflug}}},
\bauthor{\binits{D.} \bsnm{{Boclet}}},
\bauthor{\binits{P.} \bsnm{{Durouchoux}}},
\batitle{{Observation of Five Sources of Cosmic X-Rays in the 15-150 keV Energy
  Range}}.
\bjtitle{\aap}
\bvolume{1},
\bfpage{48}
(\byear{1969})
\end{barticle}
\endbibitem

\bibitem[\protect\citeauthoryear{{Rodes-Roca} et~al.}{2009}]{Rodes-Roca09}
\begin{barticle}
\bauthor{\binits{J.J.} \bsnm{{Rodes-Roca}}},
\bauthor{\binits{J.M.} \bsnm{{Torrej{\'o}n}}},
\bauthor{\binits{I.} \bsnm{{Kreykenbohm}}},
\bauthor{\binits{S.} \bsnm{{Mart{\'{\i}}nez N{\'u}{\~n}ez}}},
\bauthor{\binits{A.} \bsnm{{Camero-Arranz}}},
\bauthor{\binits{G.} \bsnm{{Bernab{\'e}u}}},
\batitle{{The first cyclotron harmonic of 4U 1538-52}}.
\bjtitle{\aap}
\bvolume{508},
\bfpage{395}--\blpage{400}
(\byear{2009}).
doi:\doiurl{10.1051/0004-6361/200912815}
\end{barticle}
\endbibitem

\bibitem[\protect\citeauthoryear{{Romano} et~al.}{2014}]{Romano14}
\begin{barticle}
\bauthor{\binits{P.} \bsnm{{Romano}}},
\bauthor{\binits{H.A.} \bsnm{{Krimm}}},
\bauthor{\binits{D.M.} \bsnm{{Palmer}}},
\bauthor{\binits{L.} \bsnm{{Ducci}}},
\bauthor{\binits{P.} \bsnm{{Esposito}}},
\bauthor{\binits{S.} \bsnm{{Vercellone}}},
\bauthor{\binits{P.A.} \bsnm{{Evans}}},
\bauthor{\binits{C.} \bsnm{{Guidorzi}}},
\bauthor{\binits{V.} \bsnm{{Mangano}}},
\bauthor{\binits{J.A.} \bsnm{{Kennea}}},
\bauthor{\binits{S.D.} \bsnm{{Barthelmy}}},
\bauthor{\binits{D.N.} \bsnm{{Burrows}}},
\bauthor{\binits{N.} \bsnm{{Gehrels}}},
\batitle{{The 100-month Swift catalogue of supergiant fast X-ray transients. I.
  BAT on-board and transient monitor flares}}.
\bjtitle{\aap}
\bvolume{562},
\bfpage{2}
(\byear{2014}).
doi:\doiurl{10.1051/0004-6361/201322516}
\end{barticle}
\endbibitem

\bibitem[\protect\citeauthoryear{{Roques} et~al.}{2003}]{Roques2003;integral}
\begin{barticle}
\bauthor{\binits{J.P.} \bsnm{{Roques}}},
\bauthor{\binits{S.} \bsnm{{Schanne}}},
\bauthor{\binits{A.} \bsnm{{von Kienlin}}},
\bauthor{\binits{J.} \bsnm{{Kn{\"o}dlseder}}},
\bauthor{\binits{R.} \bsnm{{Briet}}},
\bauthor{\binits{L.} \bsnm{{Bouchet}}},
\bauthor{\binits{P.} \bsnm{{Paul}}},
\bauthor{\binits{S.} \bsnm{{Boggs}}},
\bauthor{\binits{P.} \bsnm{{Caraveo}}},
\bauthor{\binits{M.} \bsnm{{Cass{\'e}}}},
\bauthor{\binits{B.} \bsnm{{Cordier}}},
\bauthor{\binits{R.} \bsnm{{Diehl}}},
\bauthor{\binits{P.} \bsnm{{Durouchoux}}},
\bauthor{\binits{P.} \bsnm{{Jean}}},
\bauthor{\binits{P.} \bsnm{{Leleux}}},
\bauthor{\binits{G.} \bsnm{{Lichti}}},
\bauthor{\binits{P.} \bsnm{{Mandrou}}},
\bauthor{\binits{J.} \bsnm{{Matteson}}},
\bauthor{\binits{F.} \bsnm{{Sanchez}}},
\bauthor{\binits{V.} \bsnm{{Sch{\"o}nfelder}}},
\bauthor{\binits{G.} \bsnm{{Skinner}}},
\bauthor{\binits{A.} \bsnm{{Strong}}},
\bauthor{\binits{B.} \bsnm{{Teegarden}}},
\bauthor{\binits{G.} \bsnm{{Vedrenne}}},
\bauthor{\binits{P.} \bsnm{{von Ballmoos}}},
\bauthor{\binits{C.} \bsnm{{Wunderer}}},
\batitle{{SPI/INTEGRAL in-flight performance}}.
\bjtitle{\aap}
\bvolume{411},
\bfpage{91}--\blpage{100}
(\byear{2003}).
doi:\doiurl{10.1051/0004-6361:20031501}
\end{barticle}
\endbibitem

\bibitem[\protect\citeauthoryear{{Rothschild} et~al.}{1980}]{Rothschild80}
\begin{barticle}
\bauthor{\binits{R.E.} \bsnm{{Rothschild}}},
\bauthor{\binits{D.E.} \bsnm{{Gruber}}},
\bauthor{\binits{F.K.} \bsnm{{Knight}}},
\bauthor{\binits{P.L.} \bsnm{{Nolan}}},
\bauthor{\binits{Y.} \bsnm{{Soong}}},
\bauthor{\binits{A.M.} \bsnm{{Levine}}},
\bauthor{\binits{F.A.} \bsnm{{Primini}}},
\bauthor{\binits{W.H.G.} \bsnm{{Lewin}}},
\bauthor{\binits{W.A.} \bsnm{{Wheaton}}},
\batitle{{A high sensitivity determination of the hard X-ray spectrum of SCO
  X-1}}.
\bjtitle{\nat}
\bvolume{286},
\bfpage{786}--\blpage{788}
(\byear{1980}).
doi:\doiurl{10.1038/286786a0}
\end{barticle}
\endbibitem

\bibitem[\protect\citeauthoryear{{Rothschild} et~al.}{1998}]{Rothschild98}
\begin{barticle}
\bauthor{\binits{R.E.} \bsnm{{Rothschild}}},
\bauthor{\binits{P.R.} \bsnm{{Blanco}}},
\bauthor{\binits{D.E.} \bsnm{{Gruber}}},
\bauthor{\binits{W.A.} \bsnm{{Heindl}}},
\bauthor{\binits{D.R.} \bsnm{{MacDonald}}},
\bauthor{\binits{D.C.} \bsnm{{Marsden}}},
\bauthor{\binits{M.R.} \bsnm{{Pelling}}},
\bauthor{\binits{L.R.} \bsnm{{Wayne}}},
\bauthor{\binits{P.L.} \bsnm{{Hink}}},
\batitle{{In-Flight Performance of the High-Energy X-Ray Timing Experiment on
  the Rossi X-Ray Timing Explorer}}.
\bjtitle{\apj}
\bvolume{496},
\bfpage{538}--\blpage{549}
(\byear{1998}).
doi:\doiurl{10.1086/305377}
\end{barticle}
\endbibitem

\bibitem[\protect\citeauthoryear{{Rothschild} et~al.}{1979}]{Rothschild79}
\begin{barticle}
\bauthor{\binits{R.} \bsnm{{Rothschild}}},
\bauthor{\binits{E.} \bsnm{{Boldt}}},
\bauthor{\binits{S.} \bsnm{{Holt}}},
\bauthor{\binits{P.} \bsnm{{Serlemitsos}}},
\bauthor{\binits{G.} \bsnm{{Garmire}}},
\bauthor{\binits{P.} \bsnm{{Agrawal}}},
\bauthor{\binits{G.} \bsnm{{Riegler}}},
\bauthor{\binits{S.} \bsnm{{Bowyer}}},
\bauthor{\binits{M.} \bsnm{{Lampton}}},
\batitle{{The cosmic X-ray experiment aboard HEAO-1}}.
\bjtitle{Space Science Instrumentation}
\bvolume{4},
\bfpage{269}--\blpage{301}
(\byear{1979})
\end{barticle}
\endbibitem

\bibitem[\protect\citeauthoryear{{Sacco} et~al.}{1990}]{Sacco90}
\begin{barticle}
\bauthor{\binits{B.} \bsnm{{Sacco}}},
\bauthor{\binits{G.} \bsnm{{Gerardi}}},
\bauthor{\binits{T.} \bsnm{{Mineo}}},
\bauthor{\binits{L.} \bsnm{{Scarsi}}},
\bauthor{\binits{B.} \bsnm{{Agrinier}}},
\bauthor{\binits{E.} \bsnm{{Barouch}}},
\bauthor{\binits{J.L.} \bsnm{{Masnou}}},
\bauthor{\binits{B.} \bsnm{{Parlier}}},
\bauthor{\binits{E.} \bsnm{{Costa}}},
\bauthor{\binits{E.} \bsnm{{Massaro}}},
\bauthor{\binits{M.} \bsnm{{Salvati}}},
\bauthor{\binits{M.} \bsnm{{Niel}}},
\bauthor{\binits{P.} \bsnm{{Mandrou}}},
\bauthor{\binits{C.} \bsnm{{Hote}}},
\bauthor{\binits{P.M.} \bsnm{{McCulloch}}},
\batitle{{Observation of the VELA pulsar, PSR 0833-45, at 0.2-6.0 MeV with the
  FIGARO II experiment}}.
\bjtitle{\apjl}
\bvolume{349},
\bfpage{21}--\blpage{24}
(\byear{1990}).
doi:\doiurl{10.1086/185641}
\end{barticle}
\endbibitem

\bibitem[\protect\citeauthoryear{{Sakamoto} et~al.}{2005}]{Sakamoto05}
\begin{barticle}
\bauthor{\binits{T.} \bsnm{{Sakamoto}}},
\bauthor{\binits{D.Q.} \bsnm{{Lamb}}},
\bauthor{\binits{N.} \bsnm{{Kawai}}},
\bauthor{\binits{A.} \bsnm{{Yoshida}}},
\bauthor{\binits{C.} \bsnm{{Graziani}}},
\bauthor{\binits{E.E.} \bsnm{{Fenimore}}},
\bauthor{\binits{T.Q.} \bsnm{{Donaghy}}},
\bauthor{\binits{M.} \bsnm{{Matsuoka}}},
\bauthor{\binits{M.} \bsnm{{Suzuki}}},
\bauthor{\binits{G.} \bsnm{{Ricker}}},
\bauthor{\binits{J.-L.} \bsnm{{Atteia}}},
\bauthor{\binits{Y.} \bsnm{{Shirasaki}}},
\bauthor{\binits{T.} \bsnm{{Tamagawa}}},
\bauthor{\binits{K.} \bsnm{{Torii}}},
\bauthor{\binits{M.} \bsnm{{Galassi}}},
\bauthor{\binits{J.} \bsnm{{Doty}}},
\bauthor{\binits{R.} \bsnm{{Vanderspek}}},
\bauthor{\binits{G.B.} \bsnm{{Crew}}},
\bauthor{\binits{J.} \bsnm{{Villasenor}}},
\bauthor{\binits{N.} \bsnm{{Butler}}},
\bauthor{\binits{G.} \bsnm{{Prigozhin}}},
\bauthor{\binits{J.G.} \bsnm{{Jernigan}}},
\bauthor{\binits{C.} \bsnm{{Barraud}}},
\bauthor{\binits{M.} \bsnm{{Boer}}},
\bauthor{\binits{J.-P.} \bsnm{{Dezalay}}},
\bauthor{\binits{J.-F.} \bsnm{{Olive}}},
\bauthor{\binits{K.} \bsnm{{Hurley}}},
\bauthor{\binits{A.} \bsnm{{Levine}}},
\bauthor{\binits{G.} \bsnm{{Monnelly}}},
\bauthor{\binits{F.} \bsnm{{Martel}}},
\bauthor{\binits{E.} \bsnm{{Morgan}}},
\bauthor{\binits{S.E.} \bsnm{{Woosley}}},
\bauthor{\binits{T.} \bsnm{{Cline}}},
\bauthor{\binits{J.} \bsnm{{Braga}}},
\bauthor{\binits{R.} \bsnm{{Manchanda}}},
\bauthor{\binits{G.} \bsnm{{Pizzichini}}},
\bauthor{\binits{K.} \bsnm{{Takagishi}}},
\bauthor{\binits{M.} \bsnm{{Yamauchi}}},
\batitle{{Global Characteristics of X-Ray Flashes and X-Ray-Rich Gamma-Ray
  Bursts Observed by HETE-2}}.
\bjtitle{\apj}
\bvolume{629},
\bfpage{311}--\blpage{327}
(\byear{2005}).
doi:\doiurl{10.1086/431235}
\end{barticle}
\endbibitem

\bibitem[\protect\citeauthoryear{{Santangelo} et~al.}{1998}]{Santangelo98}
\begin{barticle}
\bauthor{\binits{A.} \bsnm{{Santangelo}}},
\bauthor{\binits{S.} \bsnm{{del Sordo}}},
\bauthor{\binits{A.} \bsnm{{Segreto}}},
\bauthor{\binits{D.} \bsnm{{dal Fiume}}},
\bauthor{\binits{M.} \bsnm{{Orlandini}}},
\bauthor{\binits{S.} \bsnm{{Piraino}}},
\batitle{{BeppoSAX detection of a Cyclotron Feature in the spectrum of Cen
  X-3}}.
\bjtitle{\aap}
\bvolume{340},
\bfpage{55}--\blpage{59}
(\byear{1998})
\end{barticle}
\endbibitem

\bibitem[\protect\citeauthoryear{{Santangelo} et~al.}{1999}]{Santangelo99}
\begin{barticle}
\bauthor{\binits{A.} \bsnm{{Santangelo}}},
\bauthor{\binits{A.} \bsnm{{Segreto}}},
\bauthor{\binits{S.} \bsnm{{Giarrusso}}},
\bauthor{\binits{D.} \bsnm{{Dal Fiume}}},
\bauthor{\binits{M.} \bsnm{{Orlandini}}},
\bauthor{\binits{A.N.} \bsnm{{Parmar}}},
\bauthor{\binits{T.} \bsnm{{Oosterbroek}}},
\bauthor{\binits{T.} \bsnm{{Bulik}}},
\bauthor{\binits{T.} \bsnm{{Mihara}}},
\bauthor{\binits{S.} \bsnm{{Campana}}},
\bauthor{\binits{G.L.} \bsnm{{Israel}}},
\bauthor{\binits{L.} \bsnm{{Stella}}},
\batitle{{A BEPPOSAX Study of the Pulsating Transient X0115+63: The First X-Ray
  Spectrum with Four Cyclotron Harmonic Features}}.
\bjtitle{\apjl}
\bvolume{523},
\bfpage{85}--\blpage{88}
(\byear{1999}).
doi:\doiurl{10.1086/312249}
\end{barticle}
\endbibitem

\bibitem[\protect\citeauthoryear{{Sbarrato} et~al.}{2013}]{Sbarrato13}
\begin{barticle}
\bauthor{\binits{T.} \bsnm{{Sbarrato}}},
\bauthor{\binits{G.} \bsnm{{Tagliaferri}}},
\bauthor{\binits{G.} \bsnm{{Ghisellini}}},
\bauthor{\binits{M.} \bsnm{{Perri}}},
\bauthor{\binits{S.} \bsnm{{Puccetti}}},
\bauthor{\binits{M.} \bsnm{{Balokovi{\'c}}}},
\bauthor{\binits{M.} \bsnm{{Nardini}}},
\bauthor{\binits{D.} \bsnm{{Stern}}},
\bauthor{\binits{S.E.} \bsnm{{Boggs}}},
\bauthor{\binits{W.N.} \bsnm{{Brandt}}},
\bauthor{\binits{F.E.} \bsnm{{Christensen}}},
\bauthor{\binits{P.} \bsnm{{Giommi}}},
\bauthor{\binits{J.} \bsnm{{Greiner}}},
\bauthor{\binits{C.J.} \bsnm{{Hailey}}},
\bauthor{\binits{F.A.} \bsnm{{Harrison}}},
\bauthor{\binits{T.} \bsnm{{Hovatta}}},
\bauthor{\binits{G.M.} \bsnm{{Madejski}}},
\bauthor{\binits{A.} \bsnm{{Rau}}},
\bauthor{\binits{P.} \bsnm{{Schady}}},
\bauthor{\binits{V.} \bsnm{{Sudilovsky}}},
\bauthor{\binits{C.M.} \bsnm{{Urry}}},
\bauthor{\binits{W.W.} \bsnm{{Zhang}}},
\batitle{{NuSTAR Detection of the Blazar B2 1023+25 at Redshift 5.3}}.
\bjtitle{\apj}
\bvolume{777},
\bfpage{147}
(\byear{2013}).
doi:\doiurl{10.1088/0004-637X/777/2/147}
\end{barticle}
\endbibitem

\bibitem[\protect\citeauthoryear{{Sch{\"o}nfelder} and
  {Lichti}}{1973}]{Schonfelder73}
\begin{barticle}
\bauthor{\binits{V.} \bsnm{{Sch{\"o}nfelder}}},
\bauthor{\binits{G.} \bsnm{{Lichti}}},
\batitle{{A Balloon Borne Soft Gamma Ray Telescope}}.
\bjtitle{International Cosmic Ray Conference}
\bvolume{4},
\bfpage{2709}
(\byear{1973})
\end{barticle}
\endbibitem

\bibitem[\protect\citeauthoryear{{Sch{\"o}nfelder} and
  {Lichti}}{1974}]{Schonfelder74}
\begin{barticle}
\bauthor{\binits{V.} \bsnm{{Sch{\"o}nfelder}}},
\bauthor{\binits{G.} \bsnm{{Lichti}}},
\batitle{{Energy Spectrum and Evidence for Extragalactic Origin of Diffuse
  Gammaradiation in the MeV Range}}.
\bjtitle{\apjl}
\bvolume{191},
\bfpage{1}
(\byear{1974}).
doi:\doiurl{10.1086/181528}
\end{barticle}
\endbibitem

\bibitem[\protect\citeauthoryear{{Sch\"onfelder} et~al.}{1980}]{Schonfelder80}
\begin{barticle}
\bauthor{\binits{V.} \bsnm{{Sch\"onfelder}}},
\bauthor{\binits{F.} \bsnm{{Graml}}},
\bauthor{\binits{F.-P.} \bsnm{{Penningsfeld}}},
\batitle{{The vertical component of 1-20 MeV gamma rays at balloon altitudes}}.
\bjtitle{\apj}
\bvolume{240},
\bfpage{350}--\blpage{362}
(\byear{1980}).
doi:\doiurl{10.1086/158239}
\end{barticle}
\endbibitem

\bibitem[\protect\citeauthoryear{{Sch{\"o}nfelder}
  et~al.}{1982}]{Schonfelder82}
\begin{barticle}
\bauthor{\binits{V.} \bsnm{{Sch{\"o}nfelder}}},
\bauthor{\binits{U.} \bsnm{{Graser}}},
\bauthor{\binits{R.} \bsnm{{Diehl}}},
\batitle{{Properties and performance of the MPI balloon borne Compton
  telescope}}.
\bjtitle{\aap}
\bvolume{110},
\bfpage{138}--\blpage{151}
(\byear{1982})
\end{barticle}
\endbibitem

\bibitem[\protect\citeauthoryear{{Sch{\"o}nfelder}
  et~al.}{1993}]{Schonfelder1993;cgro}
\begin{barticle}
\bauthor{\binits{V.} \bsnm{{Sch{\"o}nfelder}}},
\bauthor{\binits{H.} \bsnm{{Aarts}}},
\bauthor{\binits{K.} \bsnm{{Bennett}}},
\bauthor{\binits{H.} \bsnm{{de Boer}}},
\bauthor{\binits{J.} \bsnm{{Clear}}},
\bauthor{\binits{W.} \bsnm{{Collmar}}},
\bauthor{\binits{A.} \bsnm{{Connors}}},
\bauthor{\binits{A.} \bsnm{{Deerenberg}}},
\bauthor{\binits{R.} \bsnm{{Diehl}}},
\bauthor{\binits{A.} \bsnm{{von Dordrecht}}},
\bauthor{\binits{J.W.} \bsnm{{den Herder}}},
\bauthor{\binits{W.} \bsnm{{Hermsen}}},
\bauthor{\binits{M.} \bsnm{{Kippen}}},
\bauthor{\binits{L.} \bsnm{{Kuiper}}},
\bauthor{\binits{G.} \bsnm{{Lichti}}},
\bauthor{\binits{J.} \bsnm{{Lockwood}}},
\bauthor{\binits{J.} \bsnm{{Macri}}},
\bauthor{\binits{M.} \bsnm{{McConnell}}},
\bauthor{\binits{D.} \bsnm{{Morris}}},
\bauthor{\binits{R.} \bsnm{{Much}}},
\bauthor{\binits{J.} \bsnm{{Ryan}}},
\bauthor{\binits{G.} \bsnm{{Simpson}}},
\bauthor{\binits{M.} \bsnm{{Snelling}}},
\bauthor{\binits{G.} \bsnm{{Stacy}}},
\bauthor{\binits{H.} \bsnm{{Steinle}}},
\bauthor{\binits{A.} \bsnm{{Strong}}},
\bauthor{\binits{B.N.} \bsnm{{Swanenburg}}},
\bauthor{\binits{B.} \bsnm{{Taylor}}},
\bauthor{\binits{C.} \bsnm{{de Vries}}},
\bauthor{\binits{C.} \bsnm{{Winkler}}},
\batitle{{Instrument description and performance of the Imaging Gamma-Ray
  Telescope COMPTEL aboard the Compton Gamma-Ray Observatory}}.
\bjtitle{\apjs}
\bvolume{86},
\bfpage{657}--\blpage{692}
(\byear{1993}).
doi:\doiurl{10.1086/191794}
\end{barticle}
\endbibitem

\bibitem[\protect\citeauthoryear{{Schwartz} et~al.}{1970}]{Schwartz70}
\begin{barticle}
\bauthor{\binits{D.A.} \bsnm{{Schwartz}}},
\bauthor{\binits{H.S.} \bsnm{{Hudson}}},
\bauthor{\binits{L.E.} \bsnm{{Peterson}}},
\batitle{{The Spectrum of Diffuse Cosmic X-Rays 7.7-113 keV}}.
\bjtitle{\apj}
\bvolume{162},
\bfpage{431}
(\byear{1970}).
doi:\doiurl{10.1086/150675}
\end{barticle}
\endbibitem

\bibitem[\protect\citeauthoryear{{Serlemitsos}
  et~al.}{1976}]{Serlemitsos1976;oso8}
\begin{barticle}
\bauthor{\binits{P.J.} \bsnm{{Serlemitsos}}},
\bauthor{\binits{R.H.} \bsnm{{Becker}}},
\bauthor{\binits{E.A.} \bsnm{{Boldt}}},
\bauthor{\binits{S.S.} \bsnm{{Holt}}},
\bauthor{\binits{S.H.} \bsnm{{Pravdo}}},
\bauthor{\binits{R.E.} \bsnm{{Rothschild}}},
\bauthor{\binits{J.H.} \bsnm{{Swank}}},
\batitle{{Cosmic X-ray observations with OSO-8}}.
\bjtitle{NASA Special Publication}
\bvolume{389},
\bfpage{67}--\blpage{79}
(\byear{1976})
\end{barticle}
\endbibitem

\bibitem[\protect\citeauthoryear{{Sguera} et~al.}{2005}]{Sguera05}
\begin{barticle}
\bauthor{\binits{V.} \bsnm{{Sguera}}},
\bauthor{\binits{E.J.} \bsnm{{Barlow}}},
\bauthor{\binits{A.J.} \bsnm{{Bird}}},
\bauthor{\binits{D.J.} \bsnm{{Clark}}},
\bauthor{\binits{A.J.} \bsnm{{Dean}}},
\bauthor{\binits{A.B.} \bsnm{{Hill}}},
\bauthor{\binits{L.} \bsnm{{Moran}}},
\bauthor{\binits{S.E.} \bsnm{{Shaw}}},
\bauthor{\binits{D.R.} \bsnm{{Willis}}},
\bauthor{\binits{A.} \bsnm{{Bazzano}}},
\bauthor{\binits{P.} \bsnm{{Ubertini}}},
\bauthor{\binits{A.} \bsnm{{Malizia}}},
\batitle{{INTEGRAL observations of recurrent fast X-ray transient sources}}.
\bjtitle{\aap}
\bvolume{444},
\bfpage{221}--\blpage{231}
(\byear{2005}).
doi:\doiurl{10.1051/0004-6361:20053103}
\end{barticle}
\endbibitem

\bibitem[\protect\citeauthoryear{{Sguera} et~al.}{2006}]{Sguera06}
\begin{barticle}
\bauthor{\binits{V.} \bsnm{{Sguera}}},
\bauthor{\binits{A.} \bsnm{{Bazzano}}},
\bauthor{\binits{A.J.} \bsnm{{Bird}}},
\bauthor{\binits{A.J.} \bsnm{{Dean}}},
\bauthor{\binits{P.} \bsnm{{Ubertini}}},
\bauthor{\binits{E.J.} \bsnm{{Barlow}}},
\bauthor{\binits{L.} \bsnm{{Bassani}}},
\bauthor{\binits{D.J.} \bsnm{{Clark}}},
\bauthor{\binits{A.B.} \bsnm{{Hill}}},
\bauthor{\binits{A.} \bsnm{{Malizia}}},
\bauthor{\binits{M.} \bsnm{{Molina}}},
\bauthor{\binits{J.B.} \bsnm{{Stephen}}},
\batitle{{Unveiling Supergiant Fast X-Ray Transient Sources with INTEGRAL}}.
\bjtitle{\apj}
\bvolume{646},
\bfpage{452}--\blpage{463}
(\byear{2006}).
doi:\doiurl{10.1086/504827}
\end{barticle}
\endbibitem

\bibitem[\protect\citeauthoryear{{Shaposhnikov} et~al.}{2010}]{Shaposhnikov10}
\begin{barticle}
\bauthor{\binits{N.} \bsnm{{Shaposhnikov}}},
\bauthor{\binits{C.} \bsnm{{Markwardt}}},
\bauthor{\binits{J.} \bsnm{{Swank}}},
\bauthor{\binits{H.} \bsnm{{Krimm}}},
\batitle{{Discovery and Monitoring of a New Black Hole Candidate XTE J1752-223
  with RXTE: Rms Spectrum Evolution, Black Hole Mass, and the Source
  Distance}}.
\bjtitle{\apj}
\bvolume{723},
\bfpage{1817}--\blpage{1824}
(\byear{2010}).
doi:\doiurl{10.1088/0004-637X/723/2/1817}
\end{barticle}
\endbibitem

\bibitem[\protect\citeauthoryear{{Share} et~al.}{1990}]{Share90}
\begin{barticle}
\bauthor{\binits{G.H.} \bsnm{{Share}}},
\bauthor{\binits{M.D.} \bsnm{{Leising}}},
\bauthor{\binits{D.C.} \bsnm{{Messina}}},
\bauthor{\binits{W.R.} \bsnm{{Purcell}}},
\batitle{{Limits on a variable source of 511 keV annihilation radiation near
  the Galactic center}}.
\bjtitle{\apjl}
\bvolume{358},
\bfpage{45}--\blpage{48}
(\byear{1990}).
doi:\doiurl{10.1086/185776}
\end{barticle}
\endbibitem

\bibitem[\protect\citeauthoryear{{Shirai} et~al.}{2008}]{Shirai08}
\begin{barticle}
\bauthor{\binits{H.} \bsnm{{Shirai}}},
\bauthor{\binits{Y.} \bsnm{{Fukazawa}}},
\bauthor{\binits{M.} \bsnm{{Sasada}}},
\bauthor{\binits{M.} \bsnm{{Ohno}}},
\bauthor{\binits{D.} \bsnm{{Yonetoku}}},
\bauthor{\binits{S.} \bsnm{{Yokota}}},
\bauthor{\binits{R.} \bsnm{{Fujimoto}}},
\bauthor{\binits{T.} \bsnm{{Murakami}}},
\bauthor{\binits{Y.} \bsnm{{Terashima}}},
\bauthor{\binits{H.} \bsnm{{Awaki}}},
\bauthor{\binits{S.} \bsnm{{Ikeda}}},
\bauthor{\binits{M.} \bsnm{{Ozawa}}},
\bauthor{\binits{T.G.} \bsnm{{Tsuru}}},
\batitle{{Detailed Hard X-Ray Measurements of Nuclear Emission from the
  Seyfert2 Galaxy NGC4388 with Suzaku}}.
\bjtitle{\pasj}
\bvolume{60},
\bfpage{263}--\blpage{276}
(\byear{2008})
\end{barticle}
\endbibitem

\bibitem[\protect\citeauthoryear{{Sidoli} and {Mereghetti}}{2002}]{Sidoli02}
\begin{barticle}
\bauthor{\binits{L.} \bsnm{{Sidoli}}},
\bauthor{\binits{S.} \bsnm{{Mereghetti}}},
\batitle{{The broad band X-ray spectrum of the black hole candidate GRS
  1758-258}}.
\bjtitle{\aap}
\bvolume{388},
\bfpage{293}--\blpage{297}
(\byear{2002}).
doi:\doiurl{10.1051/0004-6361:20020546}
\end{barticle}
\endbibitem

\bibitem[\protect\citeauthoryear{{Siegert} et~al.}{2016}]{Siegert16}
\begin{barticle}
\bauthor{\binits{T.} \bsnm{{Siegert}}},
\bauthor{\binits{R.} \bsnm{{Diehl}}},
\bauthor{\binits{J.} \bsnm{{Greiner}}},
\bauthor{\binits{M.G.H.} \bsnm{{Krause}}},
\bauthor{\binits{A.M.} \bsnm{{Beloborodov}}},
\bauthor{\binits{M.C.} \bsnm{{Bel}}},
\bauthor{\binits{F.} \bsnm{{Guglielmetti}}},
\bauthor{\binits{J.} \bsnm{{Rodriguez}}},
\bauthor{\binits{A.W.} \bsnm{{Strong}}},
\bauthor{\binits{X.} \bsnm{{Zhang}}},
\batitle{{Positron annihilation signatures associated with the outburst of the
  microquasar V404 Cygni}}.
\bjtitle{\nat}
\bvolume{531},
\bfpage{341}--\blpage{343}
(\byear{2016}).
doi:\doiurl{10.1038/nature16978}
\end{barticle}
\endbibitem

\bibitem[\protect\citeauthoryear{{Singh} et~al.}{2014}]{Singh14}
\begin{bchapter}
\bauthor{\binits{K.P.} \bsnm{{Singh}}},
\bauthor{\binits{S.N.} \bsnm{{Tandon}}},
\bauthor{\binits{P.C.} \bsnm{{Agrawal}}},
\bauthor{\binits{H.M.} \bsnm{{Antia}}},
\bauthor{\binits{R.K.} \bsnm{{Manchanda}}},
\bauthor{\binits{J.S.} \bsnm{{Yadav}}},
\bauthor{\binits{S.} \bsnm{{Seetha}}},
\bauthor{\binits{M.C.} \bsnm{{Ramadevi}}},
\bauthor{\binits{A.R.} \bsnm{{Rao}}},
\bauthor{\binits{D.} \bsnm{{Bhattacharya}}},
\bauthor{\binits{B.} \bsnm{{Paul}}},
\bauthor{\binits{P.} \bsnm{{Sreekumar}}},
\bauthor{\binits{S.} \bsnm{{Bhattacharyya}}},
\bauthor{\binits{G.C.} \bsnm{{Stewart}}},
\bauthor{\binits{J.} \bsnm{{Hutchings}}},
\bauthor{\binits{S.A.} \bsnm{{Annapurni}}},
\bauthor{\binits{S.K.} \bsnm{{Ghosh}}},
\bauthor{\binits{J.} \bsnm{{Murthy}}},
\bauthor{\binits{A.} \bsnm{{Pati}}},
\bauthor{\binits{N.K.} \bsnm{{Rao}}},
\bauthor{\binits{C.S.} \bsnm{{Stalin}}},
\bauthor{\binits{V.} \bsnm{{Girish}}},
\bauthor{\binits{K.} \bsnm{{Sankarasubramanian}}},
\bauthor{\binits{S.} \bsnm{{Vadawale}}},
\bauthor{\binits{V.B.} \bsnm{{Bhalerao}}},
\bauthor{\binits{G.C.} \bsnm{{Dewangan}}},
\bauthor{\binits{D.K.} \bsnm{{Dedhia}}},
\bauthor{\binits{M.K.} \bsnm{{Hingar}}},
\bauthor{\binits{T.B.} \bsnm{{Katoch}}},
\bauthor{\binits{A.T.} \bsnm{{Kothare}}},
\bauthor{\binits{I.} \bsnm{{Mirza}}},
\bauthor{\binits{K.} \bsnm{{Mukerjee}}},
\bauthor{\binits{H.} \bsnm{{Shah}}},
\bauthor{\binits{P.} \bsnm{{Shah}}},
\bauthor{\binits{R.} \bsnm{{Mohan}}},
\bauthor{\binits{A.K.} \bsnm{{Sangal}}},
\bauthor{\binits{S.} \bsnm{{Nagabhusana}}},
\bauthor{\binits{S.} \bsnm{{Sriram}}},
\bauthor{\binits{J.P.} \bsnm{{Malkar}}},
\bauthor{\binits{S.} \bsnm{{Sreekumar}}},
\bauthor{\binits{A.F.} \bsnm{{Abbey}}},
\bauthor{\binits{G.M.} \bsnm{{Hansford}}},
\bauthor{\binits{A.P.} \bsnm{{Beardmore}}},
\bauthor{\binits{M.R.} \bsnm{{Sharma}}},
\bauthor{\binits{S.} \bsnm{{Murthy}}},
\bauthor{\binits{R.} \bsnm{{Kulkarni}}},
\bauthor{\binits{G.} \bsnm{{Meena}}},
\bauthor{\binits{V.C.} \bsnm{{Babu}}},
\bauthor{\binits{J.} \bsnm{{Postma}}},
\bctitle{{ASTROSAT mission}},
in \bbtitle{Space Telescopes and Instrumentation 2014: Ultraviolet to Gamma
  Ray}.
\bsertitle{\procspie},
vol. \bseriesno{9144},
\byear{2014},
p. \bfpage{91441}.
doi:\doiurl{10.1117/12.2062667}
\end{bchapter}
\endbibitem

\bibitem[\protect\citeauthoryear{{Smith} et~al.}{1999}]{Smith99}
\begin{barticle}
\bauthor{\binits{I.A.} \bsnm{{Smith}}},
\bauthor{\binits{E.P.} \bsnm{{Liang}}},
\bauthor{\binits{D.} \bsnm{{Lin}}},
\bauthor{\binits{M.} \bsnm{{Moss}}},
\bauthor{\binits{A.} \bsnm{{Crider}}},
\bauthor{\binits{R.P.} \bsnm{{Fender}}},
\bauthor{\binits{P.} \bsnm{{Durouchoux}}},
\bauthor{\binits{S.} \bsnm{{Corbel}}},
\bauthor{\binits{R.} \bsnm{{Sood}}},
\batitle{{Multiwavelength Observations of GX 339-4 in 1996. I. Daily Light
  Curves and X-Ray and Gamma-Ray Spectroscopy}}.
\bjtitle{\apj}
\bvolume{519},
\bfpage{762}--\blpage{770}
(\byear{1999}).
doi:\doiurl{10.1086/307390}
\end{barticle}
\endbibitem

\bibitem[\protect\citeauthoryear{{Smith} et~al.}{2008}]{Smith08}
\begin{barticle}
\bauthor{\binits{R.K.} \bsnm{{Smith}}},
\bauthor{\binits{R.} \bsnm{{Mushotzky}}},
\bauthor{\binits{K.} \bsnm{{Mukai}}},
\bauthor{\binits{T.} \bsnm{{Kallman}}},
\bauthor{\binits{C.B.} \bsnm{{Markwardt}}},
\bauthor{\binits{J.} \bsnm{{Tueller}}},
\batitle{{The Symbiotic System SS73 17 Seen with Suzaku}}.
\bjtitle{\pasj}
\bvolume{60},
\bfpage{43}--\blpage{48}
(\byear{2008})
\end{barticle}
\endbibitem

\bibitem[\protect\citeauthoryear{{Sood} et~al.}{1996}]{Sood96}
\begin{barticle}
\bauthor{\binits{R.K.} \bsnm{{Sood}}},
\bauthor{\binits{J.} \bsnm{{Panettieri}}},
\bauthor{\binits{D.} \bsnm{{Grey}}},
\bauthor{\binits{G.} \bsnm{{Woods}}},
\bauthor{\binits{J.} \bsnm{{Hoffman}}},
\bauthor{\binits{R.K.} \bsnm{{Manchanda}}},
\bauthor{\binits{R.} \bsnm{{Staubert}}},
\bauthor{\binits{E.} \bsnm{{Kendziorra}}},
\bauthor{\binits{G.K.} \bsnm{{Rochester}}},
\batitle{{Axel: A balloon-borne X-ray astronomy experiment}}.
\bjtitle{\pasa}
\bvolume{13},
\bfpage{156}--\blpage{161}
(\byear{1996})
\end{barticle}
\endbibitem

\bibitem[\protect\citeauthoryear{{Soong} and {Rothschild}}{1983}]{Soong83}
\begin{barticle}
\bauthor{\binits{Y.} \bsnm{{Soong}}},
\bauthor{\binits{R.E.} \bsnm{{Rothschild}}},
\batitle{{Long-term, hard X-ray observations of Scorpius X-1 from HEAO 1}}.
\bjtitle{\apj}
\bvolume{274},
\bfpage{327}--\blpage{332}
(\byear{1983}).
doi:\doiurl{10.1086/161449}
\end{barticle}
\endbibitem

\bibitem[\protect\citeauthoryear{{Stanek} et~al.}{2003}]{Stanek03}
\begin{barticle}
\bauthor{\binits{K.Z.} \bsnm{{Stanek}}},
\bauthor{\binits{T.} \bsnm{{Matheson}}},
\bauthor{\binits{P.M.} \bsnm{{Garnavich}}},
\bauthor{\binits{P.} \bsnm{{Martini}}},
\bauthor{\binits{P.} \bsnm{{Berlind}}},
\bauthor{\binits{N.} \bsnm{{Caldwell}}},
\bauthor{\binits{P.} \bsnm{{Challis}}},
\bauthor{\binits{W.R.} \bsnm{{Brown}}},
\bauthor{\binits{R.} \bsnm{{Schild}}},
\bauthor{\binits{K.} \bsnm{{Krisciunas}}},
\bauthor{\binits{M.L.} \bsnm{{Calkins}}},
\bauthor{\binits{J.C.} \bsnm{{Lee}}},
\bauthor{\binits{N.} \bsnm{{Hathi}}},
\bauthor{\binits{R.A.} \bsnm{{Jansen}}},
\bauthor{\binits{R.} \bsnm{{Windhorst}}},
\bauthor{\binits{L.} \bsnm{{Echevarria}}},
\bauthor{\binits{D.J.} \bsnm{{Eisenstein}}},
\bauthor{\binits{B.} \bsnm{{Pindor}}},
\bauthor{\binits{E.W.} \bsnm{{Olszewski}}},
\bauthor{\binits{P.} \bsnm{{Harding}}},
\bauthor{\binits{S.T.} \bsnm{{Holland}}},
\bauthor{\binits{D.} \bsnm{{Bersier}}},
\batitle{{Spectroscopic Discovery of the Supernova 2003dh Associated with GRB
  030329}}.
\bjtitle{\apjl}
\bvolume{591},
\bfpage{17}--\blpage{20}
(\byear{2003}).
doi:\doiurl{10.1086/376976}
\end{barticle}
\endbibitem

\bibitem[\protect\citeauthoryear{{Staubert} et~al.}{1978}]{Staubert78}
\begin{barticle}
\bauthor{\binits{R.} \bsnm{{Staubert}}},
\bauthor{\binits{E.} \bsnm{{Kendziorra}}},
\bauthor{\binits{W.} \bsnm{{Pietsch}}},
\bauthor{\binits{C.} \bsnm{{Reppin}}},
\bauthor{\binits{J.} \bsnm{{Tr{\"u}mper}}},
\bauthor{\binits{W.} \bsnm{{Voges}}},
\batitle{{The hard X-ray spectrum of AM Herculis}}.
\bjtitle{\apjl}
\bvolume{225},
\bfpage{113}
(\byear{1978}).
doi:\doiurl{10.1086/182805}
\end{barticle}
\endbibitem

\bibitem[\protect\citeauthoryear{{Staubert} et~al.}{1980}]{Staubert80}
\begin{barticle}
\bauthor{\binits{R.} \bsnm{{Staubert}}},
\bauthor{\binits{E.} \bsnm{{Kendziorra}}},
\bauthor{\binits{W.} \bsnm{{Pietsch}}},
\bauthor{\binits{C.} \bsnm{{Reppin}}},
\bauthor{\binits{J.} \bsnm{{Tr{\"u}mper}}},
\bauthor{\binits{W.} \bsnm{{Voges}}},
\batitle{{Hard X-ray pulses from 4U 0900-40}}.
\bjtitle{\apj}
\bvolume{239},
\bfpage{1010}--\blpage{1016}
(\byear{1980}).
doi:\doiurl{10.1086/158189}
\end{barticle}
\endbibitem

\bibitem[\protect\citeauthoryear{{Strickman} et~al.}{1979}]{Strickman79}
\begin{barticle}
\bauthor{\binits{M.S.} \bsnm{{Strickman}}},
\bauthor{\binits{W.N.} \bsnm{{Johnson}}},
\bauthor{\binits{J.D.} \bsnm{{Kurfess}}},
\batitle{{The hard X-ray spectrum of the Crab Nebula}}.
\bjtitle{\apjl}
\bvolume{230},
\bfpage{15}--\blpage{19}
(\byear{1979}).
doi:\doiurl{10.1086/182953}
\end{barticle}
\endbibitem

\bibitem[\protect\citeauthoryear{{Strickman} et~al.}{1980}]{Strickman80}
\begin{barticle}
\bauthor{\binits{M.S.} \bsnm{{Strickman}}},
\bauthor{\binits{W.N.} \bsnm{{Johnson}}},
\bauthor{\binits{J.D.} \bsnm{{Kurfess}}},
\batitle{{The hard X-ray pulse profile of GX 1+4}}.
\bjtitle{\apjl}
\bvolume{240},
\bfpage{21}--\blpage{25}
(\byear{1980}).
doi:\doiurl{10.1086/183316}
\end{barticle}
\endbibitem

\bibitem[\protect\citeauthoryear{{Strickman} et~al.}{1982}]{Strickman82}
\begin{barticle}
\bauthor{\binits{M.S.} \bsnm{{Strickman}}},
\bauthor{\binits{J.D.} \bsnm{{Kurfess}}},
\bauthor{\binits{W.N.} \bsnm{{Johnson}}},
\batitle{{A transient 77 keV emission feature from the Crab pulsar}}.
\bjtitle{\apjl}
\bvolume{253},
\bfpage{23}--\blpage{27}
(\byear{1982}).
doi:\doiurl{10.1086/183729}
\end{barticle}
\endbibitem

\bibitem[\protect\citeauthoryear{{Strickman} et~al.}{1996}]{Strickman96b}
\begin{barticle}
\bauthor{\binits{M.S.} \bsnm{{Strickman}}},
\bauthor{\binits{C.D.} \bsnm{{Dermer}}},
\bauthor{\binits{J.E.} \bsnm{{Grove}}},
\bauthor{\binits{W.N.} \bsnm{{Johnson}}},
\bauthor{\binits{G.V.} \bsnm{{Jung}}},
\bauthor{\binits{J.D.} \bsnm{{Kurfess}}},
\bauthor{\binits{B.F.} \bsnm{{Phlips}}},
\bauthor{\binits{G.H.} \bsnm{{Share}}},
\bauthor{\binits{S.J.} \bsnm{{Sturner}}},
\bauthor{\binits{D.C.} \bsnm{{Messina}}},
\bauthor{\binits{S.M.} \bsnm{{Matz}}},
\batitle{{Hard X-Ray Spectroscopy and Pulsar Phase Analysis of the Bursting
  X-Ray Pulsar GRO J1744-28 with OSSE}}.
\bjtitle{\apjl}
\bvolume{464},
\bfpage{131}
(\byear{1996}).
doi:\doiurl{10.1086/310112}
\end{barticle}
\endbibitem

\bibitem[\protect\citeauthoryear{{Strong} et~al.}{1974}]{Strong74}
\begin{barticle}
\bauthor{\binits{I.B.} \bsnm{{Strong}}},
\bauthor{\binits{R.W.} \bsnm{{Klebesadel}}},
\bauthor{\binits{R.A.} \bsnm{{Olson}}},
\batitle{{A Preliminary Catalog of Transient Cosmic Gamma-Ray Sources Observed
  by the VELA Satellites}}.
\bjtitle{\apjl}
\bvolume{188},
\bfpage{1}
(\byear{1974}).
doi:\doiurl{10.1086/181415}
\end{barticle}
\endbibitem

\bibitem[\protect\citeauthoryear{{Sunyaev} et~al.}{1988}]{Sunyaev88b}
\begin{barticle}
\bauthor{\binits{R.A.} \bsnm{{Sunyaev}}},
\bauthor{\binits{I.Y.} \bsnm{{Lapshov}}},
\bauthor{\binits{S.A.} \bsnm{{Grebenev}}},
\bauthor{\binits{V.V.} \bsnm{{Efremov}}},
\bauthor{\binits{A.S.} \bsnm{{Kaniovskii}}},
\bauthor{\binits{D.K.} \bsnm{{Stepanov}}},
\bauthor{\binits{S.N.} \bsnm{{Yunin}}},
\bauthor{\binits{E.A.} \bsnm{{Gavrilova}}},
\bauthor{\binits{V.M.} \bsnm{{Loznikov}}},
\bauthor{\binits{A.V.} \bsnm{{Prudkoglyad}}},
\bauthor{\binits{V.G.} \bsnm{{Rodin}}},
\bauthor{\binits{O.P.} \bsnm{{Babushkina}}},
\bauthor{\binits{S.V.} \bsnm{{Kiselev}}},
\bauthor{\binits{A.V.} \bsnm{{Kuznetsov}}},
\bauthor{\binits{A.S.} \bsnm{{Melioranskii}}},
\bauthor{\binits{A.} \bsnm{{Smith}}},
\bauthor{\binits{A.N.} \bsnm{{Parmar}}},
\bauthor{\binits{W.} \bsnm{{Pietsch}}},
\bauthor{\binits{S.} \bsnm{{Dobereiner}}},
\bauthor{\binits{J.} \bsnm{{Engelhauser}}},
\bauthor{\binits{C.} \bsnm{{Reppin}}},
\bauthor{\binits{J.} \bsnm{{Trumper}}},
\bauthor{\binits{W.} \bsnm{{Voges}}},
\bauthor{\binits{E.} \bsnm{{Kendziorra}}},
\bauthor{\binits{M.} \bsnm{{Maisack}}},
\bauthor{\binits{B.} \bsnm{{Mony}}},
\bauthor{\binits{R.} \bsnm{{Staubert}}},
\batitle{{Detection of a Hard Component in the Spectrum of the Vulpecula X-Ray
  Nova - Preliminary KVANT Results}}.
\bjtitle{Soviet Astronomy Letters}
\bvolume{14},
\bfpage{327}
(\byear{1988})
\end{barticle}
\endbibitem

\bibitem[\protect\citeauthoryear{{Sunyaev} et~al.}{1989}]{Sunyaev89}
\begin{barticle}
\bauthor{\binits{R.A.} \bsnm{{Sunyaev}}},
\bauthor{\binits{A.S.} \bsnm{{Kaniovskii}}},
\bauthor{\binits{V.V.} \bsnm{{Efremov}}},
\bauthor{\binits{S.A.} \bsnm{{Grebenev}}},
\bauthor{\binits{A.V.} \bsnm{{Kuznetsov}}},
\bauthor{\binits{V.M.} \bsnm{{Loznikov}}},
\bauthor{\binits{A.S.} \bsnm{{Melioranskii}}},
\bauthor{\binits{J.} \bsnm{{Engelhauser}}},
\bauthor{\binits{S.} \bsnm{{Dobereiner}}},
\bauthor{\binits{W.} \bsnm{{Pietsch}}},
\bauthor{\binits{C.} \bsnm{{Reppin}}},
\bauthor{\binits{J.} \bsnm{{Trumper}}},
\bauthor{\binits{E.} \bsnm{{Kendziorra}}},
\bauthor{\binits{M.} \bsnm{{Maisack}}},
\bauthor{\binits{B.} \bsnm{{Mony}}},
\bauthor{\binits{R.} \bsnm{{Staubert}}},
\batitle{{Hard X-Ray Decline of Supernova 1987A in 1988 - the MIR / KVANT
  Data}}.
\bjtitle{Soviet Astronomy Letters}
\bvolume{15},
\bfpage{125}
(\byear{1989})
\end{barticle}
\endbibitem

\bibitem[\protect\citeauthoryear{{Sunyaev} et~al.}{1990}]{Sunyaev1990;granat}
\begin{barticle}
\bauthor{\binits{R.A.} \bsnm{{Sunyaev}}},
\bauthor{\binits{S.I.} \bsnm{{Babichenko}}},
\bauthor{\binits{D.A.} \bsnm{{Goganov}}},
\bauthor{\binits{S.R.} \bsnm{{Tabaldyev}}},
\bauthor{\binits{N.S.} \bsnm{{Iamburenko}}},
\batitle{{X-ray telescopes ART-P and ART-S for the GRANAT project}}.
\bjtitle{Advances in Space Research}
\bvolume{10},
\bfpage{233}--\blpage{237}
(\byear{1990}).
doi:\doiurl{10.1016/0273-1177(90)90147-R}
\end{barticle}
\endbibitem

\bibitem[\protect\citeauthoryear{{Sunyaev} et~al.}{1991a}]{Sunyaev91b}
\begin{barticle}
\bauthor{\binits{R.A.} \bsnm{{Sunyaev}}},
\bauthor{\binits{V.A.} \bsnm{{Arefev}}},
\bauthor{\binits{K.N.} \bsnm{{Borozdin}}},
\bauthor{\binits{M.R.} \bsnm{{Gilfanov}}},
\bauthor{\binits{V.V.} \bsnm{{Efremov}}},
\bauthor{\binits{A.S.} \bsnm{{Kaniovskii}}},
\bauthor{\binits{E.M.} \bsnm{{Churazov}}},
\bauthor{\binits{E.} \bsnm{{Kendziorra}}},
\bauthor{\binits{B.} \bsnm{{Mony}}},
\bauthor{\binits{P.} \bsnm{{Kretschmar}}},
\bauthor{\binits{M.} \bsnm{{Maisack}}},
\bauthor{\binits{R.} \bsnm{{Staubert}}},
\bauthor{\binits{S.} \bsnm{{Dobereiner}}},
\bauthor{\binits{J.} \bsnm{{Englhauser}}},
\bauthor{\binits{W.} \bsnm{{Pietsch}}},
\bauthor{\binits{C.} \bsnm{{Reppin}}},
\bauthor{\binits{J.} \bsnm{{Trumper}}},
\bauthor{\binits{G.K.} \bsnm{{Skinner}}},
\bauthor{\binits{M.R.} \bsnm{{Nottingham}}},
\bauthor{\binits{H.} \bsnm{{Pan}}},
\bauthor{\binits{A.P.} \bsnm{{Willmore}}},
\batitle{{Broadband X-Ray Spectra of Black-Hole Candidates X-Ray Pulsars and
  Low-Mass Binary X-Ray Systems - KVANT Module Results}}.
\bjtitle{Soviet Astronomy Letters}
\bvolume{17},
\bfpage{409}
(\byear{1991}a)
\end{barticle}
\endbibitem

\bibitem[\protect\citeauthoryear{{Sunyaev} et~al.}{1991b}]{Sunyaev91a}
\begin{barticle}
\bauthor{\binits{R.} \bsnm{{Sunyaev}}},
\bauthor{\binits{K.} \bsnm{{Borozdin}}},
\bauthor{\binits{M.} \bsnm{{Gilfanov}}},
\bauthor{\binits{V.} \bsnm{{Efremov}}},
\bauthor{\binits{A.} \bsnm{{Kaniovskii}}},
\bauthor{\binits{E.} \bsnm{{Churazov}}},
\bauthor{\binits{G.K.} \bsnm{{Skinner}}},
\bauthor{\binits{O.} \bsnm{{Al-Emam}}},
\bauthor{\binits{T.G.} \bsnm{{Patterson}}},
\bauthor{\binits{A.P.} \bsnm{{Willmore}}},
\bauthor{\binits{A.C.} \bsnm{{Brinkman}}},
\bauthor{\binits{J.} \bsnm{{Heise}}},
\bauthor{\binits{J.J.M.} \bsnm{{Int-Zand}}},
\bauthor{\binits{R.} \bsnm{{Jager}}},
\bauthor{\binits{W.} \bsnm{{Voges}}},
\bauthor{\binits{W.} \bsnm{{Pietsch}}},
\bauthor{\binits{S.} \bsnm{{Doebereiner}}},
\bauthor{\binits{J.} \bsnm{{Engelhauser}}},
\bauthor{\binits{J.} \bsnm{{Trumper}}},
\bauthor{\binits{C.} \bsnm{{Reppin}}},
\bauthor{\binits{E.} \bsnm{{Kendziorra}}},
\bauthor{\binits{B.} \bsnm{{Mony}}},
\bauthor{\binits{M.} \bsnm{{Maisack}}},
\bauthor{\binits{R.} \bsnm{{Staubert}}},
\batitle{{Anomalously Hard Spectrum from the Source 1E:1740.7$-$294 - Data from
  the Rontgen Observatory on the KVANT Module}}.
\bjtitle{Soviet Astronomy Letters}
\bvolume{17},
\bfpage{54}
(\byear{1991}b)
\end{barticle}
\endbibitem

\bibitem[\protect\citeauthoryear{{Tagliaferri} et~al.}{2000}]{Tagliaferri00}
\begin{barticle}
\bauthor{\binits{G.} \bsnm{{Tagliaferri}}},
\bauthor{\binits{G.} \bsnm{{Ghisellini}}},
\bauthor{\binits{P.} \bsnm{{Giommi}}},
\bauthor{\binits{L.} \bsnm{{Chiappetti}}},
\bauthor{\binits{L.} \bsnm{{Maraschi}}},
\bauthor{\binits{A.} \bsnm{{Celotti}}},
\bauthor{\binits{M.} \bsnm{{Chiaberge}}},
\bauthor{\binits{G.} \bsnm{{Fossati}}},
\bauthor{\binits{E.} \bsnm{{Massaro}}},
\bauthor{\binits{M.} \bsnm{{Maesano}}},
\bauthor{\binits{F.} \bsnm{{Montagni}}},
\bauthor{\binits{R.} \bsnm{{Nesci}}},
\bauthor{\binits{G.} \bsnm{{Nucciarelli}}},
\bauthor{\binits{E.} \bsnm{{Pian}}},
\bauthor{\binits{C.M.} \bsnm{{Raiteri}}},
\bauthor{\binits{F.} \bsnm{{Tavecchio}}},
\bauthor{\binits{G.} \bsnm{{Tosti}}},
\bauthor{\binits{A.} \bsnm{{Treves}}},
\bauthor{\binits{M.} \bsnm{{Villata}}},
\bauthor{\binits{A.} \bsnm{{Wolter}}},
\batitle{{The concave X-ray spectrum of the blazar ON 231: the signature of
  intermediate BL Lacertae objects}}.
\bjtitle{\aap}
\bvolume{354},
\bfpage{431}--\blpage{438}
(\byear{2000})
\end{barticle}
\endbibitem

\bibitem[\protect\citeauthoryear{{Takahashi} et~al.}{2008}]{Takahashi08}
\begin{barticle}
\bauthor{\binits{H.} \bsnm{{Takahashi}}},
\bauthor{\binits{Y.} \bsnm{{Fukazawa}}},
\bauthor{\binits{T.} \bsnm{{Mizuno}}},
\bauthor{\binits{A.} \bsnm{{Hirasawa}}},
\bauthor{\binits{S.} \bsnm{{Kitamoto}}},
\bauthor{\binits{K.} \bsnm{{Sudoh}}},
\bauthor{\binits{T.} \bsnm{{Ogita}}},
\bauthor{\binits{A.} \bsnm{{Kubota}}},
\bauthor{\binits{K.} \bsnm{{Makishima}}},
\bauthor{\binits{T.} \bsnm{{Itoh}}},
\bauthor{\binits{A.N.} \bsnm{{Parmar}}},
\bauthor{\binits{K.} \bsnm{{Ebisawa}}},
\bauthor{\binits{S.} \bsnm{{Naik}}},
\bauthor{\binits{T.} \bsnm{{Dotani}}},
\bauthor{\binits{M.} \bsnm{{Kokubun}}},
\bauthor{\binits{K.} \bsnm{{Ohnuki}}},
\bauthor{\binits{T.} \bsnm{{Takahashi}}},
\bauthor{\binits{T.} \bsnm{{Yaqoob}}},
\bauthor{\binits{L.} \bsnm{{Angelini}}},
\bauthor{\binits{Y.} \bsnm{{Ueda}}},
\bauthor{\binits{K.} \bsnm{{Yamaoka}}},
\bauthor{\binits{T.} \bsnm{{Kotani}}},
\bauthor{\binits{N.} \bsnm{{Kawai}}},
\bauthor{\binits{M.} \bsnm{{Namiki}}},
\bauthor{\binits{T.} \bsnm{{Kohmura}}},
\bauthor{\binits{H.} \bsnm{{Negoro}}},
\batitle{{Low/Hard State Spectra of GRO J1655-40 Observed with Suzaku}}.
\bjtitle{\pasj}
\bvolume{60},
\bfpage{69}--\blpage{84}
(\byear{2008})
\end{barticle}
\endbibitem

\bibitem[\protect\citeauthoryear{{Takahashi} et~al.}{2007}]{Takahashi07}
\begin{barticle}
\bauthor{\binits{T.} \bsnm{{Takahashi}}},
\bauthor{\binits{K.} \bsnm{{Abe}}},
\bauthor{\binits{M.} \bsnm{{Endo}}},
\bauthor{\binits{Y.} \bsnm{{Endo}}},
\bauthor{\binits{Y.} \bsnm{{Ezoe}}},
\bauthor{\binits{Y.} \bsnm{{Fukazawa}}},
\bauthor{\binits{M.} \bsnm{{Hamaya}}},
\bauthor{\binits{S.} \bsnm{{Hirakuri}}},
\bauthor{\binits{S.} \bsnm{{Hong}}},
\bauthor{\binits{M.} \bsnm{{Horii}}},
\bauthor{\binits{H.} \bsnm{{Inoue}}},
\bauthor{\binits{N.} \bsnm{{Isobe}}},
\bauthor{\binits{T.} \bsnm{{Itoh}}},
\bauthor{\binits{N.} \bsnm{{Iyomoto}}},
\bauthor{\binits{T.} \bsnm{{Kamae}}},
\bauthor{\binits{D.} \bsnm{{Kasama}}},
\bauthor{\binits{J.} \bsnm{{Kataoka}}},
\bauthor{\binits{H.} \bsnm{{Kato}}},
\bauthor{\binits{M.} \bsnm{{Kawaharada}}},
\bauthor{\binits{N.} \bsnm{{Kawano}}},
\bauthor{\binits{K.} \bsnm{{Kawashima}}},
\bauthor{\binits{S.} \bsnm{{Kawasoe}}},
\bauthor{\binits{T.} \bsnm{{Kishishita}}},
\bauthor{\binits{T.} \bsnm{{Kitaguchi}}},
\bauthor{\binits{Y.} \bsnm{{Kobayashi}}},
\bauthor{\binits{M.} \bsnm{{Kokubun}}},
\bauthor{\binits{J.} \bsnm{{Kotoku}}},
\bauthor{\binits{M.} \bsnm{{Kouda}}},
\bauthor{\binits{A.} \bsnm{{Kubota}}},
\bauthor{\binits{Y.} \bsnm{{Kuroda}}},
\bauthor{\binits{G.} \bsnm{{Madejski}}},
\bauthor{\binits{K.} \bsnm{{Makishima}}},
\bauthor{\binits{K.} \bsnm{{Masukawa}}},
\bauthor{\binits{Y.} \bsnm{{Matsumoto}}},
\bauthor{\binits{T.} \bsnm{{Mitani}}},
\bauthor{\binits{R.} \bsnm{{Miyawaki}}},
\bauthor{\binits{T.} \bsnm{{Mizuno}}},
\bauthor{\binits{K.} \bsnm{{Mori}}},
\bauthor{\binits{M.} \bsnm{{Mori}}},
\bauthor{\binits{M.} \bsnm{{Murashima}}},
\bauthor{\binits{T.} \bsnm{{Murakami}}},
\bauthor{\binits{K.} \bsnm{{Nakazawa}}},
\bauthor{\binits{H.} \bsnm{{Niko}}},
\bauthor{\binits{M.} \bsnm{{Nomachi}}},
\bauthor{\binits{Y.} \bsnm{{Okada}}},
\bauthor{\binits{M.} \bsnm{{Ohno}}},
\bauthor{\binits{K.} \bsnm{{Oonuki}}},
\bauthor{\binits{N.} \bsnm{{Ota}}},
\bauthor{\binits{H.} \bsnm{{Ozawa}}},
\bauthor{\binits{G.} \bsnm{{Sato}}},
\bauthor{\binits{S.} \bsnm{{Shinoda}}},
\bauthor{\binits{M.} \bsnm{{Sugiho}}},
\bauthor{\binits{M.} \bsnm{{Suzuki}}},
\bauthor{\binits{K.} \bsnm{{Taguchi}}},
\bauthor{\binits{H.} \bsnm{{Takahashi}}},
\bauthor{\binits{I.} \bsnm{{Takahashi}}},
\bauthor{\binits{S.} \bsnm{{Takeda}}},
\bauthor{\binits{K.-I.} \bsnm{{Tamura}}},
\bauthor{\binits{T.} \bsnm{{Tamura}}},
\bauthor{\binits{T.} \bsnm{{Tanaka}}},
\bauthor{\binits{C.} \bsnm{{Tanihata}}},
\bauthor{\binits{M.} \bsnm{{Tashiro}}},
\bauthor{\binits{Y.} \bsnm{{Terada}}},
\bauthor{\binits{S.} \bsnm{{Tominaga}}},
\bauthor{\binits{Y.} \bsnm{{Uchiyama}}},
\bauthor{\binits{S.} \bsnm{{Watanabe}}},
\bauthor{\binits{K.} \bsnm{{Yamaoka}}},
\bauthor{\binits{T.} \bsnm{{Yanagida}}},
\bauthor{\binits{D.} \bsnm{{Yonetoku}}},
\batitle{{Hard X-Ray Detector (HXD) on Board Suzaku}}.
\bjtitle{\pasj}
\bvolume{59},
\bfpage{35}--\blpage{51}
(\byear{2007}).
doi:\doiurl{10.1093/pasj/59.sp1.S35}
\end{barticle}
\endbibitem

\bibitem[\protect\citeauthoryear{{Takahashi} et~al.}{2008}]{Takahashi08b}
\begin{barticle}
\bauthor{\binits{T.} \bsnm{{Takahashi}}},
\bauthor{\binits{T.} \bsnm{{Tanaka}}},
\bauthor{\binits{Y.} \bsnm{{Uchiyama}}},
\bauthor{\binits{J.S.} \bsnm{{Hiraga}}},
\bauthor{\binits{K.} \bsnm{{Nakazawa}}},
\bauthor{\binits{S.} \bsnm{{Watanabe}}},
\bauthor{\binits{A.} \bsnm{{Bamba}}},
\bauthor{\binits{J.P.} \bsnm{{Hughes}}},
\bauthor{\binits{H.} \bsnm{{Katagiri}}},
\bauthor{\binits{J.} \bsnm{{Kataoka}}},
\bauthor{\binits{M.} \bsnm{{Kokubun}}},
\bauthor{\binits{K.} \bsnm{{Koyama}}},
\bauthor{\binits{K.} \bsnm{{Mori}}},
\bauthor{\binits{R.} \bsnm{{Petre}}},
\bauthor{\binits{H.} \bsnm{{Takahashi}}},
\bauthor{\binits{Y.} \bsnm{{Tsuboi}}},
\batitle{{Measuring the Broad-Band X-Ray Spectrum from 400eV to 40keV in the
  Southwest Part of the Supernova Remnant RXJ1713.7-3946}}.
\bjtitle{\pasj}
\bvolume{60},
\bfpage{131}--\blpage{140}
(\byear{2008})
\end{barticle}
\endbibitem

\bibitem[\protect\citeauthoryear{{Takahashi} et~al.}{2014}]{Takahashi14}
\begin{bchapter}
\bauthor{\binits{T.} \bsnm{{Takahashi}}},
\bauthor{\binits{K.} \bsnm{{Mitsuda}}},
\bauthor{\binits{R.} \bsnm{{Kelley}}},
\bauthor{\binits{F.} \bsnm{{Aharonian}}},
\bauthor{\binits{H.} \bsnm{{Akamatsu}}},
\bauthor{\binits{F.} \bsnm{{Akimoto}}},
\bauthor{\binits{S.} \bsnm{{Allen}}},
\bauthor{\binits{N.} \bsnm{{Anabuki}}},
\bauthor{\binits{L.} \bsnm{{Angelini}}},
\bauthor{\binits{K.} \bsnm{{Arnaud}}},
\bauthor{\bparticle{et} \bsnm{al.}},
\bctitle{{The ASTRO-H X-ray astronomy satellite}},
in \bbtitle{Space Telescopes and Instrumentation 2014: Ultraviolet to Gamma
  Ray}.
\bsertitle{\procspie},
vol. \bseriesno{9144},
\byear{2014},
p. \bfpage{914425}.
doi:\doiurl{10.1117/12.2055681}
\end{bchapter}
\endbibitem

\bibitem[\protect\citeauthoryear{{Takei} et~al.}{2009}]{Takei09}
\begin{barticle}
\bauthor{\binits{D.} \bsnm{{Takei}}},
\bauthor{\binits{M.} \bsnm{{Tsujimoto}}},
\bauthor{\binits{S.} \bsnm{{Kitamoto}}},
\bauthor{\binits{J.-U.} \bsnm{{Ness}}},
\bauthor{\binits{J.J.} \bsnm{{Drake}}},
\bauthor{\binits{H.} \bsnm{{Takahashi}}},
\bauthor{\binits{K.} \bsnm{{Mukai}}},
\batitle{{Suzaku Detection of Superhard X-Ray Emission from the Classical Nova
  V2491 Cygni}}.
\bjtitle{\apjl}
\bvolume{697},
\bfpage{54}--\blpage{57}
(\byear{2009}).
doi:\doiurl{10.1088/0004-637X/697/1/L54}
\end{barticle}
\endbibitem

\bibitem[\protect\citeauthoryear{{Talon} et~al.}{1993}]{Talon1993;granat}
\begin{barticle}
\bauthor{\binits{R.} \bsnm{{Talon}}},
\bauthor{\binits{G.} \bsnm{{Trottet}}},
\bauthor{\binits{N.} \bsnm{{Vilmer}}},
\bauthor{\binits{C.} \bsnm{{Barat}}},
\bauthor{\binits{J.-P.} \bsnm{{Dezalay}}},
\bauthor{\binits{R.} \bsnm{{Sunyaev}}},
\bauthor{\binits{O.} \bsnm{{Terekhov}}},
\bauthor{\binits{A.} \bsnm{{Kuznetsov}}},
\batitle{{Hard X-ray and gamma-ray observations of solar flares with the PHEBUS
  experiment}}.
\bjtitle{\solphys}
\bvolume{147},
\bfpage{137}--\blpage{155}
(\byear{1993}).
doi:\doiurl{10.1007/BF00675491}
\end{barticle}
\endbibitem

\bibitem[\protect\citeauthoryear{{Tanaka} et~al.}{2008}]{Tanaka08}
\begin{barticle}
\bauthor{\binits{T.} \bsnm{{Tanaka}}},
\bauthor{\binits{Y.} \bsnm{{Uchiyama}}},
\bauthor{\binits{F.A.} \bsnm{{Aharonian}}},
\bauthor{\binits{T.} \bsnm{{Takahashi}}},
\bauthor{\binits{A.} \bsnm{{Bamba}}},
\bauthor{\binits{J.S.} \bsnm{{Hiraga}}},
\bauthor{\binits{J.} \bsnm{{Kataoka}}},
\bauthor{\binits{T.} \bsnm{{Kishishita}}},
\bauthor{\binits{M.} \bsnm{{Kokubun}}},
\bauthor{\binits{K.} \bsnm{{Mori}}},
\bauthor{\binits{K.} \bsnm{{Nakazawa}}},
\bauthor{\binits{R.} \bsnm{{Petre}}},
\bauthor{\binits{H.} \bsnm{{Tajima}}},
\bauthor{\binits{S.} \bsnm{{Watanabe}}},
\batitle{{Study of Nonthermal Emission from SNR RX J1713.7-3946 with Suzaku}}.
\bjtitle{\apj}
\bvolume{685},
\bfpage{988}--\blpage{1004}
(\byear{2008}).
doi:\doiurl{10.1086/591020}
\end{barticle}
\endbibitem

\bibitem[\protect\citeauthoryear{{Tanaka} et~al.}{1984}]{Tanaka1984;tenma}
\begin{barticle}
\bauthor{\binits{Y.} \bsnm{{Tanaka}}},
\bauthor{\binits{M.} \bsnm{{Fujii}}},
\bauthor{\binits{H.} \bsnm{{Inoue}}},
\bauthor{\binits{N.} \bsnm{{Kawai}}},
\bauthor{\binits{K.} \bsnm{{Koyama}}},
\bauthor{\binits{Y.} \bsnm{{Maejima}}},
\bauthor{\binits{F.} \bsnm{{Makino}}},
\bauthor{\binits{K.} \bsnm{{Makishima}}},
\bauthor{\binits{M.} \bsnm{{Matsuoka}}},
\bauthor{\binits{K.} \bsnm{{Mitsuda}}},
\batitle{{X-ray astronomy satellite Tenma}}.
\bjtitle{\pasj}
\bvolume{36},
\bfpage{641}--\blpage{658}
(\byear{1984})
\end{barticle}
\endbibitem

\bibitem[\protect\citeauthoryear{{Tavecchio} et~al.}{2007}]{Tavecchio07}
\begin{barticle}
\bauthor{\binits{F.} \bsnm{{Tavecchio}}},
\bauthor{\binits{L.} \bsnm{{Maraschi}}},
\bauthor{\binits{G.} \bsnm{{Ghisellini}}},
\bauthor{\binits{J.} \bsnm{{Kataoka}}},
\bauthor{\binits{L.} \bsnm{{Foschini}}},
\bauthor{\binits{R.M.} \bsnm{{Sambruna}}},
\bauthor{\binits{G.} \bsnm{{Tagliaferri}}},
\batitle{{Low-Energy Cutoffs and Hard X-Ray Spectra in High-z Radio-loud
  Quasars: The Suzaku View of RBS 315}}.
\bjtitle{\apj}
\bvolume{665},
\bfpage{980}--\blpage{989}
(\byear{2007}).
doi:\doiurl{10.1086/519156}
\end{barticle}
\endbibitem

\bibitem[\protect\citeauthoryear{{Tawara} et~al.}{2001}]{Tawara2001;infocus}
\begin{bchapter}
\bauthor{\binits{Y.} \bsnm{{Tawara}}},
\bauthor{\binits{K.} \bsnm{{Yamashita}}},
\bauthor{\binits{Y.} \bsnm{{Ogasaka}}},
\bauthor{\binits{K.} \bsnm{{Tamura}}},
\bauthor{\binits{K.} \bsnm{{Haga}}},
\bauthor{\binits{T.} \bsnm{{Okajima}}},
\bauthor{\binits{S.} \bsnm{{Ichimaru}}},
\bauthor{\binits{S.} \bsnm{{Takahasi}}},
\bauthor{\binits{A.} \bsnm{{Gotou}}},
\bauthor{\binits{H.} \bsnm{{Kitou}}},
\bauthor{\binits{S.} \bsnm{{Fukuda}}},
\bauthor{\binits{Y.} \bsnm{{Kamata}}},
\bauthor{\binits{A.} \bsnm{{Furuzawa}}},
\bauthor{\binits{F.} \bsnm{{Akimoto}}},
\bauthor{\binits{T.} \bsnm{{Yoshioka}}},
\bauthor{\binits{K.} \bsnm{{Kondou}}},
\bauthor{\binits{Y.} \bsnm{{Haba}}},
\bauthor{\binits{T.} \bsnm{{Tanaka}}},
\bauthor{\binits{H.} \bsnm{{Kuneida}}},
\bauthor{\binits{J.} \bsnm{{Tueller}}},
\bauthor{\binits{P.J.} \bsnm{{Serlemitsos}}},
\bauthor{\binits{Y.} \bsnm{{Soong}}},
\bauthor{\binits{K.W.} \bsnm{{Chan}}},
\bauthor{\binits{S.M.} \bsnm{{Owens}}},
\bauthor{\binits{B.} \bsnm{{Barber}}},
\bauthor{\binits{E.} \bsnm{{Dereniak}}},
\bauthor{\binits{E.E.} \bsnm{{Young}}},
\bctitle{{InFOCuS Balloon-borne Hard X-ray Experiment with Multilayer
  Supermirror X-ray Telescope}},
in \bbtitle{New Century of X-ray Astronomy},
ed. by \beditor{\binits{H.} \bsnm{{Inoue}}},
\beditor{\binits{H.} \bsnm{{Kunieda}}}
\bsertitle{Astronomical Society of the Pacific Conference Series},
vol. \bseriesno{251},
\byear{2001},
p. \bfpage{594}
\end{bchapter}
\endbibitem

\bibitem[\protect\citeauthoryear{{Taylor} et~al.}{1981}]{Taylor1981;exosat}
\begin{barticle}
\bauthor{\binits{B.G.} \bsnm{{Taylor}}},
\bauthor{\binits{R.D.} \bsnm{{Andresen}}},
\bauthor{\binits{A.} \bsnm{{Peacock}}},
\bauthor{\binits{R.} \bsnm{{Zobl}}},
\batitle{{The EXOSAT mission}}.
\bjtitle{\ssr}
\bvolume{30},
\bfpage{479}--\blpage{494}
(\byear{1981}).
doi:\doiurl{10.1007/BF01246069}
\end{barticle}
\endbibitem

\bibitem[\protect\citeauthoryear{{Teegarden}}{1994}]{Teegarden94}
\begin{barticle}
\bauthor{\binits{B.J.} \bsnm{{Teegarden}}},
\batitle{{A review of recent results in gamma-ray astronomy obtained from
  high-altitude balloons}}.
\bjtitle{\apjs}
\bvolume{92},
\bfpage{363}--\blpage{368}
(\byear{1994}).
doi:\doiurl{10.1086/191979}
\end{barticle}
\endbibitem

\bibitem[\protect\citeauthoryear{{Teegarden}
  et~al.}{1985}]{Teegarden1985;gris2}
\begin{barticle}
\bauthor{\binits{B.J.} \bsnm{{Teegarden}}},
\bauthor{\binits{T.L.} \bsnm{{Cline}}},
\bauthor{\binits{N.} \bsnm{{Gehrels}}},
\bauthor{\binits{G.} \bsnm{{Porreca}}},
\bauthor{\binits{J.} \bsnm{{Tueller}}},
\bauthor{\binits{M.} \bsnm{{Leventhal}}},
\bauthor{\binits{A.F.} \bsnm{{Huters}}},
\bauthor{\binits{C.J.} \bsnm{{MacCallum}}},
\bauthor{\binits{P.D.} \bsnm{{Stang}}},
\batitle{{The Gamma-Ray Imaging Spectrometer (GRIS): A new balloon-borne
  experiment for gamma-ray line astronomy}}.
\bjtitle{International Cosmic Ray Conference}
\bvolume{3},
\bfpage{307}--\blpage{310}
(\byear{1985})
\end{barticle}
\endbibitem

\bibitem[\protect\citeauthoryear{{Tendulkar} et~al.}{2014}]{Tendulkar14}
\begin{barticle}
\bauthor{\binits{S.P.} \bsnm{{Tendulkar}}},
\bauthor{\binits{F.} \bsnm{{F{\"u}rst}}},
\bauthor{\binits{K.} \bsnm{{Pottschmidt}}},
\bauthor{\binits{M.} \bsnm{{Bachetti}}},
\bauthor{\binits{V.B.} \bsnm{{Bhalerao}}},
\bauthor{\binits{S.E.} \bsnm{{Boggs}}},
\bauthor{\binits{F.E.} \bsnm{{Christensen}}},
\bauthor{\binits{W.W.} \bsnm{{Craig}}},
\bauthor{\binits{C.A.} \bsnm{{Hailey}}},
\bauthor{\binits{F.A.} \bsnm{{Harrison}}},
\bauthor{\binits{D.} \bsnm{{Stern}}},
\bauthor{\binits{J.A.} \bsnm{{Tomsick}}},
\bauthor{\binits{D.J.} \bsnm{{Walton}}},
\bauthor{\binits{W.} \bsnm{{Zhang}}},
\batitle{{NuSTAR Discovery of a Cyclotron Line in the Be/X-Ray Binary RX
  J0520.5-6932 during Outburst}}.
\bjtitle{\apj}
\bvolume{795},
\bfpage{154}
(\byear{2014}).
doi:\doiurl{10.1088/0004-637X/795/2/154}
\end{barticle}
\endbibitem

\bibitem[\protect\citeauthoryear{{Terada} et~al.}{2006}]{Terada06}
\begin{barticle}
\bauthor{\binits{Y.} \bsnm{{Terada}}},
\bauthor{\binits{T.} \bsnm{{Mihara}}},
\bauthor{\binits{M.} \bsnm{{Nakajima}}},
\bauthor{\binits{M.} \bsnm{{Suzuki}}},
\bauthor{\binits{N.} \bsnm{{Isobe}}},
\bauthor{\binits{K.} \bsnm{{Makishima}}},
\bauthor{\binits{H.} \bsnm{{Takahashi}}},
\bauthor{\binits{T.} \bsnm{{Enoto}}},
\bauthor{\binits{M.} \bsnm{{Kokubun}}},
\bauthor{\binits{T.} \bsnm{{Kitaguchi}}},
\bauthor{\binits{S.} \bsnm{{Naik}}},
\bauthor{\binits{T.} \bsnm{{Dotani}}},
\bauthor{\binits{F.} \bsnm{{Nagase}}},
\bauthor{\binits{T.} \bsnm{{Tanaka}}},
\bauthor{\binits{S.} \bsnm{{Watanabe}}},
\bauthor{\binits{S.} \bsnm{{Kitamoto}}},
\bauthor{\binits{K.} \bsnm{{Sudoh}}},
\bauthor{\binits{A.} \bsnm{{Yoshida}}},
\bauthor{\binits{Y.} \bsnm{{Nakagawa}}},
\bauthor{\binits{S.} \bsnm{{Sugita}}},
\bauthor{\binits{T.} \bsnm{{Kohmura}}},
\bauthor{\binits{T.} \bsnm{{Kotani}}},
\bauthor{\binits{D.} \bsnm{{Yonetoku}}},
\bauthor{\binits{L.} \bsnm{{Angelini}}},
\bauthor{\binits{J.} \bsnm{{Cottam}}},
\bauthor{\binits{K.} \bsnm{{Mukai}}},
\bauthor{\binits{R.} \bsnm{{Kelley}}},
\bauthor{\binits{Y.} \bsnm{{Soong}}},
\bauthor{\binits{M.} \bsnm{{Bautz}}},
\bauthor{\binits{S.} \bsnm{{Kissel}}},
\bauthor{\binits{J.} \bsnm{{Doty}}},
\batitle{{Cyclotron Resonance Energies at a Low X-Ray Luminosity: A0535+262
  Observed with Suzaku}}.
\bjtitle{\apjl}
\bvolume{648},
\bfpage{139}--\blpage{142}
(\byear{2006}).
doi:\doiurl{10.1086/508018}
\end{barticle}
\endbibitem

\bibitem[\protect\citeauthoryear{{Tomsick} et~al.}{1999}]{Tomsick99}
\begin{barticle}
\bauthor{\binits{J.A.} \bsnm{{Tomsick}}},
\bauthor{\binits{P.} \bsnm{{Kaaret}}},
\bauthor{\binits{R.A.} \bsnm{{Kroeger}}},
\bauthor{\binits{R.A.} \bsnm{{Remillard}}},
\batitle{{Broadband X-Ray Spectra of the Black Hole Candidate GRO J1655-40}}.
\bjtitle{\apj}
\bvolume{512},
\bfpage{892}--\blpage{900}
(\byear{1999}).
doi:\doiurl{10.1086/306797}
\end{barticle}
\endbibitem

\bibitem[\protect\citeauthoryear{{Trudolyubov} et~al.}{1999}]{Trudolyubov99}
\begin{barticle}
\bauthor{\binits{S.} \bsnm{{Trudolyubov}}},
\bauthor{\binits{E.} \bsnm{{Churazov}}},
\bauthor{\binits{M.} \bsnm{{Gilfanov}}},
\bauthor{\binits{M.} \bsnm{{Revnivtsev}}},
\bauthor{\binits{R.} \bsnm{{Sunyaev}}},
\bauthor{\binits{N.} \bsnm{{Khavenson}}},
\bauthor{\binits{A.} \bsnm{{Dyachkov}}},
\bauthor{\binits{I.} \bsnm{{Tserenin}}},
\bauthor{\binits{M.} \bsnm{{Vargas}}},
\bauthor{\binits{P.} \bsnm{{Goldoni}}},
\bauthor{\binits{P.} \bsnm{{Laurent}}},
\bauthor{\binits{J.} \bsnm{{Paul}}},
\bauthor{\binits{E.} \bsnm{{Jourdain}}},
\bauthor{\binits{J.-P.} \bsnm{{Roques}}},
\bauthor{\binits{P.} \bsnm{{Mandrou}}},
\bauthor{\binits{G.} \bsnm{{Vedrenne}}},
\batitle{{Spring, 1997 GRANAT/SIGMA observations of the Galactic Center:
  discovery of the X-ray nova GRS 1737-31}}.
\bjtitle{\aap}
\bvolume{342},
\bfpage{496}--\blpage{501}
(\byear{1999})
\end{barticle}
\endbibitem

\bibitem[\protect\citeauthoryear{{Tr{\"u}mper} et~al.}{1977}]{Trumper77}
\begin{bchapter}
\bauthor{\binits{J.} \bsnm{{Tr{\"u}mper}}},
\bauthor{\binits{W.} \bsnm{{Pietsch}}},
\bauthor{\binits{C.} \bsnm{{Reppin}}},
\bauthor{\binits{B.} \bsnm{{Sacco}}},
\bctitle{{Evidence for Strong Cyclotron Emission in the Hard X-Ray Spectrum of
  Her X-1}},
in \bbtitle{Eighth Texas Symposium on Relativistic Astrophysics},
ed. by \beditor{\binits{M.D.} \bsnm{{Papagiannis}}}
\bsertitle{Annals of the New York Academy of Sciences},
vol. \bseriesno{302},
\byear{1977},
p. \bfpage{538}.
doi:\doiurl{10.1111/j.1749-6632.1977.tb37072.x}
\end{bchapter}
\endbibitem

\bibitem[\protect\citeauthoryear{{Tr{\"u}mper} et~al.}{1978}]{Trumper78}
\begin{barticle}
\bauthor{\binits{J.} \bsnm{{Tr{\"u}mper}}},
\bauthor{\binits{W.} \bsnm{{Pietsch}}},
\bauthor{\binits{C.} \bsnm{{Reppin}}},
\bauthor{\binits{W.} \bsnm{{Voges}}},
\bauthor{\binits{R.} \bsnm{{Staubert}}},
\bauthor{\binits{E.} \bsnm{{Kendziorra}}},
\batitle{{Evidence for strong cyclotron line emission in the hard X-ray
  spectrum of Hercules X-1}}.
\bjtitle{\apjl}
\bvolume{219},
\bfpage{105}--\blpage{110}
(\byear{1978}).
doi:\doiurl{10.1086/182617}
\end{barticle}
\endbibitem

\bibitem[\protect\citeauthoryear{{Tueller} et~al.}{1990}]{Tueller90}
\begin{barticle}
\bauthor{\binits{J.} \bsnm{{Tueller}}},
\bauthor{\binits{S.} \bsnm{{Barthelmy}}},
\bauthor{\binits{N.} \bsnm{{Gehrels}}},
\bauthor{\binits{B.J.} \bsnm{{Teegarden}}},
\bauthor{\binits{M.} \bsnm{{Leventhal}}},
\bauthor{\binits{C.J.} \bsnm{{MacCallum}}},
\batitle{{Observations of gamma-ray line profiles from SN 1987A}}.
\bjtitle{\apjl}
\bvolume{351},
\bfpage{41}--\blpage{44}
(\byear{1990}).
doi:\doiurl{10.1086/185675}
\end{barticle}
\endbibitem

\bibitem[\protect\citeauthoryear{{T{\"u}mer} et~al.}{1984}]{Tumer84}
\begin{barticle}
\bauthor{\binits{O.T.} \bsnm{{T{\"u}mer}}},
\bauthor{\binits{J.} \bsnm{{Long}}},
\bauthor{\binits{T.} \bsnm{{Oneill}}},
\bauthor{\binits{A.} \bsnm{{Zych}}},
\bauthor{\binits{R.S.} \bsnm{{White}}},
\bauthor{\binits{B.} \bsnm{{Dayton}}},
\batitle{{Gamma rays of 0.3-30 MeV from PSR0833$-$45}}.
\bjtitle{\nat}
\bvolume{310},
\bfpage{214}--\blpage{216}
(\byear{1984}).
doi:\doiurl{10.1038/310214a0}
\end{barticle}
\endbibitem

\bibitem[\protect\citeauthoryear{{Turner} et~al.}{1981}]{Turner1981;exosat}
\begin{barticle}
\bauthor{\binits{M.J.L.} \bsnm{{Turner}}},
\bauthor{\binits{A.} \bsnm{{Smith}}},
\bauthor{\binits{H.U.} \bsnm{{Zimmermann}}},
\batitle{{The medium energy instrument on EXOSAT}}.
\bjtitle{\ssr}
\bvolume{30},
\bfpage{513}--\blpage{524}
(\byear{1981}).
doi:\doiurl{10.1007/BF01246071}
\end{barticle}
\endbibitem

\bibitem[\protect\citeauthoryear{{Ubertini} et~al.}{1984}]{Ubertini84}
\begin{barticle}
\bauthor{\binits{P.} \bsnm{{Ubertini}}},
\bauthor{\binits{A.} \bsnm{{Bazzano}}},
\bauthor{\binits{C.} \bsnm{{La Padula}}},
\bauthor{\binits{V.F.} \bsnm{{Polcaro}}},
\bauthor{\binits{R.K.} \bsnm{{Manchanda}}},
\batitle{{Hard X-ray variability of three active galactic nuclei}}.
\bjtitle{\apj}
\bvolume{284},
\bfpage{54}--\blpage{59}
(\byear{1984}).
doi:\doiurl{10.1086/162383}
\end{barticle}
\endbibitem

\bibitem[\protect\citeauthoryear{{Ubertini} et~al.}{1991}]{Ubertini91}
\begin{barticle}
\bauthor{\binits{P.} \bsnm{{Ubertini}}},
\bauthor{\binits{A.} \bsnm{{Bazzano}}},
\bauthor{\binits{C.} \bsnm{{La Padula}}},
\bauthor{\binits{R.K.} \bsnm{{Manchanda}}},
\bauthor{\binits{V.F.} \bsnm{{Polcaro}}},
\bauthor{\binits{R.} \bsnm{{Staubert}}},
\bauthor{\binits{E.} \bsnm{{Kendziorra}}},
\bauthor{\binits{F.} \bsnm{{Perotti}}},
\batitle{{Low-flux hard X-ray observation of Cygnus X-1}}.
\bjtitle{\apj}
\bvolume{383},
\bfpage{263}--\blpage{268}
(\byear{1991}).
doi:\doiurl{10.1086/170782}
\end{barticle}
\endbibitem

\bibitem[\protect\citeauthoryear{{Ubertini} et~al.}{1992}]{Ubertini92}
\begin{barticle}
\bauthor{\binits{P.} \bsnm{{Ubertini}}},
\bauthor{\binits{A.} \bsnm{{Bazzano}}},
\bauthor{\binits{M.} \bsnm{{Cocchi}}},
\bauthor{\binits{C.} \bsnm{{La Padula}}},
\bauthor{\binits{R.K.} \bsnm{{Sood}}},
\batitle{{Hard X-ray observation of Scorpius X - 1}}.
\bjtitle{\apj}
\bvolume{386},
\bfpage{710}--\blpage{714}
(\byear{1992}).
doi:\doiurl{10.1086/171051}
\end{barticle}
\endbibitem

\bibitem[\protect\citeauthoryear{{Ubertini} et~al.}{1993}]{Ubertini93}
\begin{barticle}
\bauthor{\binits{P.} \bsnm{{Ubertini}}},
\bauthor{\binits{A.} \bsnm{{Bazzano}}},
\bauthor{\binits{M.} \bsnm{{Cocchi}}},
\bauthor{\binits{C.} \bsnm{{La Padula}}},
\bauthor{\binits{R.} \bsnm{{Sood}}},
\batitle{{Hard X-ray observation of Centaurus A}}.
\bjtitle{\aaps}
\bvolume{97},
\bfpage{105}--\blpage{108}
(\byear{1993})
\end{barticle}
\endbibitem

\bibitem[\protect\citeauthoryear{{Ubertini} et~al.}{1994}]{Ubertini94}
\begin{barticle}
\bauthor{\binits{P.} \bsnm{{Ubertini}}},
\bauthor{\binits{A.} \bsnm{{Bazzano}}},
\bauthor{\binits{M.} \bsnm{{Cocchi}}},
\bauthor{\binits{C.} \bsnm{{La Padula}}},
\bauthor{\binits{V.F.} \bsnm{{Polcaro}}},
\bauthor{\binits{R.} \bsnm{{Staubert}}},
\bauthor{\binits{E.} \bsnm{{Kendziorra}}},
\batitle{{Hard X-ray timing observation of the Crab Pulsar and Cygnus X-1}}.
\bjtitle{\apj}
\bvolume{421},
\bfpage{269}--\blpage{275}
(\byear{1994}).
doi:\doiurl{10.1086/173644}
\end{barticle}
\endbibitem

\bibitem[\protect\citeauthoryear{{Ubertini}
  et~al.}{2003}]{Ubertini2003;integral}
\begin{barticle}
\bauthor{\binits{P.} \bsnm{{Ubertini}}},
\bauthor{\binits{F.} \bsnm{{Lebrun}}},
\bauthor{\binits{G.} \bsnm{{Di Cocco}}},
\bauthor{\binits{A.} \bsnm{{Bazzano}}},
\bauthor{\binits{A.J.} \bsnm{{Bird}}},
\bauthor{\binits{K.} \bsnm{{Broenstad}}},
\bauthor{\binits{A.} \bsnm{{Goldwurm}}},
\bauthor{\binits{G.} \bsnm{{La Rosa}}},
\bauthor{\binits{C.} \bsnm{{Labanti}}},
\bauthor{\binits{P.} \bsnm{{Laurent}}},
\bauthor{\binits{I.F.} \bsnm{{Mirabel}}},
\bauthor{\binits{E.M.} \bsnm{{Quadrini}}},
\bauthor{\binits{B.} \bsnm{{Ramsey}}},
\bauthor{\binits{V.} \bsnm{{Reglero}}},
\bauthor{\binits{L.} \bsnm{{Sabau}}},
\bauthor{\binits{B.} \bsnm{{Sacco}}},
\bauthor{\binits{R.} \bsnm{{Staubert}}},
\bauthor{\binits{L.} \bsnm{{Vigroux}}},
\bauthor{\binits{M.C.} \bsnm{{Weisskopf}}},
\bauthor{\binits{A.A.} \bsnm{{Zdziarski}}},
\batitle{{IBIS: The Imager on-board INTEGRAL}}.
\bjtitle{\aap}
\bvolume{411},
\bfpage{131}--\blpage{139}
(\byear{2003}).
doi:\doiurl{10.1051/0004-6361:20031224}
\end{barticle}
\endbibitem

\bibitem[\protect\citeauthoryear{{Ueda} et~al.}{2010}]{Ueda10}
\begin{barticle}
\bauthor{\binits{Y.} \bsnm{{Ueda}}},
\bauthor{\binits{K.} \bsnm{{Honda}}},
\bauthor{\binits{H.} \bsnm{{Takahashi}}},
\bauthor{\binits{C.} \bsnm{{Done}}},
\bauthor{\binits{H.} \bsnm{{Shirai}}},
\bauthor{\binits{Y.} \bsnm{{Fukazawa}}},
\bauthor{\binits{K.} \bsnm{{Yamaoka}}},
\bauthor{\binits{S.} \bsnm{{Naik}}},
\bauthor{\binits{H.} \bsnm{{Awaki}}},
\bauthor{\binits{K.} \bsnm{{Ebisawa}}},
\bauthor{\binits{J.} \bsnm{{Rodriguez}}},
\bauthor{\binits{S.} \bsnm{{Chaty}}},
\batitle{{Suzaku Observation of GRS 1915+105: Evolution of Accretion Disk
  Structure during Limit-cycle Oscillation}}.
\bjtitle{\apj}
\bvolume{713},
\bfpage{257}--\blpage{268}
(\byear{2010}).
doi:\doiurl{10.1088/0004-637X/713/1/257}
\end{barticle}
\endbibitem

\bibitem[\protect\citeauthoryear{{Ulmer}}{1975}]{Ulmer75}
\begin{barticle}
\bauthor{\binits{M.P.} \bsnm{{Ulmer}}},
\batitle{{Observations of six binary X-ray sources with the UCSD OSO-7 X-ray
  telescope}}.
\bjtitle{\apj}
\bvolume{196},
\bfpage{827}--\blpage{835}
(\byear{1975}).
doi:\doiurl{10.1086/153473}
\end{barticle}
\endbibitem

\bibitem[\protect\citeauthoryear{{Ulmer} et~al.}{1972a}]{Ulmer72}
\begin{barticle}
\bauthor{\binits{M.P.} \bsnm{{Ulmer}}},
\bauthor{\binits{W.A.} \bsnm{{Baity}}},
\bauthor{\binits{W.A.} \bsnm{{Wheaton}}},
\bauthor{\binits{L.E.} \bsnm{{Peterson}}},
\batitle{{Observations of the VELA XR-1 by the UCSD X-Ray Telescope on OSO-7}}.
\bjtitle{\apjl}
\bvolume{178},
\bfpage{121}
(\byear{1972}a).
doi:\doiurl{10.1086/181099}
\end{barticle}
\endbibitem

\bibitem[\protect\citeauthoryear{{Ulmer} et~al.}{1972b}]{Ulmer1972;oso7}
\begin{bchapter}
\bauthor{\binits{M.P.} \bsnm{{Ulmer}}},
\bauthor{\binits{W.A.} \bsnm{{Baity}}},
\bauthor{\binits{W.A.} \bsnm{{Wheaton}}},
\bauthor{\binits{L.E.} \bsnm{{Peterson}}},
\bctitle{{UCSD X-Ray Observations of the Vela Region from OSO-7.}},
in \bbtitle{Bulletin of the American Astronomical Society}.
\bsertitle{Bulletin of the American Astronomical Society},
vol. \bseriesno{4},
\byear{1972}b,
p. \bfpage{220}
\end{bchapter}
\endbibitem

\bibitem[\protect\citeauthoryear{{Ulmer} et~al.}{1973a}]{Ulmer73}
\begin{barticle}
\bauthor{\binits{M.P.} \bsnm{{Ulmer}}},
\bauthor{\binits{W.A.} \bsnm{{Baity}}},
\bauthor{\binits{W.A.} \bsnm{{Wheaton}}},
\bauthor{\binits{L.E.} \bsnm{{Peterson}}},
\batitle{{Observations of the Binary X-ray Source SMC X-1 from OSO-7}}.
\bjtitle{Nature Physical Science}
\bvolume{242},
\bfpage{121}--\blpage{123}
(\byear{1973}a).
doi:\doiurl{10.1038/physci242121b0}
\end{barticle}
\endbibitem

\bibitem[\protect\citeauthoryear{{Ulmer} et~al.}{1973b}]{Ulmer73b}
\begin{barticle}
\bauthor{\binits{M.P.} \bsnm{{Ulmer}}},
\bauthor{\binits{W.A.} \bsnm{{Baity}}},
\bauthor{\binits{W.A.} \bsnm{{Wheaton}}},
\bauthor{\binits{L.E.} \bsnm{{Peterson}}},
\batitle{{The Spectrum and Variability of Hercules X-1 Observed by OSO-7}}.
\bjtitle{\apjl}
\bvolume{181},
\bfpage{33}
(\byear{1973}b).
doi:\doiurl{10.1086/181179}
\end{barticle}
\endbibitem

\bibitem[\protect\citeauthoryear{{Ulmer} et~al.}{1974a}]{Ulmer74b}
\begin{barticle}
\bauthor{\binits{M.P.} \bsnm{{Ulmer}}},
\bauthor{\binits{A.} \bsnm{{Sammuli}}},
\bauthor{\binits{W.A.} \bsnm{{Baity}}},
\bauthor{\binits{W.A.} \bsnm{{Wheaton}}},
\bauthor{\binits{L.E.} \bsnm{{Peterson}}},
\batitle{{Long-Term Observations of Cygnus X-2 from OSO-7}}.
\bjtitle{\apj}
\bvolume{189},
\bfpage{339}--\blpage{342}
(\byear{1974}a).
doi:\doiurl{10.1086/152808}
\end{barticle}
\endbibitem

\bibitem[\protect\citeauthoryear{{Ulmer} et~al.}{1974b}]{Ulmer74}
\begin{barticle}
\bauthor{\binits{M.P.} \bsnm{{Ulmer}}},
\bauthor{\binits{W.A.} \bsnm{{Baity}}},
\bauthor{\binits{W.A.} \bsnm{{Wheaton}}},
\bauthor{\binits{L.E.} \bsnm{{Peterson}}},
\batitle{{Observations of the 4.8-hour variations of Cygnus X-3 above 7 keV
  from the OSO-7}}.
\bjtitle{\apj}
\bvolume{192},
\bfpage{691}--\blpage{695}
(\byear{1974}b).
doi:\doiurl{10.1086/153106}
\end{barticle}
\endbibitem

\bibitem[\protect\citeauthoryear{{Vadawale} et~al.}{2015}]{Vadawale15}
\begin{barticle}
\bauthor{\binits{S.V.} \bsnm{{Vadawale}}},
\bauthor{\binits{T.} \bsnm{{Chattopadhyay}}},
\bauthor{\binits{A.R.} \bsnm{{Rao}}},
\bauthor{\binits{D.} \bsnm{{Bhattacharya}}},
\bauthor{\binits{V.B.} \bsnm{{Bhalerao}}},
\bauthor{\binits{N.} \bsnm{{Vagshette}}},
\bauthor{\binits{P.} \bsnm{{Pawar}}},
\bauthor{\binits{S.} \bsnm{{Sreekumar}}},
\batitle{{Hard X-ray polarimetry with Astrosat-CZTI}}.
\bjtitle{\aap}
\bvolume{578},
\bfpage{73}
(\byear{2015}).
doi:\doiurl{10.1051/0004-6361/201525686}
\end{barticle}
\endbibitem

\bibitem[\protect\citeauthoryear{{Vadawale} et~al.}{2016}]{Vadawale16}
\begin{botherref}
\oauthor{\binits{S.V.} \bsnm{{Vadawale}}},
\oauthor{\binits{T.} \bsnm{{Chattopadhyay}}},
\oauthor{\binits{N.P.S.} \bsnm{{Mithun}}},
\oauthor{\binits{A.R.} \bsnm{{Rao}}},
\oauthor{\binits{D.} \bsnm{{Bhattacharya}}},
\oauthor{\binits{V.} \bsnm{{Bhalerao}}},
{GRB160131A: detection of polarisation by Astrosat CZTI.}
GRB Coordinates Network
\textbf{19011}
(2016)
\end{botherref}
\endbibitem

\bibitem[\protect\citeauthoryear{{Valinia} et~al.}{1999}]{Valinia99}
\begin{barticle}
\bauthor{\binits{A.} \bsnm{{Valinia}}},
\bauthor{\binits{M.J.} \bsnm{{Henriksen}}},
\bauthor{\binits{M.} \bsnm{{Loewenstein}}},
\bauthor{\binits{K.} \bsnm{{Roettiger}}},
\bauthor{\binits{R.F.} \bsnm{{Mushotzky}}},
\bauthor{\binits{G.} \bsnm{{Madejski}}},
\batitle{{Rossi X-Ray Timing Explorer Hard X-Ray Observation of A754:
  Constraining the Hottest Temperature Component and the Intracluster Magnetic
  Field}}.
\bjtitle{\apj}
\bvolume{515},
\bfpage{42}--\blpage{49}
(\byear{1999}).
doi:\doiurl{10.1086/307022}
\end{barticle}
\endbibitem

\bibitem[\protect\citeauthoryear{{van der Horst} et~al.}{2010}]{Vanderhorst10}
\begin{barticle}
\bauthor{\binits{A.J.} \bsnm{{van der Horst}}},
\bauthor{\binits{V.} \bsnm{{Connaughton}}},
\bauthor{\binits{C.} \bsnm{{Kouveliotou}}},
\bauthor{\binits{E.} \bsnm{{G{\"o}{\v g}{\"u}{\c s}}}},
\bauthor{\binits{Y.} \bsnm{{Kaneko}}},
\bauthor{\binits{S.} \bsnm{{Wachter}}},
\bauthor{\binits{M.S.} \bsnm{{Briggs}}},
\bauthor{\binits{J.} \bsnm{{Granot}}},
\bauthor{\binits{E.} \bsnm{{Ramirez-Ruiz}}},
\bauthor{\binits{P.M.} \bsnm{{Woods}}},
\bauthor{\binits{R.L.} \bsnm{{Aptekar}}},
\bauthor{\binits{S.D.} \bsnm{{Barthelmy}}},
\bauthor{\binits{J.R.} \bsnm{{Cummings}}},
\bauthor{\binits{M.H.} \bsnm{{Finger}}},
\bauthor{\binits{D.D.} \bsnm{{Frederiks}}},
\bauthor{\binits{N.} \bsnm{{Gehrels}}},
\bauthor{\binits{C.R.} \bsnm{{Gelino}}},
\bauthor{\binits{D.M.} \bsnm{{Gelino}}},
\bauthor{\binits{S.} \bsnm{{Golenetskii}}},
\bauthor{\binits{K.} \bsnm{{Hurley}}},
\bauthor{\binits{H.A.} \bsnm{{Krimm}}},
\bauthor{\binits{E.P.} \bsnm{{Mazets}}},
\bauthor{\binits{J.E.} \bsnm{{McEnery}}},
\bauthor{\binits{C.A.} \bsnm{{Meegan}}},
\bauthor{\binits{P.P.} \bsnm{{Oleynik}}},
\bauthor{\binits{D.M.} \bsnm{{Palmer}}},
\bauthor{\binits{V.D.} \bsnm{{Pal'shin}}},
\bauthor{\binits{A.} \bsnm{{Pe'er}}},
\bauthor{\binits{D.} \bsnm{{Svinkin}}},
\bauthor{\binits{M.V.} \bsnm{{Ulanov}}},
\bauthor{\binits{M.} \bsnm{{van der Klis}}},
\bauthor{\binits{A.} \bsnm{{von Kienlin}}},
\bauthor{\binits{A.L.} \bsnm{{Watts}}},
\bauthor{\binits{C.A.} \bsnm{{Wilson-Hodge}}},
\batitle{{Discovery of a New Soft Gamma Repeater: SGR J0418 + 5729}}.
\bjtitle{\apjl}
\bvolume{711},
\bfpage{1}--\blpage{6}
(\byear{2010}).
doi:\doiurl{10.1088/2041-8205/711/1/L1}
\end{barticle}
\endbibitem

\bibitem[\protect\citeauthoryear{{van der Klis}}{1995}]{Vanderklis95}
\begin{botherref}
\oauthor{\binits{M.} \bsnm{{van der Klis}}},
{Rapid aperiodic variability in X-ray binaries.}
X-ray Binaries,
252--307
(1995)
\end{botherref}
\endbibitem

\bibitem[\protect\citeauthoryear{{van der Klis}}{1999}]{Vanderklis99}
\begin{barticle}
\bauthor{\binits{M.} \bsnm{{van der Klis}}},
\batitle{{Rossi X-ray Timing Explorer observations of kilohertz QPO}}.
\bjtitle{Nuclear Physics B Proceedings Supplements}
\bvolume{69},
\bfpage{103}--\blpage{112}
(\byear{1999}).
doi:\doiurl{10.1016/S0920-5632(98)00192-3}
\end{barticle}
\endbibitem

\bibitem[\protect\citeauthoryear{{van der Klis} et~al.}{1996}]{Vanderklis96}
\begin{barticle}
\bauthor{\binits{M.} \bsnm{{van der Klis}}},
\bauthor{\binits{J.H.} \bsnm{{Swank}}},
\bauthor{\binits{W.} \bsnm{{Zhang}}},
\bauthor{\binits{K.} \bsnm{{Jahoda}}},
\bauthor{\binits{E.H.} \bsnm{{Morgan}}},
\bauthor{\binits{W.H.G.} \bsnm{{Lewin}}},
\bauthor{\binits{B.} \bsnm{{Vaughan}}},
\bauthor{\binits{J.} \bsnm{{van Paradijs}}},
\batitle{{Discovery of Submillisecond Quasi-periodic Oscillations in the X-Ray
  Flux of Scorpius X-1}}.
\bjtitle{\apjl}
\bvolume{469},
\bfpage{1}
(\byear{1996}).
doi:\doiurl{10.1086/310251}
\end{barticle}
\endbibitem

\bibitem[\protect\citeauthoryear{{van Paradijs} et~al.}{1997}]{Vanparadijs97}
\begin{barticle}
\bauthor{\binits{J.} \bsnm{{van Paradijs}}},
\bauthor{\binits{P.J.} \bsnm{{Groot}}},
\bauthor{\binits{T.} \bsnm{{Galama}}},
\bauthor{\binits{C.} \bsnm{{Kouveliotou}}},
\bauthor{\binits{R.G.} \bsnm{{Strom}}},
\bauthor{\binits{J.} \bsnm{{Telting}}},
\bauthor{\binits{R.G.M.} \bsnm{{Rutten}}},
\bauthor{\binits{G.J.} \bsnm{{Fishman}}},
\bauthor{\binits{C.A.} \bsnm{{Meegan}}},
\bauthor{\binits{M.} \bsnm{{Pettini}}},
\bauthor{\binits{N.} \bsnm{{Tanvir}}},
\bauthor{\binits{J.} \bsnm{{Bloom}}},
\bauthor{\binits{H.} \bsnm{{Pedersen}}},
\bauthor{\binits{H.U.} \bsnm{{N{\o}rdgaard-Nielsen}}},
\bauthor{\binits{M.} \bsnm{{Linden-V{\o}rnle}}},
\bauthor{\binits{J.} \bsnm{{Melnick}}},
\bauthor{\binits{G.} \bsnm{{van der Steene}}},
\bauthor{\binits{M.} \bsnm{{Bremer}}},
\bauthor{\binits{R.} \bsnm{{Naber}}},
\bauthor{\binits{J.} \bsnm{{Heise}}},
\bauthor{\binits{J.} \bsnm{{in't Zand}}},
\bauthor{\binits{E.} \bsnm{{Costa}}},
\bauthor{\binits{M.} \bsnm{{Feroci}}},
\bauthor{\binits{L.} \bsnm{{Piro}}},
\bauthor{\binits{F.} \bsnm{{Frontera}}},
\bauthor{\binits{G.} \bsnm{{Zavattini}}},
\bauthor{\binits{L.} \bsnm{{Nicastro}}},
\bauthor{\binits{E.} \bsnm{{Palazzi}}},
\bauthor{\binits{K.} \bsnm{{Bennett}}},
\bauthor{\binits{L.} \bsnm{{Hanlon}}},
\bauthor{\binits{A.} \bsnm{{Parmar}}},
\batitle{{Transient optical emission from the error box of the {$\gamma$}-ray
  burst of 28 February 1997}}.
\bjtitle{\nat}
\bvolume{386},
\bfpage{686}--\blpage{689}
(\byear{1997}).
doi:\doiurl{10.1038/386686a0}
\end{barticle}
\endbibitem

\bibitem[\protect\citeauthoryear{{Vanderspek} et~al.}{2003}]{Vanderspek03}
\begin{botherref}
\oauthor{\binits{R.} \bsnm{{Vanderspek}}},
\oauthor{\binits{G.} \bsnm{{Crew}}},
\oauthor{\binits{J.} \bsnm{{Doty}}},
\oauthor{\binits{J.} \bsnm{{Villasenor}}},
\oauthor{\binits{G.} \bsnm{{Monnelly}}},
\oauthor{\binits{N.} \bsnm{{Butler}}},
\oauthor{\binits{T.} \bsnm{{Cline}}},
\oauthor{\binits{J.G.} \bsnm{{Jernigan}}},
\oauthor{\binits{A.} \bsnm{{Levine}}},
\oauthor{\binits{F.} \bsnm{{Martel}}},
\oauthor{\binits{E.} \bsnm{{Morgan}}},
\oauthor{\binits{G.} \bsnm{{Prigozhin}}},
\oauthor{\binits{G.} \bsnm{{Azzibrouck}}},
\oauthor{\binits{J.} \bsnm{{Braga}}},
\oauthor{\binits{R.} \bsnm{{Manchanda}}},
\oauthor{\binits{G.} \bsnm{{Pizzichini}}},
\oauthor{\binits{G.} \bsnm{{Ricker}}},
\oauthor{\binits{J.-L.} \bsnm{{Atteia}}},
\oauthor{\binits{N.} \bsnm{{Kawai}}},
\oauthor{\binits{D.} \bsnm{{Lamb}}},
\oauthor{\binits{S.} \bsnm{{Woosley}}},
\oauthor{\binits{T.} \bsnm{{Donaghy}}},
\oauthor{\binits{M.} \bsnm{{Suzuki}}},
\oauthor{\binits{Y.} \bsnm{{Shirasaki}}},
\oauthor{\binits{C.} \bsnm{{Graziani}}},
\oauthor{\binits{M.} \bsnm{{Matsuoka}}},
\oauthor{\binits{T.} \bsnm{{Tamagawa}}},
\oauthor{\binits{K.} \bsnm{{Torii}}},
\oauthor{\binits{T.} \bsnm{{Sakamoto}}},
\oauthor{\binits{A.} \bsnm{{Yoshida}}},
\oauthor{\binits{E.} \bsnm{{Fenimore}}},
\oauthor{\binits{M.} \bsnm{{Galassi}}},
\oauthor{\binits{T.} \bsnm{{Tavenner}}},
\oauthor{\binits{Y.} \bsnm{{Nakagawa}}},
\oauthor{\binits{D.} \bsnm{{Takahashi}}},
\oauthor{\binits{R.} \bsnm{{Satoh}}},
\oauthor{\binits{Y.} \bsnm{{Urata}}},
\oauthor{\binits{M.} \bsnm{{Boer}}},
\oauthor{\binits{J.-F.} \bsnm{{Olive}}},
\oauthor{\binits{J.-P.} \bsnm{{Dezalay}}},
\oauthor{\binits{C.} \bsnm{{Barraud}}},
\oauthor{\binits{K.} \bsnm{{Hurley}}},
{GRB030329 (=H2652): a long, extremely bright GRB localized by the HETE WXM and
  SXC.}
GRB Coordinates Network
\textbf{1997}
(2003)
\end{botherref}
\endbibitem

\bibitem[\protect\citeauthoryear{{Vedrenne}
  et~al.}{2003}]{Vedrenne2003;integral}
\begin{barticle}
\bauthor{\binits{G.} \bsnm{{Vedrenne}}},
\bauthor{\binits{J.-P.} \bsnm{{Roques}}},
\bauthor{\binits{V.} \bsnm{{Sch{\"o}nfelder}}},
\bauthor{\binits{P.} \bsnm{{Mandrou}}},
\bauthor{\binits{G.G.} \bsnm{{Lichti}}},
\bauthor{\binits{A.} \bsnm{{von Kienlin}}},
\bauthor{\binits{B.} \bsnm{{Cordier}}},
\bauthor{\binits{S.} \bsnm{{Schanne}}},
\bauthor{\binits{J.} \bsnm{{Kn{\"o}dlseder}}},
\bauthor{\binits{G.} \bsnm{{Skinner}}},
\bauthor{\binits{P.} \bsnm{{Jean}}},
\bauthor{\binits{F.} \bsnm{{Sanchez}}},
\bauthor{\binits{P.} \bsnm{{Caraveo}}},
\bauthor{\binits{B.} \bsnm{{Teegarden}}},
\bauthor{\binits{P.} \bsnm{{von Ballmoos}}},
\bauthor{\binits{L.} \bsnm{{Bouchet}}},
\bauthor{\binits{P.} \bsnm{{Paul}}},
\bauthor{\binits{J.} \bsnm{{Matteson}}},
\bauthor{\binits{S.} \bsnm{{Boggs}}},
\bauthor{\binits{C.} \bsnm{{Wunderer}}},
\bauthor{\binits{P.} \bsnm{{Leleux}}},
\bauthor{\binits{G.} \bsnm{{Weidenspointner}}},
\bauthor{\binits{P.} \bsnm{{Durouchoux}}},
\bauthor{\binits{R.} \bsnm{{Diehl}}},
\bauthor{\binits{A.} \bsnm{{Strong}}},
\bauthor{\binits{M.} \bsnm{{Cass{\'e}}}},
\bauthor{\binits{M.A.} \bsnm{{Clair}}},
\bauthor{\binits{Y.} \bsnm{{Andr{\'e}}}},
\batitle{{SPI: The spectrometer aboard INTEGRAL}}.
\bjtitle{\aap}
\bvolume{411},
\bfpage{63}--\blpage{70}
(\byear{2003}).
doi:\doiurl{10.1051/0004-6361:20031482}
\end{barticle}
\endbibitem

\bibitem[\protect\citeauthoryear{{Villasenor} et~al.}{2003}]{Villasenor03}
\begin{bchapter}
\bauthor{\binits{J.N.} \bsnm{{Villasenor}}},
\bauthor{\binits{R.} \bsnm{{Dill}}},
\bauthor{\binits{J.P.} \bsnm{{Doty}}},
\bauthor{\binits{G.} \bsnm{{Monnelly}}},
\bauthor{\binits{R.} \bsnm{{Vanderspek}}},
\bauthor{\binits{S.} \bsnm{{Kissel}}},
\bauthor{\binits{G.} \bsnm{{Prigozhin}}},
\bauthor{\binits{G.B.} \bsnm{{Crew}}},
\bauthor{\binits{G.R.} \bsnm{{Ricker}}},
\bctitle{{An Overview of the HETE Soft X-ray Camera}},
in \bbtitle{Gamma-Ray Burst and Afterglow Astronomy 2001: A Workshop
  Celebrating the First Year of the HETE Mission},
ed. by \beditor{\binits{G.R.} \bsnm{{Ricker}}},
\beditor{\binits{R.K.} \bsnm{{Vanderspek}}}
\bsertitle{American Institute of Physics Conference Series},
vol. \bseriesno{662},
\byear{2003},
pp. \bfpage{33}--\blpage{37}.
doi:\doiurl{10.1063/1.1579294}
\end{bchapter}
\endbibitem

\bibitem[\protect\citeauthoryear{{Villasenor} et~al.}{2005}]{Villasenor05}
\begin{barticle}
\bauthor{\binits{J.S.} \bsnm{{Villasenor}}},
\bauthor{\binits{D.Q.} \bsnm{{Lamb}}},
\bauthor{\binits{G.R.} \bsnm{{Ricker}}},
\bauthor{\binits{J.-L.} \bsnm{{Atteia}}},
\bauthor{\binits{N.} \bsnm{{Kawai}}},
\bauthor{\binits{N.} \bsnm{{Butler}}},
\bauthor{\binits{Y.} \bsnm{{Nakagawa}}},
\bauthor{\binits{J.G.} \bsnm{{Jernigan}}},
\bauthor{\binits{M.} \bsnm{{Boer}}},
\bauthor{\binits{G.B.} \bsnm{{Crew}}},
\bauthor{\binits{T.Q.} \bsnm{{Donaghy}}},
\bauthor{\binits{J.} \bsnm{{Doty}}},
\bauthor{\binits{E.E.} \bsnm{{Fenimore}}},
\bauthor{\binits{M.} \bsnm{{Galassi}}},
\bauthor{\binits{C.} \bsnm{{Graziani}}},
\bauthor{\binits{K.} \bsnm{{Hurley}}},
\bauthor{\binits{A.} \bsnm{{Levine}}},
\bauthor{\binits{F.} \bsnm{{Martel}}},
\bauthor{\binits{M.} \bsnm{{Matsuoka}}},
\bauthor{\binits{J.-F.} \bsnm{{Olive}}},
\bauthor{\binits{G.} \bsnm{{Prigozhin}}},
\bauthor{\binits{T.} \bsnm{{Sakamoto}}},
\bauthor{\binits{Y.} \bsnm{{Shirasaki}}},
\bauthor{\binits{M.} \bsnm{{Suzuki}}},
\bauthor{\binits{T.} \bsnm{{Tamagawa}}},
\bauthor{\binits{R.} \bsnm{{Vanderspek}}},
\bauthor{\binits{S.E.} \bsnm{{Woosley}}},
\bauthor{\binits{A.} \bsnm{{Yoshida}}},
\bauthor{\binits{J.} \bsnm{{Braga}}},
\bauthor{\binits{R.} \bsnm{{Manchanda}}},
\bauthor{\binits{G.} \bsnm{{Pizzichini}}},
\bauthor{\binits{K.} \bsnm{{Takagishi}}},
\bauthor{\binits{M.} \bsnm{{Yamauchi}}},
\batitle{{Discovery of the short {$\gamma$}-ray burst GRB 050709}}.
\bjtitle{\nat}
\bvolume{437},
\bfpage{855}--\blpage{858}
(\byear{2005}).
doi:\doiurl{10.1038/nature04213}
\end{barticle}
\endbibitem

\bibitem[\protect\citeauthoryear{{Vilmer}}{1994}]{Vilmer1994;granat}
\begin{barticle}
\bauthor{\binits{N.} \bsnm{{Vilmer}}},
\batitle{{Solar hard X-ray and gamma-ray observations from GRANAT}}.
\bjtitle{\apjs}
\bvolume{90},
\bfpage{611}--\blpage{621}
(\byear{1994})
\end{barticle}
\endbibitem

\bibitem[\protect\citeauthoryear{{Vink} et~al.}{2001}]{Vink01}
\begin{barticle}
\bauthor{\binits{J.} \bsnm{{Vink}}},
\bauthor{\binits{J.M.} \bsnm{{Laming}}},
\bauthor{\binits{J.S.} \bsnm{{Kaastra}}},
\bauthor{\binits{J.A.M.} \bsnm{{Bleeker}}},
\bauthor{\binits{H.} \bsnm{{Bloemen}}},
\bauthor{\binits{U.} \bsnm{{Oberlack}}},
\batitle{{Detection of the 67.9 and 78.4 keV Lines Associated with the
  Radioactive Decay of $^{44}$Ti in Cassiopeia A}}.
\bjtitle{\apjl}
\bvolume{560},
\bfpage{79}--\blpage{82}
(\byear{2001}).
doi:\doiurl{10.1086/324172}
\end{barticle}
\endbibitem

\bibitem[\protect\citeauthoryear{{Virgilli} et~al.}{2017}]{Virgilli17}
\begin{botherref}
\oauthor{\binits{E.} \bsnm{{Virgilli}}},
\oauthor{\binits{V.} \bsnm{{Valsan}}},
\oauthor{\binits{F.} \bsnm{{Frontera}}},
\oauthor{\binits{E.} \bsnm{{Caroli}}},
\oauthor{\binits{V.} \bsnm{{Liccardo}}},
\oauthor{\binits{J.B.} \bsnm{{Stephen}}},
{Expected performances of a Laue Lens with bent crystals}.
Journal of Astronomical Telescopes, Instruments and Systems, submitted
(2017)
\end{botherref}
\endbibitem

\bibitem[\protect\citeauthoryear{{von Ballmoos}
  et~al.}{2005}]{vonBallmoos2005;claire}
\begin{barticle}
\bauthor{\binits{P.} \bsnm{{von Ballmoos}}},
\bauthor{\binits{H.} \bsnm{{Halloin}}},
\bauthor{\binits{J.} \bsnm{{Evrard}}},
\bauthor{\binits{G.} \bsnm{{Skinner}}},
\bauthor{\binits{N.} \bsnm{{Abrosimov}}},
\bauthor{\binits{J.} \bsnm{{Alvarez}}},
\bauthor{\binits{P.} \bsnm{{Bastie}}},
\bauthor{\binits{B.} \bsnm{{Hamelin}}},
\bauthor{\binits{M.} \bsnm{{Hernanz}}},
\bauthor{\binits{P.} \bsnm{{Jean}}},
\bauthor{\binits{J.} \bsnm{{Kn{\"o}dlseder}}},
\bauthor{\binits{B.} \bsnm{{Smither}}},
\batitle{{CLAIRE: First light for a gamma-ray lens}}.
\bjtitle{Experimental Astronomy}
\bvolume{20},
\bfpage{253}--\blpage{267}
(\byear{2005}).
doi:\doiurl{10.1007/s10686-006-9071-0}
\end{barticle}
\endbibitem

\bibitem[\protect\citeauthoryear{{von Kienlin} et~al.}{2014}]{Vonkienlin14}
\begin{barticle}
\bauthor{\binits{A.} \bsnm{{von Kienlin}}},
\bauthor{\binits{C.A.} \bsnm{{Meegan}}},
\bauthor{\binits{W.S.} \bsnm{{Paciesas}}},
\bauthor{\binits{P.N.} \bsnm{{Bhat}}},
\bauthor{\binits{E.} \bsnm{{Bissaldi}}},
\bauthor{\binits{M.S.} \bsnm{{Briggs}}},
\bauthor{\binits{J.M.} \bsnm{{Burgess}}},
\bauthor{\binits{D.} \bsnm{{Byrne}}},
\bauthor{\binits{V.} \bsnm{{Chaplin}}},
\bauthor{\binits{W.} \bsnm{{Cleveland}}},
\bauthor{\binits{V.} \bsnm{{Connaughton}}},
\bauthor{\binits{A.C.} \bsnm{{Collazzi}}},
\bauthor{\binits{G.} \bsnm{{Fitzpatrick}}},
\bauthor{\binits{S.} \bsnm{{Foley}}},
\bauthor{\binits{M.} \bsnm{{Gibby}}},
\bauthor{\binits{M.} \bsnm{{Giles}}},
\bauthor{\binits{A.} \bsnm{{Goldstein}}},
\bauthor{\binits{J.} \bsnm{{Greiner}}},
\bauthor{\binits{D.} \bsnm{{Gruber}}},
\bauthor{\binits{S.} \bsnm{{Guiriec}}},
\bauthor{\binits{A.J.} \bsnm{{van der Horst}}},
\bauthor{\binits{C.} \bsnm{{Kouveliotou}}},
\bauthor{\binits{E.} \bsnm{{Layden}}},
\bauthor{\binits{S.} \bsnm{{McBreen}}},
\bauthor{\binits{S.} \bsnm{{McGlynn}}},
\bauthor{\binits{V.} \bsnm{{Pelassa}}},
\bauthor{\binits{R.D.} \bsnm{{Preece}}},
\bauthor{\binits{A.} \bsnm{{Rau}}},
\bauthor{\binits{D.} \bsnm{{Tierney}}},
\bauthor{\binits{C.A.} \bsnm{{Wilson-Hodge}}},
\bauthor{\binits{S.} \bsnm{{Xiong}}},
\bauthor{\binits{G.} \bsnm{{Younes}}},
\bauthor{\binits{H.-F.} \bsnm{{Yu}}},
\batitle{{The Second Fermi GBM Gamma-Ray Burst Catalog: The First Four Years}}.
\bjtitle{\apjs}
\bvolume{211},
\bfpage{13}
(\byear{2014}).
doi:\doiurl{10.1088/0067-0049/211/1/13}
\end{barticle}
\endbibitem

\bibitem[\protect\citeauthoryear{{Walter} et~al.}{2015}]{Walter15}
\begin{barticle}
\bauthor{\binits{R.} \bsnm{{Walter}}},
\bauthor{\binits{A.A.} \bsnm{{Lutovinov}}},
\bauthor{\binits{E.} \bsnm{{Bozzo}}},
\bauthor{\binits{S.S.} \bsnm{{Tsygankov}}},
\batitle{{High-mass X-ray binaries in the Milky Way. A closer look with
  INTEGRAL}}.
\bjtitle{\aapr}
\bvolume{23},
\bfpage{2}
(\byear{2015}).
doi:\doiurl{10.1007/s00159-015-0082-6}
\end{barticle}
\endbibitem

\bibitem[\protect\citeauthoryear{{Walton} et~al.}{2013}]{Walton13}
\begin{barticle}
\bauthor{\binits{D.J.} \bsnm{{Walton}}},
\bauthor{\binits{F.} \bsnm{{Fuerst}}},
\bauthor{\binits{F.} \bsnm{{Harrison}}},
\bauthor{\binits{D.} \bsnm{{Stern}}},
\bauthor{\binits{M.} \bsnm{{Bachetti}}},
\bauthor{\binits{D.} \bsnm{{Barret}}},
\bauthor{\binits{F.} \bsnm{{Bauer}}},
\bauthor{\binits{S.E.} \bsnm{{Boggs}}},
\bauthor{\binits{F.E.} \bsnm{{Christensen}}},
\bauthor{\binits{W.W.} \bsnm{{Craig}}},
\bauthor{\binits{A.C.} \bsnm{{Fabian}}},
\bauthor{\binits{B.W.} \bsnm{{Grefenstette}}},
\bauthor{\binits{C.J.} \bsnm{{Hailey}}},
\bauthor{\binits{K.K.} \bsnm{{Madsen}}},
\bauthor{\binits{J.M.} \bsnm{{Miller}}},
\bauthor{\binits{A.} \bsnm{{Ptak}}},
\bauthor{\binits{V.} \bsnm{{Rana}}},
\bauthor{\binits{N.A.} \bsnm{{Webb}}},
\bauthor{\binits{W.W.} \bsnm{{Zhang}}},
\batitle{{An Extremely Luminous and Variable Ultraluminous X-Ray Source in the
  Outskirts of Circinus Observed with NuSTAR}}.
\bjtitle{\apj}
\bvolume{779},
\bfpage{148}
(\byear{2013}).
doi:\doiurl{10.1088/0004-637X/779/2/148}
\end{barticle}
\endbibitem

\bibitem[\protect\citeauthoryear{{Webber} and {Reinert}}{1970}]{Webber70}
\begin{barticle}
\bauthor{\binits{W.R.} \bsnm{{Webber}}},
\bauthor{\binits{C.P.} \bsnm{{Reinert}}},
\batitle{{Balloon-Borne Studies of X-Ray Emission from Selected Regions in the
  Northern Sky}}.
\bjtitle{\apj}
\bvolume{162},
\bfpage{883}
(\byear{1970}).
doi:\doiurl{10.1086/150717}
\end{barticle}
\endbibitem

\bibitem[\protect\citeauthoryear{{Wei} et~al.}{2016}]{Wei16}
\begin{botherref}
\oauthor{\binits{J.} \bsnm{{Wei}}},
\oauthor{\binits{B.} \bsnm{{Cordier}}},
\oauthor{\binits{S.} \bsnm{{Antier}}},
\oauthor{\binits{P.} \bsnm{{Antilogus}}},
\oauthor{\binits{J.-L.} \bsnm{{Atteia}}},
\oauthor{\binits{A.} \bsnm{{Bajat}}},
\oauthor{\binits{S.} \bsnm{{Basa}}},
\oauthor{\binits{V.} \bsnm{{Beckmann}}},
\oauthor{\binits{M.G.} \bsnm{{Bernardini}}},
\oauthor{\binits{S.} \bsnm{{Boissier}}},
\oauthor{\binits{L.} \bsnm{{Bouchet}}},
\oauthor{\binits{V.} \bsnm{{Burwitz}}},
\oauthor{\binits{A.} \bsnm{{Claret}}},
\oauthor{\binits{Z.-G.} \bsnm{{Dai}}},
\oauthor{\binits{F.} \bsnm{{Daigne}}},
\oauthor{\binits{J.} \bsnm{{Deng}}},
\oauthor{\binits{D.} \bsnm{{Dornic}}},
\oauthor{\binits{H.} \bsnm{{Feng}}},
\oauthor{\binits{T.} \bsnm{{Foglizzo}}},
\oauthor{\binits{H.} \bsnm{{Gao}}},
\oauthor{\binits{N.} \bsnm{{Gehrels}}},
\oauthor{\binits{O.} \bsnm{{Godet}}},
\oauthor{\binits{A.} \bsnm{{Goldwurm}}},
\oauthor{\binits{F.} \bsnm{{Gonzalez}}},
\oauthor{\binits{L.} \bsnm{{Gosset}}},
\oauthor{\binits{D.} \bsnm{{G{\"o}tz}}},
\oauthor{\binits{C.} \bsnm{{Gouiffes}}},
\oauthor{\binits{F.} \bsnm{{Grise}}},
\oauthor{\binits{A.} \bsnm{{Gros}}},
\oauthor{\binits{J.} \bsnm{{Guilet}}},
\oauthor{\binits{X.} \bsnm{{Han}}},
\oauthor{\binits{M.} \bsnm{{Huang}}},
\oauthor{\binits{Y.-F.} \bsnm{{Huang}}},
\oauthor{\binits{M.} \bsnm{{Jouret}}},
\oauthor{\binits{A.} \bsnm{{Klotz}}},
\oauthor{\binits{O.} \bsnm{{La Marle}}},
\oauthor{\binits{C.} \bsnm{{Lachaud}}},
\oauthor{\binits{E.} \bsnm{{Le Floch}}},
\oauthor{\binits{W.} \bsnm{{Lee}}},
\oauthor{\binits{N.} \bsnm{{Leroy}}},
\oauthor{\binits{L.-X.} \bsnm{{Li}}},
\oauthor{\binits{S.C.} \bsnm{{Li}}},
\oauthor{\binits{Z.} \bsnm{{Li}}},
\oauthor{\binits{E.-W.} \bsnm{{Liang}}},
\oauthor{\binits{H.} \bsnm{{Lyu}}},
\oauthor{\binits{K.} \bsnm{{Mercier}}},
\oauthor{\binits{G.} \bsnm{{Migliori}}},
\oauthor{\binits{R.} \bsnm{{Mochkovitch}}},
\oauthor{\binits{P.} \bsnm{{O'Brien}}},
\oauthor{\binits{J.} \bsnm{{Osborne}}},
\oauthor{\binits{J.} \bsnm{{Paul}}},
\oauthor{\binits{E.} \bsnm{{Perinati}}},
\oauthor{\binits{P.} \bsnm{{Petitjean}}},
\oauthor{\binits{F.} \bsnm{{Piron}}},
\oauthor{\binits{Y.} \bsnm{{Qiu}}},
\oauthor{\binits{A.} \bsnm{{Rau}}},
\oauthor{\binits{J.} \bsnm{{Rodriguez}}},
\oauthor{\binits{S.} \bsnm{{Schanne}}},
\oauthor{\binits{N.} \bsnm{{Tanvir}}},
\oauthor{\binits{E.} \bsnm{{Vangioni}}},
\oauthor{\binits{S.} \bsnm{{Vergani}}},
\oauthor{\binits{F.-Y.} \bsnm{{Wang}}},
\oauthor{\binits{J.} \bsnm{{Wang}}},
\oauthor{\binits{X.-G.} \bsnm{{Wang}}},
\oauthor{\binits{X.-Y.} \bsnm{{Wang}}},
\oauthor{\binits{A.} \bsnm{{Watson}}},
\oauthor{\binits{N.} \bsnm{{Webb}}},
\oauthor{\binits{J.J.} \bsnm{{Wei}}},
\oauthor{\binits{R.} \bsnm{{Willingale}}},
\oauthor{\binits{C.} \bsnm{{Wu}}},
\oauthor{\binits{X.-F.} \bsnm{{Wu}}},
\oauthor{\binits{L.-P.} \bsnm{{Xin}}},
\oauthor{\binits{D.} \bsnm{{Xu}}},
\oauthor{\binits{S.} \bsnm{{Yu}}},
\oauthor{\binits{W.-F.} \bsnm{{Yu}}},
\oauthor{\binits{Y.-W.} \bsnm{{Yu}}},
\oauthor{\binits{B.} \bsnm{{Zhang}}},
\oauthor{\binits{S.-N.} \bsnm{{Zhang}}},
\oauthor{\binits{Y.} \bsnm{{Zhang}}},
\oauthor{\binits{X.L.} \bsnm{{Zhou}}},
{The Deep and Transient Universe in the SVOM Era: New Challenges and
  Opportunities - Scientific prospects of the SVOM mission}.
ArXiv e-prints
(2016)
\end{botherref}
\endbibitem

\bibitem[\protect\citeauthoryear{{Weidenspointner}
  et~al.}{2008}]{Weidenspointner08}
\begin{barticle}
\bauthor{\binits{G.} \bsnm{{Weidenspointner}}},
\bauthor{\binits{G.} \bsnm{{Skinner}}},
\bauthor{\binits{P.} \bsnm{{Jean}}},
\bauthor{\binits{J.} \bsnm{{Kn{\"o}dlseder}}},
\bauthor{\binits{P.} \bsnm{{von Ballmoos}}},
\bauthor{\binits{G.} \bsnm{{Bignami}}},
\bauthor{\binits{R.} \bsnm{{Diehl}}},
\bauthor{\binits{A.W.} \bsnm{{Strong}}},
\bauthor{\binits{B.} \bsnm{{Cordier}}},
\bauthor{\binits{S.} \bsnm{{Schanne}}},
\bauthor{\binits{C.} \bsnm{{Winkler}}},
\batitle{{An asymmetric distribution of positrons in the Galactic disk revealed
  by {$\gamma$}-rays}}.
\bjtitle{\nat}
\bvolume{451},
\bfpage{159}--\blpage{162}
(\byear{2008}).
doi:\doiurl{10.1038/nature06490}
\end{barticle}
\endbibitem

\bibitem[\protect\citeauthoryear{{Wheaton} et~al.}{1979}]{Wheaton79}
\begin{barticle}
\bauthor{\binits{W.A.} \bsnm{{Wheaton}}},
\bauthor{\binits{J.P.} \bsnm{{Doty}}},
\bauthor{\binits{F.A.} \bsnm{{Primini}}},
\bauthor{\binits{B.A.} \bsnm{{Cooke}}},
\bauthor{\binits{C.A.} \bsnm{{Dobson}}},
\bauthor{\binits{A.} \bsnm{{Goldman}}},
\bauthor{\binits{M.} \bsnm{{Hecht}}},
\bauthor{\binits{S.K.} \bsnm{{Howe}}},
\bauthor{\binits{J.A.} \bsnm{{Hoffman}}},
\bauthor{\binits{A.} \bsnm{{Scheepmaker}}},
\batitle{{An absorption feature in the spectrum of the pulsed hard X-ray flux
  from 4U0115 + 63}}.
\bjtitle{\nat}
\bvolume{282},
\bfpage{240}--\blpage{243}
(\byear{1979}).
doi:\doiurl{10.1038/282240a0}
\end{barticle}
\endbibitem

\bibitem[\protect\citeauthoryear{{Wilson} et~al.}{1998}]{Wilson98}
\begin{barticle}
\bauthor{\binits{C.A.} \bsnm{{Wilson}}},
\bauthor{\binits{M.H.} \bsnm{{Finger}}},
\bauthor{\binits{B.A.} \bsnm{{Harmon}}},
\bauthor{\binits{D.} \bsnm{{Chakrabarty}}},
\bauthor{\binits{T.} \bsnm{{Strohmayer}}},
\batitle{{Discovery of the 198 Second X-Ray Pulsar GRO J2058+42}}.
\bjtitle{\apj}
\bvolume{499},
\bfpage{820}--\blpage{827}
(\byear{1998})
\end{barticle}
\endbibitem

\bibitem[\protect\citeauthoryear{{Wilson-Hodge} et~al.}{2011}]{Wilson-Hodge11}
\begin{barticle}
\bauthor{\binits{C.A.} \bsnm{{Wilson-Hodge}}},
\bauthor{\binits{M.L.} \bsnm{{Cherry}}},
\bauthor{\binits{G.L.} \bsnm{{Case}}},
\bauthor{\binits{W.H.} \bsnm{{Baumgartner}}},
\bauthor{\binits{E.} \bsnm{{Beklen}}},
\bauthor{\binits{P.} \bsnm{{Narayana Bhat}}},
\bauthor{\binits{M.S.} \bsnm{{Briggs}}},
\bauthor{\binits{A.} \bsnm{{Camero-Arranz}}},
\bauthor{\binits{V.} \bsnm{{Chaplin}}},
\bauthor{\binits{V.} \bsnm{{Connaughton}}},
\bauthor{\binits{M.H.} \bsnm{{Finger}}},
\bauthor{\binits{N.} \bsnm{{Gehrels}}},
\bauthor{\binits{J.} \bsnm{{Greiner}}},
\bauthor{\binits{K.} \bsnm{{Jahoda}}},
\bauthor{\binits{P.} \bsnm{{Jenke}}},
\bauthor{\binits{R.M.} \bsnm{{Kippen}}},
\bauthor{\binits{C.} \bsnm{{Kouveliotou}}},
\bauthor{\binits{H.A.} \bsnm{{Krimm}}},
\bauthor{\binits{E.} \bsnm{{Kuulkers}}},
\bauthor{\binits{N.} \bsnm{{Lund}}},
\bauthor{\binits{C.A.} \bsnm{{Meegan}}},
\bauthor{\binits{L.} \bsnm{{Natalucci}}},
\bauthor{\binits{W.S.} \bsnm{{Paciesas}}},
\bauthor{\binits{R.} \bsnm{{Preece}}},
\bauthor{\binits{J.C.} \bsnm{{Rodi}}},
\bauthor{\binits{N.} \bsnm{{Shaposhnikov}}},
\bauthor{\binits{G.K.} \bsnm{{Skinner}}},
\bauthor{\binits{D.} \bsnm{{Swartz}}},
\bauthor{\binits{A.} \bsnm{{von Kienlin}}},
\bauthor{\binits{R.} \bsnm{{Diehl}}},
\bauthor{\binits{X.-L.} \bsnm{{Zhang}}},
\batitle{{When a Standard Candle Flickers}}.
\bjtitle{\apjl}
\bvolume{727},
\bfpage{40}
(\byear{2011}).
doi:\doiurl{10.1088/2041-8205/727/2/L40}
\end{barticle}
\endbibitem

\bibitem[\protect\citeauthoryear{{Wilson-Hodge} et~al.}{2012}]{Wilson-Hodge12}
\begin{barticle}
\bauthor{\binits{C.A.} \bsnm{{Wilson-Hodge}}},
\bauthor{\binits{G.L.} \bsnm{{Case}}},
\bauthor{\binits{M.L.} \bsnm{{Cherry}}},
\bauthor{\binits{J.} \bsnm{{Rodi}}},
\bauthor{\binits{A.} \bsnm{{Camero-Arranz}}},
\bauthor{\binits{P.} \bsnm{{Jenke}}},
\bauthor{\binits{V.} \bsnm{{Chaplin}}},
\bauthor{\binits{E.} \bsnm{{Beklen}}},
\bauthor{\binits{M.} \bsnm{{Finger}}},
\bauthor{\binits{N.} \bsnm{{Bhat}}},
\bauthor{\binits{M.S.} \bsnm{{Briggs}}},
\bauthor{\binits{V.} \bsnm{{Connaughton}}},
\bauthor{\binits{J.} \bsnm{{Greiner}}},
\bauthor{\binits{R.M.} \bsnm{{Kippen}}},
\bauthor{\binits{C.A.} \bsnm{{Meegan}}},
\bauthor{\binits{W.S.} \bsnm{{Paciesas}}},
\bauthor{\binits{R.} \bsnm{{Preece}}},
\bauthor{\binits{A.} \bsnm{{von Kienlin}}},
\batitle{{Three Years of Fermi GBM Earth Occultation Monitoring: Observations
  of Hard X-Ray/Soft Gamma-Ray Sources}}.
\bjtitle{\apjs}
\bvolume{201},
\bfpage{33}
(\byear{2012}).
doi:\doiurl{10.1088/0067-0049/201/2/33}
\end{barticle}
\endbibitem

\bibitem[\protect\citeauthoryear{{Witteborn} et~al.}{1987}]{Witteborn87}
\begin{botherref}
\oauthor{\binits{F.C.} \bsnm{{Witteborn}}},
\oauthor{\binits{J.D.} \bsnm{{Bregman}}},
\oauthor{\binits{D.M.} \bsnm{{Rank}}},
\oauthor{\binits{M.} \bsnm{{Cohen}}},
\oauthor{\binits{D.K.} \bsnm{{Lynch}}},
\oauthor{\binits{R.W.} \bsnm{{Russell}}},
\oauthor{\binits{W.} \bsnm{{Cook}}},
\oauthor{\binits{D.} \bsnm{{Palmer}}},
\oauthor{\binits{T.} \bsnm{{Prince}}},
\oauthor{\binits{S.} \bsnm{{Schindler}}},
\oauthor{\binits{E.} \bsnm{{Stone}}},
\oauthor{\binits{R.H.} \bsnm{{McNaught}}},
\oauthor{\binits{A.} \bsnm{{Beresford}}},
{Supernova 1987A in the Large Magellanic Cloud}.
\iaucirc
\textbf{4400}
(1987)
\end{botherref}
\endbibitem

\bibitem[\protect\citeauthoryear{{Yadav} et~al.}{2016a}]{Yadav16}
\begin{barticle}
\bauthor{\binits{J.S.} \bsnm{{Yadav}}},
\bauthor{\binits{R.} \bsnm{{Misra}}},
\bauthor{\binits{J.} \bsnm{{Verdhan Chauhan}}},
\bauthor{\binits{P.C.} \bsnm{{Agrawal}}},
\bauthor{\binits{H.M.} \bsnm{{Antia}}},
\bauthor{\binits{M.} \bsnm{{Pahari}}},
\bauthor{\binits{D.} \bsnm{{Dedhia}}},
\bauthor{\binits{T.} \bsnm{{Katoch}}},
\bauthor{\binits{P.} \bsnm{{Madhwani}}},
\bauthor{\binits{R.K.} \bsnm{{Manchanda}}},
\bauthor{\binits{B.} \bsnm{{Paul}}},
\bauthor{\binits{P.} \bsnm{{Shah}}},
\bauthor{\binits{C.H.} \bsnm{{Ishwara-Chandra}}},
\batitle{{Astrosat/LAXPC Reveals the High-energy Variability of GRS 1915+105 in
  the X Class}}.
\bjtitle{\apj}
\bvolume{833},
\bfpage{27}
(\byear{2016}a).
doi:\doiurl{10.3847/0004-637X/833/1/27}
\end{barticle}
\endbibitem

\bibitem[\protect\citeauthoryear{{Yadav} et~al.}{2016b}]{Yadav16a}
\begin{bchapter}
\bauthor{\binits{J.S.} \bsnm{{Yadav}}},
\bauthor{\binits{P.C.} \bsnm{{Agrawal}}},
\bauthor{\binits{H.M.} \bsnm{{Antia}}},
\bauthor{\binits{J.V.} \bsnm{{Chauhan}}},
\bauthor{\binits{D.} \bsnm{{Dedhia}}},
\bauthor{\binits{T.} \bsnm{{Katoch}}},
\bauthor{\binits{P.} \bsnm{{Madhwani}}},
\bauthor{\binits{R.K.} \bsnm{{Manchanda}}},
\bauthor{\binits{R.} \bsnm{{Misra}}},
\bauthor{\binits{M.} \bsnm{{Pahari}}},
\bauthor{\binits{B.} \bsnm{{Paul}}},
\bauthor{\binits{P.} \bsnm{{Shah}}},
\bctitle{{Large Area X-ray Proportional Counter (LAXPC) instrument onboard
  ASTROSAT}},
in \bbtitle{Society of Photo-Optical Instrumentation Engineers (SPIE)
  Conference Series}.
\bsertitle{\procspie},
vol. \bseriesno{9905},
\byear{2016}b,
p. \bfpage{99051}.
doi:\doiurl{10.1117/12.2231857}
\end{bchapter}
\endbibitem

\bibitem[\protect\citeauthoryear{{Yamagami} et~al.}{1979}]{Yamagami79}
\begin{barticle}
\bauthor{\binits{T.} \bsnm{{Yamagami}}},
\bauthor{\binits{J.} \bsnm{{Nishimura}}},
\bauthor{\binits{M.} \bsnm{{Oda}}},
\bauthor{\binits{Y.} \bsnm{{Ogawara}}},
\bauthor{\binits{M.} \bsnm{{Fujii}}},
\bauthor{\binits{Y.} \bsnm{{Tawara}}},
\bauthor{\binits{M.} \bsnm{{Yoshimori}}},
\bauthor{\binits{H.} \bsnm{{Murakami}}},
\bauthor{\binits{S.} \bsnm{{Miyamoto}}},
\batitle{{Observation of Gamma-Ray Bursts at the Balloon Altitude}}.
\bjtitle{International Cosmic Ray Conference}
\bvolume{1},
\bfpage{223}
(\byear{1979})
\end{barticle}
\endbibitem

\bibitem[\protect\citeauthoryear{{Yamamoto} et~al.}{2011}]{Yamamoto11}
\begin{barticle}
\bauthor{\binits{T.} \bsnm{{Yamamoto}}},
\bauthor{\binits{M.} \bsnm{{Sugizaki}}},
\bauthor{\binits{T.} \bsnm{{Mihara}}},
\bauthor{\binits{M.} \bsnm{{Nakajima}}},
\bauthor{\binits{K.} \bsnm{{Yamaoka}}},
\bauthor{\binits{M.} \bsnm{{Matsuoka}}},
\bauthor{\binits{M.} \bsnm{{Morii}}},
\bauthor{\binits{K.} \bsnm{{Makishima}}},
\batitle{{Discovery of a Cyclotron Resonance Feature in the X-Ray Spectrum of
  GX 304-1 with RXTE and Suzaku during Outbursts Detected by MAXI in 2010}}.
\bjtitle{\pasj}
\bvolume{63},
\bfpage{751}--\blpage{757}
(\byear{2011}).
doi:\doiurl{10.1093/pasj/63.sp3.S751}
\end{barticle}
\endbibitem

\bibitem[\protect\citeauthoryear{{Yamamoto} et~al.}{2014}]{Yamamoto14}
\begin{barticle}
\bauthor{\binits{T.} \bsnm{{Yamamoto}}},
\bauthor{\binits{T.} \bsnm{{Mihara}}},
\bauthor{\binits{M.} \bsnm{{Sugizaki}}},
\bauthor{\binits{M.} \bsnm{{Nakajima}}},
\bauthor{\binits{K.} \bsnm{{Makishima}}},
\bauthor{\binits{M.} \bsnm{{Sasano}}},
\batitle{{Firm detection of a cyclotron resonance feature with Suzaku in the
  X-ray spectrum of GRO J1008-57 during a giant outburst in 2012}}.
\bjtitle{\pasj}
\bvolume{66},
\bfpage{59}
(\byear{2014}).
doi:\doiurl{10.1093/pasj/psu028}
\end{barticle}
\endbibitem

\bibitem[\protect\citeauthoryear{{Yonetoku} et~al.}{2011a}]{Yonetoku11}
\begin{barticle}
\bauthor{\binits{D.} \bsnm{{Yonetoku}}},
\bauthor{\binits{T.} \bsnm{{Murakami}}},
\bauthor{\binits{S.} \bsnm{{Gunji}}},
\bauthor{\binits{T.} \bsnm{{Mihara}}},
\bauthor{\binits{K.} \bsnm{{Toma}}},
\bauthor{\binits{T.} \bsnm{{Sakashita}}},
\bauthor{\binits{Y.} \bsnm{{Morihara}}},
\bauthor{\binits{T.} \bsnm{{Takahashi}}},
\bauthor{\binits{N.} \bsnm{{Toukairin}}},
\bauthor{\binits{H.} \bsnm{{Fujimoto}}},
\bauthor{\binits{Y.} \bsnm{{Kodama}}},
\bauthor{\binits{S.} \bsnm{{Kubo}}},
\bauthor{\bsnm{{IKAROS Demonstration Team}}},
\batitle{{Detection of Gamma-Ray Polarization in Prompt Emission of GRB
  100826A}}.
\bjtitle{\apjl}
\bvolume{743},
\bfpage{30}
(\byear{2011}a)
\end{barticle}
\endbibitem

\bibitem[\protect\citeauthoryear{{Yonetoku} et~al.}{2011b}]{Yonetoku11a}
\begin{barticle}
\bauthor{\binits{D.} \bsnm{{Yonetoku}}},
\bauthor{\binits{T.} \bsnm{{Murakami}}},
\bauthor{\binits{S.} \bsnm{{Gunji}}},
\bauthor{\binits{T.} \bsnm{{Mihara}}},
\bauthor{\binits{T.} \bsnm{{Sakashita}}},
\bauthor{\binits{Y.} \bsnm{{Morihara}}},
\bauthor{\binits{Y.} \bsnm{{Kikuchi}}},
\bauthor{\binits{T.} \bsnm{{Takahashi}}},
\bauthor{\binits{H.} \bsnm{{Fujimoto}}},
\bauthor{\binits{N.} \bsnm{{Toukairin}}},
\bauthor{\binits{Y.} \bsnm{{Kodama}}},
\bauthor{\binits{S.} \bsnm{{Kubo}}},
\bauthor{\bsnm{{Ikaros Demonstration Team}}},
\batitle{{Gamma-Ray Burst Polarimeter (GAP) aboard the Small Solar Power Sail
  Demonstrator IKAROS}}.
\bjtitle{\pasj}
\bvolume{63},
\bfpage{625}--\blpage{638}
(\byear{2011}b).
doi:\doiurl{10.1093/pasj/63.3.625}
\end{barticle}
\endbibitem

\bibitem[\protect\citeauthoryear{{Yonetoku} et~al.}{2012}]{Yonetoku12}
\begin{barticle}
\bauthor{\binits{D.} \bsnm{{Yonetoku}}},
\bauthor{\binits{T.} \bsnm{{Murakami}}},
\bauthor{\binits{S.} \bsnm{{Gunji}}},
\bauthor{\binits{T.} \bsnm{{Mihara}}},
\bauthor{\binits{K.} \bsnm{{Toma}}},
\bauthor{\binits{Y.} \bsnm{{Morihara}}},
\bauthor{\binits{T.} \bsnm{{Takahashi}}},
\bauthor{\binits{Y.} \bsnm{{Wakashima}}},
\bauthor{\binits{H.} \bsnm{{Yonemochi}}},
\bauthor{\binits{T.} \bsnm{{Sakashita}}},
\bauthor{\binits{N.} \bsnm{{Toukairin}}},
\bauthor{\binits{H.} \bsnm{{Fujimoto}}},
\bauthor{\binits{Y.} \bsnm{{Kodama}}},
\batitle{{Magnetic Structures in Gamma-Ray Burst Jets Probed by Gamma-Ray
  Polarization}}.
\bjtitle{\apjl}
\bvolume{758},
\bfpage{1}
(\byear{2012}).
doi:\doiurl{10.1088/2041-8205/758/1/L1}
\end{barticle}
\endbibitem

\bibitem[\protect\citeauthoryear{{Yoshimori}}{1979}]{Yoshimori79b}
\begin{barticle}
\bauthor{\binits{M.} \bsnm{{Yoshimori}}},
\batitle{{Galactic gamma-ray lines resulting from interactions between low
  energy cosmic rays and the interstellar medium}}.
\bjtitle{Australian Journal of Physics}
\bvolume{32},
\bfpage{383}--\blpage{404}
(\byear{1979})
\end{barticle}
\endbibitem

\bibitem[\protect\citeauthoryear{{Yoshimori} et~al.}{1979}]{Yoshimori79}
\begin{barticle}
\bauthor{\binits{M.} \bsnm{{Yoshimori}}},
\bauthor{\binits{H.} \bsnm{{Watanabe}}},
\bauthor{\binits{K.} \bsnm{{Okudaira}}},
\bauthor{\binits{Y.} \bsnm{{Hirasima}}},
\bauthor{\binits{H.} \bsnm{{Murakami}}},
\batitle{{A balloon investigation of galactic gamma-ray lines with a high
  resolution Ge/Li/ spectrometer}}.
\bjtitle{Australian Journal of Physics}
\bvolume{32},
\bfpage{375}--\blpage{382}
(\byear{1979})
\end{barticle}
\endbibitem

\bibitem[\protect\citeauthoryear{{Yuasa} et~al.}{2008}]{Yuasa08}
\begin{barticle}
\bauthor{\binits{T.} \bsnm{{Yuasa}}},
\bauthor{\binits{K.-I.} \bsnm{{Tamura}}},
\bauthor{\binits{K.} \bsnm{{Nakazawa}}},
\bauthor{\binits{M.} \bsnm{{Kokubun}}},
\bauthor{\binits{K.} \bsnm{{Makishima}}},
\bauthor{\binits{A.} \bsnm{{Bamba}}},
\bauthor{\binits{Y.} \bsnm{{Maeda}}},
\bauthor{\binits{T.} \bsnm{{Takahashi}}},
\bauthor{\binits{K.} \bsnm{{Ebisawa}}},
\bauthor{\binits{A.} \bsnm{{Senda}}},
\bauthor{\binits{Y.} \bsnm{{Hyodo}}},
\bauthor{\binits{T.G.} \bsnm{{Tsuru}}},
\bauthor{\binits{K.} \bsnm{{Koyama}}},
\bauthor{\binits{S.} \bsnm{{Yamauchi}}},
\bauthor{\binits{H.} \bsnm{{Takahashi}}},
\batitle{{Suzaku Detection of Extended/Diffuse Hard X-Ray Emission from the
  Galactic Center}}.
\bjtitle{\pasj}
\bvolume{60},
\bfpage{207}--\blpage{222}
(\byear{2008})
\end{barticle}
\endbibitem

\bibitem[\protect\citeauthoryear{{Zdziarski} et~al.}{2000}]{Zdziarski00}
\begin{barticle}
\bauthor{\binits{A.A.} \bsnm{{Zdziarski}}},
\bauthor{\binits{J.} \bsnm{{Poutanen}}},
\bauthor{\binits{W.N.} \bsnm{{Johnson}}},
\batitle{{Observations of Seyfert Galaxies by OSSE and Parameters of Their
  X-Ray/Gamma-Ray Sources}}.
\bjtitle{\apj}
\bvolume{542},
\bfpage{703}--\blpage{709}
(\byear{2000}).
doi:\doiurl{10.1086/317046}
\end{barticle}
\endbibitem

\bibitem[\protect\citeauthoryear{{Zdziarski} et~al.}{1995}]{Zdziarski95}
\begin{barticle}
\bauthor{\binits{A.A.} \bsnm{{Zdziarski}}},
\bauthor{\binits{W.N.} \bsnm{{Johnson}}},
\bauthor{\binits{C.} \bsnm{{Done}}},
\bauthor{\binits{D.} \bsnm{{Smith}}},
\bauthor{\binits{K.} \bsnm{{McNaron-Brown}}},
\batitle{{The average X-ray/gamma-ray spectra of Seyfert galaxies from GINGA
  and OSSE and the origin of the cosmic X-ray background}}.
\bjtitle{\apjl}
\bvolume{438},
\bfpage{63}--\blpage{66}
(\byear{1995}).
doi:\doiurl{10.1086/187716}
\end{barticle}
\endbibitem

\bibitem[\protect\citeauthoryear{{Zhang} et~al.}{2014a}]{Zhang14}
\begin{barticle}
\bauthor{\binits{S.} \bsnm{{Zhang}}},
\bauthor{\binits{C.J.} \bsnm{{Hailey}}},
\bauthor{\binits{F.K.} \bsnm{{Baganoff}}},
\bauthor{\binits{F.E.} \bsnm{{Bauer}}},
\bauthor{\binits{S.E.} \bsnm{{Boggs}}},
\bauthor{\binits{W.W.} \bsnm{{Craig}}},
\bauthor{\binits{F.E.} \bsnm{{Christensen}}},
\bauthor{\binits{E.V.} \bsnm{{Gotthelf}}},
\bauthor{\binits{F.A.} \bsnm{{Harrison}}},
\bauthor{\binits{K.} \bsnm{{Mori}}},
\bauthor{\binits{M.} \bsnm{{Nynka}}},
\bauthor{\binits{D.} \bsnm{{Stern}}},
\bauthor{\binits{J.A.} \bsnm{{Tomsick}}},
\bauthor{\binits{W.W.} \bsnm{{Zhang}}},
\batitle{{High-energy X-Ray Detection of G359.89-0.08 (Sgr A-E): Magnetic Flux
  Tube Emission Powered by Cosmic Rays?}}
\bjtitle{\apj}
\bvolume{784},
\bfpage{6}
(\byear{2014}a).
doi:\doiurl{10.1088/0004-637X/784/1/6}
\end{barticle}
\endbibitem

\bibitem[\protect\citeauthoryear{{Zhang} et~al.}{2014b}]{ZhangS14}
\begin{bchapter}
\bauthor{\binits{S.} \bsnm{{Zhang}}},
\bauthor{\binits{F.J.} \bsnm{{Lu}}},
\bauthor{\binits{S.N.} \bsnm{{Zhang}}},
\bauthor{\binits{T.P.} \bsnm{{Li}}},
\bctitle{{Introduction to the hard x-ray modulation telescope}},
in \bbtitle{Space Telescopes and Instrumentation 2014: Ultraviolet to Gamma
  Ray}.
\bsertitle{\procspie},
vol. \bseriesno{9144},
\byear{2014}b,
p. \bfpage{914421}.
doi:\doiurl{10.1117/12.2054144}
\end{bchapter}
\endbibitem

\bibitem[\protect\citeauthoryear{{Zhang} et~al.}{1997}]{Zhang97}
\begin{barticle}
\bauthor{\binits{W.} \bsnm{{Zhang}}},
\bauthor{\binits{T.E.} \bsnm{{Strohmayer}}},
\bauthor{\binits{J.H.} \bsnm{{Swank}}},
\batitle{{Neutron Star Masses and Radii as Inferred from Kilohertz
  Quasi-periodic Oscillations}}.
\bjtitle{\apjl}
\bvolume{482},
\bfpage{167}--\blpage{170}
(\byear{1997}).
doi:\doiurl{10.1086/310719}
\end{barticle}
\endbibitem

\end{thebibliography}

\end{document}